PhD Thesis

VNiVERSiDAD D SALAMANCA

# Asymptoticity of QCD and massive, oriented event-shapes

## A study in the large-$\beta_0$ limit and applications to jet physics

Néstor González Gracia

University of Salamanca

Supervised by Prof. Vicent Mateu Barreda

# Acknowledgments

It is undeniable true that the work of any scientist is grounded and builds on the work of those before. The experience of veterans is key in opening up the way for those who, like this candidate, start to walk the path of scientific research. In this regard I have to thank Prof. Vicent Mateu Barreda not only for the supervision of the scientific quality of this thesis but also for his willingness to share his expertise and knowledge. I am also specially thankful to Prof. Dr. André H. Hoang, whose invitation to spend several months in the research group at the University of Vienna has been extremely fruitful for me. Special thanks to Dr. Diogo Boito and Dr. Massimiliano Procura who kindly agreed to read through this thesis.

Even with the help of a guide, a complicated path also requires traveling companions. A special thanks goes to Alejandro Bris, MSc, for the countless discussions and hours shared staring at the blackboard.

Finally, one does not walk any path at all without someone taking them to the start of the road and pushing them forward. I have to express the deepest of my gratitudes to my father, Teófilo, who's love for learning inspired me to make of knowledge a core part of my life, and to my mother, María, for her truly unconditional support and kindness.

# Table of contents

# Introduction

With the appearance and development of the broad and powerful theories of General Relativity and the Standard Model, theoretical physics came to the realization that a single theory explaining the behavior of all kinds of matter and energy may be possible. The phenomenology of three of the four fundamental interactions of nature –the strong, weak and electromagnetic interactions– has been accurately described to a great extent with the language of Quantum Field Theory (QFT) in the Standard Model, although the incorporation of an adequate description of gravity still remains a wall we have not yet climbed. General Relativity is considered to be incompatible with QFT, and thus ultimately incompatible with the Standard Model itself. Such incompatibilities are not by any means the purpose of this work, but its inevitable conclusion serves as a starting point: if there is to be a coherent explanation for the four interactions of nature, improvements in any (or both) theories have yet to come.

Improvements on a scientific model can occur in one of two main ways. On the one hand, there is the intuitive way of extending the model to describe new phenomenology. In particle physics, extensions of the Standard Model are an effort collectively known as Beyond the Standard Model Physics (BSM). BSM deals with neutrino oscillations, matter-antimatter asymmetry, the strong CP problem, dark matter and dark energy, all of them problems that the Standard Model cannot explain in its current state.

However, it is important to realize there is a second way in which progress in a model can be made: by pushing forward not the boundaries of the model itself, but the boundaries of our understanding of it. During the last century, theoretical physics has seen how, in searching for the most fundamental theory, a price of technical complexification has been paid. We went from a classical view of the world requiring only algebraic and differential calculus to theories built on the language of more and more advanced mathematical objects. Special Relativity abandoned the Euclidean space, and later General Relativity was built by employing Differential Geometry, which added tensor calculus and the properties of space into the physi-cists' toolkit. Other ingredients such as Probability theory and strategies such as perturbation theory have been progressively standardized in physics thanks to the developments in Thermodynamics and Quantum Mechanics, and they have become so common nowadays that it is hard to think there was a time not that long ago when physicists were unfamiliar with concepts such as matrices and tensors. In any case, it is clear that improvements in the understanding of these methods were key in advancing the physical knowledge of the time and developing it to the extent we find today.

The way this intertwines with the current situation of the Standard Model is simply that it is no exception to the rule: the use of perturbation theory and the machinery of Feynman diagrams implies that results for matrix elements and observables are given in the form of infinite power series. The difficulty in the computation of the coefficients of these series increases exponentially with the order of the expansion, and the state-of-the-art knowledge indicates they have in general zero convergence radius[1]. In QCD, the language of renormalon calculus and Operator Product Expansion (OPE) has been developed to deal with divergent series. Renormalon calculus is based on the study of asymptotic divergent series and deals with the problems of assigning a finite estimate and an uncertainty, usually called ambiguity, to the sum of the series. OPE adds non-perturbative corrections that cancel the ambiguities of the perturbative series.

In sum, computations within the perturbative approach to the Standard Model are carried out in the language of divergent power series that need to be supplemented with non-perturbative corrections, in which theoretical physics does not have the full mastery yet. Parts I and II of this thesis are dedicated to build on this topic through a detailed review and a number of applications and studies. In chapter 1 we introduce the concepts of asymptotic series, summation methods for divergent series and Borel summation, and we particularize them to series in QCD, in which the factorial growth of the coefficients translates into poles in the Borel plane known as renormalons. We present all these ideas formally but we also complement with several illustrative examples. We end the chapter by introducing the large-$\beta_0$ limit, a rearrangement of the usual perturbative QCD expansion in $\alpha_s$ in which the leading order contribution is an infinite tower of terms that can be computed with a single computation of one-loop difficulty. It is in this limit in which the asymptotic properties of QCD can be studied.

1. In QCD, this is supported by studies in the large-$\beta_0$ limit, to which we dedicate this thesis.

In chapters 2 to 5 we develop a formalism for the systematic study of asymptotic series in the large-$\beta_0$ limit. Chapter 2 is dedicated to set up the notation and the main ingredients necessary for the subsequent discussions, and chapters 3 and 4 develop the formalism for series without anomalous dimension and with cusp-anomalous dimension, respectively. The main results of these chapters are the integral expressions for the sum of the large-$\beta_0$ series, its renormalization factor and its anomalous dimension. A great amount of care has been put into making this formalism systematic, in the sense that all the input that is required to apply the final results to identify the renormalons of a series and obtain its Borel-sum and ambiguity is the solution of a (class of) Feynman diagrams of one-loop difficulty in terms of the bare coupling constant. Equivalently to Borel summation, in our formalism renormalons manifest as poles along the integration path of the integrals defining the sum of the series. Chapter 5 describes the usual prescription to perform such integrations with poles, and collects the main results. Finally, in chapters 6 to 9 a large number of applications are carried out, namely the short-distance $\overline{\text{MS}}$ and MSR mass-schemes, and the matching coefficients and jet functions involved in the event-shape factorization theorems for $e^+e^-\to$ hadrons in SCET and bHQET.

Part II is dedicated to the problem of asymptotic separation, a discrepancy between to prescriptions –fixed-order perturbation theory (FOPT) and contour-improved perturbation theory (CIPT)– to compute the massless QCD corrections to the $\tau$ spectral moments. The discrepancy is systematically observed in truncated fixed-order theory up to $\mathcal{O}(\alpha_s^4)$, is still present in large-$\beta_0$ studies, and translates into a major contribution to the theoretical uncertainty in the extraction of $\alpha_s(m_\tau^2)$ from $\tau$-decays. Out of the applications treated in this thesis, asymptotic separation is perhaps the one that evidences the most there is room for a better understanding in the set-up of resummed perturbative asymptotic series and non-perturbative OPE power corrections, as, although progress has been made at a technical level, the reasons for the fundamental discrepancies of FOPT and CIPT have not yet been fully unraveled. We contribute to the studies dedicated to asymptotic separation by computing and analyzing the $\tau$ spectral moment functions in FOPT and CIPT in the large-$\beta_0$ and in the gluon condensate model, i.e., a model with only one UV renormalon –instead an infinite combination of IR and UV renormalons.

In part III we change topics to explore oriented event-shapes for massive quarks in fixed-order perturbation theory. The oriented event-shape differential cross-section for $e^+e^-\to$ hadrons is computed at NLO, where oriented refers to the fact that it is differential in the angle $\theta_T$ of the thrust axis of the collision with respect to the initial beam. It is known (see references cited in the corresponding section) that only two angular structures, $1+\cos^2\theta_T$ and $1-3\cos^2\theta_T$, can arise in the distribution. Upon phase-space integration over $\theta_T$ they yield 1 and 0, respectively. In our computation we keep quarks massive and explicitly work out both the unoriented $(1+\cos^2\theta_T)$ and oriented $(1-3\cos^2\theta_T)$ contributions. For both structures we determine the analytic expressions for the coefficients of the distributional structure of the cross-section, which for massive quarks consists on the Dirac delta $\delta(e-e_{\rm min})$, the plus distribution $[1/(e-e_{\rm min})]_+$ and a non-singular function $F_e^{\rm NS}$. The unoriented part, reproducing the event-shape cross-section $\mathrm{d}\sigma/\mathrm{d}e$ after integration, was already known and provides a cross-check to our results. In both cases, we carry out the analytic expressions up to the point where an specific event-shape must be selected to obtain numeric results.

Lastly, a good amount of details and computations have been relegated to the appendices, in hopes not to deviate the focus from the main discussions but also to provide with all the complementary information and make the work as self-contained as possible. In particular, appendix A contains a brief introduction to the plus and delta distributions, and appendix F contains advanced topics related to phase-space integration such as how to build the phase-space in arbitrary $d$ dimensions, the projection of an oriented cross-section onto the thrust axis and the derivation of the curves and intersection points delimiting the Dalitz region for three particles into the three regions of constant thurst axis.

# Part I

# Formal study of the large-$\beta_0$ limit and applications to jet physics

# Chapter 1

# Asymptoticity in QCD

## 1.1 Asymptotic expansions

In quantum field theories, the standard procedure of perturbation theory gives transition amplitudes and matrix elements in the form of infinite power series in the renormalized coupling,

$$R(\alpha) \equiv 1 + \sum_{n=1}^{\infty} r_n \alpha^n, \tag{1.1}$$

where without loss of generality the $\alpha$-independent term can be normalized to 1. Despite the fact that physical values of the coupling are real and positive, the study of power series is best performed in the complex plane, so in general we let $\alpha \in \mathbb{C}$. Infinite power series such as (1.1) can be of three kinds: (1) convergent for any value of $\alpha$, hence with infinite *radius of convergence*; (2) convergent only for $|\alpha| < r < \infty$ and divergent for $|\alpha| > r$, hence with a finite radius of convergence of $r$; (3) divergent for all values of $\alpha$, which is stated as the series having zero radius of convergence. It is easy to find examples for each of the three cases:

| | | | | |
|---|---|---|---|---|
| (1) | exponential series: | $\sum_{n=0}^{\infty} \frac{z^n}{n!} = e^z$ | for any $z \in \mathbb{C}$, | (1.2) |
| (2) | geometric series: | $\sum_{n=0}^{\infty} z^n = \frac{1}{1-z}$ | for $\|z\| < 1$, | |
| (3) | factorial-rising series: | $\sum_{n=0}^{\infty} n! z^n$ | diverges for any $z \in \mathbb{C}$. | |

More generally, for a series of complex numbers $S = \sum_{n=0}^{\infty} a_n$ convergence is stated in the following terms. If the series $\sum_{n=0}^{\infty} |a_n|$ is convergent, then $S$ is convergent too and is said to be *absolutely convergent*. If the series $\sum_{n=0}^{\infty} |a_n|$ diverges, but $S$ is convergent, it is said to be *conditionally convergent*. It should be noted that all absolutely convergent series are conditionally convergent.

The procedures to determine whether a series converges or diverges are collectively known as *convergence tests*. Two of the most widely used are the *ratio test* and the *root test*. For the series $S$, the ratio and root tests define the quantities[1.1]

$$\rho_{\text{ratio}} \equiv \lim_{n \to \infty} \left| \frac{a_{n+1}}{a_n} \right|, \qquad \rho_{\text{root}} \equiv \limsup_{n \to \infty} |a_n|^{1/n}, \tag{1.3}$$

and establish that $S$ is absolutely convergent when $\rho < 1$ and divergent when $\rho > 1$. When $\rho = 1$ both the tests are inconclusive. The relation between the ratio and root tests is the following: the limit $\rho_{\text{root}}$ always exists, while $\rho_{\text{ratio}}$ may not exist; if $\rho_{\text{ratio}}$ exists, $\rho_{\text{root}} = \rho_{\text{ratio}}$. The root test is, then, more conclusive; however it is often more difficult to implement than the ratio test, so both are used in practical computations. In the particular case of a power series $\sum_{n=0}^{\infty} a_n (z-a)^n$, the convergence conditions reduce to

$$|z-a| < \frac{1}{\rho_{\text{ratio}}} \equiv r_{\text{ratio}}, \qquad |z-a| < \frac{1}{\rho_{\text{ratio}}} \equiv r_{\text{root}}. \tag{1.4}$$

The quantities $r_{\text{ratio}}$ and $r_{\text{root}}$ are the convergence radius of the power series, and define a circle in the complex plane, $U \equiv \{z \in \mathbb{C} |\, |z-a| < r\}$, where the power series converges. Remarks:

1. When $\rho_{\text{root}} = \infty$, $r_{\text{root}} = r_{\text{ratio}} = 0$ and the series diverges for all values of $z$ except for the trivial case $z = a$.

2. Convergence on the circumference $\partial U \equiv \{z \in \mathbb{C} |\, |z-a| = r\}$ is not uniform in general: the series may converge or diverge in all the points in $\partial U$ or may converge

1.1. The limit superior of a real sequence $\{x_n\}_{n=0}^{\infty}$ is defined as $\limsup_{n \mapsto \infty} x_n \equiv \lim_{n \mapsto \infty} \sup_{m \geq n} x_m$ and entails information on how the superior of the tail region of the sequence behaves. The limit superior of a sequence of real numbers always exists, due to $\mathbb{R}$ being a complete set.

in some of the points and diverge in others. For example:

- $\sum_{n=0}^{\infty} z^n$ has radius of convergence $r=1$ and diverges for all $z\in\partial U$,

- $\sum_{n=0}^{\infty} z^n/n$ has radius of convergence $r=1$, diverges for $z=1$ but converges for all other $z\in\partial U$, and

- $\sum_{n=0}^{\infty} z^n/n^2$ has radius of convergence $r=1$ and converges for all $z\in\partial U$.

When an infinite series such as (1.1) arises in the context of a physical theory and it is directly related to an observable, one may have the expectation that it should have a non-vanishing convergence radius, as cases (1) and (2) in (1.2). However, practical computations have proved this is not always the situation in QED and QCD, where series with zero radius of convergence appear in many occasions.

As discussed in textbooks on the topic and reviews such as [1, 2, 3, 4, 5] divergent series should not be immediately discarded as a failure of the theory that produces them: if they arise as the expansion of some function $f(z)$, and certain conditions are met, they can be used to construct an estimate of $f(z)$. We will specify those conditions shortly, but for the moment let us consider the following illustrative example of a function and two of its expansions:

$$\begin{aligned} f(x) &= x\,e^x\int_x^{\infty}\mathrm{d}t\,\frac{e^{-t}}{t} = x\,e^x\left[-\log(x)-\gamma_E+\sum_{n=1}^{\infty}\frac{(-1)^{n+1}x^n}{n\Gamma(n-1)}\right] \\ &\asymp \sum_{n=0}^{\infty}\frac{(-1)^n\Gamma(n+1)}{x^n}, \end{aligned} \tag{1.5}$$

where $\gamma_E\simeq 0.5772$ is the Euler-Mascheroni constant and $x>0$. The expansion in the first line converges for all $x$, while for the expansion in the second line the quotient of the $(n+1)$-th term and the $n$-th is $-n/x$, so that the series diverges for all $x$[1.2].

1.2. Because of this we cannot formally write an equality between $f(x)$ and the divergent expansion and use the symbol $\asymp$ instead. Its precise meaning will be made clear shortly.

However, the divergent series is still useful to estimate numerical values of $f(x)$. In figure 1.1 we show the behavior of both series for $x=10$, including up to $N$ terms in each of the sums. In figure 1.1($a$) it is observed that the convergent expansion starts far and initially diverges from the exact result $f(10)\simeq 0.9156$, requiring the contribution of several orders to show consistent convergence towards $f(10)$. In fact, to reproduce the exact result with three digits of precision one needs to include the first 40 terms. On the other hand, the divergent expansion starts closer to the exact answer and achieves three-digit precision after only 6 terms. In this case, after enough terms have been included, the divergent behavior sets in and the series gets uncontrollably far from the exact value.

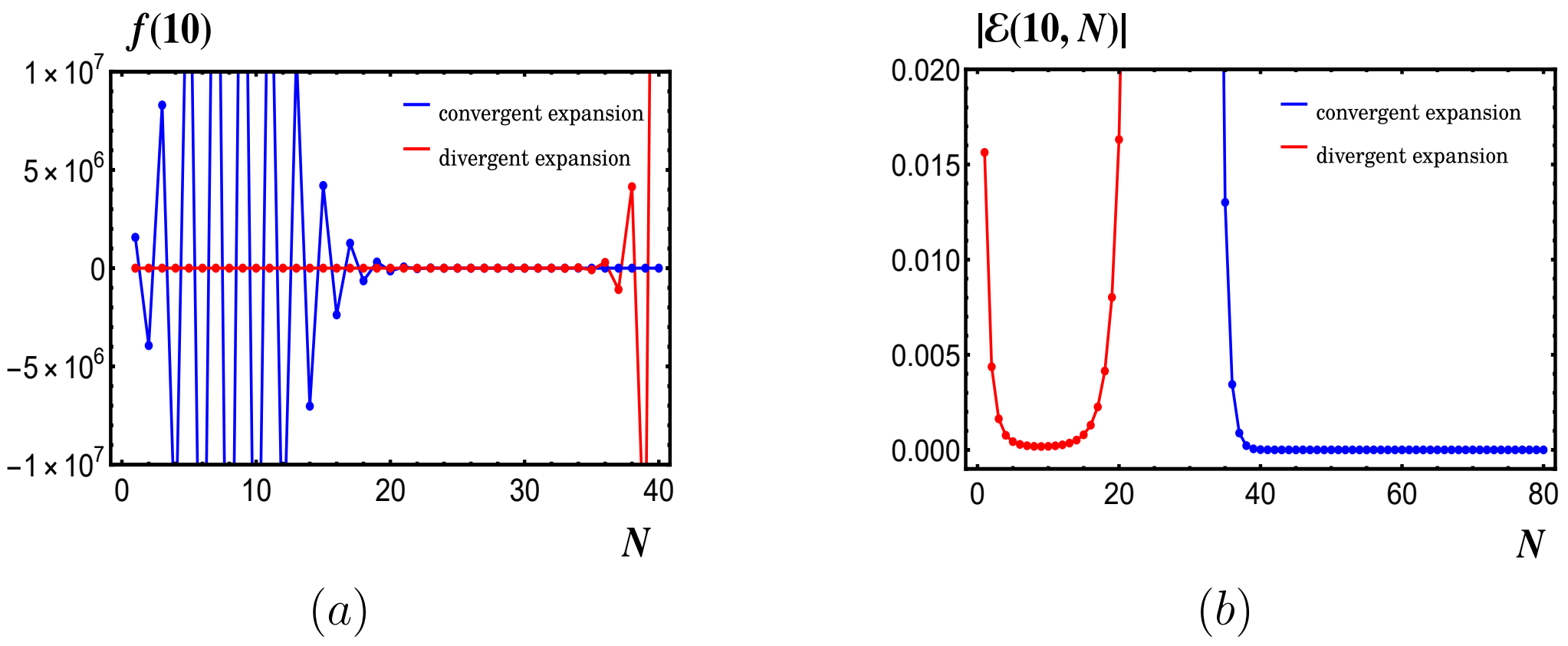

**Figure 1.1.** Comparison of the two expansions of the function $f(x)$ defined in (1.5) for a value $x=10$. Panel ($a$): partial sum up to $N$ terms of both the convergent and divergent expansions. Note that the huge scale of the $y$-axis hides that the convergent series only stays close to $f(10)\simeq 0.9156$ for $N\geq 40$. Panel ($b$): Absolute value of the truncation error made in estimating $f(x)$ by the partial sum of the first $N$ terms.

The error in using any of the two expansions up to $N$ terms to estimate $f(x)$ is defined as the difference between $f(x)$ and the truncated expansion, and denoted as $\mathcal{E}(x,N)$. Its behavior with respect to $N$ for $x=10$ is shown in figure 1.1($b$) in absolute value. When using the convergent series and enough terms are accounted for, $|\mathcal{E}(x,N)|$ monotonically decreases to 0 with $N$. In this case the error can be made arbitrarily small provided enough terms of the expansion can be computed. When using the divergent series, however, the error is not monotonic with $N$: it decreases up to $N=9$ and then increases again without boundary. In this case, for each $x$ there is a number of terms $N_{\min}(x)$ that minimizes $\mathcal{E}(x,N)$, and truncating the series at $N_{\min}(x)$ provides the best estimation. This means there is a lower boundary in the achievable precision when using the divergent series to estimate $f(x)$[1.3]. Nevertheless, as we see from 1.1(*b*), the truncation error at the best approximation $N_{\min}(10)=9$ is comparable in size to that of the convergent expansion at $N=40$.

Simply stated, the reason why the truncated divergent series offers a good approximation to $f(x)$ is because it is an *asymptotic expansion* of $f(x)$. Although, as in the previous example, in this thesis we will be mainly considering applications with $x>0$, we formulate here the formal definition of an asymptotic expansion for complex $z$. Let $D\subseteq\mathbb{C}$ be a domain in the complex plane and let $\bar{D}$ denote its closure[1.4]. Then the formal definition of an asymptotic expansion can be built as follows [1].

First, one defines the large and small "o" symbols. For two functions $f(z)$ and $g(z)$ continuous in $D$ and a point $z_0\in\bar{D}$ we write

$$f(z)\sim\mathcal{O}(g(z)) \text{ as } z\longrightarrow z_0 \tag{1.6}$$

if there exists a constant $k$ and a neighborhood $N_0$ of $z_0$ such that $|f(z)|\leq k|g(z)|$ for all $z\in N_0\cap D$. Here and in the following definitions, it is understood that $z$ tends to $z_0$ through values in $D$. Similarly,

$$f(z)\sim o(g(z)) \text{ as } z\longrightarrow z_0 \tag{1.7}$$

if for any $\varepsilon>0$ there exists a neighborhood $N_\varepsilon$ of $z_0$ such that $|f(z)|\leq\varepsilon|g(z)|$ for all $z\in N_\varepsilon\cap D$.

In other words, if $g(z)\neq 0$ in $D-\{z_0\}$, $f(z)\sim\mathcal{O}(g(z))$ when the ratio $f/g$ remains bounded as $z\longrightarrow z_0$ and $f(z)\sim o(g(z))$ when the limit of this ratio is zero. For example, one has $(z-z_0)^{n+1}\sim o((z-z_0)^n)$ as $z\longrightarrow z_0$, since

$$\lim_{z\to z_0}\frac{(z-z_0)^{n+1}}{(z-z_0)^n}=0. \tag{1.8}$$

Similarly, $z^{-n-1}\sim o(z^{-n})$ as $z\longrightarrow\infty$, due to the analogous limit vanishes.

---

1.3. For example, if one would wish to estimate $f(10)$ to 4 significant figures, the divergent series in (1.5) would not be of use, as the best approximation is found for $N=9$, for which $\sum_{n=0}^{9}(-1)^n\Gamma(n-1)x^{-n}\simeq 0.9155$.

1.4. This is, the set of points such that every ball centered at them contains points in $D$.

Second, one introduces the concept of an asymptotic sequence. The sequence of functions $\{\phi_n(z)\}_{n=0}^{\infty}$ is called an asymptotic sequence as $z \to z_0$ if every $\phi_n(z)$ is continuous in $D$ and

$$\phi_{n+1}(z) \sim o(\phi_n(z)) \text{ as } z \to z_0. \tag{1.9}$$

The formal power sequences $\{(z-z_0)^n\}_{n=0}^{\infty}$ and $\{z^{-n}\}_{n=0}^{\infty}$ are asymptotic sequences as $z \to z_0$ and $z \to \infty$, respectively.

Finally, based on the previous notions, the definition of an asymptotic expansion can be introduced. A series $\Sigma_{n=0}^{\infty} f_n \phi_n(z)$ is called an asymptotic expansion of $f(z)$ as $z \to z_0$ with respect to the sequence $\{\phi_n\}_{n=0}^{\infty}$ if[1.5]

- $\{\phi_n(z)\}_{n=0}^{\infty}$ forms an asymptotic sequence as $z \to z_0$, and (1.12)
- $f(z) - \sum_{n=0}^{N} f_n \phi_n(z) \sim o(\phi_N)$ as $z \to z_0, \quad N = 0, 1, 2...$

We then write this as

$$f(z) \asymp \sum_{n=0}^{\infty} f_n \phi_n(z), \qquad z \to z_0. \tag{1.13}$$

This definition ensures that, after truncating the sum at $N$ terms, the error $\mathcal{E}(z, N)$ vanishes at a well-defined rate as $z$ gets closer to the point of evaluation. When evaluating the expansion at $z \neq z_0$, $\mathcal{E}(z, N)$ can be made arbitrarily smaller as $N \to \infty$ if the asymptotic expansion is convergent at $z$, but has a minimum at some finite value $N_{\min}(z)$ if the asymptotic expansion diverges. In this case, although for fixed

1.5. Expansions satisfying the definition (1.12) are referred to as Poincaré asymptotic expansions to emphasize the asymptotic sequence doesn't need to be the formal power series $(x - x_0)^n$. Also, the second condition in (1.12) is often stated in the literature in terms of the big "o" as

$$f(z) - \sum_{n=0}^{N} f_n \phi_n(z) \sim \mathcal{O}(\phi_{N+1}) \text{ as } z \to z_0, \quad N = 0, 1, 2... \tag{1.10}$$

The two expressions are equivalent due to $\mathcal{O}(o(f)) = o(\mathcal{O}(f)) = o(o(f)) = o(f)$ [1]. When $f(z) \neq 0$ in $D - \{z_0\}$, both the second line of (1.12) and (1.10) are also equivalent to the limit

$$\lim_{z \to z_0} \frac{f(z) - \sum_{n=1}^{N} f_n \phi_n(z)}{\phi_N(z)} = \lim_{z \to z_0} \frac{\mathcal{E}(z, N)}{\phi_N(z)} = 0. \tag{1.11}$$

$z$ the precision in the estimation is bounded from below, the estimation improves as $z \to z_0$, since $|\mathcal{E}(z, N_{\min}(z))|$ gets smaller[1.6].

We illustrate these ideas by reconsidering the example in (1.5) under the language of asymptotic expansions. The convergent series is an expansion with respect to the sequence $\{x^n\}_{n=1}^{\infty}$, which, as mentioned, is an asymptotic sequence as $x \to 0$. Analogously, the divergent series is expanded with respect to $\{x^{-n}\}_{n=1}^{\infty}$, which is an asymptotic sequence as $x \to \infty$. Furthermore, we find that for all $N$

$$\lim_{x \to 0} \frac{\mathcal{E}^c(x, N)}{x^N} = 0, \qquad \lim_{x \to \infty} \frac{\mathcal{E}^d(x, N)}{x^{-N}} = 0, \tag{1.14}$$

where $\mathcal{E}^c(x, N)$ and $\mathcal{E}^d(x, N)$ denote the truncation errors for the convergent and divergent series, respectively[1.7]. Therefore, the convergent (divergent) series is an asymptotic expansion as $x \to 0$ ($x \to \infty$). In figure 1.2 we show the behavior with respect to $x$ of the truncation errors. In both cases (figures 1.2($a$) and 1.2($b$)) the truncation error vanishes for all values of $N$ as $x$ approaches the point at which the series is asymptotic. When evaluating at a different point, the error of the convergent series at any given $x$ gets smaller as $N$ increases (figure 1.2($a$)), while for the divergent series, there is an optimal $N$ that minimizes $\mathcal{E}^d(x, N)$ that depends on the point of evaluation. This is best seen in figure 1.2($b$), where we take the logarithm to compensate the quick decay to 0 as $x$ increases. We can see how for $x \leq 3$ taking 1 term in the expansion (blue curve) constitutes a better estimation to $f(x)$ than taking 5, 9 or 15 terms. For $3 < x \leq 7$, $N = 5$ (red curve) is the best shown estimation, and, analogously, $N = 9$ and $N = 15$ (orange and purple curves, respectively) constitute the best approximation in $7 < x \leq 12$ and $12 < x < 20$, respectively. In black dots we show the best estimation $\mathcal{E}^d(x, N_{\min}(x))$ for integer values of $0 < x \leq 20$. We observe the best estimation improves as $x$ gets closer to $\infty$.

1.6. For example, the divergent expansion (1.5), which is asymptotic as $x \to \infty$, estimates $f(10)$ to three significant figures, $f(15)$ to five significant figures and $f(20)$ to seven significant figures.

1.7. For completeness, the closed forms for $\mathcal{E}^c(x, N)$ and $\mathcal{E}^d(x, N)$, from which the limits in (1.14) are computed, are

$$\begin{aligned} \mathcal{E}^c(x, N) &= \frac{(-x)^{N+2} e^x}{(N+1)\Gamma(N+2)}\, {}_2F_2(1, N+1, N+2, N+2, -x), \\ \mathcal{E}^d(x, N) &= (-1)^{N+1} x e^x \Gamma(N+2) \Gamma(-1-N, x), \end{aligned} \tag{1.15}$$

where ${}_pF_q$ denotes the Gauss hypergeometric function and $\Gamma(a, z)$ denotes the incomplete Euler gamma function (see Appendix A for definitions).

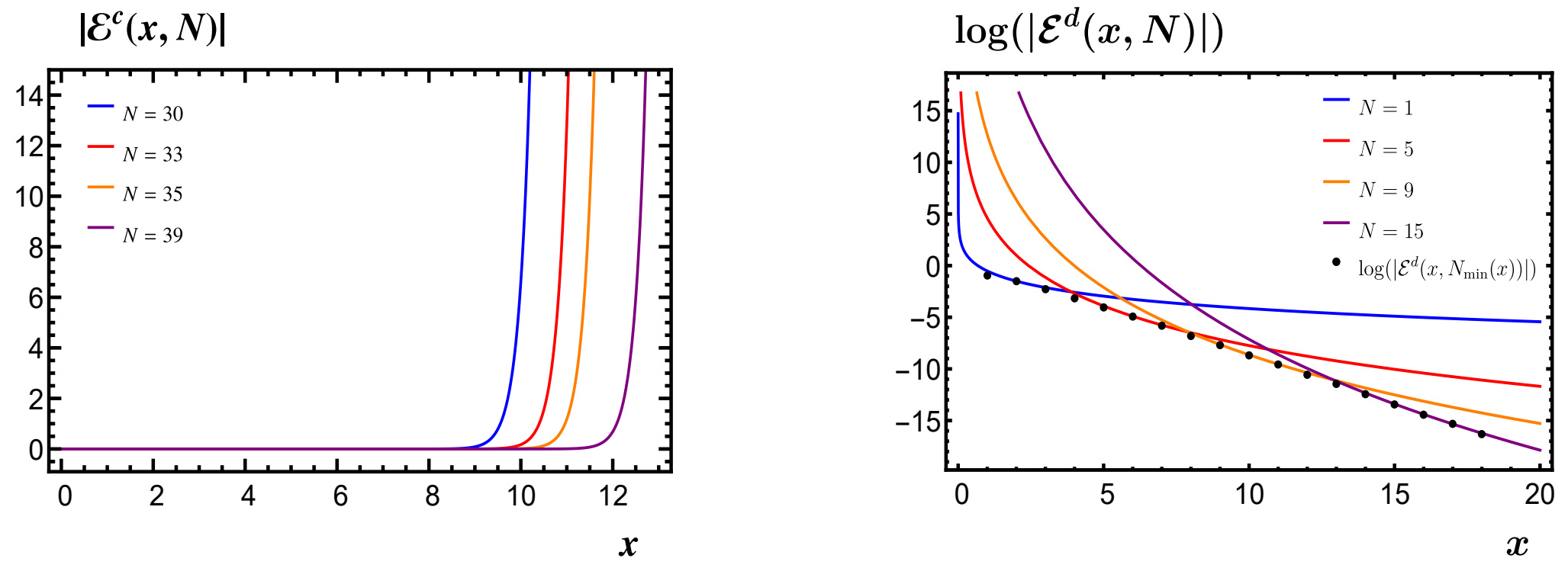

$(a)$ $(b)$

**Figure 1.2.** Dependence on the point of evaluation $x$ of the truncation errors for the convergent and divergent series in (1.5). Panel $(a)$: the truncation error for the convergent series vanishes as $x$ tends to 0, and the estimation for any $x$ improves as more terms are included. Panel $(b)$: the truncation error for the divergent series vanishes as $x$ gets closer to $\infty$. In this case for each $x$ there is an optimal value $N_{\min}(x)$ at which the truncated series provides the best approximation.

In sum, both convergent and divergent series can be of use to build an estimate of a function $f(z)$ if they constitute an asymptotic expansion of $f(z)$. This language of asymptotic expansions can be applied to QED and QCD to give meaning to their (usually divergent) perturbative power series. Before proceeding further into the particular case of QCD series let us remark some important ideas that are valid for general asymptotic expansions.

1. The coefficients of the asymptotic expansion of a function $f(z)$ as $z \to z_0$ with respect to a given sequence $\{\phi_n(z)\}_{n=0}^{\infty}$ are unique. The formal proof of this statement can be found in [1]. Of course, if $f(z)$ is analytic at $z_0$ and $\phi_n(z) = z^n$, the coefficients of the expansion are given by the derivatives of $f(z)$ at $z_0$.

2. A single function $f(z)$ can have different asymptotic expansions as $z \to z_0$, each of which can be convergent or divergent. This is due to the different choices for the asymptotic sequence functions $\phi_n(z)$.

3. Different functions can have the same asymptotic expansion. This is easy to see once one realizes there are functions whose asymptotic expansion in a given region of the complex plane vanishes at all orders. Indeed, the following asymptotic expansion,

$$e^{-1/z^{\gamma}} \asymp 0, \quad z \to 0, \text{ with respect to } \{z^n\}_{n=0}^{\infty}, \tag{1.16}$$

holds for $0<\gamma<1$ in the right-half complex plane determined by $\text{Re}(z)>0$ and for $\gamma=1$ in the positive real axis. Equivalently,

$$e^{-z}\asymp 0, \quad z\to\infty, \text{ with respect to } \{z^{-n}\}_{n=0}^{\infty}, \tag{1.17}$$

along any path in $\mathbb{C}$. Then, the family of functions $f(z)+Ae^{-1/z^{\gamma}}$ ($f(z)+Ae^{-z}$), for any constant $A$, has the same asymptotic expansion as $f(z)$ as $z\mapsto 0$ ($z\mapsto\infty$).

4. Continuing from the previous point, it follows that a series can be asymptotic to infinitely many functions, all differing by terms which have vanishing asymptotic expansions. An important consequence is that the convergent asymptotic expansions of $f(z)$ do not need to sum up to $f(z)$. For example, if $f(z)$ is analytic at $z=0$ and we denote the coefficients of its Taylor expansion at that point as $f_n$, we have the asymptotic expansion as $z\mapsto 0$ in the region $\text{Re}(z)>0$:

$$f(z)+e^{-1/z}\asymp\sum_{n=0}^{\infty} f_n z^n. \tag{1.18}$$

Summing up the series only recovers $f(z)$.

Let us suppose we are given a series $\sum_{n=0}^{\infty} f_n\phi_n(z)$ and we are asked to extract out of it the best estimate of a function $f(z)$. If $f(z)$ is known, we can verify the conditions (1.12) to decide whether the series is asymptotic to $f(z)$. If that is the case, an estimation of $f(z)$ can be constructed by computing and minimizing the truncation error $|\mathcal{E}(x,N)|$. Note that even if the series is convergent, due to remark 4 above, it may not converge to $f(z)$, as $f(z)$ can have infinitely many terms with zero asymptotic expansion. Thus, both for convergent and divergent series, $|\mathcal{E}(z,N)|$ may have a non-zero lower boundary.

A much more interesting situation is when no explicit knowledge of $f(z)$ is possessed. In this case, as asymptoticity is a property defined on a function and its expansion, the conditions in (1.12) cannot be proved and the truncation error cannot be computed explicitly. Instead, the way to proceed is by assuming the series is asymptotic to $f(z)$ and establishing a prescription to assign a finite estimate $f_e(z)$ to the sum of the series. These prescriptions are usually called summation prescriptions and they cannot rely on the explicit knowledge of $\mathcal{E}(z,N)$. As mentioned before, $f_e(z)$ may differ from $f(z)$ even when the series is convergent and the estimate is chosen as the finite value $f_\infty(z)\equiv\sum_{n=0}^{\infty} f_n\phi_n(z)$.

With regards to this thesis, the equivalence of summation prescriptions, the uniqueness of their estimates and the specific relation to $f(z)$ are topics too broad to be treated in complete generality. We thus conclude our general treatment of asymptotic expansions and focus our attention into perturbative QCD series of the form (1.1).

## 1.2 Summability of QCD series

### 1.2.1 QCD series as asymptotic expansions

In this section we apply the previous ideas on asymptotic expansions to the case of QCD perturbative series and introduce the Borel summation technique to build a finite estimate. The situation in QCD can be described by the two following remarks:

- The outcome of perturbation theory around small $\alpha$ is the power series (1.1), with no indication or information on whether the series constitutes the expansion around $\alpha=0$ of some function.

- Although some series in QCD may be convergent, the general case is that $R(\alpha)$ usually has zero radius of convergence: the coefficients $r_n$ grow factorially for sufficiently large $n$, scaling as $r_n \sim n! a^n n^b$ with real constants $a$, $b$ [4], [5].

Thus, the situation corresponds to that described at the end of section 1.1: we are only given the (usually divergent) perturbative series $R(\alpha)$ and we assume it constitutes the asymptotic expansion as $\alpha \mapsto 0$ with respect to the sequence $\{\alpha^n\}_{n=0}^{\infty}$ of an unknown function $\bar{R}(\alpha)$. We then reinterpret (1.1) as

$$\bar{R}(\alpha) \asymp R(\alpha) = 1 + \sum_{n=1}^{\infty} r_n \alpha^n. \tag{1.19}$$

When the series in the right-hand side has a non-vanishing convergence radius, $R(\alpha)$ is analytic in the corresponding convergence circle; when the radius of convergence is zero, $R(\alpha)$ is only a notational shortcut for the divergent series, which needs to be replaced by a finite estimate function $R_e(\alpha)$. In both cases, the right-hand side of the $\asymp$ symbol is finite but can differ from $\bar{R}(\alpha)$ in terms whose asymptotic expansion at $\alpha=0$ vanishes, which in QCD often take the form $e^{-q/\alpha}$ for some positive $q$. Note that under assumption (1.19), despite $\bar{R}(\alpha)$ is unknown, the truncation error is bounded as $|\mathcal{E}(\alpha, N)| \leq k_{N+1} |\alpha|^{N+1}$ for some positive $k_{N+1}$.

We now introduce two summation prescriptions: the Optimal Truncation Rule and Borel summation.

### 1.2.2 Summation methods: optimal truncation rule and Borel summation

The Optimal Truncation Rule [3], [6] constructs the finite estimate for the asymptotic expansion $R(\alpha)$ by truncating the sum at its minimal term:

$$R_{N_{\min}}(\alpha) \equiv 1 + \sum_{n=1}^{N_{\min}} r_n \alpha^n. \tag{1.20}$$

This rule is justified when the boundary on $\mathcal{E}(\alpha, N)$ is given by $k_N = k N! |a|^N N^b$, with $k > 0$. The best approximation comes when the boundary is minimal, which also corresponds to the minimal term of the series, since both follow the same pattern. Indeed, using Stirling's approximation $N! \simeq \sqrt{2\pi N}(N/e)^N$ one can verify that[1.8]

$$N_{\min} \simeq \frac{1}{|a\alpha|}, \qquad \mathcal{E}(\alpha, N_{\min}) \sim o(e^{-1/|a\alpha|}). \tag{1.23}$$

An example where the optimal truncation rule can be successfully applied is the divergent series in (1.5): the truncation error (see (1.15)) satisfies the factorial boundary and its minimal value occurs at the minimal term of the series.

---

1.8. After employing Stirling's approximation to minimize $k_N |\alpha|^N$ one finds the transcendental relation $1 + 2b + 2N \log(N|a\alpha|) = 0$, which for large $N$ vanishes at $N_{\min} \simeq 1/|a\alpha|$. With this,

$$k_{N_{\min}} |\alpha|^{N_{\min}} \simeq \sqrt{2\pi} k |a\alpha|^{-1/2-b} e^{-1/|a\alpha|}, \tag{1.21}$$

and, since $\mathcal{E}(\alpha, N_{\min}) \sim o(\alpha^{N_{\min}})$ and $\alpha^{N_{\min}} \sim o(e^{-1/|a\alpha|})$ because

$$\lim_{\alpha \to 0} \frac{\alpha^{\frac{1}{|a\alpha|}}}{e^{-1/|a\alpha|}} = 0, \quad \mathrm{Re}(\alpha) > 0, \tag{1.22}$$

we can finally write $\mathcal{E}(\alpha, N_{\min}) \sim o(o(e^{-1/|a\alpha|})) \sim o(e^{-1/|a\alpha|})$. This recovers the results in [4], [5] and [6]. Note that we could equally write $\mathcal{E}(\alpha, N_{\min}) \sim o(|a\alpha|^{-1/2-b} e^{-1/|a\alpha|})$ since the equivalent limit also vanishes.

A second summation prescription, that improves upon the truncation at the minimal term, is Borel summation. The first step in Borel summation consists on employing the integral definition of the gamma function, and the fact that $\Gamma(n+1)=n!$, to rewrite the generic series (1.19) as

$$1+\sum_{n=1}^{\infty} r_n \alpha^n = 1+\sum_{n=1}^{\infty} \frac{r_n}{(n-1)!}\alpha^n \int_0^{\infty} \mathrm{d}t\, e^{-t} t^{n-1}. \tag{1.24}$$

The integral evaluates to $\Gamma(n)$ and compensates the $(n-1)!$ in the denominator. The sum and the integral cannot be interchanged in general, but the Borel sum of the series is defined as

$$\begin{aligned} R_B(\alpha) &\equiv 1+\alpha\int_0^{\infty} \mathrm{d}t\, e^{-t} B[R(\alpha)](\alpha t), \\ B[R(\alpha)](u) &\equiv \sum_{n=0}^{\infty} \frac{r_{n+1}}{n!} u^n. \end{aligned} \tag{1.25}$$

The series $B[R(\alpha)](u)$ receives the name of the Borel transform of the original series $R(\alpha)$, and we shall sometimes employ the shorter notation $B[R]$. The $n!$ in the denominator tames the divergent factorial behavior of $r_n$ and improves the convergence of the original series. In order for $R_B(\alpha)$ to be well-defined it must occur that

- $B[R]$ has non-vanishing radius of convergence $r$ and, if $r<\infty$, can be analytically continued to the positive real axis, and

- the integral $\int_0^{\infty} \mathrm{d}t\, e^{-t} B[R(\alpha)](\alpha t)$ converges.

When this two conditions are met, the series $R(\alpha)$ is said to be Borel-summable to the Borel sum $R_B(\alpha)$.

## 1.3 Borel summation for series with non-vanishing convergence radius

When the original series has a non-vanishing convergence radius, the convergence radius of the Borel transform is infinite and $R_B(\alpha)$ returns the analytic continuation of the original series $R(\alpha)$ to a region broader than the convergence circle of $R(\alpha)$. The fact that $B[R]$ possesses an infinite radius of convergence can be seen in general from the root test for $R(\alpha)$ and $B[R]$:

$$r_R = \left[ \limsup_{n \mapsto \infty} |r_n|^{1/n} \right]^{-1}, \qquad r_{B[R]} = \left[ \limsup_{n \mapsto \infty} \left| \frac{r_{n+1}}{n!} \right|^{1/n} \right]^{-1}. \tag{1.26}$$

If $r_R$ is finite, then $r_{B[R]}$ will be infinite due to the extra factor of $1/(n!)^{1/n}$. This ensures the integral in $R_B(\alpha)$ crosses no poles and $R(\alpha)$ is Borel-summable. Moreover, since the integral and the sum in (1.24) can be reversed when $|\alpha| < r_R$, it is clear that $R_B(\alpha) = R(\alpha)$ inside the convergence circle. However, the Borel sum is not restricted to the convergence circle and in general can be evaluated for some $|\alpha| \geq r_R$. For these points, $R_B(\alpha)$ agrees with the analytic continuation of $R(\alpha)$.

Let us now illustrate these claims with an example. Since analytic continuation is usually mentioned but not explicitly performed in manipulations regarding asymptotic series and perturbative QCD, we also detail how to analytically continue the example series and compare results with Borel summation.

We take the geometric series $R(\alpha) = \sum_{n=0}^{\infty} (-2)^n \alpha^n$, which has a convergence radius $r = 1/2$. Then, for $|\alpha| < r$, $R(\alpha)$ is analytic and converges to

$$R(\alpha) = \frac{1}{1 + 2\alpha}. \tag{1.27}$$

The analytic continuation can be performed as follows. Let $U$ denote the circle of radius $1/2$ centered at $\alpha = 0$, and let $\partial U$ be its boundary. For any point $a \in U$ we can expand

$$R(\alpha) = \sum_{n=0}^{\infty} c_n (\alpha - a)^n \tag{1.28}$$

in an neighborhood of $a$ that is completely contained in $U$ –this is, if $N_a$ denotes the neighborhood and $\partial N_a$ denotes its boundary, $(N_a \cup \partial N_a) \subset U$–. Since $R(\alpha)$ is analytic at $\alpha = a$, the coefficients of its expansion are given by its derivatives and we can use Cauchy's integral formula[1.9] to compute them

$$
\begin{aligned}
c_n &= \frac{R^{(n)}(a)}{n!} = \frac{1}{2\pi i}\oint_{\partial N_a} \mathrm{d}z \frac{R(z)}{(z-a)^{n+1}} = \frac{1}{2\pi i}\sum_{m=0}^{\infty}(-2)^m \int_0^{2\pi} \mathrm{d}\theta\, i e^{i\theta} \frac{(re^{i\theta}+a)^m}{(re^{i\theta})^{n+1}} \\
&= \frac{1}{2\pi}\sum_{n=0}^{\infty}(-2)^m \sum_{k=0}^{m}\binom{m}{k}\frac{a^{m-k}}{r^{n-k}}\int_0^{2\pi}\mathrm{d}\theta\, e^{i\theta(k-n)} = \sum_{m=0}^{\infty}(-2)^m\binom{m}{n}a^{m-n} \\
&= \frac{(-2)^n}{(1+2a)^{n+1}}.
\end{aligned} \tag{1.30}
$$

To arrive to this result we first chose $N_a$ to be a circle of radius $r$ centered at $a$, with $r<1/2-|a|$ (to ensure $N_a \cup \partial N_a \subset U$). Then we expanded $R(z)$ at $z=0$, changed variables to $z \to z+a$, parametrized $z=re^{\mathrm{i}\theta}$ and used (A.17) to expand the binomial in the numerator. The resulting integral vanishes for all $k$ except $k=n$, for which it is $2\pi$. The last sum over $m$ converges for $|a|<1/2$, which is satisfied since $a \in U$. With this, the expansion of $R(\alpha)$ around $a$ is

$$
R(\alpha) = \sum_{n=0}^{\infty} \frac{(-2)^n}{(1+2a)^{n+1}}(\alpha-a)^n. \tag{1.31}
$$

This expansion converges to the previous result $R(\alpha)=1/(1+2\alpha)$ with a convergence radius

$$
r = \left[\lim_{n\mapsto\infty}\left|\frac{c_{n+1}}{c_n}\right|\right]^{-1} = \left|\frac{1}{2}+a\right|. \tag{1.32}
$$

Therefore, choosing $a \neq 0$ in $U$ sets a convergence circle for (1.31) that includes a region outside $U$. Figure 1.3 shows the convergence circles of the original series ($U$) and of (1.31) (denoted as $V$) for various choices of $a$. In figures 1.3(*a*) and 1.3(*b*) it can be observed that the convergence circle $V$ is not fully contained in $U$, effectively continuing the original series outside $U$. Figure 1.3(*c*) shows how the singularity at $\alpha=-1/2$ prevents analytic continuation to points with $\mathrm{Re}(\alpha)<-1/2$ following a path along the real negative axis, as that path crosses the singularity of the closed function.

---

1.9. Cauchy's integral formula –sometimes called Cauchy's differentiation formula– establishes that for $f(z)$ holomorphic in $U \subseteq \mathbb{C}$, the derivatives $f^{(n)}(z)$ at $a \in U$ are given by

$$
f^{(n)}(a) = \frac{n!}{2\pi i}\oint_{\gamma}\mathrm{d}z \frac{f(z)}{(z-a)^{n+1}}, \tag{1.29}
$$

where $\gamma$ is a counterclockwise-oriented circle centered at $a$ and of radius $r$ such that it is fully contained in $U$.

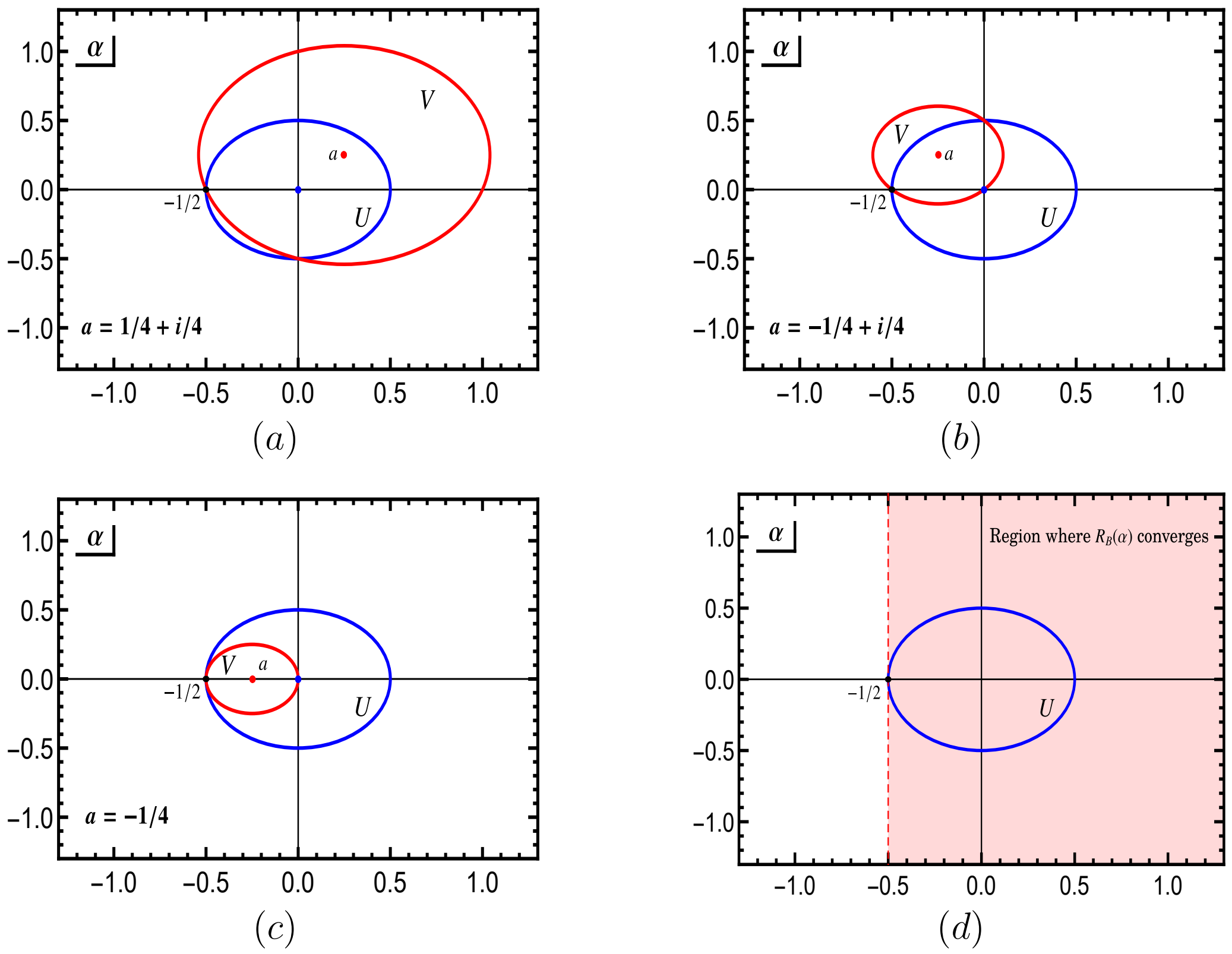

**Figure 1.3.** Panels $(a)$ to $(c)$: convergence circles of the series in (1.27) and (1.31) for three values of $a$. Panel $(d)$: convergence region of the integral defining the Borel sum $R_B(\alpha)$.

The process of analytic continuation can be repeated by picking $b \in V$, $b \notin U$ and computing the coefficients of the series in powers of $\alpha - b$. Proceeding this way, one extends the original series from $U$ to $\mathbb{C} - \{1/2\}$. Moreover, the converging function is always (1.27). This can be seen by assuming the series converges to two different functions: $R_U(\alpha)$ in $U$ and $R_V(\alpha)$ in $V$. Despite not representing the series outside their domain, $R_U(\alpha)$ and $\mathrm{R}_V(\alpha)$ can be defined in $U \cup V$. Due to the uniqueness of Taylor expansions –or, more generally, of asymptotic expansions in terms of a fixed asymptotic sequence; see remark 1 in section 1.1–, $R_U(\alpha) = R_V(\alpha)$ in $U \cap V$. Then, $R_U(\alpha) - R_V(\alpha)$ is an analytic function in $U \cup V$ that vanishes in $U \cap V$ and by the identity theorem for analytic functions[1.10], $R_U(\alpha) - R_V(\alpha) = 0$ in $U \cup V$.

Thus, in general, if a power series converges to a closed function inside a given domain in the complex plane, the closed function can be taken as its analytic continuation to the whole complex plane except in those points where it has a singularity.

1.10. The identity theorem states that if two functions $f(z)$ and $g(z)$ analytic in $U \subseteq \mathbb{C}$ are equal on a subset of $U$ which has an accumulation point (a point $a$ for which every of its neighborhoods contains at least one point different from $a$), then $f(z) = g(z)$ in $U$.

We now apply Borel summation. The Borel transform is

$$B[R(\alpha)](u) = \sum_{n=0}^{\infty} \frac{(-2)^{n+1}}{n!} u^n = -2e^{-2u}, \tag{1.33}$$

and has infinite radius of convergence. The Borel sum is then

$$R_B(\alpha) = 1 - 2\alpha \int_0^{\infty} \mathrm{d}t\, e^{-(2\alpha+1)t} = \frac{1}{1+2\alpha}, \tag{1.34}$$

where the integral converges for $\mathrm{Re}(\alpha) > -1/2$, delimiting a region that is broader than $U$, (see figure 1.3(*d*)). Figure 1.3(*d*) also shows how $R_B(\alpha)$, which only contains information on $B[R(\alpha)](u)$ for positive real values of $u$, is not immediately continued to $\mathbb{C}$.

A scenario where Borel summation proves useful is when $R(\alpha)$ is convergent but a closed form in terms of the usual functions cannot be found by direct summation. In this case the Borel sum may be easier to compute and provides an adequate closed form for the sum of the original series. For example, let us consider the series

$$R(\alpha) = \sum_{n=1}^{\infty} \frac{\Gamma(1+n, 2i\pi) - \Gamma(1+n, -2i\pi)}{2^{n+2} i\pi n} \alpha^n. \tag{1.35}$$

The two incomplete gamma functions have the same real part but imaginary parts of opposite signs, hence the subtraction is purely imaginary and the factor of $i$ in the denominator renders the series real. Moreover, the imaginary parts do not grow as $n!$, and one finds that the series has a convergence radius of $1/\pi$ through the root test:

$$\lim_{n\mapsto\infty} \left| \frac{\Gamma(1+n, 2i\pi) - \Gamma(1+n, -2i\pi)}{2^{n+2} i\pi n} \right|^{1/n} = \pi. \tag{1.36}$$

Despite convergent, finding a closed expression for the sum in (1.35) may prove challenging. One can trade the problem to finding its Borel transform and solving the inverse Borel integral. It can be seen that expanding

$$B[R(\alpha)](u) = \frac{\sin(au\pi)}{u\pi(2 - au)} \tag{1.37}$$

for small $u$ and performing the inverse Borel transform of each term returns the series in (1.35). The value the series converges to is then

$$\begin{aligned} R_B(\alpha) &= \int_0^{\infty} \mathrm{d}u\, e^{-u} B[R(\alpha)](u) \qquad (1.38) \\ &= \frac{\arctan(a\pi)}{2\pi}(1 - e^{-2/a}) + \frac{e^{-2/a}}{4\pi i}\left[\Gamma\left(0, -\frac{l}{a} - i\pi l, -\frac{l}{a} + i\pi l\right)\right.\end{aligned}$$

$$-\log\left(-\frac{2}{a}+2i\pi\right)+\log\left(-\frac{2}{a}-2i\pi\right)\Bigg].$$

Here, $\Gamma(z,a,b)$ is the doubly incomplete gamma, whose definition can be found in appendix A. In figure 1.4 we show how $R_B(\alpha)$ adequately returns the value the original series converges to when for several values of $\alpha$ in the convergence circle.

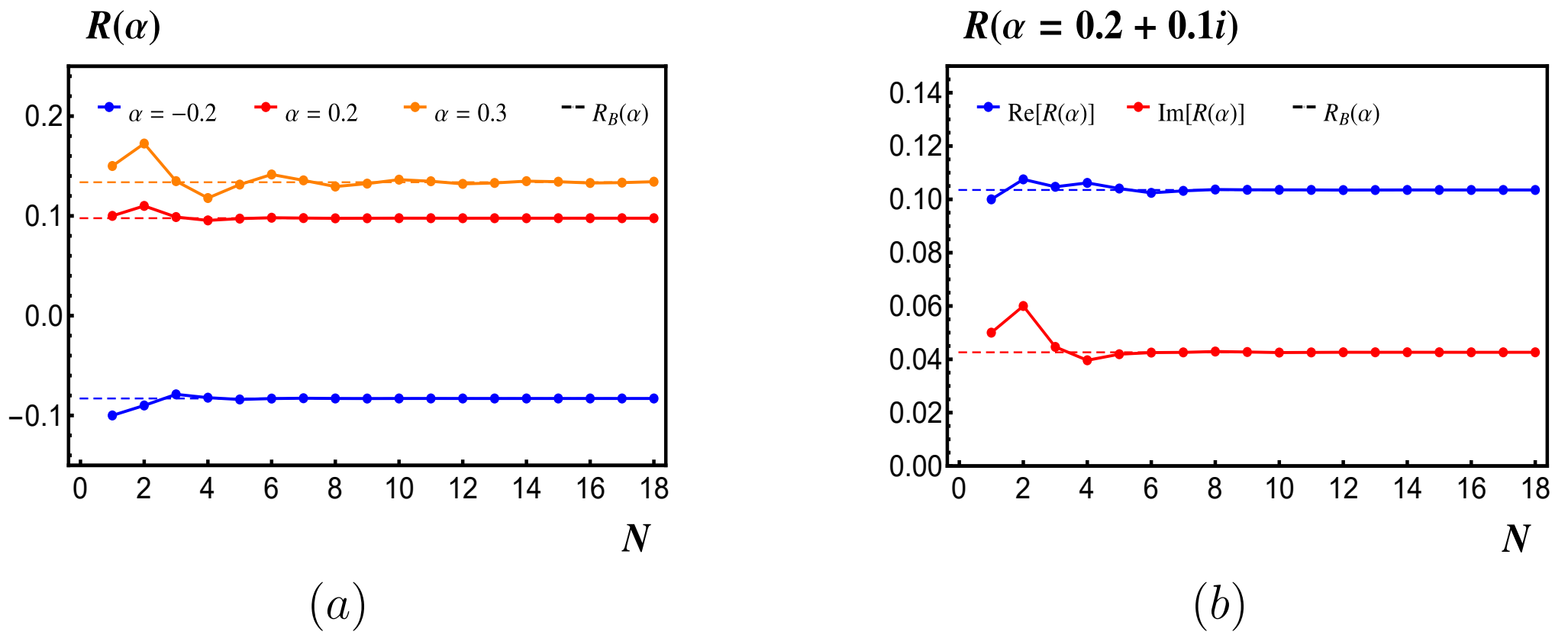

**Figure 1.4.** Comparison between the series in (1.35) truncated at $N$ terms (colored dots) and the Borel sum in (1.38) (colored, dashed lines) for values of $\alpha$ inside the convergence circle $|\alpha|<1/\pi$. In panel $(a)$ we show the series for $\alpha=-0.2, 0.2$ and $0.3$. In panel $(b)$ we show the real and imaginary parts of $R(\alpha)$ and $R_B(\alpha)$ for the complex value $\alpha=0.2+0.1i$.

# 1.4 Borel summation for series with zero convergence radius

## 1.4.1 Watson-Nevanlinna theorem

If $R(\alpha)$ has zero radius of convergence, there is no guarantee the Borel transform can be analytically continued to the positive real axis. For the moment, we focus on the case where $R(\alpha)$ is Borel-summable. In this case, the Borel sum $R_B(\alpha)$ can be interpreted as an estimate of $\bar{R}(\alpha)$, i.e., Borel summation acts as a prescription to reconstruct the function $\bar{R}(\alpha)$ from the asymptotic power series $R(\alpha)$. Since a series can be asymptotic to many different functions, the result given by Borel summation is not unique, meaning that different summation prescriptions –such as the Optimal Truncation Rule– can be applied and lead to different results. However, for the QCD series described at the beginning of the section it is believed that Borel summation constitutes the best possible estimate, in the sense of minimizing the truncation error. This idea is supported, at least partially, by the uniqueness theorems associated to Borel summation. In the context of asymptotic series, a uniqueness theorem states the conditions under which a function is uniquely determined by one of its asymptotic series. The conditions usually involve the analyticity of the function and a bound for the truncation error. In the case of QCD series, the most relevant uniqueness result is the Watson-Nevanlinna theorem.

The theorem arises from the original work of Watson [7], and was later refined by Nevanlinna [8]. Sokal rediscovered Nevanlinna's results and gave a modern formulation suitable for Quantum Field Theories [9]. Further discussions can be found in [3] and [10].

*Watson-Nevanlinna theorem.* Let $f(z)$ be analytical in the region of the complex plane $C_R = \{z \in \mathbb{C} | \mathrm{Re}(z^{-1}) > R^{-1}\}$ for some positive $R$ (see figure 1.5(*a*)) and let $\sum_{n=0}^{\infty} f_n z^n$ be asymptotic to $f(z)$ as $z \mapsto 0$. If the truncation error is bounded as

$$|\mathcal{E}(z, N-1)| \leq k\sigma^N N! |z|^N, \tag{1.39}$$

for positive $k$ and $\sigma$, then

- $B[f]$ converges for $|u| < 1/\sigma$ and can be analytically continued to the region $S_\sigma = \{u \in \mathbb{C} | \mathrm{dist}(u, \mathbb{R}^+) < 1/\sigma\}$, i.e., the region of points whose distance to the positive real axis is less than $1/\sigma$ (see figure 1.5(*b*)),
- $B[f]$ is bounded in $S_\sigma$ as $|B[f(z)](u)| \leq K \mathrm{e}^{|u|/R}$, $K > 0$, and
- $f(z)$ equals the convergent integral $1 + z\int_0^\infty \mathrm{d}t\, e^{-t} B[f(z)](zt)$ for all $z \in C_R$.

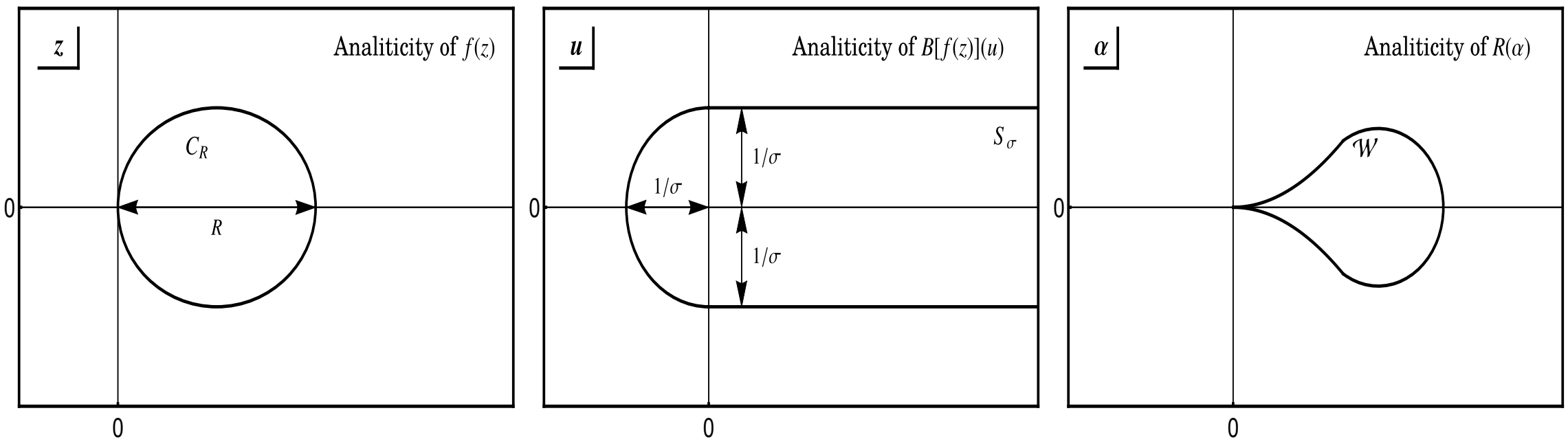

(*a*) (*b*) (*c*)

**Figure 1.5.**

The theorem also holds conversely [3]: if $B[f]$ is analytic in $S_\sigma$ and it satisfies $|B[f(z)](u)| \leq K \mathrm{e}^{|u|/R}$, the function $f(z)$ defined by the inverse Borel transform is analytic in $C_R$ and has an asymptotic expansion $\sum_{n=0}^{\infty} f_n z^n$ satisfying (1.39).

It is interesting to note that the function $e^{-1/z}$ does not satisfy the conditions of the Watson-Nevanlinna theorem. Indeed, $e^{-1/z}$ is analytic in $C_R$ but the truncation error $\mathcal{E}(z,N)=e^{-1/z}$ is unbounded. This can be seen by defining $w\equiv 1/z\equiv w_R+i w_I$, change that takes the circle $C_R$ into the infinite plane $\mathrm{Re}(w)\geq 1/R$. Then we have

$$|w^N e^{-w}|=|(w_R+i w_I)^N|e^{-w_R}\geq w_I^N e^{-w_R}, \tag{1.40}$$

where we used $w_R>0$. The right-hand side can be made arbitrarily large by increasing $w_I$, so $\mathcal{E}(z,N)$ is unbounded.

The Watson-Nevanlinna theorem is a relevant result which gives a sufficient characterization of a class of Borel summable functions. However, it has been pointed out that there are strong arguments indicating that non-perturbative QCD may not be analytical in the circle $C_R$ but instead in the wedge of zero opening angle depicted in figure 1.5(*c*) and denoted by $\mathcal{W}$ [11],[12]. A modification of this theorem that only demands $f(z)$ to be regular on $\mathcal{W}$ and continuous inside has been carried out in [13] (see also [3]). However, it comes at the price of imposing a more restrictive boundary for $|\mathcal{E}(z,N-1)|$, which translates into demanding that the series coefficients' have a more tamed divergent behavior[1.11] incompatible with the actual behavior in QCD. Nevertheless, Borel summation remains a powerful tool largely employed in QCD resummation studies.

### 1.4.2 Poles in the Borel plane and renormalons

As an example of Borel summation of a divergent series we consider the generic QCD series (1.1) with $r_n=a^n\Gamma(n)$. To take the Borel transform we multiply and divide by $\Gamma(n)$, which leads to:

$$B[R(\alpha)](u)=\sum_{n=1}^{\infty}a^n u^{n-1}, \tag{1.41}$$

which has a convergence radius $r_u=1/|a|$. For $|u|<1/|a|$ one can obtain the closed form

$$B[R(\alpha)](u)=\frac{a}{1-a u}. \tag{1.42}$$

1.11. Indeed, a balance between the analyticity properties of $f(z)$ and the boundary of the truncation error is required for $f(z)$ to be reconstructed from its divergent power expansion: requiring less analyticity of $f(z)$ allows for $f_n$ to diverge faster.

The Borel transform has a pole singularity in the positive real axis if $a>0$, which corresponds to $R(\alpha)$ being a non-alternating series. Conversely, we can note that the original series diverges faster the larger $a$ is, which translates into poles in the Borel transform being more severe the closer they are to $u=0$. We can also see that the radius of convergence of the original series can be identified as the position of the pole of its Borel transform.

When $a>0$ the series is not Borel summable, as $B[R]$ cannot be analytically continued to the positive real axis. One must then deform the integration contour, and, in doing so, the integral $R_B$ acquires a small imaginary part:

$$R_B^{a>0}(\alpha)=e^{-1/(a\alpha)}\mathrm{Ei}\left(\frac{1}{a\alpha}\right)\mp i\pi e^{-1/(a\alpha)}, \tag{1.43}$$

where the minus (plus) sign corresponds to avoiding the pole at $u=1/a$ along a clockwise (counter-clockwise) semicircle in the positive (negative) imaginary plane[1.12] and $\mathrm{Ei}(x)$ is the exponential integral function (see Appendix A for definitions). Expanding the real part in (1.43) in powers of $\alpha$ around the origin one gets back the $r_n$ coefficients. In figure 1.6(a) we compare the estimate (1.43) with the subsequent partial sums of the series. We see that between $N=8$ and 12 the series (blue dots) reaches a plateau, which is well estimated by the real part of the Borel sum $R_B^{a>0}(\alpha)$ (dashed red line). The imaginary part (gray band) is exponentially small and is often interpreted as an error in the estimation.

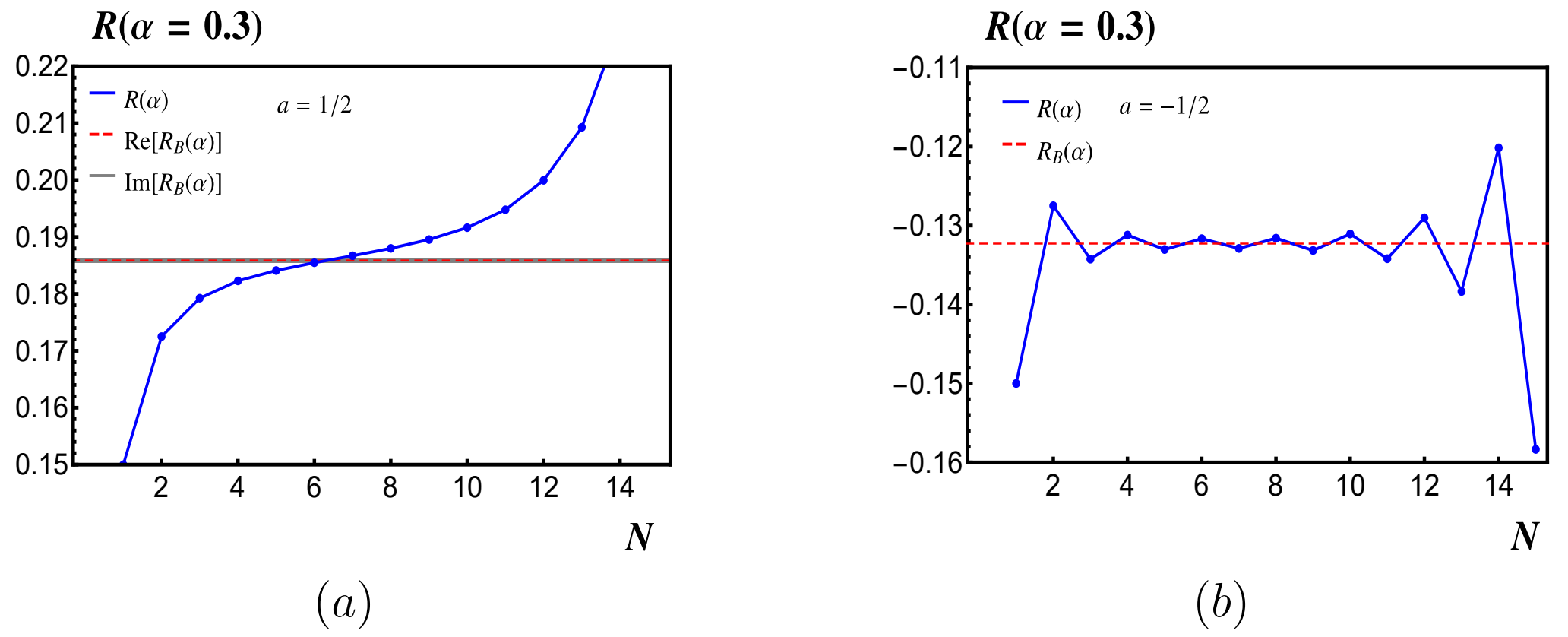

**Figure 1.6.** Borel summation for the series $R(\alpha)=\sum_{n=1}^{\infty}a^n\Gamma(n)\alpha^n$ for two values of $a$ and $\alpha=0.3$. Partial sums of the series are represented by blue dots, and the red, dashed line shows the Borel sum $R_B$. Panel $(a)$ shows a fixed sign series $(a>0)$, which translates into a singularity in the positive real axis of the Borel plane and into an imaginary part in the Borel $R_B$. Panel $(b)$ shows a sign-alternating series $(a<0)$, for which the singularity of the Borel transform lays in the negative real axis and the Borel sum is well-defined and real.

1.12. The easiest way carry out the computation is to center the semicircle of radius $\varepsilon$ at $u=1/a$ and take the limit $\varepsilon\mapsto 0$ after integration. For more details, we again refer to chapter 5.

In figure 1.6(*b*) we treat the case of alternating-sign series ($a<0$). In this case the series is Borel summable and one finds $R_B^{a<0} = \mathrm{Re}(R_B^{a>0})$ with no ambiguity. Nevertheless, we stress the original series is still divergent.

In practical QCD applications, the poles of the Borel transform of a series can be traced back to factorial divergences in the original series due to low or high values of the loop momentum in Feynman diagrams [4]. These singularities receive the name of renormalons, and are referred to as infrared (IR) renormalons when they come from low momentum and ultraviolet (UV) when they come from high loop momentum. UV renormalons are situated in the negative real axis of the Borel complex plane and IR renormalons lay in the positive real axis.

Let us conclude noting the imaginary part in (1.43) possesses an exponential of the form $e^{-q/\alpha}$, $q>0$. As mentioned earlier, $R_B(\alpha)$ and $\bar{R}(\alpha)$ can differ in terms of this exponential form, and it is believed the ambiguity signals the presence of this exponential factor for fixed-sign series.

## 1.5 A rearrangement of perturbative series: the large-flavor expansion

It is clear that in order to sum a convergent series or construct an estimate of a divergent series, one should have access to the general term of the series. In QCD, however, loop computations increase in difficulty and reaching contributions higher than three-four loops is in many cases unachievable.

A work around this problem is to carry out the series computation in a limit of full QCD known as the *large-$\beta_0$ limit*. This limit is a rearrangement of the standard perturbative expansion around small $\alpha_s$ in which each term is not a number but a whole infinite power series. The general term of the infinite sum that constitute the leading order contribution to this rearranged perturbative expansion can be usually computed at once with a single computation of one-loop difficulty. Its qualitative behavior is often expected to be that of the same as the full series, so that it can be used to learn features of the asymptotic properties of QCD.

In the remaining of this chapter we detail how to implement this limit. Let us consider again the generic series in (1.1), this time in QCD. A first important observation is that each coefficient $r_n$ can always be written as a polynomial in the number of quark flavors, $n_f$, with powers from 0 to $n-1$:

$$r_n = \sum_{m=0}^{n-1} r_{n,m} n_f^m. \tag{1.44}$$

The reason for this comes naturally if one thinks in terms of the Feynman diagrams contributing to $R(\alpha_s)$. Aside from real quark radiation, factors of $n_f$ can only appear in $R$ through diagrams with quark-loops; since each loop of any kind needs to be connected to at least two gluon lines, the power of $n_f$ in a given multi-loop structure must always be equal or lower than the power of $\alpha_s$. The attachment of the multi-loop correction into a diagram leads to additional powers of $\alpha_s$ and no extra powers of $n_f$.

As an example we can consider the quark self-energy in figure 1.7($a$). If the quantum correction, represented in the figure by the box, contains $m$ quark loops, it will be proportional to $\alpha_s^n n_f^m$, with $m \leq n$. The $n$ powers of $\alpha_s$ are given by the gluon attachments to loops of any kind. When the box is attached to the quark line more powers of $\alpha_s$ arise and the self-energy results proportional to $\alpha_s^n n_f^m$, where now $m \leq n-1$, with the equality coming from all corrections being quark bubbles attached to only two gluons.

Considering QCD amputated $n$-point functions of partonic degrees of freedom, one can see that relation (1.44) holds for any of them with the sole exception of the gluon two-point function, represented in diagram b) of figure 1.7($b$). In this case, the multi-loop structure can be directly attached to the gluon line with no additional powers of $\alpha_s$, and therefore its contributions are $\alpha_s^n n_f^m$, with $0 \leq m \leq n$. Nevertheless, we stress the gluon propagator is the only QCD diagrammatic construction in which powers of $n_f$ can run up to $n$ at $\mathcal{O}(\alpha_s^n)$, and that indeed all physical observables satisfy (1.44).

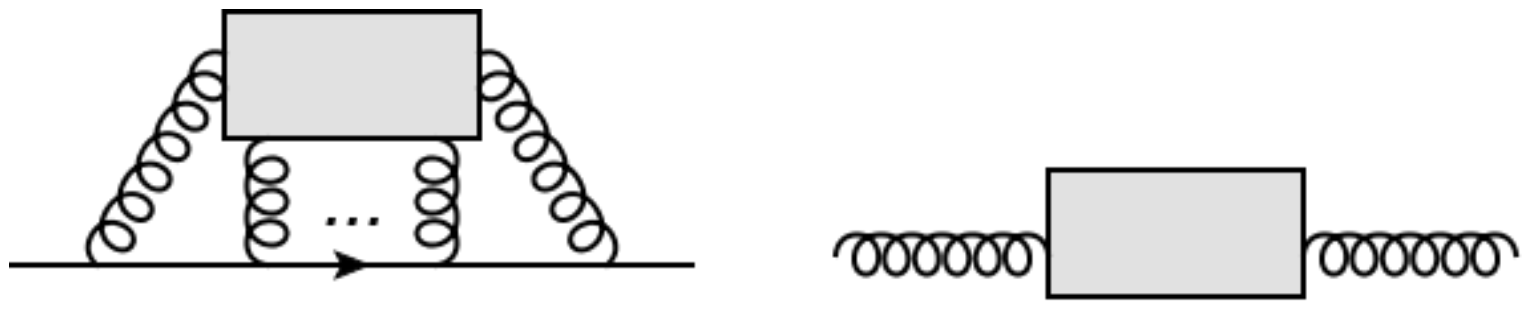

($a$) ($b$)

**Figure 1.7.** Symbolic corrections to quark ($a$) and gluon ($b$) lines. The box represents any possible quantum correction, and three dots stand for as many gluon insertions as desired. In both cases the corrections go as $\alpha_s^n n_f^m$, with $m < n$ in diagram ($a$) and $m \leq n$ in diagram ($b$).

Making explicit the dependence on $n_f$, our perturbative series for a generic observable is

$$R = 1 + \sum_{n=1}^{\infty} \alpha_s^n \sum_{m=0}^{n-1} r_{n,m} n_f^m, \tag{1.45}$$

where we drop the dependence on $\alpha_s$ for simplicity. A second observation is that we can now rearrange (1.45) by grouping together the terms attending to the relation between the powers of $\alpha_s$ and $n_f$. The terms with the maximum number of powers of $n_f$ for each order in $\alpha_s$ are of the form $\alpha_s^n n_f^{n-1}$ and constitute the first set in the rearrangement; those with next-to-maximum number of powers of $n_f$ for each order in $\alpha_s$ are proportional to $\alpha_s^n n_f^{n-2}$ and constitute the second set, etc. In sum, we make groups of the form $\alpha_s^n n_f^{n-m}$ and reorganize $R$ as

$$R = 1 + O(\alpha_s^n n_f^{n-1}) + O(\alpha_s^n n_f^{n-2}) + \ldots + O(\alpha_s^n n_f^0). \tag{1.46}$$

As we will show in a moment, this rearrangement has some important advantages. First, however, we implement (1.46) into in (1.45):

$$\begin{aligned} R &= 1 + \sum_{m=0}^{\infty} \sum_{n=m+1}^{\infty} r_{n,m} \alpha_s^n n_f^m = 1 + \sum_{m=0}^{\infty} \sum_{n=1}^{\infty} r_{n+m,m} \alpha_s^{n+m} n_f^m \\ &= 1 + \sum_{n=1}^{\infty} \sum_{m=n}^{\infty} r_{m,m-n} \alpha_s^m n_f^{m-n} = 1 + \sum_{m=1}^{\infty} \sum_{n=m}^{\infty} r_{n,n-m} \alpha_s^n n_f^{n-m}. \end{aligned} \tag{1.47}$$

To arrive to the last equality we first interchanged the sums in (1.45), switched $n \to n+m$, interchanged the sums again and finally switched $m \to m-n$. In the very last step we simply renamed $n \leftrightarrow m$ for notational convenience. For each value of $m$, the sum over $n$ groups all the contributions to $\alpha_s^n$ that go with $n_f^{n-m}$: the first term of the sum over $m$ contains all the terms of the form $\alpha_s^n n_f^{n-1}$, the second term contains those with $\alpha_s^n n_f^{n-2}$... In sum, (1.47) follows the rearrangement in (1.46).

Having our perturbative series written as in (1.46) or (1.47) has the very important advantage that the complete infinite tower of quantum corrections with $m=1$ can be analytically computed at once with a single computation of one-loop difficulty. This is the result of several features that the terms of this form exhibit:

1. Diagrammatically, all terms proportional to the highest power of $n_f$ arise from a single class of Feynman diagrams: the class of all possible insertions of one *effective gluon propagator*. The effective gluon propagator, which we also call the *effective propagator* or the *bubble chain*, is a gluon line with only quark-loop insertions (see figure 1.8).

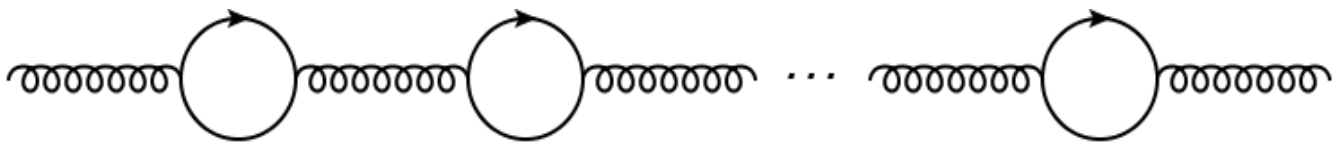

**Figure 1.8.** Effective gluon propagator. The three dots stand for any number of quark loops inserted on the gluon line.

For $n$ quark loops, the propagator itself is proportional to $\alpha_s^n n_f^n$, and its attachment gives an additional power of $\alpha_s$. One can check that with the allowed interactions in standard QCD no other structures can produce terms of the form $\alpha_s^n n_f^{n-1}$.

2. As we shall see in section 2.3, the effective propagator is computable for a generic number of quark loops, $n$. Moreover, its functional form with respect to the gluon's momentum is that of the standard gluon propagator, with the replacement $k^2 \mapsto (k^2)^{1+n\epsilon}$, where the spacetime dimension has been analytically continued to $d=4-2\epsilon$. Therefore the insertion of the effective propagator gives rise to an integral that is of one-loop difficulty, hence analytically computable for generic $n$. The whole $m=1$ contribution is then obtained by summing over the number of quark bubbles, $n$.

Submaximum contributions ($m>1$) are harder to picture, as they do not possess a unique diagrammatic counterpart and need to be considered in each case.

The fact that the $m=1$ infinite set of terms in the proposed rearrangement can be computed with relative ease, while the successive groups with $m>1$ are increasingly more complicated, makes it natural to introduce a hierarchy that can be seen as an expansion of $R$ in a parameter that is not just the coupling constant. The hierarchy

$$1 \gg \alpha_s^n n_f^{n-1} \gg \alpha_s^n n_f^{n-2} \gg \dots \tag{1.48}$$

is equivalent to considering $n_f$ large enough so that $\alpha_s n_f \sim 1$ and expanding the original series (1.45) in $1/n_f$. Indeed, (1.48) is equivalent to

$$1 \gg \frac{1}{n_f}(\alpha_s n_f)^n \gg \frac{1}{n_f^2}(\alpha_s n_f)^n \gg \dots \quad \text{or} \quad 1 \gg \frac{1}{n_f} \gg \frac{1}{n_f^2} \gg \dots \tag{1.49}$$

This is known as the *large-flavor expansion*, and under it we can truncate $R$ to leading order:

$$R = 1 + \underbrace{\sum_{n=1}^{\infty} r_{n,n-1} n_f^{n-1} \alpha_s^n}_{n_f\mathrm{LO}} + \underbrace{\mathcal{O}(\alpha_s^n n_f^{n-2})}_{n_f\mathrm{NLO}}. \tag{1.50}$$

We introduced here the shorthand notation $n_f\mathrm{N}^n\mathrm{LO}$ to refer to the different orders in the large-$n_f$ expansion. As a final comment, we remark that when computing QCD series at $n_f$LO one can in many cases ignore some of the interaction terms in the full-QCD Lagrangian and work with a simplified, abelian-like version. This is because the only particle interaction present in the effective propagator is the quark-gluon vertex, and therefore, if the diagram in which it is inserted does not contain other kinds of interactions, these will only contribute to subleading orders in the large-$n_f$ and can be safely ignored. This is the case, for example, when considering corrections to the quark propagator, where the ghost-gluon interaction and the gluon self-interactions do not play any role.

## 1.6 The large-$\beta_0$ expansion

Considering a large number of quark flavors is adequate to set up calculations from the diagrammatic point of view, as it establishes a distinctive identification between the power counting in the series expansion and the number of quark loops in the diagram. Formally, however, one needs to be careful: if $n_f$ is only allowed to take arbitrarily high positive values, the QCD-beta function will change its sign and key strong-interaction properties such as confinement and asymptotic freedom will be casted out of the theory. Indeed, the QCD-beta function is

$$\beta_{\mathrm{QCD}}(\alpha_s) \equiv \mu \frac{\mathrm{d}}{\mathrm{d}\mu} \alpha_s(\mu) = -2\epsilon\alpha_s - 2\alpha_s \sum_{n=0}^{\infty} \beta_n \left(\frac{\alpha_s}{4\pi}\right)^{n+1}, \tag{1.51}$$

where the first coefficient takes the explicit form

$$\beta_0 = \frac{11}{3} C_A - \frac{4}{3} T_F n_f, \tag{1.52}$$

with $T_F$ and $C_A$ related to the normalization of the SU(3) Lie-algebra generators in the fundamental and adjoint representation, respectively. For the standard normalization and three colors, we have $T_F = 1/2$ and $C_A = 3$, so $\beta_0 > 0$ for $n_f < 16$, which is satisfied in QCD. The limit $n_f \to \infty$ reverts the sign of $\beta_0$.

A simple solution to maintain both a perturbative expansion in $\alpha_s n_f \sim 1$ and asymptotic freedom is to consider $n_f$ as a large negative number. However, since for most important practical applications it is best to take a positive number of flavors, it is customary to implement instead the formally equivalent limit $\beta_0 \gg 1$. This defines the *large-$\beta_0$ limit* of QCD.

Let us see the equivalence of both limits. On the one hand, if we already are in the large-$n_f$ limit, $\beta_0$ gets the effective value of $-4T_F n_f/3$, a replacement often called *naive non-abelianization* [5]. With it, the leading-order contribution in (1.50) is

$$R = 1 + \sum_{n=1}^{\infty} \left( \frac{-3}{4T_F} \right)^{n-1} r_{n,n-1} \beta_0^{n-1} \alpha_s^n + O(\alpha_s^n \beta_0^{n-2}). \tag{1.53}$$

On the other hand, if we return to the original (prior to the large-$n_f$ expansion) expression for $R$, (1.45), and rewrite it in terms of the full $\beta_0$ coefficient we obtain:

$$\begin{aligned} R &= 1 + \sum_{n=1}^{\infty} \alpha_s^n \sum_{m=0}^{n-1} r_{n,m} \frac{(11C_A - 3\beta_0)^m}{(4T_F)^m} \qquad (1.54) \\ &= 1 + \sum_{n=1}^{\infty} \alpha_s^n \sum_{m=0}^{n-1} \frac{r_{n,m}}{(4T_F)^m} \sum_{i=0}^{m} \binom{m}{i} (-3\beta_0)^i (11C_A)^{m-i} = 1 + \sum_{n=1}^{\infty} \alpha_s^n \sum_{i=0}^{n-1} s_{n,i} \beta_0^i, \end{aligned}$$

with

$$s_{n,i} \equiv \sum_{m=i}^{n-1} \binom{m}{i} \frac{(-3)^i (11C_A)^{m-i}}{(4T_F)^m} r_{n,m}, \tag{1.55}$$

where we expanded $(11C_A - 3\beta_0)^m$ with the binomial expansion (A.17) and interchanged the nested sums over $m$ and $i$. Taking now the limit of large $\beta_0$, in which $\alpha_s \beta_0 \sim 1$, only the coefficient $s_{n,n-1}$ contributes, so the sum over $i$ disappears and $R$ becomes

$$\begin{aligned} R &= 1 + \sum_{n=1}^{\infty} s_{n,n-1} \beta_0^{n-1} \alpha_s^n + \mathcal{O}(\alpha_s^n \beta_0^{n-2}), \qquad (1.56) \\ s_{n,n-1} &= \left( \frac{-3}{4T_F} \right)^{n-1} r_{n,n-1}, \end{aligned}$$

in agreement with (1.53).

Of course, in analogy to (1.49), the large-$\beta_0$ limit induces a hierarchy on the terms of the perturbative series,

$$1 \gg \frac{1}{\beta_0}(\alpha_s\beta_0)^n \gg \frac{1}{\beta_0^2}(\alpha_s\beta_0)^n \gg \ldots \quad \text{or} \quad 1 \gg \frac{1}{\beta_0} \gg \frac{1}{\beta_0^2} \gg \ldots, \tag{1.57}$$

for which we also adopt the notation $\beta_0\mathrm{N}^n\mathrm{LO} \sim \mathcal{O}(1/\beta_0^{n-1})$.

# Chapter 2

# Ingredients for the study of series in the large-$\beta_0$ limit

Results (1.54) and (1.56) constitute the large-$\beta_0$ expansion of a perturbative series and its truncation at leading order. Despite being completely general, the simplicity in their derivation relays in the fact that our original expression for the generic observable $R$ was a renormalized power series in the renormalized coupling constant $\alpha_s$, where the dependence on $n_f$ was already extracted out of the coefficients at all orders in $\alpha_s$.

In practice, however, what one obtains from direct computations is the unrenormalized perturbative series in terms of the bare coupling constant $\alpha_s^0$. We will refer to this series as $A_0$ to distinguish it from the series in the previous chapter. To rewrite $A_0$ in terms of the renormalized QCD parameters, one needs to account for the appropriate multiplicative renormalization factors. Of course, this situation is analogous to any fixed-order computation in QCD, but in order to adequately expand $A_0$ in the large-$\beta_0$ limit there are some subtleties that must be observed. First, as already discussed, the $\beta_0$LO contribution to any series including the renormalization factors is an infinite tower of terms, all of which must be properly kept in order for the expansion to work. Second, the renormalization factors themselves contain different dependence on $\alpha_s$ and $\beta_0$. In particular, the quark wave-function and mass renormalization factors $Z_\psi$ and $Z_m$ start at $\mathcal{O}(\alpha_s^n \beta^{n-1})$ in the large-$\beta_0$ expansion, while the gluon wave function $Z_A$ and the coupling renormalization $Z_\alpha$ starts at $\mathcal{O}(\alpha_s^n \beta_0^n) \sim 1$. This is due to the fact that in the large-$\beta_0$ limit, $Z_A$ is defined to absorb the divergences of the effective gluon propagator in figure 1.8. As we shall see in detail, $Z_\alpha$ is completely determined by $Z_A$ in the large-$\beta_0$ limit, hence the scaling $Z_\alpha \sim 1 + \mathcal{O}[(\alpha_s\beta_0)^{n>0}]$, which can already be seen from the well-known one-loop result, $Z_\alpha^{\overline{\mathrm{MS}}} = 1 - \beta_0\alpha_s/(4\pi\epsilon)$ [14], [15].

In this chapter we start with the bare theory of QCD and work our way through regularization, and multiplicative renormalization. For clarity, we will work both at the level of the Lagrangian and the perturbative series $A_0$. The aim is double: on the one hand, we seek to find a comprehensive recipe to compute the $\beta_0$LO contribution to any perturbative series –in the renormalized theory– solely from the computation of its Feynman diagrams in the bare formalism; on the other hand, carefully working our way through also allows us to set up the relevant notation.

We remark that a complete exposition of QCD, dimensional regularization and renormalization methods is too broad to be included in this thesis. The aim of this section is to briefly go through the main important points and set up the notation and definitions that are relevant for the later applications. The reader can find excellent reviews in [14], [15] and [16].

## 2.1 The general picture of QCD renormalization

As we have mentioned, at leading order in the large-$\beta_0$ limit of QCD some of the interactions in the full theory may not play any role and can be ignored from the Lagrangian. In particular, throughout this work we will be focusing on applications in which the effective propagator will be inserted in quark and quark-like (jet) lines, so from now on we will be considering the simplified Lagrangian

$$\mathcal{L}_{\mathrm{QCD}}^{\beta_0\mathrm{LO}} \equiv \bar{\psi}^0(i\not{\partial} - m_0)\psi^0 - \frac{1}{4}(\partial_\mu A_\nu^{0a} - \partial_\nu A_\mu^{0a})^2 + g_0\bar{\psi}^0 \not{A}^0\psi^0. \tag{2.1}$$

In terms of the bare fields and parameters we denote perturbative series as

$$A_0 \equiv 1 + \sum_{n=1}^{\infty} a_n^0 \left(\frac{g_0^2}{(4\pi)^2}\right)^n = 1 + \sum_{n=1}^{\infty} a_n^0 \left(\frac{\alpha_s^0}{4\pi}\right)^n, \tag{2.2}$$

where in the second step we defined the bare strong coupling as

$$\alpha_s^0 \equiv \frac{g_0^2}{4\pi}. \tag{2.3}$$

The $a_n^0$ coefficients are formally divergent due to loop integrals. To deal with the divergences, one uses regularization and renormalization. Regularization makes the divergences explicit, while renormalization absorbs them into non-physical parameters of the theory.

The chosen regularization method is dimensional regularization, which is implemented by analytically continuing the spacetime dimension to $d=4-2\epsilon$. This affects the mass dimension[2.1] of the fields and parameters in the Lagrangian (2.1), as well as the mass dimension of the quark and gluon propagators. Indeed, the mass dimensions of the fields and parameters are

$$[\psi^0]=M^{(d-1)/2}=M^{3/2-\epsilon}, \qquad [A^0_\mu]=M^{(d-2)/2}=M^{1-\epsilon}, \qquad (2.4)$$
$$[m_0]=M, \qquad [g_0]=M^{(d-4)/2}=M^{\epsilon}.$$

In dimensional regularization the bare coupling constant $g_0$ acquires mass dimension. It is, however, convenient to expand perturbative series in terms of a dimensionless variable. To this end a mass scale $\tilde{\mu}$ –a quantity with mass dimension $M$– is introduced and one considers $g_0\tilde{\mu}^{-\epsilon}$ as the expansion parameter. In order to keep the perturbative series independent on this arbitrary scale, one needs to add a compensating factor. The perturbative series contains even powers of $g_0\tilde{\mu}^{-\epsilon}$, so the compensating factor consists on the replacement of $a_n^0$ by $\tilde{\mu}^{2n\epsilon}a_n^0(\epsilon)$, which is also dimensionless. The regularized series is then

$$A_0=1+\sum_{n=1}^{\infty}\tilde{\mu}^{2n\epsilon}a_n^0(\epsilon)\left(\frac{g_0^2\tilde{\mu}^{-2\epsilon}}{(4\pi)^2}\right)^n=1+\sum_{n=1}^{\infty}\tilde{\mu}^{2n\epsilon}a_n^0(\epsilon)\left(\frac{\alpha_s^0\tilde{\mu}^{-2\epsilon}}{4\pi}\right)^n, \qquad (2.5)$$

where the notation $a_n^0(\epsilon)$ indicates that the divergent part of $a_n^0$ is now regularized through $\epsilon$. On the other hand, the $n$ order coefficient $a_n^0$ contains $n$ loop integrals, each of which can absorb a factor of $\tilde{\mu}^{2\epsilon}$ out of the total $\tilde{\mu}^{2n\epsilon}$. These factors merge with the energy scales of the loop integral $(\Lambda_i)$ and, upon expansion around $\epsilon=0$ give rise to logarithms of the form $\log(\tilde{\mu}/\Lambda_i)$ as well as powers of $1/\epsilon$. Besides, the $\tilde{\mu}$-independence of $A_0$ after breaking it into $\tilde{\mu}$-dependent coefficients and $\tilde{\mu}$-dependent coupling constant ensures the condition

$$\tilde{\mu}\frac{\mathrm{d}}{\mathrm{d}\tilde{\mu}}A_0=0, \qquad (2.6)$$

which is known as the homogeneous *renormalization group equation* (RGE).

In sum, the regularization process can be synthesized by stating that the regularized theory consists on the bare Lagrangian plus the following rule:

$$\text{in the perturbative series:}\;\; g_0\to g_0\tilde{\mu}^{-\epsilon},\;\; \mathrm{d}^dp\to\tilde{\mu}^{2\epsilon}\mathrm{d}^dp, \qquad (2.7)$$

2.1. Here and in the following we use the so called *natural units*, which set both the speed of light $c$ and Plank's reduced constant $\hbar$ to be the dimensionless number 1. This of course leads to all the usual quantities being measured in units of mass or energy, with the former usually preferred for dimensional analysis and the latter preferred to express numeric results such as decay widths and cross sections. In particular, time and distance are measured in inverse mass units, $[t]=[x]=M^{-1}$.

where the replacement of momentum space integration is equivalent to that of the $a_n^0$ coefficients.

Once divergences are regularized, they can be removed by introducing renormalization factors in the Lagrangian. This is done by defining the renormalized fields and Lagrangian parameters as

$$\psi^0 \equiv Z_\psi^{1/2}\psi, \qquad A^0 \equiv Z_A^{1/2}A, \qquad g_0 \equiv Z_g g \qquad \text{and} \qquad m_0 \equiv Z_m m. \tag{2.8}$$

With them, the resulting Lagrangian reads

$$\mathcal{L}_{\text{QCD}}^{\beta_0\text{LO}} = Z_\psi \bar{\psi}(i\partial\!\!\!/ - Z_m m)\psi - \frac{1}{4}Z_A(\partial_\mu A_\nu^{0a} - \partial_\nu A_\mu^{0a})^2 + Z_{qg}\bar{\psi} \mathrm{i} g A\!\!\!/ \psi, \tag{2.9}$$

with $Z_{qg} \equiv Z_\psi Z_g Z_A^{1/2}$. In the renormalized theory, $g\tilde{\mu}^{-\epsilon}$ is taken as the new coupling constant, which leads to the following dimensionless definition of the renormalized strong coupling

$$\alpha_s(\tilde{\mu}) \equiv \frac{g^2\tilde{\mu}^{-2\epsilon}}{4\pi}, \tag{2.10}$$

related as follows to the ($\tilde{\mu}$-dependent) bare coupling:

$$\alpha_s^0 = \frac{g_0^2}{4\pi} = \frac{Z_g^2 g^2}{4\pi} = Z_\alpha \tilde{\mu}^{2\epsilon}\alpha_s, \qquad Z_\alpha \equiv Z_g^2. \tag{2.11}$$

The renormalization factors $Z_i$ now enter the series (2.5) and are defined to cancel the $1/\epsilon$ poles. The leading order term in the large-$\beta_0$ limit for each of them is

$$\begin{aligned} Z_\psi, Z_m &\sim 1 + O(\alpha_s^n \beta_0^{n-1}) \qquad \text{[quark self-energy]}, \\ Z_A &\sim 1 + O(\alpha_s^n \beta_0^n) \qquad \text{[gluon self-energy]}, \\ Z_{qg} &\sim 1 + O(\alpha_s^n \beta_0^{n-1}) \qquad \text{[quark-gluon vertex]}, \end{aligned} \tag{2.12}$$

as can be seen from the relevant diagrams quoted on the right. Only for $Z_A$ quantum corrections have the same large-$\beta_0$ counting as the leading term. On top of this, (2.12) determines the large-$\beta_0$ expansion for $Z_\alpha$:

$$Z_\alpha = \frac{Z_{qg}^2}{Z_\psi^2 Z_A} \sim Z_A^{-1} \sim 1 + O(\alpha_s^n \beta_0^n), \tag{2.13}$$

in agreement with the one-loop result. The moral of the story is that to adequately prepare $A_0$ to be expanded in the large-$\beta_0$ limit with regards to the renormalized coupling constant $\alpha_s$, we need to explicitly keep track of the factors of $Z_\alpha$ but we can absorb the contribution of $Z_\psi$ and $Z_m$ into the coefficients $a_n^0$. This defines the coefficients $a_n(\epsilon)$, which are polynomials in $\beta_0$ with powers up to $\beta_0^{n-1}$:

$$A_0 = 1 + \sum_{n=1}^{\infty} \tilde{\mu}^{2n\epsilon} a_n(\epsilon) Z_\alpha^n \left(\frac{\alpha_s}{4\pi}\right)^n = 1 + \sum_{n=1}^{\infty} \tilde{\mu}^{2n\epsilon} Z_\alpha^n \left(\frac{\alpha_s \beta_0}{4\pi}\right)^n \sum_{i=0}^{n-1} \frac{b_{n,i}(\epsilon)}{\beta_0^{n-i}}, \tag{2.14}$$
$$b_{n,i}(\epsilon) \equiv \sum_{m=i}^{n-1} \binom{m}{i} \frac{(-3)^i (11 C_A)^{m-i}}{(4T_F)^m} a_{n,m}(\epsilon),$$

where in the second step we proceeded as in (1.54) by writing $a_n \equiv \sum_{m=0}^{n-1} a_{n,m} n_f^m$ and switching to $\beta_0$. Also, we multiplied and divided by $\beta_0^n$ to make the following discussion clearer. Note that the coefficients $a_n$ –and, by extension, $a_{n,m}$ and $b_{n,m}$– still depend on $\epsilon$ due to the fact that $Z_\alpha$ has not been expanded yet. Also, as it will be the case for the later applications, $A_0$ may still present poles in $\epsilon$ after $Z_\alpha$ has been expanded, requiring its own renormalization factor.

It is also customary to work with the Modified Minimal Subtraction ($\overline{\text{MS}}$) scale $\mu$ instead of $\tilde{\mu}$, defined as

$$\mu^2 \equiv 4\pi e^{-\gamma_E} \tilde{\mu}^2, \tag{2.15}$$

where $\gamma_E$ is the Euler-Mascheroni constant. This scale is designed to cancel the powers of $\log(4\pi)$ and $\gamma_E$ arising in $\epsilon$ expansions.

What remains is to expand (2.14) in the large-$\beta_0$ limit. Accounting for the fact that in the sum over $i$ all the powers of $1/\beta_0$ are positive, one can clearly see that the contributions to $A_0$ at $\beta_0$LO are

- the complete $\beta_0$LO tower of terms in $Z_\alpha \sim 1 + \mathcal{O}[(\alpha_s \beta_0)^{n>0}]$,

- the $i = n-1$ slice in the $i$ sum, which is proportional to $1/\beta_0$ and contains the coefficients

$$b_{n,n-1} = \left(-\frac{3}{4T_F}\right)^{n-1} a_{n,n-1}. \tag{2.16}$$

In the next two sections we compute these two contributions and pave the way for the final form of $A_0$ in the large-$\beta_0$. First, in section 2.2 we derive a resummed expression for the leading large-$\beta_0$ terms in $Z_\alpha$, which is obtained from studying the running of $\alpha_s(\mu)$ in this limit. Then, in section 2.3 we compute the effective gluon propagator and give the explicit relation between its insertion into the relevant diagrams and the coefficients $a_{n,n-1}$. Finally, in section 2.4 we plug these results back into $A_0$ and reorganize the resulting series as a true expansion in $1/\beta_0$.

## 2.2 Coupling constant renormalization

### 2.2.1 The running of $\alpha_s$

Let us explore the running of $\alpha_s$ in the large-$\beta_0$ limit. Using $\beta_n \sim \mathcal{O}(\beta_0^n)$ for $n > 0$, the large-$\beta_0$ expansion of the QCD-beta function in (1.51) is

$$\mu\frac{\mathrm{d}}{\mathrm{d}\mu}\alpha_s(\mu) = -2\epsilon\alpha_s - \frac{\alpha_s^2\beta_0}{2\pi} + \mathcal{O}(\alpha_s^3\beta_0) = -\frac{1}{\beta_0}\left[2\epsilon\alpha_s\beta_0 + \frac{(\alpha_s\beta_0)^2}{2\pi}\right] + \mathcal{O}\left(\frac{1}{\beta_0^2}\right). \quad (2.17)$$

From now on we drop the reference to subleading terms. The above differential equation can be solved for $\epsilon = 0$ through separation of variables,

$$\frac{\mathrm{d}\alpha_s}{\alpha_s^2} = -\frac{\beta_0}{2\pi}\frac{\mathrm{d}\mu}{\mu} \quad \Longrightarrow \quad \frac{1}{\alpha_s(\mu)} - \frac{1}{\alpha_s(\mu_0)} = \frac{\beta_0}{2\pi}\log\left(\frac{\mu}{\mu_0}\right), \quad (2.18)$$

leading to the LL running of $\alpha_s$,

$$\alpha_s(\mu) = \frac{\alpha_s(\mu_0)}{1 + \alpha_s(\mu_0)\frac{\beta_0}{2\pi}\log\left(\frac{\mu}{\mu_0}\right)}, \quad (2.19)$$

which is exact in the large-$\beta_0$ limit.

In practical applications the physical energy scale of the interaction dictates the number of active quark flavors, and the unphysical renormalization scale $\mu$ –at which $\alpha_s$ is evaluated– does not activate or suppress flavors. In this case, the renormalized coupling constant remains a function of $\mu$ that is only specified by the boundary condition $\alpha_s(\mu_0)$ and the fixed value of $n_f$. Then it is useful to rewrite (2.19) in terms of the scale at which $\alpha_s$ diverges, defined as scale $\mu_{\rm div} \equiv \Lambda_{\rm QCD}$ that makes the denominator of (2.19) vanish:

$$\Lambda_{\rm QCD} = \mu_0 e^{-\frac{2\pi}{\beta_0\alpha_s(\mu_0)}}. \quad (2.20)$$

When $\Lambda_{\text{QCD}}$ is used in place of the reference scale $\mu_0$ the dependence on $\alpha_s(\mu_0)$ disappears,

$$\alpha_s(\mu) = \frac{\alpha_s(\mu_0)}{1+\alpha_s(\mu_0)\frac{\beta_0}{2\pi}\left[\log\left(\frac{\mu}{\Lambda_{\text{QCD}}}\right) - \frac{2\pi}{\beta_0\alpha_s(\mu_0)}\right]} = \frac{2\pi}{\beta_0\log\left(\frac{\mu}{\Lambda_{\text{QCD}}}\right)}. \tag{2.21}$$

### 2.2.2 The renormalization factor $Z_\alpha$

We can relate the previous results to $Z_\alpha$. From the definition (2.11) we have

$$\begin{aligned} \mu\frac{\mathrm{d}}{\mathrm{d}\mu}\alpha_s(\mu) &= \mu\frac{\mathrm{d}}{\mathrm{d}\mu}\frac{g^2\tilde{\mu}^{-2\epsilon}}{4\pi} = \mu\frac{\mathrm{d}}{\mathrm{d}\mu}\frac{Z_\alpha^{-1}g_0^2\left(\frac{\mu^2\mathrm{e}^{\gamma_E}}{4\pi}\right)^{-\epsilon}}{4\pi} \\ &= \frac{g_0^2\mathrm{e}^{-\gamma_E\epsilon}}{(4\pi)^{1-\epsilon}}\left[\mu^{1-2\epsilon}\frac{\mathrm{d}}{\mathrm{d}\mu}\frac{1}{Z_\alpha} - \frac{2\epsilon\mu^{-2\epsilon}}{Z_\alpha}\right] = Z_\alpha\frac{g_0^2\tilde{\mu}^{-2\epsilon}}{4\pi}\left[\mu Z_\alpha\frac{\mathrm{d}}{\mathrm{d}\mu}\frac{1}{Z_\alpha} - 2\epsilon\right] \\ &= -\alpha_s\left[\frac{\mathrm{d}\log(Z_\alpha)}{\mathrm{d}\log(\mu)} + 2\epsilon\right]. \end{aligned} \tag{2.22}$$

This result must agree with (2.17), so in the large-$\beta_0$ we have

$$\frac{\mathrm{d}\log(Z_\alpha)}{\mathrm{d}\log(\mu)} = \frac{\alpha_s\beta_0}{2\pi}. \tag{2.23}$$

Since $\alpha_s$ is just a function of $\mu$, we can use the chain rule on the left-hand side,

$$\frac{\mathrm{d}\log(Z_\alpha)}{\mathrm{d}\alpha_s}\frac{\mathrm{d}\alpha_s}{\mathrm{d}\log(\mu)} = -\alpha_s\frac{4\pi\epsilon+\alpha_s\beta_0}{2\pi}\frac{\mathrm{d}\log(Z_\alpha)}{\mathrm{d}\alpha_s}, \tag{2.24}$$

where we have used Eq. (2.17). We arrive to a differential equation for $Z_\alpha$

$$\frac{\mathrm{d}\log(Z_\alpha)}{\mathrm{d}\alpha_s} = -\frac{\beta_0}{4\pi\epsilon+\alpha_s\beta_0}, \tag{2.25}$$

which we integrate using the tree-level boundary condition $\log(Z_\alpha(0)) = \log(1) = 0$, obtaining

$$Z_\alpha = \frac{4\pi\epsilon}{\alpha_s\beta_0 + 4\pi\epsilon}. \tag{2.26}$$

This result effectively sums up the infinite tower of leading order contribution in the large-$\beta_0$ limit to $Z_A$, which, as we observed, is of order $\mathcal{O}(\alpha_s\beta_0)\sim 1$.

### 2.2.3 The $\beta$ parameter

In the large-$\beta_0$ limit, the $\mathcal{O}(1)$ combination

$$\beta \equiv \frac{\alpha_s \beta_0}{4\pi}, \tag{2.27}$$

arises as the natural parameter in perturbative expansions. It is useful then to write the previous results in terms of $\beta$.

First, we can derive a differential relation between $\beta$ and $\mu$ that will proof useful when computing the running of renormalized series. From (2.17) we have

$$\mu \frac{\mathrm{d}}{\mathrm{d}\mu} \beta = -2\beta(\epsilon + \beta). \tag{2.28}$$

Using Eq. (2.27) into (2.21) or, equivalently, solving (2.28) for $\epsilon = 0$ we find

$$\beta(\mu) = \frac{\beta_{\mu_0}}{1 + 2\beta_{\mu_0} \log\left(\frac{\mu}{\mu_0}\right)} = \frac{1}{2\log\left(\frac{\mu}{\Lambda_{\mathrm{QCD}}}\right)}, \tag{2.29}$$

where we defined $\beta_x \equiv \beta(x)$ to alleviate the notation.

Second, the $Z_\alpha$ factor (2.26) can also be compactly written in terms of $\beta$:

$$Z_\alpha = \frac{\epsilon}{\beta + \epsilon}. \tag{2.30}$$

Finally, from (2.29) we can find two useful relations:

$$\begin{aligned} \log\left(\frac{\beta_\mu}{\beta_{\mu_0}}\right) &= -\log\left[1 + 2\beta_{\mu_0} \log\left(\frac{\mu}{\mu_0}\right)\right] = \log\left[1 - 2\beta_\mu \log\left(\frac{\mu}{\mu_0}\right)\right], \\ \log\left(\frac{\mu}{\mu_0}\right) &= \frac{1}{2}\left(\frac{1}{\beta_\mu} - \frac{1}{\beta_{\mu_0}}\right). \end{aligned} \tag{2.31}$$

## 2.3 The effective gluon propagator

### 2.3.1 Computation

In this section we compute step-by-step the effective gluon propagator. We start by computing the amputated quark-loop (no side propagators) shown in figure 2.1.

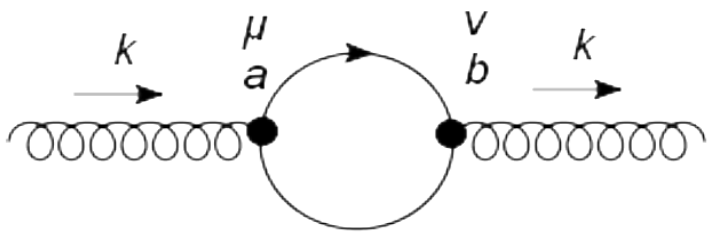

**Figure 2.1.** Amputated quark loop, which we denote as $B_{ab}^{\mu\nu}(k)$, with $k$ the gluon's momentum.

Direct application of Feynman rules gives

$$B_{ab}^{\mu\nu}(k) \equiv -g_0^2 \delta_{ab} n_f T_F \int \frac{\mathrm{d}^d\ell}{(2\pi)^d} \frac{\mathrm{tr}[\gamma^\mu \not{\ell} \gamma^\nu (\not{\ell} - \not{k})]}{\ell^2 (\ell - k)^2}, \tag{2.32}$$

where, as it is conventional, the Latin indices $a$ and $b$ denote color indices while the Greek indices $\mu$ and $\nu$ denote Lorentz indices. Before going any further, we pull out the trivial color dependence $\delta_{ab}$ by defining $B_{ab}^{\mu\nu}(k) \equiv \delta_{ab} B^{\mu\nu}(k)$ to simplify the notation.

To solve (2.32) we start by solving the trace in the numerator. Using the standard techniques of Dirac's algebra, we find

$$\begin{aligned} \mathrm{tr}[\gamma^\mu \not{\ell} \gamma^\nu (\not{\ell} - \not{k})] &= 4(2\ell^\mu \ell^\nu - \ell^\mu k^\nu - \ell^\nu k^\mu - g^{\mu\nu} \ell^2 + g^{\mu\nu} \ell \cdot k) \qquad (2.33) \\ &= 4\left[2\ell^\mu \ell^\nu - \ell^\mu k^\nu - \ell^\nu k^\mu - \frac{1}{2} g^{\mu\nu} \ell^2 + \frac{1}{2} g^{\mu\nu} k^2 - \frac{1}{2} g^{\mu\nu} (\ell - k)^2\right]. \end{aligned}$$

The terms $\ell^2$ and $(\ell - k)^2$ cancel the propagators in the denominator and lead to scaleless integrals that vanish in dimensional regularization. Using the notation for loop integrals in appendix C, the integral (2.32) becomes

$$\begin{aligned} \frac{-1}{4 g_0 n_f T_F} B^{\mu\nu}(k) &= 2 I_2(q^\mu q^\nu, Q, Q(-k)) - k^\nu I_2(q^\mu, Q, Q(-k)) \qquad (2.34) \\ &\quad - k^\mu I_2(q^\nu, Q, Q(-k)) + \frac{1}{2} g^{\mu\nu} k^2 I_2(1, Q, Q(-k)). \end{aligned}$$

Since, with the results in C.3[2.2],

$$\begin{aligned} I_2(\ell^\mu, Q, Q(-k)) &= \frac{k^\mu}{2} I_2(1, Q, Q(k)), \qquad (2.35) \\ I_2(\ell^\mu \ell^\nu, Q, Q(-k)) &= \frac{1}{4(d-1)} (-k^2 g^{\mu\nu} + d k^\mu k^\nu) I_2(1, Q, Q(k)), \end{aligned}$$

2.2. In the right-hand side of (2.35) we used the symmetry property $I_2(1, Q, Q(-k)) = I_2(1, Q, Q(k))$ that can be seen from the explicit result (C.7).

$B^{\mu\nu}(k)$ further simplifies to the transverse Lorentz structure

$$B^{\mu\nu}(k) = 2g_0^2 n_f T_F \frac{d-2}{1-d}(k^2 g^{\mu\nu} - k^\mu k^\nu) I_2(1, Q, Q(k)) = iB(k^2) k^{-2\epsilon} \mathcal{T}^{\mu\nu}, \tag{2.36}$$

$$B(k^2) = \frac{ig_0^2}{(4\pi)^{2-\epsilon}} n_f T_F P_B(\epsilon), \qquad \mathcal{T}^{\mu\nu} = (k^2 g^{\mu\nu} - k^\mu k^\nu),$$

where in the last line we defined

$$P_B(\epsilon) \equiv -8 \frac{(-1)^{-\epsilon}\Gamma(\epsilon)\Gamma^2(2-\epsilon)}{\Gamma(4-2\epsilon)}. \tag{2.37}$$

Next, the structure that repeats periodically in the $n$-chain is the bubble $B^{\mu\nu}(k)$ with its left gluon propagator. We denote this structure as $\Sigma^{\mu\nu}(k)$. The gluon propagator $G^{\mu\nu}_{ab}(k) \equiv \delta_{ab} G^{\mu\nu}(k)$ in an arbitrary gauge has the form

$$G^{\mu\nu}(k) = -\frac{i}{k^2}\left[ g^{\mu\nu} - \frac{k^\mu k^\nu}{k^2}(1-\xi) \right]. \tag{2.38}$$

The contraction $\Sigma^{\mu\nu}(k) \equiv G^{\mu\rho}(k) B_\rho{}^\nu(k)$ produces the Lorentz tensor

$$\left[ g^{\mu\rho} - \frac{k^\mu k^\rho}{k^2}(1-\xi) \right] \mathcal{T}^{\mu\nu} = \mathcal{T}^{\mu\nu}, \tag{2.39}$$

so that

$$\Sigma^{\mu\nu}(k) = B(k^2) k^{-2(\epsilon+1)} \mathcal{T}^{\mu\nu}. \tag{2.40}$$

To get the $n$-bubble chain we need to contract $\delta_{ab}\Sigma^{\mu\nu}(k)$ with itself $n$ times and then add the final gluon propagator on the right. After $n$ contractions:

- the Lorentz scalars are powered to $n$, contributing as $[B(k^2)]^n k^{-2n(\epsilon+1)}$;
- the color part is given by the Kronecker delta $\delta_{ab}$;
- the Lorentz structure $\mathcal{T}^{\mu\nu}$ contracted with itself is again transverse tensor

$$\mathcal{T}^{\mu\rho}\mathcal{T}_\rho{}^\nu = k^2(k^2 g^{\mu\nu} - k^\mu k^\nu), \tag{2.41}$$

  so each contraction simply adds a factor of $k^2$ to a total of $k^{2(n-1)}$.

All in all, contracting $\Sigma^{\mu\nu}(k)$ $n$ times yields

$$\Sigma_n^{\mu\nu}(k) \equiv [B(k^2)]^n k^{-2(\epsilon+1)} \mathcal{T}^{\mu\nu}. \tag{2.42}$$

Lastly we add the final gluon propagator. The Lorentz contraction is the same as in (2.39), so

$$\tilde{\Delta}_n^{\mu\nu}(k) \equiv \Sigma_n^{\mu\rho}(k) G_\rho^{\ \nu}(k) = -iB^n(k^2)\left[\frac{g^{\mu\nu}}{(k^2)^{1+n\epsilon}} - \frac{k^\mu k^\nu}{(k^2)^{2+n\epsilon}}\right] \tag{2.43}$$

The expression for $\tilde{\Delta}_n^{\mu\nu}(k)$ does not fully match the gluon propagator $G^{\mu\nu}(k)$ when $n=0$: the gauge-dependent term is not recovered. Including it we finally arrive to

$$\Delta_n^{\mu\nu}(k) \equiv -iB^n(k^2)\left[\frac{g^{\mu\nu}}{(k^2)^{1+n\epsilon}} - \frac{k^\mu k^\nu}{(k^2)^{2+n\epsilon}}(1-\xi\delta_{n0})\right]. \tag{2.44}$$

### 2.3.2 $\beta_0$LO contribution from shifted contribution

As mentioned above, the dependence of the effective gluon propagator (2.44) on the gluon's momentum $k$ is the same as that of the tree-level gluon propagator $G^{\mu\nu}(k)$ with the shift $k^2 \to k^{2+n\epsilon}$. Also, the extra prefactor (powered to $n$) does not depend on $k$ and will factor out in any loop integral. Due of this fact, and with the aim of making computations lighter, we define the *shifted gluon propagator* as

$$G_{n\epsilon}^{\mu\nu}(k) \equiv -i\left[\frac{g^{\mu\nu}}{(k^2)^{1+n\epsilon}} - \frac{k^\mu k^\nu}{(k^2)^{2+n\epsilon}}(1-\xi\delta_{n0})\right]. \tag{2.45}$$

We then call $D_{\rm sh}(h)$ the contribution to a given series resulting from all the possible insertions of the shifted propagator $G_h^{\mu\nu}(k)$, and we factor

$$D_{\rm sh}(h) \equiv \left(\frac{g_0}{4\pi}\right)^2 a(h,\epsilon), \tag{2.46}$$

where the notation $a(h,\epsilon)$ is meant to distinguish between the shift $h$ in the propagator and the dimensional regularization parameter $\epsilon$. The complete $\beta_0$LO contribution –obtained by inserting $\Delta_n^{\mu\nu}$ and summing over $n$– is

$$\begin{aligned} D^{\beta_0\rm LO} &= \sum_{n=1}^{\infty}\left[\frac{g_0^2}{(4\pi)^{2-\epsilon}} n_f T_F P_B(\epsilon)\right]^{n-1} D_{\rm sh}((n-1)\epsilon) \\ &= \sum_{n=1}^{\infty}[(4\pi)^\epsilon n_f T_F P_B(\epsilon)]^{n-1}\left(\frac{\alpha_s^0}{4\pi}\right)^n a((n-1)\epsilon,\epsilon), \end{aligned} \tag{2.47}$$

where in the last line we used the definition of $\alpha_s^0$ in (2.3). This gives a simple recipe for computations, as the only input needed to compute the complete $\beta_0$LO contribution is the $a(h,\epsilon)$ function. Indeed, comparing (2.14) with $a_n = \sum_{m=0}^{n-1} a_{n,m} n_f^m$ to equation (2.47) we immediately see

$$a_{n,n-1} = [(4\pi)^{\epsilon} T_F P_B(\epsilon)]^{n-1} a((n-1)\epsilon,\epsilon). \tag{2.48}$$

## 2.4 Final form

With all the results above we are finally able to expand $A_0$ in the large-$\beta_0$ limit. We plug the expression (2.30) for $Z_\alpha$ into (2.14), getting

$$\begin{aligned} A_0 &= 1 + \sum_{n=1}^{\infty} \tilde{\mu}^{2n\epsilon} \left[ \frac{\epsilon}{\beta+\epsilon} + O\left(\frac{1}{\beta_0}\right) \right]^n \beta^n \sum_{i=0}^{n-1} \frac{b_{n,i}}{\beta_0^{n-i}} \\ &= 1 + \frac{1}{\beta_0} \sum_{n=1}^{\infty} \left( \frac{\beta}{\beta+\epsilon} \right)^n \tilde{\mu}^{2n\epsilon} \epsilon^n b_{n,n-1} + O\left(\frac{1}{\beta_0^2}\right), \end{aligned} \tag{2.49}$$

where in the last line we only retained the coefficient $b_{n,n-1}$. At this stage, the large-$\beta_0$ expansion of $A_0$ has been performed and all the terms in the infinite sum over $n$ contribute equally and constitute the leading order. Having already extracted the factor $1/\beta_0$, we need to rewrite $A_0$ as a power series in $\beta$. To achieve this we use again the binomial expansion, this time in the form (A.18), to expand the $\beta$ dependence as

$$\left(\frac{\beta}{\beta+\epsilon}\right)^n = \left(\frac{\beta}{\epsilon}\right)^n (1+\beta/\epsilon)^{-n} = \sum_{i=0}^{\infty} \frac{(-1)^i \Gamma(n+i)}{\Gamma(n)\Gamma(i+1)} \left(\frac{\beta}{\epsilon}\right)^{n+i}. \tag{2.50}$$

From now on we omit the $O(1/\beta_0^2)$ and define $\delta A_0 \equiv A_0 - 1$ to lighten the notation. Plugging this result back into $A_0$ we obtain

$$\begin{aligned} \beta_0 \delta A_0 &= \sum_{n=1}^{\infty} \tilde{\mu}^{2n\epsilon} \epsilon^n b_{n,n-1} \sum_{i=0}^{\infty} \frac{(-1)^i \Gamma(n+i)}{\Gamma(n)\Gamma(i+1)} \left(\frac{\beta}{\epsilon}\right)^{n+i} \\ &= \sum_{n=1}^{\infty} \tilde{\mu}^{2n\epsilon} \epsilon^n b_{n,n-1} \sum_{i=n}^{\infty} \frac{(-1)^{i-n} \Gamma(i)}{\Gamma(n)\Gamma(i-n+1)} \left(\frac{\beta}{\epsilon}\right)^i \\ &= \sum_{i=1}^{\infty} \sum_{n=1}^{i} \tilde{\mu}^{2n\epsilon} \epsilon^n b_{n,n-1} \frac{(-1)^{i-n} \Gamma(i)}{\Gamma(n)\Gamma(i-n+1)} \left(\frac{\beta}{\epsilon}\right)^i \\ &= \sum_{n=1}^{\infty} (-\beta)^n \epsilon^{-n} \sum_{i=1}^{n} \frac{(-1)^i \Gamma(n)}{\Gamma(i)\Gamma(n-i+1)} \tilde{\mu}^{2i\epsilon} \epsilon^i b_{i,i-1}, \end{aligned}$$

where we first took $i \to i-n$, then interchanged the two sums and in the last step renamed $n \leftrightarrow i$. Finally, we relate the $b_{i,i-1}$ coefficients to the function $a(h,\epsilon)$ as explained in section 2.3.2. From (2.48) the relation is

$$\begin{aligned}\tilde{\mu}^{2i\epsilon} b_{i,i-1} &= \tilde{\mu}^{2i\epsilon}\frac{(-3)^{i-1}}{(4T_F)^{i-1}} a_{i,i-1} = \frac{(\mu^2 e^{\gamma_E})^{i\epsilon}}{(4\pi)^{i\epsilon}}\left[-\frac{3}{4}(4\pi)^{\epsilon} P_B(\epsilon)\right]^{i-1} a((i-1)\epsilon,\epsilon) \qquad (2.51)\\ &= \frac{(\mu^2 e^{\gamma_E})^{i\epsilon}}{(4\pi)^{\epsilon}}\left[-\frac{3}{4}P_B(\epsilon)\right]^{i-1} a((i-1)\epsilon,\epsilon),\end{aligned}$$

where in the last step we switched to the $\overline{\text{MS}}$ scale (2.15). With this, our final form for the $\beta_0$LO contribution to $A_0$ reads

$$\beta_0 \delta A_0 = \sum_{n=1}^{\infty}(-\beta)^n \epsilon^{-n}\sum_{i=1}^{n}\frac{(-1)^i\Gamma(n)}{\Gamma(i)\Gamma(n-i+1)}\frac{F^{\mu}(\epsilon,i\epsilon)}{i}, \qquad (2.52)$$

where we defined

$$F^{\mu}(\epsilon,u) \equiv \frac{(\mu^2 e^{\gamma_E})^u u}{(4\pi)^{\epsilon}}\left[-\frac{3}{4}\epsilon P_B(\epsilon)\right]^{\frac{u}{\epsilon}-1} a(u-\epsilon,\epsilon). \qquad (2.53)$$

We included the superscript $\mu$ in $F^{\mu}$ to make the dependence on the renormalization scale explicit, something that will prove useful in our posterior work with $A_0$. Also, since $a(u-\epsilon,\epsilon)$ contains the dependence of the series on the energy scale of the process, $\omega$, we can split:

$$\begin{aligned}F^{\mu}(\epsilon,u) &\equiv \left(\frac{\mu^2}{\omega^2}\right)^u F(\epsilon,u), \qquad (2.54)\\ F(\epsilon,u) &= \frac{e^{\gamma_E u} u}{(4\pi)^{\epsilon}}\left[-\frac{3}{4}\epsilon P_B(\epsilon)\right]^{\frac{u}{\epsilon}-1}\omega^{2u} a(u-\epsilon,\epsilon),\end{aligned}$$

where $F(\epsilon,u)$ has no $\mu$ dependence and is dimensionless.

## 2.5 Anomalous dimension and work-route

Throughout the following chapters we will be studying the large-$\beta_0$ limit of the series $A_0$. Eventually, we will make explicit the negative powers of $\epsilon$ and proceed to its renormalization by defining the renormalized series $A$ and its renormalization factor $Z_A$, as $A_0 \equiv A Z_A$. From simple inspection we can see that at $\beta_0$LO multiplicative renormalization effectively occurs by the simple addition of the $\beta_0$LO contributions to $A$ and $Z_A$, since any other terms are subleading:

$$A Z_A = \left(1+\sum_{n=0}^{\infty} A^{\beta_0 \text{N}^n\text{LO}}\right)\left(1+\sum_{n=0}^{\infty} Z_A^{\beta_0 \text{N}^n\text{LO}}\right) = 1 + A^{\beta_0\text{LO}} + Z_A^{\beta_0\text{LO}} + O\left(\frac{1}{\beta_0^2}\right), \qquad (2.55)$$

where $A^{\beta_0\mathrm{N}^n\mathrm{LO}}, Z_A^{\beta_0\mathrm{N}^n\mathrm{LO}} \sim \mathcal{O}(1/\beta_0^{n+1})$.

Due to the renormalization of $A_0$ in (2.55) and the fact that $Z_A$ is defined to cancel all the negative powers of $\epsilon$, the renormalized series $A$ is now UV finite but it has acquired a dependence on the renormalization scale $\mu$. This dependence was not present in the bare series and is described by its *anomalous dimension* $\gamma_A$. The usual definition of the anomalous dimension arises from the fact that $A_0$ obeys the homogeneous renormalization group equation (2.6), which after renormalization takes the form

$$0 = \mu\frac{\mathrm{d}}{\mathrm{d}\mu}(Z_A A) \implies \mu\frac{\mathrm{d}}{\mathrm{d}\mu}A = \left[-\frac{1}{Z_A}\mu\frac{\mathrm{d}}{\mathrm{d}\mu}Z_A\right]A \equiv \gamma_A A. \tag{2.56}$$

Both the renormalized series and the renormalization factor depend on $\mu$ explicitly –through the $F^\mu(\epsilon, u)$ function previously defined– and implicitly through the factor of $\alpha_s(\mu)$ in $\beta = \beta(\mu)$. As a consequence, $\gamma_A$, which will always have the implicit $\mu$-dependence in $\alpha_s(\mu)$, can also have explicit dependence. Of course, if $A_0$ is ultraviolet finite in the first place, no renormalization is needed and $\gamma_A = \gamma_{A_0} = 0$. It is customary then to distinguish three kinds of series according to $\gamma_A$:

1. **Finite series.** $A_0$ is UV finite –that is, in dimensional regularization, presents no negative powers of $\epsilon$–, then no renormalization is required and $A_0 = A$. In this case, the anomalous dimension identically vanishes.

2. **Usual series.** $A_0$ presents a single power $1/\epsilon$ at one loop. In this case the anomalous dimension only contains implicit $\mu$-dependence through $\alpha_s(\mu)$, and we define its fixed-order coefficients as

$$\gamma_A(\alpha_s) = \sum_{n=0}^{\infty}\gamma_n^A\left(\frac{\alpha_s}{4\pi}\right)^{n+1}. \tag{2.57}$$

3. **Cusp series.** $A_0$ diverges as $1/\epsilon^2$ at one loop. In this case, the anomalous dimension acquires linear dependence on $\log(\mu)$ and can be split as

$$\gamma_A(\alpha_s, \mu) = \gamma_A(\alpha_s) + \log\left(\frac{\mu}{\omega}\right)\Gamma_A(\alpha_s), \tag{2.58}$$

where $g_A$ is referred to as the *non-cusp* part and $\Gamma_A$ is the universal cusp anomalous dimension. We adopt the following convention for their perturbative expansion in $\alpha_s$

$$\gamma_A(\alpha_s) = \sum_{n=0}^{\infty}\gamma_n^A\left(\frac{\alpha_s}{4\pi}\right)^{n+1}, \qquad \Gamma_A(\alpha_s) = \sum_{n=0}^{\infty}\Gamma_n^A\left(\frac{\alpha_s}{4\pi}\right)^{n+1}. \tag{2.59}$$

At $\beta_0$LO the general definition of the anomalous dimension in (2.56) takes a simpler form. This is due to the fact that the the logarithm of a series is

$$\log(A) = \log\left(1 + \sum_{n=0}^{\infty} A^{\beta_0 \mathrm{N}^n\mathrm{LO}}\right) = \sum_{i=1}^{\infty} \frac{(-1)^{i+1}}{i}\left[\sum_{n=0}^{\infty} A^{\beta_0 \mathrm{N}^n\mathrm{LO}}\right]^i = A^{\beta_0\mathrm{LO}} + \mathcal{O}\left(\frac{1}{\beta_0^2}\right), \tag{2.60}$$

so that

$$\gamma_A = \mu\frac{\mathrm{d}}{\mathrm{d}\mu}\log(A) = \mu\frac{\mathrm{d}}{\mathrm{d}\mu}A + \mathcal{O}\left(\frac{1}{\beta_0^2}\right) = -\mu\frac{\mathrm{d}}{\mathrm{d}\mu}Z_A + \mathcal{O}\left(\frac{1}{\beta_0^2}\right). \tag{2.61}$$

# Chapter 3

# Series without cusp-anomalous dimension

In this chapter we deal with series with no cusp-anomalous dimension. First, in 3.1 we impose the regularity condition on $F(\epsilon,u)$ and work on our perturbative series up to a point where the McLaurin series in $\epsilon$ is made explicit and renormalization can be carried out in a simple manner. Then, in 3.2, we apply the results in appendix B and find an integral representation for the renormalized series and the renormalization factor, which we later manipulate to a more appropriate form regarding the study of its large order divergences. Finally, in 3.3 we focus on the anomalous dimension and we solve the corresponding RGE. The work developed in this chapter, as well as in the remaining of part I is published in [17].

## 3.1 Perturbative form

### 3.1.1 Form for regular $F(\epsilon, u)$

Let us assume that $F(\epsilon,u)$ is regular at the origin. Then, we can expand

$$F^{\mu}(\epsilon,u) \equiv \sum_{i=0}^{\infty}\sum_{j=0}^{\infty}\epsilon^i u^j F^{\mu}_{i,j}, \qquad F^{\mu}_{i,0} \equiv F_{i,0}. \tag{3.1}$$

Under this assumption $A_0$ in (2.52) takes the form

$$\begin{aligned}\beta_0\delta A_0 &= \sum_{n=1}^{\infty}(-\beta)^n\epsilon^{-n}\sum_{i=1}^{n}\frac{(-1)^i\Gamma(n)}{\Gamma(i+1)\Gamma(n-i+1)}\sum_{j=0}^{\infty}\sum_{k=0}^{\infty}\epsilon^j(i\epsilon)^k F^{\mu}_{j,k},\\ &= \sum_{n=1}^{\infty}(-\beta)^n\sum_{j=0}^{\infty}\sum_{k=0}^{\infty}\epsilon^{j+k-n}F^{\mu}_{j,k}\sum_{i=1}^{n}\frac{(-1)^i\Gamma(n)i^k}{\Gamma(i+1)\Gamma(n-i+1)}.\end{aligned} \tag{3.2}$$

The sum over $i$,

$$I_{n,k} \equiv \sum_{i=1}^{n}\frac{(-1)^i\Gamma(n)i^k}{\Gamma(i+1)\Gamma(n-i+1)} \quad , \tag{3.3}$$

is obviously convergent, since it is finite and does not cross any singularity. For $k=0$ it sums up to $I_{n,0}=-1/n$, and for $k\geq 1$ it exhibits the properties $I_{n,1\leq k<n}=0$, $I_{1,k}=-1$ and $I_{n,n}=(-1)^n\Gamma(n)$. In appendix B.1 we prove the closed form $I_{n,0}$ and the three stated properties by deriving a closed form for $I_{n,k\geq 1}$. For now, we use them to simplify $A_0$ by making the $n=1$, $k=0$ and $n>1$ cases explicit:

$$\beta_0\delta A_0=\beta\sum_{j=0}^{\infty}\sum_{k=0}^{\infty}\epsilon^{j+k-1}F^{\mu}_{j,k}+\sum_{n=2}^{\infty}(-\beta)^n\sum_{j=0}^{\infty}\left[-\epsilon^{j-n}\frac{F_{j,0}}{n}+\sum_{k=n}^{\infty}\epsilon^{j+k-n}F^{\mu}_{j,k}I_{n,k}\right]. \tag{3.4}$$

Note that we set the sum over $k$ to start at $k=n$, since $I_{n>1,k<n}=0$.

### 3.1.2 Form as McLaurin series in $\epsilon$

Now we turn our attention to the $\epsilon$ dependence, noting that contains positive powers. Since we are only interested in $\epsilon\mapsto 0$ and QCD renormalization has already been performed, we only need to keep the negative powers and the $\epsilon$-independent term.

To make the discarding of negative powers simple and explicit, we order $A_0$ as a power series in $\epsilon$. In the first double sum over $j$ and $k$ we change $k\mapsto k-j+1$ and switch the order of summation; in the double sum over $n$ and $j$ we take $j\mapsto n+j$; lastly, in the triple sum over $n$, $j$ and $k$ we take $k\to n+k$. The result is:

$$\beta_0\delta A_0=\beta\sum_{k=-1}^{\infty}\epsilon^k\sum_{j=0}^{k+1}F^{\mu}_{j,k-j+1}+\sum_{n=2}^{\infty}(-\beta)^n\left[-\sum_{j=-n}^{\infty}\epsilon^j\frac{F_{n+j,0}}{n}+\sum_{j=0}^{\infty}\sum_{k=0}^{\infty}\epsilon^{j+k}F^{\mu}_{j,k+n}I_{n,k+n}\right], \tag{3.5}$$

where the complete process for the first double sum over $j$ and $k$ is

$$\beta\sum_{j=0}^{\infty}\sum_{k=0}^{\infty}\epsilon^{j+k-1}F^{\mu}_{j,k}=\beta\sum_{j=0}^{\infty}\sum_{k=j-1}^{\infty}\epsilon^kF^{\mu}_{j,k-j+1}=\beta\sum_{k=-1}^{\infty}\epsilon^k\sum_{j=0}^{k+1}F^{\mu}_{j,k-j+1}. \tag{3.6}$$

Now we can easily discard the positive powers of $\epsilon$. In the first double sum over $k$ and $j$ only the cases $k=-1$ and $k=0$ have negative powers. In the second double sum over $n$ and $j$ we keep only the cases $j\leq 0$, hence we truncate the sum over $j$ at 0. Finally, in the triple sum over $n$, $j$ and $k$ we only have a contribution from the $j=k=0$ case. This leads to

$$\beta_0\delta A_0=\beta\left[\frac{F^{\mu}_{0,0}}{\epsilon}+F^{\mu}_{0,1}+F_{1,0}\right]+\sum_{n=2}^{\infty}(-\beta)^n\left[-\sum_{j=0}^{n}\epsilon^{-j}\frac{F_{n-j,0}}{n}+(-1)^n\Gamma(n)F^{\mu}_{0,n}\right], \tag{3.7}$$

where we took $j\mapsto -j$ –which reverses the sum limits– and used the last property $I_{n,n}=(-1)^n\Gamma(n)$. To completely write $A_0$ as a power series in $1/\epsilon$ we need to work on the remaining double sum over $n$ and $j$:

$$\begin{aligned}\sum_{n=2}^{\infty}\sum_{j=0}^{n}(-\beta)^n\epsilon^{-j}\frac{F_{n-j,0}}{n} &= \sum_{j=0}^{\infty}\sum_{n=\max\{2,j\}}^{\infty}(-\beta)^n\,\epsilon^{-j}\frac{F_{n-j,0}}{n} \\ &= \sum_{n=2}^{\infty}(-\beta)^n\,\frac{F_{n,0}}{n}+\frac{1}{\epsilon}\sum_{n=2}^{\infty}(-\beta)^n\,\frac{F_{n-1,0}}{n} \\ &\quad +\sum_{j=2}^{\infty}\epsilon^{-j}\sum_{n=j}^{\infty}(-\beta)^n\,\frac{F_{n-j,0}}{n},\end{aligned} \tag{3.8}$$

where in the second step made the $j=0$ and $j=1$ cases explicit to combine them with those already present in $A_0$. The maximum condition in the first line often appears when inverting the order of two sums, and arises from the fact that the functional relation between the variables of two nested sums may not have a well defined inverse. Figure 3.1 depicts the case in (3.8). The blue dots represent the points in the $(n,j)$ plane that are included in the double sum. When considering $j=j(n)$, summation occurs along the vertical arrows: each value of $n$ fixes the range of $j$. To invert the order of the sums one considers $n=n(j)$ and then summation follows the horizontal arrows. One can see that for $j=0,1$, $n$ starts from 2 while for $j\geq 2$ $n$ starts at $j$. The function $\max\{2,j\}$ encompasses these two situations.

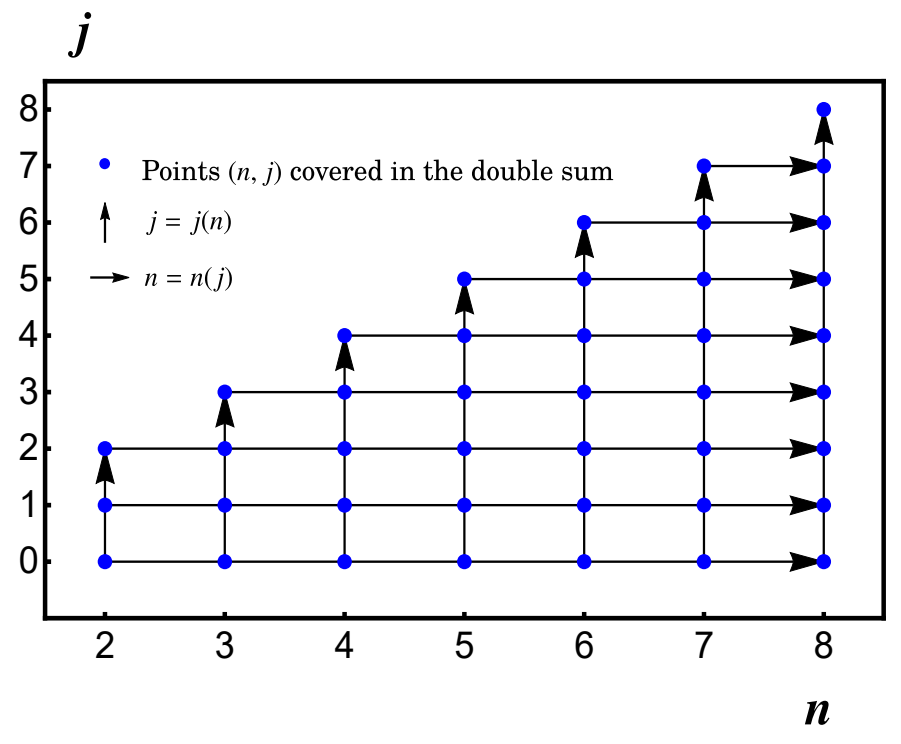

**Figure 3.1.** Points in the $(n,j)$ plane included in the double sum in (3.8). Vertical arrows indicate the order in which points are crossed when the sum over $j$ is nested inside the sum over $n$ and horizontal lines indicate the opposite situation.

Plugging (3.8) into (3.7) and grouping terms according to powers of $\epsilon$ we find

$$\begin{aligned}\beta_0\delta A_0 &= \sum_{n=1}^{\infty}(-\beta)^n\left[(-1)^n\Gamma(n)F^{\mu}_{0,n}-\frac{F_{n,0}}{n}\right]-\sum_{j=1}^{\infty}\frac{1}{\epsilon^j}\sum_{n=j}^{\infty}(-\beta)^n\,\frac{F_{n-j,0}}{n} \\ &= \sum_{n=1}^{\infty}\beta^n\left[\Gamma(n)F^{\mu}_{0,n}-\frac{(-1)^nF_{n,0}}{n}\right]-\sum_{j=1}^{\infty}\frac{1}{\epsilon^j}\sum_{n=0}^{\infty}(-\beta)^{n+j}\,\frac{F_{n,0}}{n+j},\end{aligned} \tag{3.9}$$

where we extended the sum over $n$ ($j$) in the $\epsilon$-independent (dependent) term to start from $n=1$ ($j=1$) so that it absorbs the free $F_{j,k}$ coefficients (the $1/\epsilon$ coefficients) in (3.7). In the final step we relabeled $n\mapsto n+j$ in the double sum.

We can now start to see from (3.9) the advantages of all the manipulations we carried out. For starters, the UV divergences are explicit and confined in the last sum, making the renormalization of $A_0$ trivial. More importantly, $A_0$ is in general an asymptotic series, and in (3.10) we see its divergent behavior has been confined into its first sum over $n$, which is factorially divergent through the function $\Gamma(n)$. On top of this, we see that this divergent behavior is solely determined by the function $F^\mu(0,u)$, since the sum only contains the $F^\mu_{0,n}$ coefficients. The other two sums over $n$ are indeed convergent, since $\beta^n/n\mapsto 0$ as $n\mapsto\infty$ faster that $\epsilon^n$ does in the expansion in (3.1). We also see the non-divergent part of $A_0$, as well as its $1/\epsilon$ structure, is given by the function $F(\epsilon,0)$.

By defining $A_0\equiv Z_A A$ and $\delta Z_A\equiv 1-Z_A$ we have, in the $\overline{\text{MS}}$ scheme:

$$\beta_0\delta A=\sum_{n=1}^{\infty}\left[\beta^n\Gamma(n)F^\mu_{0,n}-\frac{(-\beta)^n}{n}F_{n,0}\right]. \tag{3.10}$$

$$\beta_0\delta Z_A=-\sum_{j=1}^{\infty}\frac{1}{\epsilon^j}\sum_{n=0}^{\infty}(-\beta)^{n+j}\frac{F_{n,0}}{n+j}.$$

At this point we have the option of using the relation (B.21) to pull out the $\mu$-dependence of the $F^\mu_{0,i}$ coefficients and write them in terms of the $F_{0,i}$. This, however, would include a double sum that complicates the task of finding a closed form for $A_0$, so for now we stick with (3.10) and relegate this computation to section 3.2.3 as a consistency check.

## 3.2 Closed form

### 3.2.1 Computation

The goal now is to find a closed –that is, resummed– form for (3.10). Or course, a true closed form as a combination of usual functions is simply not possible to achieve, since, as mentioned, $A$ is in general a divergent series. A second idea would be to consider particular cases, work out the $F^\mu(\epsilon,u)$ functions and try to sum up the convergent terms for each specific case. However, the specific $F^\mu(\epsilon,u)$ functions that we shall find through the applications below are complicated enough that even the $F_{n,0}$ coefficients have a complex dependence on $n$ and this plan becomes inefficient.

Instead, what we can do to work on all three terms in $A$ not loosing generality on the $F^{\mu}_{i,j}$ coefficients is to write each term in (3.10) as an integral over the function $F^{\mu}(\epsilon, u)$. This can be reasoned in two steps. First, the factors accompanying $F^{\mu}_{0,n}$ and $F_{n,0}$ inside the sums over $n$, i.e., $\beta^n\Gamma(n)$ and $(-\beta)^n/n$, admit integral representations, which are derived in Appendix B.2. Second, the function $F^{\mu}(\epsilon, u)$ appears then from using the regularity condition (3.1) from left to right to solve the sums over $n$. Appendix B.3 contains the relevant results for this last step.

Explicitly, using the mentioned results we find for each term:

$$\sum_{n=1}^{\infty}\beta^n\Gamma(n)F^{\mu}_{0,n} = \sum_{n=1}^{\infty}F^{\mu}_{0,n}\int_0^{\infty}\mathrm{d}\tau\tau^{n-1}\mathrm{e}^{-\tau/\beta} = \int_0^{\infty}\mathrm{d}\tau\,\mathrm{e}^{-\tau/\beta}\frac{F^{\mu}(0,\tau)-F(0,0)}{\tau}, \tag{3.11}$$
$$-\sum_{n=1}^{\infty}\frac{(-\beta)^n}{n}F_{n,0} = \sum_{n=1}^{\infty}F_{n,0}\int_{-\beta}^{0}\mathrm{d}\tau\tau^{n-1} = \int_{-\beta}^{0}\mathrm{d}\tau\,\frac{F(\tau,0)-F(0,0)}{\tau},$$
$$\sum_{n=0}^{\infty}\frac{(-\beta)^{n+j}}{n+j}F_{n,0} = \sum_{n=0}^{\infty}F_{n,0}\int_{-\beta}^{0}\mathrm{d}\tau\tau^{n+j-1} = \int_{-\beta}^{0}\mathrm{d}\tau\tau^{j-1}F(\tau,0).$$

This leads to the closed form for $A$ and $Z_A$:

$$\beta_0\delta A = \int_0^{\infty}\mathrm{d}\tau\mathrm{e}^{-\tau/\beta}\,\frac{F^{\mu}(0,\tau)-F(0,0)}{\tau} + \int_{-\beta}^{0}\mathrm{d}\tau\,\frac{F(\tau,0)-F(0,0)}{\tau}, \tag{3.12}$$
$$\beta_0\delta Z_A = -\sum_{j=1}^{\infty}\frac{1}{\epsilon^j}\int_{-\beta}^{0}\mathrm{d}\tau\tau^{j-1}F(\tau,0) = -\int_{-\beta}^{0}\mathrm{d}\tau\,\frac{F(\tau,0)}{\epsilon-\tau},$$

where in $Z_A$ we resummed the $\epsilon$-dependence with the geometric series

$$\sum_{j=1}^{\infty}\frac{\tau^{j-1}}{\epsilon^j} = \frac{1}{\epsilon-\tau}. \tag{3.13}$$

Result (3.12) calls for some analysis. In place of the factorially divergent term in $A$ we obtained an integral of the function $F^{\mu}(0,u)$ over the positive real axis, with a damping exponential factor, much as the inverse Borel integral defined for asymptotic series. We expect the factorial growth of the original series gets translated into the integral by the presence of poles $F^{\mu}(0,u)$ along the integration path.

In the second integral of $A$ and the one in $Z_A$ we integrate the function $F(\epsilon,0)$ along the much smaller path $(-\beta,0)$, so that the poles of $F(\epsilon,0)$, although, as we will see, may exist, are not crossed in the integration. A second solid argument in favor of the convergence of these two integrals will be given in the light of the closed form the anomalous dimension in section 3.3.1.

While the two convergent integrals can always be computed exactly, at least with numerical methods, the inverse-Borel integral is formally divergent. We can still associate a value to such an integral through the so-called principal-value prescription. We will elaborate more on this in Chapter 5, but for the sake of the following discussion we advance that it consists on taking the computation to the complex $\tau$-plane and infinitesimally deforming the integration path to avoid the poles. This deformation is not unique in the sense that the poles can be jumped either in the upper or lower parts of the complex $\tau$ plane, and therefore the computation acquires an uncertainty –or prescription dependence– that is the size of the difference between the two contours. This uncertainty is referred to as the ambiguity of the series, and it is essentially given by the sum of the residues of the integrand at the poles along the integration path.

### 3.2.2 Removing the $\mu$-dependence from the ambiguous integral

Given the way the first integral in (3.12) is written, one may think the size of the ambiguity, i.e., the size of the residues of the first integrand, may depend on the renormalization scale $\mu$. In this section we explicitly show this is not the case.

We start by writing the ambiguous integral as

$$\int_0^\infty \mathrm{d}\tau\, \frac{\mathrm{e}^{-\tau/\beta}}{\tau}\left[\left(\frac{\mu^2}{\omega^2}\right)^\tau F(0,\tau) - F(0,0)\right] = \int_0^\infty \mathrm{d}\tau \mathrm{e}^{-\tau/\beta}\left(\frac{\mu^2}{\omega^2}\right)^\tau \frac{F(0,\tau)-F(0,0)}{\tau} + F(0,0)\int_0^\infty \mathrm{d}\tau\, \frac{\mathrm{e}^{-\tau/\beta}}{\tau}\left[\left(\frac{\mu^2}{\omega^2}\right)^\tau - 1\right]. \tag{3.14}$$

The first integral does not depend on $\mu$ since

$$\mathrm{e}^{-\tau/\beta}\left(\frac{\mu}{\omega}\right)^{2\tau} = \left(\mu \mathrm{e}^{\frac{-1}{2\beta}}\right)^{2\tau}\omega^{-2\tau} = \left(\mu \mathrm{e}^{\frac{-2\pi}{\beta_0\alpha_s}}\right)^{2\tau}\omega^{-2\tau} = \left(\frac{\Lambda_{\mathrm{QCD}}}{\omega}\right)^{2\tau}, \tag{3.15}$$

with $\Lambda_{\mathrm{QCD}}$ defined in (2.20). To compute the last integral we expand its integrand, integrate each term and then sum up the resulting series. The result is

$$\int_0^\infty \mathrm{d}\tau\, \frac{\mathrm{e}^{-\tau/\beta}}{\tau}\left[\left(\frac{\mu^2}{\omega^2}\right)^\tau - 1\right] = -\log\left[1 - \beta\log\left(\frac{\mu^2}{\omega^2}\right)\right] = \log\left(\frac{\beta_\omega}{\beta}\right), \tag{3.16}$$

where in the last step we used (2.31) for simplicity and followed the previously established notation $\beta_x = \beta(x)$ and $\beta_\mu = \beta(\mu) = \beta$. Plugging back these results into (3.12) we find

$$\beta_0\delta A = F(0,0)\log\left(\frac{\beta_\omega}{\beta}\right) + \int_0^\infty \mathrm{d}\tau \left(\frac{\Lambda_{\mathrm{QCD}}}{\omega}\right)^{2\tau}\frac{F(0,\tau)-F(0,0)}{\tau} + \int_{-\beta}^0 \mathrm{d}\tau\, \frac{F(\tau,0)-F(0,0)}{\tau} \tag{3.17}$$

In the form (3.17), the $\mu$-dependence has been extracted from the ambiguous integral and it is clear the ambiguity scales with $(\Lambda_{\rm QCD}/\omega)^{2\tau}$ with no dependence on $\mu$.

### 3.2.3 Alternate derivation: $\mu$-dependence in the perturbative series

We finish by taking the alternate path mentioned below (3.10): writing the perturbative expression in terms of the $F_{i,0}$ and $F_{0,i}$ coefficients and find a closed form involving only the function $F(\epsilon,u)$. Only the first term in (3.10) needs to be modified, and with the relation (B.21) we find

$$\begin{aligned}\sum_{n=1}^{\infty}\beta^n\Gamma(n)F^{\mu}_{0,n} &= \sum_{n=1}^{\infty}\beta^n\Gamma(n)\sum_{i=0}^{n}\frac{1}{i!}\log^i\!\left(\frac{\mu^2}{\omega^2}\right)F_{0,n-i} \\ &= \sum_{i=0}^{\infty}\frac{1}{i!}\log^i\!\left(\frac{\mu^2}{\omega^2}\right)\sum_{n=\max\{1,i\}}^{\infty}\beta^n\Gamma(n)F_{0,n-i} \\ &= \sum_{n=1}^{\infty}\beta^n\Gamma(n)F_{0,n}+\sum_{i=1}^{\infty}\frac{1}{i!}\log^i\!\left(\frac{\mu^2}{\omega^2}\right)\sum_{n=i}^{\infty}\beta^n\Gamma(n)F_{0,n-i} \\ &= \sum_{n=1}^{\infty}\beta^n\Gamma(n)F_{0,n}+\sum_{i=1}^{\infty}\frac{1}{i!}\log^i\!\left(\frac{\mu^2}{\omega^2}\right)\sum_{n=0}^{\infty}\beta^{n+i}\Gamma(n+i)F_{0,n} \\ &= \int_0^{\infty}\mathrm{d}\tau\,\mathrm{e}^{-\tau/\beta}\frac{F(0,\tau)-F(0,0)}{\tau} \\ &\quad+\sum_{i=1}^{\infty}\frac{1}{i!}\log^i\!\left(\frac{\mu^2}{\omega^2}\right)\int_0^{\infty}\mathrm{d}\tau\,\tau^{i-1}\mathrm{e}^{-\tau/\beta}F(0,\tau) \\ &= \int_0^{\infty}\mathrm{d}\tau\,\mathrm{e}^{-\tau/\beta}\frac{F(0,\tau)-F(0,0)}{\tau}+\int_0^{\infty}\mathrm{d}\tau\,\mathrm{e}^{-\tau/\beta}\left[\left(\frac{\mu^2}{\omega^2}\right)^{\tau}-1\right]\frac{F(0,\tau)}{\tau},\end{aligned} \tag{3.18}$$

where we interchanged the sums, split the $i=0$ case, relabeled $n\mapsto n+i$ in the second sum over $n$ and used the second integral representation in (B.12) and the expansion for $F(0,u)$ in (B.24). The last sum over $i$ corresponds to the exponential expansion and leads to the term in squared brackets. Note that the subtractions in both integrals are regularizing the divergences at $\tau=0$ and thus cannot be split. What we can do, however, is to combine them by adding and subtracting a $\tau=0$ term such as

$$\begin{aligned}\sum_{n=1}^{\infty}\beta^n\Gamma(n)F^{\mu}_{0,n} &= \int_0^{\infty}\mathrm{d}\tau\,\mathrm{e}^{-\tau/\beta}\frac{F(0,\tau)-F(0,0)}{\tau} \\ &\quad+\int_0^{\infty}\mathrm{d}\tau\,\mathrm{e}^{-\tau/\beta}\left[\left(\frac{\mu^2}{\omega^2}\right)^{\tau}-1\right]\frac{F(0,\tau)-F(0,0)}{\tau} \\ &\quad+F(0,0)\int_0^{\infty}\mathrm{d}\tau\frac{\mathrm{e}^{-\tau/\beta}}{\tau}\left[\left(\frac{\mu^2}{\omega^2}\right)^{\tau}-1\right]\end{aligned} \tag{3.19}$$

$$= \int_0^\infty d\tau e^{-\tau/\beta}\left(\frac{\mu^2}{\omega^2}\right)^\tau \frac{F(0,\tau)-F(0,0)}{\tau}$$
$$+F(0,0)\int_0^\infty d\tau \frac{e^{-\tau/\beta}}{\tau}\left[\left(\frac{\mu^2}{\omega^2}\right)^\tau - 1\right]$$

As discussed before, the first of these integrals does not depend on $\mu$ but on $\Lambda_{\text{QCD}}$. The second integral turns out to be (3.16) so we finally write

$$\sum_{n=1}^\infty \beta^n \Gamma(n) F^\mu_{0,n} = F(0,0)\log\left(\frac{\beta_\omega}{\beta}\right) + \int_0^\infty d\tau \left(\frac{\Lambda_{\text{QCD}}}{\omega}\right)^{2\tau} \frac{F(0,\tau)-F(0,0)}{\tau}, \qquad (3.20)$$

which upon the addition of the second term in (3.11) exactly agrees with (3.17).

## 3.3 Anomalous dimension and running

### 3.3.1 Anomalous dimension

The renormalized series $A(\mu)$ has acquired a running in $\mu$ given by the anomalous dimension in (2.61). We can work out the derivative of $A$ or $Z_A$ either in their perturbative series or in their closed integral forms. We shall work all four cases, but again we regard the demonstration through $Z_A$ as a consistency check and we postpone it to the next subsection. For the perturbative form of $A$ we find

$$\begin{aligned}
\beta_0\gamma(\alpha_s) = \mu\frac{d}{d\mu}[\beta_0\delta A] &= -2\beta^2\sum_{n=1}^\infty n\beta^{n-1}\left[\Gamma(n)F^\mu_{0,n} - \frac{(-1)^n}{n}F_{n,0}\right] \\
&\quad +\sum_{n=1}^\infty \beta^n\Gamma(n)\mu\frac{d}{d\mu}F^\mu_{0,n} \\
&= -2\beta^2\sum_{n=1}^\infty n\beta^{n-1}\left[\Gamma(n)F^\mu_{0,n} - \frac{(-1)^n}{n}F_{n,0}\right] \\
&\quad +\sum_{n=1}^\infty \beta^n\Gamma(n)2F^\mu_{0,n-1} \\
&= -2\sum_{n=1}^\infty \beta^{n+1}\Gamma(n+1)F^\mu_{0,n} + 2\beta\sum_{n=1}^\infty(-\beta)^n F_{n,0} \\
&\quad +2\sum_{n=0}^\infty \beta^{n+1}\Gamma(n+1)F^\mu_{0,n} \\
&= 2\beta\sum_{n=1}^\infty(-\beta)^n F_{n,0} + 2\beta F_{0,0} = 2\beta\sum_{n=0}^\infty(-\beta)^n F_{n,0},
\end{aligned} \qquad (3.21)$$

where we used result (2.28) with $\epsilon=0$ to take the derivative of $\beta$ and result (B.22) to take the derivative of $F^{\mu}_{0,n}$. With regards to the manipulation of the sums, we employed $n\Gamma(n)=\Gamma(n+1)$ in the first sum over $n$ and took $n\mapsto n+1$ in the third one. Finally, we observed their subtraction simply gives the $n=0$ case $2\beta F_{0,0}$ and included it in the second sum.

One can also take the $\mu$-derivative on the closed form (3.17), where the contribution from the ambiguous integral vanishes and the $\mu$-dependent term is

$$\begin{aligned}\mu\frac{\mathrm{d}}{\mathrm{d}\mu}[\beta_0\delta A] &= F(0,0)\mu\frac{\mathrm{d}}{\mathrm{d}\mu}\left[\log\left(\frac{\beta_\omega}{\beta}\right)\right]+\mu\frac{\mathrm{d}}{\mathrm{d}\mu}\left[\int_{-\beta}^{0}\mathrm{d}\tau\,\frac{F(\tau,0)-F(0,0)}{\tau}\right] \\ &= -\frac{F(0,0)}{\beta}\mu\frac{\mathrm{d}}{\mathrm{d}\mu}\beta+\frac{F(-\beta,0)-F(0,0)}{(-\beta)}\mu\frac{\mathrm{d}}{\mathrm{d}\mu}\beta \\ &= 2\beta F(0,0)+2\beta[F(-\beta,0)-F(0,0)]=2\beta F(-\beta,0),\end{aligned} \tag{3.22}$$

where to take the derivative of the integral we used Leibnitz's theorem in (A.22). Of course (3.22) can directly be obtained by summing up (3.21) with the regularity expansion (3.1).

Collecting the results, our final expression for the anomalous dimension is

$$\gamma_A(\alpha_s) = \frac{2\beta}{\beta_0}\sum_{n=0}^{\infty}(-\beta)^n F_{n,0}=\frac{2\beta}{\beta_0}F(-\beta,0), \tag{3.23}$$

where we already employed the non-cusp notation $\gamma_A(\alpha_s)$ since the expressions clearly reveal there is no explicit dependence on $\mu$. This proves indeed the equivalence between the regularity condition of $F(\epsilon,u)$ and the non-cusp character of the series.

Finally, we observe the closed form for $\gamma_A$ is not an integral representation, but instead the function $F(\epsilon,u)$ evaluated at $\epsilon=-\beta$ and $u=0$. Since the function is regular at $u=0$, the pole of $F(\epsilon,0)$ that is closest to the origin and lays in the negative real axis sets the convergence radius of $\gamma(\alpha_s)$.

### 3.3.2 Solution to the RGE equation

Since we completely know the anomalous dimension at $\beta_0$LO we can work out the $\mu$-running of $A(\mu)$. Using the definition of $\gamma_A$ in (2.57), integrating both sides from $\mu_1$ to $\mu_2$ and using (2.28) $\mathrm{d}\mu$ into $\mathrm{d}\beta$, the RGE takes the form and solution

$$\begin{aligned}A(\mu_2)-A(\mu_1) &= -\frac{1}{2}\int_{\beta_{\mu_1}}^{\beta_{\mu_2}}\mathrm{d}\beta\frac{\gamma_A(\beta)}{\beta^2}=-\frac{1}{\beta_0}\int_{\beta_{\mu_1}}^{\beta_{\mu_2}}\mathrm{d}\beta\frac{F(-\beta,0)}{\beta}, \\ &= -\frac{F_{0,0}}{\beta_0}\log\left(\frac{\beta_{\mu_2}}{\beta_{\mu_1}}\right)-\frac{1}{\beta_0}\sum_{n=1}^{\infty}(-1)^n F_{n,0}\frac{\beta^n_{\mu_2}-\beta^n_{\mu_1}}{n}.\end{aligned} \tag{3.24}$$

In the first line we show the closed integral form and in the second the perturbative form, which is obtained by expanding $F(-\beta,0)=\sum_{n=0}^{\infty}(-\beta)^n F_{n,0}$ and integrating each term.

The expression clearly shows the difference between two series is non-ambiguous, or equivalently, renormalon-free. Given the fact that $\gamma_A$ is not ambiguous, provided the values of $\beta$ are not too large, the sum can be carried out to arbitrarily high orders. This approach, employed also in Ref. [18], provides an effective way of computing the numerical integral in the first line to arbitrary precision. Furthermore, the first term in the last line corresponds to LL accuracy, while adding $n$ additional terms from the sum yields N$^n$LL accuracy.

### 3.3.3 Alternative derivation: obtaining $\gamma$ from $Z$

Here we present the alternative derivation for $\gamma_A$ based on the last equality of (2.61). This derivation explicitly shows the cancellation of $1/\epsilon$ poles, rendering an UV finite $\gamma_A$. Taking the derivative of the perturbative form we get

$$\begin{aligned}
-\mu\frac{\mathrm{d}}{\mathrm{d}\mu}[\beta_0\delta Z_A] &= -\sum_{j=1}^{\infty}\frac{1}{\epsilon^j}\sum_{n=0}^{\infty}(-1)^{n+j}(n+j)\beta^{n+j-1}[-2\beta(\epsilon+\beta)]\frac{F_{n,0}}{n+j} \\
&= 2\sum_{j=1}^{\infty}\frac{1}{\epsilon^j}\sum_{n=0}^{\infty}(-\beta)^{n+j}(\epsilon+\beta)F_{n,0}=2\sum_{n=0}^{\infty}(-\beta)^{n+1}F_{n,0},
\end{aligned} \tag{3.25}$$

where we used we used the full relation (2.28) to take the derivative of $\beta$ and only kept the finite term $j=1$ since the anomalous dimension is finite. Indeed, the infinite terms are

$$\begin{aligned}
&2\sum_{j=2}^{\infty}\frac{1}{\epsilon^{j-1}}\sum_{n=0}^{\infty}(-\beta)^{n+j}F_{n,0}+2\beta\sum_{j=1}^{\infty}\frac{1}{\epsilon^j}\sum_{n=0}^{\infty}(-\beta)^{n+j}F_{n,0} \\
=\; &-2\beta\sum_{j=1}^{\infty}\frac{1}{\epsilon^j}\sum_{n=0}^{\infty}(-\beta)^{n+j}F_{n,0}+2\beta\sum_{j=1}^{\infty}\frac{1}{\epsilon^j}\sum_{n=0}^{\infty}(-\beta)^{n+j}F_{n,0}=0.
\end{aligned} \tag{3.26}$$

Analogously, if one takes the derivative of the closed form finds

$$\mu\frac{\mathrm{d}}{\mathrm{d}\mu}\int_{-\beta}^{0}\mathrm{d}\tau\,\frac{F(\tau,0)}{\epsilon-\tau}=-\frac{F(\tau,0)}{\epsilon+\beta}\mu\frac{\mathrm{d}}{\mathrm{d}\mu}(-\beta)=2\beta F(\tau,0), \tag{3.27}$$

where the dependence on $\epsilon$ of the integrand is explicitly canceled by the derivative of $\beta$. Both expressions in (3.25) and (3.27) agree with (3.21) and (3.22).

## 3.4 Relations for the perturbative coefficients

Our perturbative result (3.10) allows to compute the renormalized series $A(\mu)$ as a perturbative series in $\beta$ at $\beta_0$LO. However, taking a closer look at the expression, one can easily see that the coefficients of this perturbative expansion are $\mu$-dependent through the $F_{0,n}^{\mu}$ coefficients, which contain powers of $L\equiv\log(\mu^2/\omega^2)$. More specifically, since $A(\mu)$ is a non-cusp series, its explicit dependence on $\mu$ takes the form of powers of $L$ that can run up to $L^n$ at each order $\beta^n$:

$$\beta_0\delta A(\mu)\equiv\sum_{n=1}^{\infty}\beta^n\sum_{i=0}^{n}c_{n,i}L^i. \tag{3.28}$$

In this last section we compare our expression (3.10) against the general form (3.28) in order to extract the scale-independent, purely numeric coefficients $c_{n,i}$. We present expressions for such coefficients in terms of $F_{n,0}$ and $F_{0,n}$, which arise from the expansions of the $F(\epsilon,u)$ function around $\epsilon=0$ or $u=0$. We also develop a recursive relation for the $c_{n,i}$, which results useful whenever $F(\epsilon,u)$ proves hard to expand. On the topic of the $F(\epsilon,u)$ function, we anticipate that in general it can take a complicated mathematical form, and therefore Taylor expanding it without additional simplifications can be very inefficient in terms of computational resources. Because of this, in each of the applications we took the time to work on $F(\epsilon,0)$ and $F(0,u)$ to bring them to forms that can be easily expanded.

### 3.4.1 Renormalized series

First, the non-logarithmic coefficients ($i=0$) can be obtained by simply setting $\mu=\omega$ in (3.10), which leads to $F_{0,n}^{\omega}=F_{0,n}$ and

$$c_{n,0}=\Gamma(n)F_{0,n}-\frac{(-1)^n}{n}F_{n,0}. \tag{3.29}$$

To obtain the remaining coefficients we take the $\mu$-derivative of (3.28). The left-hand side simply gives the anomalous dimension (3.23) and the right hand side leads to

$$\begin{aligned}\mu\frac{\mathrm{d}}{\mathrm{d}\mu}\left[\sum_{n=1}^{\infty}\beta^n\sum_{i=0}^{n}c_{n,i}L^i\right] &= \sum_{n=1}^{\infty}n\beta^{n-1}(-2\beta^2)\sum_{i=0}^{n}c_{n,i}L^i+\sum_{n=1}^{\infty}\beta^n\sum_{i=1}^{n}2i\,c_{n,i}L^{i-1} \\ &= -2\sum_{n=1}^{\infty}n\beta^{n+1}\sum_{i=0}^{n}c_{n,i}L^i+2\sum_{n=0}^{\infty}\beta^{n+1}\sum_{i=0}^{n}(i+1)\,c_{n+1,i+1}L^i \\ &= 2\beta\sum_{n=0}^{\infty}\beta^n\sum_{i=0}^{n}[(i+1)\,c_{n+1,i+1}-n\,c_{n,i}]L^i.\end{aligned} \tag{3.30}$$

where we used $\mu \mathrm{d}L/(\mathrm{d}L)=2$, realized the $i=0$ case vanishes when taking the derivative of $L$, then took $i\mapsto i+1$ and $n\mapsto n+1$ at the same time in the second sum, extended the first sum to start from $n=0$ since the contribution vanishes and combined both sums. Comparing powers of $\beta$ with (3.23) we obtain

$$(-1)^n F_{n,0}=\sum_{i=0}^{n}[(i+1)\,c_{n+1,i+1}-n\,c_{n,i}]L^i. \tag{3.31}$$

Taking a look now at the powers of $L$, we see that, since $\gamma_A$ is free of logarithms all the terms with $i\geq 1$ must vanish individually. Hence we get the two relations

$$\begin{aligned}(-1)^n F_{n,0}&=c_{n+1,1}-n\,c_{n,0},\\ 0&=(i+1)\,c_{n+1,i+1}-n\,c_{n,i},\quad i\geq 1,\end{aligned} \tag{3.32}$$

which can be best written as

$$\begin{aligned}c_{n,1}&=(n-1)\,c_{n-1,0}-(-1)^n F_{n-1,0},\\ c_{n,i}&=\frac{n-1}{i}c_{n-1,i-1},\qquad i\geq 2,\end{aligned} \tag{3.33}$$

where the latter is restricted to $i\geq 2$. Each of these relations can be further simplified. In $c_{n,1}$ we can use (3.29) and get

$$c_{n,1}=\Gamma(n)F_{0,n-1}. \tag{3.34}$$

The relation for $c_{n,i}$ can be simplified by using it recursively until the base case is reached. For arbitrary $c_{n,i}$ $(n\geq i)$ each step lowers the indices by 1 until the $i=1$ case is reached in the right hand side, so the number of steps is $i-1$. This gives

$$c_{n,i}=\frac{(n-1)!}{i!(n-i)!}c_{n-i+1,1}=\frac{\Gamma(n)}{\Gamma(i-1)}F_{0,n-i}, \tag{3.35}$$

where we have written $(n-1)(n-2)...(n-i+1)=(n-1)!/(n-i)!$ and used (3.34). Note that when using (3.34) the relation automatically includes the $j=1$ case, and hence the last equality of (3.35) contains (3.34).

Relations (3.29) and (3.35) allow to build the perturbative series $A(\mu)$ in an efficient fashion from the $F_{n,0}$ and $F_{0,n}$ coefficients. Whenever the computation of this coefficients proves difficult, one can equivalently use the recursive relations in (3.33) to accelerate the process.

### 3.4.2 Anomalous dimension

For completeness, we also provide an analogous relation for $\gamma_n^A$. The anomalous dimension does not contain logarithms and therefore its parameter-independent coefficients are just those of $\beta$. Comparing our perturbative definition in (2.57) against the expansion in (3.23) we get

$$\gamma = \sum_{n=0}^{\infty} \gamma_n^A \left(\frac{\alpha_s}{4\pi}\right)^{n+1} = \frac{1}{\beta_0} \sum_{i=0}^{\infty} \frac{\gamma_n^A}{\beta_0^n} \beta^{n+1} = \frac{2}{\beta_0} \sum_{i=0}^{\infty} \beta^{n+1} (-1)^n F_{n,0}, \tag{3.36}$$

so that the parameter-independent coefficients are simply

$$\hat{\gamma}_n^A \equiv \frac{\gamma_n^A}{\beta_0^n} = 2(-1)^n F_{n,0}. \tag{3.37}$$

# Chapter 4

# Series with cusp-anomalous dimension

## 4.1 Perturbative form

### 4.1.1 The $G(\epsilon, u)$ function

As already discussed, cases with regular $F(\epsilon, u)$ correspond to perturbative series with no cusp-anomalous dimension. In this section we explore the case where $F(\epsilon, u)$ is regular at $\epsilon \mapsto 0$ but presents a simple pole when $u \mapsto 0$. First we define

$$G^{\mu}(\epsilon, u) \equiv u F^{\mu}(\epsilon, u), \tag{4.1}$$

which enters $A_0$ as

$$\beta_0 \delta A_0 = \sum_{n=1}^{\infty} (-\beta)^n \epsilon^{-n} \sum_{i=1}^{n} \frac{(-1)^i \Gamma(n)}{\Gamma(i)\Gamma(n-i+1)} \frac{G^{\mu}(\epsilon, i\epsilon)}{i^2 \epsilon}, \tag{4.2}$$

in analogy with (2.52). The presence of an extra $i\epsilon$ alters the manipulations carried out throughout section 3.1, so we must repeat them now.

### 4.1.2 Form for regular $G(\epsilon, u)$

Expanding

$$G^{\mu}(\epsilon, u) \equiv \sum_{i=0}^{\infty} \sum_{j=0}^{\infty} G^{\mu}_{i,j} \epsilon^i u^j, \qquad G^{\mu}_{i,0} \equiv G_{i,0}, \tag{4.3}$$

we find

$$\begin{aligned}\beta_0\delta A_0 &= \sum_{n=1}^{\infty}(-\beta)^n\epsilon^{-n}\sum_{i=1}^{n}\frac{(-1)^i\Gamma(n)}{\Gamma(i)\Gamma(n-i+1)}\frac{1}{i^2\epsilon}\sum_{j=0}^{\infty}\sum_{k=0}^{\infty}G^{\mu}_{j,k}\epsilon^j(i\epsilon)^k \\ &= \sum_{n=1}^{\infty}(-\beta)^n\sum_{j=0}^{\infty}\sum_{k=0}^{\infty}\epsilon^{j+k-n-1}G^{\mu}_{j,k}\sum_{i=1}^{n}\frac{(-1)^i\Gamma(n)}{\Gamma(i+1)\Gamma(n-i+1)}i^{k-1}.\end{aligned} \tag{4.4}$$

The sum over $i$ corresponds to $I_{n,k-1}$ in (3.3) with the extended coefficient[4.1]

$$I_{n,-1} = -\frac{H_n}{n}, \tag{4.5}$$

for any value of $n$. Here, $H_i$ are the Harmonic numbers defined in (A.14). To use the properties we single out the $n=1$ case, for which $I_{1,k}=-1$ and then in the sum starting from $n=2$ we also pull out the $k=0,1$ cases, corresponding to $I_{n,-1}=-H_n/n$ and $I_{n,0}=-1/n$ respectively, and discard the $1<k<n+1$ cases, for which $I_{n,k-1}=0$. We then find

$$\begin{aligned}\beta_0\delta A_0 &= \sum_{n=1}^{\infty}(-\beta)^n\sum_{j=0}^{\infty}\sum_{k=0}^{\infty}\epsilon^{j+k-n-1}G^{\mu}_{j,k}I_{n,k-1} \\ &= \beta\sum_{j=0}^{\infty}\sum_{k=0}^{\infty}\epsilon^{j+k-2}G^{\mu}_{j,k}-\sum_{n=2}^{\infty}(-\beta)^n\sum_{j=0}^{\infty}\left[\frac{H_n}{n}\epsilon^{j-n-1}G_{j,0}+\frac{1}{n}\epsilon^{j-n}G^{\mu}_{j,1}\right] \\ &\quad+\sum_{n=2}^{\infty}(-\beta)^n\sum_{j=0}^{\infty}\sum_{k=n+1}^{\infty}\epsilon^{j+k-n-1}G^{\mu}_{j,k}I_{n,k-1}.\end{aligned} \tag{4.6}$$

The next step is to discard the positive powers of $\epsilon$. In the first term we can explicitly write all the non-positive cases since they must satisfy $j+k\leq 2$

$$\beta\sum_{j=0}^{\infty}\sum_{k=0}^{\infty}\epsilon^{j+k-2}G^{\mu}_{j,k}=\beta\left[\frac{G_{0,0}}{\epsilon^2}+\frac{G_{1,0}+G^{\mu}_{0,1}}{\epsilon}+G_{2,0}+G^{\mu}_{0,2}+G^{\mu}_{1,1}\right]. \tag{4.7}$$

In the second term the condition for non-positive powers is $j\leq n+1$ in the first sum over $j$ and $j\leq n$ in the second one, so we simply restrict the sums and reorganize them as

$$\begin{aligned}&\sum_{n=2}^{\infty}(-\beta)^n\sum_{j=0}^{\infty}\left[\frac{H_n}{n}\epsilon^{j-n-1}G_{j,0}+\frac{1}{n}\epsilon^{j-n}G^{\mu}_{j,1}\right] \\ &= \sum_{n=2}^{\infty}\frac{(-\beta)^n}{n}\left[\sum_{j=0}^{n+1}\frac{1}{\epsilon^j}H_nG_{n-j+1,0}+\sum_{j=0}^{n}\frac{1}{\epsilon^j}G^{\mu}_{n-j,1}\right].\end{aligned} \tag{4.8}$$

4.1. The closed forms in of $I_{n,k}$ are properly found and classified in (B.2).

Finally the only non-positive contribution to the last term is for $j=0$ and $k=n+1$, which using $I_{n,n}=(-1)^n\Gamma(n)$ gives the $\epsilon$-independent term

$$\sum_{n=2}^{\infty}(-\beta)^n\sum_{j=0}^{\infty}\sum_{k=n+1}^{\infty}\epsilon^{j+k-n-1}G^{\mu}_{j,k}I_{n,k-1}=\sum_{n=2}^{\infty}(-\beta)^n(-1)^n\Gamma(n)G^{\mu}_{0,n+1}. \tag{4.9}$$

Summing up all three contributions we arrive at

$$\begin{aligned}\beta_0\delta A_0 &= \beta\left[\frac{G_{0,0}}{\epsilon^2}+\frac{G_{1,0}+G^{\mu}_{0,1}}{\epsilon}+G_{2,0}+G^{\mu}_{1,1}\right]+\sum_{n=1}^{\infty}(-\beta)^n(-1)^n\Gamma(n)G^{\mu}_{0,n+1} \\ &\quad -\sum_{n=2}^{\infty}\frac{(-\beta)^n}{n}\left[\sum_{j=0}^{n+1}\frac{1}{\epsilon^j}H_nG_{n-j+1,0}+\sum_{j=0}^{n}\frac{1}{\epsilon^j}G^{\mu}_{n-j,1}\right],\end{aligned} \tag{4.10}$$

where we realized the term $G^{\mu}_{0,2}$ in (4.7) corresponds to the $n=1$ case of the sum in equation (4.9).

### 4.1.3 Form as power series in $\epsilon$

Next we write $A_0$ as a power series in $\epsilon$, and for that we need to work on the last two double sums in (4.10). The first one is

$$\begin{aligned}\sum_{n=2}^{\infty}\sum_{j=0}^{n+1}\frac{(-\beta)^nH_n}{n\epsilon^j}G_{n-j+1,0} &= \sum_{j=0}^{2}\frac{1}{\epsilon^j}\sum_{n=2}^{\infty}\frac{(-\beta)^nH_n}{n}G_{n-j+1,0} \\ &\quad +\sum_{j=3}^{\infty}\frac{1}{\epsilon^j}\sum_{n=j-1}^{\infty}\frac{(-\beta)^nH_n}{n}G_{n-j+1,0} \\ &= \sum_{n=2}^{\infty}\frac{(-\beta)^nH_n}{n}\left[G_{n+1,0}+\frac{G_{n,0}}{\epsilon}+\frac{G_{n-1,0}}{\epsilon^2}\right] \\ &\quad +\sum_{j=3}^{\infty}\frac{1}{\epsilon^j}\sum_{n=0}^{\infty}\frac{(-\beta)^{n+j-1}H_{n+j-1}}{n+j-1}G_{n,0},\end{aligned} \tag{4.11}$$

where in the first line we interchanged the sums and in the last line we took $n\mapsto n+j-1$ in the second term. On the other hand, the second double sum in (4.10) is

$$\begin{aligned}\sum_{n=2}^{\infty}\sum_{j=0}^{n}\frac{(-\beta)^n}{n\epsilon^j}G^{\mu}_{n-j,1} &= \sum_{j=0}^{2}\frac{1}{\epsilon^j}\sum_{n=2}^{\infty}\frac{(-\beta)^n}{n}G^{\mu}_{n-j,1}+\sum_{j=3}^{\infty}\frac{1}{\epsilon^j}\sum_{n=j}^{\infty}\frac{(-\beta)^n}{n}G^{\mu}_{n-j,1} \\ &= \sum_{n=2}^{\infty}\frac{(-\beta)^n}{n}\left[G^{\mu}_{n,1}+\frac{G^{\mu}_{n-1,1}}{\epsilon}+\frac{G^{\mu}_{n-2,1}}{\epsilon^2}\right] \\ &\quad +\sum_{j=3}^{\infty}\frac{1}{\epsilon^j}\sum_{n=0}^{\infty}\frac{(-\beta)^{n+j}}{n+j}G^{\mu}_{n,1},\end{aligned} \tag{4.12}$$

where we interchanged the sums and took $n\mapsto n+j$ in the second term. Note that both in (4.11) and (4.12) we pulled the $j=0,1,2$ terms out of the sum to combine them with those already present in $A_0$. Before we plug these results back into (4.10) let us sum and simplify (4.11) and (4.12). Adding up the single sums gives

$$\sum_{n=2}^{\infty}\frac{(-\beta)^n}{n}\left[H_nG_{n+1,0}+G^{\mu}_{n,1}+\frac{H_nG_{n,0}+G^{\mu}_{n-1,1}}{\epsilon}+\frac{H_nG_{n-1,0}+G^{\mu}_{n-2,1}}{\epsilon^2}\right]. \tag{4.13}$$

Since $G^{\mu}_{-1,1}=0$, the first line in (4.10) can be seen to reproduce exactly the $n=1$ case of the sum in (4.13), so its addition only extends the sum to start from $n=1$. Note that (4.13) enters (4.10) with a global negative sign, and that the first line in (4.10) also has a negative global sign since $\beta=-(-\beta)$. Going back to $A_0$ we obtain

$$\begin{aligned}\beta_0\delta A_0 \;=\; & \sum_{n=1}^{\infty}(-\beta)^n\left[(-1)^n\Gamma(n)G^{\mu}_{0,n+1}-\frac{H_n}{n}G_{n+1,0}-\frac{1}{n}G^{\mu}_{n,1}\right] \\ & -\sum_{n=1}^{\infty}\frac{(-\beta)^n}{n}\left[\frac{H_nG_{n,0}+G^{\mu}_{n-1,1}}{\epsilon}+\frac{H_nG_{n-1,0}+G^{\mu}_{n-2,1}}{\epsilon^2}\right] \\ & -\sum_{j=3}^{\infty}\frac{1}{\epsilon^j}\sum_{n=0}^{\infty}\left[\frac{(-\beta)^{n+j-1}H_{n+j-1}}{n+j-1}G_{n,0}+\frac{(-\beta)^{n+j}}{n+j}G^{\mu}_{n,1}\right].\end{aligned} \tag{4.14}$$

The last simplification consists on realizing that the $1/\epsilon^2$ term is reproduced by taking $j=2$ in the sum over $j$ –this is not the case for the $1/\epsilon$ term and the $j=1$ case–:

$$\begin{aligned}& \frac{1}{\epsilon^2}\sum_{n=0}^{\infty}\left[\frac{(-\beta)^{n+1}H_{n+1}}{n+1}G_{n,0}+\frac{(-\beta)^{n+2}}{n+2}G^{\mu}_{n,1}\right] \\ = & \frac{1}{\epsilon^2}\sum_{n=1}^{\infty}\frac{(-\beta)^nH_n}{n}G_{n-1,0}+\frac{1}{\epsilon^2}\sum_{n=1}^{\infty}\frac{(-\beta)^n}{n}G^{\mu}_{n-2,1} \\ = & \frac{1}{\epsilon^2}\sum_{n=1}^{\infty}\frac{(-\beta)^n}{n}[H_nG_{n-1,0}+G^{\mu}_{n-2,1}].\end{aligned} \tag{4.15}$$

With this, the renormalized series $A$ and its renormalization factor $Z_A$ are, in the $\overline{\mathrm{MS}}$ scheme,

$$\begin{aligned}\beta_0\delta A \;\equiv\; & \sum_{n=1}^{\infty}(-\beta)^n\left[(-1)^n\Gamma(n)G^{\mu}_{0,n+1}-\frac{H_n}{n}G_{n+1,0}-\frac{1}{n}G^{\mu}_{n,1}\right], \\ \beta_0\delta Z_A \;\equiv\; & -\frac{1}{\epsilon}\sum_{n=1}^{\infty}(-\beta)^n\frac{H_nG_{n,0}+G^{\mu}_{n-1,1}}{n} \\ & -\sum_{j=2}^{\infty}\frac{1}{\epsilon^j}\sum_{n=0}^{\infty}\left[\frac{(-\beta)^{n+j-1}H_{n+j-1}}{n+j-1}G_{n,0}+\frac{(-\beta)^{n+j}}{n+j}G^{\mu}_{n,1}\right].\end{aligned} \tag{4.16}$$

## 4.2 Closed form

### 4.2.1 Computation

We now seek to write $A$ and $Z_A$ in a closed integral form, analogous to what we did for series with no cusp-anomalous dimension. This involves finding integral representations for all the series in (4.16), see Appendix B.2. We use results (B.12) and (B.14) for the integral representations of the coefficients accompanying the $G^{\mu}_{i,j}$ and (B.24), (B.26) and (B.28) to sum up the $G^{\mu}_{i,j}$ coefficients themselves. The manipulations for each term in $A$ are

$$\begin{aligned}
\sum_{n=1}^{\infty}\beta^n\Gamma(n)G^{\mu}_{0,n+1} &= \sum_{n=1}^{\infty}G^{\mu}_{0,n+1}\int_0^{\infty}\mathrm{d}\tau\,\mathrm{e}^{-\tau/\beta}\tau^{n-1} = \int_0^{\infty}\mathrm{d}\tau\,\frac{\mathrm{e}^{-\tau/\beta}}{\tau^2}\sum_{n=2}^{\infty}\tau^n G^{\mu}_{0,n} \qquad (4.17)\\
&= \int_0^{\infty}\mathrm{d}\tau\,\mathrm{e}^{-\tau/\beta}\left[\frac{G^{\mu}(0,\tau)-G(0,0)}{\tau^2}-\frac{1}{\tau}\frac{\mathrm{d}}{\mathrm{d}u}G^{\mu}(0,u)\bigg|_{u=0}\right],\\
\sum_{n=1}^{\infty}\frac{(-\beta)^n}{n}H_n G_{n+1,0} &= \sum_{n=1}^{\infty}G_{n+1,0}\int_{-\beta}^{0}\mathrm{d}\tau\,\tau^{n-1}\log\left(1+\frac{\tau}{\beta}\right)\\
&= \int_{-\beta}^{0}\mathrm{d}\tau\,\log\left(1+\frac{\tau}{\beta}\right)\sum_{n=2}^{\infty}\tau^{n-2}G_{n,0}\\
&= \int_{-\beta}^{0}\mathrm{d}\tau\,\log\left(1+\frac{\tau}{\beta}\right)\frac{1}{\tau^2}\left[\sum_{n=1}^{\infty}\tau^n G_{n,0}-\tau G_{1,0}\right]\\
&= \int_{-\beta}^{0}\mathrm{d}\tau\log\left(1+\frac{\tau}{\beta}\right)\left[\frac{G(\tau,0)-G(0,0)}{\tau^2}-\frac{1}{\tau}\frac{\mathrm{d}}{\mathrm{d}\epsilon}G(\epsilon,0)\bigg|_{\epsilon=0}\right]\\
&= \int_{-\beta}^{0}\mathrm{d}\tau\log\left(1+\frac{\tau}{\beta}\right)\frac{G(\tau,0)-G(0,0)}{\tau^2}-\frac{\pi^2}{6}G_{1,0},\\
\sum_{n=1}^{\infty}\frac{(-\beta)^n}{n}G^{\mu}_{n,1} &= -\sum_{n=1}^{\infty}G^{\mu}_{n,1}\int_{-\beta}^{0}\mathrm{d}\tau\,\tau^{n-1} = -\int_{-\beta}^{0}\mathrm{d}\tau\,\frac{1}{\tau}\left[\sum_{n=0}^{\infty}\tau^n G^{\mu}_{n,1}-G^{\mu}_{0,1}\right]\\
&= -\int_{-\beta}^{0}\mathrm{d}\tau\frac{1}{\tau}\frac{\mathrm{d}}{\mathrm{d}u}[G^{\mu}(\tau,u)-G^{\mu}(0,u)]\bigg|_{u=0}.
\end{aligned}$$

In the last line of the computation in the center we realized the the $1/\tau^2$ pole is canceled by the combined effect of the subtraction $G(\tau,0)-G(0,0)$ and the logarithm, which vanishes at $\tau=0$, and thus we integrated the term with $1/\tau$, which is $\beta$-independent as can be seen by changing variables $\tau\to-\tau\beta$:

$$\int_{-\beta}^{0}\frac{\mathrm{d}\tau}{\tau}\log\left(1+\frac{\tau}{\beta}\right)=\frac{\pi^2}{6}. \qquad (4.18)$$

This cannot be done in the first and last computations, where all the subtractions in the squared brackets must be kept in order for the integral to be safe at $\tau=0$. Adding the three contributions the closed form for $A$ is

$$\begin{aligned}\beta_0\delta A &= \frac{\pi^2}{6}G_{1,0}+\int_0^\infty \mathrm{d}\tau\,\mathrm{e}^{-\tau/\beta}\left[\frac{G^\mu(0,\tau)-G(0,0)}{\tau^2}-\frac{1}{\tau}\frac{\mathrm{d}}{\mathrm{d}u}G^\mu(0,u)\bigg|_{u=0}\right] \\ &\quad -\int_{-\beta}^0 \mathrm{d}\tau\left\{\log\left(1+\frac{\tau}{\beta}\right)\frac{G(\tau,0)-G(0,0)}{\tau^2}-\frac{1}{\tau}\frac{\mathrm{d}}{\mathrm{d}u}[G^\mu(\tau,u)-G^\mu(0,u)]\bigg|_{u=0}\right\}.\end{aligned} \tag{4.19}$$

Turning our attention to $Z_A$, the representation of each of its four series is

$$\begin{aligned}\sum_{n=1}^\infty \frac{(-\beta)^n H_n}{n}G_{n,0} &= \int_{-\beta}^0 \mathrm{d}\tau\log\left(1+\frac{\tau}{\beta}\right)\frac{1}{\tau}\sum_{n=1}^\infty \tau^n G_{n,0} \\ &= \int_{-\beta}^0 \mathrm{d}\tau\log\left(1+\frac{\tau}{\beta}\right)\frac{G(\tau,0)-G(0,0)}{\tau} \\ &= \int_{-\beta}^0 \mathrm{d}\tau\log\left(1+\frac{\tau}{\beta}\right)\frac{G(\tau,0)}{\tau}-\frac{\pi^2}{6}G_{0,0}, \\ \sum_{n=1}^\infty \frac{(-\beta)^n}{n}G^\mu_{n-1,1} &= -\sum_{n=1}^\infty\int_{-\beta}^0 \mathrm{d}\tau\,\tau^{n-1}G^\mu_{n-1,1}=-\int_{-\beta}^0\mathrm{d}\tau\sum_{n=0}^\infty \tau^n G^\mu_{n,1} \\ &= -\int_{-\beta}^0 \mathrm{d}\tau\frac{\mathrm{d}}{\mathrm{d}u}G^\mu(\tau,u)\bigg|_{u=0}, \\ \sum_{n=0}^\infty \frac{(-\beta)^{n+j-1}H_{n+j-1}}{n+j-1}G_{n,0} &= \sum_{n=0}^\infty G_{n,0}\int_{-\beta}^0 \mathrm{d}\tau\log\left(1+\frac{\tau}{\beta}\right)\tau^{n+j-1} \\ &= \int_{-\beta}^0 \mathrm{d}\tau\log\left(1+\frac{\tau}{\beta}\right)\tau^{j-1}G(\tau,0), \\ \sum_{n=0}^\infty \frac{(-\beta)^{n+j}}{n+j}G^\mu_{n,1} &= -\sum_{n=0}^\infty G^\mu_{n,1}\int_{-\beta}^0 \mathrm{d}\tau\,\tau^{n+j-1}=\int_{-\beta}^0 \mathrm{d}\tau\,\tau^{j-1}\frac{\mathrm{d}}{\mathrm{d}u}G^\mu(\tau,u)\bigg|_{u=0}.\end{aligned} \tag{4.20}$$

Where again in the first computation we integrated the term with $G(0,0)=G_{0,0}$ due to the logarithm canceling the $1/\tau$ pole. With these results

$$\begin{aligned}\beta_0\delta Z_A &= \frac{\pi^2}{6\epsilon}G_{0,0}-\frac{1}{\epsilon}\int_{-\beta}^0 \mathrm{d}\tau\left[\log\left(1+\frac{\tau}{\beta}\right)\frac{G(\tau,0)}{\tau}-\frac{\mathrm{d}}{\mathrm{d}u}G^\mu(\tau,u)\bigg|_{u=0}\right] \\ &\quad -\sum_{j=2}^\infty\frac{1}{\epsilon^j}\int_{-\beta}^0 \mathrm{d}\tau\,\tau^{j-1}\left[\log\left(1+\frac{\tau}{\beta}\right)-\frac{\mathrm{d}}{\mathrm{d}u}G^\mu(\tau,u)\bigg|_{u=0}\right] \\ &= \frac{\pi^2}{6\epsilon}G_{0,0}+\int_{-\beta}^0\frac{\mathrm{d}\tau}{\tau-\epsilon}\left[\log\left(1+\frac{\tau}{\beta}\right)\frac{G(\tau,0)}{\tau}-\frac{\mathrm{d}}{\mathrm{d}u}G^\mu(\tau,u)\bigg|_{u=0}\right].\end{aligned} \tag{4.21}$$

Here we resummed the $\epsilon$ dependence with

$$\sum_{j=2}^{\infty}\frac{\tau^{j-1}}{\epsilon^{j}}=\frac{\tau}{\epsilon(\epsilon-\tau)}, \tag{4.22}$$

and combined both integrals.

### 4.2.2 Removing the $\mu$-dependence from the integrals

Analogously to (3.12), the closed expression for $A$ in (4.19) presents $\mu$-dependence on the integrand of the ambiguous, Borel integral. In this section we manipulate it to extract the dependence on $\mu$ from the integrand. To this end, we observe the $\mu$-dependence of the integrand arises not only through the $G^{\mu}(0,\tau)$ function but also from its derivative with respect to to $u$. Thus, the first step is to extract the dependence on $\mu$ from the derivative term. In general, we have

$$\begin{aligned}\frac{\mathrm{d}}{\mathrm{d}u}G^{\mu}(\epsilon,u)\bigg|_{u=0} &= \left[\frac{\mathrm{d}}{\mathrm{d}u}G(\epsilon,u)\right]\left(\frac{\mu^2}{\omega^2}\right)^{u}\bigg|_{u=0}+G(\epsilon,u)\left(\frac{\mu^2}{\omega^2}\right)^{u}\log\left(\frac{\mu^2}{\omega^2}\right)\bigg|_{u=0} \\ &= \frac{\mathrm{d}}{\mathrm{d}u}G(\epsilon,u)\bigg|_{u=0}+G(\epsilon,0)\log\left(\frac{\mu^2}{\omega^2}\right),\end{aligned} \tag{4.23}$$

which of course also holds for $\epsilon=0$. Following the same logic, one can also make explicit the $\mu$ dependence of the finite integral:

$$\begin{aligned}\frac{\mathrm{d}}{\mathrm{d}u}[G^{\mu}(\epsilon,u)-G^{\mu}(0,u)]\bigg|_{u=0} &= \frac{\mathrm{d}}{\mathrm{d}u}[G(\epsilon,u)-G(0,u)]\bigg|_{u=0} \\ &\quad+[G(\epsilon,0)-G(0,0)]\log\left(\frac{\mu^2}{\omega^2}\right).\end{aligned} \tag{4.24}$$

We start by working on the ambiguous integral. We remark that in any manipulation we must not split the subtractions that make the integral safe at $\tau=0$. Due to this fact we manipulate its integrand in the following way

$$\begin{aligned}&\frac{G^{\mu}(0,\tau)-G(0,0)}{\tau^2}-\frac{1}{\tau}\frac{\mathrm{d}}{\mathrm{d}u}G^{\mu}(0,u)\bigg|_{u=0} \\ =\ &\frac{G^{\mu}(0,\tau)-G(0,0)}{\tau^2}-\frac{1}{\tau}\frac{\mathrm{d}}{\mathrm{d}u}G(0,u)\bigg|_{u=0}-\frac{1}{\tau}G(0,0)\log\left(\frac{\mu^2}{\omega^2}\right)\end{aligned} \tag{4.25}$$

$$
\begin{aligned}
&= \left(\frac{\mu^2}{\omega^2}\right)^{\tau}\left[\frac{G(0,\tau)-G(0,0)}{\tau^2}-\frac{1}{\tau}\frac{\mathrm{d}}{\mathrm{d}u}G(0,u)\Big|_{u=0}\right]-\frac{1}{\tau}G(0,0)\log\left(\frac{\mu^2}{\omega^2}\right)\\
&\quad -\left[1-\left(\frac{\mu^2}{\omega^2}\right)^{\tau}\right]\left[\frac{G(0,0)}{\tau^2}+\frac{1}{\tau}\frac{\mathrm{d}}{\mathrm{d}u}G(0,u)\Big|_{u=0}\right]\\
&= \left(\frac{\mu^2}{\omega^2}\right)^{\tau}\left[\frac{G(0,\tau)-G(0,0)}{\tau^2}-\frac{1}{\tau}\frac{\mathrm{d}}{\mathrm{d}u}G(0,u)\Big|_{u=0}\right]\\
&\quad +G(0,0)\left[\frac{1}{\tau^2}\left(\frac{\mu^2}{\omega^2}\right)^{\tau}-\frac{1}{\tau}\log\left(\frac{\mu^2}{\omega^2}\right)-\frac{1}{\tau^2}\right]\\
&\quad +\frac{\mathrm{d}}{\mathrm{d}u}G(0,u)\Big|_{u=0}\left[\left(\frac{\mu^2}{\omega^2}\right)^{\tau}-1\right]\log\left(\frac{\mu^2}{\omega^2}\right),
\end{aligned}
$$

where first we used (4.23), then pushed the factor of $(\mu^2/\omega^2)^\tau$ out of the first, $\tau=0$ safe term and grouped the remaining terms into two, one proportional to $G(0,0)=G_{0,0}$ and another proportional to $(\mathrm{d}G(0,u)/\mathrm{d}u)|_{u=0}=G_{0,1}$. To solve the integral with $G_{0,0}$ we expand $(\mu^2/\omega^2)^\tau$, which starts at $i=0$, and realize the $i=0$, 1 cases are precisely the $1/\tau^2$ and $1/\tau$ subtractions, so that

$$
\begin{aligned}
&\int_0^\infty \mathrm{d}\tau\,\mathrm{e}^{-\tau/\beta}\left[\frac{1}{\tau^2}\left(\frac{\mu^2}{\omega^2}\right)^{\tau}-\frac{1}{\tau}\log\left(\frac{\mu^2}{\omega^2}\right)-\frac{1}{\tau^2}\right] \qquad (4.26)\\
&= \int_0^\infty \mathrm{d}\tau\,\mathrm{e}^{-\tau/\beta}\sum_{i=2}^{\infty}\frac{\tau^{i-2}}{\Gamma(i+1)}\log^i\left(\frac{\mu^2}{\omega^2}\right)=\sum_{i=2}^{\infty}\frac{\beta^{i-1}\Gamma(i-1)}{\Gamma(i+1)}\log^i\left(\frac{\mu^2}{\omega^2}\right)\\
&= \frac{1}{\beta}\left\{\beta\log\left(\frac{\mu^2}{\omega^2}\right)+\left[1-\beta\log\left(\frac{\mu^2}{\omega^2}\right)\right]\log\left[1-\beta\log\left(\frac{\mu^2}{\omega^2}\right)\right]\right\}\\
&= \log\left(\frac{\mu^2}{\omega^2}\right)+\left[\frac{1}{\beta}-\log\left(\frac{\mu^2}{\omega^2}\right)\right]\log\left[1-\beta\log\left(\frac{\mu^2}{\omega^2}\right)\right]\\
&= \log\left(\frac{\mu^2}{\omega^2}\right)+\frac{1}{\beta_\omega}\log\left(\frac{\beta}{\beta_\omega}\right),
\end{aligned}
$$

where in the last line we used the running of $\beta=\beta(\mu)$ in the form (2.31). The integral proportional to $G_{0,1}$ is just (3.16). With both results, the ambiguous integral becomes

$$
\begin{aligned}
&\int_0^\infty \mathrm{d}\tau\,\mathrm{e}^{-\tau/\beta}\left[\frac{G^\mu(0,\tau)-G(0,0)}{\tau^2}-\frac{1}{\tau}\frac{\mathrm{d}}{\mathrm{d}u}G^\mu(0,u)\Big|_{u=0}\right] \qquad (4.27)\\
&= \int_0^\infty \mathrm{d}\tau\left(\frac{\Lambda_{\mathrm{QCD}}}{\omega}\right)^{2\tau}\left[\frac{G(0,\tau)-G(0,0)}{\tau^2}-\frac{1}{\tau}\frac{\mathrm{d}}{\mathrm{d}u}G(0,u)\Big|_{u=0}\right]\\
&\quad +G_{0,0}\log\left(\frac{\mu^2}{\omega^2}\right)+\left[\frac{G_{0,0}}{\beta_\omega}-G_{0,1}\right]\log\left(\frac{\beta}{\beta_\omega}\right).
\end{aligned}
$$

In the finite integral, the dependence on $\mu$ occurs both through $\beta$ in the lower limit and through the $G^\mu$ functions in the integrand. This second dependence is logarithmic and can be extracted out with (4.23):

$$\int_{-\beta}^{0}\frac{\mathrm{d}\tau}{\tau}\frac{\mathrm{d}}{\mathrm{d}u}[G^\mu(\tau,u)-G^\mu(0,u)]\bigg|_{u=0} = \int_{-\beta}^{0}\frac{\mathrm{d}\tau}{\tau}\frac{\mathrm{d}}{\mathrm{d}u}[G(\tau,u)-G(0,u)]\bigg|_{u=0} + \log\left(\frac{\mu^2}{\omega^2}\right)\int_{-\beta}^{0}\mathrm{d}\tau\frac{G(\tau,0)-G(0,0)}{\tau}. \tag{4.28}$$

Combining then (4.27) and (4.28) into (4.19), our final expression for $A$ reads

$$\begin{aligned}\beta_0\delta A &= \frac{\pi^2}{6}G_{1,0}+G_{0,0}\log\left(\frac{\mu^2}{\omega^2}\right)+\left[\frac{G_{0,0}}{\beta_\omega}-G_{0,1}\right]\log\left(\frac{\beta}{\beta_\omega}\right) \\ &+\int_0^\infty \mathrm{d}\tau\left(\frac{\Lambda_{\mathrm{QCD}}}{\omega}\right)^{2\tau}\left[\frac{G(0,\tau)-G(0,0)}{\tau^2}-\frac{1}{\tau}\frac{\mathrm{d}}{\mathrm{d}u}G(0,u)\bigg|_{u=0}\right] \\ &-\int_{-\beta}^{0}\mathrm{d}\tau\left\{\log\left(1+\frac{\tau}{\beta}\right)\frac{G(\tau,0)-G(0,0)}{\tau^2}-\frac{1}{\tau}\frac{\mathrm{d}}{\mathrm{d}u}[G(\tau,u)-G(0,u)]\bigg|_{u=0}\right\} \\ &+\log\left(\frac{\mu^2}{\omega^2}\right)\int_{-\beta}^{0}\mathrm{d}\tau\frac{G(\tau,0)-G(0,0)}{\tau}.\end{aligned} \tag{4.29}$$

The logarithmic $\mu$-dependence of $\delta Z_A$ can be also made explicit by using (4.23) in the last derivative of (4.21), which removes the $\mu$ index of $G^\mu$ and introduces a term proportional to $\log(\mu^2/\omega^2)$. Defining $\delta Z_A \equiv \delta Z_{\mathrm{nc}}+\log(\mu^2/\omega^2)\delta Z_{\mathrm{cusp}}$ one finds

$$\begin{aligned}\delta Z_{\mathrm{nc}} &= \frac{\pi^2}{6\epsilon}G_{0,0}+\int_{-\beta}^{0}\frac{\mathrm{d}\tau}{\tau-\epsilon}\left[\log\left(1+\frac{\tau}{\beta}\right)\frac{G(\tau,0)}{\tau}-\frac{\mathrm{d}}{\mathrm{d}u}G(\tau,u)\bigg|_{u=0}\right], \\ \delta Z_{\mathrm{cusp}} &= -\int_{-\beta}^{0}\frac{\mathrm{d}\tau}{\tau-\epsilon}G(\tau,0).\end{aligned} \tag{4.30}$$

## 4.3 Anomalous dimension

Again we can derive the anomalous dimension both from the perturbative and closed expressions for $A(\mu)$ and $Z(\mu)$. Let us do it explicitly for the perturbative version of $A(\mu)$. The derivative of (4.16) is, step-by-step,

$$\begin{aligned}\mu\frac{\mathrm{d}}{\mathrm{d}\mu}[\beta_0\delta A(\mu)] &= \sum_{n=1}^{\infty}2(-\beta)^{n+1}[(-1)^n\Gamma(n+1)G^\mu_{0,n+1}-H_nG_{n+1,0}-G^\mu_{n,1}] \\ &+\sum_{n=1}^{\infty}(-\beta)^n\left[(-1)^n\Gamma(n)2G^\mu_{0,n}-\frac{2G_{n,0}}{n}\right] \\ &= \sum_{n=1}^{\infty}2(-\beta)^{n+1}[(-1)^n\Gamma(n+1)G^\mu_{0,n+1}-H_nG_{n+1,0}-G^\mu_{n,1}]\end{aligned}$$

$$
\begin{aligned}
&\quad +2\beta[G^{\mu}_{0,1}+G_{1,0}] \\
&\quad +\sum_{n=1}^{\infty}2(-\beta)^{n+1}\left[(-1)^{n+1}\Gamma(n+1)G^{\mu}_{0,n+1}-\frac{G_{n+1,0}}{n+1}\right] \\
&= -2\sum_{n=1}^{\infty}(-\beta)^{n}\left[H_{n-1}G_{n,0}+G^{\mu}_{n-1,1}+\frac{G_{n,0}}{n}\right] \\
&= -2\sum_{n=1}^{\infty}(-\beta)^{n}[H_nG_{n,0}+G_{n-1,1}]-2\log\left(\frac{\mu^2}{\omega^2}\right)\sum_{n=1}^{\infty}(-\beta)^{n}G_{n-1,0}.
\end{aligned}
$$

As before, we used (2.28) with $\epsilon=0$ to take the derivative with respect to $\beta$ and (B.22) to take the derivative of $G^{\mu}_{i,j}$. To simplify the resulting expression we first discarded the $n=1$ case of the second sum and took $n\mapsto n+1$. We then combined the two sums, realized the single terms correspond to the $n=0$ case and took $n\mapsto n-1$ after including them. In these simplifications the identities $H_0=0$ and $H_{n-1}+1/n=H_n$ were employed. In the very last step we used (B.21) in the form

$$
G^{\mu}_{n-1,1}=\sum_{k=0}^{1}\frac{1}{k!}\log^{k}\left(\frac{\mu^2}{\omega^2}\right)G_{n-1,1-k}=G_{n-1,1}+\log\left(\frac{\mu^2}{\omega^2}\right)G_{n-1,0}, \tag{4.31}
$$

to extract the $\mu$ dependence.

This computation may seem a nuisance anyone would like to skip, but in truth it reveals two powerful things. On the one hand, it explicitly shows how the factorial divergences of $A$, present through the term with $\Gamma(n)$, have canceled upon taking the derivative, ensuring the anomalous dimension series is convergent. On the other hand, it also shows explicitly how, out of the infinite powers of the logs that are present in $A$ through the $G^{\mu}_{0,n}$, only a single power survived. This shows indeed that $A$ is a cusp series, and proves our claim of the equivalence of such series with the $F(\epsilon,u)$ having a simple pole at $u=0$.

Following our convention in (2.58) and (2.59) for the anomalous dimension, we write down the final expressions

$$
\begin{aligned}
\gamma_A(\alpha_s) &= -\frac{2}{\beta_0}\sum_{n=1}^{\infty}(-\beta)^{n}[H_nG_{n,0}+G_{n-1,1}], \qquad (4.32)\\
\Gamma_A(\alpha_s) &= -\frac{2}{\beta_0}\sum_{n=1}^{\infty}(-\beta)^{n}G_{n-1,0}.
\end{aligned}
$$

From (4.32) we can directly derive the corresponding closed forms using again the results in Appendix B. The relevant cases are

$$\begin{aligned}
&-\sum_{n=1}^{\infty}(-\beta)^n H_n G_{n,0} = -\sum_{n=1}^{\infty} G_{n,0}\int_{-\beta}^{0}\mathrm{d}\tau\frac{\tau^n-(-\beta)^n}{\tau+\beta} = \int_{-\beta}^{0}\mathrm{d}\tau\frac{G(\tau,0)-G(-\beta,0)}{\beta+\tau}\\
&-\sum_{n=1}^{\infty}(-\beta)^n G_{n-1,1} = \beta\sum_{n=0}^{\infty}(-\beta)^n G_{n,1} = \beta\frac{\mathrm{d}}{\mathrm{d}u}G(-\beta,u)\bigg|_{u=0} \qquad (4.33)\\
&-\sum_{n=1}^{\infty}(-\beta)^n G_{n-1,0} = \beta\sum_{n=0}^{\infty}(-\beta)^n G_{n,0} = \beta G(-\beta,0).
\end{aligned}$$

With them, the closed expressions for the non-cusp and cusp parts are

$$\begin{aligned}
\gamma_A(\alpha_s) &= \frac{2}{\beta_0}\int_{-\beta}^{0}\mathrm{d}\tau\frac{G(\tau,0)-G(-\beta,0)}{\beta+\tau} + \frac{2\beta}{\beta_0}\frac{\mathrm{d}}{\mathrm{d}u}G(-\beta,u)\bigg|_{u=0}, \qquad (4.34)\\
\Gamma_A(\alpha_s) &= \frac{2\beta}{\beta_0}G(-\beta,0).
\end{aligned}$$

## 4.4 Relations for perturbative coefficients

Again we finish the chapter by finding efficient relations to obtain the $\mu$-independent perturbative coefficients of the $A(\mu)$ series and its anomalous dimension.

### 4.4.1 Renormalized series

This time the presence of a cusp term in the anomalous dimension of $A(\mu)$ translates into an extra power of the logarithm $L=\log(\mu^2/\omega^2)$ compared to the series with no cusp-anomalous dimension:

$$A(\mu) \equiv 1 + \frac{1}{\beta_0}\sum_{n=1}^{\infty}\beta^n\sum_{i=0}^{n+1}c_{n,i}L^i. \qquad (4.35)$$

The non-logarithm coefficients are obtained as before by setting $\mu=\omega$ in the first line of (4.16), which leads to

$$c_{n,0} = \Gamma(n)G_{0,n+1} - \frac{(-1)^n}{n}(H_n G_{n+1,0} + G_{n,1}) \qquad (4.36)$$

The coefficients of the logarithm are obtained by taking the $\mu$-derivative of (4.35). The left-hand side gives the anomalous dimension, and the right-hand side leads to

$$\begin{aligned} \mu\frac{\mathrm{d}}{\mathrm{d}\mu}\left[\sum_{n=1}^{\infty}\beta^n\sum_{i=0}^{n+1}c_{n,i}L^i\right] &= -2\sum_{n=1}^{\infty}n\beta^{n+1}\sum_{i=0}^{n+1}c_{n,i}L^i+2\sum_{n=1}^{\infty}\beta^n\sum_{i=1}^{n+1}i\,c_{n,i}L^{i-1} \quad (4.37)\\ &= -2\sum_{n=1}^{\infty}n\beta^{n+1}\sum_{i=0}^{n+1}c_{n,i}L^i+2\sum_{n=0}^{\infty}\beta^{n+1}\sum_{i=0}^{n+1}(i+1)c_{n+1,i+1}L^i\\ &= 2\sum_{n=1}^{\infty}\beta^{n+1}\sum_{i=0}^{n+1}[(i+1)c_{n+1,i+1}-n\,c_{n,i}]L^i+2\beta[c_{1,1}+2c_{1,2}L], \end{aligned}$$

where in the second step we took $i\mapsto i+1$ and $n\mapsto n+1$ in the second sum and extracted the $n=0$ case from the resulting sum. Now we can compare the powers of $L$ with those in (4.32): the log-free terms must equal $\gamma(\alpha_s)$, the single power terms must equal $\Gamma(\alpha_s)$ and the coefficients with higher log powers must vanish identically. These conditions lead us to the following recursive relations:

$$\begin{aligned} c_{n,1}&=(n-1)c_{n-1,0}-(-1)^n(H_nG_{n,0}+G_{n-1,1}), & i=0, \quad (4.38)\\ c_{n,2}&=\frac{n-1}{2}c_{n-1,1}-\frac{(-1)^n}{2}G_{n-1,0}, & i=1,\\ c_{n,i}&=\frac{n-1}{i}c_{n-1,i-1}, & i\geq 2, \end{aligned}$$

which hold for $n\geq 1$. We can also write the $c_{n,i}$ coefficients solely in terms of the $G_{i,j}$ using these results on themselves. When using $c_{n,0}$ in $c_{n,1}$ we obtain

$$\begin{aligned} c_{n,1} &= (n-1)\left[G_{0,n}\Gamma(n-1)+\frac{(-1)^n}{n-1}(H_{n-1}G_{n,0}+G_{n-1,1})\right]-(-1)^n(H_nG_{n,0}+G_{n-1,1})\\ &= \Gamma(n)G_{0,n}+(-1)^n(H_{n-1}-H_n)G_{n,0}=\Gamma(n)G_{0,n}-\frac{(-1)^n}{n}G_{n,0}, \quad (4.39) \end{aligned}$$

where in the last step we used $H_n-H_{n-1}=1/n$. When using $c_{n,1}$ into $c_{n,2}$ we get

$$\begin{aligned} c_{n,2} &= \frac{n-1}{2}\left[\Gamma(n-1)G_{0,n-1}+\frac{(-1)^n}{n-1}G_{n-1,0}\right]-\frac{(-1)^n}{2}G_{n-1,0} \quad (4.40)\\ &= \frac{\Gamma(n)}{2}G_{0,n-1}. \end{aligned}$$

Finally, to simplify the $i\geq 2$ result for $c_{n,i}$ we need to use it recursively. Each time both $n$ and $i$ get lowered by one unit, and the base case corresponds to $i=2$, so the number of steps is $i-2$

$$\begin{aligned} c_{n,i} &= \frac{(n-1)(n-2)...(n-i+2)}{i(i-1)...3}c_{n-i+2,2} \quad (4.41)\\ &= \frac{2(n-1)!}{i!(n-i+1)!}\frac{\Gamma(n-i+2)}{2}G_{0,n-i+1}=\frac{\Gamma(n)}{\Gamma(i+1)}G_{0,n-i+1}. \end{aligned}$$

### 4.4.2 Anomalous dimension

To better compare our perturbative definition (2.59) for the anomalous dimension with the results in (4.34) we first write it in terms of $\beta$ as

$$\gamma_A(\alpha_s) = \frac{1}{\beta_0}\sum_{n=1}^{\infty}\frac{g_{n-1}}{\beta_0^{n-1}}\beta^n, \qquad \Gamma_A(\alpha_s) = \frac{1}{\beta_0}\sum_{n=1}^{\infty}\frac{\Gamma_{n-1}}{\beta_0^{n-1}}\beta^n. \tag{4.42}$$

Then one can directly write down the coefficients as

$$\hat{\gamma}_n^A \equiv \frac{\gamma_n^A}{\beta_0^n} = 2(-1)^n[H_{n+1}G_{n+1,0} + G_{n,1}], \tag{4.43}$$
$$\hat{\Gamma}_n^A \equiv \frac{\Gamma_n^A}{\beta_0^n} = 2(-1)^n G_{n,0}.$$

# Chapter 5

# Integrals with poles in the real axis: the Principal Value prescription

As we have seen, either Borel summation or the equivalent formalism developed in chapters 3 and 4 expresses the sum of a divergent series as the integral along the positive real axis of a function that may present poles. When this is the case, a prescription to regularize the integral is required. The usual choice consists on deforming the integration path to avoid the poles either along a clockwise semicircle in the upper half of the complex plane or along a counter-clockwise semicircle in the lower half. This prescription is called the Principal Value prescription. In this chapter we develop explicit results to regularize the infinite integrals from chapters 3 and 4 and also define a prescription to compute the ambiguity associated with the choice of the deformation of the path.

## 5.1 Set up

Given a function $f(x)$ let us denote the set of its real poles by $\{x_n^{(a)}\}_{n=1}^{n_p}$, where $x_n > x_m$ if $n > m$, $a$ denotes the order of the pole and $n_p$ the number of poles. Note that a pole of order $a$ usually implies lower order poles at the same position, so all $\{x_n^{(1)}, ..., x_n^{(a)}\}$ are assumed to be included in the set of poles. Also, in general we can have $n_p \mapsto \infty$. The pole expansion of $f(x)$ is

$$f(x) \asymp \sum_{n=1}^{n_p}\sum_{k=1}^{a} \frac{f_{n,k}}{(x - x_n^{(a)})^k}. \tag{5.1}$$

The integral of $f(x)$ along the real axis can be split as

$$\begin{aligned}\int_{-\infty}^{\infty} \mathrm{d}x\, f(x) &= \int_{-\infty}^{x_1-\delta} \mathrm{d}x\, f(x) + \sum_{n=1}^{n_p-1}\int_{x_n+\delta}^{x_{n+1}-\delta} \mathrm{d}x\, f(x) \\ &\quad + \int_{x_{n_p}+\delta}^{\infty} \mathrm{d}x\, f(x) + \sum_{n=1}^{n_p}\int_{x_n-\delta}^{x_n+\delta} \mathrm{d}x\, f(x),\end{aligned} \tag{5.2}$$

where $\delta > 0$, $\delta < \min_n\{|x_{n+1} - x_n|/2\}$[5.1] and where we have dropped the pole's order dependence for simplicity. The integral is independent on the value of $\delta$ as long as it verifies the stated conditions. It is the last integral that needs a prescription, and we start by adding and subtracting the expansion $f(x)$ around each pole:

$$\sum_{n=1}^{n_p}\int_{x_n-\delta}^{x_n+\delta} \mathrm{d}x\, f(x) = \sum_{n=1}^{n_p}\sum_{k=1}^{a}\int_{x_n-\delta}^{x_n+\delta} \mathrm{d}x\left[f(x) - \frac{f_{n,k}}{(x-x_n)^k}\right] + \sum_{n=1}^{n_p}\sum_{k=1}^{a} f_{n,k}\int_{x_n-\delta}^{x_n+\delta}\frac{\mathrm{d}x}{(x-x_n)^k}. \tag{5.3}$$

Due to the pole subtraction, again only the last integral in (5.3) is divergent. To compute such an integral we define the principal value (P.V.) prescription as

$$\mathrm{P.V.}_{\pm}\left\{\int_{x_n-\delta}^{x_n+\delta}\frac{\mathrm{d}x}{(x-x_n)^k}\right\} \equiv \lim_{\varepsilon\mapsto 0}\left[\int_{x_n-\delta}^{x_n-\varepsilon}\frac{\mathrm{d}x}{(x-x_n)^k} + \int_{C_\pm(\varepsilon,x_n)}\frac{\mathrm{d}x}{(x-x_n)^k} + \int_{x_n+\varepsilon}^{x_n+\delta}\frac{\mathrm{d}x}{(x-x_n)^k}\right], \tag{5.4}$$

where $0 < \varepsilon < \delta$ and $C_\pm(\varepsilon, x_n)$ denotes either the clockwise-oriented, upper $(+)$ or counterclockwise-oriented, lower semicircunference of radius $\varepsilon$ centered at $x_n$. Let us work out each term.

## 5.2 Pure pole integrals in the P.V. prescription

### 5.2.1 Simple pole

For $k = 1$ the integrals along the real axis are

$$\int_{x_n-\delta}^{x_n-\varepsilon}\frac{\mathrm{d}x}{x-x_n} = \ln\left(\frac{\varepsilon}{\delta}\right), \qquad \int_{x_n+\varepsilon}^{x_n+\delta}\frac{\mathrm{d}x}{x-x_n} = \ln\left(\frac{\delta}{\varepsilon}\right). \tag{5.5}$$

To solve the contour integral along $C_\pm(\varepsilon, x_n)$ we first change variables to $z = x - x_n$ so that the semicircle is centered at the origin and then write $z = \varepsilon e^{i\theta}$. This leads to

$$\int_{C_\pm(\varepsilon,x_n)}\frac{\mathrm{d}x}{x-x_n} = \int_{C_\pm(\varepsilon,0)}\frac{\mathrm{d}z}{z} = i\int_{\pm\pi}^{0}\mathrm{d}\theta = \mp i\pi. \tag{5.6}$$

5.1. The conditions on $\delta$ simply ensure there is no overlap between the integrals. Note that, as long as any interval isn't integrated more that once, one can introduce as many $\delta$ parameters as desired. We choose to make a symmetric splitting to simplify the latter computations.

The sum of the integrals cancels the $\varepsilon$-dependence and the limit is finite, so

$$\text{P.V.}_{\pm}\left\{\int_{x_n-\delta}^{x_n+\delta}\frac{\mathrm{d}x}{x-x_n}\right\}=\mp\mathrm{i}\pi. \tag{5.7}$$

### 5.2.2 Higher order poles

Let us consider $k\geq 1$. Then

$$\begin{aligned}\int_{x_n-\delta}^{x_n-\varepsilon}\frac{\mathrm{d}x}{(x-x_n)^k}&=\frac{(-1)^k}{k-1}\left[\frac{1}{\varepsilon^{k-1}}-\frac{1}{\delta^{k-1}}\right],\\ \int_{x_n+\varepsilon}^{x_n+\delta}\frac{\mathrm{d}x}{(x-x_n)^k}&=\frac{1}{k-1}\left[\frac{1}{\varepsilon^{k-1}}-\frac{1}{\delta^{k-1}}\right].\end{aligned} \tag{5.8}$$

To solve the contour integral along $C_{\pm}(\varepsilon,x_n)$ we again write $z=x-x_n=\varepsilon e^{i\theta}$. This leads to

$$\int_{C_{\pm}(\varepsilon,x_n)}\frac{\mathrm{d}x}{(x-x_n)^k}=\frac{i}{\varepsilon^{k-1}}\int_{\pm\pi}^{0}\mathrm{e}^{\mathrm{i}\theta(1-k)}\mathrm{d}\theta=\frac{-1}{k-1}\frac{1+\mathrm{e}^{\mp\mathrm{i}\pi k}}{\varepsilon^{k-1}}. \tag{5.9}$$

Adding up the three integrals the $\varepsilon$-dependence cancels again and we get

$$\text{P.V.}_{\pm}\left\{\int_{x_n-\delta}^{x_n+\delta}\mathrm{d}x\frac{1}{(x-x_n)^k}\right\}=\frac{-1}{k-1}\frac{(-1)^k+1}{\delta^{k-1}}, \tag{5.10}$$

where the last result comes from the fact that both $\mathrm{e}^{\mp\mathrm{i}\pi k}=(-1)^k$.

### 5.2.3 Conclusions

Under the principal value prescription the integral when crossing a pole is

$$\text{P.V.}_{\pm}\left\{\int_{x_n-\delta}^{x_n+\delta}\mathrm{d}x\frac{1}{(x-x_n)^k}\right\}=\begin{cases}\mp\mathrm{i}\pi, & k=1\\ \dfrac{-2}{(k-1)\delta^{k-1}}, & k \text{ even}, k>1\\ 0, & k \text{ odd}, \quad k>1\end{cases}. \tag{5.11}$$

We see that simple poles only contribute to the result by adding an imaginary part, while even poles only contribute to the real part of the result and odd poles don't contribute at all. Moreover, none of the contributions depend on the pole's position in the real axis.

With result (5.11), the integral of $f$ then reads

$$\begin{aligned}\mathrm{P.V.}_{\pm}\left\{\int_{-\infty}^{\infty}\mathrm{d}x\, f(x)\right\} &= \int_{-\infty}^{x_1-\delta}\mathrm{d}x\, f(x)+\sum_{n=1}^{n_p-1}\int_{x_n+\delta}^{x_{n+1}-\delta}\mathrm{d}x\, f(x)+\int_{x_{n_p}+\delta}^{\infty}\mathrm{d}x\, f(x)\\ &+\sum_{n=1}^{n_p}\sum_{k=1}^{a}\int_{x_n-\delta}^{x_n+\delta}\mathrm{d}x\left[f(x)-\frac{f_{n,k}}{(x-x_n)^k}\right]\mp \mathrm{i}\pi\sum_{n=1}^{n_p}f_{n,1}\\ &+\sum_{n=1}^{n_p}\sum_{k=1}^{\mathrm{floor}\{a/2\}}\frac{-2f_{n,2k}}{(a-1)\delta^{2k-1}},\end{aligned} \tag{5.12}$$

where simple poles have been singled out and the sum over the remaining poles has been modified to include only those of even order. The function floor$\{x\}$ returns the largest integer lower than $x$. Note that for an infinite number of poles the sums are extended to $\infty$ and the last integral on the first line vanishes.

## 5.3 Principal value and ambiguity

When $f$ is a real function, all the terms in (5.12) are real except for the one containing the residues of the simple poles, which is purely imaginary. We want a prescription that assigns a real value to the integral of a real function, so we take the average of P.V.$_+$ and P.V.$_-$ and define

$$\mathrm{P.V.}\left\{\int_{-\infty}^{\infty}\mathrm{d}x\, f(x)\right\}\equiv\frac{1}{2}\left[\mathrm{P.V.}_{+}\left\{\int_{-\infty}^{\infty}\mathrm{d}x\, f(x)\right\}+\mathrm{P.V.}_{-}\left\{\int_{-\infty}^{\infty}\mathrm{d}x\, f(x)\right\}\right], \tag{5.13}$$

which amounts to ignoring the imaginary term in (5.12). The imaginary terms that cancel in the average are accounted for as an ambiguity for the P.V. value. The standard way of computing the ambiguity is

$$\begin{aligned}\mathrm{amb}\left\{\int_{-\infty}^{\infty}\mathrm{d}x\, f(x)\right\} &\equiv \frac{1}{2\pi\mathrm{i}}\left[\mathrm{P.V.}_{+}\left\{\int_{-\infty}^{\infty}\mathrm{d}x\, f(x)\right\}-\mathrm{P.V.}_{-}\left\{\int_{-\infty}^{\infty}\mathrm{d}x\, f(x)\right\}\right] \\ &= \sum_{n=1}^{n_p}f_{n,1}.\end{aligned} \tag{5.14}$$

The subtraction in the ambiguity leads to the sum of integrals along contours enclosing the poles, and thus the result recovers the residue theorem.

For the practical applications developed in this thesis, the real, averaged value (5.13) is usually referred to as the exact value of the integral. We implement the complete P.V. prescription (exact value and ambiguity) via algorithms in `Mathematica` [19] and `Phyton` [20] codes. We use the native built-in functions in `Mathematica` to numerically solve the integrals, while in `Python` we use the `ScyPy` [21] module, that builds on the `NumPy` [22] package which is also employed for the mathematical constants $\pi$ or $\gamma_E$. Since in our applications the function $f(x)$ has an infinite number of poles, in both cases we extend the sum of the integrals until either the (absolute value of the) integral around a pole or the integral between poles is smaller than $10^{-10}$. Algorithms in both codes give results that agree within 15 decimal places. In some cases, it is possible to derive a closed expression for the sum of the residues in (5.14), which we compare to the result given by the P.V. algorithm to find complete agreement within at least 10 decimal places.

# Chapter 6
# Applications I: $\overline{\text{MS}}$ and MSR masses

As a first application of the formalism developed in the previous chapters, we study the relation between the $\mu$-independent pole mass $m_p$ and the $\mu$-dependent masses in the $\overline{\text{MS}}$ and MSR renormalization schemes, $\overline{m}(\mu)$ and $m^{\text{MSR}}(\mu)$.

The pole and $\overline{\text{MS}}$ masses are defined by the choice of the renormalization scheme for the mass parameter in the bare Lagrangian. The conditions defining the mass scheme are imposed on the renormalized propagator, which in turn translates into conditions on the quark self-energy,

$$\Sigma(\not{p}, m) = \not{p}\Sigma_p(m^2, p^2) + \Sigma_m(m^2, p^2), \tag{6.1}$$

and also determine the quark-wavefunction renormalization $Z_\psi$. The quark self-energy has dimensions of mass and is represented at $\beta_0$LO in figure 6.1.

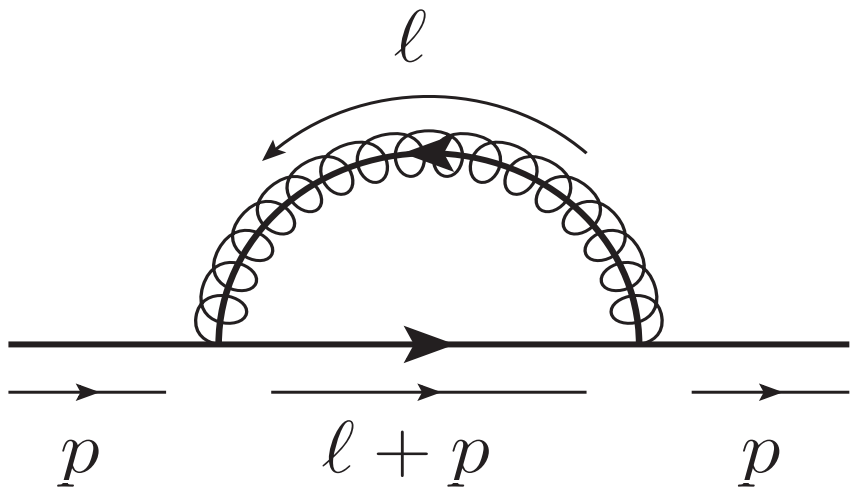

**Figure 6.1.** One-loop quark self-energy diagram for a massive or massless quark, with $\ell$ the virtual loop momentum and $p$ the quark off-shell momentum. The gluon propagator with an arrow is understood to be the shifted effective propagator.

The pole mass is defined in the so-called on-shell scheme (OS), in which one demands that the pole of the propagator occurs at the mass parameter, $m_p$, and normalizes its residue to $i$ [16]. This imposes the following conditions at $\beta_0$LO

$$\delta Z_\psi^{\text{OS}} = -\frac{\mathrm{d}}{\mathrm{d}\not{p}}\Sigma(\not{p}, m_p)\bigg|_{\not{p}=m_p}, \qquad \delta Z_m^{\text{OS}} = \Sigma_p(m_p^2, m_p^2) + \Sigma_m(m_p^2, m_p^2). \tag{6.2}$$

In the $\overline{\text{MS}}$ scheme, the renormalization factors are simply defined to cancel the divergences occurring in the propagator's renormalization. At $\beta_0$LO they are fixed as

$$\delta Z_\psi^{\overline{\text{MS}}} = -\Sigma_p(\bar{m}^2, \bar{m}^2)|_{\text{div}}, \qquad \delta Z_m^{\overline{\text{MS}}} = \Sigma_m(\bar{m}^2, \bar{m}^2)|_{\text{div}} + \Sigma_p(\bar{m}^2, \bar{m}^2)|_{\text{div}}, \tag{6.3}$$

where the notation 'div' indicates that only the negative powers of $\epsilon$ must be kept.

## 6.1 Massive quark self-energy

At $\beta_0$LO, the quark self-energy corresponds to the insertion of the shifted gluon propagator (2.45) shown in figure 6.1. Having made clear the $\not p$ and $m$ dependences of the self-energy in (6.1) to (6.3), we simply refer to the shifted version as $i\Sigma_{\rm sh}(h)$, where $h$ indicates the shift in the power of the gluon's momentum. Then, in the bare formalism we write

$$i\Sigma_{\rm sh}(h) = -g_0^2 C_F \int \frac{\mathrm{d}^d\ell}{(2\pi)^d} \frac{\gamma^\mu(\not p - \not\ell + m_0)\gamma_\mu}{(\ell^2)^{1+h}[(p-\ell)^2 - m_0^2]} \equiv -g_0^2 C_F I_\Sigma, \tag{6.4}$$

where $C_F = (C_A^2 - 1)/(2\,C_A)$. Working on the numerator,

$$\gamma^\mu(\not p - \not\ell + m_0)\gamma_\mu = (2-d)(\not p - \not\ell) + d m_0 = 2(\epsilon - 1)(\not p - \not\ell) + (4 - 2\epsilon)m_0, \tag{6.5}$$

the integral acquires the expression

$$\begin{aligned} I_\Sigma \;&=\; 2[(\epsilon-1)\not p + (2-\epsilon)m_0]\, I_2(1, Q^{1+h}, Q_{m_0}(-p)) \\ &\quad -2(\epsilon-1)\gamma_\mu\, I_2(k^\mu, Q^{1+h}, Q_{m_0}(-p)). \end{aligned} \tag{6.6}$$

Using the results in appendix C.4 and defining $J_\Sigma(x) \equiv I_2(1, Q^x, Q_{m_0}(p))$ we can rewrite the vector integral as

$$I_2(k^\mu, Q^{1+h}, Q_{m_0}(-p)) = \frac{p^\mu}{2p^2}[(p^2 - m_0^2) J_\Sigma(1+h) + J_\Sigma(h)], \tag{6.7}$$

and thus the $I_\Sigma$ becomes

$$I_\Sigma = (\epsilon - 1)\left[\left(1 + \frac{m_0^2}{p^2}\right) J^\Sigma(1+h) - \frac{1}{p^2} J_\Sigma(h)\right]\not p + 2(2-\epsilon) m_0\, J_\Sigma(1+h), \tag{6.8}$$

where we factorized $\not p$ and $m$ to match the conventional splitting in (6.1). The results for both parts of the shifted quark-self energy are

$$\begin{aligned} \Sigma_{p,\rm sh}(h) \;&=\; \frac{g_0^2}{(4\pi)^{2-\epsilon}} C_F (-1)^h (m_0^2)^{-h-\epsilon} \frac{(\epsilon-1)\Gamma(h+\epsilon)\Gamma(1-h-\epsilon)}{\Gamma(2-\epsilon)} \\ &\quad \times\left[\frac{m_0^2}{p^2}\,{}_2F_1\!\left(h, h+\epsilon-1, 2-\epsilon, \frac{p^2}{m^2}\right)\right. \\ &\quad \left. - \frac{m_0^2 + p^2}{p^2}\,{}_2F_1\!\left(h, h+\epsilon-1, 2-\epsilon, \frac{p^2}{m^2}\right)\right], \end{aligned} \tag{6.9}$$

$$\Sigma_{m,\mathrm{sh}}(h) = \frac{g_0^2}{(4\pi)^{2-\epsilon}} 2C_F(-1)^h(m^2)^{-h-\epsilon}(\epsilon-2)\Gamma(h+\epsilon)\Gamma(1-h-\epsilon) \times {}_2F_1\left(h+1, h+\epsilon, 2-\epsilon, \frac{p^2}{m_0^2}\right),$$

where ${}_2F_1$ the Gauss hypergeometric function.

Although it is not the direct object of our study, let us illustrate how to compute $Z_m^{\mathrm{OS}}$ from these results. First, applying (6.2) we sum the two expressions in (6.9) and following (2.46) we split the result to get the function $a_{Z_m^{\mathrm{OS}}}(h,\epsilon)$

$$\Sigma_{p,\mathrm{sh}}(h) + \Sigma_{m,\mathrm{sh}}(h) \equiv \left(\frac{g_0}{4\pi}\right)^2 a_{Z_m^{\mathrm{OS}}}(h,\epsilon), \tag{6.10}$$
$$a_{Z_m^{\mathrm{OS}}}(h,\epsilon) = 2C_F(-1)^h(m_p^2)^{-h-\epsilon}(4\pi)^\epsilon \frac{(1-h-\epsilon)(2\epsilon-3)\Gamma(h+\epsilon)\Gamma(1-2h-2\epsilon)}{\Gamma(3-h-2\epsilon)}.$$

From here it is immediate to obtain $F^\mu_{Z_m^{\mathrm{OS}}}(\epsilon,u)$ following (2.53),

$$F^\mu_{Z_m^{\mathrm{OS}}}(\epsilon,u) = 2C_F e^{u\gamma_E}\left(\frac{\mu^2}{m_p^2}\right)^u \frac{(1-u)(2\epsilon-3)\Gamma(1+u)\Gamma(1-2u)}{\Gamma(3-u-\epsilon)} T^{\frac{u}{\epsilon}-1}(\epsilon), \tag{6.11}$$
$$T(\epsilon) \equiv (-1)^{1+\epsilon}\frac{3}{4}\epsilon P_B(\epsilon),$$

where we defined the factor $T(\epsilon)$ since it will appear again in later applications. A quick expansion reveals $F_{Z_m^{\mathrm{OS}}}(\epsilon,u) = -3C_F + O(\epsilon) + O(u)$, so indeed it is regular at the origin. The complete $\beta_0$LO contribution to $Z_m^{\mathrm{OS}}$ can be then computed perturbatively as the sum of the two equations in (3.10) and in closed integral form as the sum of the two equations in (3.12). Note that $Z_m^{\mathrm{OS}}$ must contain both divergent and finite terms so it is not multiplicatively renormalized.

Finally, since it will be required in the later computation of the matching coefficient of SCET onto bHQET, we follow the same logic for the on-shell wavefunction renormalization factor. To compute the derivative in (6.2) one uses the identity

$$\frac{\mathrm{d}}{\mathrm{d}\not{p}} = 2\not{p}\frac{\mathrm{d}}{\mathrm{d}p^2}, \tag{6.12}$$

which is directly derived from $\not{p}^2 = p^2$. This leads to

$$\delta Z_\psi^{\mathrm{OS}} = -\frac{\mathrm{d}}{\mathrm{d}\not{p}}\Sigma(\not{p},m)\bigg|_{\not{p}=m_p} \tag{6.13}$$
$$= -\Sigma_p(m_p^2,m_p^2) - 2m_p^2\frac{\mathrm{d}}{\mathrm{d}p^2}[\Sigma_p(p^2,m_p^2)+\Sigma_m(p^2,m_p^2)]\bigg|_{p^2=m_p^2}.$$

From here it is obtained that $a_{Z_\psi^{\rm OS}}(h,\epsilon)=(1+h)a_{Z_m^{\rm OS}}(h,\epsilon)$ and thus $F^\mu_{Z_\psi^{\rm OS}}(\epsilon,u)=(1-u-\varepsilon)F^\mu_{Z_m^{\rm OS}}(\epsilon,u)$.

## 6.2 Pole and $\overline{\rm MS}$ mass relation

### 6.2.1 General expressions

To study the relation between the pole and $\overline{\rm MS}$ masses we consider the difference $\delta_{\overline{\rm MS}}(\mu)\equiv m_p-\overline{m}(\mu)$. The series $\delta_{\overline{\rm MS}}(\mu)$ gives the pole mass from the renormalized $\overline{\rm MS}$ mass as $m_p=\overline{m}(\mu)+\delta_{\overline{\rm MS}}(\mu)$, where the $\mu$-dependence on the right-hand side cancels. In addition, at $\beta_0$LO

$$m_p-\overline{m}(\mu)=\left(\frac{Z_m^{\overline{\rm MS}}}{Z_m^{\rm OS}}-1\right)\overline{m}(\mu)=(\delta Z_m^{\overline{\rm MS}}-\delta Z_m^{\rm OS})\overline{m}(\mu). \tag{6.14}$$

Since $Z_m^{\overline{\rm MS}}$ and $Z_m^{\rm OS}$ present the same divergences –see equations (6.2) and (6.3)–, the series $\delta_{\overline{\rm MS}}$ is finite and is built from $\overline{m}(\mu)F^\mu_{\overline{\rm MS}}(\epsilon,u)\equiv-\overline{m}(\mu)(\mu/\overline{m}(\mu))^{2u}F_{Z_m^{\rm OS}}(\epsilon,u)$. The final perturbative and closed expressions for the renormalized series (3.10) and (3.17) can be applied to $F_{\overline{\rm MS}}(\epsilon,u)$. The closed form is

$$\begin{aligned}\beta_0\delta_{\overline{\rm MS}}(\mu) \;=\; & \overline{m}(\mu)\Bigg\{F_{\overline{\rm MS}}(0,0)\log\left(\frac{\beta_{\overline{m}(\mu)}}{\beta}\right) \\ & +\int_0^\infty {\rm d}\tau\left(\frac{\Lambda_{\rm QCD}}{\overline{m}(\mu)}\right)^{2\tau}\frac{F_{\overline{\rm MS}}(0,\tau)-F_{\overline{\rm MS}}(0,0)}{\tau} \\ & +\int_{-\beta}^0 {\rm d}\tau\,\frac{F_{\overline{\rm MS}}(\tau,0)-F_{\overline{\rm MS}}(0,0)}{\tau}\Bigg\}.\end{aligned} \tag{6.15}$$

The value $\overline{m}(\mu)$ entering (6.15) is obtained by running from some reference scale $\mu_0$ and can also be derived from the anomalous dimension of $\delta_{\overline{\rm MS}}(\mu)$. Since $m_p$ has no $\mu$ dependence[6.1]

$$\gamma_{\overline{\rm MS}}=\mu\frac{\rm d}{{\rm d}\mu}\delta_{\overline{\rm MS}}=-\mu\frac{\rm d}{{\rm d}\mu}\overline{m}(\mu)=-2\gamma_{\overline{m}}. \tag{6.16}$$

Therefore, the $\overline{\rm MS}$ mass anomalous dimension and its running take the form (3.24)

$$\overline{m}(\mu)=\overline{m}-\frac{\overline{m}}{\beta_0}\int_{\beta_{\overline{m}}}^{\beta}{\rm d}\tau\frac{F_{\overline{\rm MS}}(-\tau,0)}{\tau}, \qquad \gamma_{\overline{m}}=-\frac{\beta}{\beta_0}F_{\overline{\rm MS}}(-\beta,0), \tag{6.17}$$

6.1. Note that the $\overline{\rm MS}$ mass anomalous dimension is defined as $\frac{\mu}{\overline{m}}\frac{{\rm d}\overline{m}}{{\rm d}\mu}=2\gamma_{\overline{m}}$, and therefore $\gamma_{\overline{m}}=-\gamma_{\overline{\rm MS}}/2$. We remind the $1/\overline{m}$ global factor is effectively 1 in the large-$\beta_0$ limit.

where we choose the reference scale $\overline{m}(\overline{m}) \equiv \overline{m}$. Note that $\overline{m}(\mu)$ can be unambiguously obtained from its initial value $\overline{m}$, in contrast to the pole mass that must be obtained from $\overline{m}(\mu)$ and carries an ambiguity.

### 6.2.2 Study of $F_{\overline{\text{MS}}}$

Before applying the results in (6.15) and (6.17), and with the aim of making the following computations algorithmically efficient, let us present a brief study on the function $F_{\overline{\text{MS}}}(\epsilon, u)$. First, the two relevant functions $F_{\overline{\text{MS}}}(\epsilon, 0)$ and $F_{\overline{\text{MS}}}(0, u)$ are

$$\begin{aligned} F_{\overline{\text{MS}}}(\epsilon, 0) &= \frac{C_F}{3} \frac{(3-2\epsilon)\Gamma(4-2\epsilon)}{\Gamma^2(2-\epsilon)\Gamma(3-\epsilon)\Gamma(\epsilon)} \qquad (6.18) \\ &= 3C_F \exp\left\{-\frac{5\epsilon}{6} + \sum_{n=2}^{\infty} \frac{\epsilon^n}{n}[\zeta_n(2^n - (-1)^n - 3) - 2^n(1 + 2\cdot 3^{-n}) + 2 + 2^{-n}]\right\}, \\ F_{\overline{\text{MS}}}(0, u) &= 6C_F e^{\frac{5u}{3}} \frac{\Gamma(1-2u)\Gamma(1+u)}{(2-u)\Gamma(1-u)} \\ &= 3C_F \exp\left\{\frac{13}{6}u + \sum_{n=2}^{\infty} \frac{u^n}{n}[2^{-n} + \zeta_n(2^n + (-1)^n - 1)]\right\}, \end{aligned}$$

with $F_{\overline{\text{MS}}}(0, 0) = F_{0,0}^{\overline{\text{MS}}} = 3C_F$. In the second equalities of each expression we have written the expression in a way that facilitates its expansion in powers of $\epsilon$ and $u$ by computer programs, such that the fixed-order coefficients can be easily obtained. The relevant equalities can be found in (A.19) and (A.20). To carry out the expansion of the exponential of a series we use the recursive relation in (A.21).

The pole structure of $F_{\overline{\text{MS}}}(\epsilon, 0)$ and $F_{\overline{\text{MS}}}(0, u)$ is summarized in table 6.1.

| | Poles | | Order | Crossed |
|---|---|---|---|---|
| $F_{\overline{\text{MS}}}(\epsilon, 0)$ | $(2n+1)/2,$ | $n = 2, 3, 4\ldots$ | 1 | No |
| $F_{\overline{\text{MS}}}(0, u)$ | $(2n+1)/2,$ | $n = 0, 1, 2\ldots$ | 1 | Yes |
| | $-n,$ | $n = 1, 2, 3\ldots$ | 1 | No |
| | 2 | | 1 | Yes |

**Table 6.1.** Pole structure of $F_{\overline{\text{MS}}}(\epsilon, 0)$ and $F_{\overline{\text{MS}}}(0, u)$. The last column indicates whether the poles are crossed in the integrals in which each function appears.

As one can see, the poles of $F_{\overline{\text{MS}}}(\epsilon, 0)$ lay in the positive real axis at all the half-integer values starting from $5/2$; none of them is crossed in the integrals of (6.15) and (6.17), which explicitly justifies for the first time why we have been treating them as finite. In particular, neither the running $\overline{m}(\mu)$ nor $\gamma_{\overline{m}}$ are ambiguous. There is, however, one subtlety with regards to this function: the expansion $\sum_{i=0}^{\infty} \epsilon^i F_{i,0}^{\overline{\text{MS}}}$ only converges to $F_{\overline{\text{MS}}}(\epsilon, 0)$ within a convergence radius of $|\epsilon| = 5/2$, which is the location of the pole that is closest to the origin. This formally restricts $\beta<2.5$ in the associated perturbative series.

On the other hand, $F_{\overline{\text{MS}}}(0,u)$ presents poles at all positive half-integers and $u=2$, which are crossed in the first integral (6.15), explicitly showing it carries an ambiguity. The integrand of this ambiguous integral is[6.2]

$$B_{\overline{\text{MS}}}(u)\equiv\frac{F_{\overline{\text{MS}}}(0,u)-F_{\overline{\text{MS}}}(0,0)}{u}. \tag{6.19}$$

The subtraction ensures there is no singularity at $u=0$, but plays no role for poles with $u>0$. Therefore, $B_{\overline{\text{MS}}}(u)$ also presents poles at all the positive half-integers and $u=2$. Since

$$\underset{u=u_0}{\text{Res}}\,[B_{\overline{\text{MS}}}(u)]=\underset{u=u_0}{\text{Res}}\,[F_{\overline{\text{MS}}}(u)]\,/\,u_0 \tag{6.20}$$

for any $u_0\neq 0$, it suffices to study $F_{\overline{\text{MS}}}(u)$ around its singularities. For $u=(2n+1)/2$, $n\in\mathbb{N}$ we can write

$$F_{\overline{\text{MS}}}\left(0,\frac{2n+1}{2}\right)=3C_F e^{\frac{5(2n+1)}{6}}\frac{\Gamma(2n+2)\Gamma(-n)}{4^{2n}(3-2n)\Gamma(n+1)}, \tag{6.21}$$

where all the divergences come solely from the factor

$$\Gamma(-x)=\frac{(-1)^{n+1}}{\Gamma(n+1)(x-n)}+O(x-n)^0. \tag{6.22}$$

Then the residues at half integers acquire the form

$$\underset{u=(2n+1)/2}{\text{Res}}[F_{\overline{\text{MS}}}(0,u)]=3C_F e^{\frac{5(2n+1)}{6}}\frac{(-1)^{n+1}\Gamma(2n+2)}{4^{2n}(3-2n)\Gamma^2(n+1)}. \tag{6.23}$$

For $u=2$ the residue can be directly extracted from the expansion to be

$$\underset{u=2}{\text{Res}}[F_{\overline{\text{MS}}}(0,u)]=-C_F e^{\frac{10}{3}}. \tag{6.24}$$

With these results we can build the pole expansion for $B_{\overline{\text{MS}}}(u)$

$$B_{\overline{\text{MS}}}(u)\asymp -C_F\left[\frac{e^{\frac{10}{3}}}{2(u-2)}+\sum_{n=0}^{\infty}\frac{(-1)^n 6 e^{\frac{5(2n+1)}{6}}\Gamma(2n+1)}{4^{2n}(3-2n)\Gamma^2(n+1)\left(u-\frac{2n+1}{2}\right)}\right]. \tag{6.25}$$

In figure 6.2(*a*) we compare the exact function $B_{\overline{\text{MS}}}(u)$ with the above expansion, finding both are hardly distinguishable. This means $B_{\overline{\text{MS}}}(u)$ is strongly dominated by its pole structure.

6.2. Except for an overall factor of $(\Lambda_{\text{QCD}}/\overline{m}(\mu))^{2u}$, which we exclude from the pole analysis since it has no singularities.

In figure 6.2(*b*) we observe how the additional factor of $(\Lambda_{\text{QCD}}/\overline{m}(\mu))^{2u}$ acts as a damping factor that gets stronger as $u$ grows, suppressing the contribution from poles the further they are from the origin. This damping makes applying the truncation-based P.V. algorithm discussed in section 5.3 possible.

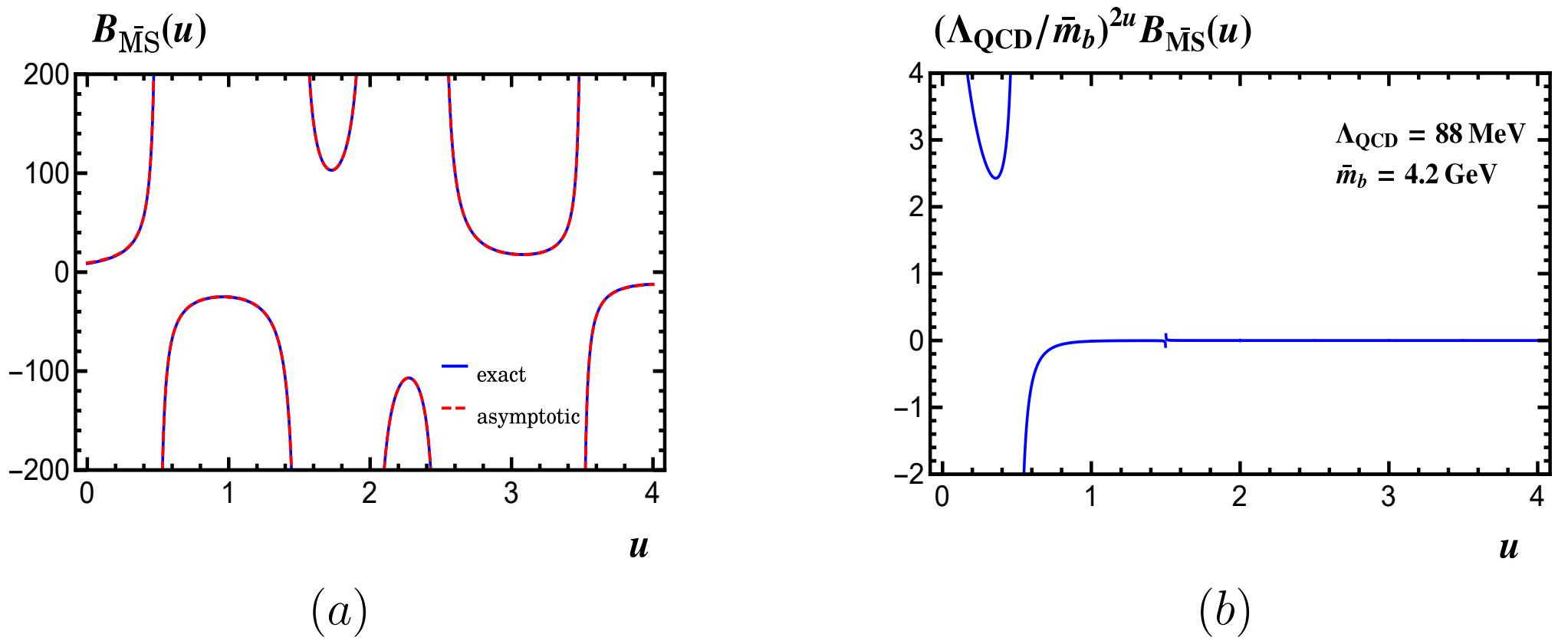

**Figure 6.2.** Panel (*a*): comparison between the function $B_{\overline{\text{MS}}}(u)$ and the pole expansion built from the expansion around its poles. Panel (*b*): damping effect of the $(\Lambda_{\text{QCD}}/\overline{m}(\mu))^{2u}$ accompanying $B_{\overline{\text{MS}}}(u)$ in the ambiguous integral.

It turns out the sum over all the residues at half-integer poles can be solved:

$$\begin{aligned}
&\sum_{n=0}^{\infty} \underset{u=(2n+1)/2}{\text{Res}}\left[\left(\frac{\Lambda_{\text{QCD}}}{\overline{m}(\mu)}\right)^{2u} B_{\overline{\text{MS}}}(0,u)\right] \qquad (6.26)\\
&= \sum_{n=0}^{\infty} 6C_F \left(\frac{e^{5/6}\Lambda_{\text{QCD}}}{\overline{m}(\mu)}\right)^{2n+1} \frac{(-1)^{n+1}\Gamma(2n+1)}{4^{2n}(3-2n)\Gamma^2(n+1)}\\
&= \frac{C_F}{2}\left(\frac{e^{5/6}\Lambda_{\text{QCD}}}{\overline{m}(\mu)}\right)\left[2-\left(\frac{e^{5/6}\Lambda_{\text{QCD}}}{\overline{m}(\mu)}\right)^2\right]\sqrt{4+\left(\frac{e^{5/6}\Lambda_{\text{QCD}}}{m_p}\right)^2},
\end{aligned}$$

which allows to obtain an expression for the full ambiguity of the $\delta_{\overline{\text{MS}}}$ series:

$$\begin{aligned}
\delta_\Lambda\{\delta_{\overline{\text{MS}}}(\mu)\} \;=\; &\frac{C_F \text{e}^{5/6}\Lambda_{\text{QCD}}}{2\beta_0}\left\{\left[2-\left(\frac{e^{5/6}\Lambda_{\text{QCD}}}{\overline{m}(\mu)}\right)^2\right]\sqrt{4+\left(\frac{e^{5/6}\Lambda_{\text{QCD}}}{\overline{m}(\mu)}\right)^2} \right. \qquad (6.27)\\
&\left. -\left(\frac{\text{e}^{5/6}\Lambda_{\text{QCD}}}{\overline{m}(\mu)}\right)^3\right\}.
\end{aligned}$$

Since, as discussed, $\overline{m}(\mu)$ is not ambiguous, the relation above corresponds to the ambiguity of computing $m_p$ from $\overline{m}(\mu)$, and at leading order does not depend on $\overline{m}(\mu)$:

$$\delta_\Lambda m_p = \frac{2\,\text{e}^{5/6}\,C_F}{\beta_0}\Lambda_{\text{QCD}} + \mathcal{O}(\Lambda_{\text{QCD}}^3). \qquad (6.28)$$

### 6.2.3 Results

We start by computing the $\overline{\text{MS}}$ anomalous dimension and running $\overline{m}(\mu)$. Figure 6.3 show the anomalous $\overline{\text{MS}}$ mass anomalous dimension $\gamma_{\overline{m}}$, which was first computed in reference [23] –see also reference [24]– and takes the explicit form

$$\gamma_{\overline{m}}(\beta) = -\frac{C_F}{3\beta_0}\frac{\beta\,(3+2\,\beta)\,\Gamma\,(4+2\,\beta)}{(2+\beta)\,\Gamma\,(1-\beta)\,\Gamma\,(2+\beta)^3}. \tag{6.29}$$

Our result reproduces the full QCD leading flavor structure up to 5 loops [25, 26, 27, 28], collected in the first column of Table 6.2. Comparing the perturbative series with the exact value obtained from its integral form, we find the set of graphs in figure 6.3.

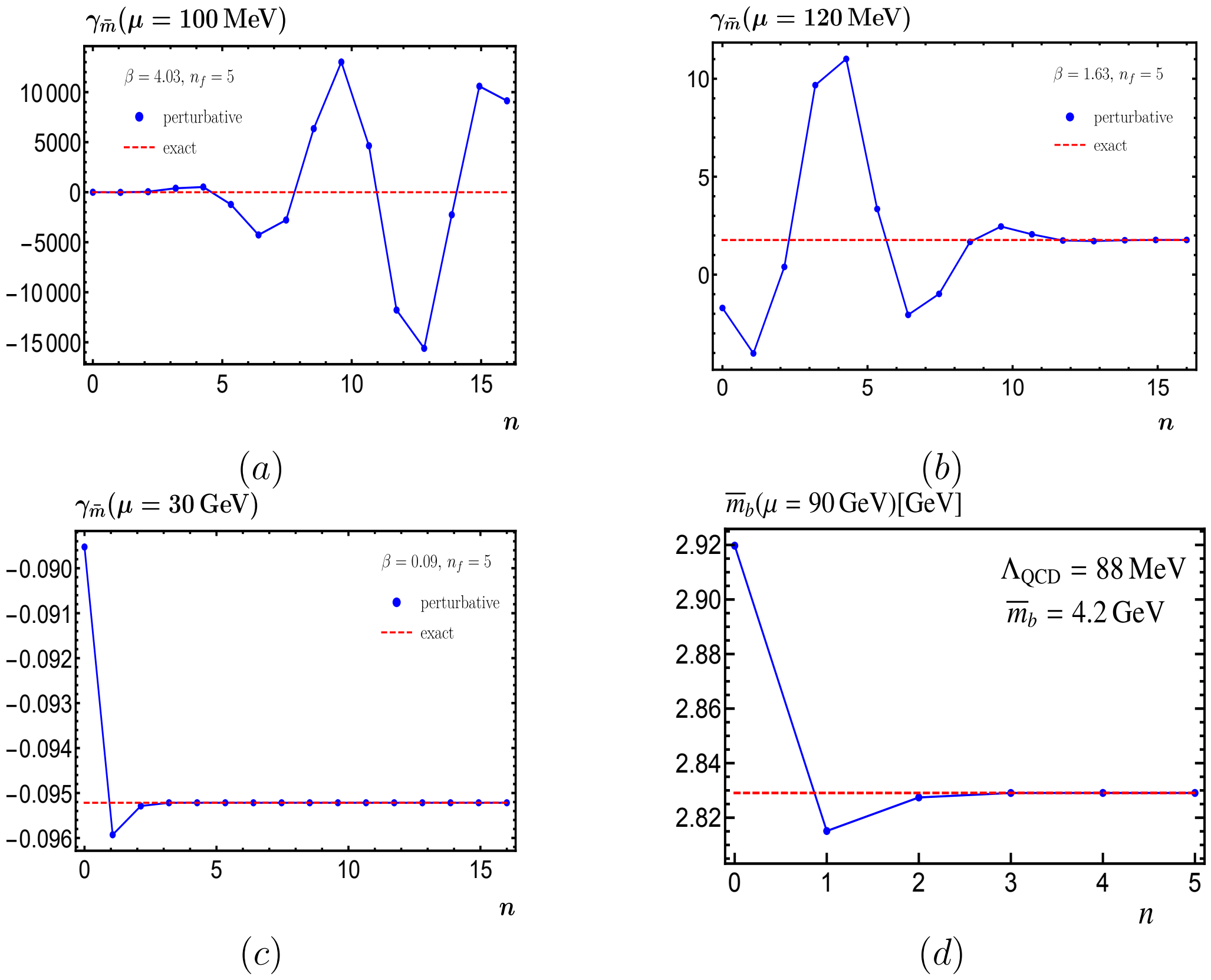

**Figure 6.3.** Panels $(a)$ to $(c)$: comparison for several values of $\mu$ of the perturbative behavior of $\gamma_{\overline{m}}$. In each case, thee exact value has been obtained by numerically solving the integral form. Panel $(d)$: fixed-order running of the $\overline{\text{MS}}$ bottom mass run from $\overline{m}_b$.

We observe that indeed, the anomalous dimension series converges for $\beta < 2.5$ (figures 6.3$(b)$ and 6.3$(c)$) but diverges for $\beta > 2.5$ (figure 6.3$(a)$). We remark convergence is achieved at sufficiently large orders even for $\beta > 1$ (figure 6.3$(b)$), while for $\beta < 1$ the convergence is greatly improved and already at NLL is very close to the exact value (figure 6.3($c$)). This behavior directly comes from the fact that the $\mu$-dependence of $\gamma_{\overline{m}}$ is entirely contained in $\beta(\mu)$, and thus the series' convergence is improven when $\mu \gg \Lambda_{\text{QCD}}^{n_f}$, as this lowers the value of $\beta$. The fixed-order running for the $\overline{\text{MS}}$ mass exhibits the same behavior and is displayed in figure 6.3($d$) in the perturbative region of $\alpha_s$ for the bottom quark mass. To evolve the $b$ quark mass we use $n_f = 5$ and $\bar{m}_b(\bar{m}_b) = 4.2\,\text{GeV}$ [29].

In figure 6.4 we compare the perturbative and closed forms of $\gamma_{\overline{m}}$ as functions of $\mu$ and $\alpha_s$. As before, the solid lines represent the exact value from the integral form and the dashed lines the perturbative series up to $n$ terms.

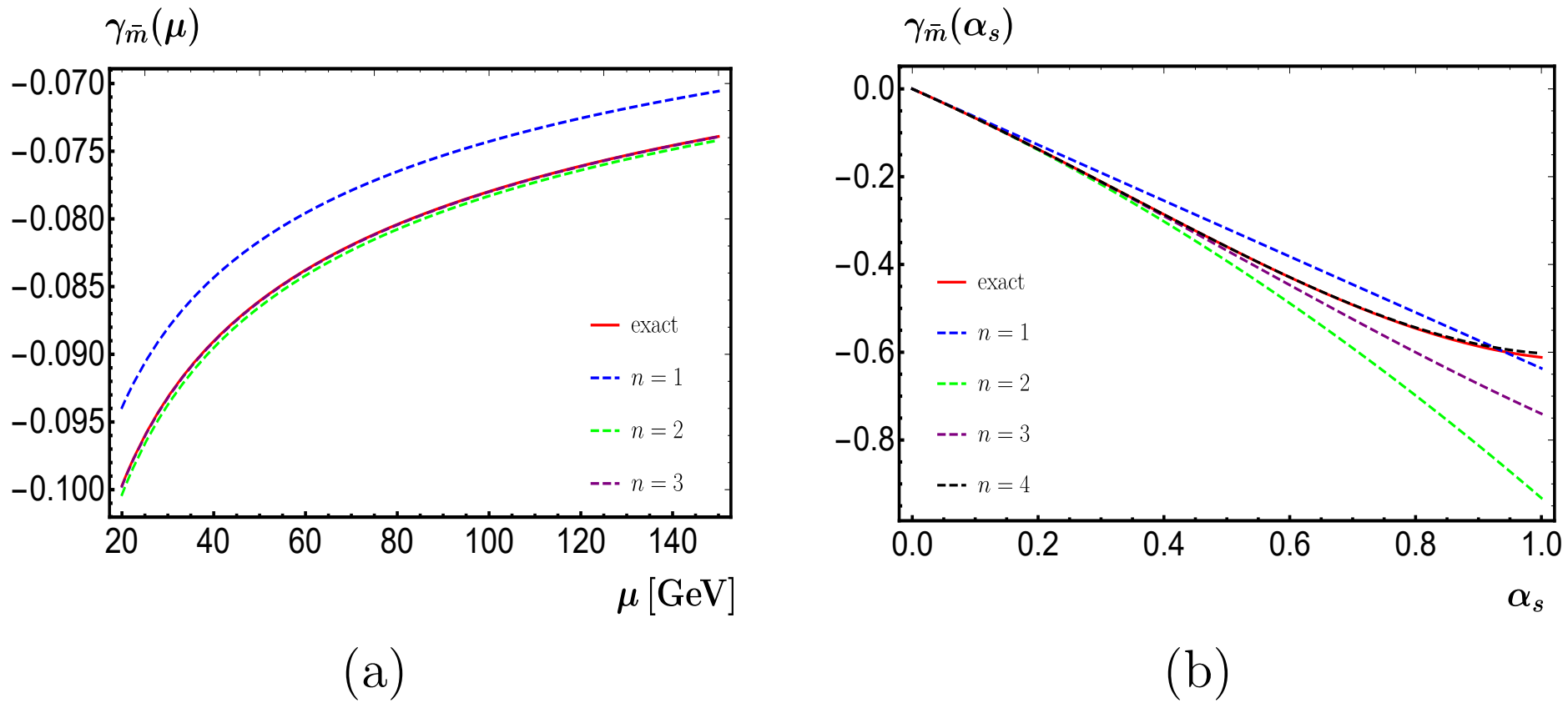

**Figure 6.4.** Anomalous dimension $\gamma_{\overline{m}}$ as function of $\mu$ (panel ($a$)) and $\alpha_s$ (panel ($b$)). The solid line represents the exact, non-ambiguous value and the dashed lines represent the partial sum of the $n$ first perturbative coefficients.

The exact running of the $\overline{\text{MS}}$ mass as function of $\mu$ can be found figure 6.7(b), which compares to the $R$-evolution of the MSR mass discussed in the next section.

Finally, we compute $m_p$ from $\overline{m}(\mu)$ with (6.15). For the $b$ and $t$ quarks the exact value and ambiguity are

$$m_b^{\text{pole}} = 4.94 \pm 0.07\,\text{GeV}, \qquad m_t^{\text{pole}} = 170.13 \pm 0.04\,\text{GeV}. \tag{6.30}$$

For the $t$ quark mass we use $n_f = 6$ and the initial value $\overline{m}_t(\overline{m}_t) = 160\,\text{GeV}$ [29]. To compute the ambiguous integral we employ the P.V. algorithm described in chapter 5, in which we integrate along the real axis until the next segment gives a contribution smaller than $10^{-10}$. The ambiguities in (6.30) have been set as the sum of the residues of the poles included this algorithm; they are not dependent on $\mu$ and agree with the closed expression (6.27) in 11 digits. The behavior of $m_b^{\text{pole}}$ and $m_t^{\text{pole}}$ as perturbative series can be seen in figure 6.5.

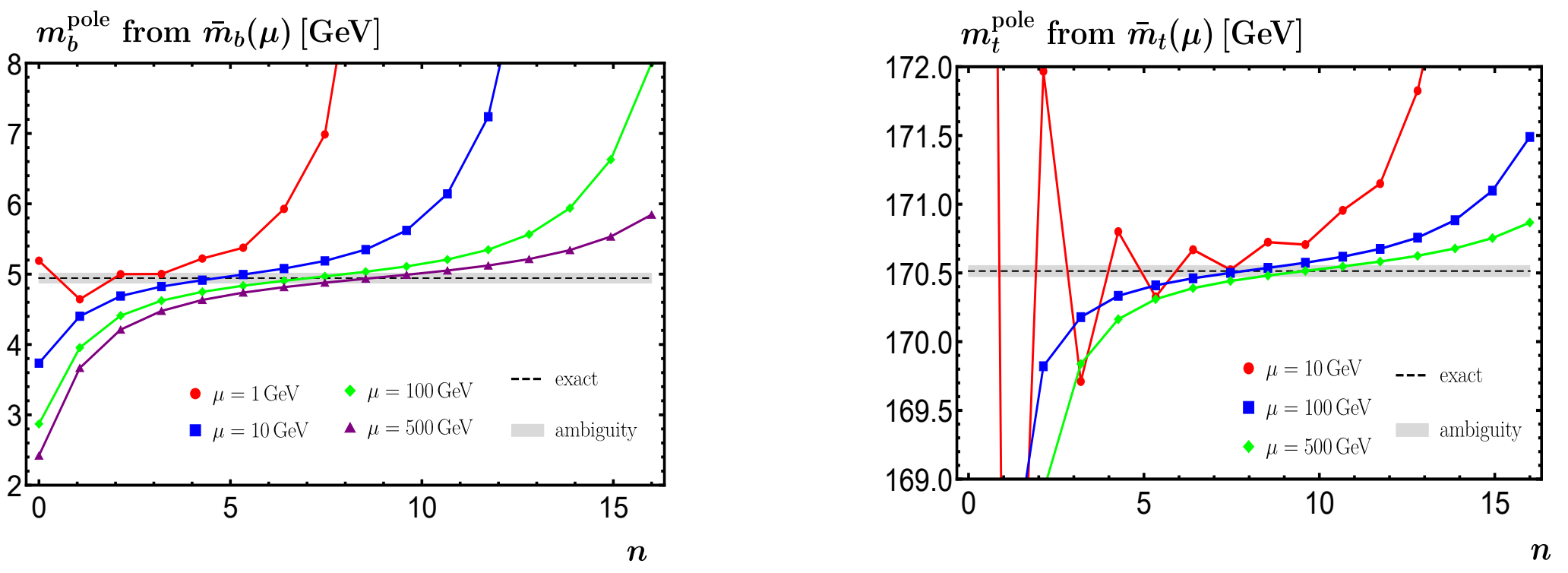

(*a*) (*b*)

**Figure 6.5.** Comparison for several values of $\mu$ of the perturbative behavior around the closed result for the pole mass in terms of the $\overline{\text{MS}}$ mass. In panel (*a*), $\overline{m}_b(\overline{m}_b) = 4.2\,\text{GeV}$, $\Lambda_{\text{QCD}}^{n_f=5} = 88\,\text{MeV}$ and the P.V. algorithm crosses the first five poles. In panel (*b*), $\overline{m}_t(\overline{m}_t) = 160\,\text{GeV}$, $\Lambda_{\text{QCD}}^{n_f=6} = 45\,\text{MeV}$ and the P.V. algorithm crosses the first three poles.

We see that the value the asymptotic series momentarily converges to does not depend on $\mu$, yet $\mu$ controls both when the convergent value is reached and when the divergent behavior sets in. In particular, when $\mu$ and the scale $\overline{m}$ are comparable in size, the perturbative series reaches the exact value sooner the smaller $\mu$ is, but in turn the divergences also start sooner. Conversely, the larger the value of $\mu$ the longer it takes to reach the exact value, but in turn the convergence lasts longer. When $\mu \ll \overline{m}$ both $\log(\mu/\overline{m})$ and $\beta(\mu)$ are large, and the series converges poorly –see the $\mu = 10$ GeV case in figure 6.5(*b*)–. This behavior is not seen for $\mu \gg \overline{m}$, where the large logarithms are tamed by the small value $\beta(\mu)$.

## 6.3 MSR mass

The MSR scheme [30, 31] is defined from the series relating the pole and $\overline{\text{MS}}$ masses. In particular, the series $\delta_{\text{MSR}}(R) = m_p - m^{\text{MSR}}(R)$ is built from the logarithm-free coefficients of $\delta_{\overline{\text{MS}}}$ as

$$\delta_{\text{MSR}}(R) \equiv \frac{R}{\beta_0}\sum_{n=1}^{\infty} c_{n,0}^{\overline{\text{MS}}}\beta_R^n = \frac{R}{\beta_0}\biggl\{\int_0^\infty \mathrm{d}\tau\, \mathrm{e}^{-\tau/\beta_R}\frac{F_{\overline{\text{MS}}}(0,\tau) - F_{\overline{\text{MS}}}(0,0)}{\tau} + \int_{-\beta_R}^0 \mathrm{d}\tau\,\frac{F_{\overline{\text{MS}}}(\tau,0) - F_{\overline{\text{MS}}}(0,0)}{\tau}\biggr\}, \tag{6.31}$$

where the coefficients $c_{n,0}^{\overline{\rm MS}}$ are computed with (3.29):

$$c_{n,0}^{\overline{\rm MS}} = \Gamma(n) F_{0,n}^{\overline{\rm MS}} - \frac{(-1)^n}{n} F_{0,n}^{\overline{\rm MS}}. \tag{6.32}$$

The closed form in the second line clearly indicates $\delta_{\rm MSR}(R)$ carries an ambiguity dominated by the renormalon at $\tau = 1/2$, which is the closest to the origin. It is also possible to use the running of $\beta$ to re-expand $\delta_{\rm MSR}$ in powers of $\beta = \beta(\mu)$. Plugging (2.29) for the running and expanding with (A.16) one obtains

$$\begin{aligned}\sum_{n=1}^{\infty} c_{n,0}^{\overline{\rm MS}} \beta_R^n &= \sum_{n=1}^{\infty}\sum_{i=0}^{\infty} c_{n,0}^{\overline{\rm MS}} \frac{(-1)^i \Gamma(n+i)}{\Gamma(n)\Gamma(i+1)} 2^i \beta^{n+i} L_R^i \\ &= \sum_{n=1}^{\infty}\sum_{i=n}^{\infty} c_{n,0}^{\overline{\rm MS}} \frac{(-1)^{i-n}\Gamma(i)}{\Gamma(n)\Gamma(i-n+1)} 2^{i-n} \beta^i L_R^{i-n} \\ &= \sum_{i=1}^{\infty} \beta^i \sum_{n=1}^{i} c_{n,0}^{\overline{\rm MS}} \frac{(-1)^{i-n}\Gamma(i)}{\Gamma(n)\Gamma(i-n+1)} 2^{i-n} L_R^{i-n} \\ &= \sum_{i=1}^{\infty} \beta^i \sum_{n=0}^{i-1} \frac{2^n(-1)^n\Gamma(i)}{\Gamma(i-n)\Gamma(n+1)} c_{i-n,0}^{\overline{\rm MS}} L_R^n,\end{aligned} \tag{6.33}$$

where we first took $i \mapsto i - n$, then switched the sums and in the last step took $n \mapsto i - n$. With this the expansion in powers of $\beta$ reads

$$\delta_{\rm MSR}(R) = \frac{R}{\beta_0} \sum_{n=1}^{\infty} \beta_R^n \sum_{n=0}^{i-1} c_{n,i}^{\rm MSR} L_R^i, \qquad c_{n,i}^{\rm MSR} \equiv \frac{2^n(-1)^n\Gamma(i)}{\Gamma(i-n)\Gamma(n+1)} c_{i-n,0}^{\overline{\rm MS}}. \tag{6.34}$$

To obtain the pole mass from $m^{\rm MSR}$ we need to compute the $R$-evolution. Taking into account that the pole mass is independent of $R$, the $R$-anomalous dimension of $\delta_{\rm MSR}$, $\gamma_R$, takes the form

$$\begin{aligned}\beta_0 \gamma_R(\beta_R) = -\beta_0 \frac{\mathrm{d}}{\mathrm{d}R} m^{\rm MSR}(R) &= -\sum_{n=1}^{\infty} c_{n,0}^{\overline{\rm MS}} \beta_R^n - R \sum_{n=1}^{\infty} n\, c_{n,0}^{\overline{\rm MS}} \beta_R^{n-1} \left( \frac{-2\beta_R^2}{R} \right) \\ &= \sum_{n=0}^{\infty} [2n\, c_{n,0}^{\overline{\rm MS}} - c_{n+1,0}^{\overline{\rm MS}}] \beta_R^{n+1} \\ &= \int_0^{\infty} \mathrm{d}\tau\, \mathrm{e}^{-\tau/\beta_R} (1-2\tau) \frac{F_{\overline{\rm MS}}(0,\tau) - F_{\overline{\rm MS}}(0,0)}{\tau}\end{aligned} \tag{6.35}$$

$$+\int_{-\beta_R}^{0}\mathrm{d}\tau\,\frac{F_{\overline{\rm MS}}(\tau,0)-F_{\overline{\rm MS}}(0,0)}{\tau}$$
$$+2\beta_R[F_{\overline{\rm MS}}(\tau,0)-F_{0,0}^{\overline{\rm MS}}].$$

From here one can observe that $\gamma_R$ is still ambiguous, but the pole at $\tau=1/2$ in the integrand of the first-line cancels, pushing the leading renormalon to $\tau=3/2$, thus making $\gamma_R$ less ambiguous than $\delta_{\rm MSR}$. On the other hand, $\gamma_R$ does not contain explicit dependence on $R$, so one can directly expand it in powers of $\beta=\beta(\mu)$:

$$\beta_0\gamma_R(\beta)=\sum_{n=0}\hat{\gamma}_R^n\;\beta^{n+1}=\sum_{n=1}^{\infty}[2n\,c_{n,0}^{\overline{\rm MS}}-c_{n+1,0}^{\overline{\rm MS}}]\beta^{n+1} \tag{6.36}$$
$$\hat{\gamma}_R^n=n![F_{0,n+1}^{\overline{\rm MS}}-2\,(1-\delta_{n,0})\,F_{0,n}^{\overline{\rm MS}}]+(-1)^n\left[\frac{F_{n+1,0}^{\overline{\rm MS}}}{n+1}+2\,(1-\delta_{n,0})\,F_{n,0}^{\overline{\rm MS}}\right].$$

The corresponding fixed-order coefficients $\hat{\gamma}_R^n$ are collected in the second column of Table 6.2. They reproduce the leading flavor structure of full QCD up to four loops [31].

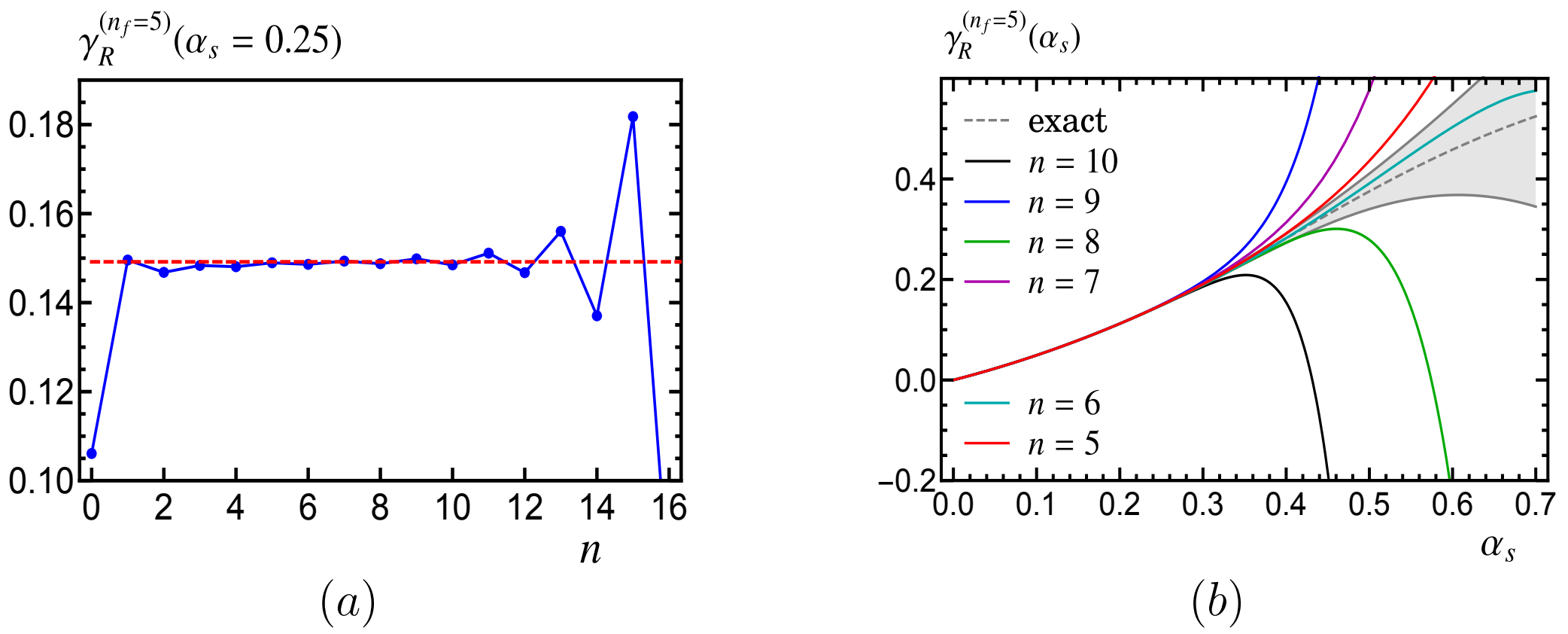

**Figure 6.6.** Panel ($a$): comparison of the R-anomalous dimension $\gamma_R\,(\alpha_s=0.25)$ computed with a partial sum including $n+1$ terms (blue dots) or through the closed integral form (red dashed line). Panel ($b$): Dependence of $\gamma_R$ with $\alpha_s$ in its exact form (gray band) and when including up to $\gamma_R^n$ with $n=5,6,7,8,9,10$ in red, cyan, magenta, green, blue and black, respectively.

In figure 6.6($a$) we compare the exact form for $\gamma_R$ (red dashed line) in (6.35) with the partial sum of (6.36) at various orders for a relatively large value of $\alpha_s$.

We observe the expected behavior: the partial sum oscillates around the exact value until $n \simeq 10$, and starts diverging after that order. In the right panel we show the dependence of the closed form for $\gamma_R$ with $\alpha_s$ (gray band) with the partial sum up to $\gamma_R^n$ for $5 \leq n \leq 10$.

The $R$-evolution of $m^{\rm MSR}(R)$ is obtained by integrating $\gamma_R$ as given in (6.35). This is achieved by switching variables from $\beta_R$ to $R$ using the second equality of the running in (2.29). In sum, one solves $R = \Lambda_{\rm QCD} e^{1/(2\beta_R)}$ and writes[6.3]

$$\begin{aligned}
m^{\rm MSR}(R_2) - m^{\rm MSR}(R_1) &= -\frac{1}{\beta_0}\sum_{n=0}^{\infty}\hat{\gamma}_R^n \int_{R_1}^{R_2} {\rm d}R\, \beta_R^{n+1} \qquad (6.37)\\
&= \frac{\Lambda_{\rm QCD}}{2\beta_0}\sum_{n=0}^{\infty}\hat{\gamma}_R^n \int_{\beta_{R_1}}^{\beta_{R_2}} {\rm d}\beta_R\, {\rm e}^{\frac{1}{2\beta_R}}\, \beta_R^{n-1}\\
&= \frac{\Lambda_{\rm QCD}}{2\beta_0}\sum_{n=1}^{\infty}\hat{\gamma}_R^{n-1}\left[\beta_{R_2}^{n-1} E_n\!\left(-\frac{1}{2\beta_{R_2}}\right) - \beta_{R_1}^{n-1} E_n\!\left(-\frac{1}{2\beta_{R_1}}\right)\right].\\
&= \Lambda_{\rm QCD}\left[{\rm Ei}\!\left(\frac{1}{2\beta_2}\right) - {\rm Ei}\!\left(\frac{1}{2\beta_1}\right)\right]\sum_{n=0} \frac{\hat{\gamma}_R^n}{2^n\,\Gamma(n-1)}\\
&\quad + \sum_{i=1} \Gamma(i)[R_2(2\beta_2)^i - R_1(2\beta_1)^i]\sum_{n=i}\frac{\hat{\gamma}_R^n}{2^n\,\Gamma(n-1)},
\end{aligned}$$

with $E_n(x) = \int_1^\infty {\rm d}t e^{-xt}/t^n$. To obtain the result in the second line integration by parts has been used to relate $I_n(\beta) \equiv \int {\rm d}\beta\, \beta^{n-1} \exp[1/(2\beta)]$ for consecutive values of $n$. Iterating this relation until $n = 0$ each $E_n\,(-x)$ can be written in terms of ${\rm Ei}(x)$ and a finite sum[6.4]. Combining this last sum with the one over $\hat{\gamma}_R^n$ one arrives at the final expression. If the sum includes up to $\hat{\gamma}_R^n$ the result is N$^n$LL accurate. It can be shown, at least numerically, that (6.37) corresponds to the subtraction $\delta_{\rm MSR}(R_2) - \delta_{\rm MSR}(R_1)$, which provides a strong cross-check for our computations.

In figure 6.7(*a*) we compare the exact value and perturbative series for the R-evolved result $m_t^{\rm MSR}(R)$. The former (cyan dashed line) is computed from the subtraction $\delta_{\rm MSR}(R_2) - \delta_{\rm MSR}(R_1)$ and the later (black joined dots) is computed with (6.37); both are evolved from $m^{\rm MSR}(\bar{m}_t) = \bar{m}_t$, which follow from the definition of $\delta_{\rm MSR}(R)$.

6.3. The integral in (6.37) is best solved in terms of the incomplete gamma function $\Gamma(a,z) = \int_z^\infty t^{a-1} {\rm e}^{-t}$. This function possesses the following relation with the exponential integral: $\Gamma(a,z) = z^a E_{1-a}(z)$, which we exploit to obtain the form in the following steps.

6.4. This result is completely analogous to the strategy followed by the `C++` public code REvolver [18].

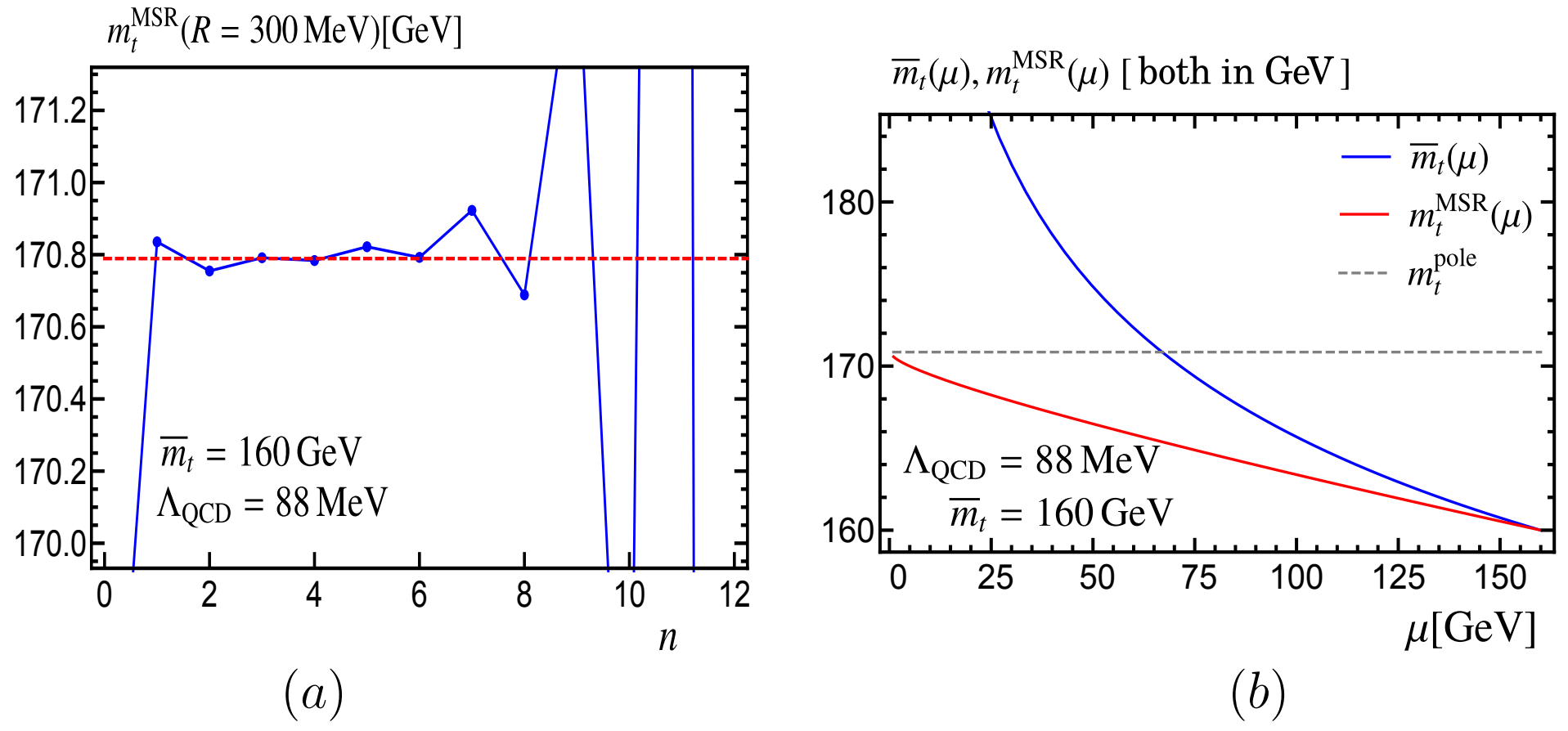

**Figure 6.7.** Panel $(a)$: MSR top quark mass $m_t^{\text{MSR}}(R)$ for $R = 300\,\text{MeV}$ and $\mu = 20\,\text{GeV}$, computed with a partial sum including $n+1$ terms or in an exact form (blue dots) and the exact value (red dashed line). Panel $(b)$: exact prediction for the top quark mass in the $\overline{\text{MS}}$ (blue), MSR (red) and pole (gray) schemes. The pole and MSR mass uncertainties are too small to be visible in the plot.

In figure 6.7$(b)$ we compare the $\overline{\text{MS}}$ and MSR evolution for the top quark mass for values of the renormalization scale smaller than $\bar{m}_t$. Both schemes coincide for $\mu = \bar{m}_t$ and, even though the $\overline{\text{MS}}$ evolution is unphysical for those scales, growing out of control and becoming even larger than the pole mass, we observe that $\bar{m}_t(\bar{m}_t/2) \simeq m_t^{\text{pole}}$. The numerical value of the MSR mass monotonically and smoothly grows as $R \to 0$ to reach $m_t^{\text{pole}}$ precisely at this limiting value. R-evolution for $R < \bar{m}_t$ is physical and should be used for processes that probe scales smaller than the quark mass itself.

In figure 6.8 the pole mass is estimated from its perturbative relation to the MSR mass. The colored dots use N$^n$LL-accurate R-evolution to compute $m_t^{\text{MSR}}(R)$, and the fixed-order relation between the pole and MSR masses including $n$ terms. The exact value and ambiguity obtained with the PV prescription are $m_t^{\text{pole}} = 170.85 \pm 0.07\,\text{GeV}$. In the left panel we fix $\mu$ and pick three values of $R$. We observe that for small $R$ the exact value of the pole mass, computed as $\bar{m}_t + \text{PV}[\delta_{\overline{\text{MS}}}(\bar{m}_t)]$, is properly estimated –within perturbative uncertainties– at low orders –since there are somewhat large logarithms for the choice $R = 2\,\text{GeV}$, the partial sum does not exactly hit the exact value for low $n$–. For larger $R$ it takes more orders to adequately predict the pole mass. The explanation for this behavior is simple: since the MSR mass at small values of $R$ is closer to the pole mass –formally both coincide if $R = 0$ –with a few matching corrections one already predicts its value. On the other hand, the large-order behavior depends only on $\mu$. In the right panel we keep $R$ fixed but change the value of $\mu$ –green dots are identical in both panels–. The three series become flat more or less at the same time –since this depends only on $R$ –but for smaller $\mu$ the divergent behavior for $n>8$ is more pronounced.

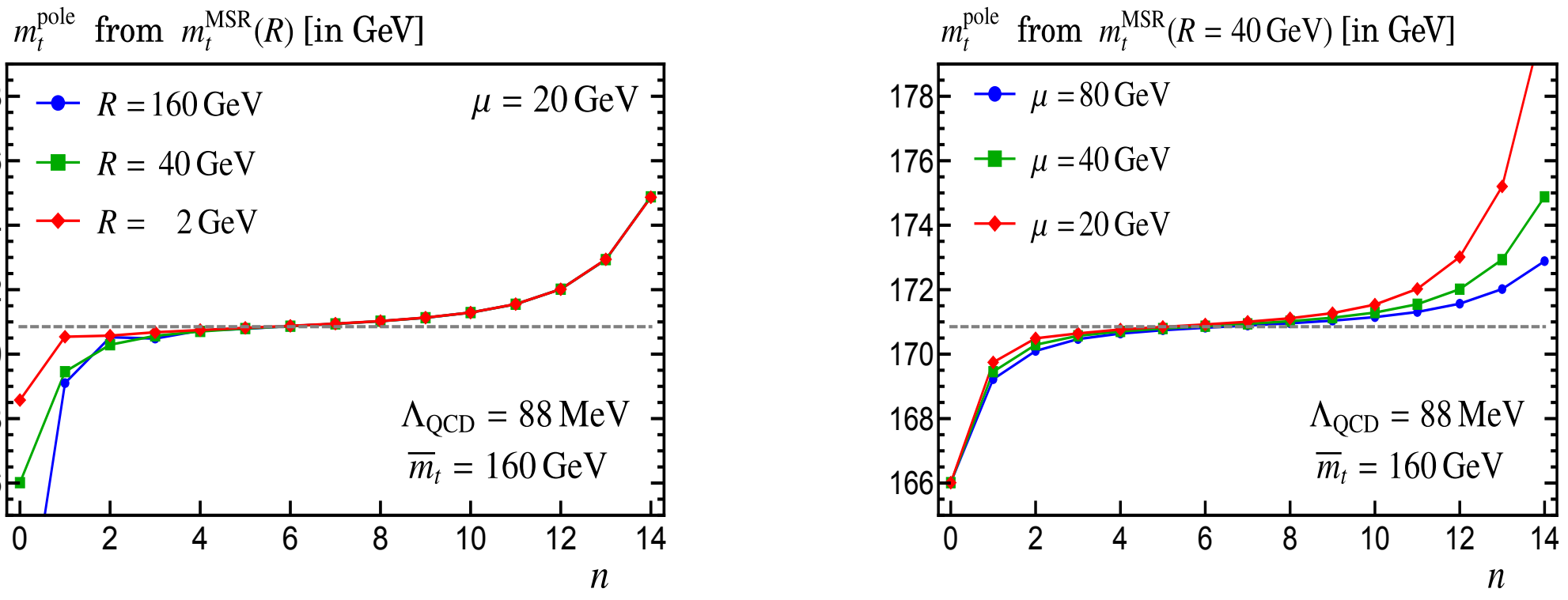

$(a)$ $(b)$

**Figure 6.8.** Determination of the top quark pole mass $m_t^{\rm pole}$ from its fixed-order relation to the MSR mass, using as boundary condition $R=\overline{m}_t$. In both panels the gray dashed line corresponds to the result computed as $\overline{m}_t + {\rm PV}[\delta_{\overline{\rm MS}}(\overline{m}_t)]$. The colored dots use $n$ terms in the fixed-order $\delta_{\rm MSR}$ series and R-evolution at N$^n$LL. In panel $(a)$ we keep $\mu=20$ GeV and show $R=2$ GeV, 40 GeV and 160 GeV in red, green and blue, respectively. In panel $(b)$ we keep $R=40$ GeV but use $\mu=20$ GeV, 40 GeV and 80 GeV in the same color ordering.

| $n$ | $\hat{\gamma}_m^n$ | $\hat{\gamma}_R^n$ |
|---|---|---|
| 0 | $-4$ | 5.33333 |
| 1 | $-3.33333$ | 14.3261 |
| 2 | 3.88889 | $-5.98376$ |
| 3 | 5.00534 | 21.989 |
| 4 | 0.442501 | $-26.3831$ |
| 5 | $-1.55019$ | 535.716 |
| 6 | $-0.670009$ | $-1397.07$ |
| 7 | 0.0812939 | 19031.9 |
| 8 | 0.123507 | $-100044$ |
| 9 | 0.022317 | $1.23114\times10^6$ |

**Table 6.2.** Numeric coefficients for the $\overline{\rm MS}$ mass and various $R$ anomalous dimensions. The hatted coefficients are defined as $f\equiv(1/\beta_0)\sum_{n=0}^{\infty}\hat{f}_n\,\beta^{n+1}$.

# Chapter 7
# Hadronic jets

The applications carried out in Chapters 8 and 9, as well as the computation of the oriented event-shape distribution for $e^+e^- \to$ hadrons in part III, belong to the field of jet physics. In the present chapter we momentarily pause the discussion on the large-$\beta_0$ formalism to introduce the relevant concepts.

## 7.1 Introduction

It is well-known that in high-energy collisions in which boosted partons are produced, the observed particles are not arranged isotropically in space, but instead conform a structure of energetic jets and background soft radiation (see figure 7.1(*a*)). Jets are sprays of particles traveling in nearly the same direction. They are produced by parton showering, a phenomenon in which a high-energy parton emits other partons as it travels, each of which emits more partons, as depicted in figure 7.1(*b*). The emission angles for energetic particles favored by QCD are small, and hence the majority of the partons tend to travel in groups of particles that are almost collinear.

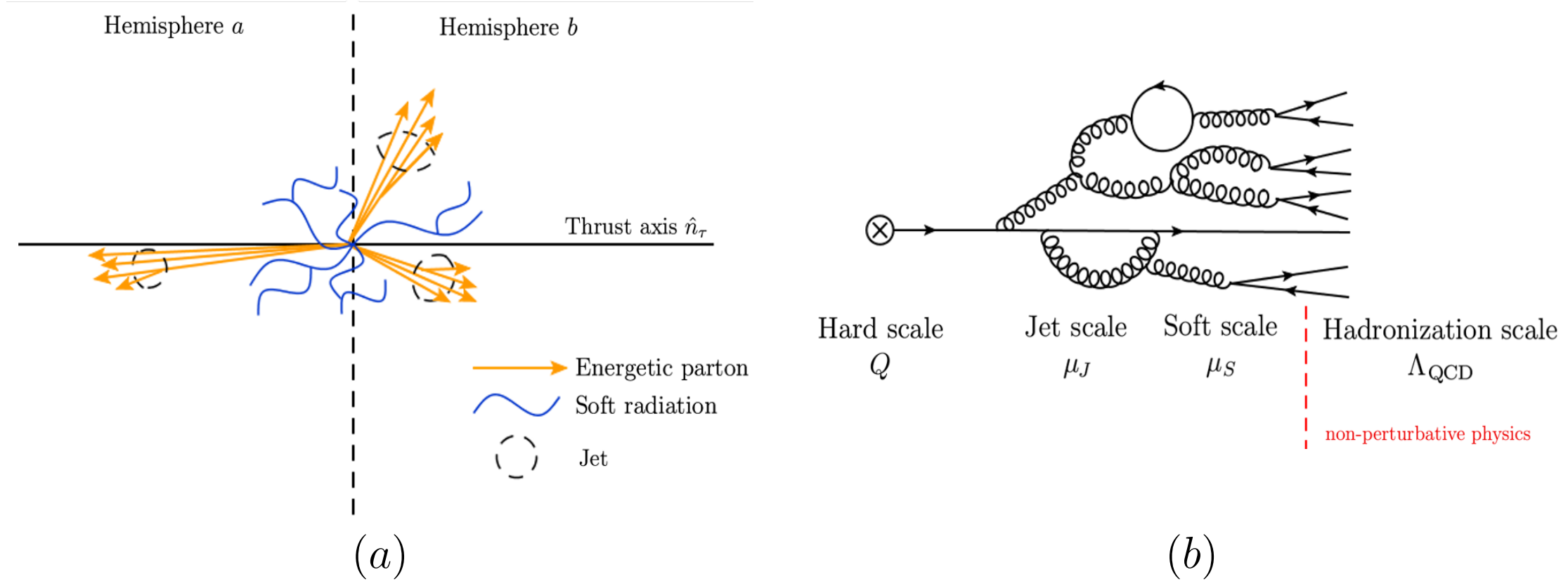

**Figure 7.1.** Panel (a): schematic representation of a high-energy collision producing partons. The particles in the final state group in a structure of jets and soft radiation. The plane perpendicular to the thrust axis divides the event in two hemispheres. Panel (b): schematic representation of parton showering. After the collision, an energetic parton is produced and emits more partons as it travels, which can also emit other partons. Since small emission angles are favored, the resulting group of particles travel in nearly the same direction. The more partons that are emitted, the softer their individual energies are, and at the non-perturbative scale $\Lambda_{\rm QCD}$ partons hadronize.

The description of jets presents a challenge both from the theoretical and experimental sides. Their complexity lies in that (1) each of them is constituted by a large number of partons; (2) they can be produced in moderate numbers at a single collision; and (3) they encompass different energy scales, like the high-energy scale $Q$ of the collision, the jet's scale $\mu_J \sim \sqrt{p_J^2}$ –which if the jet is initiated by a heavy quark is of order of its mass–, softer perturbative scales $\mu_S$ as the parton showering carries on, and finally, when the soft partons hadronize, the non-perturbative scale $\Lambda_{\rm QCD}$. The first two points make it not trivial to algorithmically resolve the number of jets of an event and consistently assign each particle to one of them, which creates problems when reconstructing the initial partons produced in the collision. The third point shows that jets test our knowledge of QCD from weak coupling $\alpha_s \ll 1$, where the dynamics are perturbative and partons travel individually, to strong coupling, where partons hadronize to the final, non-perturbative states detected in experiments [32].

To face the complexity of jet events on the theoretical side requires the combination of two main ingredients: event-shapes and effective field theories.

## 7.2 Event shapes

Event shapes are a class of observables that provide information on the jet structure of an event irrespectively from which criteria one uses to define jets (jet algorithms). Mathematically, they are functions of the momenta of the particles in the final state $(X)$, and can be studied by marginalizing the phase-space integration to include only those combinations of final-state momenta with a specific event-shape value $e$. This is done by including a delta function $\delta(e(X)-e)$ in the phase-space integration. The resulting object is called the event-shape distribution, represented as $\mathrm{d}\sigma/\mathrm{d}e$. Reviews on event shapes can be found in [32, 33, 34]. In the following we summarize the relevant information.

Those event shapes whose distribution near their minimum value $e = e_{\rm min}$ is dominated by two back-to-back narrow jets are known as two-jet event-shapes. Also, many event shapes are defined with respect to an axis called the *thrust axis* of the event, whose direction we specify with the unit vector $\hat{n}_\tau$. The thrust axis of a collision event is defined as the direction that maximizes the sum of the (absolute value of the) projections of the three-momenta of the particles in the final state,

$$\sum_{i\in X}|\hat{n}_\tau\cdot\vec{p}_i| \equiv \max_{\hat{n}}\sum_{i\in X}|\hat{n}\cdot\vec{p}_i|. \tag{7.1}$$

The components of any tri-momentum $\vec{p}$ along and normal to the thurst axis are often referred to as the longitudinal momentum $p_L \equiv |\hat{n}_\tau\cdot\vec{p}|$ and the transverse momentum $p_\perp \equiv |\vec{p}_\perp|$, respectively. Moreover, the plane perpendicular to the thrust axis divides the event in two hemispheres, which we denote as hemispheres $a$ and $b$. Unless otherwise stated, the thrust axis and all event shapes are computed in the center of mass frame. The CM frame is defined by the property $\sum_{i\in X}\vec{p}_i=\vec{0}$, and therefore the Lorentz-invariant quantity built from the square of the total momentum as $Q\equiv\sqrt{(\sum_{i\in X}p_i)^2}$ agrees with the sum of the energies of the particles and it is usually called the CM energy. The following list compiles common examples of event shapes.

1. **Thrust** [35]. Thrust is defined as the sum of the longitudinal momenta, normalized to the sum of the modulus of the momenta

$$T = 1-\tau = \frac{1}{\sum\limits_{i\in X}|\vec{p}_i|}\max_{\hat{n}}\sum_{i\in X}|\hat{n}\cdot\vec{p}_i| = \frac{1}{\sum\limits_{i\in X}|\vec{p}_i|}\sum_{i\in X}|\hat{n}_\tau\cdot\vec{p}_i|. \tag{7.2}$$

The name thrust is also employed on $\tau$, which can itself be written as

$$\tau = 1-\frac{1}{\sum\limits_{i\in X}|\vec{p}_i|}\max_{\hat{n}}\sum_{i\in X}|\hat{n}\cdot\vec{p}_i| = \frac{1}{\sum\limits_{i\in X}|\vec{p}_i|}\min_{\hat{n}}\sum_{i\in X}(|\vec{p}_i|-|\hat{n}\cdot\vec{p}_i|). \tag{7.3}$$

In an ideal dijet event all particles travel along the same direction $\hat{n}_\tau$ and then $\sum_{i\in X}|\hat{n}_\tau\cdot\vec{p}_i| = \sum_{i\in X}|\vec{p}_i|$, therefore $T=1$ and $\tau=0$, which shows $\tau$ is a two-jet event-shape.

2. **2-jettiness** [36]. 2-jettiness is similar to thrust but normalizes with the CM energy

$$T_2 = 1-\tau_2 = \frac{1}{Q}\max_{\hat{n}}\sum_{i\in X}|\hat{n}\cdot\vec{p}_i|, \qquad \tau_2 = \frac{1}{Q}\min_{\hat{n}}\sum_{i\in X}(E_i-|\hat{n}\cdot\vec{p}_i|). \tag{7.4}$$

In the dijet case one finds $T_2 = \sum_{i\in X} |\vec{p}_i|/Q$ and $\tau_2 = \sum_{i\in X}(E_i - |\vec{p}_i|)/Q$, which only take the values $T_2 = 1$ and $\tau_2 = 0$ for massless particles, so $\tau_2$ is a two-jet event-shape. Also, for massless particles one has $T_2 = T$ and $\tau_2 = \tau$, as can be seen immediately by the fact that when $p_i^2 = 0$, $Q = \sum_{i\in X} |\vec{p}_i|$.

3. **Jet broadening** [37]. Jet broadening is the total transverse momentum divided by $Q$

$$B = \frac{1}{Q}\sum_{i\in X} p_{i,\perp}.$$

4. **Hemisphere massess** [38, 39, 40]. Each hemisphere mass is defined as the total squared momentum in that hemisphere:

$$S_a = \left(\sum_{i\in a} p_i\right)^2, \qquad S_b = \left(\sum_{i\in b} p_i\right)^2. \tag{7.5}$$

The dijet case is recovered when $S_a$ and $S_b$ are both small, but, since they are independent –one can be at its minimum while the other is large– they are not two-jet event-shapes. In particular, the only way one of them can be zero is when the hemisphere is populated by a single massless particle. Two-jet event-shapes can be defined from $S_a$ and $S_b$, known as the heavy-jet mass $\rho$ and the sum of hemisphere masses $\rho_S$,

$$\rho = \frac{1}{Q^2}\max(S_a, S_b), \qquad \rho_S = \frac{S_a + S_b}{Q^2}. \tag{7.6}$$

In a dijet event both $\rho$ and $\rho_S$ are small. For completeness, the light-jet mass $\rho_L$ and the difference of hemisphere masses are also defined

$$\rho_L = \frac{1}{Q^2}\min(S_a, S_b), \qquad \rho_D = \frac{|S_a - S_b|}{Q^2}, \tag{7.7}$$

although, by the same argument as for $S_a$ and $S_b$, they are not two-jet event-shapes. In [33] a relation between the hemisphere masses and thrust/two-jettiness is given

$$\tau = 1 - \frac{Q}{\sum_{i\in X} |\vec{p}_i|}\sqrt{1 - 2\rho_S + \rho_D^2}, \tag{7.8}$$

$$\tau_2 = 1 - \sqrt{1 - 2\rho_S + \rho_D^2},$$

as well as the relations $\rho_S = \rho + \rho_L$ and $\rho_D = \rho - \rho_L$.

Using the idea that (the perpendicular plane to) an axis $\hat{n}$ divides the event into two hemispheres –which we call $n_+$ and $n_-$ to distinguish them from the hemispheres defined by $\hat{n}_\tau$, one can rewrite thrust and 2-jettiness as

$$T = \frac{2}{\sum\limits_{i\in X}|\vec{p}_i|}\max_{\hat{n}}(\hat{n}\cdot\vec{P}_{n_+}), \qquad T_2 = \frac{2}{Q}\max_{\hat{n}}(\hat{n}\cdot\vec{P}_{n_+}), \tag{7.9}$$

where $\vec{P}_{n_+}$ is the total momenta of hemisphere $+$, which depends on $\hat{n}$. To derive (7.9) one simply writes

$$\begin{aligned} \max_{\hat{n}}\sum_{i\in X}|\hat{n}\cdot\vec{p}_i| &= \max_{\hat{n}}\left(\sum_{i\in n_+}|\hat{n}\cdot\vec{p}_i| + \sum_{i\in n_-}|\hat{n}\cdot\vec{p}_i|\right) = \max_{\hat{n}}\left(\sum_{i\in n_+}\hat{n}\cdot\vec{p}_i - \sum_{i\in n_-}\hat{n}\cdot\vec{p}_i\right) \\ &= 2\max_{\hat{n}}\left(\hat{n}\cdot\sum_{i\in n_+}\vec{p}_i\right) = 2\max_{\hat{n}}(\hat{n}\cdot\vec{P}_{n_+}) = 2\max_{\hat{n}}(|\vec{P}_{n_+}|). \end{aligned} \tag{7.10}$$

We first split the sum over all particles in the final state $X$ into sums over the two hemispheres given by $\hat{n}$, then we used that we defined $\hat{n}$ so that $\hat{n}\cdot\vec{p}_i > 0$ for $i \in n_+$ and $\hat{n}\cdot\vec{p}_i < 0$ for $i \in n_-$. Finally we used the CM condition $\sum_{i\in X}\vec{p}_i = 0$, and the fact that, to obtain a maximum value, $\hat{n}$ must be proportional to $\vec{P}_{n_+}$.

Result (7.9) can be used to easily compute thrust for the cases of 2 and 3 particles. When there are 2 particles in the final state they travel back-to-back so the thrust axis lays along their common direction and each hemisphere has only one particle. Since $|\vec{p}_1| = |\vec{p}_2|$ and we can always define $\hat{n}_\tau\cdot\vec{p}_1 > 0$,

$$T = \frac{2|\vec{p}_1|}{|\vec{p}_1| + |\vec{p}_2|} = 1, \tag{7.11}$$

which is the dijet result. In the case of 3 particles there is always one particle in one of the hemispheres and two in the other. If the $n_+$ hemisphere contains one particle, then $\vec{P}_{n_+} = \vec{p}_i$; if it contains two particles, then $\vec{P}_{n_+} = \vec{p}_i + \vec{p}_j = -\vec{p}_k$, with $i \neq j \neq k$. In both cases $\vec{P}_{n_+}$ lays directly along the traveling direction of one of the particles. This leaves three possibilities for thrust,

$$T = \frac{2\max(|\vec{p}_1|, |\vec{p}_2|, |\vec{p}_3|)}{|\vec{p}_1| + |\vec{p}_2| + |\vec{p}_3|}, \tag{7.12}$$

and $\hat{n}_\tau$ is along the direction of the with highest momentum. Clearly, the discussion can be repeated for 2-jettiness, leading to

$$\begin{aligned} T_2 &= \frac{2|\vec{p}_1|}{Q} && (2\,\text{particles}), \\ T_2 &= \frac{2\max(|\vec{p}_1|, |\vec{p}_2|, |\vec{p}_3|)}{Q} && (3\,\text{particles}). \end{aligned} \tag{7.13}$$

# 7.3 Effective field theories

## 7.3.1 Effective field theories

In general, physical phenomena usually involve different energy scales. For example, a macroscopic body on Earth is subjected to gravity –at scales of the mass and radius of the Earth and its own mass–, the processes of its internal molecular and atomic structure –atomic scales– and even subatomic phenomena at the scales of the fundamental particles. Very often, the physics associated to each scale is deeply different: classical Gravitation theory may account accurately for the movement of the body under gravity, classical Thermodynamics and Solid State physics can describe with precision its atomic structure and Quantum Field Theory can address the fundamental interactions of it's atom's components. The relevant observation is that, for scales that are disparate enough, physical descriptions can be carried out by attending only to one or a few energy scales and with no knowledge from the others.

The Standard Model deals with the interactions of the fundamental particles, and thus describes phenomenology in which more than one energy scale may appear –as discussed, jets in QCD are an example–. In these cases, perturbation theory leads to an expansion in powers of the coupling constant where the coefficients are logarithms of the ratio of the energy scales involved in the process. If the scales are separated enough, their ratio and thus the logarithm becomes a big number, enough to spoil the perturbative expansion in $\alpha$. In this context, multi-scale problems are involved because the effects of heavier particles appear in low-energy processes through quantum corrections (propagators). Hence one needs to devise a method to treat them in a consistent way.

Effective Field Theories (EFTs) present a solution to this problem. They are kinematic expansions in the energy scale of interest, and their use in a multi-scale problem is capable of disentangling the physics associated to each scale. By using EFTs, the UV physics becomes local at low energies and hence gets encoded in universal (process-independent) Wilson coefficients, while the IR physics appears as operators in terms of the EFT degrees of freedom. Wilson coefficients are renormalized in the full theory, and one can pick a renormalization scalen that makes large logs disappear. Local operators develop anomalous dimensions that can be used to evolve from the high-energy renormalization scale to the low-energy typical scale, summing up large logs in the process. Each theory can be therefore independently renormalized at its own value $\mu_{\rm EFT}$, thus avoiding large logarithms. In the context of jets, there ar at least three distinct scales and one factorization theorems have been derived such that the original multi-scale problem is posed as a convolution of the results in each of the EFTs involved. In this thesis we consider two Effective Field Theories of QCD suitable for the description of jets, which we briefly address: Soft Collinear Effective Theory and Heavy Quark Effective Theory.

### 7.3.2 Soft Collinear Effective Theory

Soft Collinear Effective Theory (SCET) is a top-down effective field theory of Quantum Chromodynamics. Top-down EFTs are built by expanding a given theory, usually referred to as the full theory, into a specific kinematic region. Kinematic restrictions –this is, at the level of the momenta of the particles in the full theory–, translate into dynamical restrictions –this is, at the level of the fields in the full theory– and hence an EFT Lagrangian is derived from the original, full-theory Lagrangian. The top-down effective theory contains then less physical degrees of freedom than the full theory, and therefore describes a smaller variety of phenomenology, but this simplicity reflects on a better framework to develop computations as well as on improved precision in the results.

In the case of SCET, the expansion region is the region of *collinear* and *soft* particles, defined through their momenta. Let $Q$ be a high energy scale, usually referred to as the *hard scale*. Then, a four-momentum $p_n^\mu$ is said to be collinear to a given spatial direction $\vec{n}$ if:

1) It is energetic and highly boosted, i.e., it carries energy of order of the hard scale, $p_n^0 \sim Q$, but its spatial part is equally large, $|\vec{p}_n| \sim Q$, so that $p_n^2 = (p_n^0)^2 - |\vec{p}_n|^2 \ll Q^2$.

2) The majority of its energy flows along the direction $\vec{n}$, i.e., its component along $\vec{n}$ is large compared to its components along any other direction, so $\vec{n}\cdot\vec{p}_n \approx |\vec{p}_n|$.

In the following, we adopt the usual convention of referring to a momentum collinear to the direction $\vec{n}$ simply as $n$-collinear. On the other hand, a four-momentum $p_s^\mu$ is said to be soft if its energy and spatial parts are both much smaller than $Q$, i.e., $p_s^0 \ll Q$, $|\vec{p}_s|^2 \ll Q^2$, so that $p_s^2 \ll Q^2$. Soft momenta have no preferred spatial direction.

Both collinear and soft particles are light when compared to $Q$, and thus SCET describes the infrared limit of QCD. As real particles, this establishes a clear hierarchy between their invariant masses and the hard scale; as virtual particles, they stay close to their mass shell. This separation of energy scales allows to define a small scale parameter, such as $p^2/Q^2$, according to which the SCET Lagrangian is expanded. Operators scaling with higher powers of $p^2/Q^2$ encode smaller effects.

To better describe collinear and soft momenta the *light-cone basis* is chosen for the Minkowsky space. The light-cone basis is defined as a basis containing the vector $n^\mu$ and $\bar{n}^\mu$ that satisfy,

$$n\cdot n = n^2 = 0, \quad \bar{n}\cdot\bar{n} = \bar{n}^2 = 0, \quad n\cdot\bar{n} = C, \tag{7.14}$$

and two additional space-like four-vectors that are orthogonal both to $n$ and $\bar{n}$. The usual choice, which we follow in this thesis, is $n^\mu = (1, \vec{n})$ and $\bar{n}^\mu = (1, -\vec{n})$, which requires $|\vec{n}| = 1$ and fixes $C = 2$. Any four-vector $v^\mu$ is decomposed in the light-cone basis as

$$v^\mu = \frac{1}{2}(n\cdot v)\bar{n}^\mu + \frac{1}{2}(\bar{n}\cdot v)n^\mu + v_\perp^\mu = (v^+, v^-, \vec{v}_\perp), \tag{7.15}$$

where $v^+ \equiv n\cdot v, v^- \equiv \bar{n}\cdot v$, and $\vec{v}_\perp$ stands for the two-dimensional vector resulting from the projection of $v^\mu$ onto the orthogonal completion of the basis. The conditions for collinear and usoft momenta in this basis are

$$\begin{aligned} p_n^+, |\vec{p}_{n\perp}| \ll p_n^- \sim Q, &\quad \text{or} \quad p_n \sim Q(\lambda, 1, \lambda^2), \\ p_{\rm us}^+ \sim p_{\rm us}^- \sim |\vec{p}_{\rm us\perp}| \ll Q, &\quad \text{or} \quad p_{\rm us} \sim Q(\lambda^2, \lambda^2, \lambda^2), \end{aligned} \tag{7.16}$$

where $Q$ denotes the hard energy scale of the problem and $\lambda$ is the small parameter in which the SCET expansion is performed. Here, the subscript us denotes "ultra-soft" particles, and is used to distinguish them from the soft particles that scale as $p_{\rm us} \sim Q(\lambda, \lambda, \lambda)$. In this thesis we concern ourselves only with the particles defined in (7.16), and thus we use ultra-soft and soft synonymously.

SCET was initially proposed in the context of $B$-meson decays in 2000 [41], and it was followed in the next three years by several works [42, 43, 44, 45, 46, 47, 48], which also focused on $B$-meson physics and are considered to be the seminal papers of the theory. In the later years the community explored different applications for SCET, such as LHC high-energy collisions and jet physics [49, 50, 51, 52] and deep inelastic scattering [53, 54]. It wasn't until the years 2013-2015 that SCET became a standard in QFT books and courses: reviews such as [55, 56] were made from their respective sets of lectures, and Matthew D. Schwartz dedicated to SCET a chapter of his textbook on a modern approach to the QFT and the Standard Model [16].

At leading power in the SCET expansion, the SCET Lagrangian contains a copy of the QCD Lagrangian for the usoft quark and gluon fields and a Lagrangian for collinear quarks and gluons. The propagators and interactions belonging to the collinear sector are shown in figure 7.2.

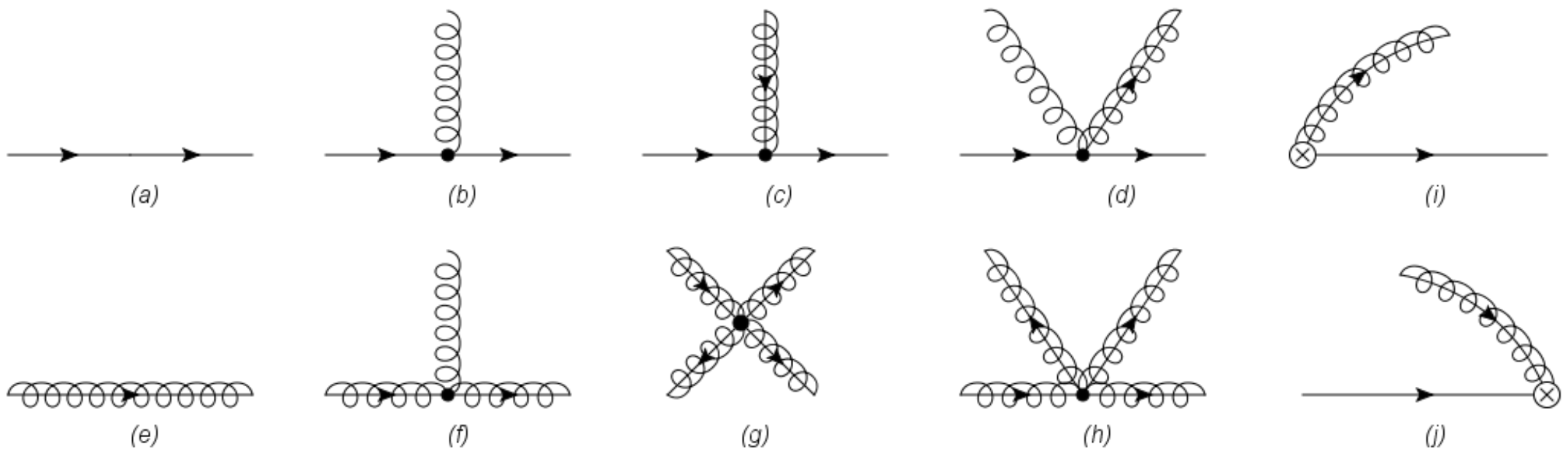

**Figure 7.2.** Interactions between collinear quarks (fermion line), collinear gluons (curly line intersected with a fermion line) and usoft gluons (curly line).

For the purposes of this thesis we will be using the Feynman rules corresponding to the following diagrams:

$$R_{(a)} = i\frac{\not{n}}{2}\frac{p^-}{p^2 - m^2}, \tag{7.17}$$

$$R_{(b)} = i g_s t^a \frac{\not{\bar{n}}}{2} n^\mu,$$

$$R_{(c)} = i g_s t^a \left[ n^\mu + \frac{\gamma_\perp^\mu(\not{p}_\perp + m)}{p^-} + \frac{(\not{p}'_\perp - m)\gamma_\perp^\mu}{p'^-} + \frac{(\not{p}'_\perp - m)(\not{p}_\perp + m)}{p^- p'^-} \right] \frac{\not{\bar{n}}}{2},$$

$$R_{(i)} = R_{(j)} = g_s t^a \frac{\bar{n}^\mu}{p^-} = g_s t^a \frac{\bar{n}^\mu}{p^-}.$$

where $\mu$ and $a$ label the quark-gluon interaction vertex, $m$ is the quark's mass and $p$ and $p'$ are the quark's incoming and outgoing momentum, respectively. In diagrams $(i)$ and $(j)$ the vertex denoted with a cross accounts for various gluon insertions in the collinear quark line that combine into a Wilson line made by collinear gluon fields –for further details see [49, 56]–.

### 7.3.3 Heavy Quark Effective Theory

Heavy Quark Effective Theory (HQET) [57, 58, 59, 60, 61] describes a heavy quark interacting with soft degrees of freedom. The quark's mass plays the role of the hard scale and its momentum is parametrized as $p^\mu = mv^\mu + r^\mu$, where $r^\mu$ has usoft scaling. A modification to a reference frame where the heavy quark is boosted is carried out in [49, 50]. The Feynman rules follow from the corresponding SCET rules with $p^\mu = mv^\mu + r^\mu$ keeping terms at leading power. For further details the reader is referred to [49].

## 7.4 $e^+e^- \to$ hadrons

Let us now focus on the collision $e^+e^- \to$ hadrons, which we will be treating in the following chapters. The event-shape distribution of $e^+e^- \to$ hadrons is given by

$$\begin{aligned} \frac{\mathrm{d}\sigma}{\mathrm{d}e} &\equiv \sum_X \frac{\mathrm{d}\sigma\,(e^+e^- \to X)}{\mathrm{d}e} \\ &= \frac{1}{2Q\sqrt{Q^2-4m_e^2}} \sum_X \int \mathrm{d}\Phi_X\, |\bar{\mathcal{M}}\,(e^+e^- \to X)|^2\, \delta\,(e(X)-e), \end{aligned} \tag{7.18}$$

where $\mathrm{d}\Phi_X$ denotes the Lorentz-invariant phase-space corresponding to the hadronic final state $X$, $Q$ denotes the center of mass energy of the collision and $m_e$ is the electron mass. The sum over $X$ is understood to include all the final states that contribute to the distribution at a given order in the electroweak and strong coupling constants $\alpha_{\text{ew}}$ and $\alpha_s$. At leading order in $\alpha_{\text{ew}}$, this process occurs first through the $s$-channel for $e^+e^- \to q\bar{q}$ mediated by a photon or a $Z$ boson, and then the final $N$-particle state $X$ is generated by parton emission from the $q\bar{q}$ pair (see figure 7.3).

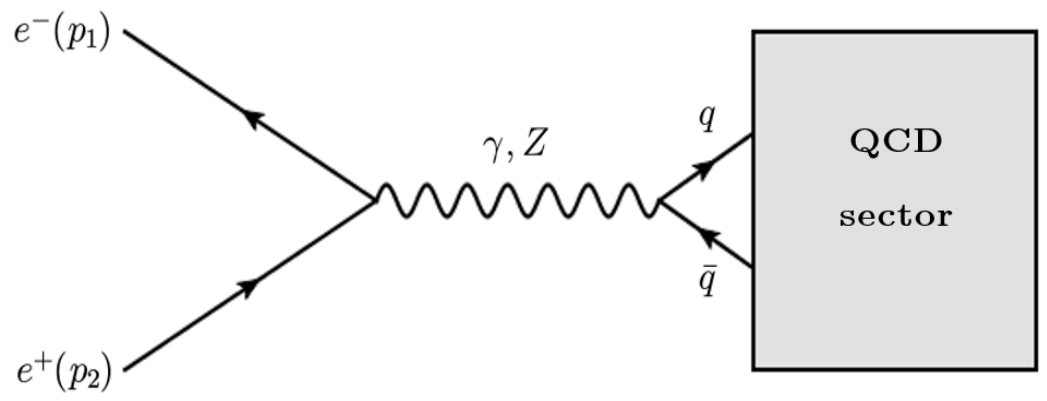

**Figure 7.3.** Class of Feynman diagrams contributing to $e^+e^- \to$ hadrons at $\mathcal{O}(\alpha_{\text{ew}})$. The box stands for any $N$-particle state produced only by QCD interaction.

With regards to this thesis' concern, and always at LO in the electroweak coupling constant, the general expression (7.18) serves as the starting point for two different sets of computations. On the one hand, we carry out its direct computation at NLO for massive quarks. The novelty of this result lies in the explicit dependence on the orientation of the thrust axis with respect to the initial $e^+e^-$ beam, i.e., $\mathrm{d}\sigma/(\mathrm{d}e\,\mathrm{d}\cos\theta_T)$, which was only done before for massless quarks [33]. This oriented cross-section is first computed for a generic event-shape up to a few phase-space, event-shape dependent integrals, and then some particular common event-shapes are studied. Being a fixed-order computation, it is developed outside the large-$\beta_0$ framework in part III.

On the other hand, the second set of computations begins from the fact that, for dijet events, the projection of the event-shape cross-section onto SCET and bHQET allows the derivation of factorization theorems that are valid at all orders in $\alpha_s$. In particular, we focus on thrust. If the initial $q\bar{q}$ pair is light, the cross-section (7.18) can be matched onto SCET to get [62], [63]

$$\begin{aligned}\frac{1}{\sigma_0}\frac{\mathrm{d}\sigma}{\mathrm{d}\tau} &= Q^2 H_Q(Q,\mu)\int\!\mathrm{d}\ell\, J_\tau(Q^2\tau - Q\ell,\mu)\, S_\tau(\ell,\mu),\\ J_\tau(s,\mu) &\equiv \int_0^s \mathrm{d}s'\, J_n(s-s',\mu)\, J_n(s',\mu).\end{aligned} \tag{7.19}$$

Here, $H_Q$, $J_n$ and $S_\tau$ are the hard, jet and soft functions, respectively. In the case the $q\bar{q}$ pair is massive (e.g. the top quark) a new scale –the quark mass $m$– becomes relevant in the problem. This mass changes the SCET jet function –secondary production of heavy quarks through gluon splitting affects also the soft function starting at $O(\alpha_s^2)$–. In the peak of the distribution the jet invariant mass and $m$ are very similar, indicating that quarks are very boosted. In this regime one can match SCET with massive quarks onto bHQET, what allows to sum up a new class of large logarithms and treat the top quark decay products in an inclusive way through the following factorization theorem [62, 63]:

$$\begin{aligned}\frac{1}{\sigma_0}\frac{\mathrm{d}\sigma_{\mathrm{bHQET}}}{\mathrm{d}\tau} &= Q^2 H(Q,\mu_m) H_m\!\left(m,\frac{Q}{m},\mu_m,\mu\right)\!\int\!\mathrm{d}\ell\, B_\tau\!\left(\frac{Q^2\tau - Q\ell}{m} - 2m,\mu\right) S_\tau(\ell,\mu),\\ B_\tau(\hat{s},\mu) &= m\int_0^{\hat{s}} \mathrm{d}\hat{s}'\, B_n(\hat{s}-\hat{s}',\mu)\, B_n(\hat{s}',\mu)\,,\end{aligned} \tag{7.20}$$

The difference between equations (7.20) and (7.19) is that the former has an additional hard factor $H_m$ that accounts for the matching between SCET and bHQET, and that the SCET jet function $J_n$ has been replaced by its bHQET counterpart $B_n$. Also, the number of active quark flavors has been reduced by one unit.

The different pieces of these factorization theorems can be computed individually and, with the exception of the soft function, they are matrix elements. The technology developed in the previous chapters can be then applied to compute their large $\beta_0$ contributions. In particular, since SCET and bHQET matrix elements have cusp-anomalous dimension, the results in chapter 4 are used. In chapter 8 we deal with the SCET factorization theorem and compute the hard function $H_Q$ and the jet function $J_n$; in chapter 9 we compute the bHQET mass-scale hard function $H_m$ and jet function $B_n$. Indications on how to compute each element are provided in their respective sections.

# Chapter 8

# Applications II: Jet factorization theorem in SCET

## 8.1 Hard function

The hard function $H_Q(Q, \mu)$ is the modulus squared of the matching coefficient $C_H(Q^2, \mu)$ between SCET and full QCD dijet operators. For its computation one needs to determine some matrix element –or Green function– of the vector current operator in both theories. Since $C_H$ does not depend on the specific process used for its computation one usually chooses the simplest possibility, which in this case is the matrix element between the vacuum and a pair of on-shell quarks. This choice implies that all diagrams in SCET as well as self-energy diagrams in QCD are scaleless and vanish [53]. The QCD computation is often referred to as the quark vector form factor. Therefore, to obtain $H_Q$ at leading order in the large-$\beta_0$ expansion one only needs to compute the diagram shown in figure 8.1 with the shifted gluon propagator.

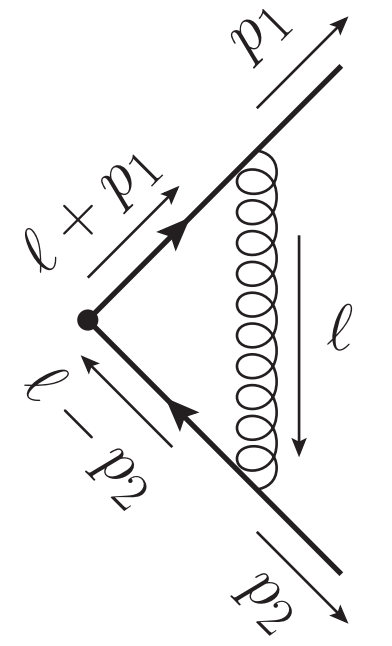

**Figure 8.1.** Vector form factor for massless quarks with on-shell momenta $p_1$ and $p_2$ and virtual loop momentum $\ell$.

The virtual correction to the bare vector form factor takes the shifted form

$$C_{H,\mathrm{sh}}(Q^2)\,\gamma^\mu \equiv i g_0^2\, C_F \int \frac{\mathrm{d}^d\ell}{(2\,\pi)^d} \frac{\gamma^\alpha(\not{\ell}+\not{p}_1)\gamma^\mu(\not{\ell}-\not{p}_2)\gamma_\alpha}{(\ell^2)^{1+h}\,(\ell+p_1)^2\,(\ell-p_2)^2} \equiv i g_0^2\, C_F I^\mu_{\mathrm{QCD}} \tag{8.1}$$

To solve the integral $I^\mu_{\mathrm{QCD}}$, we simplify the numerator to

$$\begin{aligned}\gamma^\alpha(\not{\ell}+\not{p}_1)\gamma^\mu(\not{\ell}-\not{p}_2)\gamma_\alpha \;\doteq\;& 4p_1p_2\gamma^\mu + 2(1+\epsilon)\gamma^\mu k^2 + 4(1-\epsilon)\gamma_\nu k^\nu k^\mu \\ & -2\gamma^\mu(p_1+k)^2 - 2\gamma^\mu(p_2-k)^2.\end{aligned} \tag{8.2}$$

Here the dotted equal sign $\doteq$ stands for “equal up to terms that will vanish” –either in the integration or by Dirac’s equation for massless quarks–. This makes the integral to be written as

$$\begin{aligned} I^\mu_{\mathrm{QCD}} \;\doteq\;& 2Q^2\gamma^\mu I_3(1, Q^{1+h}, Q(p_1), Q(-p_2)) + 2(1+\epsilon)\gamma^\mu I_3(1, Q^h, Q(p_1), Q(-p_2)) \\ & +4(1-\epsilon)\gamma_\nu I_3(\ell^\mu\ell^\nu, Q^{1+h}, Q(p_1), Q(-p_2)).\end{aligned} \tag{8.3}$$

The tensor integral can be further manipulated, as not all of its terms survive upon the contraction with $\gamma_\nu$ due to Dirac’s equation. In fact, from the result in (C.13) we only need to keep

$$I_3(\ell^\mu\ell^\nu, Q^{1+h}, Q(p_1), Q(-p_2)) \doteq \frac{g^{\mu\nu}}{2(1-\epsilon)}\left[A + \frac{2}{p_1{\cdot}p_2}B\right], \tag{8.4}$$

where in our case

$$A = I_3(1, Q^h, Q(p_1), Q(-p_2)), \qquad B = \frac{1}{4} I_3(1, Q^{h-1}, Q(p_1), Q(-p_2)). \tag{8.5}$$

Therefore, since all the contributions have reduced to the same integral $I_3$ with different shifts, we define $J_{\mathrm{QCD}}(h) \equiv I_3(1, Q^h, Q(p_1), Q(-p_2))$ and $Q^2 \equiv 2p_1{\cdot}p_2$ and compactly write

$$I^\mu_{\mathrm{QCD}} = 2\gamma^\mu\left[(2+\epsilon)J_{\mathrm{QCD}}(h) + Q^2 J_{\mathrm{QCD}}(h+1) + \frac{1}{Q^2}J_{\mathrm{QCD}}(h-1)\right], \tag{8.6}$$

which using the solution in (C.12) leads to

$$\begin{aligned}\delta C_{H,\mathrm{sh}}(Q^2) &= \left(\frac{g_0}{4\pi}\right)^2 2C_F(4\pi)^\epsilon(-1)^h(-Q^2)^{-h-\epsilon}[2\epsilon^3+3(h-1)\epsilon^2+(h^2-2h+3)\epsilon-2]\\ &\quad\times\frac{\Gamma(1+h+\epsilon)\Gamma^2(-h-\epsilon)}{\Gamma(3-h-2\epsilon)}.\end{aligned} \tag{8.7}$$

From here it is immediate to identify the function $a(h,\epsilon)$ and write

$$\begin{aligned}G_{C_H}(\epsilon,u) &= 2C_F\mathrm{e}^{\gamma_E u}[(u-1)\epsilon^2+(u^2-2u+3)\epsilon-2]\\ &\quad\times\frac{\Gamma(1+u)\Gamma^2(1-u)}{\Gamma(3-u-\epsilon)}T^{\frac{u}{\epsilon}-1}(\epsilon).\end{aligned} \tag{8.8}$$

With this function we can build the large-$\beta_0$ contribution to $C_H(Q^2)$ using the results in chapter 4. More interestingly, the $\beta_0$LO contribution to the hard matching coefficient can also be computed from $G_{C_H}$. This is because the squared modulus of a series in the large-$\beta_0$ limit is $|A|^2=1+2\mathrm{Re}(\delta A)+O(1/\beta_0^2)$, where $\delta A$ is the $O(1/\beta_0)$ term[8.1]. In the case of the hard matching coefficient one computes $H_Q(Q)=|C_H(Q^2+\mathrm{i}0^+)|^2$, where all the complex dependence resides in

$$(-Q^2-\mathrm{i}0^+)^{-h-\epsilon}=(Q^2\mathrm{e}^{-\mathrm{i}\pi})^{-h-\epsilon}=Q^2\mathrm{e}^{\mathrm{i}\pi(h+\epsilon)}, \tag{8.9}$$

so that the real part changes the sign of $Q^2$ and adds a cosine:

$$\begin{aligned}&H_Q(Q)=1+2\delta C_H(-Q^2)\cos\left[\pi(h+\varepsilon)\right],\\ &G_{H_Q}(\epsilon,u)=2G_{C_H}(\epsilon,u)\cos(\pi u).\end{aligned} \tag{8.10}$$

However, it is useful to notice that the hard anomalous dimension, $\gamma_H$, can be obtained directly from $C_Q(-Q^2)$, as

$$\gamma_H=\mu\frac{\mathrm{d}}{\mathrm{d}\mu}H_Q=\mu\frac{\mathrm{d}}{\mathrm{d}\mu}|C_Q|^2=2\mu\frac{\mathrm{d}}{\mathrm{d}\mu}\mathrm{Re}(\delta C_Q) \tag{8.11}$$

Next we present the relevant expansions to compute the closed forms in (4.34). In the second line we partially expand each function to easily compute the fixed-order coefficients in (4.32) recursively. We find the values $G_{0,0}^{C_Q}=-2C_F$, $G_{1,0}^{C_Q}=10C_F/3$ and

8.1. To see this one simply needs to consider $\mathrm{Re}(\delta A)\equiv\delta A_R\sim O(1/\beta_0)$ and $\mathrm{Im}(\delta A)=\delta A_I\sim O(1/\beta_0)$ and write $|A|^2=|1+\delta A_R+\mathrm{i}\delta A_I|^2=(1+\delta A_R)^2+(\delta A_I)^2=1+2\delta A_R+O(1/\beta_0^2)$.

$G_{0,1}^{C_Q} = -19C_F/3$, and the functions

$$\begin{aligned} G_{C_Q}(\epsilon,0) &= -\frac{C_F}{3\pi}\frac{\Gamma(4-2\epsilon)\sin(\pi\epsilon)}{\epsilon\Gamma^2(2-\epsilon)} \\ &= -2C_F\exp\left\{-\frac{5}{3}\epsilon-\sum_{n=2}^{\infty}\frac{\epsilon^n}{n}\left(\frac{2^n}{3^n}+2^n-1-\zeta_n[2^n-3-(-1)^n]\right)\right\}, \\ G_{C_Q}(0,u) &= -4C_F e^{5u/3}\frac{\Gamma(1+u)\Gamma(1-u)}{(1-u)(2-u)} \\ &= -2C_F\exp\left\{\frac{19}{6}u+\sum_{n=2}^{\infty}\frac{u^n}{n}[2^{-n}+1+\zeta_n((-1)^n+1)]\right\}, \\ \frac{\mathrm{d}}{\mathrm{d}u}G_{C_Q}(\epsilon,u)\bigg|_{u=0} &= -\frac{2C_F}{3}\frac{(3-2\epsilon)\Gamma(2-2\epsilon)}{(1-\epsilon)\Gamma^3(1-\epsilon)\Gamma(1+\epsilon)}\left[\frac{1}{\epsilon}\log[T(\epsilon)]+\gamma_E+H_{2-\epsilon}+\frac{\epsilon}{1-\epsilon}\right] \\ &= -2C_F\exp\left\{-\frac{5}{3}\epsilon-\sum_{n=2}^{\infty}\frac{\epsilon^n}{n}\left(\frac{2^n}{3^n}+2^n-1-\zeta_n[2^n-3-(-1)^n]\right)\right\} \\ &\quad\times\left[\sum_{n=2}^{\infty}\frac{\epsilon^{n-1}}{n}\left(\frac{2^n}{3^n}+2^n-1-\zeta_n[2^n-2-(-1)^n]\right)+H_{2-\epsilon}+\frac{5}{3}\right. \\ &\quad\left.\epsilon\exp\left\{\sum_{n=1}^{\infty}\frac{\epsilon^n}{n}\right\}\right]. \end{aligned} \tag{8.12}$$

In the last line the logarithm was manipulated as

$$\log[T(\epsilon)] = \frac{5}{3}\epsilon-\gamma_E\epsilon+\sum_{n=2}^{\infty}\frac{\epsilon^n}{n}\left(\frac{2^n}{3^n}+2^n-1-\zeta_n[2^n-2-(-1)^n]\right). \tag{8.13}$$

The pole structure of $G_{C_Q}$ is summarized in table 8.1. Again, only in the Borel-like integral along the positive real axis the singularities of the integrand are crossed. We observe that the two leading renormalons reside at $u=1,2$ and are double poles, while the following renormalons are simple poles.

| | **Poles** | **Order** | **Crossed** |
|---|---|---|---|
| $\boldsymbol{G_{C_Q}(\epsilon,0)}$ | $(2n+1)/2, \quad n=2,3,4\ldots$ | 1 | No |
| $\boldsymbol{G_{C_Q}(0,u)}$ | $-n, \quad n=2,3,4\ldots$<br>$1,2$<br>$n, \quad n=3,4,5\ldots$ | 1<br>1 and 2<br>1 | No<br>Yes<br>Yes |
| $\boldsymbol{\frac{\mathrm{d}}{\mathrm{d}u}G_{C_Q}(\epsilon,u)\Big\|_{u=0}}$ | $(2n+1)/2, \quad n=2,3,4\ldots$ | 1 | No |

**Table 8.1.** Pole structure of $G_{C_Q}(\epsilon,0)$ and $G_{C_Q}(0,u)$ and $|\mathrm{d}G_{C_Q}/\mathrm{d}u|_{u=0}$. The last column indicates whether the poles are crossed in the integrals each function appears in.

From these results we find a closed form for the cusp-anomalous dimension $\Gamma_{C_Q}$, which is related to the universal cusp-anomalous dimension as $\Gamma_{C_Q} = -2\,\Gamma_{\rm cusp}$. We find

$$\Gamma_{\rm cusp}(\beta) = -\frac{2\beta}{\beta_0} G_{C_Q}(-\beta, 0) = \frac{2\,C_F}{3\pi\beta_0}\frac{\sin(\pi\beta)\,\Gamma(4+2\beta)}{\Gamma(2+\beta)^2}. \tag{8.14}$$

Our result agrees Ref. [64], and the fixed-order coefficients reproduce the leading flavor structure of full-QCD up to $O(\alpha_s^4)$ [65, 66, 67, 68], collected in the first column of table 8.2. The convergence radius of $\Gamma_{\rm cusp}$ is set by the distance to the pole closest to the origin of the function $G_Q\,(-\beta, 0)$, which happens to be at $\beta = -2.5$. Furthermore, $\Gamma_{\rm cusp}(\beta)$ has no singularities for positive $\beta$, and the series is, as expected, free from renormalons. In figure 8.2(*a*) we compare the exact result for the cusp anomalous dimension to the partial sum of the fixed-order expansion up to 9 loops. Even though we use an unphysically large value for the strong coupling $\alpha_s = 0.9$, the fixed-order expansion converges to the exact value.

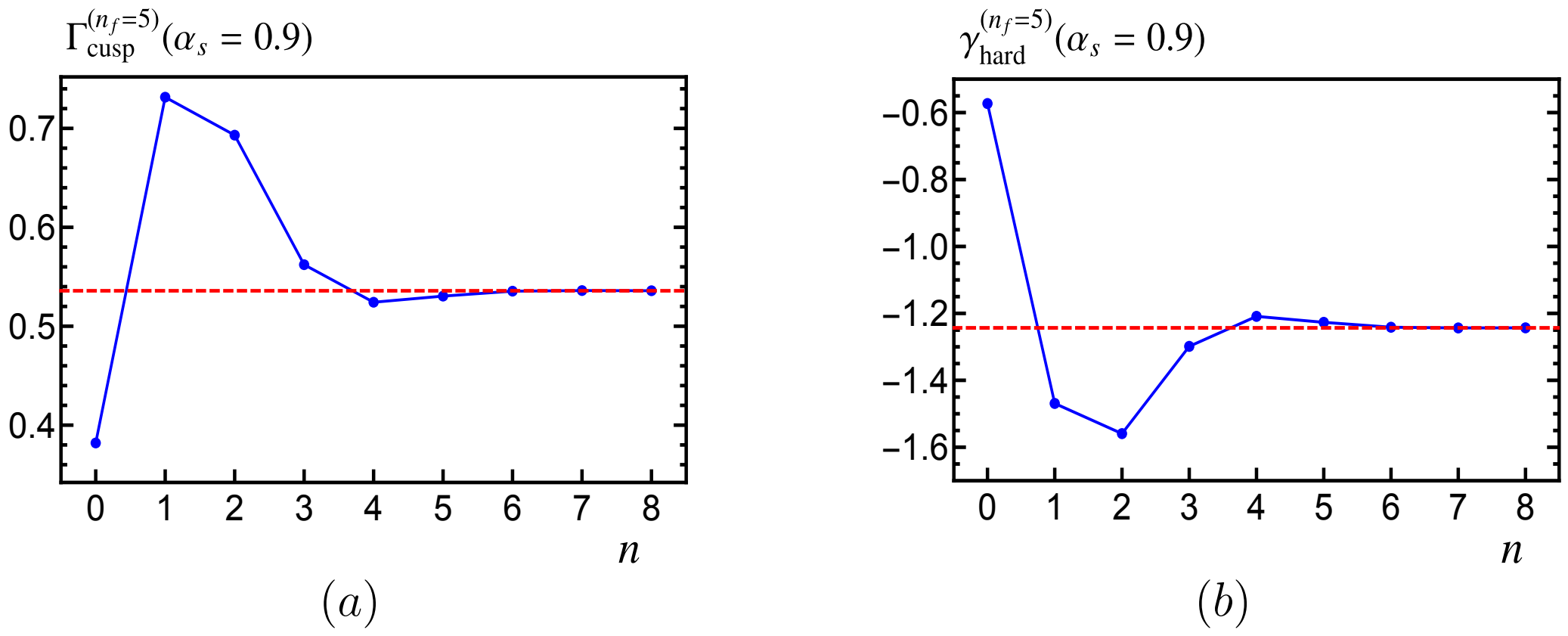

**Figure 8.2.** Comparison of the fixed-order partial sum with $n+1$ terms (blue dots) and exact results (red dashed line) for the cusp [panel (a)] and SCET non-cusp anomalous dimensions for the hard function in the large-$\beta_0$ approximation for $\alpha_s = 0.9$.

The non-cusp anomalous dimension $\gamma_H(\beta) = \gamma_{C_H}(\beta)$ is also obtained from the functions in (8.12) and results (4.32) and (4.34). We find full agreement for the fixed-order coefficients when compared to the leading flavor structure of full QCD up to $O(\alpha_s^4)$ [69, 70, 71, 68], see second column of table 8.2. To multiply out the two series in the derivative term in (8.12) we use the general result

$$\left(\sum_{i=n}^{\infty} a_i\, x^i\right)\left(\sum_{j=m}^{\infty} b_j\, x^j\right) = \sum_{j=n+m}^{\infty} x^j \sum_{i=n}^{j-m} a_i\, b_{j-i}\,, \tag{8.15}$$

where one of the two series results from expanding the exponential with the recursive formula in (A.21). The convergence radius of the series is again $\Delta\beta = 2.5$. In figure 8.2(*b*) we compare for $\alpha_s = 0.9$ the exact and fixed-order results for $\gamma_H$, finding again that the latter converges to the former already at 6 loops.

To compute $H_Q(Q)$ we need to use $G_{H_Q}(\epsilon,0) = 2G_{C_Q}(\epsilon,0)$, $G_{H_Q}(0,u) = 2G_{C_Q}(0,u)\cos(\pi u)$ and $\mathrm{d}G_{H_Q}(\epsilon,u)/\mathrm{d}u|_{u=0} = 2\mathrm{d}G_{C_Q}(\epsilon,u)/\mathrm{d}u|_{u=0}$. For the fixed-order coefficients one uses the exponential forms in (8.12) and for $G_{H_Q}(0,u)$ the Taylor expansions of $\cos(\pi u)$ and $G_Q(0,u)$ must be used in equation (8.15). Proceeding this way we fully reproduce the leading flavor structure of full QCD up to $O(\alpha_s^3)$ [72, 70, 73, 71, 74, 75, 76], see first column of table 8.3.

The integrand of the ambiguous integral, $B_H(u) \equiv [G_H(0,u) - G^H_{0,0} - uG^H_{0,1}]/u^2$, has two double poles at $u = 1, 2$ and an infinite number of simple poles at integer values of $u \geq 3$. Proceeding as explained for $\delta_{\overline{\mathrm{MS}}}$ (section 6.2.2) we obtain the residues at these poles and build the asymptotic expansion

$$B_H(u) \asymp 8C_F\left[\frac{\mathrm{e}^{5/3}}{(u-1)^2} + \frac{5\mathrm{e}^{5/3}}{3(u-1)} - \frac{\mathrm{e}^{10/3}}{2(u-2)^2} - \frac{\mathrm{e}^{10/3}}{12(u-2)} - \sum_{n=3}^{\infty}\frac{e^{5n/3}}{(n-2)\,(n-1)\,n\,(u-n)}\right]. \tag{8.16}$$

These double poles unambiguously signal the presence of anomalous dimensions with $n_f$ dependence at leading order for dimension-2 and -4 operators of the associated OPE[8.2]. The total ambiguity can be resummed and expressed in the following compact form:

$$\delta_\Lambda H_Q = \frac{C_F}{\beta_0}\left\{16x^2\log(x) - 2x^4[4\log(x) - 3] + 4x^2 - 6x^4 + 4(x^2-1)^2\log(1-x^2)\right\} \tag{8.17}$$

where $x \equiv \mathrm{e}^{5/6}\Lambda_{\mathrm{QCD}}/Q$. We have explicitly singled out the two leading contributions in the first two terms in curly brackets, while the remaining terms starts at $O(\Lambda^6_{\mathrm{QCD}})$. The logarithmic enhancement of the two leading terms appears due to the double nature of the poles and is associated to the anomalous dimension just discussed. This entails that $Q^2\,\delta_\Lambda H_Q$ retains some logarithmic dependence on $Q$. In figure 8.3(*a*) the dependence of $\delta_\Lambda H_Q$ in $Q$ is shown in red, while the ambiguity when the logarithm

8.2. Operators corresponding to simple poles might also carry anomalous dimension, but at leading-order they should not depend on $n_f$. We thank A. Hoang for clarifying this point.

is set to zero appears in blue. Despite the logarithmic enhancement, it is negligibly small even for the smallest values of $Q$ for which SCET is applicable.

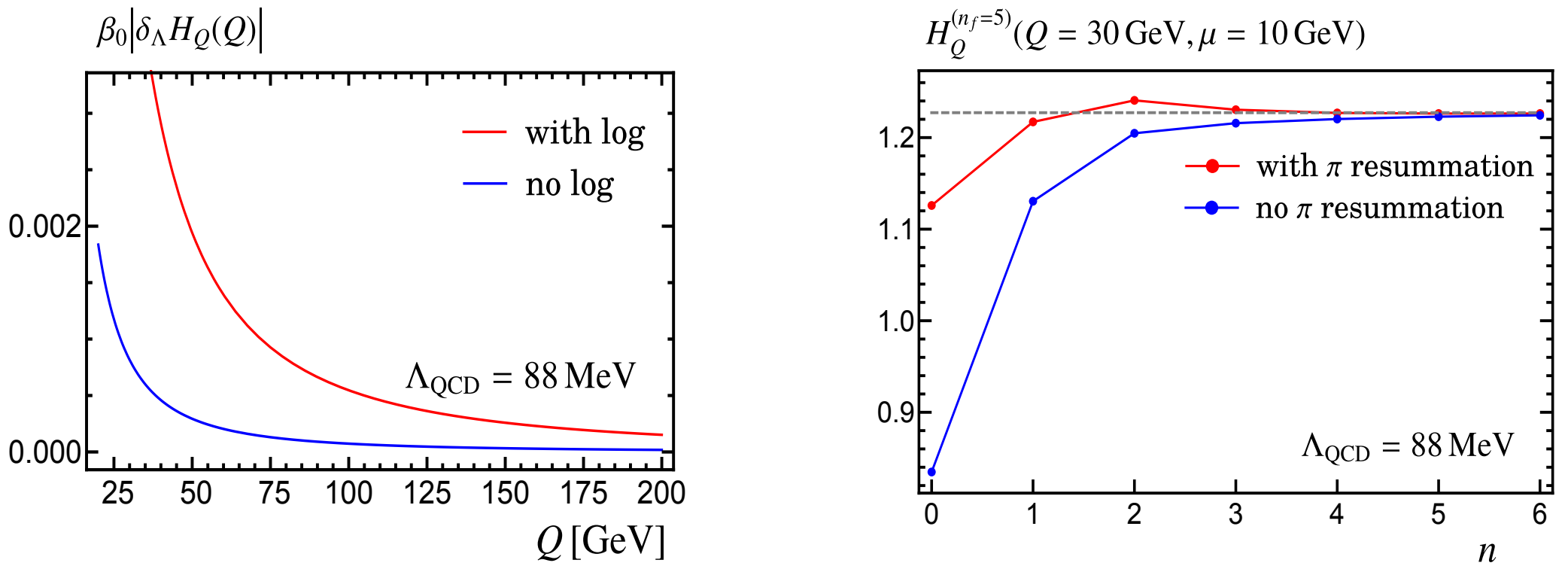

$(a)$ $(b)$

**Figure 8.3.** Left panel: dependence of the SCET hard function ambiguity (in absolute value) at leading order in $1/\beta_0$ with the center-of-mass energy $Q$. The red line is the exact result, while in blue we neglect the logarithmic term [see equation (8.17)]. Right panel: comparison of the exact result (gray dashed line) for $H_Q\,(Q = 30\,\text{GeV}, \mu = 10\,\text{GeV})$ with the fixed-order expansion including $n+1$ terms in the matching and resummation at N$^n$LL. Red dots include $\pi$-resummation (with $\mu_0 = -i\,Q$), while blue dots do not (using $\mu_0 = Q$).

We compare next the exact result for the hard function $H_Q(Q, \mu)$ obtained with equation (4.29) with its fixed-order expansion. For the latter we compute $H_Q(Q, \mu_0)$ and subsequently run to $H_Q(Q, \mu)$. A natural choice for the matching scale that avoids explicit large logarithms is $\mu_0 = Q$, but in this way $H_Q(Q, Q)$ has analytical continuation $\pi^n$ terms that can be traced as coming from the real part of $\log^n(-\mu^2/Q^2 + i0^+)$ with $Q^2$ and $\mu^2$ both positive when transitioning from Euclidean to Minkowskian regions, which therefore can also be summed up using RG evolution. In the renormalon formalism these factors come from expanding in powers of $u$ the factor $\cos(\pi u)$ appearing in $G_{H_Q}(0, u)$. To carry out this procedure one performs resummation in $C_H(Q^2, \mu)$ by computing first $C_H(Q^2, \mu_0)$ with $\mu_0 = -iQ$ and then running to from $\mu_0$ to $\mu$. From this resummed result for $C_H(Q^2, \mu)$ one obtains $H_Q\,(Q, \mu)$ simply taking the modulus squared expanded to $\mathcal{O}\,(1/\beta_0)$. We compare these two approaches in figure 8.3$(b)$ for a relatively small value of the center-of-mass energy. The red dots, which include $\pi$-resummation, converge to the exact answer much faster than the blue dots. Eventually, both approaches reproduce the exact result, and start diverging for $n > 30$. The ambiguity of $H_Q(Q, \mu)$ is too small to be visible in the plot. Since the hard factor mainly affects the norm of the event-shape distributions, this behavior might explain why the total-hadronic cross section obtained via a direct integration of the differential event-shape cross section, was undershot at low orders in the thrust and C-parameter distributions in references [77, 78]. Similarly, reference [79] found poor order-by-order convergence in peak differential cross sections for boosted tops unless they were self-normalized.

| $n$ | $\hat{\Gamma}^n_{\text{cusp}}$ | $\hat{\gamma}^n_H$ | $\hat{\gamma}^n_J$ | $\hat{\gamma}^n_S$ | $\hat{\gamma}^n_{H_m}$ | $\hat{\gamma}^n_B$ |
|---|---|---|---|---|---|---|
| 1 | 5.33333 | −16 | 8 | 0 | −10.6667 | 5.33333 |
| 2 | 8.88889 | −45.5782 | 26.6989 | −3.90981 | −17.7778 | 12.7987 |
| 3 | −1.77778 | −8.35288 | 3.26488 | 0.911559 | 3.55556 | −2.68934 |
| 4 | −11.0442 | 43.9815 | −23.6303 | 1.6396 | 22.0883 | −12.6838 |
| 5 | −5.83051 | 27.5444 | −8.49109 | −5.2811 | 11.661 | −0.549408 |
| 6 | 1.73202 | −9.98024 | 10.8212 | −5.83111 | −3.46405 | 7.56314 |
| 7 | 2.54404 | −14.9285 | 8.61429 | −1.15006 | −5.08807 | 3.6941 |
| 8 | 0.56906 | −3.87091 | 1.07053 | 0.864928 | −1.13812 | −0.295868 |
| 9 | −0.259965 | 1.00721 | −0.756306 | 0.252702 | 0.519929 | −0.512666 |
| 10 | −0.157657 | 0.398943 | 0.114127 | −0.313598 | 0.315314 | 0.155941 |

**Table 8.2.** Numeric coefficients for the cusp anomalous dimension and the non-cusp anomalous dimensions of the hard and mass-scale matching coefficients, the SCET jet function and the bHQET jet function. The hatted coefficients are defined as $f \equiv (1/\beta_0)\sum_{n=0}^{\infty} \hat{f}_n \beta^{n+1}$. For the cusp anomalous dimensions we observe $\Gamma_H = 2\Gamma_S = -2\Gamma_J = -2\Gamma_B = -4\Gamma_{\text{cusp}} = 2\Gamma_{H_m}$.

| $n$ | $H_Q$ | $\tilde{J}_n$ | $H_m$ | $\tilde{B}_n$ |
|---|---|---|---|---|
| 1 | 9.3721 | 0.560352 | 15.0532 | 7.52658 |
| 2 | 15.0279 | 10.7871 | 59.9652 | 29.0465 |
| 3 | 117.452 | 21.3687 | 267.538 | 112.621 |
| 4 | 709.259 | 50.6136 | 1656.73 | 661.256 |
| 5 | 4302.01 | 193.805 | 13629.1 | 5290.68 |
| 6 | 28592.4 | 961.578 | 138593 | 52915.6 |
| 7 | 214413 | 5740.14 | $1.67874 \times 10^6$ | 634995 |
| 8 | $1.78994 \times 10^6$ | 40068.6 | $2.36208 \times 10^7$ | $8.88995 \times 10^6$ |
| 9 | $1.66615 \times 10^7$ | 320083 | $3.78939 \times 10^8$ | $1.42239 \times 10^8$ |
| 10 | $1.70403 \times 10^8$ | $2.87865 \times 10^6$ | $6.83033 \times 10^9$ | $2.56031 \times 10^9$ |

**Table 8.3.** Numeric values for the non-logarithmic coefficients for the various SCET and bHQET coefficients $c_{n,n-1}$.

## 8.2 Jet function

The SCET jet function describes energetic radiation branching from the initiating quark, and depends on the collinear measurement function, which is the same for hemisphere masses, C-parameter and thrust. Furthermore, it is completely inclusive such that it can be computed as the discontinuity of a forward-scattering matrix element:

$$\mathcal{J}_n(s,\mu) = \frac{1}{8\pi N_c Q}\int \mathrm{d}^d x\, \mathrm{e}^{\mathrm{i}p\cdot x}\ \mathrm{Tr}\langle 0|\mathrm{T}\{\chi_n(x)\bar{\chi}_{n,Q}(0)\not{n}\}|\,0\rangle\,, \tag{8.18}$$

with $d = 4 - 2\varepsilon$, Tr a trace over spin and color, and $p^2 = s$. We use the gauge-invariant jet field $\chi_n = W_n^\dagger \xi_n$, which is the product of a collinear quark field $\xi_n$ and a path-ordered collinear Wilson line $W_n^\dagger$. Likewise, $\chi_{n,Q}$ is a jet field with total minus label momenta $\bar{n}P$ set to $Q$ with a Dirac delta function. This overly simplifies the computation of the leading $1/\beta_0$ correction as one only needs to modify the gluon propagator of the one-loop diagrams shown in figure 8.4.

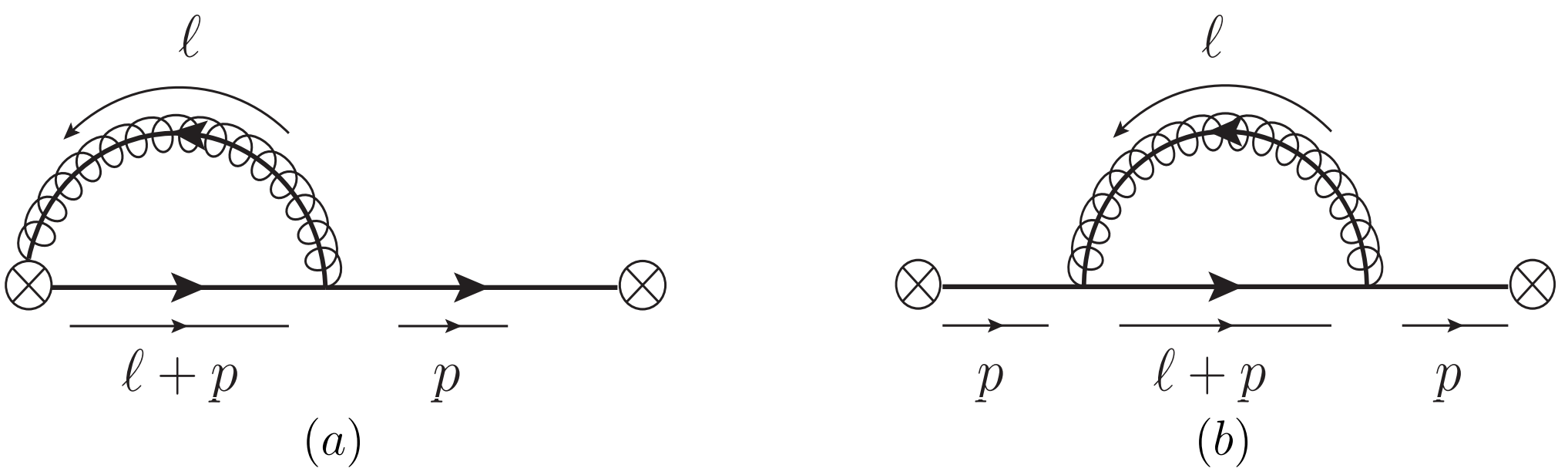

**Figure 8.4.** Feynman diagrams for the one-loop jet function for massless quarks with off-shell momentum $p^2 = s$ and virtual loop momentum $\ell$. The cross represents a quark jet-field insertion. Diagram (a) has to be multiplied by a factor of two to account for the symmetric contribution with the gluon line emitted from the right current.

The vertex diagram 8.4(a) must be doubled to account for the symmetric contribution, which is omitted along with the diagram in which the gluon is emitted and absorbed from both Wilson since it vanishes in the Feynman gauge – the light-cone vectors $n$ carried by each crossed vertex are contracted with $g_{\mu\nu}$ in the gluon propagator, thus proportional to $n^\mu n^\nu g_{\mu\nu} = n^2 = 0$–. Given that the jet functions for the three event shapes just mentioned are trivially related to one another, in this section we carry out the computation for the invariant mass of a single hemisphere. The discontinuity of $\mathcal{J}_n(s,\mu)$ -that is, $\mathcal{J}_n(s + i\,0^+,\mu) - \mathcal{J}_n(s + i\,0^-,\mu)$- yields the momentum-space jet function $J_n(s,\mu)$, with support for $s > 0$. It can be converted to the position-space jet function $\tilde{J}_n(y,\mu)$, with $y$ the variable conjugate to $s$, defined as:

$$\tilde{J}_n(y,\mu) = \int_0^\infty \mathrm{d}s\ e^{-\mathrm{i}sy}\, J_n(s,\mu)\,. \tag{8.19}$$

While $\mathcal{J}_n(s,\mu)$ is a regular function, $J_n(y,\mu)$ for massless quarks contains Dirac delta and plus distributions. For its computation we will use the following identity:

$$\mathrm{Im}[(-s-i\,0^+)^{-1-\eta}]=\frac{\pi s^{-1-\eta}\theta(s)}{\Gamma(1+\eta)\Gamma(-\eta)}\,. \tag{8.20}$$

Both $\mathcal{J}_n(s,\mu)$ and $J_n(y,\mu)$ have dimensions of an inverse squared mass, while $s$ is a real-valued mass-dimension-two variable. On the contrary, $\tilde{J}_n(y,\mu)$ is a dimensionless complex-valued regular function depending on the complex variable $y$, with mass dimension $-2$. At tree-level one has

$$\mathcal{J}_n^{\rm tree}(s)=-\frac{1}{2\,\pi}\frac{1}{s+i\,0^+}\,,\qquad J_n^{\rm tree}(s)=\delta(s)\,,\qquad \tilde{J}_n^{\rm tree}(y)=1\,, \tag{8.21}$$

and we define the corrections to the bare jet function in momentum and position space as $J_n(s,\mu)=\delta(s)+\delta J_n(s)$ and $\tilde{J}_n(y)=1+\delta\tilde{J}_n(y)$, respectively, which are computed next.

In a light-cone basis where $\vec{n}$ is given by the direction of the jet we have

$$p^-=\bar{n}\cdot p\equiv Q\,,\qquad p^+=n\cdot p\,,\qquad p_\perp=0. \tag{8.22}$$

Also, we use the notation

$$p^2=p^+p^-\equiv s\quad\text{and}\quad p^+=\frac{p^2}{p^-}=\frac{s}{Q}. \tag{8.23}$$

Inserting the shifted gluon propagator and following the SCET Feynman rules in (7.17), the vertex contribution and the self-energy correction graphs in figure 8.4 are

$$\begin{aligned}\mathcal{J}_a&=g_0^2C_F\frac{Q}{s}\not{n}\int\frac{{\rm d}^d\ell}{(2\pi)^d}\frac{\ell^--Q}{(\ell^2)^{1+h}(\ell-p)^2k^-}\equiv g_0^2C_F\frac{Q}{s}\not{n}I_{\mathcal{J}_a}, \\ \mathcal{J}_b&=g_0^2C_F\frac{Q^2}{s^2}\not{n}(1-\epsilon)\int\frac{{\rm d}^dk}{(2\pi)^d}\frac{\ell_\perp^2}{(\ell^2)^{1+h}(\ell-p)^2(Q-k^-)}\equiv g_0^2C_F\frac{Q^2}{s^2}\not{n}(1-\epsilon)I_{\mathcal{J}_b}.\end{aligned} \tag{8.24}$$

In simplifying the above expressions we used the lightcone vector properties $\not{n}\not{\bar{n}}\not{n}=4\not{n}$ and $\not{\bar{n}}\not{n}\not{\bar{n}}=4\not{\bar{n}}$. The two integrals can be put in terms of those in appendix C.6. The first one can be readily written as

$$\begin{aligned}I_{\mathcal{J}_a}&=I_2(1,Q^{1+h},Q(-p))-Q\,I_3(1,Q^{1+h},Q(-p),E_n)\\ &=I_2(1,Q^{1+h},Q(p))+Q\,I_3(1,Q^{1+h},Q(p),E_n),\end{aligned} \tag{8.25}$$

where in the second term we canceled both $\ell^-$ in the numerator and denominator and used the symmetry properties of both integrals when $p\mapsto -p$. The integral $I_{\mathcal{J}_b}$ requires a bit more manipulation:

$$\begin{aligned} I_{\mathcal{J}_b} &= \int\frac{\mathrm{d}^d\ell}{(2\pi)^d}\frac{\ell_\perp^2}{(\ell^2)^{1+h}(\ell-p)^2(Q-k^-)} = \int\frac{\mathrm{d}^d\ell}{(2\pi)^d}\frac{(p_\perp-\ell_\perp)^2}{(k^2)^{1+h}(\ell-p)^2(Q-\ell^-)}\\ &= \int\frac{\mathrm{d}^dk}{(2\pi)^d}\frac{(p-\ell)^2-(p^+-\ell^+)(Q-\ell^-)}{(\ell^2)^{1+h}(\ell-p)^2(Q-\ell^-)} = \int\frac{\mathrm{d}^dk}{(2\pi)^d}\frac{\ell^+-p^+}{(\ell^2)^{1+h}(\ell-p)^2}\\ &= n_\mu I_2(\ell^\mu, Q^{1+h}, Q(-p)) - \frac{s}{Q}I_2(1, Q^{1+h}, Q(-p))\\ &= \frac{1}{2Q}[I_2(1, Q^h, Q(p)) - s\,I_2(1, Q^{1+h}, Q(p))], \end{aligned} \tag{8.26}$$

where first we added $p_\perp = 0$ and then we expanded the squared vector in the lightcone basis. In the last step we used the Passarino-Veltman result (C.18). Identifying the two kinds of integrals $J^1_{\mathcal{J}}(h)\equiv I_2(1, Q^h, Q(p))$ and $J^2_{\mathcal{J}}(h)\equiv I_3(1, Q^h, Q(p), E_n)$ the sum both diagrams $\mathcal{J}_{\mathrm{sh}} = 2\mathcal{J}_a + \mathcal{J}_b$ yields

$$\begin{aligned} \frac{\mathrm{i}}{4\pi Q}\mathrm{tr}[\bar{n}\!\!\!/\mathcal{J}_{\mathrm{sh}}] &= \frac{2\mathrm{i}g_0^2C_F}{\pi s}\left\{2[J^1_{\mathcal{J}}(h+1) + Q\,J^2_{\mathcal{J}}(h+1)] + \frac{1-\epsilon}{2s}[J^1_{\mathcal{J}}(h) - s\,J^1_{\mathcal{J}}(h+1)]\right\}\\ &= -\left(\frac{g_0}{4\pi}\right)^2 2C_F(-1)^h(4\pi)^\epsilon(-s)^{-1-h-\epsilon}[\epsilon^2 + (h-5)\epsilon - 3h + 4]\\ &\quad\times\frac{\Gamma(h+\epsilon)\Gamma(-h-\epsilon)\Gamma(2-\epsilon)}{\pi\Gamma(1+h)\Gamma(3-h-2\epsilon)}, \end{aligned} \tag{8.27}$$

where the trace over spinnor indices is simply $\mathrm{tr}(n\!\!\!/\bar{n}\!\!\!/) = 8$. The imaginary part is taken with (8.20), yielding

$$J_{\mathrm{sh}} = -\left(\frac{g_0}{4\pi}\right)^2 2C_F(-1)^h(4\pi)^\epsilon s^{-1-h-\epsilon}\theta(s)\frac{[\epsilon^2+(h-5)\epsilon-3h+4]\Gamma(2-\epsilon)}{[h+\epsilon]\Gamma(1+h)\Gamma(3-h-2\epsilon)}, \tag{8.28}$$

$$\tilde{J}_{\mathrm{sh}} = -\left(\frac{g_0}{4\pi}\right)^2 2C_F(-1)^h(4\pi)^\epsilon(\mathrm{i}y)^{h+\epsilon}\frac{[\epsilon^2+(h-5)\epsilon-3h+4]\Gamma(2-\epsilon)\Gamma(-h-\epsilon)}{[h+\epsilon]\Gamma(1+h)\Gamma(3-h-2\epsilon)}.$$

This immediately gives the generating function $G^\mu_{\tilde{J}}(\epsilon,u) = (\mathrm{i}y\mu^2\mathrm{e}^{\gamma_E})^u G_{\tilde{J}}(\epsilon,u)$, with

$$G_{\tilde{J}}(\epsilon,u) = 2C_F\frac{[(\epsilon-3)u+4-2\epsilon]\Gamma(2-\epsilon)\Gamma(1-u)}{\Gamma(1+u-\epsilon)\Gamma(3-u-\epsilon)}T^{\frac{u}{\epsilon}-1}(\epsilon), \tag{8.29}$$

identifying $\omega^2 = -\mathrm{i}\mathrm{e}^{\gamma_E}/y$. The function $G_{\tilde{J}}(\epsilon,u)$ is regular at $\epsilon = u = 0$ with $G_{\tilde{J}}(0, 0) = G^{\tilde{J}}_{0,0} = 4C_F$. Its pole structure is summarized in

| | Poles | Order | Crossed |
|---|---|---|---|
| $G_{\tilde{J}}(\epsilon,0)$ | $(2n+1)/2, \quad n=2,3,4\ldots$ | 1 | No |
| $G_{\tilde{J}}(0,u)$ | $1,2$ | 1 | Yes |
| $\left.\frac{\mathrm{d}}{\mathrm{d}u}G_{\tilde{J}}(\epsilon,u)\right\|_{u=0}$ | $(2n+1)/2, \quad n=2,3,4\ldots$ | 1 | No |

**Table 8.4.** Pole structure of the function $G_{\tilde{J}}(\epsilon,u)$ generating the contribution to the large-$\beta_0$ SCET jet function in position space.

The Borel-like integral carries an ambiguity but this time there are only two simple poles at $u=1,2$. The relevant forms and partial expansions are

$$\begin{aligned}
G_{\tilde{J}}(\epsilon,0) &= \frac{2C_F}{3\pi}\frac{\Gamma(4-2\epsilon)\sin(\pi\epsilon)}{\epsilon\Gamma^2(2-\epsilon)} \qquad (8.30)\\
&= 4C_F\exp\left\{-\frac{5}{3}\epsilon-\sum_{n=2}^{\infty}\frac{\epsilon^n}{n}\left(\frac{2^n}{3^n}+2^n-1-\zeta_n[2^n-3-(-1)^n]\right)\right\},\\
G_{\tilde{J}}(0,u) &= \frac{2C_F e^{\left(\frac{5}{3}-\gamma_E\right)u}(4-3u)}{(1-u)(2-u)\Gamma(1+u)}\\
&= 4C_F\exp\left\{\frac{29}{12}u-\sum_{n=2}^{\infty}\frac{u^n}{n}\left[\frac{3^n}{4^n}-2^{-n}-1+(-1)^n\zeta_n\right]\right\},\\
\left.\frac{\mathrm{d}}{\mathrm{d}u}G_{\tilde{J}}(\epsilon,u)\right|_{u=0} &= \frac{4C_F}{3}\frac{(3-2\epsilon)\Gamma(2-2\epsilon)}{(1-\epsilon)\Gamma^3(1-\epsilon)\Gamma(1+\epsilon)}\left[\frac{1}{\epsilon}\log[T(\epsilon)]+\gamma_E+\frac{3-\epsilon^2}{2(1-\epsilon)(2-\epsilon)}\right]\\
&= C_F\exp\left[-\frac{1}{6}\epsilon+\sum_{n=2}^{\infty}\frac{\epsilon^n}{n}\{2+2^{-n}-2^n(1+3^{-n})+\zeta_n[2^n-3-(-1)^n]\}\right]\\
&\quad\left\{3+\epsilon^2+2(1-\epsilon)(2-\epsilon)\right.\\
&\quad\left.\times\left[\frac{5}{3}+\sum_{n=3}^{\infty}\frac{\epsilon^{n-1}}{n}\{2^n(1+3^{-n})-1+\zeta_n[2-2^n+(-1)^n]\}\right]\right\},
\end{aligned}$$

with $G_{1,0}^{\tilde{J}}=-20C_F/3$ and $G_{0,1}^{\tilde{J}}=29C_F/3$.

For the cusp anomalous dimension we find $\Gamma_{\tilde{J}}=4\Gamma_{\text{cusp}}$ and $\beta G_{\tilde{J}}(-\beta,0)/\beta_0$ reproduces the result found in equation (8.14) since $G_{\tilde{J}}(\epsilon,0)=-2G_{C_Q}(\epsilon,0)$ as a consequence of universality. For the non-cusp part $\gamma_J(\beta)$ we find full agreement with the leading flavor structure of the known full SCET results up to $O(\alpha_s^3)$ [66] - see columns 3 and 4 in table 8.2-. The soft non-cusp anomalous dimension can be computed from a consistency condition and it reads $\gamma_S = -\gamma_H - \gamma_J$. In both cases the convergence radius is $\Delta\beta = 5/2$. In figure 8.5 we compare for $\alpha_s = 0.9$ the exact and fixed-order results for the jet and soft non-cusp anomalous dimensions, respectively. The soft function has $\gamma_S^0 = 0$ and therefore the partial sum starting from the first non-vanishing order is shown. We find good convergence in both cases.

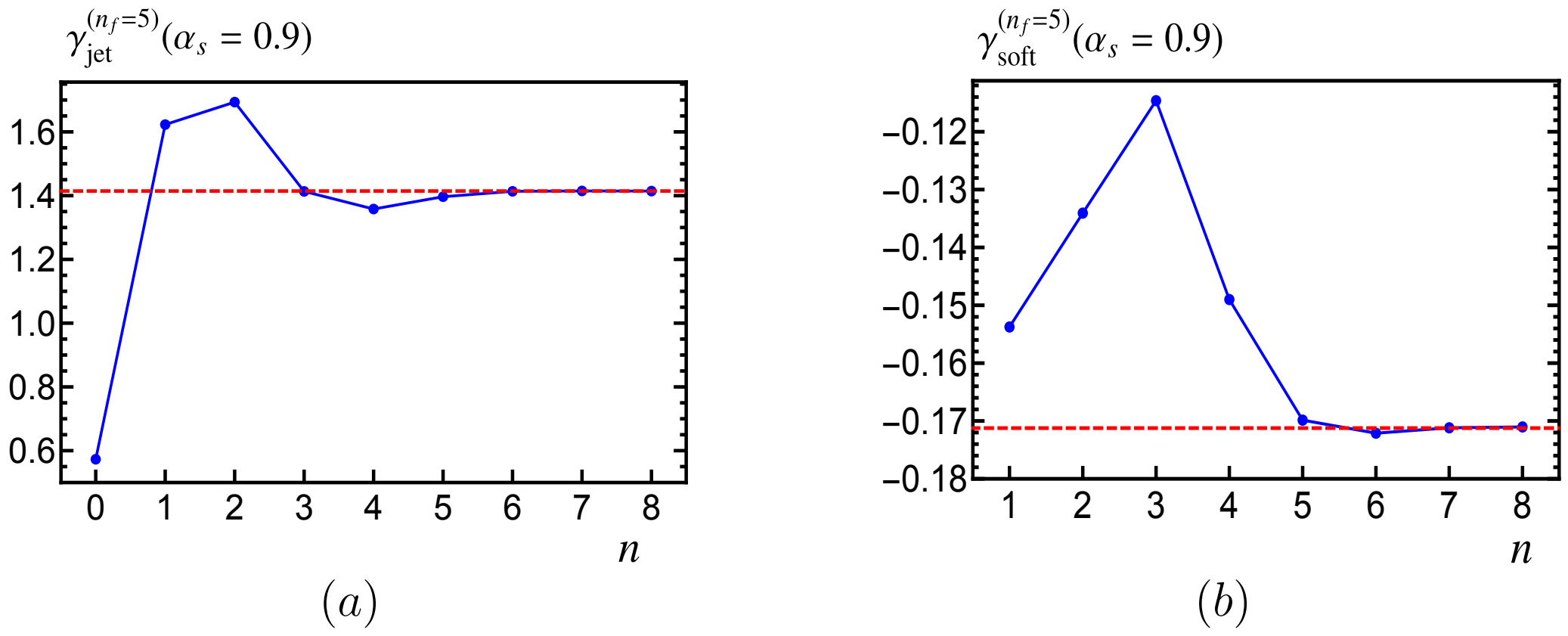

**Figure 8.5.** Comparison of the fixed-order partial sum with $n+1$ terms (blue dots) and exact results (red dashed line) for the jet –panel $(a)$– and soft functions anomalous dimensions –panel $(b)$– in the large-$\beta_0$ approximation for $\alpha_s = 0.9$. The jet anomalous dimension has been directly computed and the soft anomalous dimension is extracted from the consistency condition $\gamma_H + \gamma_J + \gamma_S = 0$.

Next we compute the non-logarithmic fixed-order coefficients $c_{n,i}^{\tilde{J}}$ of the Fourier-space jet function, which are defined as

$$\tilde{J}_n = \sum_{n=1}^{\infty} \beta^n \sum_{i=0}^{n+1} c_{n,i}^{\tilde{J}} \log^i(\mathrm{i} y e^{\gamma_E} \mu^2)\,. \tag{8.31}$$

Using the Taylor expansions of the functions in (8.30) into equations (4.36) and (4.41) we fully reproduce the leading flavor structure of the known coefficients in full SCET up to $O(\alpha_s^3)$ [80, 81, 82, 83, 84] –see second column of table 8.3–.

As mentioned before, the function $B_{\tilde{J}}(u) \equiv [G_{\tilde{J}}(0,u) - G_{0,0}^{\tilde{J}} - u\,G_{0,1}^{\tilde{J}}]/u^2$ has only two simple poles at $u = 1, 2$, with the following asymptotic expansion:

$$B_{\tilde{J}}(u) \asymp -C_F\left[\frac{e^{\frac{10}{3} - 2\gamma_E}}{2\,(u-2)} + \frac{2\,e^{\frac{5}{3} - \gamma_E}}{u-1}\right]. \tag{8.32}$$

Therefore the ambiguities in position and momentum space are

$$\delta_\Lambda \tilde{J}_n = -\frac{C_F}{\beta_0}\left[2\left(\mathrm{i}y\mathrm{e}^{\frac{5}{3}}\Lambda_{\rm QCD}^2\right)+\frac{1}{2}\left(\mathrm{i}y\mathrm{e}^{\frac{5}{3}}\Lambda_{\rm QCD}^2\right)^2\right],$$
$$\delta_\Lambda J_n = -\frac{2\,\mathrm{e}^{\frac{5}{3}}C_F}{\beta_0}\Lambda_{\rm QCD}^2\delta'(s)+O(\Lambda_{\rm QCD}^4)\simeq\frac{2\,\mathrm{e}^{\frac{5}{3}}C_F}{\beta_0}\frac{\Lambda_{\rm QCD}^2}{s}\delta(s)\,, \tag{8.33}$$

where in the second line the Dirac delta function's derivative is expressed as $\delta'(s)\simeq -\delta(s)/s$, as obtained from the identity $[s\delta(s)]'=0$. For real values of $y$ the real (imaginary) part of the jet function is free from the $u=1(2)$ renormalon, but affected by the $u=2(1)$ ambiguity. For complex $y$ both real and imaginary parts are affected by the two singularities. In the second line of equation (8.33) it is shown that when transforming back to momentum space the leading ambiguity is proportional to $\delta'(s)$. Therefore, to find an ambiguous series in momentum space one needs to consider a double cumulative (that is, the cumulative version of $\Sigma_J$) or a weighted cumulative such as $\int_0^{s'}\mathrm{d}s\,s J_n(s,\mu)$.

In figure 8.6 we compare the fixed-order expansion for the position-space jet function up to $(n+1)$-loops with the exact result, splitting in two separate panels the real and imaginary parts, for a real value of the Fourier variable $y$. The ambiguity is visible for the imaginary part only. We nevertheless observe nice convergence for different values of the matching scale.

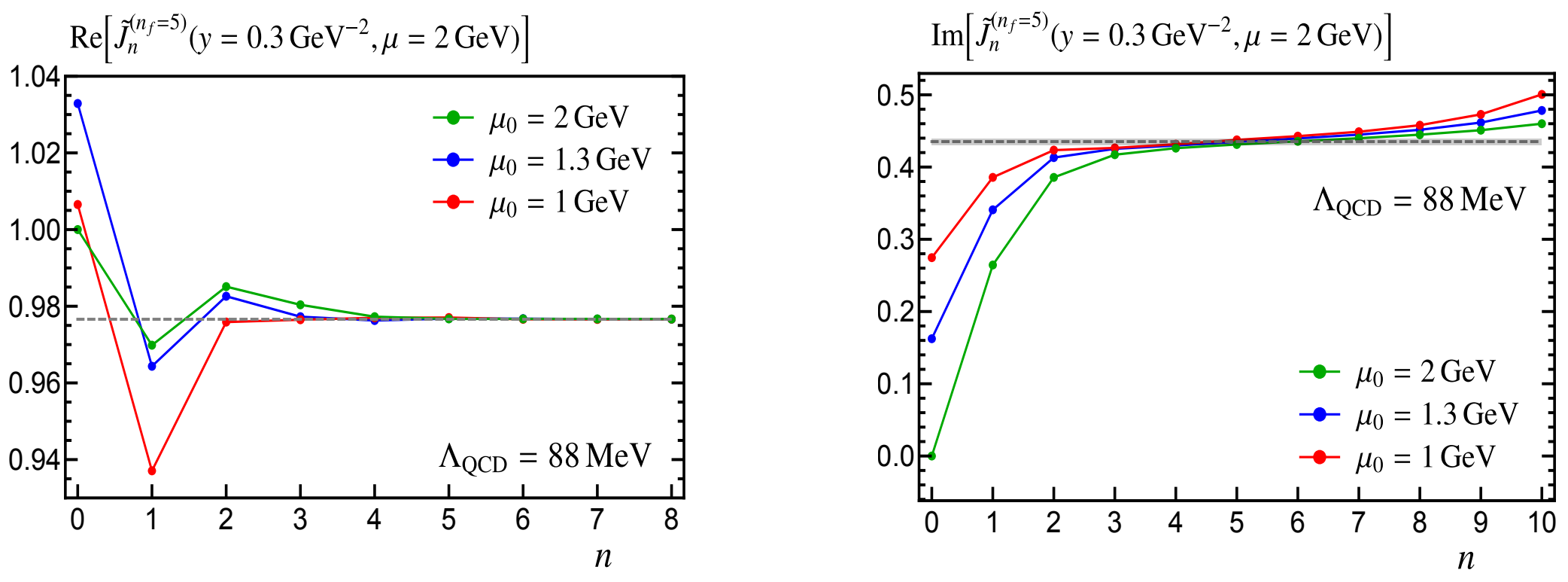

(a) (b)

**Figure 8.6.** Comparison of the fixed-order partial sum (colored dots) and exact results (red dashed line) for the real (left panel) and imaginary (right panel) part of the position-space SCET jet function $\tilde{J}_n(y,\mu)$, with $y=0.3\,\mathrm{GeV}^{-2}$ and $\mu=2\,\mathrm{GeV}$. We show fixed-order results for three matching scales: $\mu_0=2\,\mathrm{GeV}$ (green), 1.3 GeV (blue) and 1 GeV (red).

# Chapter 9
# Applications III: Jet factorization theorem in bHQET

## 9.1 Mass-scale hard factor

To determine the hard massive function $H_m$ one needs to compute the matching coefficient $C_m$ between the SCET and bHQET massive di-jet operators. As before, we compute the matrix element between the vacuum and a pair of massive, on-shell quarks. The optimal way of carrying out the computation is regularizing all infrared divergences within dimensional regularization, causing all bHQET diagrams vanish such that only SCET contributions need to be considered. The diagrams to take into account are shown in figure 9.1, where diagram 9.1$(a)$ needs to be doubled to account for the symmetric diagram not explicitly shown.

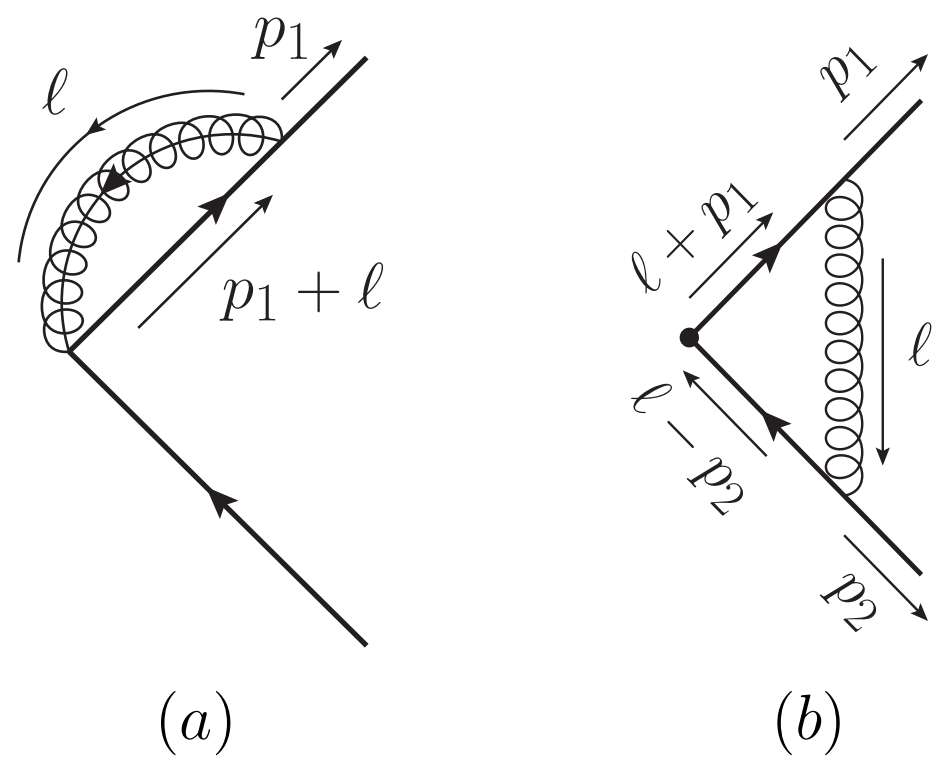

**Figure 9.1.** SCET Feynman diagrams for the computation of the bHQET to SCET matching coefficient at one-loop for a massive on-shell quark with momentum $p^2=m^2$ and virtual loop momentum $\ell$. We do not show the symmetric counterpart of diagram $(a)$, as it gives identical results.When using dimensional regularization to regularize all divergences, all bHQET diagrams vanish.

We compute the matrix element in the bare formalism and then switch to the on-shell scheme as $\langle q\bar{q}\,|\,\bar{\xi}_n^0\Gamma^\mu\xi_n^0|0\rangle = Z_\xi^{\rm OS}\langle q\bar{q}\,|\,\bar{\xi}_n^{\rm OS}\Gamma^\mu\xi_n^{\rm OS}|0\rangle$. The wavefunction renormalization for the collinear quark field in SCET is the same as in QCD, and for massive quarks the leading $1\,/\,\beta_0$ contribution does not vanish. We extract this renormalization factor from the QCD self-energy for massive quark carried out in

section 6.1. Explicitly, the contribution resulting from the insertion of the shifted propagator is

$$\begin{aligned} Z_{\xi,\text{sh}}^{\text{OS}} &= \left(\frac{g_0}{4\pi}\right)^2 2C_F(-1)^h(m_p^2)^{-h-\epsilon}(4\pi)^\epsilon \\ &\times\frac{(1+h)(1-h-\epsilon)(2\epsilon-3)\Gamma(h+\epsilon)\Gamma(1-2h-2\epsilon)}{\Gamma(3-h-2\epsilon)}. \end{aligned} \tag{9.1}$$

The diagrams in figure 9.1 are

$$\begin{aligned} D_a^{\text{mSCET}} &= -2\mathrm{i}g_0^2C_F\gamma^\mu\int\frac{\mathrm{d}^d\ell}{(2\pi)^d}\frac{\ell^-+p_1^-}{(\ell^2)^{1+h}[(\ell+p_1)^2-m^2]\ell^-}, \\ D_b^{\text{mSCET}} &= 2\mathrm{i}g_0^2C_F\gamma^\mu\int\frac{\mathrm{d}^d\ell}{(2\pi)^d}\frac{(p_1^-+\ell^-)(\ell^+-p_2^+)}{(\ell^2)^{1+h}[(\ell+p_1)^2-m^2][(\ell-p_2)^2-m^2]}. \end{aligned} \tag{9.2}$$

In digram 9.1$(a)$, the integral can be simplified as

$$\begin{aligned} I_a^{\text{mSCET}} &\equiv \int\frac{\mathrm{d}^d\ell}{(2\pi)^d}\frac{\ell^-+p_1^-}{(\ell^2)^{1+h}[(\ell+p_1)^2-m^2]\ell^-}, \\ &= I_3(1,Q^{1+h},Q_m(p_1))+p_1^-I_3(1,Q^{1+h},Q_m(p_1),E_n). \end{aligned} \tag{9.3}$$

The integral in diagram 9.1$(b)$ is scaleless and vanishes since $\ell$ is usoft

$$\begin{aligned} I_b^{\text{mSCET}} &\equiv \int\frac{\mathrm{d}^d\ell}{(2\pi)^d}\frac{(p_1^-+\ell^-)(p_2^+-\ell^+)}{(\ell^2)^{1+h}[(\ell+p_1)^2-m^2][(\ell-p_2)^2-m^2]} \\ &\sim \int\frac{\mathrm{d}^dk}{(2\pi)^d}\frac{p_1^-p_2^+}{(\ell^2)^{1+h}[\ell^+p_1^-][-\ell^-p_2^+]}=0 \end{aligned} \tag{9.4}$$

Therefore the shifted contribution to the mass-scale matching coefficient is

$$\begin{aligned} C_{m,\text{sh}}^0 &= -4\mathrm{i}g_0^2C_F[p_1^-I_3(1,Q^{1+h},Q_m(p_1),E_n)+I_2(1,Q^{1+h},Q_m(p_1))]+Z_{\xi,\text{sh}}^{\text{OS}} \\ &= \left(\frac{g_0}{4\pi}\right)^2 4C_F(4\pi)^\epsilon(-1)^h(m^2)^{-h-\epsilon}(1+h)\frac{\Gamma(-2h-2\epsilon)\Gamma(h+\epsilon)}{\Gamma(3-h-2\epsilon)} \\ &\quad\times[2\epsilon^3+(4h-5)\epsilon^2+(2h^2-8h+5)\epsilon-3h^2+4h-2], \end{aligned} \tag{9.5}$$

where the superscript refers to the fact that $C_m$ requires its renormalization $Z_{C_m}$. Nevertheless, at leading order in the large-$\beta_0$ multiplicative renormalization amounts to adding $\delta Z_{C_m}$, which, in turn, in the $\overline{\text{MS}}$ scheme amounts for simply dropping the $1/\epsilon^n$ terms. In our formalism, the generating $G(\epsilon,u)$ remains the same for the bare and renormalized series, and the renormalized series is found by applying (4.29). The the function $G_{C_m}^\mu(\epsilon,u)=(\mu^2/m^2)^uG_{C_m}(\epsilon,u)$ takes the form

$$\begin{aligned} G_{C_m}(\epsilon,u) &= 4C_F\frac{\mathrm{e}^{\gamma_Eu}u^2(1+u-\epsilon)\Gamma(u)\Gamma(-2u)}{\Gamma(3-u-\epsilon)}T^{\frac{u}{\epsilon}-1}(\epsilon) \\ &\quad\times[(2u^2-2u+1)\epsilon-3u^2+4u-2]. \end{aligned} \tag{9.6}$$

The function $G_{C_m}(\epsilon, u)$ is regular at the origin with $G_{C_m}(0,0) = G_{0,0}^{C_m} = 2C_F$. Its pole structure is summarized in table 9.1. Only $G_{C_m}(0,u)$ presents poles –at $u = 1,2$ and all the positive half-integers– that are crossed in the integrals.

| | Poles | Order | Crossed |
|---|---|---|---|
| $\boldsymbol{G_{C_m}(\epsilon,0)}$ | $(2n+1)/2, \quad n = 2,3,4...$ | 1 | No |
| $\boldsymbol{G_{C_m}(0,u)}$ | $(2n+1)/2, \quad n = 0,1,2...$<br>$-n, \quad n = 2,3,4...$<br>$1,2$ | 1<br>1<br>1 | Yes<br>No<br>Yes |
| $\left.\boldsymbol{\frac{d}{du} G_{C_m}(\epsilon,u)}\right\|_{u=0}$ | $(2n+1)/2, \quad n = 2,3,4...$ | 1 | No |

**Table 9.1.** Pole structure of the relevant evaluations of the $G_{C_m}(\epsilon, u)$ function.

The relevant evaluations of the function $G_{C_m}(\epsilon, u)$ and its partial expansions to extract the fixed order coefficients are

$$\begin{aligned}
G_{C_m}(\epsilon,0) &= \frac{C_F}{3\pi}\frac{\Gamma(4-2\epsilon)\sin(\pi\epsilon)}{\epsilon\Gamma^2(2-\epsilon)} \qquad (9.7)\\
&= C_F \exp\left\{-\frac{5}{3}\epsilon - \sum_{n=2}^{\infty}\frac{\epsilon^n}{n}\left(\frac{2^n}{3^n} + 2^n - 1 - \zeta_n[2^n - 3 - (-1)^n]\right)\right\},\\
G_{C_m}(0,u) &= 2C_F e^{\frac{5u}{3}}\frac{(3u^2-4u+2)\Gamma(2+u)\Gamma(1-2u)}{\Gamma(3-u)}\\
&= C_F \exp\left[\frac{25\,u}{6} + \sum_{n=2}^{\infty}\frac{u^n}{n}\{1-(-1)^n + 2^{-n} + \zeta_n\,[2^n - 1 + (-1)^n]\}\right]\\
&\quad \times (2 - 4\,u + 3\,u^2),\\
\left.\frac{d}{du}G_{C_m}(\epsilon,u)\right|_{u=0} &= G_{C_m}(\epsilon,0)\left[\frac{1}{\epsilon}\log[T(\epsilon)] + \gamma_E + H_{2-\epsilon} - \frac{1-2\epsilon}{1-\epsilon}\right]\\
&= 2C_F \exp\left[-\frac{2\epsilon}{3} + \sum_{n=2}^{\infty}\frac{\epsilon^n}{n}\{2 - 2^n\,(1+3^{-n}) + \zeta_n\,[2^n - 3 - (-1)^n]\}\right]\\
&\quad \left\{2\,\epsilon + (1-\epsilon)\left[\frac{13}{6} + \sum_{n=3}^{\infty}\frac{\epsilon^n}{n}\left\{2^n + \frac{2^n}{3^n} + \frac{n}{2^n} - 1\right\}\right.\right.\\
&\quad \left.\left. + \sum_{n=3}^{\infty}\frac{\epsilon^n}{n}\zeta_n[2 - n + (-1)^n - 2^n]\right]\right\},
\end{aligned}$$

with $G_{1,0}^{C_m} = -10C_F/3$ and $G_{0,1}^{C_m} = 13C_F/3$.

The second equality of each expression is expressed as a Taylor polynomial times the exponential of another series, in a form amenable to be re-expanded by a computer program with the algorithms already explained. We again see $G_{C_Q}(\epsilon,$ $0)=-2G_{\tilde{J}}(\epsilon,0)=-G_{C_m}(\epsilon,0)$ as consequence of the universality of $\Gamma_{\rm cusp}$; indeed, from $\Gamma_{C_m}(\beta)=2\beta G_{C_m}(-\beta,0)/\beta_0$ we recover $\Gamma_{C_m}=2\Gamma_{\rm cusp}$. For the $H_m$ factor one gets an additional factor of two: $G_{H_m}=2G_{C_m}$.

These results can be used to obtain a closed form for $\gamma_{H_m}$ and its fixed-order coefficients, whose series has convergence radius $\Delta\beta=5/2$. We find full agreement with the leading flavor structure of the known full bHQET results up to $O(\alpha_s^3)$ [63, 85, 86], collected in the fifth column of table 8.2. In figure 9.2 we compare the exact form of the bHQET non-cusp anomalous dimension for $\alpha_s=0.9$ with its fixed-order partial sum including up to $\gamma_n^{H_m}$. Even for this enormous value of the strong coupling, the expanded result converges to the exact answer.

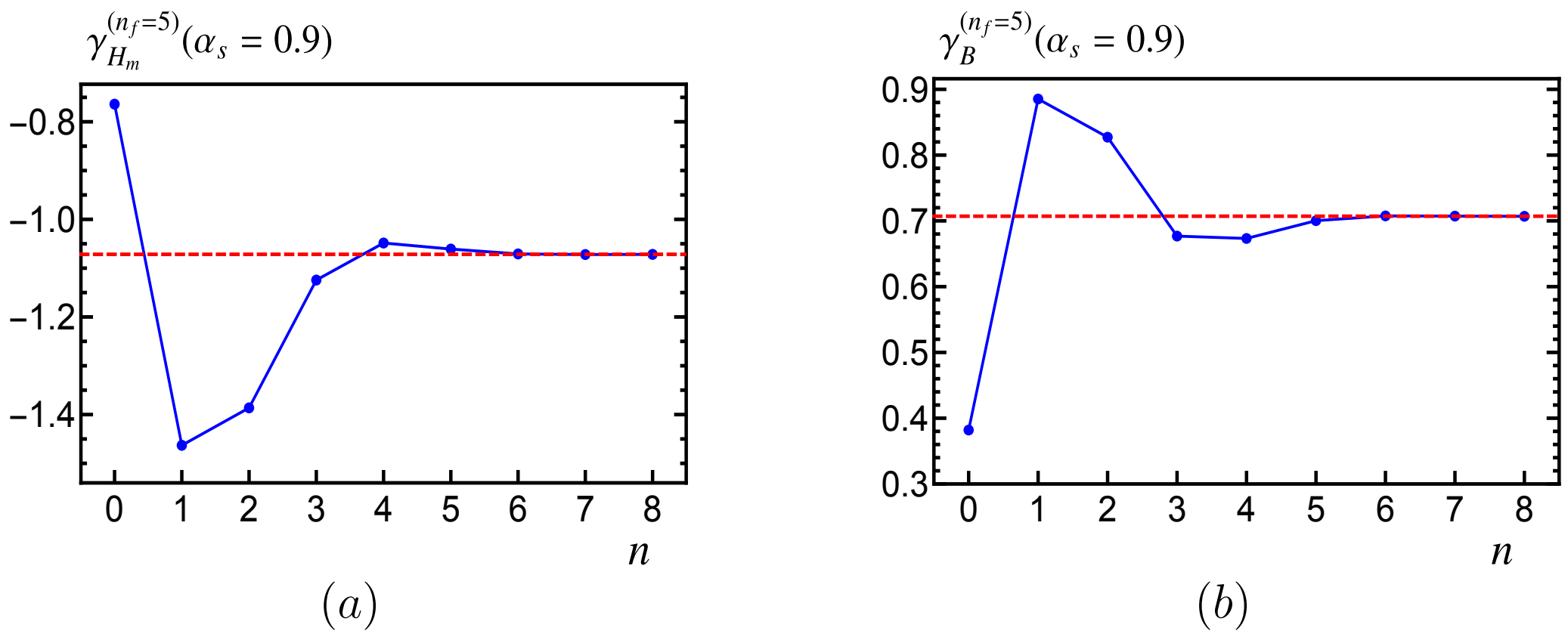

**Figure 9.2.** Comparison of the fixed-order partial sum with $n+1$ terms (blue dots) and exact results (red dashed line) for the non-cusp bHQET anomalous dimensions for the mass-scale hard function $H_m$ -panel $(a)$- and jet function $B$ -panel $(b)$- in the large-$\beta_0$ approximation for $\alpha_s=0.9$.

With the results above we reproduce the leading flavor structure of the full bHQET coefficients up to two loops [63, 86], see column 3 in table 8.3. As mentioned, the Borel transform $B_{C_m}(u)\equiv[G_{C_m}(0,u)-(G_{C_m})_{0,0}-u(G_{C_m})_{0,1}]/u^2$ has poles at $u=1,2$ and all positive half-integers, for which the leading renormalon $u=1/2$ is $O(\Lambda_{\rm QCD})$. Its asymptotic expansion can be written in terms of a finite and an infinite sum

$$B_{C_m}(u) \equiv C_F\left\{\sum_{k=1}^{5}\frac{(-1)^k e^{\frac{5k}{6}}\,[8+k\,(3\,k-8)]\,\Gamma\left(\frac{k}{2}+2\right)}{k\,\Gamma\left(3-\frac{k}{2}\right)\Gamma(k+1)}\,\frac{1}{u-\frac{k}{2}}\right. \tag{9.8}$$

$$\left.-\sum_{k=3}^{\infty}\frac{e^{\frac{5k}{3}+\frac{5}{6}}\,[3+4\,k\,(3\,k-1)]\,\Gamma\left(k+\frac{5}{2}\right)}{(2\,k+1)\,\Gamma\left(\frac{5}{2}-k\right)\Gamma(2\,k+2)}\,\frac{1}{u-\frac{2\,k+1}{2}}\right\}.$$

The leading ambiguity for $H_m$, which includes a factor of two, is three times as large as the pole mass ambiguity with the same sign:

$$\delta_\Lambda H_m = -\frac{6\,e^{\frac{5}{6}}\,C_F}{\beta_0}\frac{\Lambda_{\rm QCD}}{m}\,. \tag{9.9}$$

Therefore, the combination $H_m/m_p^3$ is free from the leading ambiguity. To illustrate how this renormalon cancellation works in practice we refer to figure 9.3, where the dimensionless quantity $(\bar m_t/m_t^{\rm pole})^3\,H_m(m_t,\mu)$ is considered, showing its exact value as a gray dashed line in both panels.

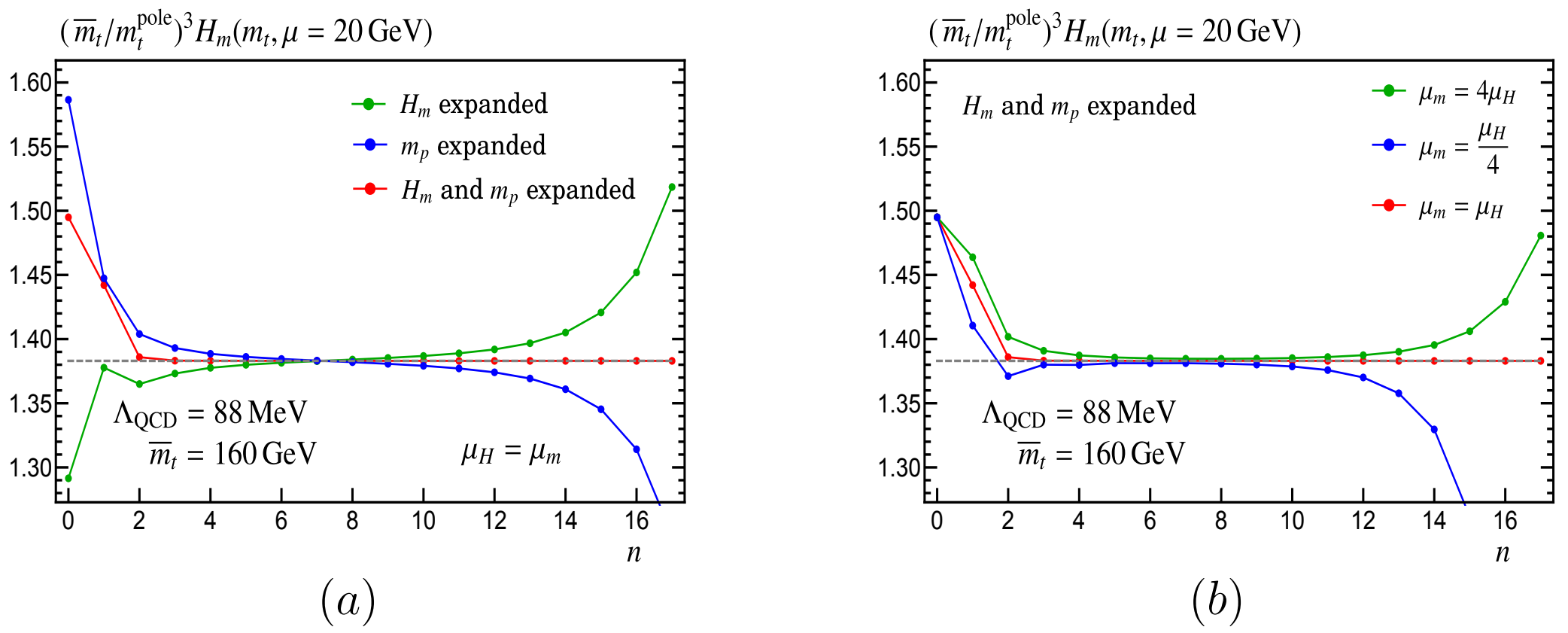

**Figure 9.3.** Comparison of the fixed-order partial sum (colored dots) and exact results (red dashed line) for the dimensionless and renormalon-free quantity $(\bar m_t/m_t^{\rm pole})^3\,H_m(m_t,\mu)$, with $\mu = 20$ GeV. In panel (a) we expand only $H_m$ (green dots), only $m_p$ (blue dots), or both (red dots), using $\alpha_s(\bar m)$ as the expansion parameter in all cases. In panel (b) we expand $H_m$ in terms of $\alpha_s(\bar m)$ and $m_p$ in powers of $\alpha_s(\mu_m)$ with $\mu_m = 4\,\bar m_t$ (green dots), $\mu_m = \bar m_t$ (red dots) and $\mu_m = \bar m_t/4$ (blue dots).

In these plots we highlight some important aspects of consistently expanding perturbative series to achieve an effective renormalon cancellation. For the perturbative expansions we use $\mu_H = \bar m_t$ as the matching scale, and therefore for a correct cancellation one needs to expand $(\bar m_t/m_t^{\rm pole})^3$ in powers of $\alpha_s(\mu_H)$. In figure 9.3(*a*) we either keep $m_p$ exact expanding in powers of $\alpha_s(\mu_H)$ only $H_m(m_t,\mu_H)$ (green dots); keep $H_m(m_t,\mu)$ exact expanding only $(\bar m_t/m_t^{\rm pole})^3$ (blue dots); or expand both (red dots), and only in the last case we see a better-behaved perturbative expansion –keeping both factors exact generates the gray line–. In figure 9.3(*b*), we expand $H_m$ and $(\bar m_t/m_t^{\rm pole})^3$ in powers of $\alpha_s(\mu_H)$ and $\alpha_s(\mu_m)$, respectively, with $\mu_m = \eta\times\mu_H$ and $\eta = 4, 1/4, 1$ shown as green, blue, and red dots, respectively. As expected, only when $\eta = 1$ the asymptotic behavior is removed. In both panels we observe that the removal of the leading renormalon ambiguity makes the series approach the exact result after fewer orders have been included in the partial sum. The $H_m$ hard factor mainly affects the norm of the distribution. In reference [79] this renormalon was not properly accounted for, and this fact, together with the lack of $\pi$ summation in $H_Q$, might explain the poor convergence found in peak cross sections for boosted top pairs unless the curves were self-normalized.

To determine the $C_m$ matching coefficient, on top of computing the relevant Feynman diagrams and accounting for the wave-function renormalization, one needs to renormalize the SCET current multiplying the bare result with $Z_{C_H}$ as computed in section 8.1. At leading order in $1/\beta_0$ this amounts to adding $\delta Z_{C_H}$ and therefore the renormalized $C_m$ at this order is obtained simply by dropping the divergent terms, denoted by $\delta Z^{(0)}_{C_m}$, from the bare computation. The renormalization factor for $C_m$ is then $\delta Z_{C_m} = \delta Z^{(0)}_{C_m} + \delta Z_{C_H}$ and accordingly, the $H_m$ anomalous dimension is $\gamma_{H_m} = \gamma_H + 2\gamma^{(0)}_{C_m}$.

## 9.2 bHQET jet function

The last matrix element we consider in this set of applications is the bHQET jet function for hemisphere masses. The relevant Feynman diagrams coincide with those of the SCET jet function shown in figure 8.4, but the actual computation uses bHQET Feynman rules in which the heavy-quark momentum $p = mv + r$ with $v^2 = 1$ has a residual component $r$ with soft scaling. We follow the same computational strategy as for the determination of the SCET jet function in section 8.2 based on the following forward-scattering matrix element:

$$m\mathcal{B}_n(\hat{s}, \mu) = \frac{1}{8\,\pi\,N_c}\int \mathrm{d}^d x\; \mathrm{e}^{\mathrm{i}k\cdot x}\; \mathrm{Tr}\langle 0|\mathrm{T}\{h_v(x)\,\bar{h}_v(0)\,W_n(0)\,W_n^\dagger(x)\}|\,0\rangle\,, \tag{9.10}$$

with $h_v$ an HQET massive quark field, $\hat{s} = 2\,v\cdot k$ and $W_n^{(\dagger)}$ the collinear Wilson line already introduced in section 8.2. Its discontinuity along the branch cut $\hat{s} > 0$ defines the momentum-space jet function $B_n(\hat{s}, \mu)$, with support for $\hat{s} > 0$. Both functions have dimensions of an inverse squared mass, although the kinematic variable $\hat{s} \equiv (s - m^2)/m$ they depend on has dimensions of energy. Since the fixed-order expansion of $B_n(\hat{s}, \mu)$ contains Dirac delta and plus distributions –even though $\mathcal{B}_n(\hat{s}, \mu)$ is a regular function–, it is convenient to consider its Fourier transform

$$\tilde{B}_n(x, \mu) \equiv m\int_0^\infty \mathrm{d}\hat{s}\; \mathrm{e}^{-\mathrm{i}\hat{s}x}\, B_n(\hat{s}, \mu)\,, \tag{9.11}$$

which due to the prefactor $m$ is dimensionless and depends on $x$, the variable conjugate to $\hat{s}$ with dimensions of an inverse energy. At lowest order one has

$$m\mathcal{B}_n^{\rm tree}(\hat{s}) = \frac{1}{2\,\pi}\frac{i}{\hat{s}+i\,0^+}\,, \qquad mB_n^{\rm tree}(\hat{s}) = \delta(\hat{s})\,, \qquad \tilde{B}_n^{\rm tree}(y) = 1\,. \tag{9.12}$$

Quantum corrections to the bare jet function, defined as $B_n(\hat{s}) = \delta(\hat{s}) + \delta B_n(\hat{s})$ and $\tilde{B}_n(x) = 1 + \delta\,\tilde{B}_n(x)$, are computed next at one-loop with a shifted gluon propagator.

Defining $a \equiv \hat{s} + \mathrm{i}\delta$

$$\begin{aligned}\mathcal{B}_a &= g_0^2 C_F \frac{2}{a} v^- \int \frac{\mathrm{d}^d\ell}{(2\pi)^d}\frac{1}{(\ell^2)^{1+h}[v\cdot(\ell+k)]\ell^-},\\ \mathcal{B}_b &= g_0^2 C_F \frac{4}{a^2}\int \frac{\mathrm{d}^d\ell}{(2\pi)^d}\frac{1}{(\ell^2)^{1+h}[v\cdot(\ell+k)]}.\end{aligned} \tag{9.13}$$

The two integrals can be put in terms of the results in appendix C.8:

$$I_{\mathcal{B}_a} = I_3(1, Q^{1+h}, E_v(v\cdot r), E_n) \qquad I_{\mathcal{B}_b} = I_2(1, Q^{1+h}, E_v(v\cdot r)),$$

so that $\mathcal{B}_{\rm sh} = 2\mathcal{B}_a + \mathcal{B}_b$ is

$$\begin{aligned}\frac{\mathrm{i}}{2\pi m}\mathcal{B}_{\rm sh} &= \frac{2\mathrm{i}g_0^2 C_F}{a\pi m}\left[v^- I_3(1, Q^{1+h}, E_v(v\cdot r), E_n) + \frac{1}{a}I_2(1, Q^{1+h}, E_v(v\cdot r))\right] \\ &= \left(\frac{g_0}{4\pi}\right)^2 \frac{4C_F}{\pi m}(-1)^h(-a)^{-1-2(h+\epsilon)}(4\pi)^\epsilon \frac{(1-h-\epsilon)\Gamma(-h-\epsilon)\Gamma(2h+2\epsilon-1)}{\Gamma(1+h)}.\end{aligned} \tag{9.14}$$

The imaginary part is again taken with (8.20), yielding

$$\mathrm{Im}[(-a)^{-1-2(h+\epsilon)}] = \frac{-\pi\hat{s}^{-1-2(h+\epsilon)}\theta(\hat{s})}{\Gamma(2h+2\epsilon)\Gamma(1-2h-2\epsilon)}, \tag{9.15}$$

so that

$$\begin{aligned}B_{\rm sh} &= \mathrm{Im}\left[\frac{\mathrm{i}}{2\pi m}\mathcal{B}_{\rm sh}\right] = \left(\frac{g_0}{4\pi}\right)^2 \frac{4C_F}{m}\hat{s}^{-1-2(h+\epsilon)}\frac{(-1)^{h+1}(4\pi)^\epsilon\Gamma(2-h-\epsilon)}{\Gamma(1+h)\Gamma(2-2h-2\epsilon)(h+\epsilon)}. \\ \tilde{B}_{\rm sh} &= \left(\frac{g_0}{4\pi}\right)^2 \frac{4C_F}{m}(\mathrm{i}x)^{2(h+\epsilon)}\frac{(-1)^h(4\pi)^\epsilon\Gamma(2-h-\epsilon)}{2(h+\epsilon)^2(1-2h-2\epsilon)\Gamma(1+h)}.\end{aligned} \tag{9.16}$$

From this result we easily identify the generating function $G_{\tilde{B}}(\epsilon, u)$

$$G_{\tilde{B}}(\epsilon,u)=2C_F\frac{\mathrm{e}^{-u\gamma_E}\Gamma(2-u)}{(1-2u)\Gamma(1+u-\epsilon)}T^{\frac{u}{\epsilon}-1}(\epsilon), \tag{9.17}$$

which is regular at the origin with $G_{\tilde{B}}(\epsilon,u)=G_{0,0}^{\tilde{B}}=2C_F$. Again we give the relevant evaluations of this function:

$$\begin{aligned} G_{\tilde{B}}(\epsilon,0) &= \frac{C_F}{3\pi}\frac{\Gamma(4-2\epsilon)\sin(\pi\epsilon)}{\epsilon\Gamma^2(2-\epsilon)} \qquad (9.18)\\ &= C_F\exp\left\{-\frac{5}{3}\epsilon-\sum_{n=2}^{\infty}\frac{\epsilon^n}{n}\left(\frac{2^n}{3^n}+2^n-1-\zeta_n[2^n-3-(-1)^n]\right)\right\},\\ G_{\tilde{B}}(0,u) &= 2C_F\mathrm{e}^{\left(\frac{5}{3}-2\gamma_E\right)u}\frac{\Gamma(2-u)}{(1-2u)\Gamma(1+u)}\\ &= 2C_F\exp\left\{\frac{8}{3}u+\sum_{n=2}^{\infty}\frac{u^n}{n}[2^n-1+\zeta_n[1-(-1)^n]\right\},\end{aligned}$$

$$\begin{aligned} \frac{\mathrm{d}}{\mathrm{d}u}G_{\tilde{B}}(\epsilon,u)\bigg|_{u=0} &= G_{\tilde{B}}(\epsilon,0)\left[\frac{1}{\epsilon}\log[T(\epsilon)]+\gamma_E-H_{1-\epsilon}+\frac{2-\epsilon}{1-\epsilon}\right]\\ &= 2\,C_F\exp\left[-\frac{5}{3}\epsilon+\sum_{n=2}^{\infty}\frac{\epsilon^n}{n}\left\{1-2^n-\frac{2^n}{3^n}+\zeta_n[2^n-3-(-1)^n]\right\}\right]\\ &\quad\times\left[\frac{8}{3}+\sum_{n=3}^{\infty}\frac{\epsilon^{n-1}}{n}\left\{2^n+\frac{2^n}{3^n}-1+\zeta_n\,[2+n-2^n+(-1)^n]\right\}\right].\end{aligned}$$

with $G_{1,0}^{\tilde{B}}=-10C_F/3$ and $G_{0,1}^{\tilde{B}}=16C_F/3$. Once again we observe $G_{\tilde{B}}(\epsilon,0)=G_{C_m}(\epsilon, 0)=2G_{\tilde{J}}(\epsilon,0)=-G_{C_Q}(\epsilon,0)$ and recover the cusp anomalous dimension as $\Gamma_B= 2\beta G_{\tilde{B}}(-\beta,0)/2=2\Gamma_{\text{cusp}}$.

When these results are applied to (4.32) and (4.34), the non-cusp part $\gamma_B(\beta)$ is obtained in closed form and as a perturbative series, finding for the latter a convergence radius of $\Delta\beta=5/2$. We find full agreement with the leading flavor structure of full bHQET results up to $O(\alpha_s^3)$ [63, 85, 86], see last column of table 8.2. From the bHQET consistency condition one can deduce the following relation between various non-cusp anomalous dimensions: $\gamma_{H_m}=-2\,(\gamma_S+\gamma_B)=2\,(\gamma_H+\gamma_J-\gamma_B)$, where in the second equality we use the SCET consistency condition to eliminate $\gamma_S$. This relation is exactly verified by our computations at $\mathcal{O}\,(1/\beta_0)$. In figure 9.2 we compare the exact result for $\gamma_B$ with the fixed-order partial sum including up to $\gamma_n^B$. We use $\alpha_s=0.9$, but nevertheless find that after adding 6 or more terms the perturbative result nicely agrees with the exact computation.

Finally, we compute the non-logarithmic fixed-order coefficients $c^B_{n,0}$ of the Fourier-space bHQET jet function, which we define as

$$\tilde{B}_n = \sum_{n=1} \beta^n \sum_{i=0}^{n+1} c^B_{n,i} \log^i(\mathrm{i}x e^{\gamma_E}\mu). \tag{9.19}$$

They are obtained from relation (4.36) and correctly reproduce the leading flavor structure of the full theory up to two loops [63, 85], see last column of table 8.3. The Borel transform $B_{\tilde{B}}(u) \equiv [G_{\tilde{B}}(0,u) - (G_{\tilde{B}})_{0,0} - u(G_{\tilde{B}})_{0,1}]/u^2$ has simple poles at $u=1/2$ –related to the pole mass ambiguity– and integer values of $u \geq 2$. Its asymptotic expansion is therefore expressed as an isolated term plus an infinite sum:

$$B_{\tilde{B}}(u) \asymp -2\,C_F \left[ \frac{2\, e^{\left(\frac{5}{6}-\gamma_E\right)}}{u-\frac{1}{2}} + \sum_{n=2}^{\infty} \frac{(-1)^n\,(n-1)\; e^{n\left(\frac{5}{3}-2\gamma_E\right)}}{n\,(1-2\,n)(n!)^2} \,\frac{1}{u-n} \right]. \tag{9.20}$$

The full ambiguity can be written in closed form in terms of Bessel functions of the first kind and generalized hypergeometric functions. The leading ambiguity is, as expected, proportional to $\Lambda_{\rm QCD}$ and takes the following form in position and momentum space:

$$\delta_\Lambda \tilde{B}(x) = -\frac{4\,C_F e^{5/6}}{\beta_0}(\mathrm{i}x\Lambda_{\rm QCD})\,, \qquad \delta_\Lambda B(x) = \frac{4\,C_F e^{5/6}}{\beta_0}\frac{\Lambda_{\rm QCD}}{\hat{s}}\,\delta(\hat{s})\,. \tag{9.21}$$

In position space, except for the factor $\mathrm{i}x$, the ambiguity is exactly twice that of $\delta_{\overline{\rm MS}}(\overline{m})$. If one writes the position-space version of the factorization theorem shown in equation (7.20), the bHQET jet function *effectively* appears in the combination $\hat{B}_n(x, m_p, \mu) \equiv \tilde{B}_n(x,\mu)\, e^{-2\mathrm{i}x m_p}$, which is free from the leading renormalon ambiguity. Therefore, in order to have a stable perturbative expansion for $\hat{B}_n$ one needs to express $m_p$ in a short-distance scheme and expand consistently in powers of $\alpha_s(\mu)$ with $\mu \simeq 1/x \simeq \hat{s}$. Since the heavy quark mass is no longer a dynamic scale of the effective theory, one should use a low-scale short-distance scheme whose relation to the pole mass is not proportional to the mass itself, such as $m^{\rm MSR}$. This avoids having renormalon-subtractions that are much larger than the corresponding fixed-order corrections for $\tilde{B}_n$, thus breaking the bHQET power counting. Furthermore, if one chooses $R \sim \mu$ there will be no large logarithms in the renormalon subtraction series. In figure 9.4 we show how the leading renormalon effectively cancels for the case of the top quark, both for the real and imaginary parts in panels ($a$) and ($b$), respectively. Since the real part of $\tilde{B}_n$ does not have the $u = 1/2$ ambiguity if $x$ is real, we choose a complex number, but for simplicity keep the renormalization scale $\mu$ real. When truncating the fixed-order series for $\tilde{B}_n$ and $\delta_{\rm MSR}$ at any finite order one gets residual dependence on $R$ and $\mu$. In the figures we use $n+1$ terms in the fixed-order partial sum, plus N$^n$LL resummation both in $\tilde{B}_n$ and $m_t^{\rm MSR}$. For this numerical analysis we choose $\mu_0 \neq R$. If one expands in powers of $\alpha_s(\mu_0)$ only $\tilde{B}_n$ (green dots) or $m_p$ (blue dots) the asymptotic behavior remains in the series, and it takes more orders to approach the exact value (shown as a gray dashed line). The way in which the series diverges in each case seems to be almost exactly opposite. When both functions are expanded consistently (red dots), the series appears much better behaved for large $n$, and converges to the exact result already at three loops.

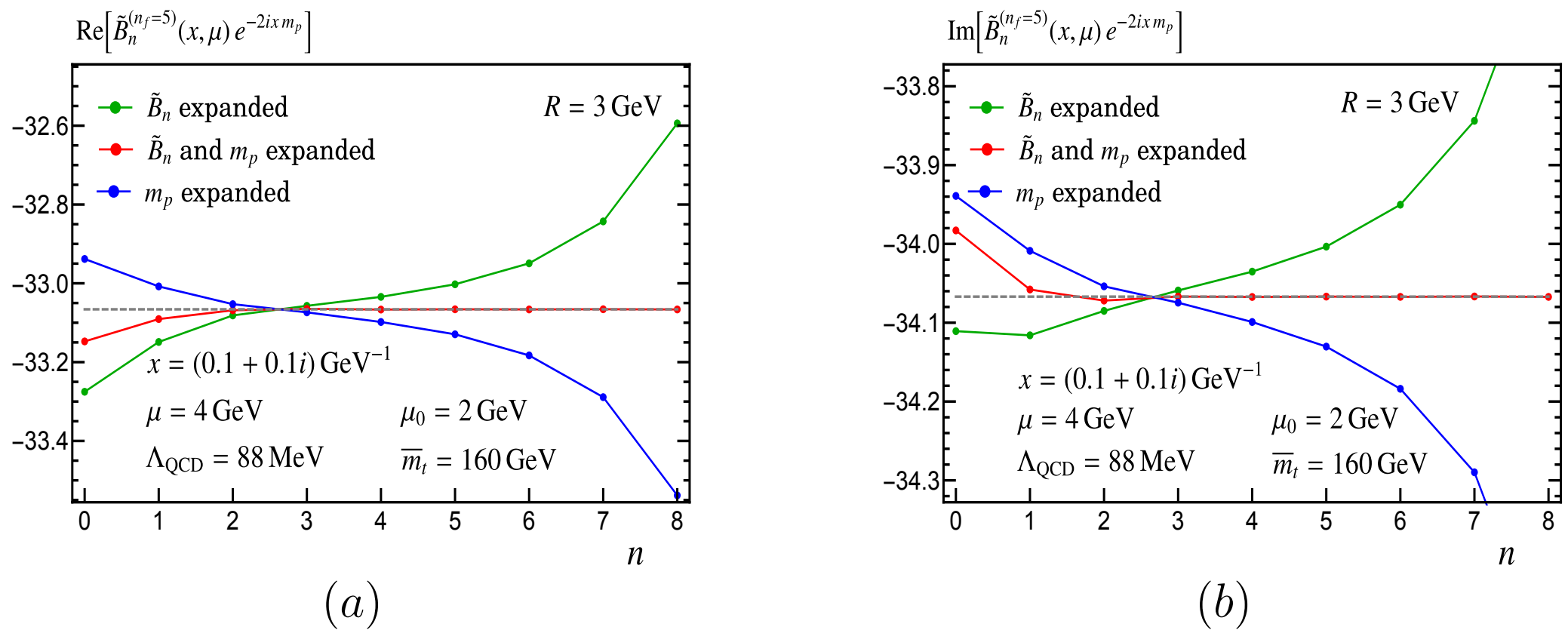

**Figure 9.4.** Comparison of the fixed-order partial sum (colored dots) and exact results (red dashed line) for the real (left panel) and imaginary (right panel) parts of the position-space bHQET jet function with renormalon subtraction $\tilde{B}_n(x,\,\mu)\,e^{-2ixm_p}$ in the MSR scheme. The plot uses $x=(0.1+0.1\,i)$ GeV$^{-1}$ and $\mu = 4$ GeV. The fixed-order results employ $\mu_0 = 2$ GeV and $R = 3$ GeV, and include $n+1$ terms in the matching condition plus N$^n$LL resummation. We show fixed-order results expanding $\tilde{B}_n$ only (green), expanding $m_p$ only (blue), and expanding both (red).

# Part II

# Asymptotic separation

# Chapter 10
# Theoretical framework of $\tau$-decays

## 10.1 Introduction

One of the most important tasks in modern particle physics is to test to great accuracy the predictions of the Standard Model. Improvements on the accuracy of the observable predictions translate into improvements on the accuracy of the fundamental parameters, such as couplings and masses, while discrepancies may reveal as clues enlightening the path towards beyond SM physics.

In the case of the strong interaction, confinement presents complications to the extraction of fundamental parameters from the predictions of perturbation theory. At high energies, where $\alpha_s$ lays well inside the perturbative regime and QCD asymptotic series may present acceptable behavior, collisions produce large numbers of particles. The inclusive decay of the $\tau$ into hadrons,

$$R_\tau \equiv \frac{\Gamma(\tau^- \to \text{hadrons}\,\nu_\tau)}{\Gamma(\tau^- \to e^-\bar{\nu}_e\nu_\tau)}, \tag{10.1}$$

stands out as one of the most appropriate observables for the study of QCD under clean conditions, with the extraction of $\alpha_s$ as one of its main applications. With a mass of $m_\tau = 1776.86 \pm 0.12$ MeV [29], the $\tau$ is the only lepton heavy enough to decay into hadrons, a decay which follows the $W$-boson channel whose lowest order diagram is depicted in figure 10.1.

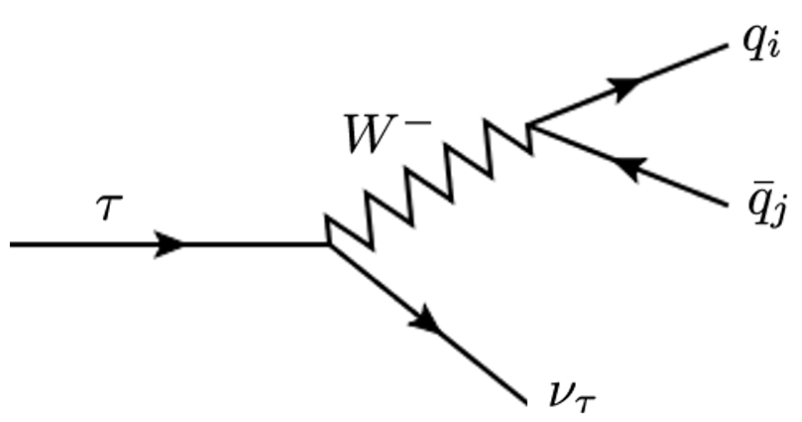

**Figure 10.1.** $\mathcal{O}(\alpha_s)$ diagram for $\tau \to \text{hadrons}\,\nu_\tau$.

The quark-antiquark pair after the $W^-$ decay is boosted and hadronizes in combinations of mesons due to the strong interaction. Accounting for this, the weak and strong interaction conservation laws dictate that the initial $q\bar{q}$ pair is either $\bar{u}d$ or $\bar{u}s$[10.1]. Thus, $\bar{u}d$ initiates a strangeness-conserved decay, while for $\bar{u}s$ strangeness is violated in 1 unit.

On the theoretical side, $R_\tau$ is written as the integral [87]

$$\begin{aligned} R_\tau &= 12\pi\int_0^{M_\tau^2}\frac{\mathrm{d}s}{m_\tau^2}\left(1-\frac{s}{m_\tau^2}\right)^2\left[\left(1+\frac{2s}{m_\tau^2}\right)\mathrm{Im}\,\Pi^{(1)}(s)+\mathrm{Im}\,\Pi^{(0)}(s)\right] \\ &= 6\pi\oint_{|s|=m_\tau^2}\frac{\mathrm{d}s}{m_\tau^2}\left(1-\frac{s}{m_\tau^2}\right)^2\left[\left(1+\frac{2s}{m_\tau^2}\right)\Pi^{(1+0)}(s)-\frac{2s}{m_\tau^2}\Pi^{(0)}(s)\right], \end{aligned} \tag{10.2}$$

where $\Pi^{(J)}(s)$ arise from the two-point correlation functions

$$\begin{aligned} &\Pi^C_{\mu\nu,ij}(p)\equiv i\int \mathrm{d}x\, e^{ipx}\langle 0|T\{J^C_{\mu,ij}(x)J^{\dagger C}_{\nu,ij}(0)\}|0\rangle, \\ &J^V_{\mu,ij}(x)=[\bar{q}_j\gamma_\mu q_i](x), \qquad J^A_{\mu,ij}(x)=[\bar{q}_j\gamma_\mu\gamma_5 q_i](x), \end{aligned} \tag{10.3}$$

where $|0\rangle$ denotes the physical vacuum, $C=V,A$ are the indices for the vector and axialvector currents, and $i$ and $j$ stand for $u$, $d$ and $s$ flavors. The correlators can be decomposed in terms of the scalar correlators $\Pi^{(J)}_{ij}$ as

$$\Pi^C_{\mu\nu,ij}(p)=(p_\mu p_\nu-g_{\mu\nu}p^2)\Pi^{(1)}_{ij,C}(p^2)+p_\mu p_\nu\Pi^{(0)}_{ij,C}(p^2), \tag{10.4}$$

where the superscripts denote $J=1$ (transversal) and $J=0$ (longitudinal) angular momentum in the hadronic rest frame. The appropriate flavor contributions to each correlator are

$$\Pi^{(J)}=|V_{ud}|^2\left[\Pi^{(J)}_{ud,V}+\Pi^{(J)}_{ud,A}\right]+|V_{us}|^2\left[\Pi^{(J)}_{us,V}+\Pi^{(J)}_{us,A}\right], \tag{10.5}$$

with $V_{ij}$ the elements of the Cabibbo-Kobayashi-Maskawa (CKM) quark-mixing matrix. The imaginary part of the correlators $\Pi^C_{\mu\nu,ij}$ is proportional to the spectral functions for hadrons with the same quantum numbers [87, 88], thus the first line of (10.1) computes the ratio $R_\tau$ as the moment $(1-s/m_\tau^2)^2(1+2s/m_\tau^2)$ of such spectral functions. To arrive to the second line of (10.1), in which $\Pi^{(1+0)}\equiv\Pi^{(1)}(s)+\Pi^{(0)}(s)$ and the circle $|s|=m_\tau^2$ runs counter-clockwise, the analyticity of $\Pi^{(J)}(s)$ in the complex plane except the positive real axis –where its imaginary part presents discontinuities– is employed[10.2].

10.1. Since only strong interactions are accounted for after the initial $q\bar{q}$ pair, the production of a $c$ quark would lead to a charmed meson, the lightest of which is already heavier than the $\tau$.

10.2. For further details on this manipulation, see Ref. [87].

Both in theoretical and experimental studies, the ratio $R_\tau$ is separated in its nonstrange and strange contributions, and the latter is further divided in a vector and axialvector current contributions: $R_\tau = R_{\tau,V} + R_{\tau,A} + R_{\tau,S}$. Experimentally, the nonstrange contribution $R_{\tau,V} + R_{\tau,A}$ is resolved into vector and axialvector parts with good accuracy. For example, for final states with only pions, isospin invariance dictates that an even number of pions corresponds to vector final states, and an odd number corresponds to axialvector states[10.3].

On the theoretical side, the scale $m_\tau^2$, at which the correlators in the contour representation (10.1) are evaluated, is large enough that the non-perturbative effects are small. The operator product expansion is used to separate perturbative and non-perturbative effects. Specifically, $R_{\tau,C}$ is parametrized as [88, 90]

$$R_{\tau,C} = \frac{N_c}{2} S_{\text{EW}} |V_{ud}|^2 \left( 1 + \delta^{(0)} + \delta'_{\text{EW}} + \sum_{D \geq 2} \delta^{(D)}_{ud,C} \right). \tag{10.6}$$

Here, $S_{\text{EW}} = 1.0198 \pm 0.0006$ [91] and $\delta'_{\text{EW}} = 0.0010 \pm 0.0010$ [92] are electroweak corrections, $\delta^{(0)}$ is the dimension-0, perturbative QCD correction for massless quarks and $\delta^{(D)}_{ud,C}$ denotes operator product expansion corrections in powers of $m_\tau^{-D}$ of quark mass dimension ($D = 2$) and higher. The massless correction $\delta^{(0)}$, which is identical for both currents, is the focus of the this part of the thesis, while mass corrections start at $D = 2$. The precise relation between the representations (10.2) and (10.6) will be made clear in the next section.

In this context, studies on the $\tau$ decay width in $R_\tau$ and moments of the $\tau$ hadronic spectral functions have allowed to determine $\alpha_s$ with a precision competitive to the world average $\alpha_s(m_Z^2) = 0.1179 \pm 0.0009$ [29]. The analysis of the data collected by the ALEPH collaboration at LEP [88, 89] yielded $\alpha_s(m_\tau^2) = 0.332 \pm 0.005_{\text{exp}} \pm 0.011_{\text{theo}}$, which after evolution to $m_Z^2$ led to $\alpha_s(m_Z^2) = 0.1199 \pm 0.0006_{\text{exp}} \pm 0.0012_{\text{theo}} \pm 0.0005_{\text{evol}} = 0.1199 \pm 0.0015_{\text{tot}}$ [93, 94].

The main contribution to the theoretical uncertainty comes from the error in perturbative QCD corrections to the moments of the spectral functions, which in turn is dominated by the fact that the two expansion methods known as fixed-order perturbation theory (FOPT) and contour-improved perturbation theory (CIPT) yield numerical differences for $\delta^{(0)}$. This problem, referred to as asymptotic separation, has been the object of various recent studies [90, 95, 96, 97, 98].

10.3. Other combinations can also be resolved into vector and axialvector contributions such as the vector state $K^- K^0$. An exception are the $K\overline{K}$ states, which are not eigenstates of $G$-parity and thus contribute to both vector and axialvector channels. These induces a small uncertainty in the experimental determination of $R_{\tau,V}$ and $R_{\tau,A}$ [88, 89].

## 10.2 The perturbative correction $\delta^{(0)}$

The second line of (10.2) presents the advantage that the correlators are only required at $|s|=m_\tau^2$, instead of along the real positive axis where its imaginary parts presents poles and where small values of $s$ lay in the non-perturbative regime. To avoid the dependence on the renormalization scale and the subtraction schemes of the correlators, it is advantageous to write $R_\tau$ in terms of the Adler functions

$$D^{(1+0)}(s) \equiv -s\frac{\mathrm{d}}{\mathrm{d}s}\Pi^{(1+0)}(s), \qquad D^{(0)}(s) \equiv \frac{s}{m_\tau^2}\frac{\mathrm{d}}{\mathrm{d}s}[s\Pi^{(0)}(s)], \tag{10.7}$$

which are physical in the sense that satisfy homogeneous renormalization group equations. Changing variables to $x = s/M_\tau^2$ one obtains

$$R_\tau = -i\pi\oint_{|x|=1}\frac{\mathrm{d}x}{x}(1-x)^3\Big\{3\,(1+x)D^{(1+0)}(m_\tau^2 x) + 4D^{(0)}(m_\tau^2 x)]\Big\}, \tag{10.8}$$

where integration by parts has been used with

$$\begin{aligned}(1-x)^2(1+2x) &= -\frac{1}{2}\frac{\mathrm{d}}{\mathrm{d}x}[(1-x)^3(1+x)], \\ (1-x)^2 &= -\frac{1}{3}\frac{\mathrm{d}}{\mathrm{d}x}(1-x)^3.\end{aligned} \tag{10.9}$$

On the other hand, the perturbative correction $\delta^{(0)}$ is computed in the massless limit, in which vector and axialvector contributions coincide and $s\Pi^{(0)}_{ij,C}(s)=0$. Thus to inspect $\delta^{(0)}$ it is enough to keep only the $D^{(1+0)}$ term in (10.8):

$$R_{\tau,C} = -3i\pi|V_{ud}|^2\oint_{|x|=1}\frac{\mathrm{d}x}{x}(1-x)^3\,(1+x)D_V^{(1+0)}(m_\tau^2 x) = \frac{N_c}{2}|V_{ud}|^2(1+\delta^{(0)}), \tag{10.10}$$

where $D_V^{(1+0)}$ is built from the corresponding vector correlator for $u$ and $d$ quarks. To account for the parton-model result in the right-hand side one defines the reduced Adler function,

$$\frac{12\pi^2}{N_c}D_V^{(1+0)}(s) \equiv 1 + \hat{D}(s), \tag{10.11}$$

getting

$$\frac{1}{2i\pi}\oint_{|x|=1}\frac{\mathrm{d}x}{x}(1-x)^3\,(1+x)[1+\hat{D}(m_\tau^2 x)] = 1+\delta^{(0)}. \tag{10.12}$$

The first integral in the left-hand side can be solved by parameterizing $x$ in the unit circle as $x = e^{i\varphi}$,

$$\frac{1}{2i\pi}\oint_{|x|=1}\frac{\mathrm{d}x}{x}(1-x)^3\,(1+x) = \frac{1}{2i\pi}\int_0^{2\pi}\mathrm{d}\varphi\, i(1-e^{i\varphi})^3\,(1+e^{i\varphi}) = 1, \qquad (10.13)$$

leading to the following form for $\delta^{(0)}$:

$$\delta^{(0)} = \frac{1}{2i\pi}\oint_{|x|=1}\frac{\mathrm{d}x}{x}(1-x)^3\,(1+x)\hat{D}(m_\tau^2 x). \qquad (10.14)$$

The vector correlator $\Pi_V^{(1+0)}$ and thus Adler function $\hat{D}(s)$ are known perturbatively up to $\mathcal{O}(\alpha_s^4)$ [99, 100, 101]. The perturbative expansion of the correlator $\Pi_V^{(1+0)}$ is

$$\Pi_V^{(1+0)}(s) = -\frac{N_c}{12\pi^2}\sum_{n=0}^{\infty}\left(\frac{\alpha_s}{\pi}\right)^n\sum_{k=0}^{n+1}c_{n,k}L^k. \qquad (10.15)$$

with $L \equiv \log(-s/\mu^2)$, and so, taking the derivative with respect to $s$ as in (10.7) yields

$$\begin{aligned}\hat{D}(s) &= \sum_{n=1}^{\infty}\left(\frac{\alpha_s(\mu^2)}{\pi}\right)^n\sum_{k=1}^{n+1}k\,c_{n,k}L^{k-1} \qquad (10.16)\\ &= \sum_{n=1}^{\infty}\left(\frac{\alpha_s(m_\tau^2)}{\pi}\right)^n\sum_{k=1}^{n+1}k\,c_{n,k}\log^{k-1}\left(\frac{-s}{m_\tau^2}\right)\\ &= \sum_{n=1}^{\infty}c_{n,1}\left(\frac{\alpha_s(-s)}{\pi}\right)^n.\end{aligned}$$

In the first line we simply took the derivative of the correlator, and in the second and third lines we used the homogeneous RGE to expand in powers of $\alpha_s(m_\tau^2)$ and $\alpha_s(-s)$, respectively. The expansion in powers of $\alpha(-s)$ cancels all the logarithms and only $c_{n,1}$ contribute.

From here two possibilities open up due to the fact that one can use either the second or third lines of (10.16) into (10.14) to build the perturbative correction $\delta^{(0)}$. In the first case the approach is called fixed order perturbation theory and in the second case it is called contour improved perturbation theory.

1. **Fixed order perturbation theory** (FOPT, in the following abbreviated to FO). The logarithms are summed up with the choice $\mu^2 = m_\tau^2$, which leads to

$$\delta_{\mathrm{FO}}^{(0)} = \frac{1}{2i\pi}\sum_{n=1}^{\infty}\frac{\alpha_s^n(m_\tau^2)}{\pi^n}\sum_{k=1}^{n+1}k\,c_{n,k}\oint_{|x|=1}\frac{\mathrm{d}x}{x}(1-x)^3(1+x)\log^{k-1}(-x). \qquad (10.17)$$

**2. Contour improved perturbation theory** (CIPT, in the following abbreviated to CI). The logarithms are summed up with the choice $\mu^2 = -s = -x m_\tau^2$, which leads to

$$\delta_{\rm CI}^{(0)} = \frac{1}{2i\pi}\sum_{n=1}^{\infty}\frac{c_{n,1}}{\pi^n}\oint\limits_{|x|=1}\frac{{\rm d}x}{x}(1-x)^3(1+x)\alpha_s^n(-x m_\tau^2)\,. \qquad (10.18)$$

In FO, the complex-valuedness of the Adler function is encoded in the logarithms, while in CI it is encoded in the strong coupling.

Finally, to better inspect the individual moments of the weight function $W(x) = (1-x)^3(1+x)$, we define the moments of $\delta^{(0)}$:

$$\delta_{{\rm FO},l}^{(0)} \equiv \frac{1}{2i\pi}\sum_{n=1}^{\infty}\frac{\alpha_s^n(m_\tau^2)}{\pi^n}\sum_{k=1}^{n+1}k c_{n,k}\oint\limits_{|x|=1}\frac{{\rm d}x}{x}(-x)^l \log^{k-1}(-x), \qquad (10.19)$$

$$\delta_{{\rm CI},l}^{(0)} \equiv \frac{1}{2i\pi}\sum_{n=1}^{\infty}\frac{c_{n,1}}{\pi^n}\oint\limits_{|x|=1}\frac{{\rm d}x}{x}(-x)^l\alpha_s^n(-x m_\tau^2).$$

In each case, the linear combination to recover $W(x)$ is $\delta^{(0)} = \delta_0^{(0)} + 2\delta_1^{(0)} - 2\delta_3^{(0)} - \delta_4^{(0)}$.

## 10.3 Recurrence relations for the Adler function

Although we will be addressing the computation of the FO and CI series in the large-$\beta_0$ limit, we can nevertheless derive in full generality the constraints imposed on the $c_{n,m}$ coefficients of the Adler function due to its homogeneous RGE. These constraints take the form of recurrence relations. The RGE equation for $\hat{D}(s)$ is

$$\mu\frac{\rm d}{{\rm d}\mu}\hat{D}(s) = \left[2\frac{\partial}{\partial L} - \beta(\alpha_s)\frac{\partial}{\partial\alpha_s}\right]\hat{D}(s) = 0, \qquad (10.20)$$

and gives the following two terms:

$$2\frac{\partial}{\partial L}\hat{D}(s) = 2\sum_{n=1}^{\infty}\left(\frac{\alpha_s}{\pi}\right)^n\sum_{k=1}^{n+1}m(m-1)\,c_{n,k}L^{k-2} \qquad (10.21)$$

$$-\beta(\alpha_s)\frac{\partial}{\partial\alpha_s}\hat{D}(s) = -2\alpha_s\sum_{j=0}^{\infty}\beta_j\left(\frac{\alpha_s}{4\pi}\right)^{j+1}\sum_{n=1}^{\infty}\frac{n\alpha_s^{n-1}}{\pi^n}\sum_{k=1}^{n+1}k c_{n,k}L^{k-1}.$$

In the second term we used the expansion of the beta function in (1.51). The idea is to work on each term to explicitly to write it as a polynomial in $\alpha_s$ and $L$, upon the addition of both contributions, each power must vanish identically. In the first term it is enough to discard the $k=1$ case and take $k\to k+1$:

$$2\frac{\partial}{\partial L}\hat{D}(s)=2\sum_{n=1}^{\infty}\left(\frac{\alpha_s}{\pi}\right)^n\sum_{k=1}^{n}(k+1)k\,c_{n,k+1}L^{k-1}. \tag{10.22}$$

In the second term we manipulate the sums the following way

$$\begin{aligned}
-\beta(\alpha_s)\frac{\partial}{\partial\alpha_s}\hat{D}(s) &= 2\sum_{n=1}^{\infty}\sum_{k=0}^{\infty}\frac{n\alpha_s^{n+j+1}\,\beta_j}{4^{j+1}\pi^{n+j+1}}\sum_{k=1}^{n+1}k\,c_{n,k}L^{k-1} \\
&= 2\sum_{n=1}^{\infty}\sum_{j=n+1}^{\infty}\frac{n\alpha_s^{j}\,\beta_{k-n-1}}{4^{j-n}\pi^{j}}\sum_{k=1}^{n+1}k\,c_{n,k}L^{k-1} \\
&= 2\sum_{j=2}^{\infty}\left(\frac{\alpha_s}{4\pi}\right)^j\sum_{n=1}^{j-1}4^n n\,\beta_{j-n-1}\sum_{k=1}^{n+1}k\,c_{n,k}L^{k-1} \\
&= \frac{1}{2}\sum_{j=2}^{\infty}\left(\frac{\alpha_s}{4\pi}\right)^j\sum_{k=1}^{j}k\,L^{k-1}\sum_{n=k}^{j}4^n(n-1)\,\beta_{j-n}\,c_{n-1,k}.
\end{aligned} \tag{10.23}$$

First we performed $j\to j-n-1$, then we swapped the $n$ sum first with the $j$ sum and then with the $k$ sum, and finally took $n\to n-1$. When swapping the $n$ and $k$ sums it is best to extend the $n$ sum to start from $n=0$, since that only includes a vanishing contribution. Adding both derivatives together and comparing powers of $\alpha_s$ and $L$ yields $c_{1,2}=0$ and

$$c_{n,k}=-\frac{1}{4^{n+1}k}\sum_{j=k-1}^{n}4^j(j-1)\beta_{n-j}c_{j-1,k-1},\qquad n\ge 2,\quad 2\le k\le n+1. \tag{10.24}$$

The particular case of $c_{n,n+1}$ can be seen to vanish, since the recurrence relation (10.24) reduces to

$$c_{n,n+1}=-\frac{\beta_0(n-1)}{4(n+1)}c_{n-1,n}, \tag{10.25}$$

for which the initial case is $c_{1,2}=0$. This can be accounted for in (10.24), where it corresponds to eliminating the $j=k-1$ case, so the final expression for $c_{n,k}$ is

$$c_{n,k}=-\frac{1}{4^{n+1}k}\sum_{j=k}^{n}4^j(j-1)\beta_{n-j}c_{j-1,k-1},\qquad n\ge 2,\quad 2\le k\le n. \tag{10.26}$$

This relation provides a way of computing the coefficients $c_{n,k}$ recursively from the coefficients of the beta function.

# Chapter 11

# The gluon condensate renormalon model in the large-$\beta_0$ limit

In this chapter we study the FO and CI series in the gluon condensate model in the large-$\beta_0$. This model has the advantage that the Adler function in powers of $a(-s)$ has a single renormalon at $u=2$, and thus a form simple enough that analytical expressions can be found for the spectral moments in FO and CI.

## 11.1 The model

In the large-$\beta_0$ limit we take $a(\mu^2)\equiv\alpha_s(\mu^2)\beta_0/(4\pi)$ as the coupling in perturbative expansions and define the notation $a_\mu\equiv a(\mu^2)$ with the particular cases $a\equiv a(m_\tau^2)$ and $a_-\equiv a(-s)$. The LO running of $\alpha_s$ in (2.19) is exact in the large-$\beta_0$ and we can write

$$\frac{a(-xm_\tau^2)}{a}=\frac{1}{1+a\log(-x)}. \tag{11.1}$$

We work under a Borel model for the reduced Adler that contains a single IR renormalon at $u=2$. The Borel transform of $\hat{D}(s)$ setting $\mu^2=-s$ with respect to $a_-$ (this is, the third line in (10.16)) is

$$B[\hat{D}(s)](u)=\frac{1}{2-u}. \tag{11.2}$$

The motivation for this model is that it is known that the Borel transform of the complete Adler function is a combination of non-analytic structures $1/(p_{\rm IR}-u)^\gamma$ and $1/(p_{\rm UV}+u)^\gamma$ –with $p_{\rm IR}$ and $p_{\rm UV}$ being positive integers and $\gamma$ being real– describing IR and UV renormalons, respectively, and functions that are analytic in the positive real axis of $u$ [4, 97].

The coefficients $c_{n,k}$ of the Adler function in these conditions can be easily identified. First, expanding $B[\hat{D}]$ for small $u$ and performing the inverse Borel transform for each term,

$$\hat{D}(s)=\sum_{n=0}^{\infty}\frac{1}{2^{n+1}}\int_0^{\infty}\mathrm{d}u\,e^{-u/a_-}u^n=\sum_{n=1}^{\infty}\frac{\Gamma(n)}{2^n}a_-^n, \tag{11.3}$$

one arrives, by comparison with the third line of (10.16), to

$$c_{n,1}=\left(\frac{\beta_0}{4}\right)^n\frac{\Gamma(n)}{2^n}. \tag{11.4}$$

The remaining coefficients can be computed recursively from the $c_{n,1}$ with the result (10.26). In the large-$\beta_0$ limit, all $\beta_{n\geq 1}$ are set to zero and the recurrence relation simplifies to

$$c_{n,k}=-\frac{\beta_0(n-1)}{4k}c_{n-1,k-1},\qquad n\geq 2,\quad 2\leq k\leq n. \tag{11.5}$$

It is useful at this stage to define $\bar{c}_{n,k}\equiv(4/\beta_0)^n c_{n,k}$, for which (11.5) reduces to

$$\begin{aligned}\bar{c}_{n,k} &= -\frac{n-1}{k}\bar{c}_{n-1,k-1}=\left[-\frac{n-1}{k}\right]\left[-\frac{n-2}{k-1}\right]\ldots\left[-\frac{n-(k-1)}{k-(k-2)}\right]\bar{c}_{n-k+1,1} \\ &= \frac{(-1)^{k+1}\Gamma(n)}{\Gamma(k+1)\Gamma(n-k+1)}\bar{c}_{n-k+1,1},\end{aligned} \tag{11.6}$$

where in the second step we used it recursively on itself $k-1$ times. Substituting $\bar{c}_{n,1}=\Gamma(n)/2^n$ the coefficients are

$$\bar{c}_{n,k}=\frac{(-1)^{k+1}\Gamma(n)}{2^{n-k+1}\Gamma(k+1)}. \tag{11.7}$$

## 11.2 Approaches to $\delta^{(0)}$

### 11.2.1 Fixed order perturbation theory

From the first line of (10.19) and from (11.7), the momenta of the FO series are

$$\begin{aligned}\delta^{(0)}_{\mathrm{FO},l} &= \sum_{n=1}^{\infty}a^n d_{n,l}^{\mathrm{FO}}\equiv\sum_{n=1}^{\infty}a^n\sum_{k=1}^{n}k\,\bar{c}_{n,k}I_{k-1,l}, \\ I_{k,l} &\equiv \frac{1}{2i\pi}\oint_{|x|=1}\frac{\mathrm{d}x}{x}(-x)^l\log^k(-x).\end{aligned} \tag{11.8}$$

where we have eliminated the $k=n+1$ case due to $\bar{c}_{n,n+1}=0$. The $I_{k,l}$ integrals can be solved by the parametrization $x=e^{i\varphi}$ in the unit circle, where, due to the minus sign accompanying $x$, we let $\varphi\in[0,2\pi]$. The integrand contains the factors

$$\begin{aligned}&\log(-e^{i\varphi})=i\mathrm{Arg}(e^{i(\varphi-\pi)})=i(\varphi-\pi),\\&(-e^{i\varphi})^l=e^{i(\varphi-\pi)l},\end{aligned}\tag{11.9}$$

where we took the principal value of the logarithm since $\varphi-\pi\in[-\pi,\pi]$. All in all,

$$I_{k,l}=\frac{i^k}{2\pi}\int_0^{2\pi}\mathrm{d}\varphi e^{i(\varphi-\pi)l}(\varphi-\pi)^k=\frac{i^k}{2\pi}\int_{-\pi}^{\pi}\mathrm{d}\varphi e^{il\varphi}\varphi^k=\frac{\Gamma(k+1,\mathrm{i}l\pi,-\mathrm{i}l\pi)}{2i\pi(-l)^{k+1}},\tag{11.10}$$

where $\Gamma(a,z_1,z_2)$ is the doubly incomplete gamma function and the result holds for integer $k$ and $l$. The $l=0$ case is

$$I_{k,0}=\frac{i^k}{2\pi}\int_{-\pi}^{\pi}\mathrm{d}\varphi\varphi^k=\frac{(i\pi)^k[1+(-1)^k]}{2(k+1)}=\frac{(i\pi)^k}{k+1}\cos\left(\frac{\pi k}{2}\right).\tag{11.11}$$

It is worth noticing that the doubly incomplete gamma in (11.10) can be written as the subtraction of two incomplete gamma functions as $\Gamma(z,a,b)=\Gamma(z,a)-\Gamma(z,b)$. This gamma subtraction cancels the real part, and the dividing factor of $i$ renders $I_{k,l}$ real. The same occurs for $I_{k,0}$, where the term in square brackets only allows for even values of $k$, for which $i^k$ is real.

From these results the coefficients $d_{n,l}^{\mathrm{FO}}$ can be given explicit closed forms. In the case $l=0$, the sum over $k$ can be directly solved, yielding

$$d_{n,0}^{\mathrm{FO}}=\frac{\Gamma(n+1,2i\pi,-2i\pi)}{2^{n+2}in\pi}.\tag{11.12}$$

It is also possible to find a closed form for the coefficients when $l\neq 0$. Indeed, making use of the integral representation of the gamma function one finds

$$\begin{aligned}d_{n,l}^{\mathrm{FO}}&=\frac{\Gamma(n)}{2^{n+2}i\pi}\sum_{k=1}^{n}\left(\frac{2}{l}\right)^k\frac{1}{\Gamma(k)}\int_{-i\pi l}^{i\pi l}\mathrm{d}t\,t^{k-1}e^{-t}=\int_{-i\pi l}^{i\pi l}\mathrm{d}t\,\frac{e^{t(2-l)/l}\Gamma\left(n,\frac{2t}{l}\right)}{2^{n+1}i\pi l}\\&=\frac{(-1)^l2^{-n}\Gamma(n,-2i\pi,2i\pi)-l^{-n}\Gamma(n,-i\pi l,i\pi l)}{2i\pi(l-2)},\end{aligned}\tag{11.13}$$

where in the second step we interchanged the sum and the integral, as both are finite. The $l=2$ case can be found either by solving the last integral in (11.13) or as the $l\to 2$ limit of $d_{n,l}^{\mathrm{FO}}$:

$$d_{n,2}^{\mathrm{FO}}=\frac{\Gamma(n,2i\pi)+\Gamma(n,-2i\pi)}{2^{n+1}}+\frac{\Gamma(n+1,-2i\pi,2i\pi)}{2^{n+2}i\pi}.\tag{11.14}$$

In this case, the sum of incomplete gamma functions in the first term cancels the imaginary part. The real part is factorially increasing, therefore one expects $\delta^{(0)}_{\rm FO,2}$ to have zero radius of convergence.

### 11.2.2 Contour improved perturbation theory

In CIPT, the momenta of $\delta^{(0)}$ are,

$$\delta^{(0)}_{{\rm CI},l} = \sum_{n=1}^{\infty} \frac{\Gamma(n)}{2^n} a^n H_{n,l}(a), \tag{11.15}$$
$$H_{n,l}(a) \equiv \frac{1}{2i\pi} \oint\limits_{|x|=1} \frac{{\rm d}x}{x} (-x)^l \left[\frac{a(-x m_\tau^2)}{a}\right]^n .$$

In the large-$\beta_0$ the LO running of $\alpha_s$ (2.19) is exact and we can write

$$H_{n,l}(a) = \frac{1}{2i\pi} \oint\limits_{|x|=1} \frac{{\rm d}x}{x} \frac{(-x)^l}{[1+a\log(-x)]^n} = \frac{1}{2\pi}\int_{-\pi}^{\pi} {\rm d}\varphi \frac{e^{i\varphi l}}{(1+ia\varphi)^n}, \tag{11.16}$$

where first we evolved to $a(m_\tau^2) = a$ and then we parametrized $x$ in the unit circle using the ideas in (11.9). The integral $H_{n,l}(a)$ is well defined in the region of the complex $a$-plane given by ${\rm Im}(a) < 1/\pi$. For $l = 0$ one has

$$\begin{aligned} H_{n,0}(a) &= \frac{(1-ia\pi)^{1-n} - (1+ia\pi)^{1-n}}{2i\pi a(n-1)} = {}_2F_1\!\left(\frac{n}{2}, \frac{n+1}{2}, \frac{3}{2}, -a^2\pi^2\right) \\ &= \frac{\sin[(n-1)\arctan(a\pi)]}{a\pi(n-1)(1+a^2\pi^2)^{\frac{1}{2}(n-1)}}, \end{aligned} \tag{11.17}$$

for $n \geq 2$ and $H_{1,0}(a) = \arctan(\pi a)/(\pi a)$. We collected three representations of $H_{n,0}(a)$, all of which hold in the entirety of its domain of definition. The first one corresponds to the one used in [95] and is obtained by direct integration; the second one is found by expanding the binomial, integrating term by term and summing up the results to the hypergeometric function ${}_2F_1$; finally, the third representation can be derived by writing the complex binomials in the first representation in polar coordinates.

For $l \neq 0$ the integral can also be solved, yielding different expressions for ${\rm Re}(a) > 0$ and ${\rm Re}(a) < 0$:

$$H_{n,l}(a) = \frac{e^{-l/a}}{2\pi i l}\left(-\frac{l}{a}\right)^n \Gamma\!\left(1-n, -\frac{l}{a} - il\pi, -\frac{l}{a} + il\pi\right) + \theta[{\rm Re}(a)]\frac{e^{-l/a}}{l\Gamma(n)}\left(\frac{l}{a}\right)^n, \tag{11.18}$$

### 11.2.3 IR subtracted CIPT

In Ref. [98] a model for a renormalon-free gluon condensate matrix element is proposed and applied to the CI series –it does not contribute in FO except for $l=2$–. The subtraction is implemented perturbatively at a scale $R$ on the Adler function, and in the large-$\beta_0$ limit takes the form (recall $a_- = a(-s)$):

$$\hat{D}^{\rm RF}(s,R^2) \equiv \frac{c_0(R^2)}{s^2} + \sum_{n=1}^{\infty}\frac{\Gamma(n)}{2^n}a_-^n - \frac{R^4}{s^2}\sum_{n=1}^{\infty}\frac{\Gamma(n)}{2^n}a_R^n, \tag{11.19}$$
$$c_0(R^2) \equiv R^4\ {\rm P.V.}\bigg\{\int_0^\infty {\rm d}u \frac{{\rm e}^{-u/a_R}}{2-u}\bigg\} = R^4 e^{-2/a_R}{\rm Ei}\bigg(\frac{2}{a_R}\bigg).$$

Once can verify the $R$-subtraction successfully removes the gluon condensate renormalon by taking the Borel transform of $\hat{D}^{\rm RF}$:

$$\begin{aligned}\hat{D}_B^{\rm RF}(s\,,R^2) &= \frac{c_0(R^2)}{s^2} + \int_0^\infty {\rm d}u\left[\frac{{\rm e}^{-u/a_-}}{2-u} - \frac{R^4}{s^2}\frac{{\rm e}^{-u/a_R}}{2-u}\right] \\ &= \frac{c_0(R^2)}{s^2} + \int_0^\infty {\rm d}u\,\frac{{\rm e}^{-u/a_-}}{2-u}\left[1 - \frac{R^4}{s^2}\bigg(\frac{-s}{R^2}\bigg)^u\right],\end{aligned} \tag{11.20}$$

where in the last line we evolved $a_R$ to $a(-s)$. This result evidences how the $u=2$ renormalon is removed from the Adler function.

To build the subtracted CI series, we first reexpand $\hat{D}^{\rm RF}(s)$ in terms of the coupling $a(-s)$, where the coefficients are polynomials in $L_R \equiv \log(-s\,/\,R^2)$. The relevant term is

$$\begin{aligned}\sum_{n=1}^{\infty}\frac{\Gamma(n)}{2^n}a_R^n &= \sum_{n=1}^{\infty}\frac{\Gamma(n)}{2^n}\sum_{k=0}^{\infty}\frac{a_-^{n+k}\Gamma(n+k)L_R^k}{\Gamma(n)\Gamma(k+1)} = \sum_{k=0}^{\infty}\sum_{n=k+1}^{\infty}\frac{a_-^n\,\Gamma(n)L_R^k}{2^{n-k}\Gamma(k+1)} \\ &= \sum_{n=1}^{\infty}a_-^n\frac{\Gamma(n)}{2^n}\sum_{k=0}^{n-1}\frac{2^kL_R^k}{\Gamma(k+1)} = \frac{s^2}{R^4}\sum_{n=1}^{\infty}\frac{\Gamma(n,2L_R)}{2^n}a_-^n,\end{aligned} \tag{11.21}$$

where in the second step we took $n\to n-k$ and in the third we interchanged the order of the sums. In the last step, the finite sum over $k$ is solved exactly. This leads to the closed form

$$\hat{D}^{\rm RF}(s,R^2) = \frac{c_0(R^2)}{s^2} + \sum_{n=1}^{\infty}\frac{\Gamma(n)}{2^n}\tilde{\gamma}(n,2L_R)a_-^n, \tag{11.22}$$

where $\gamma(a,z)$ is the lower incomplete gamma function and $\tilde{\gamma}(a,z)=\gamma(a,z)/\Gamma(a)$. In terms of $x=s/m_\tau^2$, the momenta of the subtracted CI series reads

$$\begin{aligned}
\delta^{(0)}_{\mathrm{RFCI},l} &= \frac{1}{2i\pi}\oint_{|x|=1}\frac{\mathrm{d}x}{x}(-x)^l\hat{D}^{\mathrm{RF}}(m_\tau^2 x, R^2) \\
&= \frac{c_0(R^2)}{2i\pi m_\tau^4}\oint_{|x|=1}\frac{\mathrm{d}x}{x^3}(-x)^l+\frac{1}{2i\pi}\sum_{n=1}^{\infty}\frac{\Gamma(n)}{2^n}a^n\oint_{|x|=1}\frac{\mathrm{d}x}{x}(-x)^l\frac{\tilde{\gamma}(n,2L_R)}{[1+a\log(-x)]^n} \\
&= \frac{c_0(R^2)}{m_\tau^4}\delta_{l,2}+\sum_{n=1}^{\infty}\frac{\Gamma(n)}{2^n}a^n H^{\mathrm{RF}}_{n,l}(a,R^2),
\end{aligned} \tag{11.23}$$

with

$$H^{\mathrm{RF}}_{n,l}(a,R^2)\equiv\frac{1}{2i\pi}\oint_{|x|=1}\frac{\mathrm{d}x}{x}(-x)^l\frac{\tilde{\gamma}(n,2\log(-m_\tau^2x/R^2))}{[1+a\log(-x)]^n}. \tag{11.24}$$

The scale $R$ can be freely adjusted, and in the following analysis we set $R=m_\tau$, yielding:

$$\begin{aligned}
\delta^{(0)}_{\mathrm{RFCI},l} &= e^{-2/a}\mathrm{Ei}\left(\frac{2}{a}\right)\delta_{l,2}+\sum_{n=1}^{\infty}\frac{\Gamma(n)}{2^n}a^n H^{\mathrm{RF}}_{n,l}(a,R^2), \\
H^{\mathrm{RF}}_{n,l}(a,m_\tau^2) &\equiv H^{\mathrm{RF}}_{n,l}(a)=\frac{1}{2\pi}\int_{-\pi}^{\pi}\mathrm{d}\varphi\, e^{i\varphi l}\frac{\tilde{\gamma}(n,2i\varphi)}{(1+ia\varphi)^n}.
\end{aligned} \tag{11.25}$$

As in the unsubtracted case, the integrand of $H^{\mathrm{RF}}_{n,l}(a)$ does not cross any singularity as long as $\mathrm{Im}(a)<1/\pi$. In the region of the complex $a$-plane where $H^{\mathrm{RF}}_{n,l}(a)$ is defined, no representation in terms of the usual functions that does not involve other sums or integrals has been found. For $l=0$ we find the alternate representation

$$H^{\mathrm{RF}}_{n,0}(a)=H_{n,0}(a)+\frac{1}{4\pi}\sum_{k=1}^{n}\left(\frac{2}{a}\right)^n\frac{\Gamma\left(k-n,\frac{2}{a}-2i\pi,\frac{2}{a}+2i\pi\right)\Gamma\left(1-k+n,\frac{2}{a}\right)}{\Gamma(k)\Gamma(1-k+n)}, \tag{11.26}$$

valid in the integral's domain of definition.

### 11.2.4 Numerical analysis

The FO, CI and subtracted CI series built in the previous section are now plotted as functions of the truncation order in figure 11.1.

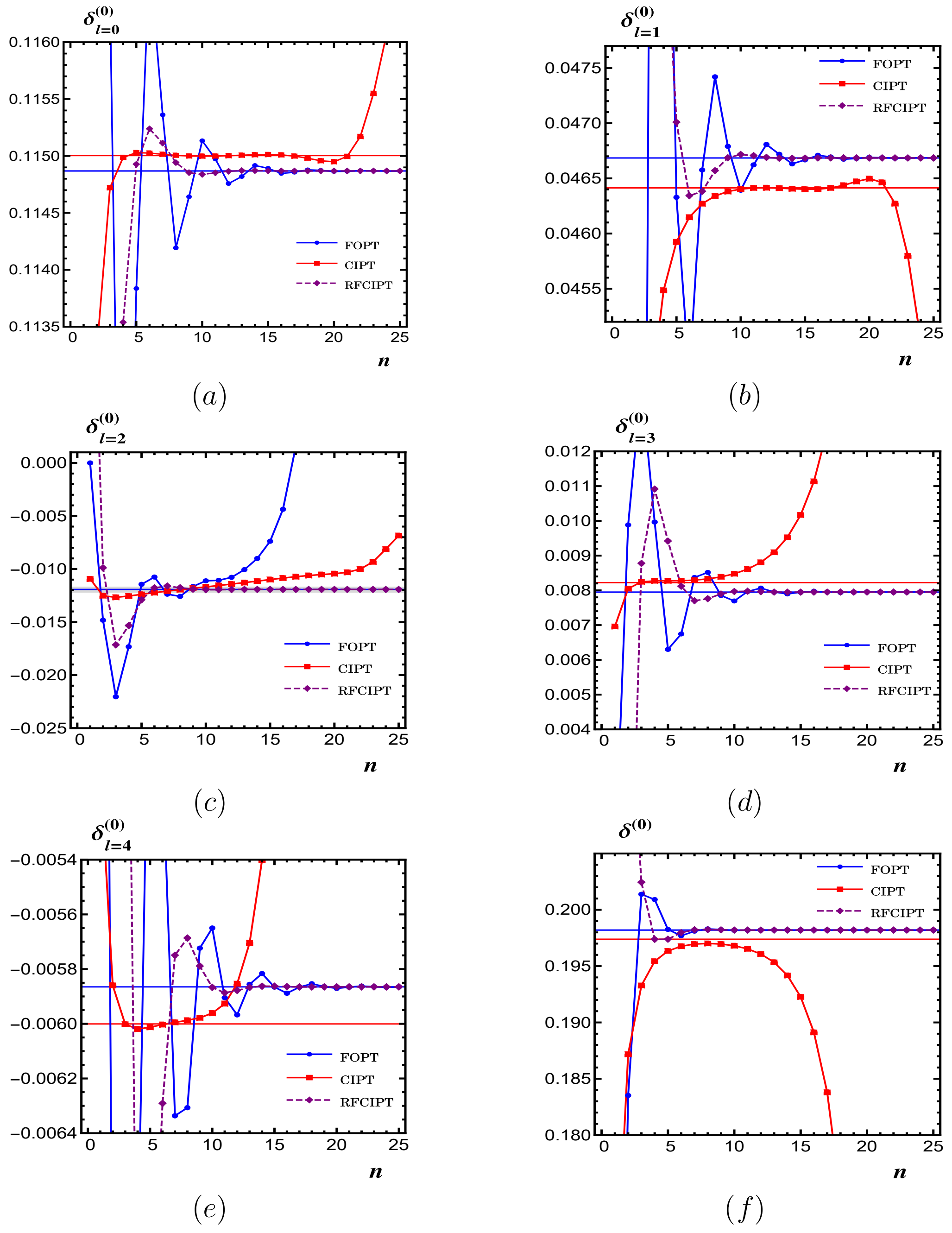

**Figure 11.1.** Fixed-order, contour-improved and subtracted contour-improved series in the gluon condensate renormalon model in the large-$\beta_0$ limit at $\alpha(m_\tau^2) = 0.34$. Plots $(a)$ – $(e)$ show the individual momentum contributions for $l = 0, 1, 2, 3$ and 4, respectively, while plot $(e)$ shows the total contribution $\delta^{(0)} = \delta_0^{(0)} + 2\delta_1^{(0)} - 2\delta_3^{(0)} - \delta_4^{(0)}$. The colored dots represent the truncated perturbative series, and the straight lines represent the exact sum of the series (when convergent) and the Borel sum (when divergent). In plot $(c)$ the gray band represents the ambiguity of the FO series for $l = 2$.

As mentioned above, the FO series converges for $l \neq 2$, and for $l = 2$ presents an IR renormalon at $u = 2/a$, which adds an ambiguity to the Borel sum that goes as $e^{-2/a}$. The CI series diverges for all values of the momentum, but, remarkably, the poles of its Borel transform do not lay in the real axis, hence it is resummed with no ambiguity. It is also remarkable that for $l=2,3$ and 4, the inverse Borel integral does not converge at infinity, and that the exact values are found by analytically continuing that of $l=1$. The CI series also approaches its Borel sum faster than the FO series, and both series sum up to different values. This is a behavior consistently observed in previous studies. On the other hand, after $R$-subtraction, the CI series converges to the FO value.

In the next section we discuss the convergence of the three series in the complex $a$ plane, and compute the explicit expressions for their Borel transforms and Borel sums. It is from these expressions that the exact (and ambiguity) values in figure 11.1 are made, which are in agreement with the numerical results in [97].

## 11.3 Borel sum and convergence

In Ref. [97] it is shown that the correct Borel representations of the spectral moments in the FO and CI schemes are

$$\delta^{(0)}_{\mathrm{FO},l,B} = \frac{1}{2\pi i}\int_0^\infty \mathrm{d}u \oint_{|x|=1}\frac{\mathrm{d}x}{x}(-x)^l B[\hat{D}^{\mathrm{FOPT}}](u)\, e^{-u/a} \tag{11.27}$$
$$\delta^{(0)}_{\mathrm{CI},l,B} = \frac{1}{2\pi i}\int_0^\infty \mathrm{d}u \oint_{\mathcal{C}_x}\frac{\mathrm{d}x}{x}(-x)^l\left(\frac{a(-x m_\tau^2)}{a}\right) B[\hat{D}]\left(\frac{a(-x m_\tau^2)}{a}u\right) e^{-u/a}$$

where $\mathcal{C}_x$ is a deformation of the unit circle into the real-negative complex plane, $B[\hat{D}](u)$ is the Borel transform of the reduced Adler function with respect to $a(-s)$, which in the gluon condensate model is $B[\hat{D}](u) = 1\,/\,(2-u)$, and $B[\hat{D}^{\mathrm{FOPT}}] = B[\hat{D}]e^{-u/a(-x)+u/a}$, which in the large-$\beta_0$ limit reduces to

$$B[\hat{D}^{\mathrm{FOPT}}] = B[\hat{D}]e^{-u\log(-x)}. \tag{11.28}$$

The Borel representations in (11.27) are derived by attending to the non-analytic structures (poles and cuts) of the strong coupling and generic UV and IR renormalon models for $B[\hat{D}](u)$. Here we reproduce the discussion in the large-$\beta_0$ limit and the gluon condensate model, in which they take the form

$$\delta^{(0)}_{\mathrm{FO},l,B} = \frac{1}{2\pi i}\int_0^\infty \mathrm{d}u \,\frac{e^{-\bar{u}/a}}{2-u}\oint_{|x|=1}\frac{\mathrm{d}x}{x}(-x)^{l+u}\,, \tag{11.29}$$
$$\delta^{(0)}_{\mathrm{CI},l,B} = \frac{1}{2\pi i}\int_0^\infty \mathrm{d}u\, e^{-u/a}\oint_{\mathcal{C}_x}\frac{\mathrm{d}x}{x}\frac{(-x)^l}{2+2a\log(-x)-u}.$$

The path along $|x|$ has endpoints at $x=1\pm i0$ but can be deformed freely as long as the non-analytic structures of the strong coupling (a cut along the positive real axis and the Landau pole) and of the Adler function are not crossed.

In FO, the contour integral only affects the norm of the $u=2$ pole and not its position in the complex $u$ plane, hence the Borel transform of $B[\delta^{(0)}_{\mathrm{FO},l}]$ has a convergence radius of 2 in $u$. The integration along values $u>2$ is carried out as usual by analytically continuing the function[11.1]. Therefore, one can keep the unit circle path, making integration trivial (see the results after the current discussion). In CI, however, there is a non trivial interplay between the contour integral and the non-analytic structures of $B[\delta^{(0)}_{\mathrm{CI},l}]$ in the complex plane of $u$. The contour integral presents a cut along the positive real axis (due to $\log(-x)$), a pole at $x=0$ if $l=0$ and a second pole at $\tilde{x}=-e^{\frac{u-2}{2a}}$. With the integration over $u$, the pole at $\tilde{x}$ moves further into the negative real $u$ axis. The important observation is that $B[\hat{D}](u)$ sums to $1/(2-u)$ with a convergence radius of 2, and therefore $B[\hat{D}]\Big(\frac{a(-xm_\tau^2)}{a}u\Big)$ has a convergence radius of

$$\left|\frac{a(-xm_\tau^2)}{a}u\right|<2 \implies u<2|1+a\log(-x)|<2. \tag{11.30}$$

One can observe then that for $\bar{u}<2$, this is, inside the radius of convergence of $B[\hat{D}]\Big(\frac{a(-xm_\tau^2)}{a}u\Big)$, the pole at $\tilde{x}$ is enclosed by the unit circle path. When $B[\hat{D}]$ is analytically continued to the region $\bar{u}>2$ the pole moves into the real negative axis and outside the unit circle. Thus to perform analytic continuation of the contour integrand and not only of its integrand the contour must be deformed to always enclose $\tilde{x}$. This leads to define $\mathcal{C}_x$ into the negative infinite of the real axis.

With these considerations the contour integrals can be solved. The results in [97] applied to the gluon condensate model in the large-$\beta_0$ read

$$\begin{aligned}
&\frac{1}{2\pi i}\frac{1}{2-u}\oint_{|x|=1}\frac{\mathrm{d}x}{x}(-x)^{l+u}=\frac{(-1)^{l+1}\sin(\pi u)}{\pi(l-u)(2-u)}, \\
&\frac{1}{2\pi i}\oint_{\mathcal{C}_x}\frac{\mathrm{d}x}{x}\frac{(-x)^l}{2+2a\log(-x)-\bar{u}}\equiv C(l,u,a),
\end{aligned} \tag{11.31}$$

where

$$\begin{aligned}
C(0,u,a)&=\frac{1}{4\pi i a}\left[\log\!\left(\frac{1}{a}-\frac{u}{2a}+i\pi\right)-\log\!\left(\frac{1}{a}-\frac{u}{2a}-i\pi\right)\right], \\
C(l,u,a)&=\frac{e^{\frac{l}{2a}(u-2)}}{2a}\left[1-\frac{1}{2i\pi}\Gamma\!\left(0,\frac{u}{2a}(l-2)-i\pi l,\frac{u}{2a}(l-2)+i\pi l\right)\right].
\end{aligned} \tag{11.32}$$

11.1. Let us remark that $u=2$ is a pole and not a cut, so analytic continuation is well defined.

The first result in (11.31) shows the FO spectral moments $l\neq 2$ have no poles in the Borel plane, since the zero of the sine function cancels the simple pole in the denominator. The FO series is then convergent to the value that is returned by performing the $u$ integration. For $l=2$ there is a zero of order two in the denominator at $u=2$ which the sine function only reduces to a simple pole. In this case the perturbative FO series diverges and it the $u$ integral is regularized with the principal value prescription. The result sums the FO series with an ambiguity of order $e^{-2/a}$ and returns the approximate value of its plateau. In the case of CI, the poles of the functions $C(l,\bar{u},a)$ lay outside the positive real axis of the $\bar{u}$ complex plane, hence the $\bar{u}$ integration returns the Borel sum of the (divergent) CI series with no ambiguity.Remarkably, the momentum $(-x)^l$ transforms after contour integration into the exponential $e^{\frac{l(\bar{u}-2)}{2a}}$, which for $l\geq 2$ compensates the suppression effect of the Borel exponential $e^{-\bar{u}/a}$ and renders the $\bar{u}$ integral formally infinite. In this case, the result from $l<2$ is analytically continued.

In the next sections we study the FO, CI and RFCI series solely from the perturbative expressions derived in section 11.2. Doing this allows to address the relevant aspects –convergence radius, Borel transform and Borel sum– in the complex plane of $a$. Our results reproduce (11.29) and (11.31) for real, positive $a$.

### 11.3.1 Fixed order perturbation theory

From the explicit results for $d_{n,l}^{\rm FO}$ in (11.12), (11.13) and (11.14), one can study the convergence of the FO series analytically. Since convergence is dictated by the large $n$ coefficients, it is enough to take the leading contribution in the large $n$ expansion of $d_{n,l}^{\rm FO}$, which can be extracted from the corresponding asymptotic expansion for $\Gamma(n,z)$:

$$\Gamma(a,z)=\Gamma(a)-\frac{z^a}{a}-z^a\sum_{j=0}^{\infty}\frac{(-1)^j}{a^{j+1}}\sum_{k=1}^{\infty}\frac{k^j(-z)^k}{k!}=\Gamma(a)-\frac{e^{-z}z^a}{a}-\frac{e^{-z}z^{a+1}}{a^2}+\mathcal{O}\bigg(\frac{1}{a^3}\bigg). \tag{11.33}$$

The double incomplete gamma functions appearing in the FO and CI series always have the form $\Gamma(a,x-ik\pi,x+ik\pi)$, where $x$ is any complex number and $k$ is a positive integer. The asymptotic expansion of such gamma function for large $a$ up to $1/a^2$ reads

$$\begin{aligned}\Gamma(a,x-ik\pi,x+ik\pi) \;\asymp\; & \frac{(-1)^k e^{-x}}{a}[(x+ik\pi)^a-(x-ik\pi)^a] \\ & +\frac{(-1)^k e^{-x}}{a^2}[(x+ik\pi)^{a+1}-(x-ik\pi)^{a+1}].\end{aligned} \tag{11.34}$$

With (11.33) and (11.34) we derive the following asymptotic expressions for the series coefficients

$$d_{n,l}^{\mathrm{FO}} \asymp \frac{(-1)^{n+l+1}\pi^n}{n^2}\cos\!\left(\frac{n\pi}{2}\right), \qquad d_{n,2}^{\mathrm{FO}} \asymp \frac{\Gamma(n)}{2^n}, \tag{11.35}$$

where we kept the leading contribution, which in the cases $l \neq 2$ is $1/n^2$. The root test can be easily applied now[11.2]:

$$\begin{aligned} \lim_{n\to\infty} |d_{n,l}^{\mathrm{FO}}|^{1/n} &= \pi \lim_{n\to\infty} n^{-2/n} = \pi, \\ \lim_{n\to\infty} |d_{n,2}^{\mathrm{FO}}|^{1/n} &= \frac{\pi}{2}\lim_{n\to\infty} \Gamma(n) = \infty, \end{aligned} \tag{11.36}$$

Therefore we conclude the FO series converges absolutely in the circle $|a| < 1/\pi$ in the complex plane of $a$ for $l \neq 2$ and has zero radius of convergence for $l = 2$.

Since in the subtracted CI scheme the coefficients $d_{n,l}^{\mathrm{RFCI}}$ are given as a contour integral, it is interesting to repeat the convergence analysis for the FO series using the integral representation of $d_{n,l}^{\mathrm{FO}}$ that is given by the incomplete gamma functions. This approach also allows to find the sum of the series, and, for $l = 2$, its Borel sum.

For $l = 0$ we have

$$\begin{aligned} \delta_{\mathrm{FO},0}^{(0)} &= \sum_{n=1}^{\infty} \int_{-2i\pi}^{2i\pi} \mathrm{d}t\, \frac{e^{-t}}{4i\pi n}\left(\frac{at}{2}\right)^n = \frac{i}{4\pi}\int_{-2i\pi}^{2i\pi} \mathrm{d}t\, e^{-t}\log\!\left(1 - \frac{at}{2}\right) \\ &= \frac{\arctan(a\pi)}{2\pi} - \frac{f(a,2)}{2} - \frac{e^{-2/a}}{4i\pi}\left[\log\!\left(-\frac{2}{a} + 2i\pi\right) - \log\!\left(-\frac{2}{a} - 2i\pi\right)\right], \end{aligned} \tag{11.37}$$

where

$$f(a,l) \equiv \frac{e^{-l/a}}{2\pi}\left[2\arctan(a\pi) + i\Gamma\!\left(0, -\frac{l}{a} - i\pi l, -\frac{l}{a} + i\pi l\right)\right]. \tag{11.38}$$

In the second step we have interchanged the sum and the integral, which is valid for $|a| < 2/|t|$. Since $t$ is integrated in $(-2i\pi, 2i\pi)$, only for $|a| < 1/\pi$ the singularity of the logarithm is not crossed. In the third step we solved the integral for complex $a$ inside the convergence circle, and therefore (11.37) sums up the series $\delta_{\mathrm{FO},0}^{(0)}$.

For $l > 0$, $l \neq 2$, it is advantageous to use the recurrence relation obtained through integration by parts in the integral definition of the incomplete gamma function

$$\Gamma(a-1, z_1, z_2) = \frac{\Gamma(a, z_1, z_2) - e^{-z_1} z_1^a + e^{-z_2} z_2^a}{a-1}, \tag{11.39}$$

11.2. The cosines can be safely ignored as $|\cos(n\pi/2)|=1$ for even $n$ and $|\cos(n\pi/2)|=0$ for odd $n$.

to write the coefficients of the series as[11.3]

$$d_{n,l}^{\text{FO}} = \frac{(-1)^l 2^{-n}\Gamma(n+1,-2i\pi,2i\pi) - l^{-n}\Gamma(n+1,-i\pi l,i\pi l)}{2i\pi n(l-2)}. \tag{11.40}$$

Then the sum of the series is

$$\begin{aligned}
\delta_{\text{FO},l}^{(0)} &= \frac{1}{2i\pi(l-2)}\sum_{n=1}^{\infty}\left[(-1)^l\int_{-2i\pi}^{2i\pi}\mathrm{d}t\,\frac{e^{-t}}{n}\left(\frac{at}{2}\right)^n - \int_{-il\pi}^{il\pi}\mathrm{d}t\,\frac{e^{-t}}{n}\left(\frac{at}{l}\right)^n\right] \qquad (11.41)\\
&= \frac{i}{2\pi(l-2)}\left[(-1)^l\int_{-2i\pi}^{2i\pi}\mathrm{d}t\,e^{-t}\log\left(1-\frac{at}{2}\right) - \int_{-il\pi}^{il\pi}\mathrm{d}t\,e^{-t}\log\left(1-\frac{at}{l}\right)\right]\\
&= \frac{(-1)^l f(a,2) - f(a,l)}{l-2} + \frac{(-1)^l e^{-2/a}}{2i\pi(l-2)}\left[\log\left(-\frac{2}{a}+2i\pi\right) - \log\left(-\frac{2}{a}-2i\pi\right)\right]\\
&\quad -\frac{e^{-l/a}}{2i\pi(l-2)}\left[\log\left(-\frac{l}{a}+il\pi\right) - \log\left(-\frac{l}{a}-il\pi\right)\right],
\end{aligned}$$

where again the condition for the convergence of the sums is $|a|<1/\pi$ and the solution of the integral holds inside the convergence circle. For $l=2$ it is enough to inspect the first term,

$$\sum_{n=1}^{\infty}\frac{\Gamma(n,2i\pi)}{2^{n+1}}a^n = \sum_{n=1}^{\infty}\frac{1}{2}\int_{2i\pi}^{\infty}\mathrm{d}t\,e^{-t}\,t\left(\frac{at}{2}\right)^n = \frac{1}{2}\int_{2i\pi}^{\infty}\mathrm{d}t\,\frac{e^{-t}}{2-at}, \tag{11.42}$$

to see $\delta_{\text{FO},2}^{(0)}$ always diverges: the sum only converges when $|a|<2/|t|$ but this time $t$ is integrated to $\infty$, so the radius of convergence is zero. This is translated into a pole at $t=2/a$ in the resummed expression. Borel resummation can be used to give an estimate of $\delta_{\text{FO},2}^{(0)}$. For completeness, we compute the Borel transforms for all $l$. In order for Borel summation to hold in the complex plane of $a$, the Borel prescription $a^n = (a^n/\Gamma(n+1))\int_0^{\infty}\mathrm{d}t\,e^{-t}t^n$ must not be followed by the usual change of variable $at\equiv u$ if one wants to keep the integration limits at 0 and $\infty$. Following this idea we find

$$B[\delta_{\text{FO},l}^{(0)}](u) = \frac{(-1)^{l+1}\,a\sin(au\pi)}{\pi(2-au)(l-au)}. \tag{11.43}$$

We observe that for $l\neq 2$ the zeros of the sine cancel those of the denominator, hence the corresponding Borel transform does not present poles. The $u$ integration of $B[\delta_{\text{FO},l}^{(0)}](u)$ returns the explicit sums in (11.37) and (11.45). For $l=2$ there is a simple pole at $u=2/a$, which requires a prescription to avoid the pole when $a$ is a

11.3. The exponential terms in the recurrence relation cancel in the subtraction of the two doubly incomplete gammas. The advantage of the form in (11.40) is its close resemblance with the $l=0$ case, which allows to reuse the solution of the integral in (11.37).

real, positive number. Applying the principal value prescription in chapter 5, for $a \in \mathbb{R}^+$ one finds

$$\begin{aligned}\delta^{(0)}_{\mathrm{FO},2,B} &= ie^{-2/a}\Big[\arctan(a\pi)+\mathrm{arccot}(a\pi)-\frac{\pi}{2}\Big] \\ &+\frac{e^{-2/a}}{2ai\pi}\Big[(1+ia\pi)\mathrm{Ei}\Big(\frac{2}{a}+2i\pi\Big)-(1-ia\pi)\mathrm{Ei}\Big(\frac{2}{a}-2i\pi\Big)\Big],\end{aligned} \tag{11.44}$$

with an ambiguity (imaginary part) $\delta[\delta^{(0)}_{\mathrm{FO},2}]=e^{-2/a}$ that is the size of the residue of the integrand at $u=2$. For $a \notin \mathbb{R}^+$ the inverse Borel transform requires no prescription and yields

$$\begin{aligned}\delta^{(0)}_{\mathrm{FO},2,B} &= \frac{ie^{-2/a}}{2\pi a}\Gamma\Big(0,-\frac{2}{a}-2i\pi,-\frac{2}{a}+2i\pi\Big) \\ &-\frac{e^{-2/a}}{2}\Big[\Gamma\Big(0,-\frac{2}{a}+2i\pi\Big)+\Gamma\Big(0,-\frac{2}{a}-2i\pi\Big)\Big].\end{aligned} \tag{11.45}$$

Our resummed results for the FO spectral momenta (11.37), (11.41) and (11.44), obtained from the perturbative series, agree with the corresponding results from the resummed Adler function in (11.31), providing a cross-check. Including also (11.45) for $l=2$, our results also sum the series for any other $a$ inside the circle $1/\pi$.

### 11.3.2 Contour improved perturbation theory

The convergence behavior of the CI series can be computed from the explicit results (11.33) and (11.34) for its coefficients. For $l=0$, the coefficients of the series satisfy

$$\left|\frac{\Gamma(n)}{2^n}H_{n,0}\right| < \left|\frac{\Gamma(n-1)}{2^n a\pi(1+a^2\pi^2)^{\frac{1}{2}(n-1)}}\right|, \tag{11.46}$$

and the factorial rising immediately leads to a divergent series

$$\lim_{n\to\infty}\left|\frac{\Gamma(n-1)}{2^n a\pi(1+a^2\pi^2)^{\frac{1}{2}(n-1)}}\right|^{1/n} = \frac{1}{2}\lim_{n\to\infty}[\Gamma(n-1)]^{1/n}=\infty, \tag{11.47}$$

which makes $\delta^{(0)}_{\mathrm{CI},0}$ diverge for all values of $a$. For $l>0$ it is enough to employ (11.34) to see

$$\Gamma\Big(1-n,-\frac{l}{a}-il\pi,-\frac{l}{a}+il\pi\Big)\asymp\frac{(-1)^l\sin[(n-1)\arctan(a\pi)]}{a\pi(n-1)(1+a^2\pi^2)^{\frac{1}{2}(n-1)}}, \tag{11.48}$$

so that $H_{n,l}(a)$ does not compensate the $\Gamma(n)$ rising and thus $\delta^{(0)}_{\mathrm{CI},l}$ diverges for all values of $a$. The same conclusion can be reached from the contour integral representation of $\delta^{(0)}_{\text{CI},l}$:

$$\delta^{(0)}_{\text{CI},l} = \sum_{n=1}^{\infty} \frac{\Gamma(n)}{2^n} a^n H_{n,l}(a) = \sum_{n=1}^{\infty} \frac{\Gamma(n)}{2\pi} \int_{-\pi}^{\pi} \mathrm{d}\varphi\, e^{i\varphi l} \left[\frac{a}{2(1+ia\varphi)}\right]^n. \tag{11.49}$$

Since for $\text{Im}(a) < 1/\pi$ the integration path does not cross any of the singularities of the integrand, the integral is finite, and then one can interchange it with the sum as long as the latter is finite. However, it is clear that due to the factorial rising of $\Gamma(n)$ the sum diverges for all values of $\varphi$ in the integration range.

An estimation for the sum of the series can be found by Borel summation. In the $l=0$ case, the coefficients are given by the functions in (11.17) and the Borel transform can be directly performed

$$\begin{aligned} B[\delta^{(0)}_{\text{CI},0}](u) &= \sum_{n=1}^{\infty} e^{-u} \frac{a^n u^{n-1}}{2^n} H_{n,0}(a) \\ &= \frac{\arctan(a\pi)}{2\pi} + \frac{1}{4\pi i}\left[\log\left(2 - \frac{au}{1+ia\pi}\right) - \log\left(2 - \frac{au}{1-ia\pi}\right)\right]. \end{aligned} \tag{11.50}$$

We observe that the poles of $B[\delta^{(0)}_{\text{CI},l}]$ are not in the positive real axis but at $u = 2/a \pm 2i\pi$. The inverse Borel integral can then be computed without any ambiguity:

$$\begin{aligned} \delta^{(0)}_{\text{CI},0,B} &= \frac{\arctan(a\pi)}{2\pi} + \frac{e^{-2/a}}{4\pi i}\Gamma\left(0, -\frac{2}{a} - 2i\pi, -\frac{2}{a} + 2i\pi\right) \\ &+ \frac{1}{4i\pi}\left[\log\left(-\frac{2}{a} + 2i\pi\right) - \log\left(-\frac{2}{a} - 2i\pi\right)\right] \\ &+ \frac{1}{4i\pi}\left[\log\left(\frac{a}{-1+ai\pi}\right) - \log\left(\frac{a}{-1-ai\pi}\right)\right]. \end{aligned} \tag{11.51}$$

This result Borel-sums the series $\delta^{(0)}_{\text{CI},0}$ and agrees with the corresponding modified-contour integral in (11.29), providing an important cross-check.

For $l > 0$ the form of $H_{n,l}(a)$ is complicated enough that we cannot carry over the Borel sum analytically. Instead we are forced to work with the contour integral representation of $H_{n,l}(a)$:

$$B[\delta^{(0)}_{\text{CI},l}](u) = \sum_{n=1}^{\infty} \frac{u^{n-1}}{2^{n+1} i\pi} \int_{|x|=1} \frac{\mathrm{d}x}{x} \frac{(-x)^l}{[1 + a\log(-x)]^n} = \frac{1}{2i\pi} \oint_{\mathcal{C}_x} \frac{\mathrm{d}x}{x} \frac{(-x)^l}{2 - u + 2a\log(-x)}. \tag{11.52}$$

where we set $a > 0$ to keep the discussion simple and took $ua \to u$. The sum over $n$ is carried out with the original contour $|x| = 1$ and as expected has a convergence radius of $u = 2$. We observed the pole at $\tilde{x}(u) = -e^{\frac{2-u}{2a}}$ discussed at the beginning of the section emerges also from the perturbative series, and thus modified the contour to $\mathcal{C}_x$. The reasoning behind the contour $\mathcal{C}_x$ was discussed at the beginning of the section and in [97], and for clarity we review it here in the light of the explicit expressions. For $u<2$ –this is, inside the convergence radius of $B[\delta^{(0)}_{\mathrm{CI},l}]$– the pole at $\tilde{x}(u)$ lays inside the unit circle $|x|=1$. To carry out the integration over $u$, $B[\delta^{(0)}_{\mathrm{CI},l}]$ needs to be analytically continued to $u>2$, but as $u$ grows, $\tilde{x}(u)$ moves forward into the negative real axis, exiting the unit circle at $u=2$. The adequate prescription to sum the series is to deform the unit circle so that the pole at $\tilde{x}(u)$ is always enclosed. Thus, after summing the series and before performing any integration, we deform the unit circle into the contour $\mathcal{C}_x$, which can always be done since there are no other non-analytic structures along the negative real axis. The modified contour $\mathcal{C}_x$ is represented in figure 11.2.

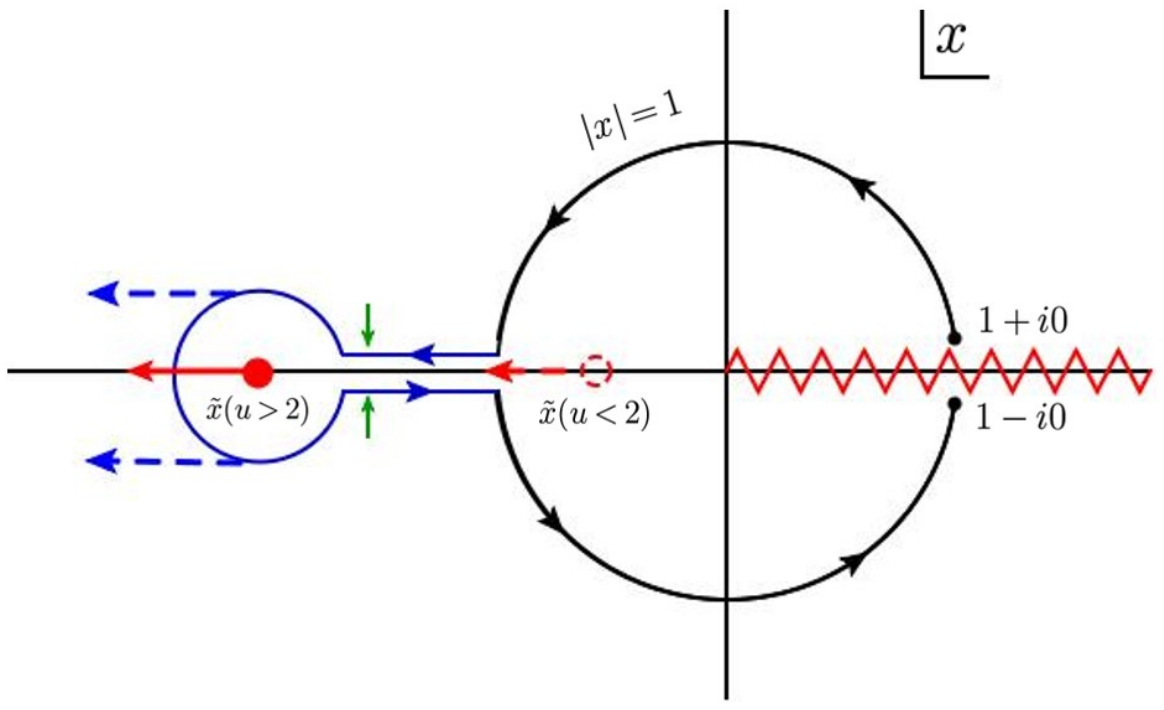

**Figure 11.2.** Modified contour $\mathcal{C}_x$ designed to enclose the pole $\tilde{x}(u)$ inside the contour integral for all values in the integration range of $u$. The non-analytic structures of the integrand are shown in red: the branch cut of $\log(-x)$ along the positive real axis is represented by the horizontal zigzag line, the dashed circle corresponds to the position of the pole at $\tilde{x}$ when $u<0$, and the full red circle represents the pole at any other $u>2$. In practice, since $\tilde{x}(u)\to-\infty$ as $u\to\infty$, the modification of the contour indicated by the blue lines is extended to $-\infty$. The green arrows indicate the limit in which the blue circle and the left opening of the $|x|=1$ circle are closed so that the residue theorem can be applied.

To compute the integral along $\mathcal{C}_x$ we simply realize that the integrand is analytic along the segment of the negative real axis that connects the origin with $\tilde{x}(u)$, so that the limit in which the blue circle and the left opening of the $|x|=1$ circle are closed can be taken. Note that an analogous limit cannot be imposed to close the the endpoints at $x\pm i0$, due to the cut of $\log(-x)$. The integral along $\mathcal{C}_x$ is then the sum of three contributions: (1) the original $|x|=1$ circuit (black circle), which needs to be explicitly evaluated; (2) the horizontal path along the real negative axis (blue horizontal lines), which vanishes due to being run twice in opposite directions; and (3) the closed circular path enclosing the pole (blue circle), which can be computed with the residue theorem. The Borel transform then reads

$$\begin{aligned} B[\delta^{(0)}_{\mathrm{CI},l}](u) &= \frac{e^{\frac{l}{2a}(u-2)}}{2a}\theta(u-2)+\frac{1}{2i\pi}\oint_{|x|=1}\frac{\mathrm{d}x}{x}\frac{(-x)^l}{2-u+2a\log(-x)} \qquad (11.53)\\ &= \frac{e^{\frac{l}{2a}(u-2)}}{2a}\theta(u-2)+\frac{1}{2\pi}\int_{-\pi}^{\pi}\mathrm{d}\varphi\,\frac{e^{i\varphi l}}{2-u+2ai\varphi}, \end{aligned}$$

where the integral over $\varphi$ evaluates to

$$\begin{aligned} \frac{1}{2\pi}\int_{-\pi}^{\pi}\mathrm{d}\varphi\,\frac{e^{i\varphi l}}{2-u+2ai\varphi} &= \frac{e^{\frac{l}{2a}(u-2)}}{2a}\theta(2-u) \qquad (11.54)\\ &\quad -\frac{e^{\frac{l}{2a}(u-2)}}{4a\pi i}\Gamma\Big(0,\frac{l}{2a}(u-2)-il\pi,\frac{l}{2a}(u-2)+il\pi\Big). \end{aligned}$$

When adding the residue, the complementary Heaviside functions add up to 1 and one is left with

$$B[\delta^{(0)}_{\mathrm{CI},l}](u)=\frac{e^{\frac{l}{2a}(u-2)}}{2a}\bigg[1-\frac{1}{2\pi i}\Gamma\Big(0,\frac{l}{2a}(u-2)-il\pi,\frac{l}{2a}(u-2)+il\pi\Big)\bigg], \qquad (11.55)$$

in total agreement with [97]. We again observe the exponential factor $e^{\frac{l}{2a}(u-2)}$, which combined with the Borel exponential amounts for a total of $e^{-\frac{2-l}{2a}u}$. This renders the inverse Borel integral infinite if $l\geq 2$. The result for $l<2$ is

$$\begin{aligned} \delta^{(0)}_{\mathrm{CI},l,B} &= \frac{(-1)^l e^{-2/a}}{l-2}\bigg[1-\frac{\Gamma(0,-\frac{2}{a}-2i\pi,-\frac{2}{a}+2i\pi)}{2i\pi}\bigg] \qquad (11.56)\\ &\quad -\frac{e^{-l/a}}{l-2}\bigg[1-\frac{\Gamma(0,-\frac{l}{a}-il\pi,-\frac{l}{a}+il\pi)}{2i\pi}\bigg], \end{aligned}$$

and can be analytically continued to $l>2$. The sum $\delta^{(0)}_{\mathrm{CI},l,B}$ also reduces to $\delta^{(0)}_{\mathrm{CI},0,B}$ in (11.51) for $l=0$.

### 11.3.3 Subtracted CI series

We again work with the contour integral representation of $H^{\mathrm{RF}}_{n,l}(a)$ in terms of $x$:

$$\begin{aligned} \delta^{(0)}_{\mathrm{RFCI},l} &= \frac{1}{2i\pi}\sum_{n=1}^{\infty}\frac{a^n}{2^n}\int_{|x|=1}\frac{\mathrm{d}x}{x}\frac{(-x)^l\gamma(n,2\log(-x))}{[1+a\log(-x)]^n} \qquad (11.57)\\ &= \frac{1}{2i\pi}\sum_{n=1}^{\infty}\int_{|x|=1}\frac{\mathrm{d}x}{x}\int_0^1\frac{\mathrm{d}t}{t}\bigg[\frac{at\log(-x)}{1+a\log(-x)}\bigg]^n(-x)^{l-2t}\\ &= \frac{1}{2i\pi}\int_{|x|=1}\frac{\mathrm{d}x}{x}\int_0^1\frac{\mathrm{d}t}{t}\frac{at\log(-x)}{1+a(1-t)\log(-x)}(-x)^{l-2t}, \end{aligned}$$

where we employed the integral representation of $\gamma(n, 2\log(-x))$ and the change of variables $t \to 2\log(-x)t$. The main difference with the unsubtracted case (11.49) is that the factorial rising of $\Gamma(n)$, responsible for the CI series zero convergence radius, is replaced by the lower incomplete gamma function $\gamma(n, 2i\varphi)$, whose interval of integration is finite. In the last step we performed the sum over $n$. This sum converges for all real $a$, as can be readily seen from the condition $|a t\log(-x)|/|1 + a t\log(-x)| < 1$ accounting for the integration ranges of $t$ and $x$. Inspecting the result of the sum in the complex plane of $a$ yields a pole at $a = -1/[(1-t)\log(-x)]$, which for $x$ in the $|x| = 1$ circumference is the closest to the origin when $\log(-x) = \pm i\pi$. This establishes that $\delta^{(0)}_{\mathrm{RFCI},l}$ also converges in the $|a| < 1/\pi$ circle in the complex plane.

On the other hand, if one swaps the $x$ and $t$ integration, the integrand presents a pole at $\tilde{x}(t) = -e^{-\frac{1}{a(1-t)}}$. For real $a \neq 0$, $\tilde{x}(t)$ is a monotonic continuous function of $t$, hence the pole moves with the $t$ integration along a path with endpoints at $\tilde{x}_1 = -e^{-1/a}$ and

$$x_2 = \begin{cases} 0, & a > 0 \\ -\infty, & a < 0 \end{cases}. \tag{11.58}$$

The function $-e^{-1/a}$ has an horizontal asymptote at $a = -1$, and lies above it for $a > 0$ and below for $a < 0$. The consequence is that $\tilde{x}(t)$ always lays inside the $|x| = 1$ circle for $a > 0$ and outside when $a < 0$, as depicted in figure 11.3. Thus, no contour modification is required in the subtracted CI scheme.

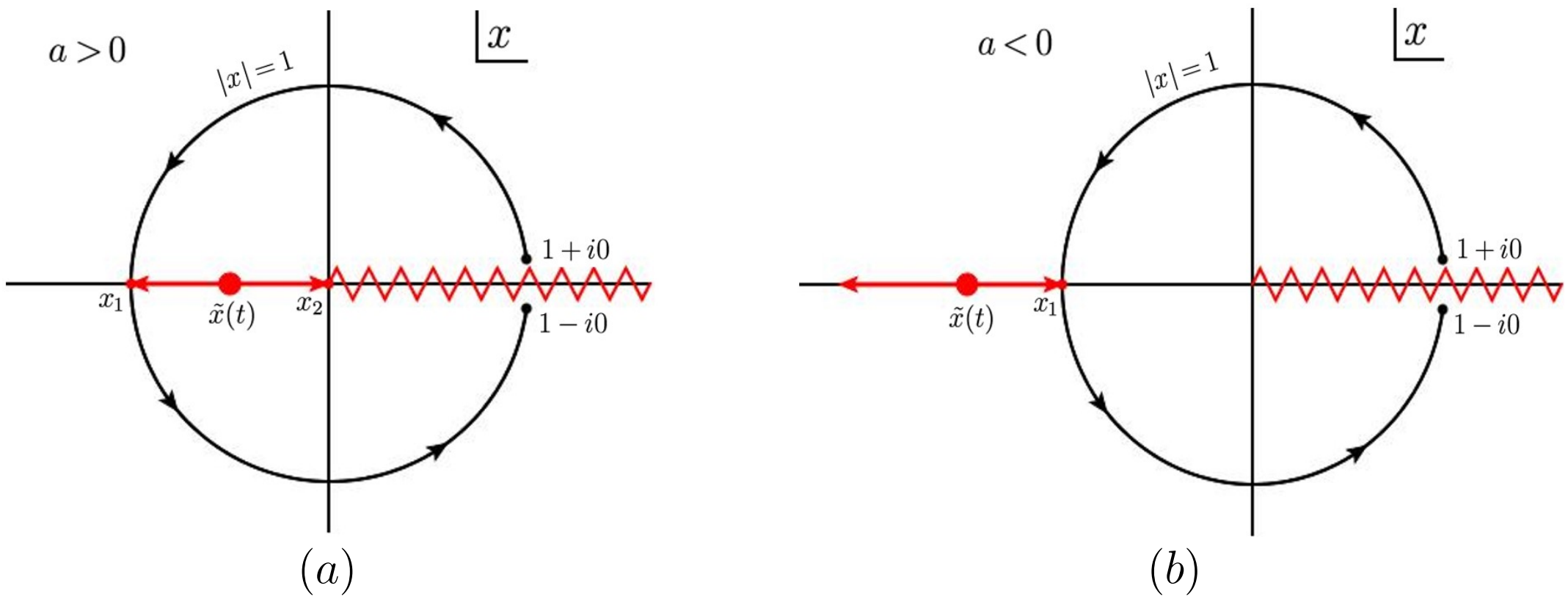

**Figure 11.3.** Non-analytic structures of the CI series in the renormalon free-scheme. For $a > 0$ (panel $(a)$), the pole at $\tilde{x}(t)$ moves from $x_1 = e^{-1/a}$ to $x_2 = 0$, and therefore is always enclosed in the unit circle, while for $a < 0$ (panel $(b)$), $x_2 = -\infty$ and the pole lays always outside.

Carrying out both integrations we find $\delta^{(0)}_{\mathrm{RFCI},l} = \delta^{(0)}_{\mathrm{FO},l}$ for $l \neq 2$. For $l = 2$ and

$a>0$, the subtracted CI series converges to

$$\delta^{(0)}_{\mathrm{RFCI},2} = \frac{e^{-2/a}}{2a\pi}\left[(i+a\pi)\mathrm{Ei}\left(\frac{2}{a}-2i\pi\right)-(i-a\pi)\mathrm{Ei}\left(\frac{2}{a}+2i\pi\right)-2a\pi\mathrm{Ei}\left(\frac{2}{a}\right)\right]. \tag{11.59}$$

### 11.3.4 Asymptotic separation

Asymptotic separation is defined as the difference between the FO and CI series. Using the expressions for the sum of the series in (11.41) and (11.51) we find for $\mathrm{Re}(a)>0$

$$\delta^{(0)}_{\mathrm{FO},l,B} - \delta^{(0)}_{\mathrm{CI},l,B} = \frac{(-1)^l}{l-2}e^{-2/a}, \tag{11.60}$$

which agrees with the corresponding result in [97]. For $l=0$, since our results for the Borel sums hold in the $1/\pi$ circle of the complex $a$ plane, we can interestingly determine the asymptotic expansion for $a<0$ vanishes. We recall the observation that the asymptotic separation at a given scale $s$ is of order $e^{-2/a(s)} = (\Lambda^2_{\mathrm{QCD}}/s)^2$, therefore identical to the gluon condensate contribution.

## 11.4 FOPT and (RF)CIPT as rearrangements

The FO expansion is a power series in $a$, while the CI and RFCI expansions are series in the functions $a^n H_n(a)$ and $a^n H_n^{\mathrm{RF}}(a)$, respectively. To connect both prescriptions we define the $S$ and $T$ matrices in the following way

$$a^n H_{n,l}(a) \equiv \sum_{k=n}^{\infty} s^l_{n,k} a^k, \qquad a^n \equiv \sum_{k=n}^{\infty} t^{(\mathrm{RF})}_{n,k}\, a^k H^{(\mathrm{RF})}_{k,l}(a).$$

Neither $S$ nor $T$ depend on $a$, and of course $T=S^{-1}$. The entries of the $S$ matrix can be determined directly from the expansion of the corresponding $H^{(\mathrm{RF})}_{n,l}(a)$ functions or by expanding binomial in the contour integral form and integrating term by term. The analytic derivation of the $T$ matrix as the inverse of $S$, however, remains a complicated task, and in this thesis we can only present the $l=0$ case for the unsubtracted CI prescription. The results are

$$
\begin{aligned}
s^0_{n,k} &= \frac{(i\pi)^{k-n}\Gamma(k)}{\Gamma(n)\Gamma(k-n+2)}\frac{1+(-1)^{k-n}}{2}, \qquad (11.61)\\
s^{0,\mathrm{RF}}_{n,k} &= \frac{(-1)^{k-n}\Gamma(k)}{2\Gamma(2+k-n)\Gamma(n)}\left[\frac{\Gamma(k+1,2i\pi,-2i\pi)}{2^{k-n+1}i\pi\Gamma(n)}\right.\\
&\quad \left.-\frac{\tilde{\Gamma}(n,2i\pi)-1+(-1)^{k-n}(\tilde{\Gamma}(n,-2i\pi)-1)}{(i\pi)^{n-k}}\right],\\
t^0_{n,k} &= \frac{(i\pi)^{k-n}\Gamma(k)}{\Gamma(n)\Gamma(k-n+1)}(2-2^{k-n})B_{k-n}.
\end{aligned}
$$

The relevant observation is that, while the expansion defining the $S$ matrix has a convergence radius of $1/\pi$ –as it arises from the binomial $(1+ia\pi)^{-n}$–, the inverse relation presents worse convergence (see figure 11.4).

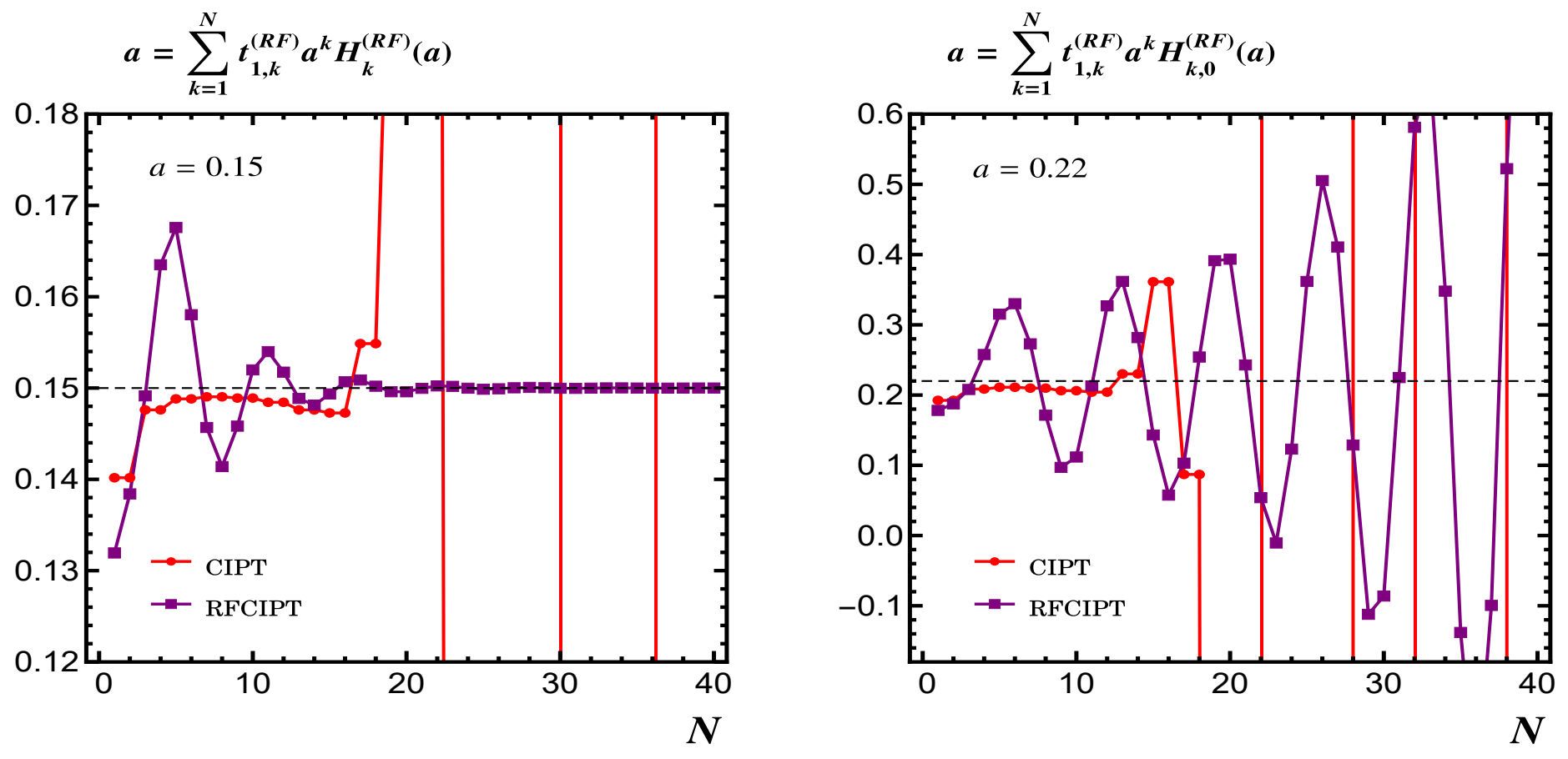

$(a)$ $(b)$

**Figure 11.4.** A single power of $a$ written as a series in the functions $H^{(\mathrm{RF})}_{n,0}$. Panel $(a)$ shows $a=0.15$, and panel $(b)$ $a=0.2$. In CI both values of $a$ yield a divergent series whose flat segment do not correspond to $a$. In RFCI, numerical evidence supports that the series converges in for $a=0.15$ and diverges for $a=0.2$.

Indeed, expressing a single power $a^n$ in terms of the CI functions $a^k H_{k,0}(a)$ yields a series that diverges for all values of $a$. This can be seen by using the asymptotic approximation of the even-index Bernoulli numbers for large $n$:

$$B_{2n} \asymp (-1)^{n+1} 4\sqrt{\pi n}\left(\frac{n}{\pi \mathrm{e}}\right)^{2n}. \qquad (11.62)$$

Since the odd-index Bernoulli numbers except for $B_1 = -1/2$ vanish, we can combine both even and odd indices as

$$\begin{aligned} B_n &\asymp \frac{1+(-1)^n}{2}(-1)^{\frac{n}{2}+1} 4\sqrt{\frac{\pi n}{2}}\left(\frac{n}{2\pi \mathrm{e}}\right)^n \\ &= \frac{1+(-1)^n}{(2\pi)^n}(-1)^{\frac{n}{2}+1}\Gamma(n+1). \end{aligned} \tag{11.63}$$

where the factor of $[1+(-1)^n]/2$ is 1 for even $n$ and 0 for odd $n$. The second line is obtained from Stirling's formula,

$$\Gamma(n+1) \sim \sqrt{2\pi n}\left(\frac{n}{e}\right)^n. \tag{11.64}$$

These approximations lead to

$$t^0_{n,k} \asymp \frac{\Gamma(k)}{\Gamma(n)}[2^{n-k+1}-1][1+(-1)^{k-n}], \tag{11.65}$$

for all fixed $n$ and large $k$, from which the root test can be seen to be infinite.

In the subtracted CI series we studied the $T$ matrix expansion up to $N = 800$, and found support for convergence for $0 < a < 0.2$. We see then the $R$ subtraction cures the renormalon but for $a > 0.2$ RFCI needs all terms in FO to properly converge. In both cases the described behavior holds for all the inspected $a^n$ powers.

# Part III

# Oriented event-shapes

# Chapter 12

# Event-shape distribution for $e^+e^- \rightarrow$ Hadrons at NLO

In the last part of this thesis we compute the fixed-order oriented event-shape distribution for the process $e^+e^- \rightarrow$ hadrons. The computation is carried out for massive stable quarks at LO in the electroweak coupling constant and at NLO in the strong coupling constant.

## 12.1 Introduction

### 12.1.1 The oriented event-shape distribution

The general definition of the event-shape distribution of $e^+e^- \rightarrow$ hadrons was given in (7.18). The oriented distribution is built by writing the phase space in (7.18) in terms of the angle $\theta_T$ between the thrust axis and the $e^+e^-$ beam and integrating over all the variables but $\theta_T$. At NLO, the massive, oriented event-shape distribution, can be written in full generality according to the distribution[12.1] structure [33, 34]

$$\begin{aligned}\frac{1}{\sigma_C^0}\frac{\mathrm{d}\sigma_C}{\mathrm{d}e\,\mathrm{d}\cos\theta_T} &= R^C\,\delta(e-e_{\mathrm{min}}) + \frac{\alpha_s C_F}{\pi} A_e^C\,\delta(e-e_{\mathrm{min}}) \\ &\quad + \frac{\alpha_s C_F}{\pi} B^C \left[\frac{1}{e-e_{\mathrm{min}}}\right]_+ + \frac{\alpha_s C_F}{\pi} F_e^C + \mathcal{O}(\alpha_s^2),\end{aligned} \tag{12.1}$$

12.1. Appendix A.4 contains the basic notions on distributions and focuses on the two distributions that are relevant for this thesis: Dirac delta distribution and plus distribution.

where the index $C=V,A$ denotes the type of current and the subscript $e$ denotes dependence on the specific event-shape under consideration. The Born-level cross section $\sigma_C^0$ stands for the LO total cross-section for massless quarks and acts as the customary normalization, and the symbol $[1/(e-e_{\min})]_+$ stands for the plus distribution defined in (A.26). In the following we compute the oriented distribution for massive quarks at NLO and identify the delta coefficients $R_C$ and $A_e^C$, the plus coefficient $B^C$ and the non-singular distribution $F_e^C$. This represents a generalization of the oriented, massless results in [33] and of the unoriented, massive results in [34].

The diagrams contributing to the distribution up to NLO order are depicted in figure 12.1.

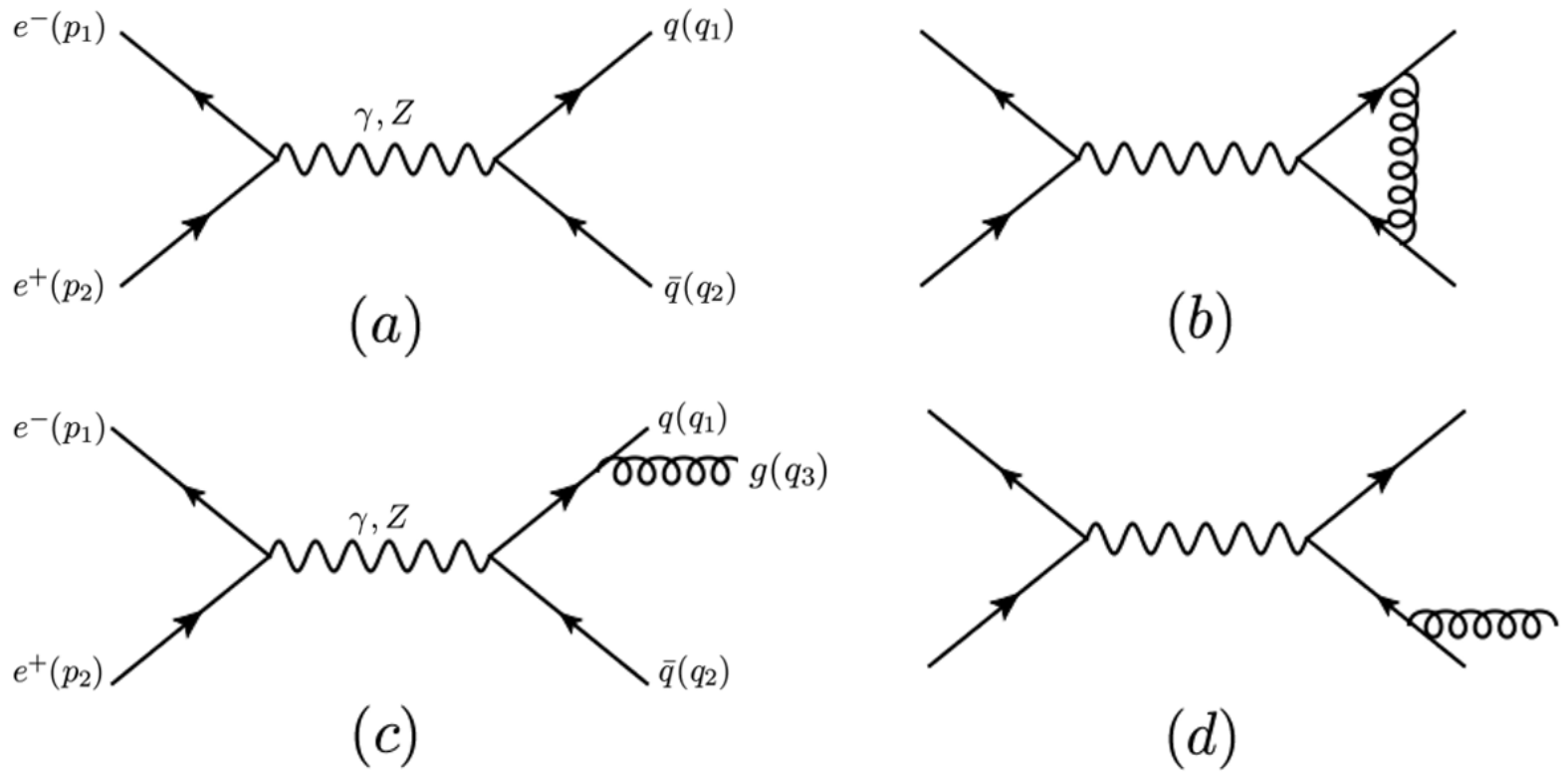

**Figure 12.1.** Diagrams contributing to the event-shape distribution to $O(\alpha_s)$.

Diagrams 12.1$(a)$ and 12.1$(b)$ correspond to the two-particle final state $q\bar{q}$, and diagrams 12.1$(c)$ and 12.1$(d)$ correspond to the three-particle final state $q\bar{q}g$. Each pair must be summed coherently, and QCD renormalization is accounted for through the on-shell quark wavefunction renormalization in diagram 12.1$(a)$, leading to:

$$|Z_\psi^{\rm OS}\mathcal{M}_a+\mathcal{M}_b|^2=\underbrace{|\mathcal{M}_a|^2}_{\rm LO}+\underbrace{\frac{\alpha_s C_F}{2\pi}Z_\psi^{\rm OS(1)}|\mathcal{M}_a|^2+2{\rm Re}(\mathcal{M}_a^\dagger\mathcal{M}_b)}_{\text{NLO, virtual}}+\mathcal{O}(\alpha_s^2), \tag{12.2}$$
$$|\mathcal{M}_c+\mathcal{M}_d|^2=\underbrace{|\mathcal{M}_c|^2+|\mathcal{M}_d|^2+2{\rm Re}(\mathcal{M}_c^\dagger\mathcal{M}_d)}_{\text{NLO, real}}$$

where

$$Z_\psi^{\rm OS}=1+\frac{\alpha_s C_F}{4\pi}Z_\psi^{\rm OS(1)},\qquad Z_\psi^{\rm OS(1)}=-\frac{3}{\epsilon}+3L_m-4,\qquad L_m\equiv\log\frac{m^2}{\mu^2}. \tag{12.3}$$

The term $|\mathcal{M}_a|^2$ in the first line of (12.2) gives the LO distribution, while the remaining terms constitute the NLO contribution. Also, the $\mathcal{O}(\alpha_s)$ terms in the first line of (12.2) are collectively known as the *virtual gluon radiation*, and those in the second line are known as the *real gluon radiation*.

In the virtual contribution, the loop diagram 12.1($b$) contains both UV and IR divergences, which in dimensional regularization are not distinguished and with the choice $d=4-2\epsilon$ show as $1/\epsilon$ poles. UV divergences are canceled by the $Z_\psi^{\mathrm{OS}}$ correction to diagram 12.1($a$). The remaining divergences are thus of IR origin and cancel out in the total distribution when adding the real radiation contribution given by diagrams 12.1($c$) and 12.1($d$), which are themselves infrared divergent in the soft, i.e., when the energy of the gluon tends to 0.

The most illustrative way to explicitly show the cancellation of UV and IR divergences is to compute all four diagrams in dimensional regularization with $4-2\epsilon$ dimensions. The real radiation contribution is then split in a hard and a soft part,

$$\begin{aligned}\frac{\mathrm{d}\sigma_C^{\mathrm{real}}}{\mathrm{d}e\mathrm{d}\cos\theta_T} &= \left[\frac{\mathrm{d}\sigma_C^{\mathrm{real}}}{\mathrm{d}e\mathrm{d}\cos\theta_T} - \frac{\mathrm{d}\sigma_C^{\mathrm{soft}}}{\mathrm{d}e\mathrm{d}\cos\theta_T}\right] + \frac{\mathrm{d}\sigma_C^{\mathrm{soft}}}{\mathrm{d}e\mathrm{d}\cos\theta_T} \\ &\equiv \frac{\mathrm{d}\sigma_C^{\mathrm{hard}}}{\mathrm{d}e\mathrm{d}\cos\theta_T} + \frac{\mathrm{d}\sigma_C^{\mathrm{soft}}}{\mathrm{d}e\mathrm{d}\cos\theta_T},\end{aligned} \tag{12.4}$$

where the soft distribution is the leading term in the soft limit expansion of the real radiation contribution. Due to the soft limit subtraction the hard distribution is finite, and in the total distribution,

$$\frac{\mathrm{d}\sigma_C}{\mathrm{d}e\mathrm{d}\cos\theta_T} = \frac{\mathrm{d}\sigma_C^{\mathrm{LO}}}{\mathrm{d}e\mathrm{d}\cos\theta_T} + \frac{\mathrm{d}\sigma_C^{\mathrm{virt}}}{\mathrm{d}e\mathrm{d}\cos\theta_T} + \frac{\mathrm{d}\sigma_C^{\mathrm{hard}}}{\mathrm{d}e\mathrm{d}\cos\theta_T} + \frac{\mathrm{d}\sigma_C^{\mathrm{soft}}}{\mathrm{d}e\mathrm{d}\cos\theta_T} + \mathcal{O}(\alpha_s^2), \tag{12.5}$$

the IR divergences from the virtual and soft contributions cancel out. We show in section 12.3.4 that this cancellation occurs already at the level of the differential distribution in $\cos\theta_T$ so that (12.5) is finite.

The separation in (12.5) allows to identify the different contributions to the distribution coefficients $A_e^C$, $B^C$ and $F_e^C$. In order to adequately see the contributions from each cross-section to these coefficients, we recall one of the results discussed in appendix F: in a $a+b\to 1+2+\cdots+N$ scattering process, the number of independent kinematic variables –or, equivalently, the number of phase-space integrations– is $3N-5$.

In the case of the LO and virtual contributions the final state is the $q\bar{q}$ pair so we have $N=2$ and there is only one phase-space integration left, which in the CM frame is over the angle between the initial $e^+e^-$ beam and the quark (see the phase space (F.44)). This angle corresponds to the thrust angle $\theta_T$, since the thrust axis lays in the traveling direction of the $q\bar{q}$ pair. Moreover, for two particles in the final state the dijet event-shape $e$ automatically takes its minimum value, so both contributions are proportional to $\delta(e-e_{\min})$, contributing to the delta coefficients $R^C$ and $A_e^C$, and not being event-shape dependent.

For the real radiation, we have $N=3$ and hence there are 4 phase-space integrations, which can be chosen to be over the energies of the quark and the antiquark and the angles they form with the beam pipe (see the phase space (F.70)). At this stage the soft limit expansion is performed. It is carried out with (A.26), in which the energy of the gluon is the expansion variable. In the soft part, the divergences are regularized by $\epsilon$, while the hard part remains finite. Then, in each contribution, the angular integrations are marginalized in favor of $\cos\theta_T$, and the energies are marginalized in favor of the event-shape. In the soft part, the combination of the soft-limit delta and plus distributions from expansion (A.26) and the event-shape delta $\delta(e-e(X))$ produces, upon integration, terms proportional to $\delta(e-e_{\min})$, to $[1/(e-e_{\min})]_+$ and regular functions. Thus the soft cross-section contributes to all $A_e^C$, $B_{\rm plus}^C$, and $F_e^{{\rm NS},C}$. The hard part, on the other hand, contains no distribution other than $\delta(e-e(X))$, which is used to integrate over one of the energies. Therefore, the hard part only contributes to the non-singular function $F_e^C$. As a result, the contributions to the coefficients in (12.1) are

$$A_e^C \equiv A^{{\rm virt},C} + A_e^{{\rm soft},C} \quad , \quad F_e^C \equiv F_e^{{\rm soft},C} + F_e^{{\rm hard},C}, \qquad B^C = B^{{\rm soft},C}. \tag{12.6}$$

## 12.1.2 Electroweak-QCD factorization

An idea that greatly simplifies the computation of the event-shape distribution is that the squared, spin-averaged matrix element admits the following factorization at leading order in the electroweak coupling [32]:

$$|\bar{\mathcal{M}}(e^+e^- \to X)|^2 \equiv \frac{1}{4}\sum_{s(e^\pm)}\sum_{s(X)}|\bar{\mathcal{M}}(e^+e^-\to X)|^2 = \sum_{C=V,A} L_{\mu\nu}^C H_C^{\mu\nu}(X), \tag{12.7}$$

$$H_C^{\mu\nu}(X) \equiv \sum_{s(X)} \langle 0|J_C^{\dagger\mu}|X\rangle\langle X|J_C^\nu|0\rangle,$$

where the sums extend over spins and polarizations for the initial and final states, denoted $s(e^{\pm})$ and $s(X)$ respectively, the factor $1/4$ averages over initial state spins, and the vector and axial currents,

$$J_C^{\mu} = \bar{q}_f^a \, \Gamma_C^{\mu} \, q_f^a, \qquad C = V, A, \ \text{with} \ \begin{cases} \Gamma_V^{\mu} = \gamma^{\mu} \\ \Gamma_A^{\mu} = \gamma^{\mu} \gamma^5 \end{cases}, \tag{12.8}$$

describe the production of the $q\bar{q}$ pair. In (12.8) a sum over colors $a$ is understood, so that the currents describe any possible $q\bar{q}$ pair allowed by energy conservation. The vector and axial lepton tensors $L_{\mu\nu}^C \equiv L_{\mu\nu} L_C$ are defined as the factors accompanying the vector and axial hadron squared matrix elements –note that there are no crossed vector-axial terms in the squared matrix element–, respectively, and are independent on the final state $X$. For $Q \gg m_e$, one can set $m_e = 0$ and the leptons tensors have the form:

$$\begin{aligned} L_V &= \frac{\alpha_{\text{ew}}^2 \, (4\pi)^2}{Q^4} \left[ Q_q^2 - \frac{2\, Q^2 \, Q_q \, v_e \, v_q}{Q^2 - m_Z^2} + \frac{Q^4 \, (v_e^2 + a_e^2) \, v_q^2}{(Q^2 - m_Z^2)^2} \right], \\ L_A &= \frac{\alpha_{\text{ew}}^2 \, (4\pi)^2}{(Q^2 - m_Z^2)^2} (a_e^2 + v_e^2) \, a_q^2, \\ L_{\mu\nu} &= p_{1\mu} \, p_{2\nu} + p_{1\nu} \, p_{2\mu} - \frac{Q^2}{2} g_{\mu\nu}, \end{aligned} \tag{12.9}$$

where $p_1$ and $p_2$ are the momenta of the electron and the positron, $Q_f$ is the electric charge of the fermion $f$, and $v_f$ and $a_f$ are vector and axial charges given by

$$v_f = \frac{T_3^f - 2 Q_f \sin^2 \theta_W}{2 \sin \theta_W \cos \theta_W}, \qquad a_f = \frac{T_3^f}{2 \sin \theta_W \cos \theta_W}. \tag{12.10}$$

Here, $T_3^f$ is the third component of the weak isospin and $\theta_W$ is the weak mixing angle. The factor of $1/4$ and the spin sums $s(e^{\pm})$ in the left-hand side of (12.7), as well as the coupling constants from the electroweak vertices, have been absorbed into $L_C$. The diagrammatic representation of this factorization is shown in figure 12.2, where the box represents the different QCD structures in figure 12.1.

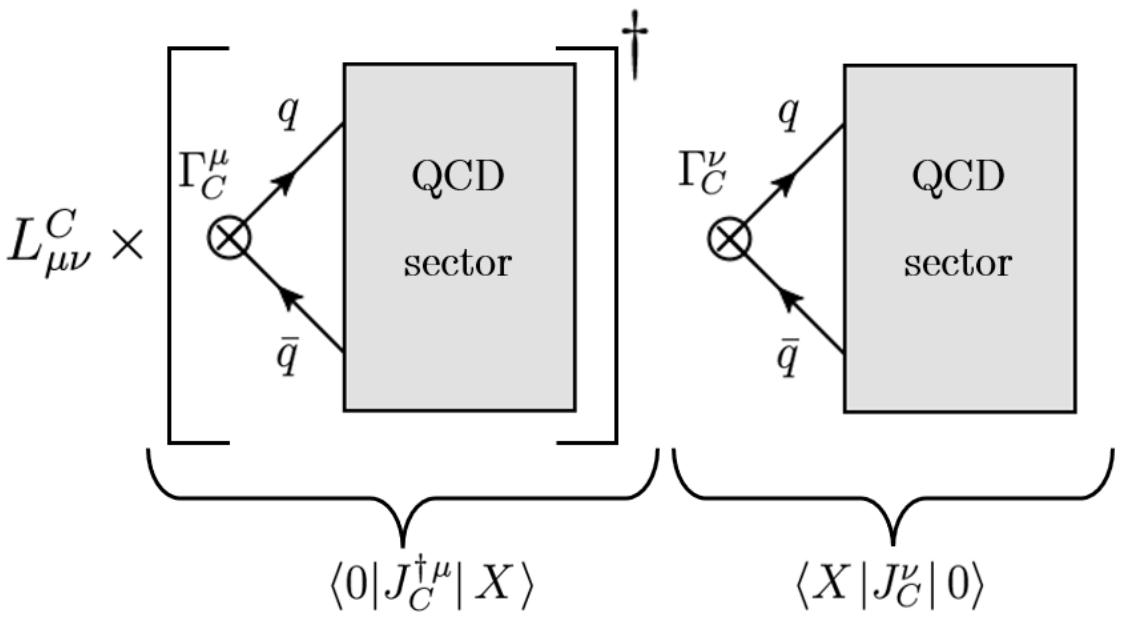

**Figure 12.2.** Schematic representation of the factorization of the squared matrix element $|\bar{\mathcal{M}}(e^+ e^- \to X)|^2$.

Before proceeding with the computation of the oriented distribution let us provide some insight on the factorization result (12.7). To do this we naively compute the lowest-order contribution to $|\mathcal{M}(e^+e^- \to X)|^2$, which is the purely electroweak process depicted in figure 12.1(a). According to the QCD Feynman rules, the matrix elements when the process is mediated by a photon ($\gamma$) and by a $Z$ boson ($Z$) of momentum $k = p_1 + p_2 = q_1 + q_2$ are

$$\begin{aligned}
\mathcal{M}_{a,\gamma} &= -\frac{\mathrm{i}e^2 Q_q}{Q^2} \bar{v}(p_2)\gamma^\mu u(p_1)\bar{u}(q_1)\gamma_\mu v(q_2), \qquad (12.11)\\
\mathcal{M}_{a,Z} &= \frac{\mathrm{i}e^2}{Q^2 - m_Z^2}\Big[ v_e v_q \bar{v}(p_2)\gamma^\mu u(p_1)\bar{u}(q_1)\gamma_\mu v(q_2) - v_e a_q \bar{v}(p_2)\gamma^\mu u(p_1)\bar{u}(q_1)\gamma^5\gamma_\mu v(q_2)\\
&\quad - a_e v_q \bar{v}(p_2)\gamma^5\gamma^\mu u(p_1)\bar{u}(q_1)\gamma_\mu v(q_2) + a_e a_q \bar{v}(p_2)\gamma^5\gamma^\mu u(p_1)\bar{u}(q_1)\gamma^5\gamma_\mu v(q_2)\Big],
\end{aligned}$$

where the photon propagator is written in the 't Hooft-Feynman gauge and in the axial case we discarded the $k_\mu k_\nu$ terms since they lead to $\bar{v}(p_2)(\not{p}_1 + \not{p}_2)u(p_1) = 0$ since we have set $m_e = 0$. The next step is to compute the total squared matrix element, which at $\mathcal{O}(\alpha_s^0)$ is

$$|\mathcal{M}_a|^2 = |\mathcal{M}_{a,\gamma} + \mathcal{M}_{a,Z}|^2 = (\mathcal{M}_{a,\gamma}^\dagger + \mathcal{M}_{a,Z}^\dagger)(\mathcal{M}_{a,\gamma} + \mathcal{M}_{a,Z}), \qquad (12.12)$$

which may seem to contain an intimidating number of terms: there are 4 different spinnor structures in $\mathcal{M}_a$ and therefore a total of 16 in $|\mathcal{M}_a|^2$. However, all of them arise from the Lorentz and spinnor structure

$$G_{i,j} \equiv \bar{v}(p_2)\Gamma_i^\mu u(p_1)\bar{u}(q_1)\Gamma_{j\mu} v(q_2), \quad i,j = V, A, \qquad (12.13)$$

with $\Gamma_i^\mu$ given by (12.8). The most general term in $|\mathcal{M}_a|^2 = \mathcal{M}_a \mathcal{M}_a^\dagger$ is then proportional to[12.2]

$$G_{i,j,k,l} \equiv G_{i,j} G_{k,l}^\dagger = \bar{v}(p_2)\Gamma_i^\mu u(p_1)\bar{u}(q_1)\Gamma_{j\mu} v(q_2)\bar{v}(q_2)\Gamma_k^\nu u(q_1)\bar{u}(p_1)\Gamma_{l\nu} v(p_2). \qquad (12.14)$$

When the spinnor sums are carried out they lead to the following traces

$$\sum_{s(e^\pm)}\sum_{s(q\bar{q})} G_{i,j,k,l} = \mathrm{tr}[\not{p}_2 \Gamma_i^\mu \not{p}_1 \Gamma_l^\nu]\,\mathrm{tr}[(\not{q}_1 + m)\Gamma_{j\mu}(\not{q}_2 - m)\Gamma_{k\nu}]. \qquad (12.15)$$

12.2. In taking the Dirac adjoint we are using $[\bar{v}\gamma^\mu u]^\dagger = \bar{u}\gamma^\mu v$ and $[\bar{v}\gamma^\mu\gamma^5 u]^\dagger = \bar{u}\gamma^\mu\gamma^5 v$.

Since traces containing $\gamma^5$ and four or less $\gamma^\mu$ matrices vanish, the two $\Gamma$ structures in each trace must be of the same kind, which fixes $i=l$ and $j=k$. Then, it is enough to write the total matrix element as

$$\begin{aligned} -\mathrm{i}\mathcal{M}_a \;=\; & \left[e^2\left(\frac{v_e v_q}{Q^2-m_Z^2}-\frac{Q_q}{Q^2}\right)\bar{v}(p_2)\gamma_\mu u(p_1)-\frac{e^2 a_e v_q}{Q^2-m_Z^2}\bar{v}(p_2)\gamma^5\gamma_\mu u(p_1)\right]\bar{u}(q_1)\gamma^\mu v(q_2) \\ & +\frac{e^2}{Q^2-m_Z^2}\left[a_e a_q \bar{v}(p_2)\gamma^5\gamma_\mu u(p_1)-v_e a_q \bar{v}(p_2)\gamma_\mu u(p_1)\bar{u}(q_1)\right]\bar{u}(q_1)\gamma^5\gamma^\mu v(q_2) \end{aligned} \tag{12.16}$$

to see each term can only multiply its own conjugated. Defining for simplicity

$$\tilde{L}_i^{\mu\nu}\equiv \mathrm{tr}[\not{p}_2\Gamma_i^\mu\not{p}_1\Gamma_i^\nu], \qquad H_{i,a}^{\mu\nu}\equiv \mathrm{tr}[(\not{q}_1+m)\Gamma_i^\mu(\not{q}_2-m)\Gamma_i^\nu], \tag{12.17}$$

the squared matrix element is

$$\begin{aligned} |\bar{\mathcal{M}}_a|^2\equiv\frac{1}{4}\sum_{s(e^\pm)}\sum_{s(q\bar{q})}|\mathcal{M}_a|^2 \;=\; & \frac{1}{4}\left[e^4\left(\frac{v_e v_q}{Q^2-m_Z^2}-\frac{Q_q}{Q^2}\right)^2\tilde{L}_{V\mu\nu}+\frac{e^4 a_e^2 v_q^2}{(Q^2-m_Z^2)^2}\tilde{L}_{A\mu\nu}\right]H_{V,a}^{\mu\nu} \\ & +\frac{e^4}{4(Q^2-m_Z^2)^2}\left[a_e^2 a_q^2\tilde{L}_{A\mu\nu}+v_e^2 a_q^2\tilde{L}_{V\mu\nu}\right]H_{A,a}^{\mu\nu}. \end{aligned} \tag{12.18}$$

As a consequence of electrons being massless, the lepton traces are the same for both $i=V$ and $i=A$,

$$\mathrm{tr}[\not{p}_2\gamma^\mu\not{p}_1\gamma^\nu]=\mathrm{tr}[\not{p}_2\gamma^5\gamma^\mu\not{p}_1\gamma^5\gamma^\nu]=4(p_1^\mu p_2^\nu+p_1^\nu p_2^\mu)-2Q^2 g^{\mu\nu}, \tag{12.19}$$

so we can write $\tilde{L}_V^{\mu\nu}=\tilde{L}_A^{\mu\nu}\equiv\tilde{L}^{\mu\nu}$ and simplify expression (12.18):

$$|\bar{\mathcal{M}}_a|^2=\left[e^4\left(\frac{v_e v_q}{Q^2-m_Z^2}-\frac{Q_q}{Q^2}\right)^2+\frac{e^4 a_e^2 v_q^2}{(Q^2-m_Z^2)^2}\right]L_{\mu\nu}H_{V,a}^{\mu\nu}+\frac{e^4(a_e^2 a_q^2+v_e^2 a_q^2)}{4(Q^2-m_Z^2)^2}L_{\mu\nu}H_{A,a}^{\mu\nu}, \tag{12.20}$$

where $\tilde{L}^{\mu\nu}/4=L^{\mu\nu}$ in (12.9). Finally, using $e=\sqrt{4\pi\alpha_{\mathrm{ew}}}$, the scalar prefactors can be shown to be $L_V$ and $L_A$ also in (12.9), which leaves the final factorized expression for the squared matrix element at leading order

$$|\bar{\mathcal{M}}_a|^2=L_V L_{\mu\nu}H_{V,a}^{\mu\nu}+L_A L_{\mu\nu}H_{A,a}^{\mu\nu}=L_{\mu\nu}\sum_{C=V,A}L_C H_{C,a}^{\mu\nu}. \tag{12.21}$$

Note that, as mentioned and despite notation, $L_{\mu\nu}^V$ and $L_{\mu\nu}^A$ do not arise only from vector and axial lepton vertices, respectively. They are defined instead as the electroweak tensor that multiplies the vector and axial part of the hadron squared matrix element, respectively. Note also that, having grouped all the scalar factors in the $q\bar{q}$ pair production vertex into $L_V$ and $L_A$, in further computations these vertices only contribute to the QCD sector with the simple Feynman rules $\Gamma_V^\mu = \gamma^\mu$ and $\Gamma_A^\mu = \gamma^5\gamma^\mu$, as (12.8) expresses.

Finally, since at $\mathcal{O}(\alpha_{\mathrm{ew}})$ the QCD part of the process always starts with the production of a $q\bar{q}$ pair, the factorization in (12.21) holds at any order in $\alpha_s$. This gives the following computation strategy for the spin-averaged squared matrix-element $|\bar{\mathcal{M}}(e^+e^- \to X)|^2$.

1. Compute the QCD matrix-element corresponding to the final state $X$ and the current $J_C^\mu$, i.e.,

$$H_C^\mu \equiv \langle X | J_C^\mu | 0 \rangle, \tag{12.22}$$

   at a given order in $\alpha_s$, where $J_C^\mu$ is given in (12.8).

2. Take its modulus squared and sum over the spins of the particles in the final state, getting $H_C^{\mu\nu} = \sum_{s(X)} H_C^{\dagger\mu} H_C^\nu$.

3. Contract $H_C^{\mu\nu}$ with the lepton tensor according to (12.7):

$$|\bar{\mathcal{M}}(e^+e^- \to X)|^2 = \sum_{C=V,A} L_{\mu\nu}^C H_C^{\mu\nu}. \tag{12.23}$$

At NLO, this strategy allows to reinterpret (12.2) as

$$|Z_\psi^{\mathrm{OS}} \mathcal{M}_a + \mathcal{M}_b|^2 = \sum_{C=V,A} L_C L_{\mu\nu} \left[ H_{C,a}^{\mu\nu} + \frac{\alpha_s C_F}{2\pi} Z_\psi^{\mathrm{OS}(1)} H_{C,a}^{\mu\nu} + H_{C,ab}^{\mu\nu} \right] + \mathcal{O}(\alpha_s^2), \tag{12.24}$$
$$|\mathcal{M}_c + M_d|^2 = \sum_{C=V,A} L_C L_{\mu\nu} [H_{C,c}^{\mu\nu} + H_{C,d}^{\mu\nu} + H_{C,cd}^{\mu\nu}],$$

where

$$H_{C,ab}^{\mu\nu} \equiv 2\mathrm{Re}\left( \sum_{s(q\bar{q})} H_{C,a}^{\dagger\mu} H_{C,b}^\nu \right), \qquad H_{C,cd}^{\mu\nu} \equiv 2\mathrm{Re}\left( \sum_{s(q\bar{q})} H_{C,c}^{\dagger\mu} H_{C,d}^\nu \right). \tag{12.25}$$

## 12.2 Leading order

The leading order contribution to the event-shape distribution comes entirely from diagram 12.1(a). The momenta of the involved particles are

$$p_i^\mu = (E_i, \vec{p}_i), \qquad q_i^\mu = (U_i, \vec{q}_i), \qquad p_i^2 = 0, \qquad q_i^2 = m^2, \tag{12.26}$$

for $i=1,2$. All the kinematic invariants, which will arise in the contraction of the lepton tensor and the hadron matrix element in (12.7), can be written in terms of the CM energy $Q^2\equiv(p_1+p_2)^2=(q_1+q_2)^2$ and the angle $\theta$ between the incoming electron and the outgoing quark[12.3]:

$$\begin{aligned}
p_1{\cdot}p_2 &= \frac{1}{2}[(p_1+p_2)^2]=\frac{Q^2}{2},\\
p_1{\cdot}q_1 &= \frac{Q^2}{4}-\frac{Q}{4}\sqrt{Q^2-4m^2}\cos\theta=\frac{Q^2}{4}(1-\beta\cos\theta)\\
p_1{\cdot}q_2 &= p_1{\cdot}(p_1+p_2-q_1)=\frac{Q^2}{4}(1+\beta\cos\theta),\\
p_2{\cdot}q_1 &= (q_1+q_2-p_1){\cdot}q_1=p_1{\cdot}q_2,\\
p_2{\cdot}q_2 &= (q_1+q_2-p_1){\cdot}q_2=p_1{\cdot}q_1.\\
q_1{\cdot}q_2 &= \frac{1}{2}[(q_1+q_2)^2-2m^2]=\frac{Q^2}{2}-m^2=\frac{Q^2}{4}(1+\beta^2),
\end{aligned} \tag{12.28}$$

where we defined the reduced mass $\hat{m}=m/Q$ and the quark velocity $\beta=\sqrt{1-4\hat{m}^2}$. Finally, we need to account for the equations of motion for spinnors and their Dirac conjugates in the massive case, which are

$$\begin{aligned}
\not{p}u(p) &= m u(p), & \bar{u}(p)\not{p} &= \bar{u}(p)m,\\
\not{p}v(p) &= -m v(p), & \bar{v}(p)\not{p} &= -\bar{v}(p)m.
\end{aligned} \tag{12.29}$$

The QCD matrix elements defined in (12.22) are simply

$$H^\mu_{C,a}=\bar{u}(q_1)\Gamma^\mu_i v(q_2), \qquad H^{\dagger\mu}_{C,a}=\bar{v}(q_2)\Gamma^\mu_i u(q_1), \tag{12.30}$$

and their traces are[12.4]

$$\begin{aligned}
H^{\mu\nu}_{V,a} &= \mathrm{tr}[(\not{q}_1+m)\gamma^\mu(\not{q}_2-m)\gamma_\nu]=\mathrm{tr}[\not{q}_1\gamma^\mu\not{q}_2\gamma_\nu]-m^2\mathrm{tr}[\gamma^\mu\gamma^\nu]\\
&= 4(q_1^\mu q_2^\nu+q_1^\nu q_2^\mu)-2(Q^2-2m^2)g^{\mu\nu}-4m^2g^{\mu\nu}\\
&= 4(q_1^\mu q_2^\nu+q_1^\nu q_2^\mu)-2Q^2g^{\mu\nu},\\
H^{\mu\nu}_{A,a} &= \mathrm{tr}[(\not{q}_1+m)\gamma^5\gamma^\mu(\not{q}_2-m)\gamma^5\gamma^\nu]=\mathrm{tr}[(\not{q}_1+m)\gamma^\mu(\not{q}_2+)\gamma^\nu]\\
&= \mathrm{tr}[\not{q}_1\gamma^\mu\not{q}_2\gamma^\nu]+m^2\mathrm{tr}[\gamma^\mu\gamma^\nu]=4(q_1^\mu q_2^\nu+q_1^\nu q_2^\mu)-2(Q^2-4m)g^{\mu\nu}\\
&= 4(q_1^\mu q_2^\nu+q_1^\nu q_2^\mu)-2Q^2\beta^2g^{\mu\nu}.
\end{aligned} \tag{12.31}$$

---

12.3. In the CM frame, the energies and momenta of the final-state quark and antiquark are

$$U_1=U_2=|\vec{p}_1|=|\vec{p}_1|=\frac{Q}{2} \quad\text{and}\quad |\vec{q}_1|=|\vec{q}_2|=\frac{1}{2}\sqrt{Q^2-4m^2}=\frac{Q}{2}\beta. \tag{12.27}$$

12.4. The trace of an odd number of gamma matrices vanish. Also, to push the two $\gamma^5$ together in the axial trace and cancel them as $(\gamma^5)^2=\mathbb{I}$ we have used the anticommutation relation $\{\gamma^5,\gamma^\mu\}=0$ in the convenient form $\gamma^5\gamma^\mu(\not{q}+m)=\gamma^\mu(\not{q}-m)\gamma^5$.

The axial result depends explicitly on the quark's mass but the vector coupling is not, and both agree in the limit $m\to 0$ –which corresponds to $\beta\to 1-$. When the Lorentz structure $L_{\mu\nu}$ from the lepton tensor is contracted with (12.31) gives

$$\frac{1}{Q^4 L_V}|\bar{\mathcal{M}}_{a,V}|^2 = L_{\mu\nu}H^{\mu\nu}_{V,a} = \frac{1}{2}(3-\beta^2-3\epsilon)(1+\cos^2\theta)+\frac{1}{2}(1-\beta^2-\epsilon)(1-3\cos^2\theta),$$
$$\frac{1}{Q^4 L_A}|\bar{\mathcal{M}}_{a,A}|^2 = L_{\mu\nu}H^{\mu\nu}_{A,a} = \frac{1}{2}\beta^2(2-3\epsilon)(1+\cos^2\theta)-\frac{1}{2}\beta^2\epsilon(1-3\cos^2\theta). \quad (12.32)$$

The most general expression for the 2-particle phase-space in $d$ dimensions can be found in (F.123). In the present case, and accounting for the appropriate flux factor, one obtains[12.5]

$$\frac{1}{2Q^2}\int \mathrm{d}\Phi_2 = \frac{\beta}{32\pi Q^2\Gamma(1-\epsilon)}\left(\frac{Q^2\beta^2}{16\pi}\right)^{-\epsilon}\int_{-1}^{1}\mathrm{d}\cos\theta\,(1-\cos^2\theta)^{-\epsilon}. \quad (12.33)$$

The final step is to write (12.32) and (12.33) in terms of the thrust axis. Since in the CM frame the thrust axis lays in the common direction of the quark and the antiquark, this is simply done by identifying $\theta=\theta_T$. Then the LO oriented event-shape distribution in $d=4-2\epsilon$ dimensions reads

$$\begin{aligned}\frac{1}{L_V}\frac{\mathrm{d}\sigma_V^{\mathrm{LO}}}{\mathrm{d}\cos\theta_T} &= \frac{Q^2\beta}{64\pi\Gamma(1-\epsilon)}\left(\frac{Q^2\beta^2}{16\pi}\right)^{-\epsilon}(1-\cos^2\theta)^{-\epsilon} \quad (12.34)\\ &\times[(3-\beta^2-3\epsilon)(1+\cos^2\theta_T)+(1-\beta^2-\epsilon)(1-3\cos^2\theta_T)],\\ \frac{1}{L_A}\frac{\mathrm{d}\sigma_A^{\mathrm{LO}}}{\mathrm{d}\cos\theta_T} &= \frac{Q^2\beta^3}{64\pi\Gamma(1-\epsilon)}\left(\frac{Q^2\beta^2}{16\pi}\right)^{-\epsilon}(1-\cos^2\theta)^{-\epsilon}\\ &\times[(2-3\epsilon)(1+\cos^2\theta_T)-\epsilon(1-3\cos^2\theta_T)].\end{aligned}$$

The two structures in $\cos^2\theta_T$ allow to easily integrate over the angle, since

$$\int_{-1}^{1}\mathrm{d}\cos\theta\,(1+\cos^2\theta)=\frac{8}{3}, \qquad \int_{-1}^{1}\mathrm{d}\cos\theta\,(1-3\cos^2\theta)=0, \quad (12.35)$$

and therefore we refer to them as the *unoriented* and *oriented* structures, respectively[12.6]. These structures reappear in the NLO computation, so we adopt the notation

$$S_{\mathrm{un}}\equiv 1+\cos^2\theta_T, \qquad S_{\mathrm{or}}\equiv 1-3\cos^2\theta_T \quad (12.36)$$

12.5. Here we use for the Källén function $\lambda(Q^2,0,0)=Q^4$ and $\lambda(Q^2,m^2,m^2)=Q^2(Q^2-4m^2)=Q^4\beta^2$.

12.6. Upon integration over $\cos\theta_T$ the angular information is lost, and the result is exactly the term proportional to $1+\cos^2\theta_T$. This is the reason why we refer to it as the unoriented contribution.

to prevent expressions from becoming too large.

From result (12.34) we can compute the total vector and axialvector cross-sections at LO by simply setting $\epsilon = 0$ and performing the integration over $\theta_T$:

$$\frac{1}{L_V}\sigma_V^{\mathrm{LO}} = \frac{Q^2\beta\,(3-\beta^2)}{24\pi}, \qquad \frac{1}{L_A}\sigma_A^{\mathrm{LO}} = \frac{Q^2\beta^3}{12\pi}. \tag{12.37}$$

Instead of the lepton tensors $L_V$ and $L_A$, it is customary to choose the Born-level cross-sections as normalization, which is the LO cross section for massless quarks. They can be directly found from (12.37) by taking the massless limit $\beta \to 1$:

$$\frac{1}{L_V}\sigma_V^0 = \frac{1}{L_A}\sigma_A^0 = \frac{Q^2}{12\pi}, \tag{12.38}$$

leading to the total Born-level cross-section

$$\sigma^0 = \sigma_V^0 + \sigma_A^0 = \frac{Q^2}{12\,\pi}(L_V + L_A) = \frac{4\pi\alpha_{\mathrm{ew}}^2}{3Q^2}\left[Q_q^2 - \frac{2Q^2 Q_q v_e v_q}{Q^2 - m_Z^2} + \frac{Q^4(a_e^2 + v_e^2)(a_q^2 + v_q^2)}{(Q^2 - m_Z^2)^2}\right] \tag{12.39}$$

which agrees with [32] and [33]. The distributions in (12.34) become

$$\begin{aligned}\frac{1}{\sigma_V^0}\frac{\mathrm{d}\sigma_V^{\mathrm{LO}}}{\mathrm{d}\cos\theta_T} &= \frac{3\beta(1-\cos^2\theta_T)^{-\epsilon}}{16\Gamma(1-\epsilon)}\left(\frac{Q^2\beta^2}{16\pi}\right)^{-\epsilon}[(3-\beta^2-3\epsilon)S_{\mathrm{un}} + (1-\beta^2-\epsilon)S_{\mathrm{or}}],\\ \frac{1}{\sigma_A^0}\frac{\mathrm{d}\sigma_A^{\mathrm{LO}}}{\mathrm{d}\cos\theta_T} &= \frac{3\beta^3(1-\cos^2\theta_T)^{-\epsilon}}{16\Gamma(1-\epsilon)}\left(\frac{Q^2\beta^2}{16\pi}\right)^{-\epsilon}[(2-3\epsilon)S_{\mathrm{un}} - \epsilon S_{\mathrm{or}}],\end{aligned} \tag{12.40}$$

and the total distributions in (12.37) become

$$R_V^{\mathrm{LO}} = \frac{1}{\sigma_V^0}\sigma_V^{\mathrm{LO}} = \frac{\beta(3-\beta^2)}{2}, \qquad R_A^{\mathrm{LO}} = \frac{1}{\sigma_A^0}\sigma_A^{\mathrm{LO}} = \beta^3. \tag{12.41}$$

Upon expansion in $\epsilon$, the $\mathcal{O}(\epsilon^0)$ terms in (12.40) constitute the LO oriented distribution, while the $\mathcal{O}(\epsilon)$ terms contribute to NLO (see (12.2)). The LO contribution is

$$\begin{aligned}\frac{1}{\sigma_V^0}\frac{\mathrm{d}\sigma_V^{\mathrm{LO}}}{\mathrm{d}\cos\theta_T} &= R_V^0 = \frac{3\beta}{16}[(3-\beta^2)S_{\mathrm{un}} + (1-\beta^2)S_{\mathrm{or}}],\\ \frac{1}{\sigma_A^0}\frac{\mathrm{d}\sigma_A^{\mathrm{LO}}}{\mathrm{d}\cos\theta_T} &= R_A^0 = \frac{3\beta^3}{8}S_{\mathrm{un}},\\ \frac{1}{\sigma_V^0}\frac{\mathrm{d}\sigma_V^0}{\mathrm{d}\cos\theta_T} &= \frac{1}{\sigma_A^0}\frac{\mathrm{d}\sigma_A^0}{\mathrm{d}\cos\theta_T} = \frac{3}{8}S_{\mathrm{un}},\end{aligned} \tag{12.42}$$

where we identified the delta coefficients $R_V^0$ and $R_A^0$ in (12.1). For completeness, and to provide a small cross-check for our computations, we can sum the vector and axial contributions

$$\frac{1}{\sigma^0}\frac{\mathrm{d}\sigma^{\mathrm{LO}}}{\mathrm{d}\cos\theta_T} = \frac{3\beta}{16}\left[\frac{(3-\beta^2)\sigma_V^0+2\beta^2\sigma_A^0}{\sigma^0}S_{\mathrm{un}}+\frac{(1-\beta^2)\sigma_V^0}{\sigma_0}S_{\mathrm{or}}\right] \tag{12.43}$$
$$\frac{1}{\sigma^0}\frac{\mathrm{d}\sigma^0}{\mathrm{d}\cos\theta_T} = \frac{3}{8}S_{\mathrm{un}}.$$

The oriented differential distribution $\mathrm{d}\sigma^0/\mathrm{d}\cos\theta_T$ reproduces the result in [33]. Integrating the first line of (12.43) over the thrust angle one obtains the total cross-section at LO, which can be put in terms of the quark's mass:

$$\begin{aligned}\sigma^{\mathrm{LO}} &= \frac{4\pi\alpha^2}{3}\frac{(Q^2+2m^2)\sqrt{Q^2-4m^2}}{Q^5} \\ &\times\left[Q_q^2-\frac{2Q^2Q_qv_ev_q}{Q^2-m_Z^2}+\frac{Q^4(a_e^2+v_e^2)}{(Q^2-m_Z^2)^2}\left(\frac{Q^2-4m^2}{Q^2+2m^2}a_q^2+v_q^2\right)\right],\end{aligned} \tag{12.44}$$

which in turn corresponds to the sum of the vector and axial contributions in (12.41). This expression is the massive equivalent of the Born-level cross section $\sigma^0$ and reproduces the form in (12.39) in the massless limit.

## 12.3 Next-to-leading order

### 12.3.1 Virtual radiation

The virtual radiation of the gluon consists on the $\mathcal{O}(\alpha_s)$ wavefunction correction to diagram 12.1($a$) and the crossed product between diagrams 12.1($a$) and 12.1($b$) in the first line of (12.24). Diagram 12.1($a$) and its modulus square were computed in $4-2\epsilon$ dimensions in section 12.2; here we compute diagram 12.1($b$) and its crossed product with diagram 12.1($a$). Being a quantum correction to diagram 12.1($a$), the kinematics described in (12.26) and (12.28), as well as the identification $\theta=\theta_T$, also holds for diagram 12.1($b$). In dimensional regularization, the matrix element corresponding to diagram $b$ is

$$H_{C,b}^{\mu} = -4\pi i\alpha_s C_F\tilde{\mu}^{2\epsilon}\bar{u}(q_1)I_C^{\mu}v(q_2), \tag{12.45}$$
$$I_C^{\rho} = \int\frac{\mathrm{d}^{4-2\epsilon}k}{(2\pi)^{4-2\epsilon}}\frac{\gamma^{\mu}(\not{q}_1+\not{k}+m)\Gamma_C^{\rho}(\not{k}-\not{q}_2+m)\gamma_{\mu}}{k^2[(q_1+k)^2-m^2][(q_2-k)^2-m^2]}.$$

The vector and axial integrals are computed in [34]:

$$\tilde{\mu}^{2\epsilon} I^{\rho}_{V,b} = \frac{i}{(4\pi)^2}\left[\tilde{A}\gamma^{\rho} + \tilde{B}\frac{(q_1-q_2)^{\rho}}{2m}\right],$$
$$\tilde{\mu}^{2\epsilon} I^{\rho}_{A,b} = \frac{i}{(4\pi)^2}\left[\tilde{C}\gamma^{\rho} + \tilde{D}\frac{(q_1+q_2)^{\rho}}{2m}\right]\gamma^5 \quad , \tag{12.46}$$

where the coefficients are[12.7]

$$\tilde{A} \equiv \left(\frac{1}{\epsilon} - L_m\right)\left(1 - \frac{1+\beta^2}{\beta}\tilde{L}_{\beta}\right) + \frac{1+\beta^2}{\beta}\left(\pi^2 - 2\mathrm{Li}_{2\beta} - \frac{1}{2}\tilde{L}_{\beta}\right) - 3\beta\tilde{L}_{\beta}, \tag{12.47}$$
$$\tilde{B} \equiv \frac{\beta^2-1}{\beta}\tilde{L}_{\beta}, \qquad \tilde{C} \equiv \tilde{A} + 2\frac{\beta^2-1}{\beta}\tilde{L}_{\beta}, \qquad \tilde{D} \equiv \frac{1-\beta^2}{\beta}\left[2\beta + (2+\beta^2)\tilde{L}_{\beta}\right],$$

with

$$\tilde{L}_{\beta} \equiv \log\left(\frac{\beta-1}{\beta+1}\right), \qquad \mathrm{Li}_{2\beta} = \mathrm{Li}_2\left(\frac{\beta-1}{\beta+1}\right). \tag{12.48}$$

Here, $\mathrm{Li}_2$ represents the di-logarithm[12.8]. Note that, since $\beta \leq 1$, the coefficients in (12.47) are complex numbers due to $\tilde{L}_{\beta}$. The crossed term between diagrams 12.1($a$) and 12.1($b$) takes the form

$$H^{\mu\nu}_{V,ab} = 2\frac{\alpha_s C_F}{4\pi}\mathrm{Re}\left[\sum_{s(q\bar{q})} \bar{v}(q_2)\gamma^{\mu}u(q_1)\bar{u}(q_1)\left(\tilde{A}\gamma^{\nu} + \tilde{B}\frac{(q_1-q_2)^{\nu}}{2m}\right)v(q_2)\right], \tag{12.49}$$
$$H^{\mu\nu}_{A,ab} = 2\frac{\alpha_s C_F}{4\pi}\mathrm{Re}\left[\sum_{s(q\bar{q})} \bar{v}(q_2)\gamma^{\mu}\gamma^5 u(q_1)\bar{u}(q_1)\left(\tilde{C}\gamma^{\nu} + \tilde{D}\frac{(q_1+q_2)^{\nu}}{2m}\right)\gamma^5 v(q_2)\right].$$

The sum over spins in these terms leads to the traces previously computed in (12.31) as well as to the following two additional traces

$$\sum_{s(q\bar{q})} \bar{v}(q_2)\gamma^{\mu}u(q_1)\bar{u}(q_1)v(q_2) = \mathrm{tr}[(\not{q}_2 - m)\gamma^{\mu}(\not{q}_1 + m)] = 4m(q_2-q_1)^{\mu}, \tag{12.50}$$
$$\sum_{s(q\bar{q})} \bar{v}(q_2)\gamma^{\mu}\gamma^5 u(q_1)\bar{u}(q_1)\gamma^5 v(q_2) = \mathrm{tr}[(\not{q}_2 - m)\gamma^{\mu}\gamma^5(\not{q}_1 + m)\gamma^5] = 4m(q_1+q_2)^{\mu}.$$

With them, the crossed terms become

$$H^{\mu\nu}_{V,ab} = 2\frac{\alpha_s C_F}{4\pi}\left[\mathrm{Re}(\tilde{A})[4(q_1^{\mu}q_2^{\nu} + q_1^{\nu}q_2^{\mu}) - 2Q^2 g^{\mu\nu}] - 2\mathrm{Re}(\tilde{B})(q_2-q_1)^{\mu}(q_2-q_1)^{\nu}\right],$$
$$H^{\mu\nu}_{A,ab} = 2\frac{\alpha_s C_F}{4\pi}\left[\mathrm{Re}(\tilde{C})[4(q_1^{\mu}q_2^{\nu} + q_1^{\nu}q_2^{\mu}) - 2\beta^2 Q^2 g^{\mu\nu}] + 2\mathrm{Re}(\tilde{D})(q_2+q_1)^{\mu}(q_2+q_1)^{\nu}\right], \tag{12.51}$$

12.7. In [34], the coefficients are simply denoted $A$, $B$, $C$ and $D$, but we choose to keep the tilde to make a redefinition when taking the modulus square.

12.8. The formal definition of the poly-logarithm is $\mathrm{Li}_n(z) \equiv \int_0^z ds \mathrm{Li}_{n-1}(s)/s$, with $\mathrm{Li}_1(z) = -\log(1-z)$.

and when contracted with $L^C_{\mu\nu}$ one finds

$$\begin{aligned}\frac{1}{Q^4 L_V} L^V_{\mu\nu} H^{\mu\nu}_{V,ab} &= \frac{\alpha_s C_F}{4\pi}\Big\{[\mathrm{Re}(\tilde{A})(3-3\epsilon-\beta^2) - \mathrm{Re}(\tilde{B})\beta^2]S_{\mathrm{un}} \\ &\quad +[\mathrm{Re}(\tilde{A})(1-\epsilon-\beta^2) - \mathrm{Re}(\tilde{B})\beta^2]S_{\mathrm{or}}\Big\}, \\ \frac{1}{Q^4 L_A} L^A_{\mu\nu} H^{\mu\nu}_{A,ab} &= \frac{\alpha_s C_F}{4\pi}\mathrm{Re}(\tilde{C})\beta^2[(2-3\epsilon)S_{\mathrm{un}} - \epsilon S_{\mathrm{or}}].\end{aligned} \tag{12.52}$$

Then the virtual contribution to the cross-section is found by summing (12.32) and (12.52) as in (12.24)

$$\begin{aligned}\frac{1}{Q^4 L_V}|\bar{\mathcal{M}}_{ab,V}|^2 &= \frac{\alpha_s C_F}{4\pi}\Big\{[A(3-3\epsilon-\beta^2) - B\beta^2]S_{\mathrm{un}} + [A(1-\epsilon-\beta^2) - B\beta^2]S_{\mathrm{or}}\Big\} \\ \frac{1}{Q^4 L_A}|\bar{\mathcal{M}}_{ab,A}|^2 &= \frac{\alpha_s C_F}{4\pi} C\beta^2\Big[(2-3\epsilon)S_{\mathrm{un}} - \epsilon S_{\mathrm{or}}\Big]\end{aligned} \tag{12.53}$$

where the coefficients without the tilde are

$$A \equiv \mathrm{Re}(\tilde{A}) + Z_\psi^{\mathrm{OS}(1)}, \qquad B \equiv \mathrm{Re}(B), \qquad C \equiv \mathrm{Re}(\tilde{C}) + Z_\psi^{\mathrm{OS}(1)}. \tag{12.54}$$

To extract the real part it is enough to change $\tilde{L}_\beta \to L_\beta$ in the tilded coefficients, as

$$\mathrm{Re}(\tilde{L}_\beta) \equiv L_\beta = \log\left(\frac{1-\beta}{1+\beta}\right). \tag{12.55}$$

Combining these matrix elements with $\tilde{\mu}^{2\epsilon}$ times[12.9] the two-particle phase-space in (12.33) and expanding to $\mathcal{O}(\epsilon)$ we find the structure

$$\begin{aligned}\frac{1}{\sigma_C^0}\frac{\mathrm{d}\sigma_C^{\mathrm{virt}}}{\mathrm{d}e\mathrm{d}\cos\theta_T} &= \frac{\alpha_s C_F}{\pi} A^{\mathrm{virt},C}\delta(e-e_{\min}) \\ &\equiv \frac{\alpha_s C_F}{\pi}\left[A_{\mathrm{fin}}^{\mathrm{virt},C} + A_{\mathrm{div}}^{\mathrm{virt},C}\left(L_\theta - \frac{1}{\epsilon}\right)\right]\delta(e-e_{\min}),\end{aligned} \tag{12.57}$$

12.9. The virtual contribution to the cross-section is of the form

$$\frac{1}{Q^2}\int \mathrm{d}\Phi_2 |\bar{\mathcal{M}}_{ab}|^2 \propto Q^{-2-2\epsilon}, \tag{12.56}$$

where $|\bar{\mathcal{M}}_{ab}|^2$ is adimensional due to the factor of $\tilde{\mu}^{2\epsilon}$ included in the loop integral in (12.46). The extra factor of $\tilde{\mu}^{2\epsilon}$ we include with the phase-space compensates $Q^{-2\epsilon}$ and ensures an adimensional expansion in $\epsilon$. An alternative method, not followed in this work, is to normalize by the Born-level cross-section in $4-2\epsilon$ dimensions, which renders $\sigma_C/\sigma_C^{0,4-2\epsilon}$ adimensional.

where we extracted the terms that cancel in the sum with the real radiation, which are those proportional to $1/\epsilon$ (soft divergences) and to $L_\theta \equiv \log(1-\cos^2\theta_T)$ (collinear divergences). Explicitly, the divergent terms are:

$$A_{\rm div}^{{\rm virt},V} = \frac{3\beta}{16}\left(1+\frac{\beta^2+1}{2\beta}L_\beta\right)\Big[(3-\beta^2)S_{\rm un}+(1-\beta^2)S_{\rm or}\Big], \tag{12.58}$$
$$A_{\rm div}^{{\rm virt},A} = \frac{3\beta^3}{8}\left(1+\frac{\beta^2+1}{2\beta}L_\beta\right)S_{\rm un},$$

where one can observe no dependence on the hard scale $Q$ and the renormalization scale $\mu$. The finite part is:

$$\begin{aligned}
A_{\rm fin}^{{\rm virt},V} &= \frac{3\beta}{16}\left[(L_m+L_Q)\left(1+\frac{\beta^2+1}{2\beta}L_\beta\right)-g_{\rm virt}(\beta)\right][(3-\beta^2)S_{\rm un}+(1-\beta^2)S_{\rm or}] \\
&\quad +\frac{3\beta^3}{16}\left[\left(1+\frac{\beta^2-1}{2\beta}L_\beta\right)S_{\rm un}+\left(1+\frac{\beta^2+1}{2\beta}L_\beta\right)S_{\rm or}\right], \qquad (12.59)\\
A_{\rm fin}^{{\rm virt},A} &= \frac{3\beta^3}{8}\left[(L_m+L_Q)\left(1+\frac{\beta^2+1}{2\beta}L_\beta\right)-g_{\rm virt}(\beta)\right]S_{\rm un} \\
&\quad +\frac{3\beta^3}{16}\left[\left(1+\frac{3(\beta^2-1)}{2\beta}L_\beta\right)S_{\rm un}+\left(1+\frac{\beta^2+1}{2\beta}L_\beta\right)S_{\rm or}\right],
\end{aligned}$$

where we have kept separated the dependence on $\mu$ and where

$$L_Q \equiv \log\left(\frac{Q^2}{\mu^2}\right), \tag{12.60}$$
$$g_{\rm virt}(\beta) \equiv \frac{1+\beta^2}{2\beta}\left(2{\rm Li}_{2\beta}+\frac{1}{2}L_\beta^2-L_\beta-L_\beta\log\left(\frac{\beta^2}{4}\right)-\pi^2\right)+\beta L_\beta-\log\left(\frac{\beta^2}{4}\right)+1.$$

The sum of the virtual and real contributions into the full coefficients $A_e^V$ and $A_e^A$ will be carried out in section 12.3.4.

### 12.3.2 Real radiation I: general set-up

The real gluon radiation encodes all the terms in the second line of (12.24), whose diagrammatic counterparts are in figures 12.1$(c)$ and 12.1$(d)$. The momenta of the involved particles are defined outgoing as in (12.26), with the addition of $q_3=(U_3,\vec{q}_3)$, $q_3^2=m_3^2=0$ for the gluon. Also, the momentum of the virtual quark is represented by $q$ when the gluon is emitted by the quark and $\bar{q}$ when it is emitted by the antiquark, both also defined outgoing, i.e.,

$$q=q_1+q_3, \qquad \bar{q}=q_2+q_3. \tag{12.61}$$

As discussed, there are 4 non-fixed kinematic variables, which can be chosen as the energies of the quark and the antiquark, $U_1$ and $U_2$, and their angles with respect the incoming electron, denoted $\theta_1$ and $\theta_2$. For convenience, instead of explicitly using the CM energies $U_i$ we will use the dimensionless variables

$$x_i \equiv \frac{2U_i}{Q}, \qquad i = 1, 2, 3. \tag{12.62}$$

In terms of the $x_i$, the on-shell relation for the quark, the antiquark and the gluon leads to

$$\begin{aligned} |\vec{q}_i| &= \sqrt{U_i^2 - m^2} = \frac{Q}{2} x_i \sqrt{1 - \frac{4\hat{m}^2}{x_i^2}} = \frac{Q}{2} x_i \beta_i, \qquad i = 1, 2, \\ |\vec{q}_3| &= U_3 = \frac{Q}{2} x_3, \end{aligned} \tag{12.63}$$

where as before $\hat{m} = m/Q$, and we defined $\beta_i \equiv \sqrt{1 - 4\hat{m}^2/x_i^2}$. The fact that only two out of the three energies and two out of the three angles are independent is reflected in the $x_i$ variables through the conditions

$$\begin{aligned} & x_1 + x_2 + x_3 = 2, \\ & 0 = |\vec{q}_1|\cos\theta_1 + |\vec{q}_2|\cos\theta_2 + |\vec{q}_3|\cos\theta_3 = \sqrt{x_1^2 - 4\hat{m}^2}\cos\theta_1 + \sqrt{x_1^2 - 4\hat{m}^2}\cos\theta_2 + x_3^2\cos\theta_3, \end{aligned} \tag{12.64}$$

which arise, respectively, from conservation of energy and the total three-momentum in the incoming beam direction. Finally, we can write all the Lorentz-invariant structures in terms of $Q$, $x_1$, $x_2$, $\theta_1$ and $\theta_2$:

$$\begin{aligned} p_1 \cdot p_2 &= \frac{1}{2}[(p_1 + p_2)^2] = \frac{Q^2}{2}, \\ p_1 \cdot q_1 &= E_1 U_1 - |\vec{p}_1||\vec{q}_1|\cos\theta_1 = \frac{Q^2}{4} x_1 (1 - \beta_1 \cos\theta_1) \\ p_1 \cdot q_2 &= \frac{Q^2}{4} x_2 (1 - \beta_2 \cos\theta_2), \\ p_1 \cdot q_3 &= E_1 U_3 (1 - \cos\theta_3) = \frac{Q^2}{4}\Big[2 - (1 - \beta_1\cos\theta_1) x_1 - (1 - \beta_2 \cos\theta_2) x_2\Big], \\ p_2 \cdot q_1 &= E_2 U_1 + |\vec{p}_2||\vec{q}_1|\cos\theta_1 = \frac{Q^2}{4} x_1 (1 + \beta_1 \cos\theta_1) \\ p_2 \cdot q_2 &= \frac{Q^2}{4} x_2 (1 + \beta_2 \cos\theta_2), \\ p_2 \cdot q_3 &= E_2 U_3 (1 + \cos\theta_3) = \frac{Q^2}{4}\Big[2 - (1 + \beta_1\cos\theta_1) x_1 - (1 + \beta_2 \cos\theta_2) x_2\Big], \\ q_1 \cdot q_2 &= \frac{1}{2}[(q_1 + q_2)^2 - 2m^2] = \frac{1}{2}[(p_1 + p_2 - q_3)^2 - 2m^2] = Q^2\left(\frac{x_1 + x_2 - 1}{2} - \hat{m}^2\right), \end{aligned} \tag{12.65}$$

$$\begin{aligned} q_1\cdot q_3 &= \frac{1}{2}[(q_1+q_3)^2-m^2]=\frac{1}{2}[(p_1+p_2-q_2)^2-m^2]=\frac{Q^2}{2}(1-x_2),\\ q_2\cdot q_3 &= \frac{1}{2}[(q_2+q_3)^2-m^2]=\frac{1}{2}[(p_1+p_2-q_1)^2-m^2]=\frac{Q^2}{2}(1-x_1). \end{aligned}$$

For the products involving the positron momentum $p_2$ and the particles in the final state, $q_1$, $q_2$ and $q_3$, we have used the fact that the angle any of the latter forms with the former is $\pi-\theta_i$, which changes the sign of the cosine function –equivalently, one can use $p_2=q_1+q_2`q_3-p_1$–. Finally, for the invariants involving the virtual quark's momenta, we simply use (12.61) and (12.65) to get

$$q^2=Q^2(1-x_2+\hat{m}^2),\qquad \bar{q}^2=(1-x_1+\hat{m}^2). \tag{12.66}$$

Now we focus on the hadron matrix elements,

$$\begin{aligned} H^\mu_{V,c} &= \frac{g_s t_a}{Q^2(1-x_2)}\bar{u}(q_1)\gamma^\rho(\not{q}+m)\gamma^\mu\varepsilon_\rho(q_3)v(q_2), \qquad (12.67)\\ H^\mu_{V,d} &= \frac{-g_s t_a}{Q^2(1-x_1)}\bar{u}(q_1)\gamma^\mu(\not{\bar{q}}-m)\gamma^\rho\varepsilon_\rho(q_3)v(q_2),\\ H^\mu_{A,c} &= \frac{g_s t_a}{Q^2(1-x_2)}\bar{u}(q_1)\gamma^\rho(\not{q}+m)\gamma^\mu\gamma^5\varepsilon_\rho(q_3)v(q_2),\\ H^\mu_{A,d} &= \frac{-g_s t_a}{Q^2(1-x_1)}\bar{u}(q_1)\gamma^\mu\gamma^5(\not{\bar{q}}-m)\gamma^\rho\varepsilon_\rho(q_3)v(q_2), \end{aligned}$$

and their Dirac adjoints[12.10]

$$\begin{aligned} H^{\dagger\mu}_{V,c} &= \frac{g_s t_a^\dagger}{Q^2(1-x_2)}\bar{v}(q_2)\gamma^\mu(\not{q}+m)\gamma^\rho\varepsilon^*_\rho(q_3)u(q_1), \qquad (12.69)\\ H^{\dagger\mu}_{V,d} &= \frac{-g_s t_a^\dagger}{Q^2(1-x_1)}\bar{v}(q_2)\gamma^\rho(\not{\bar{q}}-m)\gamma^\mu\varepsilon^*_\rho(q_3)u(q_1),\\ H^{\dagger\mu}_{A,c} &= \frac{-g_s t_a^\dagger}{Q^2(1-x_2)}\bar{v}(q_2)\gamma^5\gamma^\mu(\not{q}+m)\gamma^\rho\varepsilon^*_\rho(q_3)u(q_1),\\ H^{\dagger\mu}_{A,d} &= \frac{-g_s t_a^\dagger}{Q^2(1-x_1)}\bar{v}(q_2)\gamma^\rho(\not{\bar{q}}-m)\gamma^5\gamma^\mu\varepsilon^*_\rho(q_3)u(q_1). \end{aligned}$$

12.10. Here we are using that for any number of $\gamma^\mu$ matrices $[\bar{u}(p_1)\gamma^{\mu_1}...\gamma^{\mu_n}v(p_2)]^\dagger=\bar{v}(p_2)\gamma^{\mu_n}...\gamma^{\mu_1}u(p_1)$, while when there is a $\gamma^5$ at the beginning or at the end of the string, the adjoint picks a minus sign, $[\bar{u}(p_1)\gamma^{\mu_1}...\gamma^{\mu_n}\gamma^5 v(p_2)]^\dagger=-\bar{v}(p_2)\gamma^5\gamma^{\mu_n}...\gamma^{\mu_1}u(p_1)$. This leads to the two cases

$$\begin{aligned} &[\bar{u}(p_1)\gamma^\mu(\not{p}+m)\gamma^\nu v(p_2)]^\dagger=\bar{v}(p_2)\gamma^\nu\not{p}\gamma^\mu u(p_1)+m\bar{v}(p_2)\gamma^\nu\gamma^\mu u(p_1)=\bar{v}(p_2)\gamma^\nu(\not{p}+m)\gamma^\mu u(p_1), \qquad (12.68)\\ &[\bar{u}(p_1)\gamma^\mu(\not{p}+m)\gamma^\nu\gamma^5 v(p_2)]^\dagger=-\bar{v}(p_2)\gamma^5\gamma^\nu\not{p}\gamma^\mu u(p_1)-m\bar{v}(p_2)\gamma^5\gamma^\nu\gamma^\mu u(p_1)=-\bar{v}(p_2)\gamma^5\gamma^\nu(\not{p}+m)\gamma^\mu u(p_1). \end{aligned}$$

Summing over quark and antiquark spins and using $\sum_\lambda \varepsilon_{\rho,\lambda}(q_3)\varepsilon^*_{\sigma,\lambda}(q_3) = -g_{\rho\sigma}$ to sum over gluon polarizations, the contributions to the squared total matrix element lead to the traces

$$
\begin{aligned}
H^{\mu\nu}_{V,c} &= \frac{-g_s^2 C_F}{Q^4(1-x_2)^2}\mathrm{tr}[(\not{q}_2 - m)\gamma^\mu(\not{q}+m)\gamma^\rho(\not{q}_1+m)\gamma_\rho(\not{q}+m)\gamma^\nu], \qquad (12.70)\\
H^{\mu\nu}_{V,d} &= \frac{-g_s^2 C_F}{Q^4(1-x_1)^2}\mathrm{tr}[(\not{q}_2 - m)\gamma^\rho(\bar{\not{q}}-m)\gamma^\mu(\not{q}_1+m)\gamma^\nu(\bar{\not{q}}-m)\gamma_\rho],\\
H^{\mu\nu}_{V,cd} &= \frac{2g_s^2 C_F}{Q^4(1-x_1)(1-x_2)}\mathrm{tr}[(\not{q}_2 - m)\gamma^\mu(\not{q}+m)\gamma^\rho(\not{q}_1+m)\gamma^\nu(\bar{\not{q}}-m)\gamma_\rho],\\
H^{\mu\nu}_{A,c} &= \frac{-g_s^2 C_F}{Q^4(1-x_2)^2}\mathrm{tr}[(\not{q}_2 + m)\gamma^\mu(\not{q}+m)\gamma^\rho(\not{q}_1+m)\gamma_\rho(\not{q}+m)\gamma^\nu],\\
H^{\mu\nu}_{A,d} &= \frac{-g_s^2 C_F}{Q^4(1-x_1)^2}\mathrm{tr}[(\not{q}_2 - m)\gamma^\rho(\bar{\not{q}}-m)\gamma^\mu(\not{q}_1-m)\gamma^\nu(\bar{\not{q}}-m)\gamma_\rho],\\
H^{\mu\nu}_{A,cd} &= \frac{2g_s^2 C_F}{Q^4(1-x_1)(1-x_2)}\mathrm{tr}[(\not{q}_2 + m)\gamma^\mu(\not{q}+m)\gamma^\rho(\not{q}_1+m)\gamma^\nu(\bar{\not{q}}+m)\gamma_\rho].
\end{aligned}
$$

The results of the traces in (12.70) are too long to provide them explicitly. We solve them with `Tracer` [102] and show the results for the squared matrix elements, i.e., after contraction with the lepton tensor $L_{\mu\nu}$. We obtain the following angular structure

$$
\frac{|M_{c,C}+M_{d,C}|^2}{8\pi^2 Q^2 L_C} = \frac{\alpha_s C_F}{\pi}[f^C_{0,0} + f^C_{2,0}\cos^2\theta_1 + f^C_{0,2}\cos^2\theta_2 + f^C_{1,1}\cos\theta_1\cos\theta_2], \qquad (12.71)
$$

where $f^C_{i,j} = f^C_{i,j}(x_1, x_2)$ are given in appendix E.1. Note that by symmetry one has $f^C_{i,j}(x_1,x_2) = f^C_{j,i}(x_2,x_1)$, as can be explicitly verified

On the other hand, the 3-particle phase-space can be written in terms of any pair of final state particles. In $d = 4-2\,\epsilon$ dimensions it has the generic form that can be found in (F.136), which for a pair of massless particles in the initial state reduces to

$$
\begin{aligned}
\frac{1}{2Q^2}\int \mathrm{d}\Phi_3 \;=\; & \frac{1}{512\pi^4\Gamma(1-2\epsilon)}\left(\frac{Q^2}{8\pi}\right)^{-2\epsilon}\int \mathrm{d}x_i \mathrm{d}x_j[(x_i^2-4\hat{m}_i^2)(x_j^2-4\hat{m}_j^2)]^{-\epsilon}\Theta_{ij}\\
&\times\int\frac{\mathrm{d}\cos\theta_i\,\mathrm{d}\cos\theta_j\,\theta(h_{ij})}{(h_{ij})^{1/2+\epsilon}}, \qquad (12.72)
\end{aligned}
$$

where

$$
\begin{aligned}
\Theta_{ij} &= \theta(2-x_i-x_j)\theta(\sin^2\tilde{\theta}_{ij}), \qquad (12.73)\\
h_{ij} &= \sin^2\tilde{\theta}_{ij} + 2\cos\tilde{\theta}_{ij}\cos\theta_i\cos\theta_j - \cos^2\theta_i - \cos^2\theta_j,\\
\cos\tilde{\theta}_{ij} &= \frac{2+x_ix_j-2(x_i+x_j)+2(\hat{m}_i^2+\hat{m}_j^2-\hat{m}_k^2)}{\sqrt{x_i^2-4\hat{m}_i^2}\,\sqrt{x_j^2-4\hat{m}_j^2}}, \qquad i\neq j\neq k.
\end{aligned}
$$

In (12.72) the integrals over $x_i$ extend along $(2\hat{m}_i, \infty)$ and those over $\cos\theta_i$ extend along $(-1, 1)$; the Heaviside functions $\Theta_{ij}$ and $\theta(h_{ij})$ restrict these integrations to the physical region. Appendix F contains the details on the integration boundaries, and in our case $\hat{m}_1 = \hat{m}_2 = \hat{m}$ and $\hat{m}_3 = 0$.

When combining the matrix elements in (12.71) with $\tilde{\mu}^{4\epsilon}$ times[12.11] the three-particle phase-space in (12.72) one obtains the multiple-differential cross-section for real gluon radiation $\mathrm{d}\sigma_C^{\mathrm{real}}/(\mathrm{d}x_1\mathrm{d}x_2\mathrm{d}\cos\theta_1\mathrm{d}\cos\theta_2)$. To match with the contribution from virtual radiation in (12.57) one needs to take the following steps.

1. Project the cross-section onto the thrust axis, i.e., write the differential cross-section in $\mathrm{d}\cos\theta_T$ and integrate over the remaining angle, obtaining $\mathrm{d}\sigma_C^{\mathrm{real}}/(\mathrm{d}x_1\mathrm{d}x_2\mathrm{d}\cos\theta_T)$.

2. Take the limit in which the gluon is soft, since it is the one in which the IR divergences occur. This limit occurs when $x_3 = 0$ and $x_1 = x_2 = 1$ simultaneously, and in $4 - 2\epsilon$ dimensions $\epsilon$ regulates the divergences. By adding and subtracting the IR-divergences, $\mathrm{d}\sigma_C^{\mathrm{real}}/(\mathrm{d}x_1\mathrm{d}x_2\mathrm{d}\cos\theta_T)$ is split into an IR-safe (hard cross-section) part and a purely IR-divergent part (soft cross-section).

3. Integrate over $x_1$ and $x_2$ to finally get $\mathrm{d}\sigma_C^{\mathrm{real}}/\mathrm{d}\cos\theta_T$. The IR-safe combines with the finite part of the virtual cross-section (12.58) and the IR-divergent part cancels against its divergent and $\log(1 - \cos^2\theta_T)$ parts.

### 12.3.3 Real-radiation II: projection onto thrust and soft limit

Let us begin with the projection onto the thrust axis. It is shown in result (7.12) and the corresponding discussion that for three particles the thrust axis lays in the traveling direction of the particle with the highest $|\vec{p}_i|$. In our case, this can be written as

$$T = \max\left\{\sqrt{x_1^2 - 4\hat{m}^2}, \sqrt{x_2^2 - 4\hat{m}^2}, x_3\right\}. \tag{12.75}$$

12.11. The real contribution to the cross-section is of the form

$$\frac{1}{Q^2}\int \mathrm{d}\Phi_3 |M_{c,C} + M_{d,C}|^2 \propto Q^{-2-4\epsilon}, \tag{12.74}$$

where $|M_{c,C} + M_{d,C}|^2$ has dimensions of $Q^{-2}$. The extra factor of $\tilde{\mu}^{2\epsilon}$ we include with the phase-space groups with that coming from $\alpha_s^0$ and compensates $Q^{-4\epsilon}$ and ensures an adimensional expansion in $\epsilon$. It also makes the virtual and real contributions to have the same mass dimension, so that they can be summed.

Appendix F.6 contains the specific details of the projection of a differential cross-section $\mathrm{d}\sigma/(\mathrm{d}x_1\mathrm{d}x_2\mathrm{d}\cos\theta_1\mathrm{d}\cos\theta_2)$ with polynomial dependence on $\cos\theta_1$ and $\cos\theta_2$ onto the thrust axis. The general idea is to take $\mathrm{d}\Phi_3\to\mathrm{d}\Phi_3^T\equiv\mathrm{d}\Phi_3\,\delta_T$ [33], where

$$\begin{aligned}\delta_T &\equiv \delta_T^1+\delta_T^2+\delta_T^3,\\ \delta_T^i &\equiv \Theta_i^T\,\delta(\cos\theta_i-\cos\theta_T)\,\mathrm{d}\cos\theta_T,\\ \Theta_i^T &\equiv \theta\Big(\sqrt{x_i^2-4\hat{m}_i^2}-\sqrt{x_j^2-4\hat{m}_j^2}\Big)\,\theta\Big(\sqrt{x_i^2-4\hat{m}_i^2}-\sqrt{x_k^2-4\hat{m}_k^2}\Big).\end{aligned} \tag{12.76}$$

When $\delta_T$ is introduced, the function $\delta(\cos\theta_i-\cos\theta_T)$ simply changes the notation $\cos\theta_i\to\cos\theta_T$ for the particle determining the thrust axis. The remaining angle is then integrated over with the ideas in sections F.3.2.5 and F.4.4.

The introduction of $\delta_T$ also divides the differential distribution into three regions specified by the projectors $\Theta_i^T$, in a way that we can define

$$\frac{1}{\sigma_C^0}\frac{\mathrm{d}\sigma_C^{\mathrm{real}}}{\mathrm{d}x_1\mathrm{d}x_2\mathrm{d}\cos\theta_T}\equiv\sum_{i=1}^{3}\frac{1}{\sigma_C^0}\frac{\mathrm{d}\sigma_C^i}{\mathrm{d}x_1\mathrm{d}x_2\mathrm{d}\cos\theta_T}\Theta_i^T. \tag{12.77}$$

The relevant result is that the cross-section in each region can be put as[12.12]

$$\begin{aligned}\frac{1}{\sigma_C^0}\frac{\mathrm{d}\sigma_C^i}{\mathrm{d}x_1\mathrm{d}x_2\mathrm{d}\cos\theta_T} &= F(x_1,x_2,\epsilon)\frac{1}{\sigma_C^0}\frac{\mathrm{d}\bar{\sigma}_C^i}{\mathrm{d}x_1\mathrm{d}x_2\mathrm{d}\cos\theta_T}\\ F(x_1,x_2,\epsilon) &= \frac{\alpha_s C_F}{\pi}\frac{3\mathrm{e}^{2\epsilon\gamma_E}(1-\cos\theta_T)^{-\epsilon}}{4^{3-\epsilon}\Gamma^2(1-\epsilon)}\left(\frac{Q^2}{\mu^2}\right)^{-2\epsilon}[(1-x_1)(1-x_2)(1-x_3)-\hat{m}^2x_3^2]^{-\epsilon}\end{aligned} \tag{12.79}$$

where the differential distributions $\bar{\sigma}_C^i$ take the form (F.189) in $4-2\epsilon$ dimensions. This result is given for three massive particles, and in our case ($m_1=m_2=m$, $m_3=0$) it reduces to

$$\begin{aligned}\frac{\mathrm{d}\bar{\sigma}_C^1}{\mathrm{d}x_1\mathrm{d}x_2\mathrm{d}\cos\theta_T} &= \Big[3f_{0,0}+f_{2,0}+f_{2,0}+\cos\tilde{\theta}_{12}f_{1,1}+\frac{\epsilon}{1-\epsilon}\sin^2\tilde{\theta}_{12}f_{0,2}\Big]S_{\mathrm{un}}\\ &\quad+\Big[f_{0,0}-f_{2,0}-f_{0,2}-\cos\tilde{\theta}_{12}f_{1,1}+\frac{2-\epsilon}{1-\epsilon}\sin^2\tilde{\theta}_{12}f_{0,2}\Big]S_{\mathrm{or}},\\ \frac{\mathrm{d}\bar{\sigma}_C^2}{\mathrm{d}x_1\mathrm{d}x_2\mathrm{d}\cos\theta_T} &= \Big[3f_{0,0}+f_{2,0}+f_{2,0}+\cos\tilde{\theta}_{12}\,f_{1,1}+\frac{\epsilon}{1-\epsilon}\sin^2\tilde{\theta}_{12}f_{2,0}\Big]S_{\mathrm{un}}\end{aligned} \tag{12.80}$$

12.12. In appendix F.6.2, the projection onto thrust is carried out for a cross-section of the form $\mathrm{d}\sigma/(\mathrm{d}x_1\mathrm{d}x_2\mathrm{d}\cos\theta_1\mathrm{d}\cos\theta_2)$, with a matrix element $|M|^2=\sum_{i,j}f_{i,j}\cos^i\theta_1\cos^j\theta_2$. For the real radiation, the matrix element in (12.71) is multiplied by the constant factor $32\pi^2Q^2L_C\alpha_sC_F/(4\pi)$, and the cross-section includes the normalization $1/\sigma_C^0$. Both contributions are accounted for in $F(x_1,x_2,\epsilon)$:

$$F(x_1,x_2,\epsilon)\equiv\frac{32\pi^2Q^2L_C\,\alpha_sC_F}{4\pi\sigma_C^0}F^{\mathrm{app}}(x_1,x_2,\epsilon), \tag{12.78}$$

where $F^{\mathrm{app}}(x_1,x_2,\epsilon)$ is defined in (F.171). The current dependence cancels in the quotient $L_C/\sigma_C^0$ by (12.38).

$$
\begin{aligned}
&+\left[f_{0,0}-f_{2,0}-f_{0,2}-\cos\tilde{\theta}_{12}\,f_{1,1}+\frac{2-\epsilon}{1-\epsilon}\sin^2\tilde{\theta}_{12}f_{2,0}\right]S_{\rm or},\\
\frac{{\rm d}\bar{\sigma}_C^3}{{\rm d}x_1{\rm d}x_2{\rm d}\cos\theta_T} &= S_{\rm un}\left\{3\,f_{0,0}+f_{2,0}+f_{2,0}+\cos\tilde{\theta}_{12}\,f_{1,1}+\frac{\epsilon}{1-\epsilon}\frac{\sin^2\tilde{\theta}_{12}}{x_3^2}\right.\\
&\left.\times\left[(x_2^2-4\hat{m}^2)f_{2,0}+(x_1^2-4\hat{m}^2)f_{0,2}+\sqrt{x_1^2-4\hat{m}^2}\,\sqrt{x_2^2-4\hat{m}^2}\,f_{1,1}\right]\right\}\\
&+S_{\rm or}\left\{f_{0,0}-f_{2,0}-f_{2,0}-\cos\tilde{\theta}_{12}f_{1,1}+\frac{2-\epsilon}{1-\epsilon}\frac{\sin^2\tilde{\theta}_{12}}{x_3^2}\right.\\
&\left.\times\left[(x_2^2-4\hat{m}^2)\,f_{2,0}+(x_1^2-4\hat{m}^2)f_{0,2}+\sqrt{x_1^2-4\hat{m}^2}\,\sqrt{x_2^2-4\hat{m}^2}\,f_{1,1}\right]\right\}.
\end{aligned}
$$

Note that for simplicity we dropped the superscript $C$ in the coefficients $f_{i,j}^C$. By symmetry one has that $\bar{\sigma}_C^1(x_1,x_2)=\bar{\sigma}_C^2(x_2,x_1)$ and $\bar{\sigma}_C^3(x_1,x_2)=\bar{\sigma}_C^3(x_2,x_1)$.

The next step is to extract the limit in which the gluon is soft, which is characterized by $x_3=0$ and $x_1=x_2=1$. In order for that we change variables to $x_1=1-(1-z)y$ and $x_2=1-zy$. This implies $y=x_3$ and makes the soft limit immediate to implement by expanding around $y=0$, since this automatically gives $x_1=x_2=1$ independently on the value of $z$. Let us start by writing the oriented event-shape distribution[12.13]:

$$
\begin{aligned}
\frac{1}{\sigma_C^0}\frac{{\rm d}\sigma_C^{\rm real}}{{\rm d}e{\rm d}\cos\theta_T} &= \int{\rm d}x_1{\rm d}x_2\frac{1}{\sigma_C^0}\frac{{\rm d}\,\sigma_C^{\rm real}(x_1,x_2)}{{\rm d}x_1{\rm d}x_2{\rm d}\cos\theta_T}\delta[e-\hat{e}(x_1,x_2)]. \qquad (12.81)\\
&= \sum_{i=1}^{3}\int{\rm d}y{\rm d}z\,y\,F(y,z,\epsilon)\frac{1}{\sigma_C^0}\frac{{\rm d}\bar{\sigma}_C^i(y,z)}{{\rm d}x_1{\rm d}x_2{\rm d}\cos\theta_T}\Theta_i^T\delta[e-\hat{e}(y,z)],
\end{aligned}
$$

where no expansion has been performed so far and where the $y$ and $z$ integrals extend to

$$
\int_{2\hat{m}}^{\infty}{\rm d}x_1\int_{2\hat{m}}^{\infty}{\rm d}x_2\,\theta(2-x_i-x_j)\theta(\sin^2\tilde{\theta}_{12})=\int_{(1-\beta)/2}^{(1+\beta)/2}{\rm d}z\int_0^{y_+(z)}{\rm d}y\,y, \qquad (12.82)
$$

$$
y_+(z)=1-\frac{\hat{m}^2}{z(1-z)}.
$$

12.13. Making the dependence on $x_1$ and $x_2$ in $\sigma$ and $e$ explicit makes it easier to follow the subsequent computations. The dependence on $\cos\theta_T$ is kept hidden for the sake of simplicity.

The complete discussion on the integration limits can be found in appendix F.3.3 and F.5. For simplicity, from now on we omit the integration limits. Coming back to the expansion in $\epsilon$, the first observation is that the expansion of $F(y,z,\epsilon)$ has an overall factor of $y^{-2\epsilon}$,

$$\begin{aligned} F(y,z,\epsilon) &= y^{-2\epsilon}\left\{\frac{\alpha_s C_F}{\pi}\frac{3e^{2\epsilon\gamma_E}(1-\cos\theta_T)^{-\epsilon}}{4^{3-\epsilon}\Gamma^2(1-\epsilon)}\left(\frac{Q^2}{\mu^2}\right)^{-2\epsilon}[(1-z)z-\hat{m}]^{-\epsilon}+\mathcal{O}(y)\right\} \\ &\equiv y^{-2\epsilon}\{F(z,\epsilon)+\mathcal{O}(y)\}. \end{aligned} \tag{12.83}$$

Next, the expansion of any $\mathrm{d}\bar{\sigma}_C^i/(\mathrm{d}x_1\mathrm{d}x_2\mathrm{d}\cos\theta_T)$ starts at $\mathcal{O}(1/y^2)$, so, when picking up the factor $y$ from the Jacobian, only the leading term remains divergent. We refer to this divergent term as the soft-limit contribution to the real-radiation cross-section. Explicitly,

$$\begin{aligned} yF(y,z,\epsilon)\frac{1}{\sigma_C^0}\frac{\mathrm{d}\bar{\sigma}_C^i(y,z)}{\mathrm{d}x_1\mathrm{d}x_2\mathrm{d}\cos\theta_T} &\equiv y^{1-2\epsilon}\left[\frac{F(z,\epsilon)}{y^2}\frac{1}{\sigma_C^0}\frac{\mathrm{d}\bar{\sigma}_C^{\mathrm{soft},i}(z)}{\mathrm{d}x_1\mathrm{d}x_2\mathrm{d}\cos\theta_T}+\mathcal{O}\left(\frac{1}{y}\right)\right] \\ &\equiv y^{-1-2\epsilon}\frac{1}{\sigma_C^0}\frac{\mathrm{d}\sigma_C^{\mathrm{soft},i}(z)}{\mathrm{d}x_1\mathrm{d}x_2\mathrm{d}\cos\theta_T}+\mathcal{O}(y^0), \end{aligned} \tag{12.84}$$

with the soft cross-section defined as

$$\frac{1}{\sigma_C^0}\frac{\mathrm{d}\sigma_C^{\mathrm{soft},i}(z)}{\mathrm{d}x_1\mathrm{d}x_2\mathrm{d}\cos\theta_T}=F(\epsilon,z)\frac{1}{\sigma_C^0}\frac{\mathrm{d}\bar{\sigma}_C^{\mathrm{soft},i}(z)}{\mathrm{d}x_1\mathrm{d}x_2\mathrm{d}\cos\theta_T}. \tag{12.85}$$

Now that we have isolated the IR divergences, we can subtract and add them,

$$\begin{aligned} \frac{1}{\sigma_C^0}\frac{\mathrm{d}\sigma_C^{\mathrm{real}}}{\mathrm{d}e\mathrm{d}\cos\theta_T} &= \sum_{i=1}^{3}\int\mathrm{d}y\mathrm{d}z\,y\frac{1}{\sigma_C^0}\frac{\mathrm{d}\sigma_C^{\mathrm{hard},i}(y,z)}{\mathrm{d}x_1\mathrm{d}x_2\mathrm{d}\cos\theta_T}\Theta_i^T\delta[e-\hat{e}(y,z)] \\ &+\sum_{i=1}^{3}\int\mathrm{d}y\mathrm{d}z\frac{y^{-1-2\epsilon}}{\sigma_C^0}\frac{\mathrm{d}\sigma_C^{\mathrm{soft},i}(z)}{\mathrm{d}x_1\mathrm{d}x_2\mathrm{d}\cos\theta_T}\Theta_i^T\delta[e-\hat{e}(y,z)], \\ \frac{1}{\sigma_C^0}\frac{\mathrm{d}\sigma_C^{\mathrm{hard},i}(y,z)}{\mathrm{d}x_1\mathrm{d}x_2\mathrm{d}\cos\theta_T} &\equiv \frac{1}{\sigma_C^0}\frac{\mathrm{d}\sigma_C^i(y,z)}{\mathrm{d}x_1\mathrm{d}x_2\mathrm{d}\cos\theta_T}-\frac{y^{-2-2\epsilon}}{\sigma_C^0}\frac{\mathrm{d}\sigma_C^{\mathrm{soft},i}(z)}{\mathrm{d}x_1\mathrm{d}x_2\mathrm{d}\cos\theta_T}, \end{aligned} \tag{12.86}$$

splitting the real cross-section into a IR-finite part, which we call the hard cross-section, and the IR-divergent soft cross-section[12.14]. Since it is IR and UV finite, the hard part can be computed in 4 dimensions, meaning that after integration it entirely contributes to the non-singular coefficient $F_e$. The soft part, being proportional to $y^{-1-2\epsilon}$, needs to be expanded around $\epsilon=0$ with the distribution identity (A.26), contributing to $F_e$ and $A_e^C$ and giving the entire plus coefficient $B_{\mathrm{plus}}^C$.

12.14. The names hard and soft are not completely rigorous, since the $\mathcal{O}(1)$ term in the soft expansion (12.84) also belongs to the soft limit, as is it independent on $y$. The way we have carried out the subtraction, this $\mathcal{O}(1)$ term contributes to the hard cross-section.

### 12.3.4 Real radiation III: soft part

The last step is to perform the integration on $y$ and $z$, which in our case we carry out by marginalizing the distribution to a given generic event-shape whose value is $\hat{e}(y, z)$. Let us focus on the soft part first, which in each thrust region can be written as

$$\frac{1}{\sigma_C^0}\frac{\mathrm{d}\sigma_C^{\mathrm{soft},i}(z)}{\mathrm{d}x_1\mathrm{d}x_2\mathrm{dcos}\,\theta_T} \equiv \frac{\alpha_s C_F}{\pi} P(Q,\epsilon)\, M_C^i(z,\epsilon), \tag{12.87}$$
$$P(Q,\epsilon) \equiv \frac{4^\epsilon e^{2\epsilon\gamma_E}(1-\cos\theta_T)^{-\epsilon}}{\Gamma^2(1-\epsilon)}\left(\frac{Q^2}{\mu^2}\right)^{-2\epsilon}$$

where the functions $M_C^i$ corresponding to each region, as well as their power expansions in $\epsilon$,

$$M_C^i(z,\epsilon) \equiv M_C^{i,(0)}(z) + \epsilon M_C^{i,(1)}(z) + \mathcal{O}(\epsilon^2), \tag{12.88}$$

can be found in appendix E.2. The expansion of the prefactor $P(Q,\epsilon)$ is

$$P(Q,\epsilon) = 1 + \Big(\log(4) - 2L_Q - L_\theta\Big)\epsilon + \mathcal{O}(\epsilon^2) \equiv 1 + P_1\epsilon + \mathcal{O}(\epsilon^2). \tag{12.89}$$

For the expansion of $y^{-1-2\epsilon}$ we employ the distribution identity (A.26) in the form

$$y^{-1-2\epsilon} = -\frac{1}{2\epsilon}\delta(y) + \left[\frac{1}{y}\right]_+ + \mathcal{O}(\epsilon). \tag{12.90}$$

With these ideas the soft distribution takes the form

$$\begin{aligned}\frac{1}{\sigma_C^0}\frac{\mathrm{d}\sigma_C^{\mathrm{soft}}(z)}{\mathrm{dedcos}\theta_T} &= \frac{\alpha_s C_F}{\pi} P(Q,\epsilon)\sum_{i=1}^{3}\int \mathrm{d}z\mathrm{d}y\, y^{-1-2\epsilon} M_C^i(z,\epsilon)\Theta_i^T(y,z)\delta[e-\hat{e}(y,z)] \\ &= \frac{\alpha_s C_F}{\pi}\Bigg\{-\frac{\delta(e-e_{\mathrm{min}})}{2}\int \mathrm{d}z\left[\left(P_1+\frac{1}{\epsilon}\right)M_C^{1,(0)}(z) + M_C^{1,(1)}(z)\right] \\ &\quad + \sum_{i=1}^{3}\int \mathrm{d}z\mathrm{d}y\, M_C^{i,(0)}(z)\left[\frac{1}{y}\right]_+\delta[e-\hat{e}(y,z)]\Bigg\}.\end{aligned} \tag{12.91}$$

The terms proportional to $\delta(y)$ have been integrated over $y$, setting $y=0$, which causes several simplifications:

- The event-shape takes its minimum value, $\hat{e}(0,z) = e_{\mathrm{min}}$, and is therefore independent on $z$.

- The projector $\Theta_3^T(0,z)=\theta(-\beta)$ vanishes, and so does its contribution. In regions 1 and 2, before setting $y=0$ one needs to realize

$$\Theta_1^T(y,z)=\theta(z-1/2)\Theta_1^T(y,z), \qquad \Theta_2^T(y,z)=\theta(1/2-z)\Theta_2^T(y,z). \tag{12.92}$$

  When $y=0$, $\Theta_1^T(0,z)=\Theta_2^T(0,z)=\theta(\beta)=1$ and, due to $M_C^{1,(i)}=M_C^{2,(i)}$ for $i=0$, 1, the contribution from the regions 1 and 2 sums as

$$\begin{aligned}&\frac{1}{2\epsilon}\sum_{i=1}^{3}\int \mathrm{d}z\mathrm{d}y\,\delta(y)\,[M_C^{1,(0)}(z,\epsilon)+\epsilon M_C^{1,(1)}(z,\epsilon)]\Theta_i^T(y,z)\delta[e-\hat{e}(y,z)] \quad (12.93)\\ =\;&\frac{\delta(e-e_{\min})}{2\epsilon}\int \mathrm{d}z\;[M_C^{1,(0)}(z,\epsilon)+\epsilon M_C^{1,(1)}(z,\epsilon)][\theta(z-1/2)+\theta(1/2-z)].\end{aligned}$$

  Since $z$ is integrated from $(1-\beta)/2$ to $(1+\beta)/2$, which always crosses $1/2$, the sum of the last Heaviside functions just renders 1. As punchline for further computations, one can assume $\Theta_1^T(0,z)+\Theta_2^T(0,z)=1$.

These ideas, together with the expansion in $\epsilon$, produce the $z$ integral proportional to $\delta(e-e_{\min})$. The terms with the plus distribution $[1/y]_+$ are kept in the double integral over $y$ and $z$. We seek to write (12.91) in the form (12.1), where the delta, plus and non-singular structures in $e-e_{\min}$ are explicit. Only the double integral in $y$ and $z$ remains, so we focus on it and denote it as

$$I\equiv\sum_{i=1}^{3}\int \mathrm{d}z\mathrm{d}y\,M_C^{i,(0)}(z)\,\Theta_i^T(y,z)\left[\frac{1}{y}\right]_+\delta[e-\hat{e}(y,z)]. \tag{12.94}$$

The $y$ integral in $I$ can be solved by following the first idea discussed in appendix A.4.2. It consists on adding and subtracting the term

$$K_e\equiv\sum_{i=1}^{3}\int \mathrm{d}z\mathrm{d}y\,M_C^{i,(0)}(z)\,\Theta_i^T(y,z)\left[\frac{1}{y}\right]_+\delta[e-\bar{e}(y,z)], \tag{12.95}$$

where $\bar{e}(y,z)$ is the linear $y$-expansion of $\hat{e}(y,z)$ near its minimum, $\bar{e}(y,z)=e_{\min}+y f_e(z)$. Note that $f_e(z)=\mathrm{d}\hat{e}(y,z)/\mathrm{d}y\big|_{y=0}>0$ since $\hat{e}(0,z)$ is a minimum. Since the integrand of $K_e$ reduces to that of the double integral in (12.91) for $y=0$, when both terms are subtracted the integrand is not singular at $y=0$ and the plus distribution can be treated as the function $1/y$,

$$I-K_e=\sum_{i=1}^{3}\int \mathrm{d}z\mathrm{d}y\frac{M_C^{i,(0)}(z)}{y}\,\Theta_i^T(y,z)\Big\{\delta[e-\hat{e}(y,z)]-\delta[e-\bar{e}(y,z)]\Big\}, \tag{12.96}$$

contributing only to the non-singular term. On the other hand, in the added $K_e$ term, the integral over $y$ can be solved with the delta function,

$$\delta[e-e_{\min}-f_e(z)y]=\frac{1}{f_e(z)}\delta\bigg(y-\frac{e-e_{\min}}{f_e(z)}\bigg), \tag{12.97}$$

to give

$$\begin{aligned} K_e &= \sum_{i=1}^{3}\int \mathrm{d}z\,\frac{M_C^{i,(0)}(z)}{f_e(z)}\,\Theta_i^T\bigg(\frac{e-e_{\min}}{f_e(z)},z\bigg)\bigg[\frac{f_e(z)}{e-e_{\min}}\bigg]_+\theta\bigg(y_+(z)-\frac{e-e_{\min}}{f_e(z)}\bigg) \\ &= \sum_{i=1}^{3}\int \mathrm{d}z\,M_C^{i,(0)}(z)\,\Theta_i^T\bigg(\frac{e-e_{\min}}{f_e(z)},z\bigg)\theta\bigg(y_+(z)-\frac{e-e_{\min}}{f_e(z)}\bigg) \\ &\quad\times\bigg\{-\log[f_e(z)]\delta(e-e_{\min})+\bigg[\frac{1}{e-e_{\min}}\bigg]_+\bigg\}, \end{aligned} \tag{12.98}$$

where in the second step we used (A.39) to extract $f_e(z)$ out of the plus distribution. In the term proportional to the delta we set $e=e_{\min}=0$, which leads to $\theta(y_+(z))=1$ due to $y_+(z)>0$ in the integration interval of $z$. Specifically, the term is

$$-\delta(e-e_{\min})\int \mathrm{d}z\log[f_e(z)]M_C^{1,(0)}(z), \tag{12.99}$$

were we used again $\Theta_3^T(0,z)=0$ and $\Theta_1^T(0,z)+\Theta_2^T(0,z)=1$. The term with the plus distribution is of the form $f(e-e_{\min})[1/(e-e_{\min})]_+$, with

$$f(e-e_{\min})\equiv\Theta_i^T\bigg(\frac{e-e_{\min}}{f_e(z)},z\bigg)\theta\bigg(y_+(z)-\frac{e-e_{\min}}{f_e(z)}\bigg), \tag{12.100}$$

and therefore hides a non-singular term, as evidenced by (A.35). It can then be put as

$$\begin{aligned} &\int \mathrm{d}z\,M_C^{1,(0)}(z)\bigg[\frac{1}{e-e_{\min}}\bigg]_+ \\ &+\sum_{i=1}^{3}\int \mathrm{d}z\,\frac{M_C^{i,(0)}(z)}{e-e_{\min}}\bigg[\Theta_i^T\bigg(\frac{e-e_{\min}}{f_e(z)},z\bigg)\theta\bigg(y_+(z)-\frac{e-e_{\min}}{f_e(z)}\bigg)-\Theta_i^T(0,z)\bigg]. \end{aligned} \tag{12.101}$$

Adding both contributions, $K_e$ is

$$
\begin{aligned}
K_e &\equiv \delta(e-e_{\min})K_e^\delta + \left[\frac{1}{e-e_{\min}}\right]K^{\text{plus}} + K_e^{\text{NS}}, \qquad (12.102)\\
K_e^\delta &= -\int \mathrm{d}z \log[f_e(z)] M_C^{1,(0)}(z),\\
K^{\text{plus}} &= \int \mathrm{d}z\, M_C^{1,(0)}(z),\\
K_e^{\text{NS}} &= \sum_{i=1}^{3} \int \mathrm{d}z \frac{M_C^{i,(0)}(z)}{e-e_{\min}}\left[\Theta_i^T\!\left(\frac{e-e_{\min}}{f_e(z)}, z\right)\theta\!\left(y_+(z) - \frac{e-e_{\min}}{f_e(z)}\right) - \Theta_i^T(0,z)\right].
\end{aligned}
$$

Accounting for all the results, the soft part of the cross-section reads

$$
\begin{aligned}
\frac{1}{\sigma_C^0}\frac{\mathrm{d}\sigma_C^{\text{soft}}}{\mathrm{d}e\mathrm{d}\cos\theta_T} &= \frac{\alpha_s C_F}{\pi}\left\{A_e^{\text{soft},C}\delta(e-e_{\min}) + B^{\text{soft},C}\left[\frac{1}{e-e_{\min}}\right]_+ + F_e^{\text{soft}}\right\}, \qquad (12.103)\\
A_e^{\text{soft},C} &= -\frac{1}{2}\int \mathrm{d}z \left\{\left(\frac{1}{\epsilon} - L_\theta + \log(4) - 2L_Q + 2\log[f_e(z)]\right)M_C^{1,(0)}(z) + M_C^{1,(1)}(z)\right\},\\
B^{\text{soft},C} &= \int \mathrm{d}z\, M_C^{1,(0)}(z),\\
F_e^{\text{soft}} &= \sum_{i=1}^{3} \int \mathrm{d}z\mathrm{d}y \frac{M_C^{i,(0)}(z)}{y}\,\Theta_i^T(y,z)\delta[e-\hat{e}(y,z)] - \frac{B^{\text{soft},C}}{e-e_{\min}}.
\end{aligned}
$$

The expressions for $A_e^{\text{soft},C}$ and $B^{\text{soft},C}$ can be immediately read off in (12.91) and (12.102). To arrive to the expression for $F_e^{\text{soft}}$, we realize $K_e^{\text{NS}}$ can be written as a double integral that matches that of (12.96):

$$
\begin{aligned}
K_e^{\text{NS}} &= \sum_{i=1}^{3} \int \mathrm{d}z\mathrm{d}y \frac{M_C^{i,(0)}(z)}{y}\Theta_i^T(y,z)\left[\frac{1}{f_e(z)}\delta\!\left(y - \frac{e-e_{\min}}{f_e(z)}\right) - \frac{y}{e-e_{\min}}\delta(y)\right]\\
&= \sum_{i=1}^{3} \int \mathrm{d}z\mathrm{d}y \frac{M_C^{i,(0)}(z)}{y}\Theta_i^T(y,z)\left[\delta[e-\bar{e}(y,z)] - \frac{y}{e-e_{\min}}\delta(y)\right]. \qquad (12.104)
\end{aligned}
$$

When combined with (12.96), the terms proportional to $\delta[e-\bar{e}(y,z)]$ cancel and one is left with

$$
F_e^{\text{soft}} = \sum_{i=1}^{3} \int \mathrm{d}z\mathrm{d}y \frac{M_C^{i,(0)}(z)}{y}\,\Theta_i^T(y,z)\left\{\delta[e-\hat{e}(y,z)] - \frac{y}{e-e_{\min}}\delta(y)\right\}, \qquad (12.105)
$$

whose second term integrates to $B^{\text{soft},C}/(e-e_{\min})$.

The coefficients $B^{\text{soft},C}$ and part of the $A_e^{\text{soft},C}$ are independent on the event-shape, so we also present their explicit values. The coefficients of the plus distribution for the vector and axial currents are

$$B^{\text{soft},V} = -\frac{3\beta}{8}\left[1+\frac{1+\beta^2}{2\beta}L_\beta\right]\left[(3-\beta^2)\,S_{\text{un}}+(1-\beta^2)\,S_{\text{or}}\right], \tag{12.106}$$
$$B^{\text{soft},A} = -\frac{3\beta^3}{4}\left[1+\frac{1+\beta^2}{2\beta}L_\beta\right]S_{\text{un}}\,.$$

For the delta coefficients we split

$$A_e^{\text{soft},C} \equiv A_{\text{fin},e}^{\text{soft},C} + A_{\text{fin}}^{\text{soft},C} + A_{\text{div}}^{\text{soft},C}\left(L_\theta - \frac{1}{\epsilon}\right), \tag{12.107}$$

where the event-shape dependent part is defined as

$$A_{\text{fin},e}^{\text{soft},C} \equiv -\int \log[f_e(z)]M_C^{1,(0)}(z). \tag{12.108}$$

On the one hand, the divergent part is

$$A_{\text{div}}^{\text{soft},C} = \frac{1}{2}\int \mathrm{d}z\, M_C^{1,(0)}(z) = \frac{1}{2}B^{\text{soft},C} \tag{12.109}$$

and exactly cancels the virtual contribution in (12.58). On the other hand, the finite, event-shape independent part takes the form

$$\begin{aligned} A_{\text{fin}}^{\text{soft},V} &= \frac{3\beta}{16}\left[2(\log(2)-L_Q)\left(1+\frac{\beta^2+1}{2\beta}L_\beta\right)-g_{\text{soft}}(\beta)\right]\left[(3-\beta^2)S_{\text{un}}+(1-\beta^2)S_{\text{or}}\right] \\ &\quad -\frac{3\beta}{16}\left[\left(3+\frac{\beta^2+9}{2\beta}L_\beta\right)S_{\text{un}}+\left(1+\frac{3-\beta^2}{2\beta}L_\beta\right)S_{\text{or}}\right], \\ A_{\text{fin}}^{\text{soft},V} &= \frac{3\beta^3}{8}\left[2(\log(2)-L_Q)\left(1+\frac{\beta^2+1}{2\beta}L_\beta\right)-g_{\text{soft}}(\beta)\right]S_{\text{un}} \\ &\quad -\frac{3\beta^3}{16}\left[\left(3+\frac{3\beta^2+7}{2\beta}L_\beta\right)S_{\text{un}}+\left(1+\frac{\beta^2+1}{2\beta}L_\beta\right)S_{\text{or}}\right], \end{aligned} \tag{12.110}$$

with

$$g_{\text{soft}}(\beta) \equiv \frac{\beta^2+1}{2\beta}\left(2\text{Li}_{2\beta}+\frac{1}{2}L_\beta^2+\log(\beta^2)L_\beta\right)+\log(\beta^2). \tag{12.111}$$

The solution to the integrals of $M_C^{1,(1)}(z)$ can be found in appendix E.2.

### 12.3.5 Real radiation IV: hard part

The hard cross-section is obtained by subtracting the soft limit of the real distribution (see (12.86)). The subtraction renders it finite, so it can be directly computed in 4 dimensions. To compactly express the results, and since the oriented contribution is the same in all regions, we split the contributions to each thrust region as

$$\begin{aligned}\frac{1}{\sigma_C^0}\frac{\mathrm{d}\sigma_C^{\mathrm{hard}}(y,z)}{\mathrm{d}x_1\mathrm{d}x_2\,\mathrm{d}\cos\theta_T} &\equiv \frac{\alpha_s C_F}{\pi}\sum_{i=1}^{3}\frac{1}{\sigma_C^0}\frac{\mathrm{d}\bar{\sigma}_C^{\mathrm{hard},i}(y,z)}{\mathrm{d}x_1\mathrm{d}x_2\,\mathrm{d}\cos\theta_T}\Theta_i^T \\ &\equiv \frac{\alpha_s C_F}{\pi}\left[h_{\mathrm{hard}}^C(y,z)S_{\mathrm{un}} + \sum_{i=1}^{3}h_{\mathrm{hard}}^{C,i}(y,z)\Theta_i^T S_{\mathrm{or}}\right],\end{aligned} \tag{12.112}$$

where the unoriented contribution is the same in all three regions:

$$\begin{aligned} h_{\mathrm{hard}}^V(y,z) &\equiv \frac{3}{16}\frac{y[1-2z(1-z)]-2-4\hat{m}^2}{yz(1-z)}, \\ h_{\mathrm{hard}}^A(y,z) &\equiv \frac{3}{16}\frac{y[1+2\hat{m}^2-2z(1-z)]-2+8\hat{m}^2}{16yz(1-z)}.\end{aligned} \tag{12.113}$$

The oriented ones take the form

$$\begin{aligned} h_{\mathrm{hard}}^{V,1}(y,z) &= \frac{3[yz^2(1-z)(1-y)-\hat{m}^2z(2-y(1-2z^2))+2\hat{m}^4(y(1-z)+4z)]}{8yz^2(1-z)[(1-y(1-z))^2-4\hat{m}^2]}, \\ h_{\mathrm{hard}}^{V,2}(y,z) &= \frac{3\big[yz(1-y)(1-z)-\hat{m}^2(2+y(1-2z(2-z)))+2\hat{m}^4\big(4+\frac{yz}{1-z}\big)\big]}{8yz(1-z)[(1-yz)^2-4\hat{m}^2]}, \\ h_{\mathrm{hard}}^{V,3}(y,z) &= -\frac{3}{4y}, \\ h_{\mathrm{hard}}^{A,1}(y,z) &= \frac{3[(1-y)z^2+\hat{m}^2z(y^2(1-z)-2z(1-y)-2)+2\hat{m}^4]}{8z^2[(1-y(1-z))^2+4\hat{m}^2]}, \\ h_{\mathrm{hard}}^{A,2}(y,z) &= \frac{3\,[(1-y)(1-z)^2+\hat{m}^2(1-z)(2z-(2-y)(2+yz))+2\hat{m}^4]}{8(1-z)^2[(1-y\,z)^2-4\hat{m}^2]}, \\ h_{\mathrm{hard}}^{A,3}(y,z) &= \frac{3\,[-2z(1-z)+\hat{m}^2(2+y)]}{8yz(1-z)}.\end{aligned} \tag{12.114}$$

With this, the distribution $F_e^{\mathrm{hard},C}$ can be written as

$$\begin{aligned} F_e^{\mathrm{hard},C} &= S_{\mathrm{un}}\int \mathrm{d}z\mathrm{d}y\,h_{\mathrm{hard}}^C(y,z)\delta[e-\hat{e}(y,z)] \\ &\quad + S_{\mathrm{or}}\sum_{i=1}^{3}\int \mathrm{d}z\mathrm{d}y\,h_{\mathrm{hard}}^{C,i}(y,z)\Theta_i^T\delta[e-\hat{e}(y,z)],\end{aligned} \tag{12.115}$$

where the integrals depend on the specific event-shape under consideration.

## 12.4 Total coefficients

Lastly, we collect the complete expressions for the coefficients of the cross-section at NLO. For the coefficient of the delta we have $R_C^0$ at LO and $A_e^C = A_e^{\text{soft},C} + A^C$ at NLO:

$$
\begin{aligned}
R_V^0 &= \frac{3\beta}{16}[(3-\beta^2)S_{\text{un}} + (1-\beta^2)S_{\text{or}}], \qquad (12.116)\\
R_A^0 &= \frac{3\beta^3}{8} S_{\text{un}},\\
A^V &= \frac{3\beta}{16}\left\{\left[\log(\hat{m})\left(1+\frac{\beta^2+1}{2\beta}L_\beta\right) - g(\beta)\right]\left[(3-\beta^2)S_{\text{un}} + (1-\beta^2)S_{\text{or}}\right] - \beta L_\beta S_{\text{un}}\right\},\\
A^A &= \frac{3\beta^3}{8}\left[\log(\hat{m})\left(1+\frac{\beta^2+1}{2\beta}L_\beta\right) - g(\beta) - \frac{2-\beta^2}{2\beta}L_\beta\right]S_{\text{un}},\\
A_e^{\text{soft},C} &= -\int \log[f_e(z)] M_C^{1,(0)}(z),
\end{aligned}
$$

where

$$
g(\beta) \equiv \frac{\beta^2+1}{2\beta}(4\text{Li}_{2\beta} + L_\beta^2 + 2L_\beta - \pi^2) + 2. \qquad (12.117)
$$

We stress that the event-shape dependent term in $A_e^C$ comes entirely from the soft limit of the real radiation while the event-shape independent part takes contributions from both virtual and the soft limit of real radiation.

The plus-distribution is preceded by the coefficient $B^C = B^C_{\text{soft}}$ that is event-shape independent and only takes a contribution from the soft limit of real radiation:

$$
\begin{aligned}
B^V &= -\frac{3\beta}{8}\left[1+\frac{1+\beta^2}{2\beta}L_\beta\right]\left[(3-\beta^2)\,S_{\text{un}} + (1-\beta^2)\,S_{\text{or}}\right], \qquad (12.118)\\
B^A &= -\frac{3\beta^3}{4}\left[1+\frac{1+\beta^2}{2\beta}L_\beta\right]S_{\text{un}}\,.
\end{aligned}
$$

Finally, the non-singular distribution $F^C = F_e^{\text{soft},C} + F_e^{\text{hard},C}$ takes contributions from the soft and hard real radiation:

$$
\begin{aligned}
F_e^{\text{soft},C} &= \sum_{i=1}^{3}\int \text{d}z\text{d}y \frac{M_C^{i,(0)}(z)}{y}\,\Theta_i^T(y,z)\delta[e-\hat{e}(y,z)] - \frac{B^C}{e-e_{\text{min}}}, \qquad (12.119)\\
F_e^{\text{hard},C} &= S_{\text{un}}\int \text{d}z\text{d}y\, h^C_{\text{hard}}(y,z)\delta[e-\hat{e}(y,z)]\\
&\quad + S_{\text{or}}\sum_{i=1}^{3}\int \text{d}z\text{d}y\, h^{C,i}_{\text{hard}}(y,z)\Theta_i^T\delta[e-\hat{e}(y,z)].
\end{aligned}
$$

We also observe that for the axialvector current the oriented contributions to both $A_e^A$ and $B^A$ vanish, while $F_e^A$ does present oriented and unoriented contributions.

## 12.5 Numerical analysis

Our results contain the analytic expressions of all the event-shape independent contributions to the oriented cross-section at NLO. They are only dependent on the ratio $m/Q$ between the quark mass and the CM energy. The event-shape dependent contributions require the event shape measurement function $\hat{e}(y,z)$ and its linear expansion coefficient $f_e(z)$ since they are given as phase-space integrals involving these functions. They can be found in [34] for a large number of event-shapes.

The event-shape independent coefficients $R^C$ and $B^C$ are shown in figure 12.3 as a function of the reduced mass $\hat{m}$.

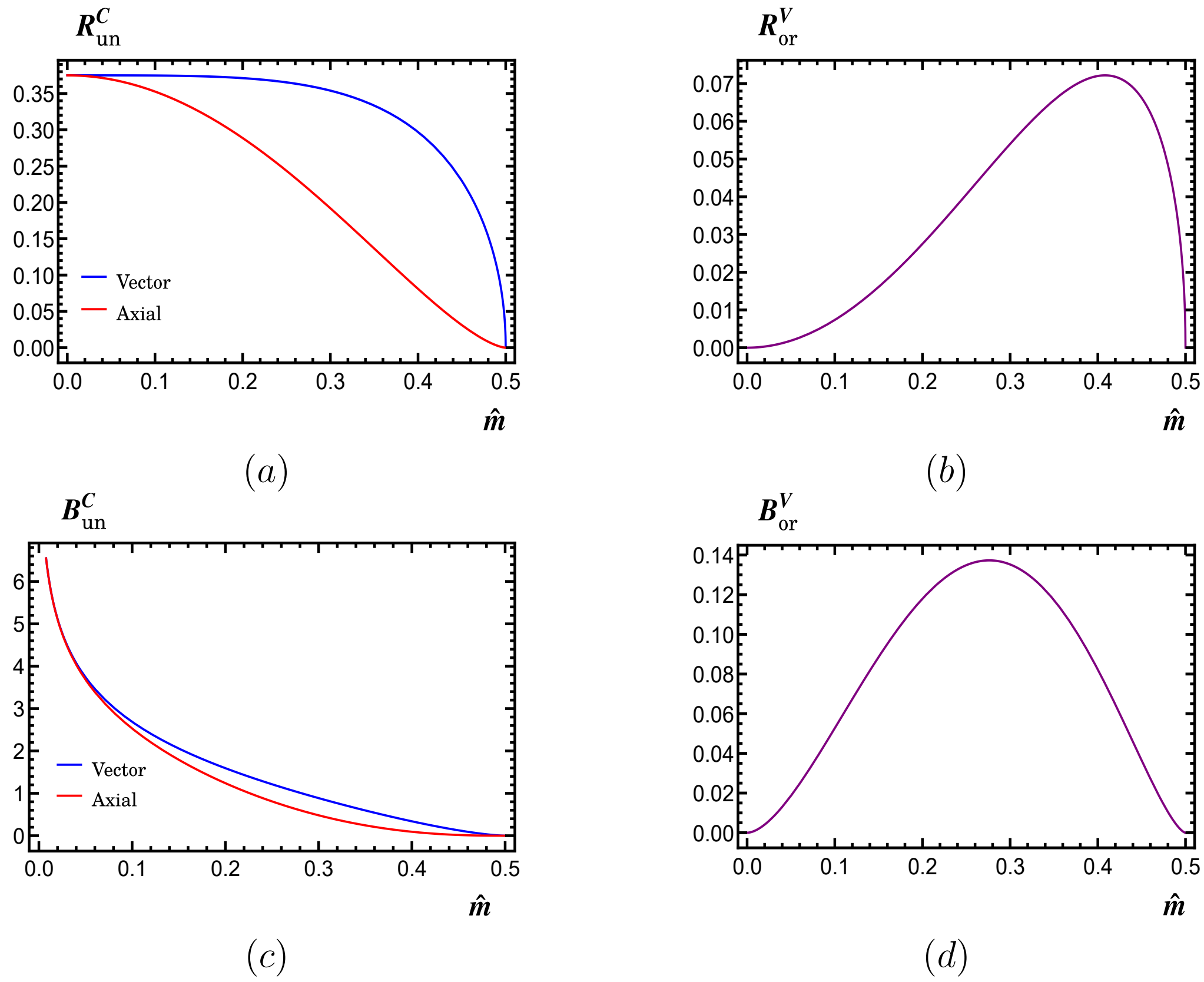

**Figure 12.3.** Coefficients $R^C$ (delta, LO) and $B^C$ (plus distribution, NLO) of the oriented event-shape cross section. Panels $(a)$ and $(c)$ contain the contributions to the unoriented cross-section, and panels $(b)$ and $(d)$ those to the oriented cross-section.

We observe the axialvector current has a bigger contribution to the unoriented cross-section for all $\hat{m}$ except in the massless limit, where both currents agree. In the massless limit $R_{\rm or}^V$ vanishes and the oriented event-shape cross-section starts at $O(\alpha_s)$. We see the mass corrections for the vector current start at tree-level. The plus distribution coefficient reproduces that found in [34] by universality as $B^C = \lim\limits_{e\to e_{\min}} (e - e_{\min}) F_e^C$. The coefficients of the oriented cross-section have no contribution from the axialvector current, and that of the vector is of considerably smaller size.

In figure 12.4 we show the NLO delta coefficient $A_e^C$, which is dependent on the event-shape. We have selected thrust and heavy jet mass and used

$$f_\tau(z) = \frac{2z + \beta - 1}{2\beta^2}, \qquad f_\rho(z) = z, \tag{12.120}$$

which are directly obtained from the functions

$$\begin{aligned} \tau(y, z) &= \sqrt{(1 - y(1 - z))^2 - 4\hat{m}^2} - \sqrt{(1 - yz)^2 - 4\hat{m}^2} - y, \\ \rho(y, z) &= yz + \hat{m}^2. \end{aligned} \tag{12.121}$$

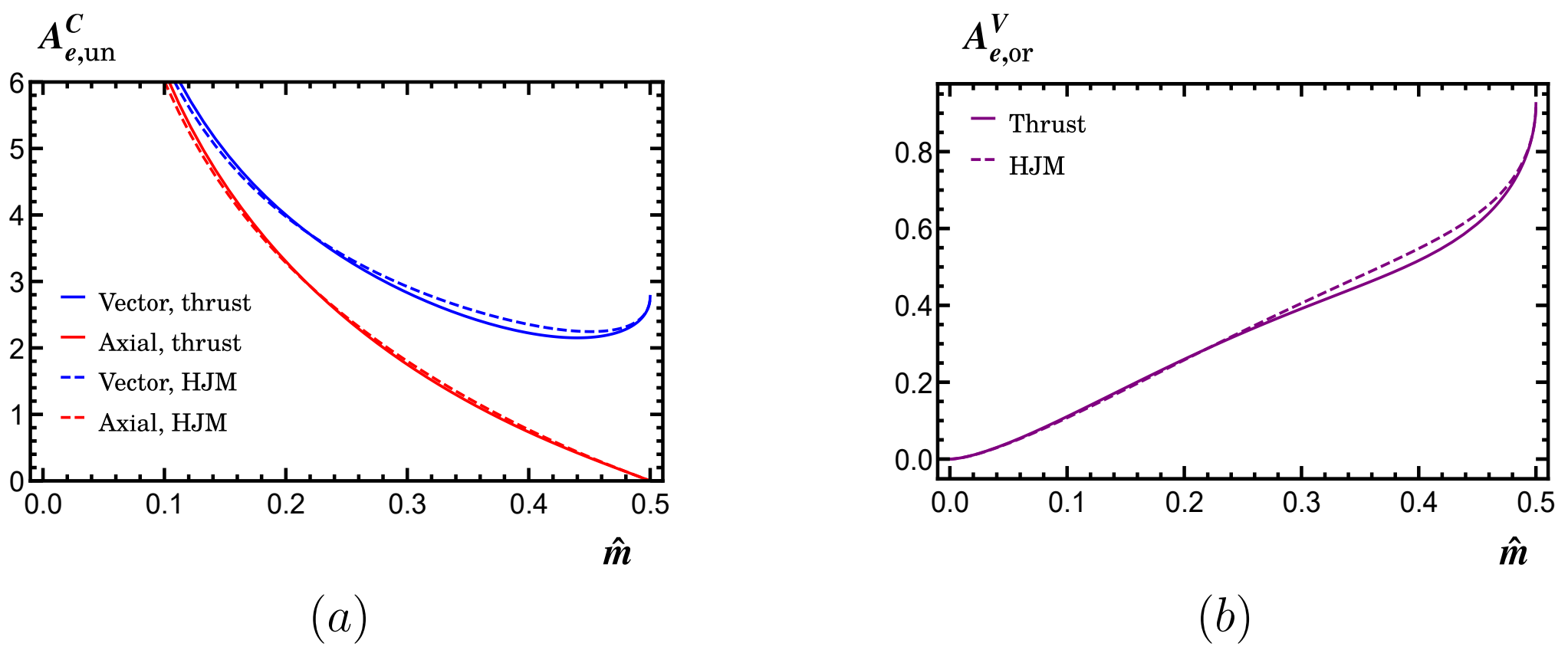

**Figure 12.4.** Coefficient $A_e^C$ of the delta of the NLO event-shape oriented distribution for the event-shapes thrust ($\tau$) and heavy jet mass ($\rho$). Panel ($a$) contains the contributions to the unoriented cross-section, and panel ($b$) those to the oriented cross-section

As before, the axialvector contribution is bigger in the unoriented cross-section and vanishes in the oriented. For both event-shapes, the $A_{e,\rm un}^C$ diverges in the massless limit, which signals that for $\hat{m} = 0$ the cross-section at NLO contains a term with $[\ln(e - e_{\min}) / (e - e_{\min})]_+$ ($e_{\min} = 0$ when $m = 0$) The contribution from the oriented cross-section is also again smaller, vanishes in the massless limit and is maximal at $\hat{m} = 0.5$.

In figure 12.5 we show the event-shape oriented cross-section as a function of the event-shape for thrust ($\tau$) and heavy jet mass ($\rho$).

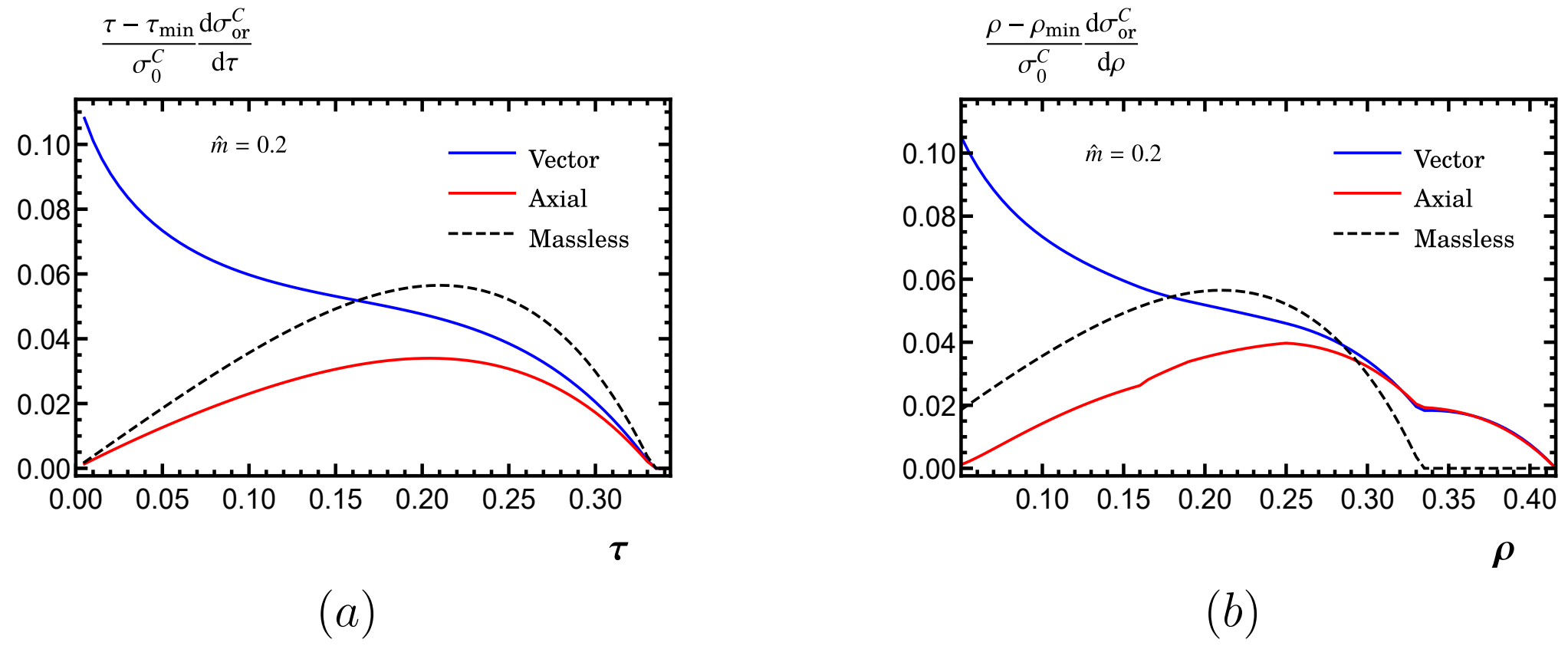

**Figure 12.5.** Oriented cross-sections for thrust (panel $(a)$) and heavy jet mass (panel $(b)$).

The chosen event shapes have a range $(\tau_{\min}, \tau_{\max}) = (0, 1/3)$ and $(\rho_{\min}, \rho_{\max}) = (\hat{m}^2, (5-4r)/3)$, where $r \equiv \sqrt{1-3\hat{m}^2}$. To compute the non-singular distribution $F_e^C$ we have applied the numerical algorithm developed in [34], which integrates the double phase-space integration by solving the delta function for $\hat{e}(y,z)$ and figuring out the integration limits, in both cases numerically, along with the integration itself.

Finally, in figure 12.5 we integrated over the event-shape and show the cross-section as a function of the reduced mass $\hat{m}$. The axialvector current vanishes at the threshold level $\hat{m} = 0.5$ but, interestingly, the vector current is enhanced. This opens up the possibility of determining the top quark's mass from threshold scans.

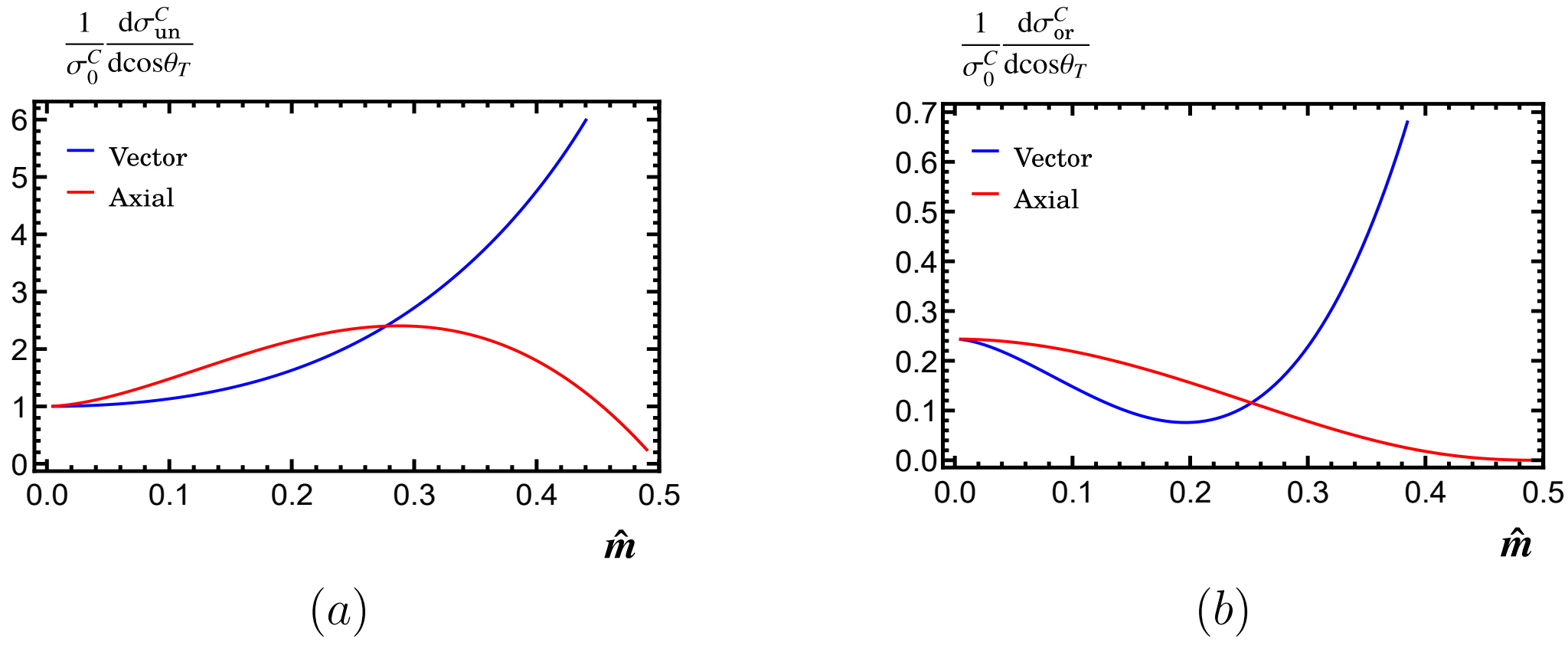

**Figure 12.6.** Total (panel $(a)$) and oriented (panel$(b)$) cross-sections.

# Conclusions and further work

In this thesis we have developed a formalism to sum perturbative series in the large-$\beta_0$ limit of QCD and systematically find their renormalons, ambiguity and (no cusp and cusp) anomalous dimension. Its application to the short-distance masses $\overline{m}$ and $m^{\rm MSR}$ and the pieces in the SCET and bHQET factorization theorems $H_Q$, $H_m$ $J_n$ and $B_n$ yielded their exact values and ambiguities, the complete set of their renormalons and their perturbative coefficients. We found matrix elements are all divergent while anomalous dimensions are all convergent. In all cases, the re-expansion of our closed results in powers of $\alpha_s$ finds agreement with the state-of-the-art perturbative coefficients. In particular, it is observed that $\pi$ resummation in $H_Q$ and the leading renormalon in $H_m$, not accounted for in [79], might improve the convergence found in peak cross sections for boosted top pairs. Although $\gamma_s$ is found from a consistence condition, the soft function in the factorization theorems yet remains to be addressed. Doing so might constitute a future study that opens up the possibility to a formulation of the resummed complete SCET and bHQET factorization theorems.

Regarding the problem of asymptotic separation in the moments of the spectral $\tau$ functions, we have studied the gluon condensate model in the large-$\beta_0$ limit, in which FOPT and CIPT present very different behavior. We found the moments $l \neq 2$ in FOPT converge when the value of the expansion parameter $a$ lays in a circle of radius $1/\pi$ centered at the origin of the complex plane. However, all moments in CIPT diverge with zero convergence radius. Asymptotic separation has been analytically computed as the difference $\delta^{(0)}_{{\rm FO},l} - \delta^{(0)}_{{\rm CI},l,B}$ for complex $a$ when $l = 0$ and for $a > 0$ for $l \neq 0$. All these results have been derived from analytic expressions of the $\delta^{(0)}_{{\rm FO},l}$ and $\delta^{(0)}_{{\rm CI},l}$ perturbative series and recover the results in [97], which are derived from the resummed expressions of the Adler function. The renormalon-free scheme in [98] has also been implemented in our model, yielding a cancellation of the $u = 2$ renormalon which makes RFCIPT converge towards the value in FOPT. We showed explicitly the RFCIPT converges for real $a$ and sums up to FOPT. We have made the important observation that the expansion of any of the powers in $a^n$ in the functions $H_{n,l}(a)$ of CIPT has zero convergence radius. The formal proof of this statement from explicit expressions for this expansion has been worked out for the $l=0$ case and checked numerically for $l>0$. The conclusion is that any truncated series in the FOPT scheme is transformed to a divergent series in the CIPT scheme. For the renormalon-free CIPT we found numerical evidence that the expansion in $H_{n,l}^{\mathrm{RF}}(a)$ has a finite convergence radius that is in any case smaller than $a(m_\tau^2)$. At the moment this thesis is written, this issue is still being explored and further work is being done on understanding the expansions in $H_{n,l}^{(\mathrm{RF})}(a)$. We remark that neither CIPT nor its renormalon-free counterpart are true power series in $a$, but instead they are expansions on the sequence of functions $H_{n,l}^{(\mathrm{RF})}(a)$.

Lastly, we explored the oriented event-shape cross-section for $e^+e \longrightarrow$ hadrons for massive quarks at NLO. The coefficients $R^C$, $A_e^C$, $B^C$ and $F_e^{\mathrm{NS}}$ of its distribution structure have been computed for an arbitrary event-shape for both the unoriented and oriented structures (see [33]). The unoriented results serve as a cross-check with [34], and the oriented results are novel. We find that mass corrections make the oriented contributions start already at tree-level for the vector current, something that was not seen for massless quarks or for the axial current. The observable dependence in the analytic expressions is encoded in the event-shape measurement function $\hat{e}(y,z)$ and the linear coefficient of its expansion at $y=0$, $f_e(z)$. These functions can be computed for any event-shape ([34] lists large number of them) and plugged back into our results to get numerical results for the distribution. This makes it easy to implement and study different event-shapes in different schemes to test sensitivity to the quark's mass. In this thesis we illustrated this by choosing two event-shapes, thrust and heavy jet mass –both in their defining original schemes–, but currently a more complete study, which shows some of the event-shapes can be analytically computed, is being carried out and prepared for publication.

# Appendix A
# Formulae from charts

## A.1 Special functions

The computations carried out in this thesis involve non-elemental functions. In this appendix we collect their definitions and some of its most basic properties. For a complete list of properties of these functions, such as relation to equivalent functions or asymptotic series representations, the reader is referred to [103] and [104].

### A.1.1 Gamma functions and Pochammer symbol

The gamma function $\Gamma(z)$ is the generalization of the factorial $n!$ to complex numbers, as it is the complex-valued function that satisfies $\Gamma(z+1)=z\Gamma(z)$. Its definition for $\mathrm{Re}(z>0)$ is

$$\Gamma(z)\equiv\int_0^\infty \mathrm{d}z e^{-t}t^{z-1}, \tag{A.1}$$

and extends to $\mathrm{Re}(z)\leq 0$ by analytic continuation. $\Gamma(z)$ is analytical in the complex plane except in the negative integers, where it presents poles. The lower and upper incomplete gamma functions $\Gamma(z,a)$ and $\gamma(z,a)$ are defined by restricting the integration interval as

$$\Gamma(z,a)\equiv\int_a^\infty \mathrm{d}z\, t^{z-1}e^{-t}, \qquad \gamma(z,a)\equiv\int_0^a \mathrm{d}z\, t^{z-1}e^{-t}, \tag{A.2}$$

and satisfy $\Gamma(z)=\gamma(z,a)+\Gamma(z,a)$. They also satisfy the recurrence relations:

$$\begin{aligned}\Gamma(z+1,a)&=z\Gamma(z,a)+e^{-a}a^z,\\ \gamma(z+1,a)&=z\gamma(z,a)-e^{-a}a^z.\end{aligned} \tag{A.3}$$

Finally, the doubly incomplete gamma function $\Gamma(z,a,b)$ is

$$\Gamma(z,a,b) \equiv \int_a^b \mathrm{d}z\, t^{z-1} e^{-t}, \tag{A.4}$$

and can be written as the subtraction $\Gamma(z,a,b) = \Gamma(z,a) - \Gamma(z,b)$. Its recurrence relation can be written immediately from (A.3):

$$\Gamma(z+1,a,b) = z\Gamma(z,a) + e^{-a}a^z - e^{-b}b^z. \tag{A.5}$$

The regularized version of each incomplete gamma function is defined by dividing by $\Gamma(z)$ and denoted with a tilde:

$$\tilde{\Gamma}(z,a) \equiv \frac{\Gamma(z,a)}{\Gamma(z)}, \qquad \tilde{\gamma}(z,a) \equiv \frac{\gamma(z,a)}{\Gamma(z)}, \qquad \tilde{\Gamma}(z,a,b) \equiv \frac{\Gamma(z,a,b)}{\Gamma(z)}. \tag{A.6}$$

The regularized versions are sometimes used in numerical computations as their factorial growth is tamed by the denominator.

Related to $\Gamma(z)$, the Pochammer symbol is defined as

$$(a)_b \equiv \frac{\Gamma(a+b)}{\Gamma(a)}, \tag{A.7}$$

for $a, b \in \mathbb{C}$.

### A.1.2 Exponential integral

The incomplete exponential integral is defined for $x > 0$ as

$$\mathrm{Ei}(x) \equiv -\mathrm{P.V.}\int_{-x}^{\infty} \frac{e^{-t}}{t}, \quad \mathrm{Ei}(-x) \equiv -\int_{x}^{\infty} \frac{e^{-t}}{t}, \tag{A.8}$$

where the integral in $\mathrm{Ei}(x)$ includes the principal value prescription to regularize the pole at $t = 0$.

### A.1.3 Hypergeometric function

The generalized Hypergeometric function is defined as the series

$$ {}_pF_q(a_1,...a_p,b_1,...b_q,z) = \sum_{k=0}^{\infty} \frac{\prod_{j=1}^{p}(a_j)_k\, z^k}{\prod_{j=1}^{q}(b_j)_k\, k!}, \tag{A.9} $$

and its regularized version is

$$ {}_p\tilde{F}_q(a_1,...a_p,b_1,...b_q,z) = \frac{{}_pF_q(a_1,...a_p,b_1,...b_q,z)}{\prod_{j=1}^{q}\Gamma(b_j)}. \tag{A.10} $$

### A.1.4 Bernoulli polynomials and numbers

Bernoulli polynomials arise from the expansion

$$ \frac{te^{xt}}{e^t-1} \equiv \sum_{n=0}^{\infty} B_n(x)\frac{t^n}{n!}, \tag{A.11} $$

and the Bernoulli numbers are $B_n \equiv B_n(0)$. A prominent property of Bernoulli numbers is that all odd Bernoulli numbers except $n=1$ vanish:

$$ B_{2n+1} = \begin{cases} 1, & n=1 \\ 0, & n>1 \end{cases}. \tag{A.12} $$

### A.1.5 Harmonic numbers

The harmonic numbers $H_n$ are defined as

$$ H_n \equiv \sum_{i=1}^{n} \frac{1}{i}. \tag{A.13} $$

They admit the sum representations

$$ H_n = \sum_{i=1}^{n} \binom{n}{i} \frac{(-1)^{i+1}}{i} = \sum_{i=1}^{n} \frac{(-1)^{i+1}\Gamma(n+1)}{\Gamma(i)\Gamma(n-i+1)i^2} \tag{A.14} $$

as well as the integral representation

$$H_n = \int_0^1 \mathrm{d}\tau \frac{1-\tau^n}{1-\tau}. \tag{A.15}$$

# A.2 Expansions

## A.2.1 Binomial expansion

In its most general expression, the expansion of the binomial $(x+a)^\alpha$ accounts for $x, a, \alpha \in \mathbb{C}$ and takes the form

$$(x+a)^\alpha = \sum_{i=0}^{\infty} \binom{\alpha}{i} x^i a^{\alpha-i} = \sum_{i=0}^{\infty} \frac{\Gamma(\alpha+1)}{\Gamma(i+1)\Gamma(\alpha-i+1)} x^i a^{\alpha-i}. \tag{A.16}$$

However, there are more convenient expressions for particular cases:

- When $\alpha = n$, for $n \in \mathbb{N}$:

$$(x+a)^n = \sum_{i=0}^{n} \binom{n}{i} x^i a^{n-i} = \sum_{i=0}^{n} \frac{\Gamma(n+1)}{\Gamma(i+1)\Gamma(n-i+1)} x^i a^{n-i}. \tag{A.17}$$

- When $\alpha = -r$, for $r \in \mathbb{R}^+$:

$$(x+a)^{-r} = \sum_{i=0}^{\infty} (-1)^i \binom{r+i-1}{i} x^i a^{-r-i} = \sum_{i=0}^{\infty} \frac{(-1)^i \Gamma(r+i)}{\Gamma(r)\Gamma(i+1)} x^i a^{-r-i}. \tag{A.18}$$

## A.2.2 Other expansions

Monomial and gamma expansions:

$$1+z = \exp\left\{ -\sum_{n=1}^{\infty} \frac{(-1)^n z^n}{n} \right\} . \tag{A.19}$$

$$\Gamma(1+z) = \exp\left\{ -\gamma_E z + \sum_{n=2}^{\infty} (-1)^n \frac{\zeta_n}{n} z^n \right\}, \tag{A.20}$$

To carry out the expansion of the exponential of a series we use the following recursive relation, which is derived imposing $\exp\{f(x)\}' = f'(x)\exp\{f(x)\}$:

$$\exp\left\{\sum_{i=1}^{\infty} a_i x^i\right\} = 1 + \sum_{i=1}^{\infty} b_i x^i, \qquad b_i = \frac{1}{i}\sum_{j=0}^{i-1} (j+1) b_{i-j-1} a_{j+1}. \tag{A.21}$$

with $b_0 = 1$.

## A.3 Leibniz integral rule

Let $f(x,t)$ be a function such that both $f(x,t)$ and its partial derivative $f_x(x,t)$ are continuous in $t$ and $x$ in some region of the $(x,t)$-plane, including $a(x) \leq t \leq b(x)$, $x_0 \leq x \leq x_1$. Also suppose that the functions $a(x)$ and $b(x)$ are both continuous and both have continuous derivatives for $x_0 \leq x \leq x_1$. Then,

$$\frac{\mathrm{d}}{\mathrm{d}x}\left[\int_{a(x)}^{b(x)} f(x,t)\mathrm{d}t\right] = f(x,b(x))\frac{\mathrm{d}}{\mathrm{d}x}b(x) - f(x,a(x))\frac{\mathrm{d}}{\mathrm{d}x}a(x) + \int_{a(x)}^{b(x)} \frac{\partial}{\partial x} f(x,t)\mathrm{d}t. \tag{A.22}$$

## A.4 Plus distribution

### A.4.1 Definition

Distribution theory generalizes the notion of a function, and, in its most rigorous sense, it constitutes a branch of mathematics. With regards to this thesis' scope, it is enough to consider that distributions are objects $\rho(x)$ that, instead of being defined point-wise as maps between number sets, are defined by their integral action on a test function[A.1],

$$\int_{a(t)}^{b(t)} \mathrm{d}x \rho(x) f(x) \equiv F_\rho[f(t)], \tag{A.23}$$

A.1. Test functions are well-behaved functions. Specifically, they are usually defined as infinitely differentiable functions with compact support.

where $f(x)$ is the test function and $F_\rho$ is the defining prescription. The canonical example is Dirac's delta distribution $\delta$, which can be defined as

$$\int_{-\infty}^{\infty} \mathrm{d}x\, \delta(x) f(x) \equiv f(0). \tag{A.24}$$

In the traditional sense of a function, there is no function $\delta(x)$ that satisfies (A.24)[A.2]. As a consequence, $\delta(x)$ is not defined point-wise on a set of numbers but instead through its integral action on a test function.

A second example is the plus distribution $[\log^n(x)/x]_+$, which arises in the expansion of $x^{-1+\epsilon}$ around $\epsilon = 0$,

$$x^{-1+\epsilon} = \frac{1}{\epsilon}\delta(x) + \sum_{n=0}^{\infty} \frac{\epsilon^n}{n!}\left[\frac{\log^n(x)}{x}\right]_+ = \frac{1}{\epsilon}\delta(x) + \left[\frac{1}{x}\right]_+ + \mathcal{O}(\epsilon), \tag{A.26}$$

and is defined as

$$\int_0^t \mathrm{d}x \left[\frac{\log^n(x)}{x}\right]_+ f(x) \equiv \int_0^t \frac{f(x) - f(0)}{x} \log^n(x) + f(0)\frac{\log^n(t)}{n+1}. \tag{A.27}$$

To prove (A.26) we consider $x^{-1+\epsilon}$ as a distribution and manipulate it to

$$\int_0^A \mathrm{d}x\, x^{-1+\epsilon} f(x) = \int_0^A \mathrm{d}x\, x^{-1+\epsilon}[f(x) - f(0)] + f(0)\frac{A^\epsilon}{\epsilon}, \tag{A.28}$$

where we subtracted and added $x^{-1-\epsilon} f(0)$ and solved the integral corresponding to the added term. Now we expand both terms in the right-hand side at $\epsilon = 0$

$$\int_0^A \mathrm{d}x\, x^{-1+\epsilon}[f(x) - f(0)] = \int_0^A \mathrm{d}x\, \frac{f(x) - f(0)}{x} \sum_{n=0}^{\infty} \frac{\epsilon^n \log^n(x)}{n!}, \tag{A.29}$$

$$f(0)\frac{A^\epsilon}{\epsilon} = f(0)\frac{1}{\epsilon}\sum_{n=0}^{\infty}\frac{\epsilon^n \log^n(A)}{n!} = f(0)\left[\frac{1}{\epsilon} + \sum_{n=0}^{\infty}\frac{\epsilon^n \log^{n+1}(A)}{(n+1)!}\right],$$

A.2. In order for the definition to hold for all test functions, the only possibility is that $\delta(x) = 0$ everywhere except at $x = 0$, where $\delta(0) \neq 0$. If $\delta(0)$ were a finite number, the integration would yield zero. If $\delta(0)$ were infinity, one would have $a\delta(x) = \delta(x)$ for all $x$ and real, finite $a$, as 0 and $\infty$ both satisfy $a\cdot 0 = 0$ and $a\cdot\infty = \infty$. However, this is not compatible with the original definition, which states

$$\int_{-\infty}^{\infty} \mathrm{d}x\, [a\delta(x)] f(x) = a f(0). \tag{A.25}$$

Therefore, $\delta(x)$ cannot be defined as a point-wise function.

and combine them to get

$$\begin{aligned}\int_0^A dx\, x^{-1+\epsilon} f(x) &= \frac{f(0)}{\epsilon} + \sum_{n=0}^{\infty} \frac{\epsilon^n}{n!} \left\{ \int_0^A dx \left[ \frac{f(x)-f(0)}{x} \ln^n(x) \right] + f(0) \frac{\log^{n+1}(A)}{n+1} \right\} \\ &= \frac{1}{\epsilon} \int_0^A dx\, f(x)\delta(x) + \sum_{n=0}^{\infty} \frac{\epsilon^n}{n!} \left\{ \int_0^A dx\, f(x) \left[ \frac{\log^n(x)}{x} \right]_+ \right\} \\ &= \int_0^A dx\, f(x) \left[ \frac{1}{\epsilon}\delta(x) + \sum_{n=0}^{\infty} \frac{\epsilon^n}{n!} \left[ \frac{\log^n(x)}{x} \right]_+ \right]. \end{aligned} \tag{A.30}$$

This corresponds to (A.26).

### A.4.2 Properties

We first present a property that allows to treat the plus distribution $[\ln^n(x)/x]_+$ as the corresponding function $\log^n(x)/x$. Indeed, if $g(x)$ is a function such that $g(0)=0$, from (A.27) we immediately have

$$\int_0^t dx \left[ \frac{\log^n(x)}{x} \right]_+ [f(x) - f(g(x))] = \int_0^t \frac{f(x) - f(g(x))}{x} \ln^n(x). \tag{A.31}$$

This shows the plus distribution is equivalent to the function $\log^n(x)/x$ whenever the test function is such that $f(0)=0$. This idea proves useful when the test function includes a Dirac delta. Let us take $f(x)\delta(g(x))$ as our test function, where $g(0)=g_0$ and $g(x) = g_0 + g_0' x + \mathcal{O}(x^2)$ when expanded around $x=0$. In other words, let us consider

$$I \equiv \int_0^t dx \left[ \frac{\log^n(x)}{x} \right]_+ f(x)\delta(g(x)). \tag{A.32}$$

Before applying any distribution property, we add and subtract the term

$$K \equiv \int_0^t dx \left[ \frac{\log^n(x)}{x} \right]_+ f(x)\delta(g_0 - g_0' x), \tag{A.33}$$

whose test function equals $f(x)\delta(g(x))$ for $x=0$. Then we have

$$\begin{aligned} I &= \int_0^t dx \left[ \frac{\log^n(x)}{x} \right]_+ \Big[ f(x)\delta(g(x)) - f(x)\delta(g_0 + g_0' x) \Big] \\ &\quad + \int_0^t dx \left[ \frac{\log^n(x)}{x} \right]_+ f(x)\delta(g_0 + g_0' x) \\ &= \int_0^t dx \frac{f(x)\delta(g(x)) - f(0)\delta(g_0)}{x} \ln^n(x) + \left[ \frac{\log^n(x_0)}{x_0} \right]_+ f(x_0)\theta(t - x_0)\theta(x_0), \end{aligned} \tag{A.34}$$

where $x_0 = -g_0/g_0'$ is the root of $\delta(g_0 + g_0' x)$. By following this idea, we ended up with the integrand of a standard function and a distribution term where no integral over $x$ remains.

As a consequence of these idea we can see the important fact that a term of the form $f(x)[\log^n(x)/x]_+$ hides a non-singular contribution:

$$f(x)\left[\frac{\log^n(x)}{x}\right]_+ = f(0)\left[\frac{\log^n(x)}{x}\right]_+ + \frac{f(x)-f(0)}{x}\ln^n(x), \tag{A.35}$$

This be readily proven by writing

$$f(x)\left[\frac{\log^n(x)}{x}\right]_+ = f(0)\left[\frac{\log^n(x)}{x}\right]_+ - \Big[f(x)-f(0)\Big]\left[\frac{\log^n(x)}{x}\right]_+ \tag{A.36}$$

and treating the distribution in the subtracted term as $\ln^n(x)/x$ as discussed.

Next the work out the extraction of a constant $c > 0$ out of the plus and delta distributions. For the delta distribution, there is the well-known property

$$\delta(cx) = \frac{\delta(x)}{c}. \tag{A.37}$$

For the plus distribution the equivalent result is

$$\left[\frac{\log^n(cx)}{cx}\right]_+ = \frac{1}{c}\left\{\delta(x)\frac{\log^{n+1}(c)}{n+1} + \sum_{m=0}^{n}\binom{n}{m}\log^m(c)\left[\frac{\log^{n-m}(x)}{x}\right]_+\right\}, \tag{A.38}$$

which leads to the particular cases

$$\begin{aligned} \left[\frac{1}{cx}\right]_+ &= \frac{1}{c}\delta(x)\log(c) + \frac{1}{c}\left[\frac{1}{x}\right]_+, \\ \left[\frac{\log(cx)}{cx}\right]_+ &= \frac{1}{2c}\delta(x)\log^2(c) + \frac{1}{c}\left[\frac{\log(x)}{x}\right]_+ + \frac{1}{c}\left[\frac{1}{x}\right]_+. \end{aligned} \tag{A.39}$$

These results can be derived from the expansion of $x^{-1+\epsilon}$. Consider the action of $(cx)^{-1+\epsilon}$ around $\epsilon = 0$ as a distribution, where $x$ is the integration variable:

$$(cx)^{-1+\epsilon} = \frac{1}{\epsilon}\delta(cx) + \sum_{n=0}^{\infty}\frac{\epsilon^n}{n!}\left[\frac{\log^n(cx)}{cx}\right]_+. \tag{A.40}$$

One also has

$$
\begin{aligned}
c^{-1+\epsilon}x^{-1+\epsilon} &= \left(\sum_{n=0}^{\infty}\frac{\epsilon^n}{n!}\frac{\log^n(c)}{c}\right)\left(\frac{1}{\epsilon}\delta(x)+\sum_{m=0}^{\infty}\frac{\epsilon^m}{m!}\left[\frac{\log^m(x)}{x}\right]_+\right) \qquad (A.41)\\
&= \delta(x)\sum_{n=0}^{\infty}\frac{\epsilon^{n-1}}{n!}\frac{\log^n(c)}{c}+\sum_{n=0}^{\infty}\sum_{m=0}^{\infty}\frac{\epsilon^{n+m}}{n!m!}\frac{\log^n(c)}{c}\left[\frac{\log^m(x)}{x}\right]_+ .
\end{aligned}
$$

Now we work on both terms separately to organize them as an expansion in $\epsilon^n/n!$. The first term is simply

$$
\delta(x)\sum_{n=0}^{\infty}\frac{\epsilon^{n-1}}{n!}\frac{\log^n(c)}{c} = \frac{\delta(x)}{c\epsilon}+\frac{\delta(x)}{c}\sum_{n=0}^{\infty}\frac{\epsilon^n}{n!}\frac{\log^{n+1}(c)}{n+1}, \qquad (A.42)
$$

where we took $n\to n+1$ and extracted the first case $(n=-1)$ from the sum. The second term is

$$
\begin{aligned}
\sum_{n=0}^{\infty}\sum_{m=0}^{\infty}\frac{\epsilon^{n+m}}{n!m!}\frac{\log^n(c)}{c}\left[\frac{\log^m(x)}{x}\right]_+ &= \sum_{n=0}^{\infty}\sum_{m=n}^{\infty}\frac{\epsilon^m}{n!(m-n)!}\frac{\log^n(c)}{c}\left[\frac{\log^{m-n}(x)}{x}\right]_+ \qquad (A.43)\\
&= \sum_{n=0}^{\infty}\frac{\epsilon^n}{n!}\sum_{m=0}^{n}\binom{n}{m}\frac{\log^m(c)}{c}\left[\frac{\log^{n-m}(x)}{x}\right]_+ ,
\end{aligned}
$$

where we first took $m\to m-n$, then reversed the order of the sums and finally we interchanged $n\leftrightarrow m$. We now combine both results into

$$
c^{-1+\epsilon}x^{-1+\epsilon} = \frac{\delta(x)}{c\epsilon}+\frac{1}{c}\sum_{n=0}^{\infty}\frac{\epsilon^n}{n!}\left\{\delta(x)\frac{\log^{n+1}(c)}{n+1}+\sum_{m=0}^{n}\binom{n}{m}\ln^m(c)\left[\frac{\log^{n-m}(x)}{x}\right]_+\right\}. \qquad (A.44)
$$

Comparing powers of $\epsilon$ between (A.40) and (A.44), the delta property (A.37) is found for the lowest order, and for $n\geq 0$ one gets (A.38).

# Appendix B

# Auxiliary results for the large-$\beta_0$ formalism

## B.1 Study of the sum in sections 3.1 and 4.1

In this appendix we study the sum $I_{n,k}$ appearing in (3.3) and (4.4),

$$I_{n,k} = \sum_{i=1}^{n} \frac{(-1)^i \Gamma(n) i^k}{\Gamma(i+1)\Gamma(n-i+1)}, \tag{B.1}$$

which is defined for $n \geq 1$ and $k \geq -1$. This sum has the following closed forms

$$\begin{aligned} I_{n,-1} &= -H_n/n, \\ I_{n,0} &= -1/n, \\ I_{n,k\geq 1} &= -{}_kF_{k-1}(\{1-n,2,\ldots,2\},\{1,\ldots,1\},1), \end{aligned} \tag{B.2}$$

where $H_n$ is the $n$-th harmonic number and ${}_pF_q$ represents the Gauss Hypergeometric function, which carries $p$ ($q$) elements in its first (second) argument and a final complex variable (see Appendix A for definitions). The first two forms can be readily proved:

$$I_{n,-1} = \frac{1}{n}\sum_{i=1}^{n} \frac{(-1)^i \Gamma(n+1)}{\Gamma(i)\Gamma(n-i+1) i^2} = -\frac{H_n}{n}, \qquad I_{n,0} = \frac{1}{n}\sum_{i=1}^{n} \binom{n}{i}(-1)^i = -\frac{1}{n}, \tag{B.3}$$

where in the first one we used the definition of the harmonic numbers in (A.14) and in the second one we used the binomial expansion (A.17). For $I_{n,k\geq 1}$, although it seems not-summable in softwares such as `Mathematica`, for particular values of $k$ one obtains

$$\begin{aligned} I_{n,1} &= -{}_1F_0(\{1-n\},\{\},1), \\ I_{n,2} &= -{}_2F_1(\{1-n,2\},\{1\},1), \\ I_{n,3} &= -{}_3F_2(\{1-n,2,2\},\{1,1\},1)\ldots \end{aligned} \tag{B.4}$$

and the result in (B.2) is the generalization to arbitrary, integer $k$. We can now prove the two properties claimed in section 3.1:

$$\begin{aligned}&(1)\;\; I_{n,1<k<n}=0,\\&(2)\;\; I_{n,n}=(-1)^n\Gamma(n).\end{aligned} \tag{B.5}$$

Both properties come from the following identity for the hypergeometric function:

$${}_kF_{k-1}(\{-a,a_2...,a_k\},\{a_2-n_2,...,a_k-n_k\},1)=\frac{a!\delta_{a,\sigma}}{\prod_{i=2}^k(1-a_i)_{n_i}}, \tag{B.6}$$

which holds for $n_i\in\mathbb{N}$ and $\sigma\le a$, where $\sigma\equiv\sum_{i=2}^k n_i$. Here $(x)_y$ is the Pochammer symbol. Note also that since $n_i$ are natural numbers, $a\ge 0$. In our case, $a=1-n$, $a_{i\ge2}=2$ and $n_{i\ge2}=1$, so that $\sigma=k-1$ and the product in the denominator is $\prod_{i=2}^k(-1)=(-1)^{k+1}$. All the conditions hold for $n\ge1$ $(a\ge0)$ and $k\le n$ $(\sigma\le a)$. We write then

$${}_kF_{k-1}(\{1-n,2,...,2\},\{1,...,1\},1)=(-1)^{k+1}(n-1)!\delta_{n-1,k-1}. \tag{B.7}$$

Due to the Kronecker delta, the above expression vanishes for $k<n$, which proves property (1); for $k=n$ it yields $I_{n,n}=(-1)^n\Gamma(n)$, as stated in property (2).

In [17] we developed an alternate proof for (1) based on the sum form. We first use $i^k/\Gamma(i+1)=i^{k-1}/\Gamma(i)$, take $i\mapsto i+1$ and expand $(i+1)^{k-1}$ with (A.17), getting

$$\begin{aligned}I_{n,k}&=\sum_{i=0}^{n-1}\frac{(-1)^{i+1}\Gamma(n)(i+1)^{k-1}}{\Gamma(i+1)\Gamma(n-i)}=-1+\sum_{j=0}^{k-1}\frac{\Gamma(k)}{\Gamma(j+1)\Gamma(k-j)}J_{n,j},\\J_{n,j}&\equiv\sum_{i=1}^{n-1}\frac{(-1)^{i+1}\Gamma(n)i^j}{\Gamma(i+1)\Gamma(n-i)}.\end{aligned} \tag{B.8}$$

We pulled out the $i=0$ case before expanding the binomial to avoid a $0^0$ indetermination. The binomial expansion, and therefore the above result, is valid for $k\ge1$; for $k=0$ one has result (2) in (B.2). The $J_{n,j}$ sum is defined in a way that we can find a recurrence relation:

$$J_{n,j}=(n-1)\sum_{i=0}^{n-2}\frac{(-1)^i\Gamma(n-1)(i+1)^{j-1}}{\Gamma(i+1)\Gamma(n-i-1)}=(n-1)\left[1-\sum_{k=0}^{j-1}\binom{j-1}{k}J_{n-1,k}\right]. \tag{B.9}$$

Again we first used $i^j/\Gamma(i+1) = i^{j-1}/\Gamma(i)$, took $i \mapsto i+1$, pulled out the $i=0$ case and expanded the binomial $(i+1)^{j-1}$ for $j \geq 1$, so (B.9) holds for $n, j \geq 1$. For $j=0$ one has $J_{1,0}=0$ and $J_{n>1,0}=1$ from the second line of (B.8). Also, $J_{1,j\geq 1}=0$ due to the $n-1$ prefactor in (B.9). With these values we can analyze a few iterations of the recurrence relation:

$$\begin{aligned} J_{n,1} &= (n-1)(1-J_{n-1,0}) = 0, && \text{unless } n=2, \\ J_{n,2} &= (n-1)[1-J_{n-1,0}-J_{n-1,1}] = 0, && \text{unless } n=2,3, \\ J_{n,3} &= (n-1)[1-J_{n-1,0}-2J_{n-1,1}-J_{n-1,2}] = 0, && \text{unless } n=2,3,4. \end{aligned} \tag{B.10}$$

Summarizing all the observations

$$J_{n,j} = 0 \text{ for } \begin{cases} n=1, & \text{or,} \\ n>1, & 0<j<n-1. \end{cases} \tag{B.11}$$

Going back to $I_{n,k}$ in (B.8), if we set $k<n$ the sum only includes $J_{n,j}$ with $j<n-1$, so only the $j=0$ contributes and we have $I_{n,k<n} = -1 + J_{n,0}$, which vanishes for $n>1$.

## B.2 Integral representations

For series with no cusp-anomalous dimension, the two following integral representations are needed:

$$\begin{aligned} \frac{(-\beta)^n}{n} &= -\int_{-\beta}^{0} \mathrm{d}\tau \tau^{n-1}, \\ \beta^n \Gamma(n) &= \int_0^\infty \mathrm{d}\tau \tau^{n-1} \mathrm{e}^{-\tau/\beta}. \end{aligned} \tag{B.12}$$

While the first relation is straightforward, the second one requires the integral definition of the gamma function and a bit of manipulation:

$$\beta^n \Gamma(n) = \beta^n \int_0^\infty \mathrm{d}t t^{n-1} \mathrm{e}^{-t} = \beta^i \int_0^\infty \frac{\mathrm{d}\tau}{\beta} \left(\frac{\tau}{\beta}\right)^{n-1} \mathrm{e}^{-\tau/\beta} = \int_0^\infty \mathrm{d}\tau \tau^{n-1} \mathrm{e}^{-\tau/\beta}, \tag{B.13}$$

where in the second step we changed variables to $\tau \equiv t\beta$ assuming $\beta > 0$.

For series with cusp-anomalous dimension, one needs, on top of (B.12), the following integral representations for sequences involving harmonic numbers

$$(-\beta)^n H_n = -\int_{-\beta}^{0} \mathrm{d}\tau \frac{\tau^n - (-\beta)^n}{\tau + \beta}, \qquad \text{(B.14)}$$
$$\frac{(-\beta)^n}{n} H_n = \int_{-\beta}^{0} \mathrm{d}\tau \tau^{n-1} \log\left(1 + \frac{\tau}{\beta}\right),$$

which they both hold for $n > 0$. We provide proofs of both relations. The first one arises immediately from the integral representation of $H_n$ in (A.15):

$$\begin{aligned}(-\beta)^n H_n &= (-\beta)^n \int_0^1 \mathrm{d}\tau \frac{1-\tau^n}{1-\tau} = (-\beta)^{n-1} \int_{-\beta}^{0} \mathrm{d}x \frac{1-(-x/\beta)^n}{1+x/\beta} \\ &= -\int_{-\beta}^{0} \mathrm{d}x \frac{x^n - (-\beta)^n}{x+\beta},\end{aligned} \qquad \text{(B.15)}$$

where $x \equiv -\beta\tau$.

The proof of the second relation in (B.14) requires a bit more detail. First, we start by writing the integral in the right-hand side as

$$\begin{aligned}\int_{-\beta}^{0} \mathrm{d}\tau \tau^{n-1} \log\left(1 + \frac{\tau}{\beta}\right) &= -(-\beta)^n \int_0^1 \mathrm{d}x \log(1-x) x^{n-1} \\ &= -(-\beta)^n \int_0^1 \mathrm{d}y \log(y) (1-y)^{n-1} \\ &= (-\beta)^n \sum_{i=0}^{n-1} \frac{(-1)^{i+1}\Gamma(n)}{\Gamma(i+1)\Gamma(n-i)} \int_0^1 \mathrm{d}y \log(y) y^i,\end{aligned} \qquad \text{(B.16)}$$

where we performed the changes of variable $\tau \equiv -\beta x$ and $x \equiv 1 - y$ and expanded the binomial with (A.17). For the next step we use that the integrand can be written as a total derivative with

$$\log(y) y^i = \frac{\mathrm{d}}{\mathrm{d}i} y^i, \qquad \text{(B.17)}$$

and so the resulting integral can be solved:

$$\int_0^1 \mathrm{d}y \log(y) y^i = \frac{\mathrm{d}}{\mathrm{d}i} \int_0^1 \mathrm{d}y y^i = \frac{\mathrm{d}}{\mathrm{d}i}\left(\frac{1}{i+1}\right) = -\frac{1}{(i+1)^2}. \qquad \text{(B.18)}$$

Plugging this back into (B.16), the sum over $i$ can be manipulated to the defining sum of the harmonic numbers in (A.14) by simply taking $i \mapsto i-1$:

$$\sum_{i=0}^{n-1} \frac{(-1)^i \Gamma(n)}{\Gamma(i+1)\Gamma(n-i)(i+1)^2} = \frac{1}{n} \sum_{i=1}^{n} \frac{(-1)^{i-1}\Gamma(n+1)}{\Gamma(i)\Gamma(n-i+1)i^2} = \frac{H_n}{n}. \tag{B.19}$$

## B.3 Relations for a regular function $F(\epsilon, u)$

The other side of the coin when deriving the closed forms non-cusp and cusp series is summing up the regularity expansion (3.1) for $F(\epsilon, u)$ and $G(\epsilon, u)$. Since this expansion is carried out in the first step of each computation and undergoes several manipulations throughout the subsequent steps, its final form does not fully match (3.1). In this section we relate these final forms to the original regularity expansion and in turn to the full $F(\epsilon, u)$ or $G(\epsilon, u)$ functions. We stress that, since the only condition we employ is regularity at the origin, we only use the notation $F(\epsilon, u)$, but all the results of the section also hold for $G(\epsilon, u)$.

### B.3.1 Relation between $F^{\mu}_{0,i}$ and $F_{0,i}$

Let us start by pulling out the $\mu$-dependence of the $F^{\mu}_{i,j}$ coefficients in order to relate them with $F_{i,j}$. Expanding the definition (2.54) and comparing the $O(\epsilon^i)$ term we get

$$\begin{aligned}
\sum_{j=0}^{\infty} u^j F^{\mu}_{i,j} &= \sum_{k=0}^{\infty} \frac{u^k}{k!} \log^k\left(\frac{\mu^2}{\omega^2}\right) \sum_{j=0}^{\infty} u^j F_{i,j} = \sum_{k=0}^{\infty}\sum_{j=0}^{\infty} \frac{u^{j+k}}{k!} \log^k\left(\frac{\mu^2}{\omega^2}\right) F_{i,j} \\
&= \sum_{k=0}^{\infty}\sum_{j=k}^{\infty} \frac{u^j}{k!} \log^k\left(\frac{\mu^2}{\omega^2}\right) F_{i,j-k} = \sum_{j=0}^{\infty} u^j \sum_{k=0}^{j} \frac{1}{k!} \log^k\left(\frac{\mu^2}{\omega^2}\right) F_{i,j-k},
\end{aligned} \tag{B.20}$$

where in the second line we changed variables $j \mapsto j-k$ and swapped the summation order. Comparing now the powers of $u$ we find

$$F^{\mu}_{i,j} = \sum_{k=0}^{j} \frac{1}{k!} \log^k\left(\frac{\mu^2}{\omega^2}\right) F_{i,j-k}. \tag{B.21}$$

It is also useful to compute the $\mu$-derivative of (B.21):

$$\mu\frac{\mathrm{d}}{\mathrm{d}\mu}F^{\mu}_{i,j} = 2\sum_{k=1}^{j}\frac{1}{(k-1)!}\log^{k-1}\left(\frac{\mu^2}{\omega^2}\right)F_{i,j-k} = 2\sum_{k=0}^{i-1}\frac{1}{k!}\log^{k}\left(\frac{\mu^2}{\omega^2}\right)F_{i,j-k-1} = 2F^{\mu}_{i,j-1}. \tag{B.22}$$

### B.3.2 Expansions for the function

Now we focus on the expansions of $F$. First we take it to a form where it is immediate to put either $\epsilon=0$ or $u=0$ –or both–

$$F(\epsilon,u) = \sum_{i=0}^{\infty}\sum_{j=0}^{\infty}\epsilon^i u^j F_{i,j} = \sum_{i=0}^{\infty}\left(\epsilon^i F_{i,0} + \sum_{j=1}^{\infty}\epsilon^i u^j F_{i,j}\right) = F_{0,0} + \sum_{i=1}^{\infty}\epsilon^i F_{i,0} + \sum_{j=1}^{\infty}u^j F_{0,j} + \sum_{i=1}^{\infty}\sum_{j=1}^{\infty}\epsilon^i u^j F_{i,j}. \tag{B.23}$$

From here one can clearly see that $F(0,0)=F_{0,0}$ and thus

$$F(\epsilon,0) - F(0,0) = \sum_{i=1}^{\infty}\epsilon^i F_{i,0}, \qquad F(0,u) - F(0,0) = \sum_{j=1}^{\infty}u^j F_{0,j}. \tag{B.24}$$

### B.3.3 Expansions for the derivatives

Finally we derive analogous relations for the $\epsilon$ and $u$ derivatives of $F$. As before, we first take them to a form that is immediate to evaluate at $\epsilon=0$ and $u=0$.

For the $\epsilon$-derivative we have

$$\frac{\mathrm{d}}{\mathrm{d}\epsilon}F(\epsilon,u) = F_{1,0} + \sum_{i=2}^{\infty} i\epsilon^{i-1}F_{i,0} + \sum_{j=1}^{\infty}u^j F_{1,j} + \sum_{i=2}^{\infty}\sum_{j=1}^{\infty} i\epsilon^{i-1}u^j F_{i,j}, \tag{B.25}$$

so that the expansions become

$$\frac{\mathrm{d}}{\mathrm{d}\epsilon}F(\epsilon,u)\bigg|_{\epsilon=0} = F_{1,0} + \sum_{j=1}^{\infty} u^j F_{1,j} = \sum_{j=0}^{\infty} u^j F_{1,j}, \tag{B.26}$$

$$\frac{\mathrm{d}}{\mathrm{d}\epsilon}F(\epsilon,u)\bigg|_{u=0} = F_{1,0} + \sum_{i=2}^{\infty} i\epsilon^{i-1} F_{i,0} = \sum_{i=1}^{\infty} i\epsilon^{i-1} F_{i,0}$$

For the $u$-derivative we proceed in the same way,

$$\frac{\mathrm{d}}{\mathrm{d}u}F(\epsilon,u) = F_{0,1} + \sum_{i=1}^{\infty} \epsilon^i F_{i,1} + \sum_{j=2}^{\infty} j u^{j-1} F_{0,j} + \sum_{j=2}^{\infty}\sum_{i=1}^{\infty} j\epsilon^i u^{j-1} F_{i,j}, \tag{B.27}$$

and the expansions are

$$\frac{\mathrm{d}}{\mathrm{d}u}F(\epsilon,u)\bigg|_{\epsilon=0} = F_{0,1} + \sum_{j=2}^{\infty} j u^{j-1} F_{0,j} = \sum_{j=1}^{\infty} j u^{j-1} F_{0,j}, \tag{B.28}$$

$$\frac{\mathrm{d}}{\mathrm{d}u}F(\epsilon,u)\bigg|_{u=0} = F_{0,1} + \sum_{i=1}^{\infty} \epsilon^i F_{i,1} = \sum_{i=0}^{\infty} \epsilon^i F_{i,1}.$$

Finally, we find the coefficient results

$$\frac{\mathrm{d}}{\mathrm{d}\epsilon}F(\epsilon,0)\bigg|_{\epsilon=0} = F_{1,0}, \qquad \frac{\mathrm{d}}{\mathrm{d}u}F(0,u)\bigg|_{u=0} = F_{0,1}. \tag{B.29}$$

# Appendix C
# Loop integrals

## C.1 Notation

Schematically, one-loop integrals have the form

$$I_n \;=\; \int \frac{\mathrm{d}^d k}{(2\pi)^d} \frac{T(k)}{P_1 \ldots P_n}, \tag{C.1}$$

where $T(k)$ represents any tensor structure made from the loop momentum $k$ and any of the external momenta, and $P_i$ represent the denominators of the propagators constituting the loop, each of them a function of $k$ and the external momenta. The spacetime dimension $d = 4 - 2\epsilon$ is chosen here as the divergence regulator.

In this work, instead of writing the full integral (C.1) we employ the shorthand notation $I_n(T(k), P_1, \ldots, P_n)$,where instead of the explicit expressions for each $P_i$ we write one of the three functions:

- Quadratic massless propagator: $Q \equiv k^2 + \mathrm{i}\delta$.
- Quadratic massive propagator: $Q_m \equiv k^2 - m^2 + \mathrm{i}\delta$.
- Eikonal $n$-collinear propagator: $E_n \equiv k^- + \mathrm{i}\delta$

To cover for all the variety of integrals throughout this work, it is enough to consider the three previous denominator functions with the following set of notational rules:

1. If the propagator depends on an external momentum $p$ we write $Q(p)$ and take the convention that $p$ is added to $k$:

$$Q(p) = (k+p)^2 + \mathrm{i}\delta, \qquad Q_m(-p) = (k-p)^2 - m^2 + \mathrm{i}\delta, \qquad E_n(s) = k^- + s + \mathrm{i}\delta.$$

2. When the loop momentum of an eikonal propagator appears multiplied by a number we will employ the notation

$$E_{an} = a k^- + \mathrm{i}\delta, \qquad E_{an}(s) = a k^- + s + \mathrm{i}\delta.$$

3. If the propagator appears powered to any number we simply take the power of the corresponding function:

$$Q^x(p) = [(k+p)^2 + \mathrm{i}\delta]^x, \qquad Q^y_m(p) = [(k+p)^2 - m^2 + \mathrm{i}\delta]^y, \qquad E^z_n = [k^- + \mathrm{i}\delta]^z.$$

## C.2 Master integrals

We have solved the loop integrals listed in sections C.3 to C.8 by reducing them to the following master integral:

$$M_n(T(k), A) \equiv \int \frac{\mathrm{d}^d k}{(2\pi)^d} \frac{T(k)}{[k^2 - A + \mathrm{i}\delta]^n}. \tag{C.2}$$

To arrive to this form one uses the so-called parametrization tricks to combine the propagators $P_i$. The two most widely used parametrization formulas are

$$\frac{1}{P_1^n P_2^m} = \frac{\Gamma(n+m)}{\Gamma(n)\Gamma(m)} \int_0^1 \mathrm{d}\lambda \frac{\lambda^{n-1}(1-\lambda)^{m-1}}{[P_2 + (P_1 - P_2)\lambda]^{n+m}}, \qquad \text{[Feynman param.]} \tag{C.3}$$

$$\frac{1}{P_1^n P_2^m} = \frac{2^m \Gamma(n+m)}{\Gamma(n)\Gamma(m)} \int_0^\infty \mathrm{d}\lambda \frac{\lambda^{m-1}}{[P_1 + 2P_2\lambda]^{n+m}}. \qquad \text{[Georgi param.]}$$

Typically, one uses Feynman parametrization to combine quadratic propagators such as the QCD propagators $Q$ and $Q_m$, and Georgi parametrization when one or more of them are linear in the loop momentum, such as the SCET and bHQET propagators $E_n$. Both parametrization formulas introduce additional integrals into $I_n$, which are exchanged with the loop-momentum integral over $k$, giving the form in (C.2). More explicitly, after parametrization one arrives to the form

$$I_n = \int \mathrm{d}\lambda_1 \ldots \int \mathrm{d}\lambda_{n-1} f(\lambda_1, \ldots, \lambda_n) M_n(T(k), A). \tag{C.4}$$

The parameter $A$ in the master integral, known as the scale of the integral, is a function of the external momenta and the introduced parameters.

For the scalar cases $T(k) = 1, k^2$ we have employed the solutions:

$$\begin{aligned} M_n(1, A) &= \frac{\mathrm{i}\pi^{d/2}}{(2\pi)^d} \frac{(-1)^n \Gamma\left(n - \frac{d}{2}\right)}{\Gamma(n)} (A - \mathrm{i}\delta)^{\frac{d}{2} - n} \\ &= \frac{\mathrm{i}}{(4\pi)^2} \frac{(-1)^n (4\pi)^\epsilon \Gamma(n - 2 + \epsilon)}{\Gamma(n)} (A - \mathrm{i}\delta)^{2 - n - \epsilon} \\ M_n(k^2, A) &= \frac{\mathrm{i}}{(4\pi)^{d/2}} \frac{d}{2} \frac{(-1)^{n+1} \Gamma\left(n - \frac{d}{2} - 1\right)}{\Gamma(n)} (A - \mathrm{i}\delta)^{\frac{d}{2} + 1 - n} \\ &= \frac{\mathrm{i}}{(4\pi)^2} \frac{(-1)^{n+1} (4\pi)^\epsilon \Gamma(n - 1 + \epsilon)}{\Gamma(n)} (A - \mathrm{i}\delta)^{1 - n - \epsilon}. \end{aligned} \tag{C.5}$$

When $T(k)$ takes tensor form, i.e., $T(k) = k^\mu$ or $k^\mu k^\nu$, the Passarino-Veltman reduction method can be applied to relate $I_n(T(k), P_1, \ldots P_n)$ to the scalar cases $I_n(1, P_1, \ldots P_n)$ and $I_n(k^2, P_1, \ldots P_n)$ and then proceed with (C.3) and (C.5). These results also illustrate the very common statement that scaleless integrals vanish: since one is only interested in the limit $\delta \mapsto 0$, when setting $A = 0$ in (C.5) the master integral vanishes.

As a final comment, we also note that we one can freely add or subtract $\mathrm{i}\delta$ in the numerator of the integrals, since the additional term will vanish when $\delta = 0$. This results specially convenient to simplify some of the integrals before embarking into the parametrization process, such as in the following example:

$$\begin{aligned} I_3(k^2, Q, Q(p_1), Q(p_2)) &= \int \frac{\mathrm{d}^d k}{(2\pi)^d} \frac{k^2}{[k^2 + \mathrm{i}\delta][(k + p_1)^2 + \mathrm{i}\delta][(k + p_2)^2 + \mathrm{i}\delta]} \\ &= \int \frac{\mathrm{d}^d k}{(2\pi)^d} \frac{k^2 + \mathrm{i}\delta}{[k^2 + \mathrm{i}\delta][(k + p_1)^2 + \mathrm{i}\delta][(k + p_2)^2 + \mathrm{i}\delta]} \\ &= \int \frac{\mathrm{d}^d k}{(2\pi)^d} \frac{1}{[(k + p_1)^2 + \mathrm{i}\delta][(k + p_2)^2 + \mathrm{i}\delta]} \\ &= I_2(1, Q(p_1), Q(p_2)), \end{aligned} \tag{C.6}$$

which, in turn, is equal to $I_2(1,Q,Q(p_1-p_2))$ or $I_2(1,Q,Q(p_2-p_1))$, as seen by the change of variables $k\mapsto k-p_2$ or $k\mapsto k-p_1$.

## C.3 Integrals in the effective gluon-propagator

For $p^2\neq 0$:

$$\begin{aligned} I_2(1,Q^x,Q^y(p)) &= \int\frac{\mathrm{d}^dk}{(2\pi)^d}\frac{1}{[k^2+\mathrm{i}\delta]^x[(k+p)^2+\mathrm{i}\delta]^y} \\ &= \frac{\mathrm{i}}{(4\pi)^2}(-1)^{x+y}(-p^2-\mathrm{i}\delta)^{2-x-y}\left(\frac{-p^2-\mathrm{i}\delta}{4\pi}\right)^{-\epsilon} \\ &\quad\times\frac{\Gamma(x+y+\epsilon-2)\Gamma(2-x-\epsilon)\Gamma(2-y-\epsilon)}{\Gamma(x)\Gamma(y)\Gamma(4-x-y-2\epsilon)}, \end{aligned} \tag{C.7}$$

$$\begin{aligned} I_2(k^\mu,Q^x,Q^y(p)) &= \frac{p^\mu}{2p^2}\Big[I_2(1,Q^x,Q^{y-1}(p)-p^2I_2(1,Q^x,Q^y(p))- \\ &\quad -I_2(1,Q^{x-1},Q^y(p))\Big], \end{aligned} \tag{C.8}$$

$$I_2(k^\mu k^\nu,Q^x,Q^y(p)) = \frac{1}{d-1}\left(A-\frac{1}{p^2}B\right)g^{\mu\nu}+\frac{1}{d-1}\frac{1}{p^2}\left(\frac{d}{p^2}B-A\right)p^\mu p^\nu, \tag{C.9}$$

$$\begin{aligned} A &= I_2(1,Q^{x-1},Q^y(p)), \\ B &= \frac{1}{4}\Big[I_2(1,Q^x,Q^{y-2}(p))+I_2(1,Q^{x-2},Q^y(p))+p^4I_2(1,Q^x,Q^y(p)) \\ &\quad +2p^2I_2(1,Q^{x-1},Q^y(p))-2p^2I_2(1,Q^{x-1},Q^{y-1}(p))\Big]. \end{aligned}$$

## C.4 Integrals in the quark-self energy

For $p^2\neq m^2$:

$$\begin{aligned} I_2(1,Q^x,Q_m(p)) &= \int\frac{\mathrm{d}^dk}{(2\pi)^d}\frac{1}{[k^2+\mathrm{i}\delta]^x[(k+p)^2-m^2+\mathrm{i}\delta]} \\ &= \frac{-\mathrm{i}}{(4\pi)^2}(-1)^x\left(\frac{4\pi}{m^2}\right)^{\epsilon}(m^2)^{1-x}\frac{\Gamma(x+\epsilon-1)\Gamma(2-\epsilon-x)}{\Gamma(2-\epsilon)} \\ &\quad\times{}_2F_1\left(x+\epsilon-1,x,2-\epsilon,\frac{p^2}{m^2}\right), \end{aligned} \tag{C.10}$$

$$\begin{aligned} I_2(k^\mu,Q^x,Q_m^y(p)) &= \frac{p^\mu}{2p^2}\Big[I_2(1,Q^x,Q_m^{y-1}(p))+(m^2-p^2)I_2(1,Q^x,Q_m^y(p)) \\ &\quad -I_2(1,Q^{x-1},Q_m^y(p))\Big]. \end{aligned} \tag{C.11}$$

## C.5 Integrals in the QCD matrix element

For $p_1^2 = p_2^2 = 0$

$$
\begin{aligned}
I_3(1, Q^x, Q^y(p_1), Q^z(p_2)) &= \int \frac{\mathrm{d}^d k}{(2\pi)^d} \frac{1}{[k^2 + \mathrm{i}\delta]^x [(k+p_1)^2 + \mathrm{i}\delta]^y [(k+p_2)^2 + \mathrm{i}\delta]^z} \\
&= \frac{\mathrm{i}(-1)^{x+y+z}}{(4\pi)^2} (2p_1{\cdot}p_2 - \mathrm{i}\delta)^{2-x-y-z} \left( \frac{2p_1{\cdot}p_2 - \mathrm{i}\delta}{4\pi} \right)^{-\epsilon} \\
&\quad \times \frac{\Gamma(x+y+z+\epsilon-2)\Gamma(2-x-z-\epsilon)\Gamma(2-x-y-\epsilon)}{\Gamma(y)\Gamma(z)\Gamma(4-x-y-z-2\epsilon)},
\end{aligned}
\tag{C.12}
$$

$$
\begin{aligned}
I_3(k^\mu k^\nu, Q^x, Q^y(p_1), Q^z(p_2)) &= \frac{g^{\mu\nu}}{2(1-\epsilon)} \left[ A - \frac{2}{p_1{\cdot}p_2} B \right] + \frac{p_1^\mu p_1^\nu + p_2^\mu p_2^\nu}{(p_1{\cdot}p_2)^2} C \\
&\quad - \frac{p_1^\mu p_2^\nu + p_1^\nu p_2^\mu}{2(1-\epsilon)} \left[ \frac{A}{p_1{\cdot}p_2} - \frac{2(2-\epsilon)}{(p_1{\cdot}p_2)^2} B \right],
\end{aligned}
\tag{C.13}
$$

$$
\begin{aligned}
A &= I_3(1, Q^{x-1}, Q^y(p_1), Q^z(p_2)), \\
B &= \frac{1}{4} \Big[ I_3(1, Q^x, Q^{y-1}(p_1), Q^{z-1}(p_2)) - I_3(1, Q^{x-1}, Q^{y-1}(p_1), Q^z(p_2)) \\
&\quad - I_3(1, Q^{x-1}, Q^y(p_1), Q^{z-1}(p_2)) + I_3(1, Q^{x-2}, Q^y(p_1), Q^z(p_2)) \Big], \\
C &= \frac{1}{4} \Big[ I_3(1, Q^x, Q^{y-2}(p_1), Q^z(p_2)) + I_3(1, Q^{x-2}, Q^y(p_1), Q^z(p_2)) \\
&\quad - 2 I_3(1, Q^{x-1}, Q^{y-1}(p_1), Q^z(p_2)) \Big].
\end{aligned}
$$

## C.6 Integrals in the SCET jet function

For $p^2 \neq 0$,

$$
\begin{aligned}
I_3(1, Q^x, Q^y(p), E_n^z) &= \int \frac{\mathrm{d}^d k}{(2\pi)^d} \frac{1}{[k^2 + \mathrm{i}\delta]^x [(k+p)^2 + \mathrm{i}\delta]^y [k^- + \mathrm{i}\delta]^z} \\
&= \frac{\mathrm{i}}{(4\pi)^2} (-1)^{x+y+z} (-p^2 - \mathrm{i}\delta)^{2-x-y} \left( \frac{-p^2 - \mathrm{i}\delta}{4\pi} \right)^{-\epsilon} (p^-)^{-z} \\
&\quad \times \frac{\Gamma(x+y+\epsilon-2)\Gamma(2-x-z-\epsilon)\Gamma(2-y-\epsilon)}{\Gamma(x)\Gamma(y)\Gamma(4-x-y-z-2\epsilon)},
\end{aligned}
\tag{C.14}
$$

$$
I_3(k^-, Q^x, Q^y(p), E_n^z) = I_3(1, Q^x, Q^y(p), E_n^{z-1}),
\tag{C.15}
$$

$$
\begin{aligned}
I_2(1, Q^x, Q^y(p)) &= \int \frac{\mathrm{d}^d k}{(2\pi)^d} \frac{1}{[k^2 + \mathrm{i}\delta]^x [(k+p)^2 + \mathrm{i}\delta]^y} \\
&= \frac{\mathrm{i}}{(4\pi)^2} (-1)^{x+y} (-p^2 - \mathrm{i}\delta)^{2-x-y} \left( \frac{-p^2 - \mathrm{i}\delta}{4\pi} \right)^{-\epsilon} \\
&\quad \times \frac{\Gamma(x+y+\epsilon-2)\Gamma(2-x-\epsilon)\Gamma(2-y-\epsilon)}{\Gamma(x)\Gamma(y)\Gamma(4-x-y-2\epsilon)},
\end{aligned}
\tag{C.16}
$$

$$
I_2(1, Q^x, E_n^y(s)) \equiv \int \frac{\mathrm{d}^d k}{(2\pi)^d} \frac{1}{[k^2 + \mathrm{i}\delta]^x [k^- + s + \mathrm{i}\delta]^y} = 0,
\tag{C.17}
$$

$$
\begin{aligned}
I_2(k^\mu, Q^x, Q^y(p)) &= \frac{p^\mu}{2p^2} \Big[ I_2(1, Q^x, Q^{y-1}(p) - p^2 I_2(1, Q^x, Q^y(p)) \\
&\quad - I_2(1, Q^{x-1}, Q^y(p)) \Big].
\end{aligned}
\tag{C.18}
$$

## C.7 Integrals in the massive SCET matrix element

For $p^2 = m^2$:

$$
\begin{aligned}
I_3(1, Q^x, Q_m^y(p), E_n^z) &= \int \frac{\mathrm{d}^d k}{(2\pi)^d} \frac{1}{[k^2 + \mathrm{i}\delta]^x [(k+p)^2 - m^2 + \mathrm{i}\delta]^y [k^- + \mathrm{i}\delta]^z} \\
&= \frac{\mathrm{i}}{(4\pi)^2} (-1)^{x+y+z} (p^-)^{-z} (m^2 - \mathrm{i}\delta)^{2-x-y} \left( \frac{m^2 - \mathrm{i}\delta}{4\pi} \right)^{-\epsilon} \\
&\quad \times \frac{\Gamma(x+y+\epsilon-2)\Gamma(4-2x-y-z-2\epsilon)}{\Gamma(y)\Gamma(4-x-y-z-2\epsilon)},
\end{aligned}
\tag{C.19}
$$

$$
\begin{aligned}
I_3(1, Q^x, Q_m^y(p), E_{\bar{n}}^z) &\equiv \int \frac{\mathrm{d}^d k}{(2\pi)^d} \frac{1}{[k^2 + \mathrm{i}\delta]^x [(k+p)^2 - m^2 + \mathrm{i}\delta]^y [k^+ + \mathrm{i}\delta]^z} \\
&= \frac{\mathrm{i}}{(4\pi)^2} (-1)^{x+y+z} (p^+)^{-z} (m^2 - \mathrm{i}\delta)^{2-x-y} \left( \frac{m^2 - \mathrm{i}\delta}{4\pi} \right)^{-\epsilon} \\
&\quad \times \frac{\Gamma(x+y+\epsilon-2)\Gamma(4-2x-y-z-2\epsilon)}{\Gamma(y)\Gamma(4-x-y-z-2\epsilon)},
\end{aligned}
\tag{C.20}
$$

$$
I_3(k^-, Q^x, Q_m^y(p), E_n^z) = I_3(1, Q^x, Q_m^y(p), E_n^{z-1}),
\tag{C.21}
$$

$$
\begin{aligned}
I_2(1, Q^x, Q_m^y(p)) &= \int \frac{\mathrm{d}^d k}{(2\pi)^d} \frac{1}{[k^2 + \mathrm{i}\delta]^x [(k+p)^2 - m^2 + \mathrm{i}\delta]^y} \\
&= \frac{\mathrm{i}}{(4\pi)^2} (-1)^{x+y} (m^2 - \mathrm{i}\delta)^{2-x-y} \left( \frac{m^2 - \mathrm{i}\delta}{4\pi} \right)^{-\epsilon} \\
&\quad \times \frac{\Gamma(x+y+\epsilon-2)\Gamma(4-2x-y-2\epsilon)}{\Gamma(y)\Gamma(4-x-y-2\epsilon)}.
\end{aligned}
\tag{C.22}
$$

## C.8 Integrals in the bHQET jet function

For $v^2=1$:

$$
\begin{aligned}
I_2(1,Q^x,E_v^y(s)) &= \int \frac{\mathrm{d}^dk}{(2\pi)^d}\frac{1}{[k^2+\mathrm{i}\delta]^x[v\cdot k+s+\mathrm{i}\delta]^y} \\
&= \frac{\mathrm{i}}{(4\pi)^2}(-1)^{x+y}(-2s-\mathrm{i}\delta)^{4-2x-y}\left(\frac{-2s-\mathrm{i}\delta}{\sqrt{4\pi}}\right)^{-2\epsilon} \\
&\quad\times\frac{2^y\Gamma(2-x-\epsilon)\Gamma(2x+y+2\epsilon-4)}{\Gamma(x)\Gamma(y)},
\end{aligned}
\tag{C.23}
$$

$$
I_3(1,Q,E_{\bar{n}}(s),E_n) \equiv \frac{\mathrm{d}^dk}{(2\pi)^d}\frac{1}{[k^2+\mathrm{i}\delta]^x[k^++s+\mathrm{i}\delta]^y[k^-+\mathrm{i}\delta]^z}=0,
\tag{C.24}
$$

$$
\begin{aligned}
I_3(1,Q^x,E_v^y(s),E_n^z) &= \int \frac{\mathrm{d}^dk}{(2\pi)^d}\frac{1}{[k^2+\mathrm{i}\delta]^x[v\cdot k+s+\mathrm{i}\delta]^y[k^-+\mathrm{i}\delta]^z} \\
&= \frac{\mathrm{i}}{(4\pi)^2}(-1)^{x+y+z}(4\pi)^{\epsilon}(-2s-\mathrm{i}\delta)^{4-2x-y-z-2\epsilon}(v^-)^{-z} \\
&\quad\times\frac{2^y\Gamma(2x+y+z+2\epsilon-4)\Gamma(2-x-z-\epsilon)}{\Gamma(x)\Gamma(y)},
\end{aligned}
\tag{C.25}
$$

$$
I_2(1,Q^x,E_{\bar{n}}^y(s)) \equiv \int \frac{\mathrm{d}^dk}{(2\pi)^d}\frac{1}{[k^2+\mathrm{i}\delta]^x[k^++s+\mathrm{i}\delta]^y}=0,
\tag{C.26}
$$

$$
I_2(1,Q^x,E_n^y(s)) \equiv \int \frac{\mathrm{d}^dk}{(2\pi)^d}\frac{1}{[k^2+\mathrm{i}\delta]^x[k^-+s+\mathrm{i}\delta]^y}=0.
\tag{C.27}
$$

# Appendix D

# $T$ matrix in contour-improved theory

In this appendix we detail the computation of the $T$ matrix defined in (11.4) in contour-improved theory for a momentum $l=0$. For simplicity in the following we drop the superscript 0.

The $T$ matrix is defined as $T=S^{-1}$, where the coefficients $s_{n,k}$ are in defined (11.61). To finite order, the $S$ matrix has the form

$$S=\begin{pmatrix} -1 & 0 & \frac{\pi^2}{3} & 0 & -\frac{\pi^4}{5} & 0 & \frac{\pi^6}{7} & 0 & \dots \\ 0 & -1 & 0 & \pi^2 & 0 & -\pi^4 & 0 & \pi^6 & \dots \\ 0 & 0 & -1 & 0 & 2\,\pi^2 & 0 & -3\,\pi^4 & 0 & \dots \\ 0 & 0 & 0 & -1 & 0 & \frac{10\,\pi^2}{3} & 0 & -7\,\pi^4 & \dots \\ 0 & 0 & 0 & 0 & -1 & 0 & 5\,\pi^2 & 0 & \dots \\ 0 & 0 & 0 & 0 & 0 & -1 & 0 & 7\,\pi^2 & \dots \\ 0 & 0 & 0 & 0 & 0 & 0 & 0 & 0 & \dots \\ \vdots & \vdots & \vdots & \vdots & \vdots & \vdots & \vdots & \vdots & \vdots \end{pmatrix}, \tag{D.1}$$

and the result of inverting the finite order matrix in (D.1) is

$$T=\begin{pmatrix} -1 & 0 & -\frac{\pi^2}{3} & 0 & -\frac{7\,\pi^4}{15} & 0 & -\frac{31\,\pi^6}{21} \\ 0 & -1 & 0 & -\pi^2 & 0 & -\frac{7\,\pi^4}{3} & 0 \\ 0 & 0 & -1 & 0 & -2\,\pi^2 & 0 & -7\,\pi^4 \\ 0 & 0 & 0 & -1 & 0 & -\frac{10\,\pi^2}{3} & 0 \\ 0 & 0 & 0 & 0 & -1 & 0 & -5\,\pi^2 \\ 0 & 0 & 0 & 0 & 0 & -1 & 0 \end{pmatrix}. \tag{D.2}$$

The task is now to determine a closed expression for the coefficients $t_{n,k}$. We start by listing three properties that can be found by inspection.

1. The $n$-dimensional $T$ is completely determined by the $n$-dimensional $S$. This can be seen consistently by computing $T_{n\times n}$ solely from $S_{n\times n}$ and checking $T_{n\times n}\subset T_{m\times m}$ for $m>n$.

2. We denote by $\mathrm{Diag}_{S,n}$ the sequence of elements in the diagonal containing powers of $\pi^n$ in $S$. Then

$$\mathrm{Diag}_{S,n}=\left\{g_n\pi^n\frac{\Gamma(n+m)}{\Gamma(m)}\right\}_{m=1}^{\infty}, \tag{D.3}$$

where the prefactors $g_n$ can be deduced from $s_{n,k}$ in (11.61). Indeed, using that the element $s_{n,k}=s_{n,n+(k-n)}$ is the $n$-th element of the diagonal $k-n$ we have

$$s_{n,k}=-\frac{\Gamma(k)}{\Gamma(n)\Gamma(2+k-n)}\frac{(\mathrm{i}\pi)^{k-n}}{2}[1+(-1)^{k-n}]=g_{k-n}\pi^{k-n}\frac{\Gamma(k)}{\Gamma(n)}, \tag{D.4}$$

which leads to

$$g_n=-\frac{\mathrm{i}^n[1+(-1)^n]}{2\Gamma(2+n)}. \tag{D.5}$$

3. The diagonals of the $T$ matrix also obey the same structure

$$\mathrm{Diag}_{T,n}=\left\{f_n\pi^n\frac{\Gamma(n+m)}{\Gamma(m)}\right\}_{m=1}^{\infty}, \tag{D.6}$$

however this time we can't yet derive an expression for the prefactors $f_n$. The first values are

$$\{f_{2n}\}_{n=0}=\left\{-1,-\frac{1}{6},-\frac{7}{360},-\frac{31}{15120},-\frac{127}{604800},-\frac{73}{3421440},\ldots\right\}, \tag{D.7}$$

and $f_{2n+1}=0$ for any $n\in\mathbb{N}$.

One can now use these three properties, plus the relation $ST = \mathbb{I}$ to derive and solve a recursive relation for $t_{n,k}$. We start by writing the product of row $n$ of $S$ with column $k$ of $T$,

$$\delta_{n,k} = \sum_{i=1}^{k} s_{n,i} t_{i,k} = \sum_{i=n}^{k} s_{n,i} t_{i,k}, \tag{D.8}$$

where the sum extends to $k$ due to property 1 and where on the second step we used $s_{n,i<n} = 0$. Then we pull the $k = n$ case out of the sum and use $s_{n,n} = -1$ to solve for $t_{n,k}$

$$t_{n,k} = -\delta_{n,k} + \sum_{i=n+1}^{k} s_{n,i} t_{i,k}. \tag{D.9}$$

The relation expresses each coefficient of the $k$ column of $T_{k\times k}$ as a combination of the coefficients immediately below. When $n = k$, there are no coefficients below $t_{k,k}$ and one obtains $t_{k,k} = -1$. For any other $k > n$ we simply have

$$t_{n,k} \;=\; \sum_{i=n+1}^{k} s_{n,i} t_{i,k}, \qquad k > n. \tag{D.10}$$

Now we use properties 2 and 3 to write

$$t_{n,k} = f_{k-n} \pi^{k-n} \frac{\Gamma(k)}{\Gamma(n)}, \qquad s_{n,k} = g_{k-n} \pi^{k-n} \frac{\Gamma(k)}{\Gamma(n)}, \tag{D.11}$$

Plugging this back into the recursive relation we obtain

$$\begin{aligned} t_{n,n+m} = f_m \pi^m \frac{\Gamma(n+m)}{\Gamma(n)} &= \sum_{i=n+1}^{n+m} s_{n,i} t_{i,n+m} = \sum_{i=1}^{m} s_{n,n+i} t_{n+i,n+m} \\ &= \sum_{i=1}^{m} g_i \pi^i \frac{\Gamma(n+i)}{\Gamma(n)} f_{m-i} \pi^{m-i} \frac{\Gamma(n+m)}{\Gamma(n+i)} \\ &= \frac{\Gamma(n+m)}{\Gamma(n)} \pi^m \sum_{i=1}^{m} g_i f_{m-i}, \end{aligned} \tag{D.12}$$

which leads to $f_m = \sum_{i=1}^{m} g_i f_{m-i}$ or equivalently

$$f_{m+1} = \sum_{i=1}^{m+1} g_i f_{m+1-i} = -\sum_{i=1}^{m+1} \frac{\mathrm{i}^i [1 + (-1)^i]}{2\Gamma(2+i)} f_{m+1-i}. \tag{D.13}$$

This recursive relation can be solved with the condition $f_0 = -1$,

$$f_n = -\frac{(2\mathrm{i})^n}{n!} B_n(1/2) = -\frac{\mathrm{i}^n(2-2^n)}{\Gamma(n+1)} B_n, \tag{D.14}$$

where $n \geq 0$ and $B_n(x)$ are the Bernoulli polynomials. In the second equality we used that for $x = 1/2$, the Bernoulli polynomials can be put in terms of the Bernoulli numbers $B_n$,

$$B_n(1/2) = (2^{1-n} - 1) B_n. \tag{D.15}$$

Bernouli numbers are 0 for all odd $n$ except $n = 1$, for which the factor $2 - 2^n$ makes $f_1 = 0$. The sequence $\{f_n\}$ reproduces that in (D.7) for even $n$. With this we obtain our final closed expression for $t_{n,k}$ in (11.61).

# Appendix E
# Real radiation coefficients

## E.1 Matrix element coefficients

Here we collect the coefficients $f^C_{i,j}$ of the real-radiation matrix element, which are defined in (12.71). We write them in terms of $x_1$ and $x_2$ and expand up to $\mathcal{O}(\epsilon^2)$. For an easier comparison with the massless case in [33], terms are organized according to the power of the reduced quark mass $\hat{m} = m/Q$.

$$\begin{aligned}
f^V_{0,0} &= \frac{x_1^2+x_2^2}{(1-x_1)(1-x_2)} - \frac{3(x_1^2+x_2^2)-4(x_1+x_2)+2x_1x_2+4}{(1-x_1)(1-x_2)}\epsilon + \frac{2(2-x_1-x_2)^2}{(1-x_1)(1-x_2)}\epsilon^2 \\
&\quad +2\hat{m}^2\left[\frac{4x_1-5}{(1-x_1)^2} - \frac{1}{(1-x_2)^2} + \frac{4x_1-2}{(1-x_1)(1-x_2)} + \frac{4(2-x_1-x_2)^2}{(1-x_1)(1-x_2)}\epsilon\right] \qquad \text{(E.1)} \\
&\quad -\frac{8\hat{m}^4(2-x_1-x_2)^2}{(1-x_1)^2(1-x_2)^2} \\
f^V_{1,1} &= 2(2\hat{m}^2-\epsilon)\frac{\sqrt{x_1^2-4\hat{m}^2}\sqrt{x_2^2-4\hat{m}^2}}{(1-x_1)(1-x_2)}, \\
f^V_{2,0} &= \frac{x_1^2(1-\epsilon)}{(1-x_1)(1-x_2)} + 2\hat{m}^2\left[\frac{x_1^2}{(1-x_1)^2} + \frac{2+4\epsilon}{(1-x_1)(1-x_2)}\right] + \frac{8\hat{m}^4}{(1-x_1)^2}, \\
f^V_{0,2} &= f^V_{2,0} \quad (1\leftrightarrow 2), \\
f^A_{0,0} &= \frac{x_1^2+x_2^2}{(1-x_1)(1-x_2)} - \frac{3(x_1^2+x_2^2)-4(x_1+x_2)+2x_1x_2+4}{(1-x_1)(1-x_2)}\epsilon + \frac{2(2-x_1-x_2)^2}{(1-x_1)(1-x_2)}\epsilon^2 \\
&\quad -2\hat{m}^2\left[\frac{4x_1^2-14x_1+9}{(1-x_1)^2} - \frac{1}{(1-x_2)^2} - \frac{2x_1(4-x_1)+2x_2(1-x_2)}{(1-x_1)(1-x_2)}\right. \\
&\quad \left. -4\left(\frac{2x_1^2-9x_1+6}{(1-x_1)^2} - \frac{1}{(1-x_2)^2} - \frac{2x_1(6-x_1)+x_2(1-x_2)+1}{(1-x_1)(1-x_2)}\right)\epsilon\right] \\
&\quad +8\hat{m}^4\frac{(2-x_1-x_2)^2(1-2\epsilon)}{(1-x_1)^2(1-x_2)^2}, \\
f^A_{1,1} &= 2[2\hat{m}^2(2\epsilon-1)-\epsilon]\frac{\sqrt{x_1^2-4\hat{m}^2}\sqrt{x_2^2-4\hat{m}^2}}{(1-x_1)(1-x_2)}, \\
f^A_{2,0} &= \frac{x_1^2(1-\epsilon)}{(1-x_1)(1-x_2)} - 2\hat{m}^2\left[\frac{x_1^2(3-2x_1-x_2)+2(1-x_1)}{(1-x_1)^2(1-x_2)} - \frac{2(x_1^2+1)\epsilon}{(1-x_1)(1-x_2)}\right] \\
&\quad +8\hat{m}^4\left[\frac{3-2x_1-x_2}{(1-x_1)^2(1-x_2)} - \frac{2\epsilon}{(1-x_1)(1-x_2)}\right], \\
f^A_{0,2} &= f^A_{2,0} \quad (1\leftrightarrow 2).
\end{aligned}$$

## E.2 Soft distribution coefficients

The functions $M_C^i(z,\epsilon)$ in (12.87) describe the soft cross-section in each of the thrust regions ($i=1,2,3$). Using the auxiliary function

$$h(z,\epsilon)\equiv\frac{3}{16}\frac{[z(1-z)-\hat{m}^2]^{1-\epsilon}}{z^2(1-z)^2}, \tag{E.2}$$

they can be put as

$$\begin{aligned}
M_V^1(z,\epsilon) &= h(z,\epsilon)\Big\{(2+4\hat{m}^2-3\epsilon)S_{\rm un}+(4\hat{m}^2-\epsilon)S_{\rm or}\Big\}, \\
M_V^3(z,\epsilon) &= \frac{h(z,\epsilon)}{1-\epsilon}\Big\{[2+4\hat{m}^2+3\epsilon^2+(2z(1-z)-6\hat{m}^2-5)\epsilon]S_{\rm un} \\
&\quad +[4z(1-z)+\epsilon^2-(2z(1-z)+2\hat{m}^2+1)]S_{\rm or}\Big\}, \\
M_A^1(z,\epsilon) &= h(z,\epsilon)(1-4\hat{m}^2)\Big\{(2-3\epsilon)S_{\rm un}-\epsilon S_{\rm or}\Big\}, \\
M_A^3(z,\epsilon) &= \frac{h(z,\epsilon)}{1-\epsilon}\Big\{[2-8\hat{m}^2+(3-12\hat{m}^2)\epsilon^2+(2z(1-z)-18\hat{m}^2-5)\epsilon]S_{\rm un} \\
&\quad +[4z(1-z)-4\hat{m}^2+(1-4\hat{m}^2)\epsilon^2-(2z(1-z)-6\hat{m}^2+1)]S_{\rm or}\Big\}.
\end{aligned} \tag{E.3}$$

When expanded in $\epsilon$ as $M_C^i(z,\epsilon)=M_C^{i,(0)}(z,\epsilon)+\epsilon M_C^{i,(1)}(z,\epsilon)+\mathcal{O}(\epsilon^2)$ the following logarithm appears

$$L_z=z(1-z)-\hat{m}^2, \tag{E.4}$$

and the coefficients take the form

$$\begin{aligned}
M_V^{1,(0)}(z,\epsilon) &= 4h(z,0)\Big\{\frac{1+2\hat{m}^2}{2}S_{\rm un}+\hat{m}^2S_{\rm or}\Big\}, \\
M_V^{1,(1)}(z,\epsilon) &= -h(z,0)\Big\{[(2+4\hat{m}^2)L_z+3]S_{\rm un}+[4\hat{m}^2L_z+1]S_{\rm or}\Big\}, \\
M_V^{3,(0)}(z,\epsilon) &= 4h(z,0)\Big\{\frac{1+2\hat{m}^2}{2}S_{\rm un}+z(1-z)S_{\rm or}\Big\}, \\
M_V^{3,(1)}(z,\epsilon) &= -h(z,0)\Big\{[(2+4\hat{m}^2)L_z-2z(1-z)+3+2\hat{m}^2]S_{\rm un} \\
&\quad +[4z(1-z)L_z-2z(1-z)+2\hat{m}^2+1]S_{\rm or}\Big\}, \\
M_A^{1,(0)}(z,\epsilon) &= 2h(z,0)(1-4\hat{m}^2)S_{\rm un}, \\
M_A^{1,(1)}(z,\epsilon) &= -h(z,0)(1-4\hat{m}^2)\Big\{[2L_z+3]S_{\rm un}+S_{\rm or}\Big\}, \\
M_A^{3,(0)}(z,\epsilon) &= 2h(z,0)\Big\{[1-4\hat{m}^2]S_{\rm un}+2[z(1-z)-\hat{m}^2]S_{\rm or}\Big\}, \\
M_A^{3,(1)}(z,\epsilon) &= -h(z,0)\Big\{[(2-8\hat{m}^2)L_z-2z(1-z)+3-10\hat{m}^2]S_{\rm un} \\
&\quad +[4(z(1-z)-\hat{m}^2)L_z-2z(1-z)-2\hat{m}^2+1]S_{\rm or}\Big\}.
\end{aligned} \tag{E.5}$$

Lastly, in the computation of the event-shape independent part of the delta coefficient $A_e^{\rm soft}$, one has to integrate over the $\mathcal{O}(\epsilon)$ term in the above expansion for the thurst region 1. The values of these integrals, written solely in terms of $\beta = \sqrt{1-4\hat{m}^2}$, are

$$\begin{aligned}
\int_{(1-\beta)/2}^{(1+\beta)/2} \mathrm{d}z\, M_V^{1,(1)}(z) &= \frac{3\beta}{8} g_{\rm soft}(\beta)\Big[(3-\beta^2)S_{\rm un} + (1-\beta^2)S_{\rm or}\Big] \\
&\quad + \frac{3\beta}{8}\bigg[\bigg(3 + \frac{\beta^2+9}{2\beta}L_\beta\bigg)S_{\rm un} + \bigg(1 + \frac{3-\beta^2}{2\beta}L_\beta\bigg)S_{\rm or}\bigg], \\
\int_{(1-\beta)/2}^{(1+\beta)/2} \mathrm{d}z\, M_A^{1,(1)}(z) &= \frac{3\beta^3}{4} g_{\rm soft}(\beta) S_{\rm un} + \frac{3\beta^3}{8}\bigg[\bigg(3 + \frac{3\beta^2+7}{2\beta}L_\beta\bigg)S_{\rm un} \\
&\quad + \bigg(1 + \frac{\beta^2+1}{2\beta}L_\beta\bigg)S_{\rm or}\bigg],
\end{aligned} \tag{E.6}$$

where $g_{\rm soft}(\beta)$ can be found in (12.111).

# Appendix F
# Two- and three-particle phase space

In this appendix we develop the tools required to perform the different phase-space integrations in part III of this thesis.

In section F.1 we start with the usual definition of the $N$-particle phase-space and introduce the notation and some general results and properties. We also discuss the choice of reference frame for the decay and scattering processes. In sections F.2 and F.3 we detail the derivation of the two- and three-particle phase-space, and in the case of the three-particle phase-space we derive the analytic form of the curves determining the integration region for the case $m_1 = m_2 = m$, $m_3 = 0$.

The remaining sections are dedicated to more advanced topics. In section F.4 we generalize the two- and three-particle phase-space to $d$ dimensions and particularize for the dimensional regularization choice $d = 4 - 2\epsilon$. Section F.5 explores the limit in which the massless particle 3 is soft and finally, section F.6 details how to express the two- and three-particle cross sections onto the thrust axis.

## F.1 Decay widths, cross-sections and computation strategy

### F.1.1 General definitions and formulas

We consider processes with $N$ particles in the final state, which arise either from the decay of a single particle or from of the collision of two particles, i.e.,

$$a \to 1 + \cdots + N, \qquad\qquad a + b \to 1 + \cdots + N. \tag{F.1}$$

The decay width and cross-section of these processes are given by Fermi's golden rule as an integration of the squared matrix element $|\mathcal{M}|^2$ over the $N$-particle Lorentz-invariant phase-space,

$$\Gamma(a\to 1+\cdots+N)\equiv\frac{1}{2m_a}\int \mathrm{d}\Phi_N|\mathcal{M}(a\to 1+\cdots+N)|^2, \tag{F.2}$$

$$\sigma(a+b\to 1+\cdots+N)\equiv\frac{1}{2\lambda^{1/2}(s,m_a^2,m_b^2)}\int \mathrm{d}\Phi_N|\mathcal{M}(a+b\to 1+\cdots+N)|^2, \tag{F.3}$$

where $s$ is the square of the initial state total momentum, i.e., $s=p_a^2$ in the case of the decay and $s=(p_a+p_b)^2$ in the case of the scattering, and $m_a$ and $m_b$ denote the masses of the particles $a$ and $b$, respectively. The totally-symmetric Källén function is defined by[F.1]

$$\lambda(x,y,z)\equiv x^2+y^2+z^2-2xy-2xz-2yz=(x+y-z)^2-4xy. \tag{F.4}$$

In momentum space, the phase space for $N$ particles has the form

$$\int \mathrm{d}\Phi_N\equiv\prod_{j=1}^{N}\int\frac{\mathrm{d}^4p_j}{(2\pi)^3}\delta(p_j^2-m_j^2)\theta(p_j^0)(2\pi)^4\delta^{(4)}\left(P^\mu-\sum_{i=1}^{N}p_i^\mu\right), \tag{F.5}$$

where $P^\mu$ is the total momentum of the initial state, $p_i^\mu$ are the momenta of the particles in the final state, and the squared momenta is given by the Minkowskian metric, $p^2=(p^0)^2-|\vec{p}|^2$, according to which the modulus of the spatial part is positive definite, $|\vec{p}|\geq 0$. The mass parameters are also positive definite, $m_i^2\geq 0$, and we implement the shorthand notation of having only one integral sign but as many integrals as differentials. These integrals extend to all momentum space, but the Heaviside step and Dirac delta functions restrict the integration region to the physical region, with the former imposing positive $p_i^0$ and the latter ruling out off-shell momenta and those combinations violating four-momentum conservation. Lastly, we observe that due to $m_a,m_b\geq 0$, one has $s>0$, as can be seen from the CM frame, where either $s=m_a^2$ or $s=m_a^2+m_b^2+2p_a^0p_b^0$.

The form in (F.4) contains $4N$ integration variables and $N+4$ constraints imposed by the on-shell delta functions and the momentum conserving delta, which amounts for a total of $3N-4$ integrations. Moreover, all these delta integrations can be carried out without the explicit knowledge of the integrand $|\mathcal{M}|^2$, since they are solved by using the identity

$$\int_a^b \mathrm{d}x\, f(x)\delta(x-x_0)=f(x_0)\theta(b-x_0)\theta(x_0-a), \tag{F.6}$$

F.1. Moreover, the factors $1/(2m_a)$ and $1/(2\lambda^{1/2}(s,m_a,m_b))$ in front of the integrals are referred to as the flux factors.

where the Heaviside step functions ensure the integral vanishes unless $x_0 \in [a, b]$ and where no knowledge of the integrand $f(x)$ is required. After all delta integrations are solved, the conditions of on-shell external momenta and momentum conservation are imposed on $|\mathcal{M}|^2$. This is the reason why the computation of decay widths and cross-sections is usually split into the following two separate steps:

1. Reduction, ignoring $|\mathcal{M}|^2$, of the phase space by solving the $N+4$ delta integrations. In some cases it is possible to employ geometric arguments to argue the independence of $|\mathcal{M}|^2$ on one or more of the remaining $3N-4$ coordinates, further simplifying the phase space[F.2].

2. Computation of $|\mathcal{M}|^2$ from Feynman diagrams imposing on-shell external momenta and momentum conservation to express it in terms of the phase-space integration variables left in the previous step.

Once this two steps are completed, (F.2) and (F.3) allow to obtain the differential and total decay width and cross section.

### F.1.2 *N*-particle phase space

Before facing particular cases, let us simplify the $N$-particle phase space as much as possible. First, we can use the on-shell delta function of each momentum to solve one of the four $\mathrm{d}^4 p_i$ integrals. We show this manipulation explicitly, as it illustrates some of the techniques later required. The usual choice is to solve for $p_i^0$ –because that leaves the Euclidean integral $\mathrm{d}^3\vec{p}_i$–, so we explicitly write the integration in terms of this variable:

$$\int \frac{\mathrm{d}^4 p_i}{(2\pi)^3}\delta(p_i^2 - m^2)\theta(p_i^0) = \int \frac{\mathrm{d}^3\vec{p}_i}{(2\pi)^3}\int_{-\infty}^{\infty}\mathrm{d}p_i^0\, \delta\big[(p_i^0)^2 - |\vec{p}_i|^2 - m^2\big]\theta(p_i^0). \tag{F.7}$$

Instead of having the form in (F.6), the argument of the delta function is a function of the integration variable, so the next step is to apply the property

$$\delta(f(x)) = \sum_i \frac{\delta(x - x_i)}{|f'(x_i)|} = \sum_i \frac{\theta(f'(x_i)) - \theta(-f'(x_i))}{f'(x_i)}\delta(x - x_i), \tag{F.8}$$

F.2. For example, due to momentum conservation, both the decay and the scattering with two particles in the final state happen in a plane, so $\mathcal{M}$ cannot depend on the perpendicular components.

where the $x_i$ are the roots of $f(x_i)$ and where the second equality arises from expressing the absolute value as

$$|x| = x\theta(x) - x\theta(-x) = x[\theta(x) - \theta(-x)]. \tag{F.9}$$

In the second equality of (F.8) we have also used that the Heaviside functions are mutually exclusive and therefore each fraction can be inverted individually[F.3]. In our case, the roots and the derivatives of the argument are

$$\begin{aligned} p^0 &= \pm\sqrt{|\vec{p}_i|^2 + m_i^2}, \\ \frac{\mathrm{d}}{\mathrm{d}p^0}\big[(p_i^0)^2 - |\vec{p}_i|^2 - m^2\big]\bigg|_{p^0 = \pm\sqrt{|\vec{p}_i|^2 + m^2}} &= \pm 2\sqrt{|\vec{p}_i|^2 + m_i^2}, \end{aligned} \tag{F.11}$$

so that, making use that they have definite sign ($\sqrt{|\vec{p}_i|^2 + m^2} > 0$), we rewrite[F.4]

$$\delta(p_i^2 - m^2) = \frac{1}{2\sqrt{|\vec{p}_i|^2 + m^2}}\Big[\delta\Big(p_i^0 - \sqrt{|\vec{p}_i|^2 + m_i^2}\Big) + \delta\Big(p_i^0 + \sqrt{|\vec{p}_i|^2 + m_i^2}\Big)\Big]. \tag{F.12}$$

Then,

$$\int \frac{\mathrm{d}^4 p_i}{(2\pi)^3}\delta(p_i^2 - m^2)\theta(p_i^0) = \int \frac{\mathrm{d}^3\vec{p}_i}{(2\pi)^3}\frac{\theta(E_i)}{2E_i} \tag{F.13}$$

where $E_i \equiv \sqrt{|\vec{p}_i|^2 + m^2}$ and $\theta(p^0)$ gets evaluated to $\theta\big(\sqrt{|\vec{p}_i|^2 + m^2}\big)$ after the delta integration with (F.6), picking the positive $p_0$ term in (F.12) via

$$\theta(x)\theta(x) = \theta(x), \qquad \theta(x)\theta(-x) = 0. \tag{F.14}$$

Also, both Heaviside functions arising from the integration are 1 due to the infinite integration interval. Performing this manipulation for each integral in the first line of (F.4) gives the following form for the phase space

$$\int \mathrm{d}\Phi_N = \prod_{j=1}^{N}\int \frac{\mathrm{d}^3\vec{p}_j\,\theta(E_j)}{2(2\pi)^3 E_j}(2\pi)^4\delta^{(4)}\Bigg(P^\mu - \sum_{i=1}^{N} p_i^\mu\Bigg). \tag{F.15}$$

F.3. Explicitly,

$$\Big[\frac{a}{b}\theta(x) + \frac{c}{d}\theta(-x)\Big]^{-1} = \frac{bd}{ad\theta(x) + cb\theta(-x)} = \frac{b}{a}\theta(x) + \frac{d}{c}\theta(-x) = \Big(\frac{a}{b}\Big)^{-1}\theta(x) + \Big(\frac{c}{d}\Big)^{-1}\theta(-x). \tag{F.10}$$

F.4. In the first equality we also used $\theta(ax) = \theta(x)$ if $a > 0$.

To further simplify it, we now turn our attention to the four-dimensional delta function, which can be split as

$$\delta^{(4)}\left(P^\mu-\sum_{i=1}^N p_i^\mu\right)=\delta\left(E-\sum_{i=1}^N E_i\right)\delta^{(3)}\left(\vec{P}-\sum_{i=1}^N \vec{p}_i\right), \tag{F.16}$$

where we defined $P^\mu \equiv (E, \vec{P})$. The spatial part can be used to immediately solve one of the three-dimensional integrals in (F.15), which for simplicity we choose to be the $N$-th. Since the integration over $\mathrm{d}^3\vec{p}_N$ is carried out for all possible values of $\vec{p}_N$, the delta function always crosses 0 and therefore both Heaviside functions in (F.6) are automatically 1. Then the phase space is

$$\int \mathrm{d}\Phi_N=\frac{1}{2^{4(N-1)}\pi^{3N-4}}\int \mathrm{d}^3\vec{p}_1\ldots\mathrm{d}^3\vec{p}_{N-1}\frac{\theta(E_1)\ldots\theta(E_N^*)}{E_1\ldots E_N^*}\delta\left(E-\sum_{i=1}^{N-1}E_i-E_N^*\right), \tag{F.17}$$

where $E_N$, which was the integrand of the integral over $\vec{p}_N$, gets evaluated in the tri-momentum conservation condition, giving

$$E_N^* \equiv E_N\Big|_{\vec{P}=\sum_{i=1}^{N-1}\vec{p}_i}=\left[\left|\vec{P}-\sum_{i=1}^{N-1}\vec{p}_i\right|^2+m_N^2\right]^{1/2}. \tag{F.18}$$

The next step is to match the integration variables with the ones in the argument of the delta, for which we change to spherical coordinates[F.5]:

$$\mathrm{d}^3\vec{p}_i=|\vec{p}_i|^2\sin\theta_i\mathrm{d}|\vec{p}_i|\mathrm{d}\theta_i\mathrm{d}\varphi_i=|\vec{p}_i|^2\mathrm{d}|\vec{p}_i|\mathrm{d}\cos\theta_i\mathrm{d}\varphi_i, \qquad i=1,2,\ldots,N-1. \tag{F.19}$$

Note that the appropriate ranges of the three scalar integration variables to cover all momentum space are

$$|\vec{p}_i|\in(0,\infty), \qquad \theta_i\in(0,\pi) \text{ or } \cos\theta_i\in(-1,1), \qquad \varphi_i\in(0,2\pi). \tag{F.20}$$

To be consistent $E_N^*$ needs also to be expressed in terms of the spherical coordinates. This can be achieved by using the sum identity

$$\left(\sum_{i=0}^n x_i\right)^2=\sum_{i=0}^n x_i^2+2\sum_{i=0}^n x_i\sum_{j=i+1}^n x_j, \tag{F.21}$$

F.5. Convention dictates that the minus sign arising from $\mathrm{d}\cos\theta_i=-\sin\theta_i\mathrm{d}\theta_i$ gets compensated by inverting the order of integration in $\mathrm{d}\cos\theta_i$, which is taken to go from $-1$ to 1 instead of from $\cos(0)=1$ to $\cos(\pi)=-1$, so it is not explicitly written in the differential transformation. As a consequence, we perform the abuse of notation of writing $\mathrm{d}\cos\theta_i=\sin\theta_i\mathrm{d}\theta_i$, which is correct as long as the differential is integrated over.

which is also valid for vectors, on (F.18). Defining $\vec{p}_0 \equiv -\vec{P}$ one has

$$\left|\vec{P}-\sum_{i=1}^{N-1}\vec{p}_i\right|^2=\left|\sum_{i=0}^{N-1}\vec{p}_i\right|^2=|\vec{P}|^2+\sum_{i=1}^{N-1}|\vec{p}_i|^2+2\sum_{i=0}^{N-1}\sum_{j=i+1}^{N-1}|\vec{p}_i||\vec{p}_j|\cos\theta_{ij}. \tag{F.22}$$

In a generic reference frame, tri-momenta can be written in spherical coordinates as

$$\vec{p}_i=|\vec{p}_i|(\sin\theta_i\cos\varphi_i,\sin\theta_i\sin\varphi_i,\cos\theta_i), \tag{F.23}$$

and therefore the angle $\theta_{ij}$ between $\vec{p}_i$ and $\vec{p}_j$ takes the form

$$\cos\theta_{ij}=\frac{\vec{p}_i\cdot\vec{p}_j}{|\vec{p}_i||\vec{p}_j|}=\sin\theta_i\sin\theta_j(\cos\varphi_i\cos\varphi_j+\sin\varphi_i\sin\varphi_j)+\cos\theta_i\cos\theta_j. \tag{F.24}$$

In the general case of $N$ particles and a reference frame with generic orientation all the integration variables (moduli and angles) appear in the integrand through the crossed products in $E_N^*$. One can then see that some choices for the axes orientations can eliminate the dependence on some angles, rendering the corresponding integrals to be immediately solvable[F.6]. A way of reducing the complexity of the general case is to perform a change of variable from the moduli to the energies by differentiating the on-shell relation, i.e.,

$$E_i^2-|\vec{p}_i|^2=m_i^2 \implies \frac{|\vec{p}_i|^2\mathrm{d}|\vec{p}_i|}{E_i}=\sqrt{E_i^2-m_i^2}\,\mathrm{d}E_i, \tag{F.25}$$

where now $E_i\in(m_i,\infty)$. With this we arrive at

$$\begin{aligned}\int\mathrm{d}\Phi_N &= \frac{1}{2^{4(N-1)}\pi^{3N-4}}\int\mathrm{d}E_1\mathrm{d}\cos\theta_1\mathrm{d}\varphi_1\ldots\mathrm{d}E_{N-1}\mathrm{d}\cos\theta_{N-1}\mathrm{d}\varphi_{N-1}\\ &\quad\times\frac{\sqrt{E_1^2-m_1^2}\ldots\sqrt{E_{N-1}^2-m_{N-1}^2}}{E_N^*}\theta(E_1)\ldots\theta(E_N^*)\delta\left(E-\sum_{i=1}^{N-1}E_i-E_N^*\right),\end{aligned} \tag{F.26}$$

where by the on-shell relation and $P^2=s$,

$$\begin{aligned}E_N^* &= \Bigg[E^2-s+\sum_{i=1}^{N-1}(E_i^2-m_i^2)+m_N^2\\ &\quad+2\sum_{i=1}^{N-1}\left(\sqrt{E^2-s}\,|\vec{p}_i|\cos\theta_{i0}+\sqrt{E_i^2-m_i^2}\sum_{j=i+1}^{N-1}\sqrt{E_j^2-m_j^2}\cos\theta_{ij}\right)\Bigg]^{1/2}.\end{aligned} \tag{F.27}$$

F.6. For example, choosing the $z$ axis along one of the $\vec{p}_i$ immediately eliminates the dependence on $\theta_i$ and $\varphi_i$. In this case one can immediately solve $\int\mathrm{d}\cos\theta_i\int\mathrm{d}\varphi_i=4\pi$.

The next step is to choose one of the remaining integration variables, use (F.8) to take the energy delta function to a form that is linear on it and integrate over with (F.6). Due to the complexity of (F.27) we will not carry on with the generic computation. Nevertheless, there is an important comment to be made when transforming the delta function with (F.8). The Heaviside and delta functions in (F.26) impose the conditions $E_i \geq 0$ and $E - \sum_{i=1}^{m} E_i \geq 0$ for any $m = 1, \ldots, N-1$. However, due to the quadratic character of the on-shell relation and the square root in (F.27), when finding the roots of the argument of the delta one has to take the square,

$$\left(E - \sum_{i=1}^{N-1} E_i\right)^2 = (E_N^*)^2 \implies E_N^* = \pm\left|E - \sum_{i=1}^{N-1} E_i\right|, \tag{F.28}$$

and the extra solution $E - \sum_{i=1}^{m} E_i \leq 0$ seems to appear: the function $\theta(E_N^*)$ cancels the negative solution in (F.28), yet the absolute value remains. After the manipulation the energy delta function simply takes the linear form $\delta(x_i - \tilde{x}_i)$, for $x = E_i$, $\theta_i$, $\varphi_i$, $i = 1, \ldots, N-1$, which does not cancel the extra solution. It is a known fact that when squaring terms in an equation involving square roots spurious solutions might be introduced. Therefore one always needs to check which of the multiple solutions obtained after squaring are indded solution of the original equation[F.7]. In our case, where the variables $E_1$ to $E_{N-1}$ are integrated over, the way of discarding the spureous solutions is by enforcing the energy comes out positive. This is achieved by introducing a Heaviside function as

$$\delta\left(E - \sum_{i=1}^{N-1} E_i - E_N^*\right) = \sum_j \frac{\delta(x - \tilde{x}_i)}{\left|\frac{\mathrm{d}}{\mathrm{d}x}(E - \sum_{i=1}^{N-1} E_i - E_N^*)\big|_{x=\tilde{x}_i}\right|} \theta\left(E - \sum_{i=1}^{N-1} E_i\right). \tag{F.31}$$

The last simplification of the general, $N$-particle case comes from making use of the freedom to choose the reference frame in which we perform the computation. First, we set the origin in the CM frame of the initial state, so that $P^\mu \equiv (\sqrt{s}, \vec{0})$. With this choice one has $E \to \sqrt{s}$ and $\vec{P} \to \vec{0}$ in all the previous expressions. Second,

F.7. As a simple example, consider the the equation $x - a = 0$ and proceed to solve it by taking squares:

$$x - a = 0 \implies x^2 = a^2 \implies x = \pm\sqrt{a^2} = \pm|a|. \tag{F.29}$$

The unique solution to the first equation on (F.29) is $x = a$, yet after taking squares the equation admits both $x = a$ and $x = -a$ as solutions. The relevant point is to realize that by taking squares one is implicitly assuming the initial solution $x = a$

$$x = a \implies x \cdot x = a \cdot x \overset{x=a}{\implies} x^2 = a^2. \tag{F.30}$$

Therefore, the solution $x = -a$ must be discarded.

with regards to the orientation of the axes, the key idea is to remember that in 3 dimensions two non-parallel vectors $\vec{v}_1$ and $\vec{v}_2$ define three mutually perpendicular axes: one axis is aligned with one of the vectors, namely $\vec{v}_1$, a second axis is aligned with the component of $\vec{v}_2$ that is perpendicular to $\vec{v}_1$, and the remaining axis is fixed to be mutually perpendicular to the two previous ones. This idea leads to appropriate choices for each the decay and the scattering process, which are depicted in figure F.1.

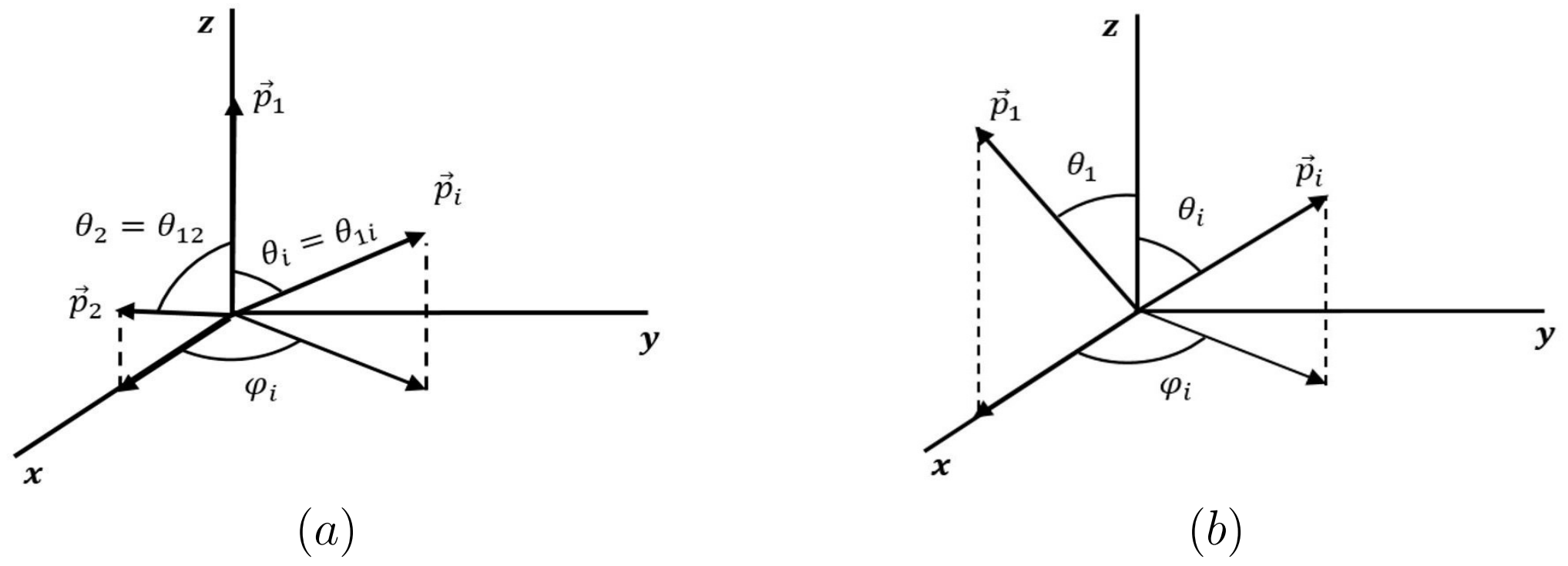

**Figure F.1.** Convenient reference frames for decay (a) and scattering (b) processes.

In the decay (figure F.1(*a*)), the CM frame sets $\vec{p}_a = \vec{0}$, so the initial particle is at rest and does not define any direction. We are free then to choose the axis orientation in terms of the particles in the final state:

1. We choose to align the $z$-axis with $\vec{p}_1$, making $\theta_1 = \varphi_1 = 0$ and $\theta_{1i} = \theta_i$.

2. We choose to align the $x$-axis with the component of $\vec{p}_2$ that is perpendicular to the $z$-axis, which makes $\varphi_2 = 0$[F.8].

Under these choices neither $E_N^*$ nor $|\mathcal{M}|^2$ can depend on $\theta_1$, $\varphi_1$ and $\varphi_2$, so we can integrate them over –they give a factor of $4\pi$ from particle 1 and $2\pi$ from particle 2– to finally get

$$\begin{aligned}\int \mathrm{d}\Phi_N &= \frac{\delta_{N,2} + 2\pi(1-\delta_{N,2})}{2^{4N-6}\pi^{3N-5}} \qquad \text{(F.32)}\\ &\times \int \mathrm{d}E_1 \mathrm{d}E_2 \mathrm{d}\cos\theta_2 \mathrm{d}E_3 \mathrm{d}\cos\theta_3 \mathrm{d}\varphi_3 \ldots \mathrm{d}E_{N-1} \mathrm{d}\cos\theta_{N-1} \mathrm{d}\varphi_{N-1}\\ &\times \frac{\sqrt{E_1^2 - m_1^2} \ldots \sqrt{E_{N-1}^2 - m_{N-1}^2}}{E_N^*} \theta(E_1) \ldots \theta(E_N^*) \delta\left(\sqrt{s} - \sum_{i=1}^{N-1} E_i - E_N^*\right),\end{aligned}$$

F.8. We choose $\vec{p}_1$ and $\vec{p}_2$ as the defining vectors for our axis orientation due to the fact that they will be present in all of the $N$-particle phase spaces –with the exception of $N = 2$–.

where the $\delta_{N,2}$ comes from the fact that for $N=2$ the integrals over the variables of particle 2 were solved in (F.17) with the tri-momentum conserving delta, so that the $2\pi$ contribution from $\varphi_2$ must not be included.

For the scattering process (figure F.1(*b*)) the CM frame choice sets $\vec{p}_a = -\vec{p}_b$, so particles $a$ and $b$ travel along the same line. This defines a direction, referred to as the direction of the incoming beam. We then define our reference frame:

1. We choose the direction of the incoming beam to be the $z$-axis. This imposes $\theta_{0i} = \theta_i$.

2. We choose the $x$-axis along the component of $\vec{p}_1$ that is perpendicular to the $z$-axis, so that $\varphi_1 = 0$.

Integrating (F.26) over $\varphi_1$ –to a factor of $2\pi$– the final phase space is

$$\int \mathrm{d}\Phi_N = \frac{1}{2^{4N-5}\pi^{3N-5}} \int \mathrm{d}E_1 \mathrm{d}\cos\theta_1 \mathrm{d}E_2 \mathrm{d}\cos\theta_2 \mathrm{d}\varphi_2 \ldots \mathrm{d}E_{N-1} \mathrm{d}\cos\theta_{N-1} \mathrm{d}\varphi_{N-1} \times \frac{\sqrt{E_1^2 - m_1^2} \ldots \sqrt{E_{N-1}^2 - m_{N-1}^2}}{E_N^*} \theta(E_1) \ldots \theta(E_N^*) \delta\left(\sqrt{s} - \sum_{i=1}^{N-1} E_i - E_N^*\right) \tag{F.33}$$

In both cases, the functional form of $E_N^*$ in the CM frame is

$$E_N^* = \sqrt{\sum_{i=1}^{N-1} (E_i^2 - m_i^2) + 2\sum_{i=1}^{N-1} \sum_{j=i+1}^{N-1} \sqrt{E_i^2 - m_i^2} \sqrt{E_j^2 - m_j^2} \cos\theta_{ij} + m_N^2}, \tag{F.34}$$

where the angles $\theta_{ij}$ have dependence on the polar and azimuthal angles of particles $i$ and $j$ (see (F.24)). Note that the choice of reference frame sets two angles to zero in the case of the decay and one in the case of the scattering, leaving a total of $3N-6$ and $3N-5$ integrations left, respectively.

Finally, one may argue that the functions $\theta(E_i)$ are redundant since by definition $E_i$ is positive and is integrated from $m_i \geq 0$ to infinity. This is not the case for $E_N^*$, that is not integrated over. The remaining $\theta(E_i)$ functions are necessary to prove the invariance of $\mathrm{d}\Phi_N$ on the minimal set of coordinates chosen to parametrize the process (see section F.3.2.3).

## F.2 The two-particle phase space

### F.2.1 Decay

From (F.32) and (F.34), the 2 particle phase space for a decay is

$$\int \mathrm{d}\Phi_2 = \frac{1}{4\pi}\int \mathrm{d}E_1 \frac{\sqrt{E_1^2 - m_1^2}}{E_2^*}\theta(E_1)\theta(E_2^*)\delta(\sqrt{s} - E_1 - E_2^*), \tag{F.35}$$

$$E_2^* = \sqrt{E_1^2 - m_1^2 + m_2^2}.$$

The remaining delta function can only be used to solve the integral over $E_1$, for which we use (F.8). The only root of the argument and the derivative are, respectively

$$\tilde{E}_1 = \frac{s + m_1^2 - m_2^2}{2\sqrt{s}}, \tag{F.36}$$

$$\frac{\mathrm{d}}{\mathrm{d}E_1}(\sqrt{s} - E_1 - E_2^*)\bigg|_{E_1 = \tilde{E}_1} = -\frac{2s\theta(s - m_1^2 + m_2^2)}{s - m_1^2 + m_2^2} - \frac{2(m_1^2 - m_2^2)\theta(-s + m_1^2 - m_2^2)}{-s + m_1^2 - m_2^2}.$$

The absolute value of the derivative only removes the global minus sign, since both terms are positive due to the Heaviside functions[F.9]. Therefore, by (F.8),

$$\delta(\sqrt{s} - E_1 - E_2^*) = (s - m_1^2 + m_2^2)\left[\frac{\theta(s - m_1^2 + m_2^2)}{2s} - \frac{\theta(-s + m_1^2 - m_2^2)}{2(m_1^2 - m_2^2)}\right] \times\theta(\sqrt{s} - E_1)\delta(E_1 - \tilde{E}_1), \tag{F.37}$$

where we have added the Heaviside function mentioned in (F.31) to discard the non-physical solutions introduced by squaring both sides in (F.36). With result (F.37) we have

$$\int \mathrm{d}\Phi_2 = \frac{1}{4\pi}\int \mathrm{d}E_1 \frac{\sqrt{E_1^2 - m_1^2}}{E_2^*}\left[\frac{(s - m_1^2 + m_2^2)\theta(s - m_1^2 + m_2^2)}{2s} + \frac{(m_1^2 - m_2^2 - s)\theta(-s + m_1^2 - m_2^2)}{2(m_1^2 - m_2^2)}\right]\theta(E_1)\theta(E_2^*)\theta(\sqrt{s} - E_1)\delta(E_1 - \tilde{E}_1), \tag{F.38}$$

F.9. In the second term note that since $s > 0$, $-s + m_1^2 - m_2^2 > 0$ implies $m_1^2 - m_2^2 > s > 0$.

where now the integral over $E_1$ can be explicitly carried out. In order for argument of the delta to cross zero one must have $\tilde{E}_1 \geq m_1$, which leads to[F.10]

$$s + m_1^2 - m_2^2 - 2m_1\sqrt{s} \geq 0. \tag{F.39}$$

Therefore, after integration one has

$$\left.\frac{\sqrt{E_1^2 - m_1^2}}{E_2^*}\right|_{E_1 = \tilde{E}_1} = \frac{\lambda^{1/2}(s, m_1^2, m_2^2)}{|s - m_1^2 + m_2^2|} \tag{F.40}$$

and all the Heaviside functions combine into $\theta(s - m_1^2 + m_2^2)\theta(s + m_1^2 - m_2^2 - 2m_1\sqrt{s})$. The function $\theta(s - m_1^2 + m_2^2)$ cancels the second contribution in (F.38) and the phase space takes the form

$$\int \mathrm{d}\Phi_2 = \frac{\lambda^{1/2}(s, m_1^2, m_2^2)}{8\pi s}\theta(s - m_1^2 + m_2^2)\theta(s + m_1^2 - m_2^2 - 2m_1\sqrt{s}). \tag{F.41}$$

The Heaviside theta functions can be safely ignored since there are no integrals left and they only impose momentum conservation on $|\mathcal{M}|^2$, so we can compactly write

$$\int \mathrm{d}\Phi_2 = \frac{\lambda^{1/2}(s, m_1^2, m_2^2)}{8\pi s}. \tag{F.42}$$

### F.2.2 Scattering

From (F.33) and (F.34), the 2 particle phase space for a scattering process is

$$\int \mathrm{d}\Phi_2 = \frac{1}{8\pi}\int \mathrm{d}E_1 \mathrm{d}\cos\theta_1 \frac{\sqrt{E_1^2 - m_1^2}}{E_2^*}\theta(E_1)\theta(E_2^*)\delta(\sqrt{s} - E_1 - E_2^*) \tag{F.43}$$
$$E_2^* = \sqrt{E_1^2 - m_1^2 + m_2^2}.$$

Since the delta function only depends on $E_1$, we can use the computation in the decay case and immediately write

$$\int \mathrm{d}\Phi_2 = \frac{\lambda^{1/2}(s, m_1^2, m_2^2)}{16\pi s}\int_{-1}^{1} \mathrm{d}\cos\theta_1. \tag{F.44}$$

F.10. The superior limit condition is $\tilde{E}_1 < \infty$, which is always satisfied.

Evidently, integrating (F.44) leads to (F.42). Also, in the CM frame $\theta_1 = -\theta_2$, so (F.44) is invariant under the choice of what particle remains in the final integration.

# F.3 The three-particle phase space

## F.3.1 Decay

### F.3.1.1 Standard form

From (F.32) and (F.34), the 3 particle phase space for a decay is

$$\begin{aligned} \int \mathrm{d}\Phi_3 &= \frac{1}{32\pi^3}\int \mathrm{d}E_1 \mathrm{d}E_2 \mathrm{d}\cos\theta_2 \frac{\sqrt{E_1^2 - m_1^2}\sqrt{E_2^2 - m_2^2}}{E_3^*} \\ &\quad \times \theta(E_1)\theta(E_2)\theta(E_3^*)\delta(\sqrt{s} - E_1 - E_2 - E_3^*), \\ E_3^* &= \sqrt{E_1^2 + E_2^2 + 2\sqrt{E_1^2 - m_1^2}\sqrt{E_2^2 - m_2^2}\cos\theta_2 + m_3^2 - m_1^2 - m_2^2}, \end{aligned} \tag{F.45}$$

where we used that, in the reference frame of figure F.1(*a*), $\theta_{12} = \theta_2$. The delta function can be used to solve the integral either over $E_1$, $E_2$ or $\cos\theta_2$; we choose $\cos\theta_2$ due to its simpler dependence. The root of the argument of for $\cos\theta_2$ is

$$\cos\tilde{\theta}_2 \equiv \frac{s + 2E_1E_2 - 2(E_1 + E_2)\sqrt{s} + m_1^2 + m_2^2 - m_3^2}{2\sqrt{E_1^2 - m_1^2}\sqrt{E_2^2 - m_2^2}}. \tag{F.46}$$

When the derivative of the argument is evaluated at $\cos\tilde{\theta}_2$ one finds

$$\left| \frac{\mathrm{d}}{\mathrm{d}\cos\theta_2}(\sqrt{s} - E_1 - E_2 - E_3^*)\bigg|_{\theta_2 = \tilde{\theta}_2} \right| = \frac{\sqrt{E_1^2 - m_1^2}\sqrt{E_2^2 - m_2^2}}{|\sqrt{s} - E_1 - E_2|}. \tag{F.47}$$

Thus,

$$\delta(\sqrt{s} - E_1 - E_2 - E_3^*) = \frac{\sqrt{s} - E_1 - E_2}{\sqrt{E_1^2 - m_1^2}\sqrt{E_2^2 - m_2^2}}\theta(\sqrt{s} - E_1 - E_2)\delta(\cos\theta_2 - \cos\tilde{\theta}_2), \tag{F.48}$$

and substituting back into the phase space we get

$$\int d\Phi_3 = \frac{1}{32\pi^3}\int dE_1 dE_2 d\cos\theta_2 \frac{\sqrt{s}-E_1-E_2}{E_3^*} \times\theta(E_1)\theta(E_2)\theta(E_3^*)\theta(\sqrt{s}-E_1-E_2)\delta(\cos\theta_2-\cos\tilde{\theta}_2). \tag{F.49}$$

When the delta integration is performed, since $\cos\theta_2$ is integrated along $(-1,1)$, the arising Heaviside step functions are

$$\theta(1-\cos\tilde{\theta}_2)\theta(\cos\tilde{\theta}_2+1)=\theta(1-|\cos\tilde{\theta}_2|)=\theta(1-\cos^2\tilde{\theta}_2)=\theta(\sin^2\tilde{\theta}_2), \tag{F.50}$$

On the other hand, the integrand evaluated at the delta condition is

$$E_3^*|_{\theta_2=\tilde{\theta}_2}=\sqrt{(\sqrt{s}-E_1-E_2)^2}=|\sqrt{s}-E_1-E_2|, \quad \theta(E_3^*)=\theta(|\sqrt{s}-E_1-E_2|)=1, \tag{F.51}$$

so the final form is

$$\int d\Phi_3=\frac{1}{32\pi^3}\int dE_1 dE_2 \Theta_{12}, \quad \Theta_{12}\equiv\theta(E_1)\theta(E_2)\theta(\sqrt{s}-E_1-E_2)\theta(\sin^2\tilde{\theta}_2). \tag{F.52}$$

The Heaviside functions grouped in $\Theta_{12}$ delimit the integration region in the $(E_1,E_2)$ plane. This integration region is found in section F.3.3.

#### F.3.1.2 Adimensionalization: the $x_i$ variables

There is a convenient parametrization of (F.52) in terms of the adimensional variables

$$x_i=\frac{2E_i}{\sqrt{s}}, \qquad \hat{m}_i=\frac{m_i}{\sqrt{s}}, \qquad i=1,2,3, \tag{F.53}$$

which have a range of $x_i\in(2\hat{m}_i,\infty)$ and satisfy the conditions

$$x_1+x_2+x_3=2, \quad \sqrt{x_1^2-4\hat{m}_1^2}\cos\theta_1+\sqrt{x_2^2-4\hat{m}_2^2}\cos\theta_2+\sqrt{x_3^2-4\hat{m}_3^2}\cos\theta_3=0, \tag{F.54}$$

where the last line comes from momentum conservation in the $z$-axis. In terms of these variables, the phase space reads

$$\begin{aligned}\int \mathrm{d}\Phi_3 &= \frac{s}{128\pi^3}\int \mathrm{d}x_1\mathrm{d}x_2\Theta_{12}, \\ \Theta_{12} &\equiv \theta(x_1)\theta(x_2)\theta(2-x_1-x_2)\theta(\sin^2\tilde{\theta}_{12}), \\ \cos\tilde{\theta}_{12} &= \frac{2+x_1x_2-2(x_1+x_2)+2(\hat{m}_1^2+\hat{m}_2^2-\hat{m}_3^2)}{\sqrt{x_1^2-4\hat{m}_1^2}\sqrt{x_2^2-4\hat{m}_2^2}}.\end{aligned} \tag{F.55}$$

Note we returned to the notation $\tilde{\theta}_2 = \tilde{\theta}_{12}$ for an easier generalization.

#### F.3.1.3 General form

Let us prove that the phase space in (F.56) is invariant under the choice of which two particles are left in the final integration, i.e., let us prove that

$$\begin{aligned}\int \mathrm{d}\Phi_3 &= \frac{s}{128\pi^3}\int \mathrm{d}x_i\mathrm{d}x_j\Theta_{ij}, \\ \Theta_{ij} &\equiv \theta(x_i)\theta(x_j)\theta(2-x_i-x_j)\theta(\sin^2\tilde{\theta}_{ij}), \\ \cos\tilde{\theta}_{ij} &\equiv \frac{2+x_ix_j-2(x_i+x_j)+2(\hat{m}_i^2+\hat{m}_j^2-\hat{m}_k^2)}{\sqrt{x_i^2-4\hat{m}_i^2}\sqrt{x_j^2-4\hat{m}_j^2}}, \quad i\neq j\neq k,\end{aligned} \tag{F.56}$$

is invariant under any choice for $i$ and $j$. To do that, the first line of (F.54) can be used to change variables, namely from $x_2$ to $x_3$: $\mathrm{d}x_2 = -\mathrm{d}x_3$. With this change the first three Heaviside functions in $\Theta_{12}$ transform as:

$$\theta(x_1)\theta(x_2)\theta(2-x_1-x_2)|_{x_2=2-x_1-x_3} \;=\; \theta(x_1)\theta(2-x_1-x_3)\theta(x_3). \tag{F.57}$$

To transform $\theta(\sin^2\tilde{\theta}_{12})$ we use the relation

$$\begin{aligned}\frac{\sqrt{x_2^2-4\hat{m}_2^2}}{\sqrt{x_3^2-4\hat{m}_3^2}}\cos\tilde{\theta}_{12} &= \frac{2+x_1(2-x_1-x_3)-2(2-x_3)+2(\hat{m}_1^2+\hat{m}_2^2-\hat{m}_3^2)}{\sqrt{x_1^2-4\hat{m}_1^2}\sqrt{x_3^2-4\hat{m}_3^2}} \\ &= \cos\tilde{\theta}_{13} + \frac{x_3^2-x_2^2+4(\hat{m}_2^2-\hat{m}_3^2)}{\sqrt{x_1^2-4\hat{m}_1^2}\sqrt{x_3^2-4\hat{m}_3^2}},\end{aligned} \tag{F.58}$$

which implies

$$\begin{aligned}\frac{x_2^2-4\hat{m}_2^2}{x_3^2-4\hat{m}_3^2}\sin^2\tilde{\theta}_{12} &= \frac{x_2^2-4\hat{m}_2^2}{x_3^2-4\hat{m}_3^2} - \frac{x_2^2-4\hat{m}_2^2}{x_3^2-4\hat{m}_3^2}\cos^2\tilde{\theta}_{12} \\ &= \frac{x_2^2-4\hat{m}_2^2}{x_3^2-4\hat{m}_3^2} - \left[\cos\tilde{\theta}_{13} + \frac{x_3^2-x_2^2+4(\hat{m}_2^2-\hat{m}_3^2)}{\sqrt{x_1^2-4\hat{m}_1^2}\sqrt{x_3^2-4\hat{m}_3^2}}\right]^2 = \sin^2\tilde{\theta}_{13}.\end{aligned} \tag{F.59}$$

where we used (F.24) in the linear term in $\cos\tilde{\theta}_{13}$. Therefore

$$\theta(\sin^2\tilde{\theta}_{12}) = \theta\left(\frac{x_3^2 - 4\hat{m}_3^2}{x_2^2 - 4\hat{m}_2^2}\sin^2\tilde{\theta}_{12}\right) = \theta(\sin^2\tilde{\theta}_{13}) \tag{F.60}$$

The same procedure can be done to put the phase space in terms of any pair $x_i$, $x_j$, so the generalization (F.56) holds.

## F.3.2 Scattering

### F.3.2.1 Standard form

From (F.33) and (F.34), the 3 particle phase space for a scattering process is

$$\begin{aligned} &\int d\Phi_3 = \frac{1}{2^7\pi^4}\int dE_1 d\cos\theta_1 dE_2 d\cos\theta_2 d\varphi_2 \frac{\sqrt{E_1^2 - m_1^2}\sqrt{E_2^2 - m_2^2}}{E_3^*} \\ &\times\theta(E_1)\theta(E_2)\theta(E_3^*)\delta(\sqrt{s} - E_1 - E_2 - E_3^*), \\ &E_3^* = \sqrt{E_1^2 + E_2^2 + 2\sqrt{E_1^2 - m_1^2}\sqrt{E_2^2 - m_2^2}\cos\theta_{12} + m_3^2 - m_1^2 - m_2^2}, \\ &\cos\theta_{12} = \sin\theta_1\sin\theta_2\cos\varphi_2 + \cos\theta_1\cos\theta_2, \end{aligned} \tag{F.61}$$

where we used (F.24) adapted to the reference frame in figure F.1($b$) for the expression of $\cos\theta_{12}$. In this case the delta function depends on all $E_1$, $E_2$, $\theta_1$, $\theta_2$ and $\varphi_2$ variables, and we observe that the dependence on $\varphi_2$ is simpler, since it only appears once through its cosine. We choose then to solve for $\varphi_2$, but, as an intermediate step that allows to reuse the computations carried out in the decay case, we first rewrite the delta function for $\cos\theta_{12}$:

$$\begin{aligned} &\delta(\sqrt{s} - E_1 - E_2 - E_3^*) = \frac{\sqrt{s} - E_1 - E_2}{\sqrt{E_1^2 - m_1^2}\sqrt{E_2^2 - m_2^2}}\theta(\sqrt{s} - E_1 - E_2)\delta(\cos\theta_{12} - \cos\tilde{\theta}_{12}), \\ &\cos\tilde{\theta}_{12} \equiv \frac{s + 2E_1E_2 - 2(E_1 + E_2)\sqrt{s} + m_1^2 + m_2^2 - m_3^2}{2\sqrt{E_1^2 - m_1^2}\sqrt{E_2^2 - m_2^2}}. \end{aligned} \tag{F.62}$$

With this

$$\begin{aligned} \int d\Phi_3 \;=\; &\frac{1}{128\pi^4}\int dE_1 d\cos\theta_1 dE_2 d\cos\theta_2 d\varphi_2 \frac{\sqrt{s} - E_1 - E_2}{E_3^*} \\ &\times\theta(E_1)\theta(E_2)\theta(E_3^*)\theta(\sqrt{s} - E_1 - E_2)\delta(\cos\theta_{12} - \cos\tilde{\theta}_{12}). \end{aligned} \tag{F.63}$$

Now we can easily solve for $\varphi_2$:

$$\begin{aligned} &\delta(\sin\theta_1\sin\theta_2\cos\varphi_2 + \cos\theta_1\cos\theta_2 - \cos\tilde{\theta}_{12}) = \frac{\delta(\cos\varphi_2 - \cos\tilde{\varphi}_2)}{|\sin\theta_1\sin\theta_2|}, \\ &\cos\tilde{\varphi}_2 = \frac{\cos\tilde{\theta}_{12} - \cos\theta_1\cos\theta_2}{\sin\theta_1\sin\theta_2}. \end{aligned} \tag{F.64}$$

We find then

$$\int \mathrm{d}\Phi_3 = \frac{1}{128\pi^4}\int \mathrm{d}E_1\mathrm{d}\cos\theta_1\mathrm{d}E_2\mathrm{d}\cos\theta_2\mathrm{d}\varphi_2\frac{\sqrt{s}-E_1-E_2}{E_3^*\sin\theta_1\sin\theta_2} \\ \times\theta(E_1)\theta(E_2)\theta(E_3^*)\theta(\sqrt{s}-E_1-E_2)\delta(\cos\varphi_2-\cos\tilde{\varphi}_2), \tag{F.65}$$

where we removed the absolute value in $|\sin\theta_1\sin\theta_2|$ since the sine is positive in the integration region $(0,\pi)$. We can now integrate over $\cos\varphi_2$, for which there are several subtleties. First, the differential can be transformed with $\mathrm{d}\cos\varphi_2 = \mathrm{d}\varphi_2\sin\varphi_2$, but $\sin\varphi_2 = \pm\sqrt{1+\cos^2\varphi_2}$, with the positive root occurring for $(0,\pi)$ and the negative root for $(\pi, 2\pi)$. Therefore, the adequate way of performing the integral is by splitting the integration region into the two intervals

$$\int_0^{2\pi}\mathrm{d}\varphi_2 = \int_0^{\pi}\mathrm{d}\varphi_2+\int_{\pi}^{2\pi}\mathrm{d}\varphi_2=\int_{-1}^{1}\frac{\mathrm{d}\cos\varphi_2}{\sqrt{1-\cos^2\varphi_2}}+\int_1^{-1}\frac{\mathrm{d}\cos\varphi_2}{-\sqrt{1-\cos^2\varphi_2}} \\ = 2\int_{-1}^{1}\frac{\mathrm{d}\cos\varphi_2}{\sqrt{1-\cos^2\varphi_2}}. \tag{F.66}$$

Second, in order for the delta function to cross zero we need to demand $|\cos\tilde{\varphi}_2|\leq 1$, which leads to

$$h_{12}\equiv\sin^2\tilde{\theta}_{12}+2\cos\tilde{\theta}_{12}\cos\theta_1\cos\theta_2-\cos^2\theta_1-\cos^2\theta_2\geq 0, \tag{F.67}$$

so the resulting Heaviside function is $\theta(h_{12})$. Finally, when evaluating the integrand at the condition of the delta we get

$$E_3^*|_{\varphi_2=\tilde{\varphi}_2}=|\sqrt{s}-E_1-E_2|, \qquad \sqrt{1-\cos^2\tilde{\varphi}_2}=\frac{\sqrt{h_{12}}}{|\sin\theta_1\sin\theta_2|}. \tag{F.68}$$

Plugging back these ideas, the phase space is

$$\int\mathrm{d}\Phi_3=\frac{1}{64\pi^4}\int\frac{\mathrm{d}E_1\mathrm{d}E_2\mathrm{d}\cos\theta_1\mathrm{d}\cos\theta_2}{\sqrt{h_{12}}}\theta(E_1)\theta(E_2)\theta(\sqrt{s}-E_1-E_2)\theta(h_{12}). \tag{F.69}$$

#### F.3.2.2 Adimensionalization: the $x_i$ variables

As before we write the phase space in terms of the adimensional variables defined in (F.53). The expression is

$$\int\mathrm{d}\Phi_3=\frac{s}{256\pi^4}\int\frac{\mathrm{d}x_1\mathrm{d}x_2\mathrm{d}\cos\theta_1\mathrm{d}\cos\theta_2}{\sqrt{h_{12}}}\theta(x_1)\theta(x_2)\theta(2-x_1-x_2)\theta(h_{12}), \tag{F.70}$$

$$h_{12}=\sin^2\tilde{\theta}_{12}+2\cos\tilde{\theta}_{12}\cos\theta_1\cos\theta_2-\cos^2\theta_1-\cos^2\theta_2,$$

$$\cos\tilde{\theta}_{12}=\frac{2+x_1x_2-2(x_1+x_2)+2(\hat{m}_1^2+\hat{m}_2^2-\hat{m}_3^2)}{\sqrt{x_1^2-4\hat{m}_1^2}\sqrt{x_2^2-4\hat{m}_2^2}}.$$

### F.3.2.3 General form

The form (F.70) holds for any pair of particles, so that we can write

$$\int \mathrm{d}\Phi_3 = \frac{s}{256\pi^4}\int \frac{\mathrm{d}x_i \mathrm{d}x_j\, \mathrm{d}\cos\theta_i \mathrm{d}\cos\theta_j}{\sqrt{h_{ij}}}\theta(x_i)\theta(x_j)\theta(2-x_i-x_j)\theta(h_{ij}), \tag{F.71}$$

$$h_{ij} \equiv \sin^2\tilde{\theta}_{ij} + 2\cos\tilde{\theta}_{ij}\cos\theta_i\cos\theta_j - \cos^2\theta_i - \cos^2\theta_j,$$

$$\cos\tilde{\theta}_{ij} \equiv \frac{2 + x_i x_j - 2(x_i + x_j) + 2(\hat{m}_i^2 + \hat{m}_j^2 - \hat{m}_k^2)}{\sqrt{x_i^2 - 4\hat{m}_i^2}\sqrt{x_j^2 - 4\hat{m}_j^2}}, \quad i \neq j \neq k.$$

Let us prove this by changing variables to $x_2 = 2 - x_1 - x_3$. The invariance of the three functions $\theta(x_i)\theta(x_j)\theta(2-x_i-x_j)$ was proved in the decay case, so we need only to transform the differential $\mathrm{d}\cos\theta_2$ and $h_{12}$. From the second line of (F.54) we find

$$\mathrm{d}\cos\theta_2 = -\frac{\sqrt{x_3^2 - 4\hat{m}_3^2}}{\sqrt{x_2^2 - 4\hat{m}_2^2}}\mathrm{d}\cos\theta_3. \tag{F.72}$$

Accounting for this Jacobian, the factor $1/\sqrt{h_{12}}$ in the integrand transforms as

$$\begin{aligned}\frac{x_2^2 - 4\hat{m}_2^2}{x_3^2 - 4\hat{m}_3^2}h_{12} \;&=\; \frac{x_2^2 - 4\hat{m}_2^2}{x_3^2 - 4\hat{m}_3^2}\sin^2\tilde{\theta}_{12} + 2\left[\frac{\sqrt{x_2^2 - 4\hat{m}_2^2}}{\sqrt{x_3^2 - 4\hat{m}_3^2}}\cos\tilde{\theta}_{12}\right]\cos\theta_1\left[\frac{\sqrt{x_2^2 - 4\hat{m}_2^2}}{\sqrt{x_3^2 - 4\hat{m}_3^2}}\cos\theta_2\right]\\ &\quad - \frac{x_2^2 - 4\hat{m}_2^2}{x_3^2 - 4\hat{m}_3^2}\cos^2\theta_1 - \frac{x_2^2 - 4\hat{m}_2^2}{x_3^2 - 4\hat{m}_3^2}\cos^2\theta_2.\end{aligned} \tag{F.73}$$

The term with $\sin^2\tilde{\theta}_{12}$ corresponds to $\sin^2\tilde{\theta}_{13}$ by (F.59), and the term with $\cos\tilde{\theta}_{12}$ is put in terms of $\cos\tilde{\theta}_{13}$ with (F.58). The term with $\cos\theta_2$ is simply

$$\frac{\sqrt{x_2^2 - 4\hat{m}_2^2}}{\sqrt{x_3^2 - 4\hat{m}_3^2}}\cos\theta_2 = -\cos\theta_3 - \frac{\sqrt{x_1^2 - 4\hat{m}_1^2}}{\sqrt{x_3^2 - 4\hat{m}_3^2}}\cos\theta_1, \tag{F.74}$$

and its square reads

$$\frac{x_2^2 - 4\hat{m}_2^2}{x_3^2 - 4\hat{m}_3^2}\cos^2\theta_2 \;=\; \cos^2\theta_3 + \frac{x_1^2 - 4\hat{m}_1^2}{x_3^2 - 4\hat{m}_3^2}\cos^2\theta_1 + 2\frac{\sqrt{x_1^2 - 4\hat{m}_1^2}}{\sqrt{x_3^2 - 4\hat{m}_3^2}}\cos\theta_1\cos\theta_3. \tag{F.75}$$

All in all we obtain

$$\begin{aligned}\frac{x_2^2-4\hat{m}_2^2}{x_3^2-4\hat{m}_3^2}h_{12} &= \sin^2\tilde{\theta}_{13}-2\left[\cos\tilde{\theta}_{13}+\frac{x_3^2-x_2^2+4(\hat{m}_2^2-\hat{m}_3^2)}{\sqrt{x_1^2-4\hat{m}_1^2}\sqrt{x_3^2-4\hat{m}_3^2}}\right]\cos\theta_1 \\ &\quad\times\left[\cos\theta_3+\frac{\sqrt{x_1^2-4\hat{m}_1^2}}{\sqrt{x_3^2-4\hat{m}_3^2}}\cos\theta_1\right]-\frac{x_2^2-4\hat{m}_2^2}{x_3^2-4\hat{m}_3^2}\cos^2\theta_1-\cos^2\theta_3 \\ &\quad-\frac{x_1^2-4\hat{m}_1^2}{x_3^2-4\hat{m}_3^2}\cos^2\theta_1-2\frac{\sqrt{x_1^2-4\hat{m}_1^2}}{\sqrt{x_3^2-4\hat{m}_3^2}}\cos\theta_1\cos\theta_3 \\ &= \sin^2\tilde{\theta}_{13}+2\cos\tilde{\theta}_{13}\cos\theta_1\cos\theta_3-\cos^2\theta_1-\cos^2\theta_3=h_{13},\end{aligned} \tag{F.76}$$

where we used (F.24) to simplify both terms in the square brackets. This also shows the remaining Heaviside function transforms invariantly:

$$\theta(h_{12})=\theta\left(\frac{x_3^2-4\hat{m}_3^2}{x_2^2-4\hat{m}_2^2}h_{13}\right)=\theta(h_{13}). \tag{F.77}$$

All these results prove the generalization (F.71) holds.

### F.3.2.4 Angular region of integration

The function $h_{ij}$ appearing in (F.71) depends on all the integration variables. However, its dependence on the cosines is simpler than that on $x_i$ and $x_j$: with respect to the cosines it is a second-degree polynomial. This hints it is simpler to integrate first over the angles and then over $x_i$ and $x_j$, and in fact it is possible to find a convenient expression.

We then see $h_{ij}$ as a second-degree polynomial in the cosines, where $\cos\tilde{\theta}_{ij}$ is a parameter and where the angles $\theta_i$ and $\theta_j$ run along the interval $(0,\pi)$. The first observation is that $h_{ij}(\theta_i,\theta_j)$ is always negative if $|\cos\tilde{\theta}_{ij}|>1$. Because of this

$$\begin{aligned}&\int \mathrm{d}\Phi_3=\frac{s}{256\pi^4}\int \mathrm{d}x_i\mathrm{d}x_j\,\Theta_{ij}\int\frac{\mathrm{d}\cos\theta_i\mathrm{d}\cos\theta_j}{\sqrt{h_{ij}}}\theta(h_{ij}), \\ &\Theta_{ij}=\theta(x_i)\theta(x_j)\theta(2-x_i-x_j)\theta(\sin^2\tilde{\theta}_{ij}),\end{aligned} \tag{F.78}$$

where we implemented the condition $|\cos\tilde{\theta}_{ij}|\leq 1$ by adding $\theta(\sin^2\tilde{\theta}_{ij})$, which only restricts the integration region for $x_i$ and $x_j$ and therefore we moved it into the $\mathrm{d}x_i\mathrm{d}x_j$ integral, which now has the same form as the that in the decay case. The function $\theta(h_{ij})$ restricts the integration region for the cosines, is symmetric in $i$ and $j$ and can be factorized as

$$h_{ij}=(\cos\theta_j^+-\cos\theta_j)(\cos\theta_j-\cos\theta_j^-), \tag{F.79}$$

where

$$\cos\theta_j^{\pm}(\cos\theta_i, x_i, x_j) \;=\; \cos\tilde{\theta}_{ij}\cos\theta_i \pm \sin\tilde{\theta}_{ij}\sin\theta_i = \cos(\theta_i \mp \tilde{\theta}_{ij}). \tag{F.80}$$

Note we reversed one of the monomials in (F.79) to correctly reproduce the negative $\cos^2\theta_j$ term of $h_{ij}$. Positive $h_{ij}(\theta_i, \theta_j)$ occurs when $|\cos\tilde{\theta}_{12}| \leq 1$ for the $\cos\theta_j$ values in between those in (F.80). Then the explicit angular integration is

$$\begin{aligned}\int \mathrm{d}\Phi_3 \;&=\; \frac{s}{256\pi^4}\int \mathrm{d}x_i \mathrm{d}x_j\, \Theta_{ij} \\ &\times \int_{-1}^{1} \mathrm{d}\cos\theta_i \int_{\cos\theta_j^-}^{\cos\theta_j^+} \frac{\mathrm{d}\cos\theta_j}{\sqrt{(\cos\theta_j^+ - \cos\theta_j)(\cos\theta_j - \cos\theta_j^-)}}.\end{aligned} \tag{F.81}$$

These phase-space integrals cannot be solved without the explicit knowledge of the matrix element of the process under consideration. In the next section we solve the angular integral for a matrix element with polynomial dependence on the cosines. Then, in section F.3.3, we make explicit the integration path imposed by $\Theta_{ij}$ in the integral over $x_i$ and $x_j$. This is carried out for three massless particles and for the case $m_1 = m_2$, $m_3 = 0$. For this purposes we define

$$\begin{aligned}&\int \mathrm{d}\Phi_3^E \equiv \int \mathrm{d}x_i \mathrm{d}x_j\, \Theta_{ij}, \\ &\int \mathrm{d}\Phi_3^\theta \equiv \int_{-1}^{1} \mathrm{d}\cos\theta_i \int_{\cos\theta_j^-}^{\cos\theta_j^+} \frac{\mathrm{d}\cos\theta_j}{\sqrt{(\cos\theta_j^+ - \cos\theta_j)(\cos\theta_j - \cos\theta_j^-)}}.\end{aligned} \tag{F.82}$$

#### F.3.2.5 Solution of the angular integral for polynomial dependence

Let us solve the angular part in (F.82) for the case where the angular dependence of the matrix element is of the form $\cos^m\theta_i \cos^n\theta_j$, for $m, n \geq 0$. To settle up notation, let us consider contributions to the matrix element of the form

$$|\mathcal{M}|^2 \equiv \sum_{m,n} M_{m,n}, \qquad M_{m,n} = f_{m,n}(x_i, x_j)\cos^m\theta_i \cos^n\theta_j, \qquad i \neq j. \tag{F.83}$$

Written in general, the type of integral we face is

$$I_{m,n}(c) \equiv \int \mathrm{d}\Phi_3^\theta \cos^m\theta_i \cos^n\theta_j = \int_{-1}^{1} \mathrm{d}x\, x^m \int_{a(x)}^{b(x)} \mathrm{d}y \frac{y^n}{\sqrt{[b(x) - y][y - a(x)]}}, \tag{F.84}$$

where

$$\begin{aligned}a(x) &= cx - (1 - c^2)^{1/2}(1 - x^2)^{1/2}, \\ b(x) &= cx + (1 - c^2)^{1/2}(1 - x^2)^{1/2}.\end{aligned} \tag{F.85}$$

Here, $c = \cos\tilde{\theta}_{12}$ as can be read from (F.80). Again in general, the inner integral is

$$J_n(a,b) \equiv \int_a^b \mathrm{d}y \frac{y^n}{\sqrt{(b-y)(y-a)}} = \int_0^{b-a} \mathrm{d}x \frac{(y+a)^n}{\sqrt{y(b-a-y)}} \tag{F.86}$$
$$= \int_0^1 \mathrm{d}z \frac{(b-a)[z(b-a)+a]^n}{\sqrt{z(b-a)(b-a-z(b-a))}}$$
$$= a^n \int_0^1 \mathrm{d}z z^{-1/2}(1-z)^{-1/2}\left[z\left(\frac{b}{a}-1\right)+1\right]^n = \pi a^n\, {}_2F_1\left(-n, 1/2, 1, 1-\frac{b}{a}\right),$$

where we performed the changes of variable $y \mapsto y+a$ and $y = z(b-a)$ and in the last step used that $b(x) > a(x)$. However, the closed hypergeometric form is not optimal to perform the $x$ integration. It is preferable then to solve $J_n(a,b)$ by expanding the integrand in the second step of (F.86) as

$$J_n(a,b) = \sum_{i=0}^{n} \binom{n}{i} a^{n-i} \int_0^{b-a} \mathrm{d}y \frac{y^i}{\sqrt{y(b-a-y)}} \tag{F.87}$$
$$= \sum_{i=0}^{n} \binom{n}{i} a^{n-i}(b-a)^i \int_0^1 \mathrm{d}z z^{i-1/2}(1-z)^{-1/2}$$
$$= \pi \sum_{i=0}^{n} \binom{n}{i} a^{n-i}(b-a)^i \frac{\Gamma(1+2i)}{4^i \Gamma^2(i+1)},$$

where again $y = z(b-a)$. Of course, summing up (F.87) gives (F.86) back. Using that in our case, $a$ and $b$ are given by (F.85) we have

$$a^{n-i} = [cx - (1-c^2)^{1/2}(1-x^2)^{1/2}]^{n-i} \tag{F.88}$$
$$= \sum_{j=0}^{n-i} \binom{n-i}{j} (-1)^{n-i-j} c^j (1-c^2)^{\frac{n-i-j}{2}} x^j (1-x^2)^{\frac{n-i-j}{2}}$$
$$(b-a)^i = 2^i (1-c^2)^{i/2}(1-x^2)^{i/2}.$$

Defining $J_n(a(x), b(x)) \equiv J_n(x,c)$ we arrive to

$$J_n(x,c) \equiv \pi \sum_{i=0}^{n} \binom{n}{i} \frac{\Gamma(1+2i)}{2^i \Gamma^2(1+i)} \tag{F.89}$$
$$\times \sum_{j=0}^{n-i} \binom{n-i}{j} (-1)^{n-i-j} c^j (1-c^2)^{\frac{n-j}{2}} x^j (1-x^2)^{\frac{n-j}{2}}.$$

whose first values of are

$$J_0(x,c) = \pi, \qquad J_1(x,c) = \pi c x, \qquad J_2(x,c) = \frac{\pi}{2}[(3x^2-1)c^2 + 1 - x^2]. \tag{F.90}$$

The remaining integration over $x$ is immediate and one reaches the final form

$$
\begin{aligned}
I_{m,n}(c) &= \frac{\pi}{2}\sum_{i=0}^{n}\sum_{j=0}^{n-i}(-1)^{n-i-j}[1+(-1)^{j+m}] \qquad \text{(F.91)}\\
&\times\frac{\Gamma(n+1)\Gamma(1+2i)\Gamma\left(\frac{1+j+m}{2}\right)\Gamma\left(1+\frac{n-j}{2}\right)}{2^i\Gamma^3(1+i)\Gamma(j+1)\Gamma(n+1-i-j)\Gamma\left(\frac{3+n+m}{2}\right)}c^j(1-c^2)^{\frac{n-j}{2}},
\end{aligned}
$$

whose first values in $m$ and $n$ take simple forms

$$
\begin{aligned}
I_{0,0}(c) &= 2\pi, & I_{1,0}(c) &= I_{0,1}(c) = 0, \qquad \text{(F.92)}\\
I_{1,1}(c) &= \frac{2\pi}{3}c, & I_{2,0}(c) &= I_{0,2}(c) = \frac{2\pi}{3}.
\end{aligned}
$$

Moreover, in general, for integrals involving no crossed terms one has

$$
I_{n,0}(c) = I_{0,n}(c) = \frac{[1+(-1)^n]\pi}{n+1}, \qquad \text{(F.93)}
$$

which can be directly proven by setting $n=0$ in (F.91) (which sets $i=j=0$). The symmetry condition $I_{n,0}(c) = I_{0,n}(c)$ is found by arguing that if there is no dependence on $\theta_1$ one can choose to integrate over $\cos\theta_1$ first, finding the same form for the roots from $\theta(h_{12})$ since $h_{12}=h_{21}$.

As a cross-check, if the matrix element has no angular dependence the angular integrals can be solved,

$$
\begin{aligned}
\int \mathrm{d}\Phi_3 &= \frac{s}{256\pi^4}\int \mathrm{d}x_1\mathrm{d}x_2\theta(\sin^2\tilde{\theta}_{12})\theta(x_1)\theta(x_2)\theta(2-x_1-x_2)I_{0,0}(\cos\tilde{\theta}_{12}) \qquad \text{(F.94)}\\
&= \frac{s}{128\pi^3}\int \mathrm{d}x_1\mathrm{d}x_2\theta(\sin^2\tilde{\theta}_{12})\theta(x_1)\theta(x_2)\theta(2-x_1-x_2),
\end{aligned}
$$

in agreement with the decay result (F.56).

### F.3.3 Dalitz region for the energy integrals

The dependence of the matrix element on $x_1$ and $x_2$ can be complicated enough to prevent us from making any sensible guess –take for example the fact that after solving the angular integral one gets powers of $\cos\tilde{\theta}_{12}$–. Because of this, in this section we will only focus on finding the Dalitz region of the process, i.e., the region of the $(x_1, x_2)$ plane where $|\mathcal{M}|^2$ must be integrated.

### F.3.3.1 Three massless particles

Let us start by putting $m_1 = m_2 = m_3 = 0$ to get

$$\int \mathrm{d}\Phi_3^E = \int_0^\infty \int_0^\infty \mathrm{d}x_1 \mathrm{d}x_2\, \theta(\sin^2\tilde{\theta}_{12})\theta(2 - x_1 - x_2). \tag{F.95}$$

Our first objective is to find the set of points in the integration region that satisfy the positivity condition $\sin^2\tilde{\theta}_{12} \geq 0$. The squared sine function and its roots for $x_2$ are

$$\sin^2\tilde{\theta}_{12} = \frac{4(1 - x_1)(1 - x_2)(x_1 + x_2 - 1)}{x_1^2 x_2^2}, \qquad \tilde{x}_2 = 1, 1 - x_1. \tag{F.96}$$

To find out whether the positive region is outside the boundaries or in between them we also compute

$$\lim_{x_2 \mapsto \infty} \sin^2\tilde{\theta}_{12} = \frac{4(x_1 - 1)}{x_1^2}, \tag{F.97}$$

which is positive if $x_1 > 1$. So

- For $x_1 \in (0, 1)$ the positive region for $\sin^2\tilde{\theta}_{12}$ lays in between the two roots, which are both in the positive half of the $x_2$ axis. In this case $x_2$ must be integrated in $(1 - x_1, 1)$.

- For $x_1 \in (1, \infty)$ the positive region for $\sin^2\tilde{\theta}_{12}$ lays outside the two roots, and in this case only $x_1 = 1$ lays on the integration path. Therefore, $x_2$ must be integrated in $(1, \infty)$.

With these ideas we have

$$\int \mathrm{d}\Phi_3^E = \left[\int_0^1 \mathrm{d}x_1 \int_{1-x_1}^1 \mathrm{d}x_2 + \int_1^\infty \mathrm{d}x_1 \int_1^\infty \mathrm{d}x_2\right] \theta(2 - x_1 - x_2) = \int_0^1 \mathrm{d}x_1 \int_{1-x_1}^1 \mathrm{d}x_2 \tag{F.98}$$

where in the last equality the condition $2 - x_1 - x_2 \geq 0$ simply discards the second integral.

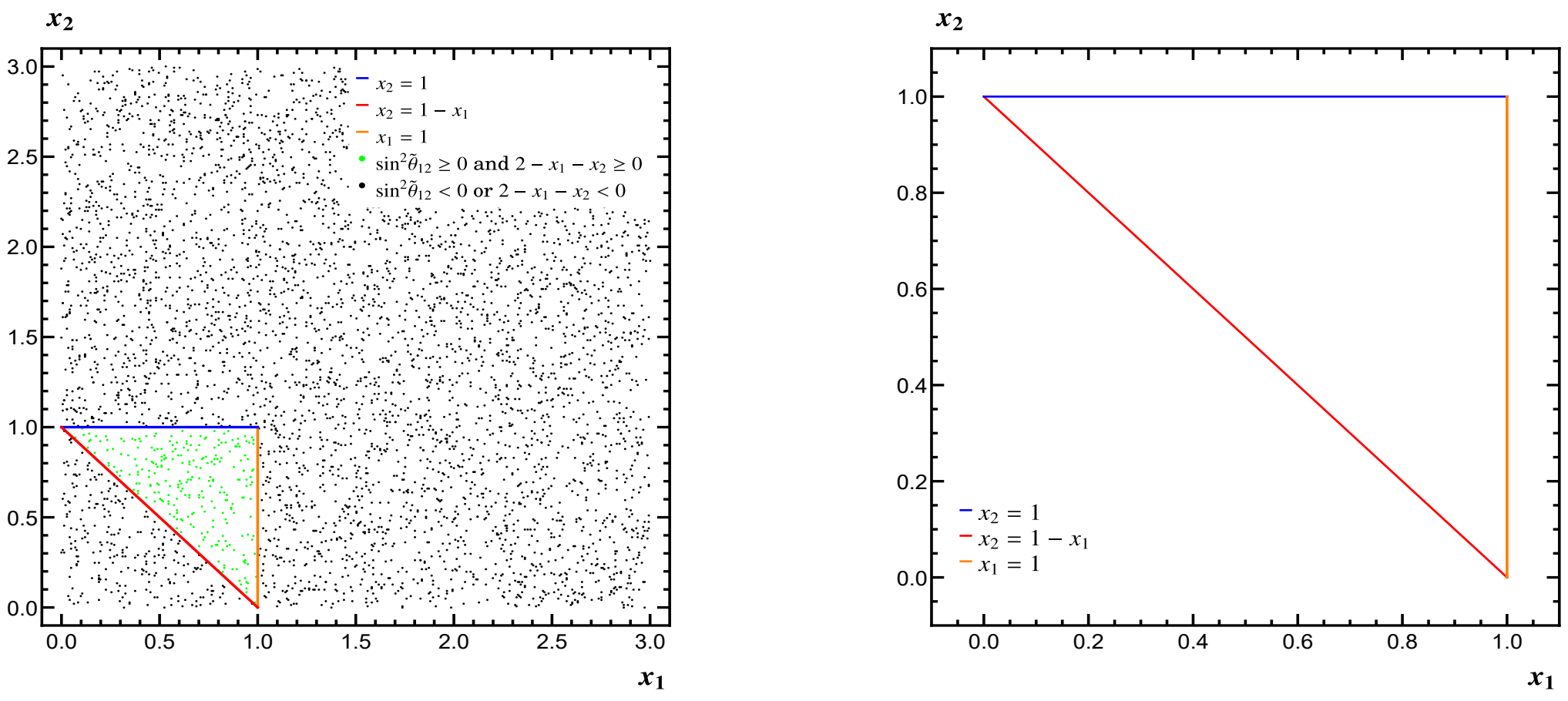

$(a)$ $(b)$

**Figure F.2.** Dalitz plot for three massless particles. Panel $(a)$: $5 \times 10^3$ random points $(x_1, x_2)$ have been generated in the square $0 \leq x_i \leq 3$. Those satisfying the two positivity conditions are colored in green, and those not satisfying one or both of them are colored in black. The analytically-found boundaries delimit the physical region. Panel $(b)$: detail of the physical region.

#### F.3.3.2 Two particles of equal mass and one massless particle

Let us now put $m_1 = m_2 \equiv m$ and $m_3 = 0$, getting

$$\int \mathrm{d}\Phi_3^E = \int_{2\hat{m}}^{\infty} \int_{2\hat{m}}^{\infty} \mathrm{d}x_1 \mathrm{d}x_2 \, \theta(\sin^2\tilde{\theta}_{12}) \theta(2 - x_1 - x_2). \tag{F.99}$$

As before, our first objective is to find the set of points in the integration region that satisfy $\sin^2\tilde{\theta}_{12} \geq 0$. In this case the squared sine function and its roots for $x_2$ are

$$\sin^2\tilde{\theta}_{12} = \frac{4(1 - x_1)(1 - x_2)(x_1 + x_2 - 1) - 4\hat{m}^2(2 - x_1 - x_2)^2}{(x_1^2 - 4\hat{m}^2)(x_2^2 - 4\hat{m}^2)}, \tag{F.100}$$

$$\tilde{x}_2^{\pm} = \frac{2\hat{m}^2(2 - x_1) + (1 - x_1)\left(2 - x_1 \pm \sqrt{x_1^2 - 4\hat{m}^2}\right)}{2(1 + \hat{m}^2 - x_1)}.$$

It is also useful to take the difference between the roots

$$\tilde{x}_2^+ - \tilde{x}_2^- = \frac{(1 - x_1)\sqrt{x_1^2 - 4\hat{m}^2}}{1 + \hat{m}^2 - x_1}, \tag{F.101}$$

which is positive when $x_1 \in (2\hat{m}, 1) \cup (1+\hat{m}^2, \infty)$ and negative when $x_1 \in (1, 1+\hat{m}^2)$. Since $x_1 \geq 2\hat{m}$, the square root in $\tilde{x}_2$ is always real, and also $\hat{m} \in (0, 1/2)$[F.11]. In this case:

- For $x_1 \in (2\hat{m}, \infty)$,it is satisfied that $\tilde{x}_2^+ > 2\hat{m}$, so the $\tilde{x}_2^+$ root is always crossed in the integration path.

- With regards to $\tilde{x}_2^-$, when $x_1 < 1+\hat{m}^2$ one has $\tilde{x}_2^- \geq 2\hat{m}$, while for $x_1 > 1+\hat{m}^2$ it is satisfied that $\tilde{x}_2^- < 2\hat{m}$. Only in the first case $\tilde{x}_2^-$ is crossed in the integration. Note that since $1+\hat{m}^2 > 1 > 2\hat{m}$ this value is always crossed in the $x_1$ integration.

Again we study the limit of the squared sine,

$$\lim_{x_2 \mapsto \infty} \sin^2\tilde{\theta}_{12} = \frac{4(x_1 - 1 - \hat{m}^2)}{x_1^2 - 4\hat{m}^2}, \tag{F.102}$$

which is positive when $x_1 > 1+\hat{m}^2$ and negative when $x_1 < 1+\hat{m}^2$. With all these we extract the following conclusions:

- For $x_1 \in (2\hat{m}, 1)$ the positive region of the sine lays in between the roots. In this case both $\tilde{x}_2^\pm > 2\hat{m}$ and $\tilde{x}_2^- < \tilde{x}_2^+$. Therefore, $x_2$ must be integrated along $(\tilde{x}_2^-, \tilde{x}_2^+)$.

- For $x_1 \in (1, 1+\hat{m}^2)$ the positive region of the sine lays in between the roots. In this case both $\tilde{x}_2^\pm > 2\hat{m}$ and $\tilde{x}_2^+ < \tilde{x}_2^-$. Therefore, $x_2$ must be integrated along $(\tilde{x}_2^+, \tilde{x}_2^-)$.

- For $x_1 \in (1+\hat{m}^2, \infty)$ the positive region of the sine lays outside the roots. In this case $\tilde{x}_2^+ > 2\hat{m}$ and $\tilde{x}_2^- < 2\hat{m}$. Therefore, $x_2$ must be integrated along $(\tilde{x}_2^+, \infty)$.

All in all this leads to

$$\begin{aligned}\int_{2\hat{m}}^{\infty}\int_{2\hat{m}}^{\infty} \mathrm{d}x_1 \mathrm{d}x_2\, \theta(\sin^2\tilde{\theta}_{12}) &= \int_{2\hat{m}}^{1} \mathrm{d}x_1 \int_{\tilde{x}_2^-}^{\tilde{x}_2^+} \mathrm{d}x_2 + \int_{1}^{1+\hat{m}^2} \mathrm{d}x_1 \int_{\tilde{x}_2^+}^{\tilde{x}_2^-} \mathrm{d}x_2 \\ &+ \int_{1+\hat{m}^2}^{\infty} \mathrm{d}x_1 \int_{\tilde{x}_2^+}^{\infty} \mathrm{d}x_2.\end{aligned} \tag{F.103}$$

Finally we have to add the positivity condition $2 - x_1 - x_2 \geq 0$. One can see that when $x_1 > 1$ (second and third integrals) both $x_2^\pm > 1$, so $x_1 + x_2 > 2$. As conclusion, this last Heaviside function effectively cancels the last two integrals and we reach

$$\int \mathrm{d}\Phi_3^E = \int_{2\hat{m}}^{\infty}\int_{2\hat{m}}^{\infty} \mathrm{d}x_1 \mathrm{d}x_2\, \theta(\sin^2\tilde{\theta}_{12})\theta(2 - x_1 - x_2) = \int_{2\hat{m}}^{1} \mathrm{d}x_1 \int_{\tilde{x}_2^-}^{\tilde{x}_2^+} \mathrm{d}x_2. \tag{F.104}$$

F.11. Since the third particle is massless, it has to receive some amount of kinetic energy, leading to $\sqrt{s} = T_1 + T_2 + 2m + E_3 > 2m$ due to $T_1, T_2, E_3 \geq 0$ and therefore to $1 > 2\hat{m}$ and $\hat{m} < 1/2$.

As a final cross-check, if we let $\hat{m}=0$ in (F.104), $x_1$ is integrated from 0 to 1 and

$$\tilde{x}_2^-|_{\hat{m}=0} = \frac{(1-x_1)(2-x_1-x_1)}{2(1-x_1)} = 1-x_1, \qquad \text{(F.105)}$$
$$\tilde{x}_2^+|_{\hat{m}=0} = \frac{(1-x_1)(2-x_1+x_1)}{2(1-x_1)} = 1,$$

so we recover the limits of the massless case (F.98).

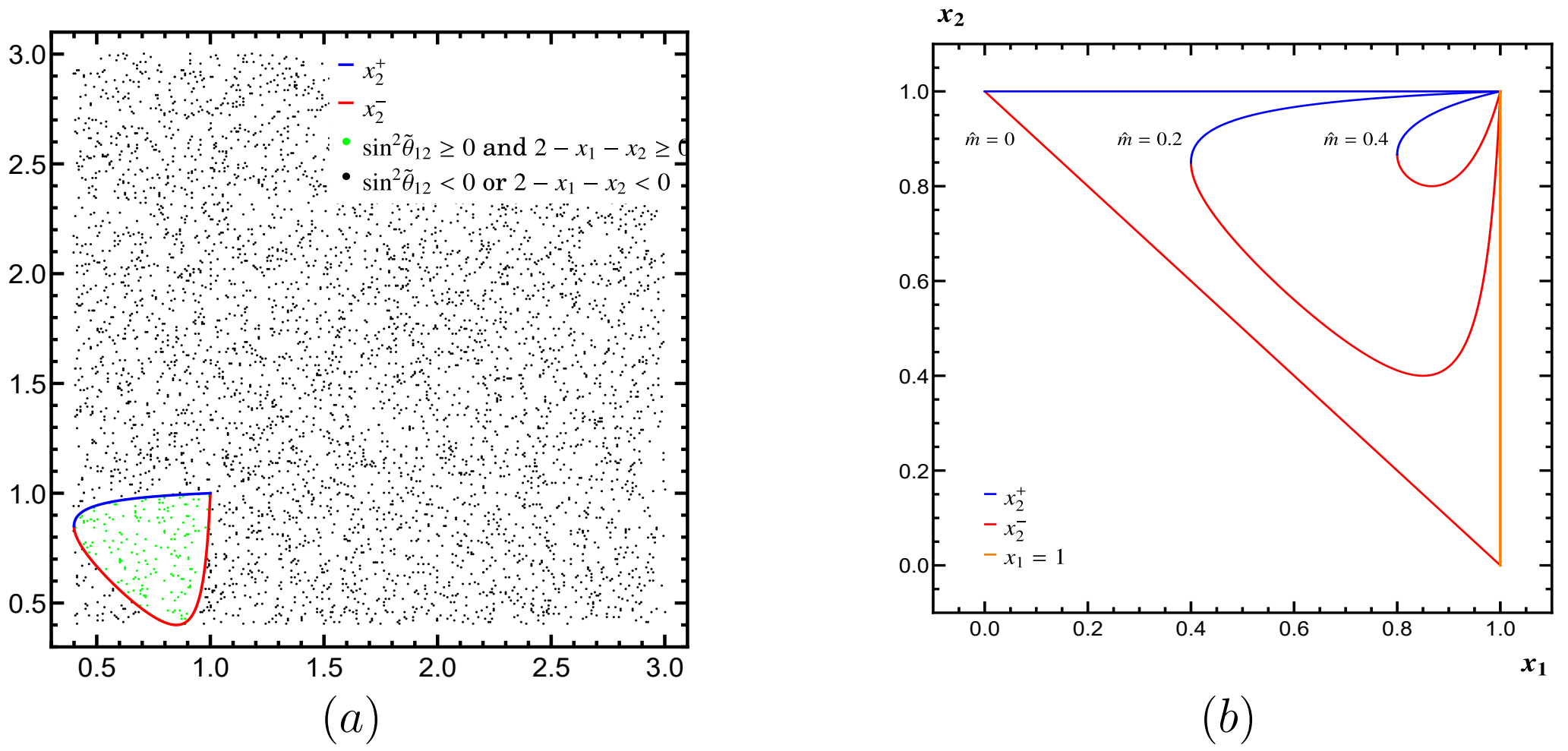

**Figure F.3.** Dalitz plot for two particles of equal mass and one massless particle. Panel $(a)$: $5\times10^3$ random points $(x_1, x_2)$ have been generated in the square $2\hat{m}_i \leq x_i \leq 3$. Those satisfying the two positivity conditions are colored in green, and those not satisfying one or both of them are colored in black. The analytically-found boundaries delimit the physical region. Panel $(b)$: detail of the physical region for $\hat{m}=0, 0.2$ and 0.4.

# F.4 Generalization to $d$ dimensions

In this section we generalize the previous results for the two and three-particle phase-space from 4 to $d$ dimensions. Having performed the four-dimensional case in such amount of detail, this generalization comes straightforward after the brief discussion in section F.4.1. Because of this, we rely heavily on the subtleties explained in sections F.2 and F.3 and here we include only the most relevant steps.

## F.4.1 General ideas

To generalize the $N$-particle phase space to $d$ spacetime dimensions we go back to the original definition in (F.4) and let $4\to d$:

$$\int \mathrm{d}\Phi_N = \prod_{j=1}^{N}\int \frac{\mathrm{d}^d p_j}{(2\pi)^{d-1}}\delta(p_j^2-m_j^2)\theta(p_j^0)(2\pi)^d\delta^{(d)}\left(P^\mu - \sum_{i=1}^{N} p_i^\mu\right). \qquad \text{(F.106)}$$

The extra dimensions are added to the vector part of the four-momenta, i.e., $p^2=(p^0)^2-|\vec{p}|^2$ with $|\vec{p}|^2=(p^1)^2+\dots+(p^{d-1})^2$. As for 4 dimensions, the zeroth component of each $\mathrm{d}^d p_i$ can be integrated over by using the on-shell delta functions, and splitting the momentum conserving delta function as

$$\delta^{(d)}\left(P^\mu-\sum_{i=1}^N p_i^\mu\right)=\delta\left(E_N-\sum_{i=1}^N E_i\right)\delta^{(d-1)}\left(\vec{P}^\mu-\sum_{i=1}^N \vec{p}_i^\mu\right) \tag{F.107}$$

allows to integrate $\mathrm{d}^{d-1}\vec{p}_N$ by using its vector part. The resultant expression, already in the center of mass frame, is

$$\begin{aligned}
&\int \mathrm{d}\Phi_N=\frac{\pi^N}{(2\pi)^{d(N-1)}}\int \mathrm{d}^{d-1}\vec{p}_1\dots\mathrm{d}^{d-1}\vec{p}_{N-1}\frac{\theta(E_1)\dots\theta(E_N^*)}{E_1\dots E_N^*}\delta\left(\sqrt{s}-\sum_{i=1}^{N-1}E_i-E_N^*\right),\\
&E_i=\sqrt{|\vec{p}_i|^2+m_i^2},\\
&E_N^*=\sqrt{\sum_{i=1}^{N-1}|\vec{p}_i|^2+2\sum_{i=0}^{N-1}\sum_{j=i+1}^{N-1}|\vec{p}_i||\vec{p}_j|\cos\theta_{ij}+m_N^2}.
\end{aligned} \tag{F.108}$$

To proceed we need to write the differentials in terms of the moduli $|\vec{p}_i|$ or the energies $E_i$, for which we employ spherical coordinates. In an $n$-dimensional Euclidean space, spherical coordinates consist on the modulus and $n-1$ angles, out of which $n-2$ are polar angles and 1 is azimuthal. A vector whose Cartesian components are $\vec{v}=(v_1,\dots,v_n)$ is written in spherical coordinates as

$$\begin{aligned}
v_1&=v\cos\theta_1,\\
v_2&=v\sin\theta_1\cos\theta_2,\\
v_3&=v\sin\theta_1\sin\theta_2\cos\theta_3,\\
&\vdots\\
v_{n-1}&=v\sin\theta_1\sin\theta_2\dots\sin\theta_{n-2}\cos\theta_{n-1},\\
v_n&=v\sin\theta_1\sin\theta_2\dots\sin\theta_{n-2}\sin\theta_{n-1},
\end{aligned} \tag{F.109}$$

where $v\equiv|\vec{v}|$, $\theta_i\in(0,\pi)$ with $i=1,\dots,n-2$ are the polar angles and $\theta_{n-1}\in(0,2\pi)$ is the azimuthal angle. The Jacobian of the transformation in (F.109) gives the differential volume form, which reads

$$\mathrm{d}^n\vec{v}=\mathrm{d}v\,\mathrm{d}\Omega_n\,v^{n-1},\quad \mathrm{d}\Omega_n\equiv\mathrm{d}\theta_1\mathrm{d}\theta_2\dots\mathrm{d}\theta_{n-2}\mathrm{d}\theta_{n-1}\sin^{n-2}\theta_1\sin^{n-3}\theta_2\dots\sin\theta_{n-2}. \tag{F.110}$$

To integrate over these coordinates the following results are useful

$$\int_0^\pi \sin^m\theta\mathrm{d}\theta=\frac{\pi^{1/2}\Gamma\left(\frac{m+1}{2}\right)}{\Gamma\left(\frac{m+2}{2}\right)},\qquad \int \mathrm{d}\Omega^{(n)}=\frac{2\pi^{n/2}}{\Gamma(n/2)}. \tag{F.111}$$

Coming back to the phase space, from (F.110) and each particle's on-shell relation we can write the phase space integral in terms of the energies and the spherical angles. The differentiation of the $d$-dimensional on-shell relation

$$\frac{|\vec{p}_i|^{d-2}\mathrm{d}|\vec{p}_i|}{E_i} = (E_i^2 - m_i^2)^{\frac{d-3}{2}}\mathrm{d}E_i, \tag{F.112}$$

from which we obtain

$$\begin{aligned}\int \mathrm{d}\Phi_N &= \frac{\pi^N}{(2\pi)^{d(N-1)}}\int \mathrm{d}E_1\mathrm{d}\Omega_{d-1}^{(1)}...\mathrm{d}E_{N-1}\mathrm{d}\Omega_{d-1}^{(n)} \\ &\times\frac{[(E_1^2-m_1^2)...(E_{N-1}^2-m_{N-1}^2)]^{\frac{d-3}{2}}}{E_N^*}\theta(E_1)...\theta(E_N^*)\delta\left(\sqrt{s}-\sum_{i=1}^{N-1}E_i - E_N^*\right),\end{aligned} \tag{F.113}$$

where $\mathrm{d}\Omega_{d-1}^{(i)}$ encodes the spherical angles of the $i$-th particle. The adequate expression for $E_N^*$ is (F.34).

Next, we split the $(d-1)$-dimensional euclidean space into two mutually normal sub-spaces: a subspace of dimension 3 and one of dimension $(d-4)$, such that all 3-vectors appearing in the phase space belong to the 3-dimensional subspace. Let us go through this explicitely. All the $\vec{p}_i$ belong to the same 3-dimensional space, and in the scattering case so does the tri-momentum of the initial beam. We choose the beam's direction as axis 1 –the axis with respect $\theta_1$ is measured–. We choose axis 2 in the direction of the component of $\vec{p}_1$ that is perpendicular to axis 1, i.e., we set $\theta_2^{(1)} = 0$. This sets $\vec{p}_1$ in the $1-2$ plane. Finally, we choose axis 3 so that the remaining $\vec{p}_i$ vectors are contained in the three-dimensional space determined by axis 1, 2 and 3. This sets $\theta_3^{(i)} = \theta_4^{(i)} = \cdots = \theta_{d-3}^{(i)} = \pi$. With these ideas the vectors are described by the spherical coordinates

$$\begin{aligned}\vec{p}_1 &= |\vec{p}_1|(\cos\theta_1^{(1)}, \sin\theta_1^{(1)}, 0, ..., 0), \\ \vec{p}_i &= |\vec{p}_i|(\cos\theta_1^{(i)}, \sin\theta_1^{(i)}\cos\theta_2^{(i)}, \sin\theta_1^{(i)}\sin\theta_2^{(i)}, 0, ..., 0), \quad i \geq 2.\end{aligned} \tag{F.114}$$

This corresponds to the situation depicted in figure F.4($b$). If we lack the external direction given by the beam we set axis 1 in the direction of $\vec{p}_1$, axis 2 in the direction of the component of $\vec{p}_2$ perpendicular to axis 1, and axis 3 so that the remaining $\vec{p}_i$ vectors are contained in the three-dimensional space determined by axis 1, 2 and 3. In this case

$$\begin{aligned}\vec{p}_1 &= |\vec{p}_1|(1, 0, ..., 0), \\ \vec{p}_2 &= |\vec{p}|(\cos\theta_1^{(i)}, \sin\theta_1^{(i)}\cos\theta_2^{(i)}, 0, ..., 0), \\ \vec{p}_i &= |\vec{p}_i|(\cos\theta_1^{(i)}, \sin\theta_1^{(i)}\cos\theta_2^{(i)}, \sin\theta_1^{(i)}\sin\theta_2^{(i)}, 0, ..., 0), \quad i > 2.\end{aligned} \tag{F.115}$$

This situation is depicted in figure F.4($a$).

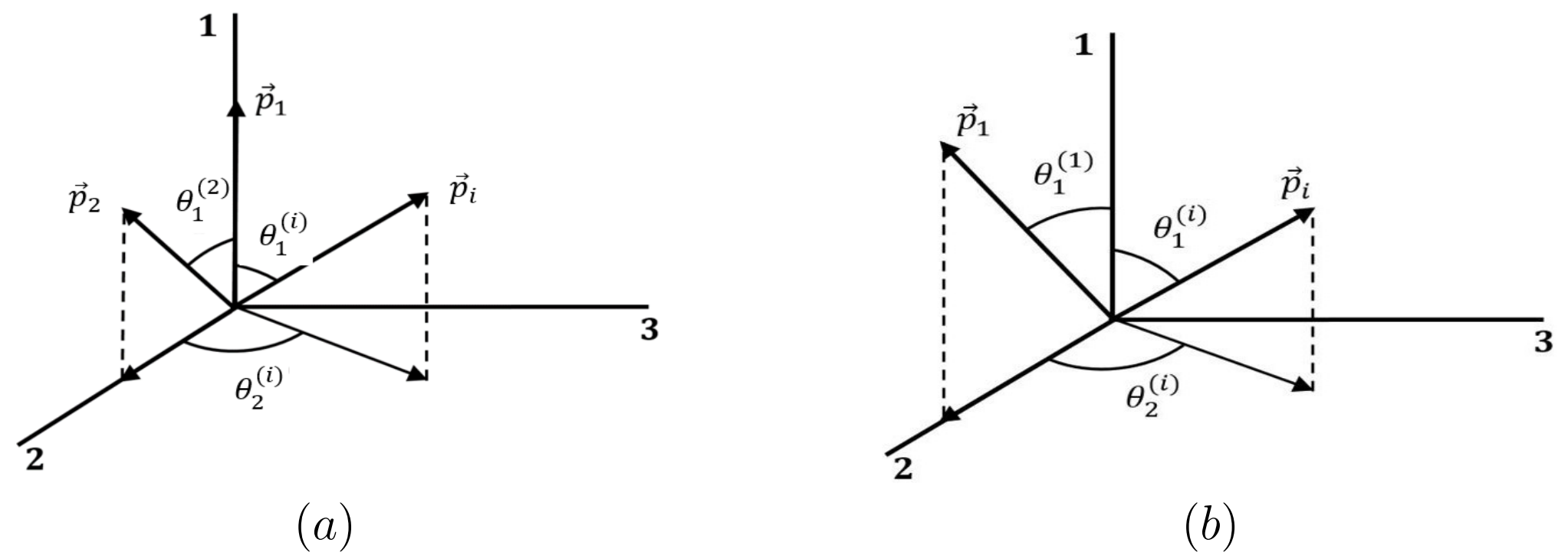

**Figure F.4.** Choices for axis orientation in $d$ spacetime dimensions for momenta of the particles in the final state of the decay of one particle (panel $(a)$) and the scattering of two particles (panel $(b)$). The three-dimensional space defined by axis $1, 2$ and $3$ contains the momenta of all the particles involved.

With the above choice of reference frame, the matrix element does not depend on angles $\theta_{j\geq 3}^{(i)}$, so they can be integrated over. In general we have

$$\begin{aligned}\int \mathrm{d}\Omega_{d-1} &= \int_0^\pi \mathrm{d}\theta_1 \sin^{d-3}\theta_1 \int_0^\pi \mathrm{d}\theta_2 \sin^{d-4}\theta_2 \ldots \int_0^\pi \mathrm{d}\theta_{d-3}\sin\theta_{d-3}\int_0^{2\pi}\mathrm{d}\theta_{d-2} \\ &= \frac{2\pi^{\frac{d-3}{2}}}{\Gamma\left(\frac{d-3}{2}\right)}\int_0^\pi \mathrm{d}\theta_1 \sin^{d-3}\theta_1 \int_0^\pi \mathrm{d}\theta_2 \sin^{d-4}\theta_2,\end{aligned} \tag{F.116}$$

appearing $N-1$ times, with an extra factor of

$$\int_0^\pi \mathrm{d}\theta_2^{(1)} \sin^{d-4}\theta_2^{(1)} = \frac{\pi^{1/2}\Gamma\left(\frac{d-3}{2}\right)}{\Gamma\left(\frac{d-2}{2}\right)}, \tag{F.117}$$

from particle 1. Together with the prefactor in (F.113) they give

$$\begin{aligned}P(N) &\equiv \frac{\pi^N}{(2\pi)^{d(N-1)}}\left[\frac{2\pi^{\frac{d-3}{2}}}{\Gamma\left(\frac{d-3}{2}\right)}\right]^{N-1}\frac{\pi^{1/2}\Gamma\left(\frac{d-3}{2}\right)}{\Gamma\left(\frac{d-2}{2}\right)} \\ &= \frac{2^{d-4}\pi^{2-N}}{(4\pi)^{\frac{(d-1)(N-1)}{2}}}\frac{\Gamma^{3-N}\left(\frac{d-3}{2}\right)}{\Gamma(d-3)},\end{aligned} \tag{F.118}$$

so the phase space is

$$\begin{aligned}\int \mathrm{d}\Phi_N &= P(N)\int \mathrm{d}E_1\ldots\mathrm{d}E_{N-1}\mathrm{d}\theta_1^{(1)}\mathrm{d}\theta_1^{(2)}\mathrm{d}\theta_2^{(2)}\ldots\theta_1^{(N-1)}\mathrm{d}\theta_2^{(N-1)} \\ &\times \sin^{d-3}\theta_1^{(1)}\sin^{d-3}\theta_1^{(2)}\sin^{d-4}\theta_2^{(2)}\ldots\sin^{d-3}\theta_1^{(N-1)}\sin^{d-4}\theta_2^{(N-1)} \\ &\times \frac{\left[(E_1^2-m_1^2)\ldots(E_{N-1}^2-m_{N-1}^2)\right]^{\frac{d-3}{2}}}{E_N^*}\theta(E_1)\ldots\theta(E_N^*)\delta\left(\sqrt{s}-\sum_{i=1}^{N-1}E_i-E_N^*\right).\end{aligned} \tag{F.119}$$

### F.4.2 Two particles

Since there is only one particle involved let us drop the superscript (1). From (F.119), our starting expression for the two-particle phase space in $d$ dimensions is:

$$\int \mathrm{d}\Phi_2 = P(2)\int \mathrm{d}E_1 \mathrm{d}\theta_1 \sin^{d-3}\theta_1 \frac{(E_1^2-m_1^2)^{\frac{d-3}{2}}}{E_2^*}\theta(E_1)\theta(E_2^*)\delta(\sqrt{s}-E_1-E_2^*),$$
$$P(2) = \frac{\pi^{1/2}}{(4\pi)^{\frac{d-1}{2}}\Gamma\left(\frac{d-2}{2}\right)}, \qquad E_2^* = \sqrt{E_1^2-m_1^2+m_2^2}. \tag{F.120}$$

We now integrate over $E_1$ using the delta function. Using (F.36) and the discussion below the only thing we need to compute is the integrand evaluated at the delta condition:

$$\left.\frac{(E_1^2-m_1^2)^{\frac{d-3}{2}}}{E_2^*}\right|_{E_1=\tilde{E}_1} = \frac{\lambda^{\frac{d-3}{2}}(s,m_1^2,m_2^2)}{(4s)^{\frac{d-4}{2}}|s-m_1^2+m_2^2|}. \tag{F.121}$$

The phase space is then

$$\int \mathrm{d}\Phi_2 = \frac{16\pi s\lambda^{\frac{d-3}{2}}(s,m_1^2,m_2^2)}{(16\pi s)^{\frac{d}{2}}\Gamma\left(\frac{d-2}{2}\right)}\int_0^\pi \mathrm{d}\theta_1 \sin^{d-3}\theta_1, \tag{F.122}$$

which leads to the following result for the scattering process

$$\begin{aligned}\int \mathrm{d}\Phi_2 &= \frac{\lambda^{\frac{d-3}{2}}(s,m_1^2,m_2^2)}{(16\pi s)^{\frac{d-2}{2}}\Gamma\left(\frac{d-2}{2}\right)}\int_{-1}^1 \mathrm{d}\cos\theta_1\,(1-\cos^2\theta_1)^{\frac{d-4}{2}} \\ &= \frac{\lambda^{\frac{1}{2}}(s,m_1^2,m_2^2)}{16\pi s}\frac{1}{\Gamma(1-\epsilon)}\left[\frac{\lambda(s,m_1^2,m_2^2)}{16\pi s}\right]^{-\epsilon}\int_{-1}^1 \mathrm{d}\cos\theta_1\,(1-\cos^2\theta_1)^{-\epsilon},\end{aligned} \tag{F.123}$$

and for the decay process

$$\int \mathrm{d}\Phi_2 = \frac{\pi^{\frac{1}{2}}\lambda^{\frac{d-3}{2}}(s,m_1^2,m_2^2)}{(16\pi s)^{\frac{d-2}{2}}\Gamma\left(\frac{d-2}{2}\right)} = \frac{\lambda^{1/2}(s,m_1^2,m_2^2)}{8\pi s}\frac{\Gamma(1-\epsilon)}{\Gamma(2-2\epsilon)}\left[\frac{\lambda(s,m_1^2,m_2^2)}{4\pi s}\right]^{-\epsilon}. \tag{F.124}$$

In both cases, the results are expressed in arbitrary $d$ dimensions and for the usual dimensional regularization choice $d=4-2\epsilon$, and reduce to the four dimensional results (F.42) and (F.44) for $d=4$ or $\epsilon=0$.

### F.4.3 Three particles

Again from (F.119) we start with

$$
\begin{aligned}
\int \mathrm{d}\Phi_3 &= P(3)\int \mathrm{d}E_1\mathrm{d}E_2\mathrm{d}\theta_1^{(1)}\mathrm{d}\theta_1^{(2)}\mathrm{d}\theta_2^{(2)}\sin^{d-3}\theta_1^{(1)}\sin^{d-3}\theta_1^{(2)}\sin^{d-4}\theta_2^{(2)} \\
&\quad\times\frac{[(E_1^2-m_1^2)(E_2^2-m_2^2)]^{\frac{d-3}{2}}}{E_3^*}\theta(E_1)\theta(E_2)\theta(E_3^*)\delta(\sqrt{s}-E_1-E_2-E_3^*) \\
&= P(3)\int \mathrm{d}E_1\mathrm{d}E_2\mathrm{d}\theta_1^{(1)}\mathrm{d}\theta_1^{(2)}\mathrm{d}\theta_2^{(2)}\sin^{d-3}\theta_1^{(1)}\sin^{d-3}\theta_1^{(2)}\sin^{d-4}\theta_2^{(2)} \\
&\quad\times\frac{[(E_1^2-m_1^2)(E_2^2-m_2^2)]^{\frac{d-4}{2}}(\sqrt{s}-E_1-E_2)}{E_3^*} \\
&\quad\times\theta(E_1)\theta(E_2)\theta(E_3^*)\theta(\sqrt{s}-E_1-E_2)\delta(\cos\theta_{12}-\cos\tilde{\theta}_{12}), \\
&\quad P(3)=\frac{2^{d-2}}{(4\pi)^d\Gamma(d-3)},
\end{aligned}
\tag{F.125}
$$

were in the second step we implemented the delta change in (F.62).

#### F.4.3.1 Decay

In the decay process we have $\theta_1^{(2)}=\theta_{12}$ and we can integrate over $\theta_1^{(1)}$ and $\theta_2^{(2)}$. We then have

$$
\begin{aligned}
\int \mathrm{d}\Phi_3 &= \frac{2\pi P(3)}{d-3}\int \mathrm{d}E_1\mathrm{d}E_2\mathrm{d}\theta_{12}\sin^{d-3}\theta_{12}\frac{(\sqrt{s}-E_1-E_2)[(E_1^2-m_1^2)(E_2^2-m_2^2)]^{\frac{d-4}{2}}}{E_3^*} \\
&\quad\times\theta(E_1)\theta(E_2)\theta(E_3^*)\theta(\sqrt{s}-E_1-E_2)\delta(\cos\theta_{12}-\cos\tilde{\theta}_{12}).
\end{aligned}
\tag{F.126}
$$

To solve the remaining angular integral we proceed as

$$
\begin{aligned}
\int_0^\pi \mathrm{d}\theta\,\sin^{d-3}\theta\,\delta(\cos\theta-\cos\tilde{\theta}_{12}) &= \int_{-1}^1 \mathrm{d}\cos\theta\frac{(1-\cos^2\theta)^{\frac{d-3}{2}}}{(1-\cos^2\theta)^{1/2}}\delta(\cos\theta-\cos\tilde{\theta}_{12}) \\
&= \sin^{d-4}\tilde{\theta}_{12}\,\theta(\sin^2\tilde{\theta}_{12}),
\end{aligned}
\tag{F.127}
$$

which leads to

$$
\int \mathrm{d}\Phi_3=\frac{\pi}{2(2\pi)^d\Gamma(d-2)}\int \mathrm{d}E_1\mathrm{d}E_2[(E_1^2-m_1^2)(E_2^2-m_2^2)\sin^2\tilde{\theta}_{12}]^{\frac{d-4}{2}}\Theta_{12}.
\tag{F.128}
$$

This result recovers (F.52) for $d=4$. In terms of the $x_i$ variables the final result reads

$$\begin{aligned}\int \mathrm{d}\Phi_3 &= \frac{1}{16\pi^2}\frac{1}{\Gamma(d-2)}\left(\frac{s}{8\pi}\right)^{d-3}\int \mathrm{d}x_1\mathrm{d}x_2[(x_1^2-4\hat{m}_1^2)(x_2^2-4\hat{m}_2^2)\sin^2\tilde{\theta}_{12}]^{\frac{d-4}{2}}\Theta_{12}\\ &= \frac{s}{128\pi^3}\frac{1}{\Gamma(2-2\epsilon)}\left(\frac{s}{8\pi}\right)^{-2\epsilon}\int \mathrm{d}x_1\mathrm{d}x_2[(x_1^2-4\hat{m}_1^2)(x_2^2-4\hat{m}_2^2)\sin^2\tilde{\theta}_{12}]^{-\epsilon}\Theta_{12},\end{aligned} \tag{F.129}$$

which again recovers (F.94) for $d=4$ or $\epsilon=0$.

#### F.4.3.2 Scattering

In the scattering case the integrand has dependence on all the angles in (F.125) through

$$\cos\theta_{12}=\cos\theta_1^{(1)}\cos\theta_1^{(2)}+\sin\theta_1^{(1)}\sin\theta_1^{(2)}\cos\theta_2^{(2)}. \tag{F.130}$$

This leads to the delta manipulation

$$\begin{aligned}\delta(\cos\theta_{12}-\cos\tilde{\theta}_{12}) &= \frac{1}{|\sin\theta_1^{(1)}\sin\theta_1^{(2)}|}\delta\left(\cos\theta_2^{(2)}-\frac{\cos\tilde{\theta}_{12}-\cos\theta_1^{(1)}\cos\theta_1^{(2)}}{\sin\theta_1^{(1)}\sin\theta_1^{(2)}}\right)\\ &= \frac{1}{\sin\theta_1^{(1)}\sin\theta_1^{(2)}}\delta(\cos\theta_2^{(2)}-\cos\tilde{\theta}_2^{(2)}),\\ \cos\tilde{\theta}_2^{(2)} &= \frac{\cos\tilde{\theta}_{12}-\cos\theta_1^{(1)}\cos\theta_1^{(2)}}{\sin\theta_1^{(1)}\sin\theta_1^{(2)}},\end{aligned} \tag{F.131}$$

where we removed the absolute value due to the positivity of $\sin\theta$ in $(0,\pi)$. The phase space is

$$\begin{aligned}\int \mathrm{d}\Phi_3 &= P(3)\int \mathrm{d}E_1\mathrm{d}E_2\mathrm{d}\theta_1^{(1)}\mathrm{d}\theta_1^{(2)}\frac{\mathrm{d}\cos\theta_2^{(2)}}{\sin\theta_2^{(2)}}\sin^{d-4}\theta_1^{(1)}\sin^{d-4}\theta_1^{(2)}\sin^{d-4}\theta_2^{(2)}\\ &\quad\times\frac{[(E_1^2-m_1^2)(E_2^2-m_2^2)]^{\frac{d-4}{2}}(\sqrt{s}-E_1-E_2)}{E_3^*}\delta(\cos\theta_2^{(2)}-\cos\tilde{\theta}_2^{(2)})\\ &\quad\times\theta(E_1)\theta(E_2)\theta(E_3^*)\theta(\sqrt{s}-E_1-E_2)\\ &= P(3)\int \mathrm{d}E_1\mathrm{d}E_2\mathrm{d}\cos\theta_1^{(1)}\mathrm{d}\cos\theta_1^{(2)}\frac{[(E_1^2-m_1^2)(E_2^2-m_2^2)]^{\frac{d-4}{2}}}{(h_{12})^{\frac{5-d}{2}}}\\ &\quad\times\theta(E_1)\theta(E_2)\theta(\sqrt{s}-E_1-E_2)\theta(h_{12}),\end{aligned} \tag{F.132}$$

where we used that

$$E_3^*|_{\theta_2^{(2)}=\tilde{\theta}_2^{(2)}} = |\sqrt{s} - E_1 - E_2| \tag{F.133}$$

$$\sin\tilde{\theta}_2^{(2)} = \sqrt{1 - \left(\frac{\cos\tilde{\theta}_{12} - \cos\theta_1^{(1)}\cos\theta_1^{(2)}}{\sin\theta_1^{(1)}\sin\theta_1^{(2)}}\right)^2} = \frac{\sqrt{h_{12}}}{\sin\theta_1^{(1)}\sin\theta_1^{(2)}},$$

and $h_{12}$ is defined by the angles $\theta_1^{(1)}$ and $\theta_1^{(2)}$ as in (F.67). In terms of the $x_i$ variables and for $d = 4 - 2\epsilon$ the result reads

$$\begin{aligned}\int \mathrm{d}\Phi_3 &= \frac{16}{s^3}\left(\frac{s}{8\pi}\right)^d \frac{1}{\Gamma(d-3)} \\ &\quad \times \int \mathrm{d}x_1 \mathrm{d}x_2\, \mathrm{d}\cos\theta_1 \mathrm{d}\cos\theta_2 \frac{[(x_1^2 - 4\hat{m}_1^2)(x_2^2 - 4\hat{m}_2^2)]^{\frac{d-4}{2}}}{(h_{12})^{\frac{5-d}{2}}} \Theta_{12}\theta(h_{12}) \\ &= \frac{s}{256\pi^4\Gamma(1-2\epsilon)}\left(\frac{s}{8\pi}\right)^{-2\epsilon} \\ &\quad \times \int \mathrm{d}x_1 \mathrm{d}x_2\, \mathrm{d}\cos\theta_1 \mathrm{d}\cos\theta_2 \frac{[(x_1^2 - 4\hat{m}_1^2)(x_2^2 - 4\hat{m}_2^2)]^{-\epsilon}}{(h_{12})^{\frac{1}{2}+\epsilon}} \Theta_{12}\theta(h_{12}),\end{aligned} \tag{F.134}$$

where for simplicity we dropped the superscripts in the angles by defining $\theta_1^{(1)} \equiv \theta_1$ and $\theta_1^{(2)} \equiv \theta_2$. The integration region is the same as for 4 dimensions: the derivation for the angular integral F.3.2.4 and the derivation for the $x_1$, $x_2$ integrals can be found in section F.3.3. Results (F.132) and (F.134) agree with their respective 4-dimensional counterparts (F.69) and (F.70) for $d = 4$ or $\epsilon = 0$.

We finish the section by proving the invariance of the $d$-dimensional phase space under the choice of the two particles defining the four final integrations. The 4-dimensional part has been proven in section F.3.2.3; all that remains is the factor powered to $\epsilon$, which under the change $x_2 = 2 - x_1 - x_3$ transforms to

$$(x_1^2 - 4\hat{m}_1^2)(x_2^2 - 4\hat{m}_2^2)h_{12} = (x_1^2 - 4\hat{m}_1^2)(x_3^2 - 4\hat{m}_3^2)h_{13}, \tag{F.135}$$

as follows directly from (F.76). This way we can write in complete generality

$$\begin{aligned}\int \mathrm{d}\Phi_3 &= \frac{16}{s^3}\left(\frac{s}{8\pi}\right)^d \frac{1}{\Gamma(d-3)} \int \mathrm{d}x_i \mathrm{d}x_j\, [(x_i^2 - 4\hat{m}_i^2)(x_j^2 - 4\hat{m}_j^2)]^{\frac{d-4}{2}} \Theta_{ij} \\ &\quad \times \int \frac{\mathrm{d}\cos\theta_i \mathrm{d}\cos\theta_j\, \theta(h_{ij})}{(h_{ij})^{\frac{5-d}{2}}} \\ &= \frac{s}{256\pi^4} \frac{1}{\Gamma(1-2\epsilon)}\left(\frac{s}{8\pi}\right)^{-2\epsilon} \int \mathrm{d}x_i \mathrm{d}x_j\, [(x_i^2 - 4\hat{m}_i^2)(x_j^2 - 4\hat{m}_j^2)]^{-\epsilon} \Theta_{ij} \\ &\quad \times \int \frac{\mathrm{d}\cos\theta_i \mathrm{d}\cos\theta_j\, \theta(h_{ij})}{(h_{ij})^{\frac{1}{2}+\epsilon}}.\end{aligned} \tag{F.136}$$

### F.4.4 Solving the angular integral for polynomial dependence

We solve the double angular integrals in (F.136) for a matrix element with polynomial dependence on the cosines (see(F.83)). Written in general, the angular integrals acquire the form

$$I^t_{m,n}(c) \equiv \int_{-1}^{1} \mathrm{d}x\, x^m \int_{a(x,c)}^{b(x,c)} \mathrm{d}y \frac{y^n}{[(b-y)(y-a)]^t}, \tag{F.137}$$

which is a generalization of (F.84). The integration limits $a(x,c)$ and $c(x,c)$ are given in (F.85). We start by solving the inner integral by expanding the integrand

$$\begin{aligned} J^t_n(a,b) &\equiv \int_a^b \mathrm{d}y \frac{y^n}{[(b-y)(y-a)]^t} = \int_0^{b-a} \mathrm{d}y \frac{(y+a)^n}{y^t(b-a-y)^t} \\ &= \sum_{i=0}^{n} \binom{n}{i} a^{n-i} \int_0^{b-a} \mathrm{d}y \frac{y^i}{y^t(b-a-y)^t} \\ &= \sum_{i=0}^{n} \binom{n}{i} a^{n-i} (b-a)^{i+1-2t} \int_0^1 \mathrm{d}z\, z^{i-t}(1-z)^{-t} \\ &= \sum_{i=0}^{n} \binom{n}{i} a^{n-i} (b-a)^{i+1-2t} \frac{\Gamma(1-t)\Gamma(1+i-t)}{\Gamma(2+i-2t)}, \end{aligned} \tag{F.138}$$

where the integral solved in the last step converges for $t<1$ and $i-t>-1$, conditions that simply reduce to $t<1$ for $i$ being a positive integer. To solve the second integral we use

$$\begin{aligned} a^{n-i} &= [cx-(1-c^2)^{1/2}(1-x^2)^{1/2}]^{n-i} \\ &= \sum_{j=0}^{n-i} \binom{n-i}{j} (-1)^{n-i-j} c^j (1-c^2)^{\frac{n-i-j}{2}} x^j (1-x^2)^{\frac{n-i-j}{2}}, \\ (b-a)^{i+1-2t} &= 2^{i+1-2t}(1-c^2)^{\frac{i+1-2t}{2}}(1-x^2)^{\frac{i+1-2t}{2}}, \end{aligned} \tag{F.139}$$

which, using again the notation $J^t_n(a(x,c),b(x,c)) \equiv J^t_n(x,c)$, leads to

$$\begin{aligned} J^t_n(x,c) &= \sum_{i=0}^{n} \binom{n}{i} \frac{2^{i+1-2t}\Gamma(1-t)\Gamma(1+i-t)}{\Gamma(2+i-2t)} \\ &\times \sum_{j=0}^{n-i} \binom{n-i}{j} (-1)^{n-i-j} c^j (1-c^2)^{\frac{n-j-2t+1}{2}} x^j (1-x^2)^{\frac{n-j-2t+1}{2}}. \end{aligned} \tag{F.140}$$

The integral over $x$ is immediate and solves to

$$\int_{-1}^{1} \mathrm{d}x\, x^{m+j}(1-x^2)^{\frac{n-j-2t+1}{2}} = \frac{1+(-1)^{m+j}}{2}\frac{\Gamma\big(\frac{1+j+m}{2}\big)\Gamma\big(\frac{n-j-2t+3}{2}\big)}{\Gamma\big(\frac{n+m-2t+4}{2}\big)}, \tag{F.141}$$

which converges for $j-n+2t<3$ and $j+m>-1$, both conditions automatically satisfied. With this the final expression is

$$\begin{aligned} I^t_{m,n}(c) &= \sum_{i=0}^{n} \binom{n}{i} \frac{2^{i-2t}\Gamma(1-t)\Gamma(1+i-t)}{\Gamma(2+i-2t)} \\ &\quad \sum_{j=0}^{n-i}[1+(-1)^{m+j}]\binom{n-i}{j}(-1)^{n-i-j}c^j(1-c^2)^{\frac{n-j-2t+1}{2}} \\ &\quad \times \frac{\Gamma\big(\frac{1+j+m}{2}\big)\Gamma\big(\frac{n-j-2t+3}{2}\big)}{\Gamma\big(\frac{n+m-2t+4}{2}\big)}. \end{aligned} \tag{F.142}$$

For $t=1/2+\epsilon$ the first values of $J^t_n$ and $I^t_{m,n}$ are

$$\begin{aligned} J_0^{1/2+\epsilon} &= (1-c^2)^{-\epsilon}(1-x)^{-\epsilon}\frac{4^{-\epsilon}\Gamma^2\big(\frac{1}{2}-\epsilon\big)}{\Gamma(1-2\epsilon)}, & J_1^{1/2+\epsilon} &= x\,c\,J_0^{1/2+\epsilon}, \\ J_2^{1/2+\epsilon} &= \frac{c^2-1+x^2[1-c^2(3-2\epsilon)]}{2(\epsilon-1)}J_0^{1/2+\epsilon}, & I_{0,0}^{1/2+\epsilon} &= \frac{2\pi(1-c^2)^{-\epsilon}}{1-2\epsilon}, \\ I_{1,0}^{1/2+\epsilon} &= I_{0,1}^{1/2+\epsilon} = 0, & I_{1,1}^{1/2+\epsilon} &= \frac{c}{3-2\epsilon}I_{0,0}^{1/2+\epsilon}, \\ I_{2,0}^{1/2+\epsilon} &= I_{0,2}^{1/2+\epsilon} = \frac{1}{3-2\epsilon}I_{0,0}^{1/2+\epsilon}. \end{aligned} \tag{F.143}$$

### F.4.5 From oriented to unoriented

As a cross-check we show one can recover (F.129) from (F.134) by solving the angular integrals. Integrating over $\theta_1$ and $\theta_2$ with (F.83), the phase space in (F.134) reads

$$\begin{aligned} \int \mathrm{d}\Phi_3 &= \frac{s}{256\pi^4\Gamma(1-2\epsilon)}\left(\frac{s}{8\pi}\right)^{-2\epsilon}\int \mathrm{d}x_1\mathrm{d}x_2\,[(x_1^2-4\hat{m}_1^2)(x_2^2-4\hat{m}_2^2)]^{-\epsilon}\Theta_{12}I_{0,0}^{1/2+\epsilon} \\ &= \frac{s}{128\pi^3\Gamma(2-2\epsilon)}\left(\frac{s}{8\pi}\right)^{-2\epsilon}\int \mathrm{d}x_1\mathrm{d}x_2\,[(x_1^2-4\hat{m}_1^2)(x_2^2-4\hat{m}_2^2)\sin^2\tilde{\theta}_{12}]^{-\epsilon}\Theta_{12}, \end{aligned} \tag{F.144}$$

which is the decay phase space in (F.129).

## F.5 Soft-limit parametrization of the 3-particle phase space

Here we discuss the limit in which one of the three particles in the final state is soft, i.e., the simultaneous limit $x_1, x_2 \to 1$, $x_3 \to 0$. To this end we change variables to $x_1 = 1-(1-z)y$ and $x_2 = 1-zy$. This change implies $y = x_3$ and makes the soft limit immediate to implement, since $y=0$ implies $x_1 = x_2 = 1$ independently on the value of $z$. The inverse and the Jacobian of this change of variables are

$$y = 2 - x_1 - x_2, \qquad z = \frac{1-x_2}{2-x_1-x_2}, \tag{F.145}$$
$$\begin{vmatrix} \frac{\partial x_1}{\partial y} & \frac{\partial x_1}{\partial z} \\ \frac{\partial x_2}{\partial y} & \frac{\partial x_2}{\partial z} \end{vmatrix} = \begin{vmatrix} z-1 & y \\ -z & -y \end{vmatrix} = y(1-z) + zy = y.$$

### F.5.1 Transforming the physical region

In section F.3.3 we established that, for $m_1 = m_2 = m$ and $m_3 = 0$, the physical region is the area of the $(x_1, x_2)$ plane enclosed by the curves $x_2 = x_2^{\pm}(x_1)$ and the vertical lines $x_1 = 2\hat{m}$ and $x_1 = 1$. More formally, we can write

$$\mathcal{D} \equiv \{(x_1, x_2) \in \mathbb{R}^2 | 2\hat{m} \leq x_1 \leq 1, x_2^-(x_1) \leq x_2 \leq x_2^+(x_1)\}. \tag{F.146}$$

The purpose of this section is to determine the form of $\mathcal{D}$ in $(y, z(y))$ and the $(z, y(z))$ planes.

The given transformation expresses each $x_1$ and $x_2$ as a combination of both $y$ and $z$, and so, when transforming the curves $x_2 = x_2^{\pm}(x_1)$ one can choose to solve for $y = y^{\pm}(z)$ or $z = z^{\pm}(y)$. To find the limits of the remaining variable it needs to be seen as a function in the $(x_1, x_2)$ plane, i.e., $y = y(x_1, x_2)$ or $z = z(x_1, x_2)$, and search for their local maximum and minimum in $\mathcal{D}$. As can be seen in (F.145), the functional form $z = z(x_1, x_2)$ involves a fraction, which makes it harder to search for the local extreme. To avoid this complication we simply choose $z$ as the dependent variable. Then we have that the curves delimiting the physical region are

$$\tilde{z}^{\pm}(y) = \frac{1 - y \pm \sqrt{(1-y)(1-4\hat{m}^2 - y)}}{2(1-y)}. \tag{F.147}$$

Now we find the minimum and maximum values of $y$, which come from, respectively, the maximum and minimum values of $x_1 + x_2$:

$$y_{\min} = 2 - \max_{(x_1,x_2)\in\mathcal{D}} \{x_1 + x_2\}, \qquad y_{\max} = 2 - \min_{(x_1,x_2)\in\mathcal{D}} \{x_1 + x_2\}. \tag{F.148}$$

Let us start with $y_{\min}$. For each $x_1$, the maximum value of $x_2$ is $x_2^+(x_1)$. Then we are interested in the value of $x_1$ that maximizes $x_1 + x_2^+(x_1)$. By computing the derivative we find that it satisfies

$$\frac{\mathrm{d}}{\mathrm{d}x_1}[x_1 + x_2^+(x_1)] > 0, \qquad \forall x_1 \in (2\hat{m}, 1), \tag{F.149}$$

and thus $x_1 + x_2^+(x_1)$ is a monotonically increasing function of $x_1$ in the interval $x_1 \in (2\hat{m}, 1)$. Its maximum value corresponds to the higher boundary of the interval, which is

$$\max_{x_1 \in (2\hat{m},1)} \{x_1 + x_2^+(x_1)\} = x_1 + x_2^+(x_1)|_{x_1=1} = 2 \implies y_{\min} = 2 - 2 = 0. \tag{F.150}$$

Now we find $y_{\max}$, for which we need the minimum value of the sum $x_1 + x_2$. For each $x_1$, the minimum value of $x_2$ is $x_2^-(x_1)$. In this case the derivative satisfies[F.12]

$$\begin{aligned} &\frac{\mathrm{d}}{\mathrm{d}x_1}[x_1 + x_2^-(x_1)] < 0, \qquad \forall x_1 \in \left(2\hat{m}, \frac{1}{2} + 2\hat{m}^2\right), \\ &\frac{\mathrm{d}}{\mathrm{d}x_1}[x_1 + x_2^-(x_1)] > 0, \qquad \forall x_1 \in \left(\frac{1}{2} + 2\hat{m}^2, 1\right), \end{aligned} \tag{F.151}$$

F.12. The intervals are always well defined, since $1 \geq \frac{1}{2} + 2\hat{m}^2 > 2\hat{m}$ for $\hat{m} \in (0, 1/2)$. The first inequality is immediate: the maximum value of $\hat{m}$ is $1/2$, for which $1/2 + 2\hat{m}^2 = 1$. The second inequality can be seen by finding the roots of the subtraction function $P(\hat{m}) = 2\hat{m}^2 - 2\hat{m} + 1/2$: the only root occurs at $\hat{m} = 1/2$. Since, for example $P(0) = 1/2 > 0$, we have $P(\hat{m}) > 0$ for any $\hat{m} < 1/2$, and in particular, for $\hat{m} \in (0, 1/2)$.

which means $x_1 + x_2^-(x_1)$ presents a minimum at $x_1 = 1/2 + 2\hat{m}^2$. The maximum value of $y$ is then

$$\begin{aligned} y_{\max} &= 2 - [x_1 + x_2^-(x_1)]_{x_1 = \frac{1}{2} + 2\hat{m}^2} \\ &= \frac{(1-4\hat{m}^2)\left(3 - 4\hat{m}^2 + \sqrt{(1-4\hat{m}^2)^2}\right)}{4 - 8\hat{m}^2} = \begin{cases} 1 - 4\hat{m}^2 \\ \dfrac{1-4\hat{m}^2}{2-4\hat{m}^2} \end{cases}, \end{aligned} \tag{F.152}$$

where the two solutions come from the square root of $(1-4\hat{m}^2)^2$. To determine which of the two solutions is higher we simply realize the denominator of the second one satisfies $2 - 4\hat{m}^2 \geq 1$, with the equality at $\hat{m} = 1/2$. Thus the second solution is smaller and we have

$$y_{\max} = 1 - 4\hat{m}^2. \tag{F.153}$$

With all this we can finally write

$$\mathcal{D} = \{(y,z) \in \mathbb{R}^2 | 0 \leq y \leq 1 - 4\hat{m}^2, z^-(y) \leq z \leq z^+(y)\}. \tag{F.154}$$

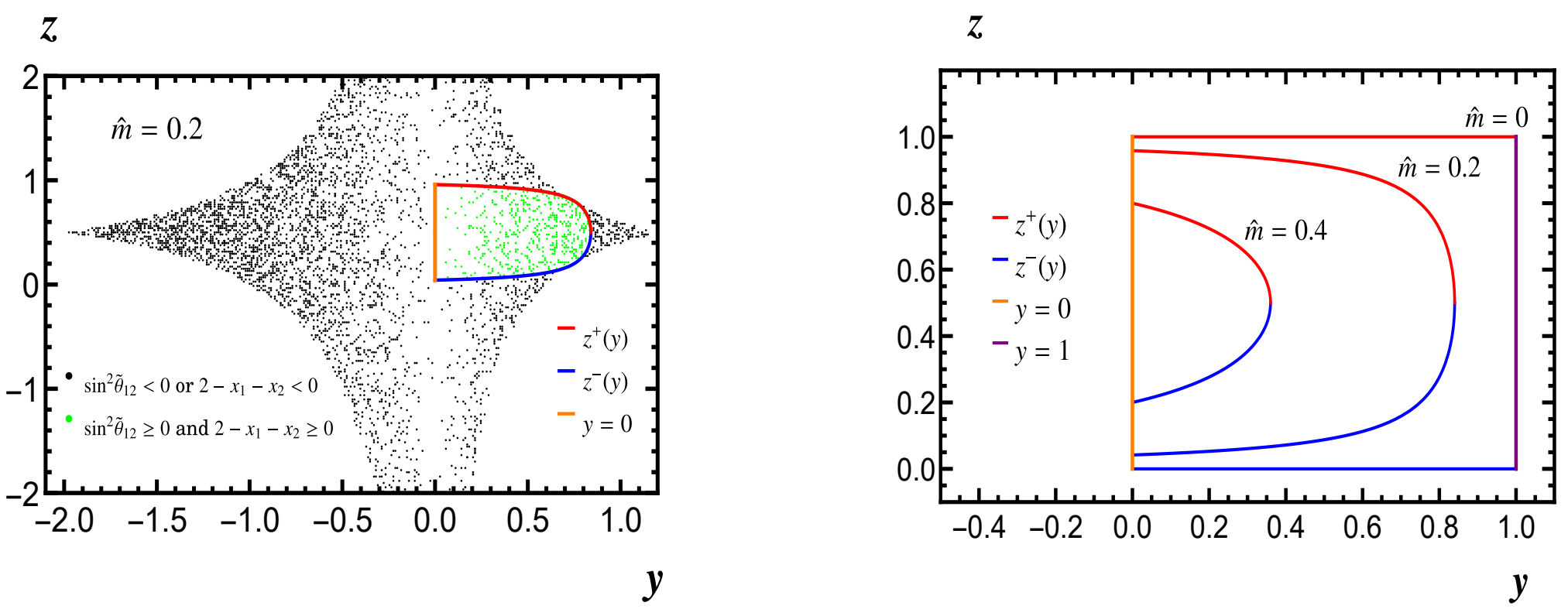

(*a*) (*b*)

**Figure F.5.** Dalitz plot for two particles of equal mass and one massless particle. Panel (*a*): $5 \times 10^3$ random points $(x_1, x_2)$ have been generated in the square $0 \leq x_i \leq 3$ and then transformed to $(y, z)$. Those satisfying the two positivity conditions are colored in green, and those not satisfying one or both of them are colored in black. The analytically-found boundaries delimit the physical region. Panel (*b*): detail of the physical region for $\hat{m} = 0$, 0.2 and 0.4.

It is now possible to invert these results and find the curve in the $(z, y(z))$ plane. Fist, applying the transformation to the curves $x_2 = x_2^{\pm}(x_1)$ and solving for $y$ one finds

$$y^-(z) = 0, \qquad y^+(z) = 1 - \frac{\hat{m}^2}{z(1-z)}. \tag{F.155}$$

The curve $y^+(z)$ has an inverted 'U' shape and crosses $y=0$ at

$$z_{\rm min} = \frac{1}{2}\left(1-\sqrt{1-4\hat{m}^2}\right), \qquad z_{\rm min} = \frac{1}{2}\left(1+\sqrt{1-4\hat{m}^2}\right). \tag{F.156}$$

This leads to the following parametrization for the physical region

$$\mathcal{D} = \{(y,z)\in \mathbb{R}^2 | z_{\rm min} \le z \le z_{\rm max}, 0 \le z \le y^+(z)\}. \tag{F.157}$$

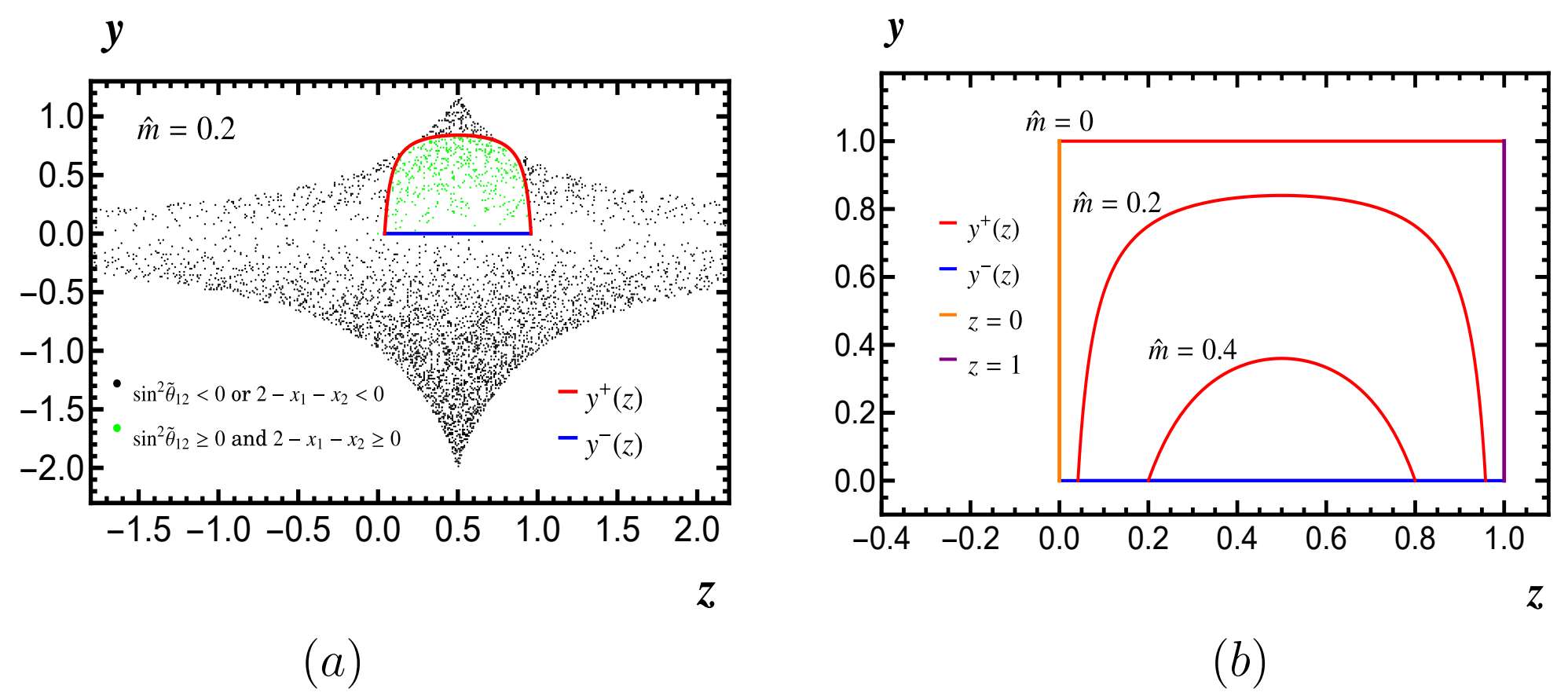

**Figure F.6.** Dalitz plot for two particles of equal mass and one massless particle. Panel $(a)$ $5\times 10^3$ random points $(x_1,x_2)$ have been generated in the square $0\le x_i \le 3$ and then transformed to $(z,y)$. Those satisfying the two positivity conditions are colored in green, and those not satisfying one or both of them are colored in black. The analytically-found boundaries delimit the physical region. Panel $(b)$: detail of the physical region for $\hat{m}=0$, 0.2 and 0.4.

As conclusion, the phase-space integral reads

$$\int {\rm d}\Phi_3^E = \int_0^{1-4\hat{m}^2} {\rm d}y\, y \int_{z^-(y)}^{z^+(y)} {\rm d}z = \int_{\frac{1}{2}\left(1-\sqrt{1-4\hat{m}^2}\right)}^{\frac{1}{2}\left(1+\sqrt{1-4\hat{m}^2}\right)} {\rm d}z \int_0^{y^+(z)} {\rm d}y\, y. \tag{F.158}$$

### F.5.2 Transforming the $(x_1, x_2)$ plane and finding the physical region

Here we present an alternate way of deriving the physical region in terms of the variables $y$ and $z$. We start from expression (F.99), in which the conditions in the Heaviside functions have not been imposed and therefore the integration carries over all the positive $(x_1,x_2)$ plane with $x_{1,2}\ge 2\hat{m}$. We transform this expression into $y$ and $z$ and then we will proceed to impose the conditions and limit the integration region.

As mentioned in the previous subsection, working with $z$ as our independent variable lead to unnecessary complications, so we choose $y$ as the independent variable. In this case, the limits of integration for $z$ are found from the boundary conditions for $x_1$ and $x_2$:

$$
\begin{aligned}
x_1 &= 1 + y\,(z-1) \geq 2\hat{m} \Longrightarrow z \geq \frac{y-1+2\hat{m}}{y},\\
x_2 &= 1 - yz \geq 2\hat{m} \Longrightarrow z \leq \frac{1-2\hat{m}}{y}.
\end{aligned}
\tag{F.159}
$$

The limits for $y = 2 - x_1 - x_2$ are easily found since at this stage since $x_1$ and $x_2$ are independent variables taking values in $(2\hat{m}, \infty)$: $y$ ranges from $2 - 4\hat{m}$ to $-\infty$. With these we have

$$
\begin{aligned}
\int \mathrm{d}\Phi_3^E &= \int_{-\infty}^{2-4\hat{m}} y\mathrm{d}y \int_{(y-1+2\hat{m})/y}^{(1-2\hat{m})/y} \mathrm{d}z\,\theta(\sin^2\tilde{\theta}_{12})\theta(y)\\
&= \int_{0}^{2-4\hat{m}} y\mathrm{d}y \int_{(y-1+2\hat{m})/y}^{(1-2\hat{m})/y} \mathrm{d}z\,\theta(\sin^2\tilde{\theta}_{12}),
\end{aligned}
\tag{F.160}
$$

where the squared sine function is

$$
\sin^2\tilde{\theta}_{12} = \frac{4y^2[\hat{m}^2 - z(1-y)(1-z)]}{[4\hat{m}^2 - (1-y\,(1-z))^2][4\hat{m}^2 - (1-yz)^2]}.
\tag{F.161}
$$

We start by computing its roots for $z$

$$
z^{\pm}(y) = \frac{1 - y \pm \sqrt{(1-y)(1-4\hat{m}^2-y)}}{2(1-y)},
\tag{F.162}
$$

result that agrees with (F.147). The study of the roots is as follows.

- For $y \in (0, 1-4\hat{m}^2) \cup (1, 2-4\hat{m})$ the square root is real. For $y \in (1-4\hat{m}^2, 1)$ the square root is complex. In this last case the sine must maintain its sign, so we pick $y = 1 - 2\hat{m}^2 \in (1-4\hat{m}^2, 1)$ and get

  $$\sin^2\tilde{\theta}_{12} \leq 0, \quad \forall y \in (1-4\hat{m}^2, 1).$$

  Note the integration obviously crosses the $(1-4\hat{m}^2, 1)$ interval, since $2 - 4\hat{m} > 1 > 1 - 2\hat{m}^2$.

- For $y \in (0, 1-4\hat{m}^2)$, $z^\pm > 0$ and $z^+ \geq z^-$, reaching equality when $y = 1-4\hat{m}^2$. In this case we have

$$\frac{y-1+2\hat{m}}{y} \leq z^- \leq z^+ \leq \frac{1-2\hat{m}}{y}. \tag{F.163}$$

- For $y \in (1, 2-4\hat{m})$ we have $z^- > z^+$, so

$$z^+ < \frac{y-1+2\hat{m}}{y} \leq \frac{1-2\hat{m}}{y} < z^-, \tag{F.164}$$

  where the equality between the integration limit occurs at $y = 2-4\hat{m}$.

Now we study the sine function to determine whether it is positive outside the roots or in between them. It smoothly tends to 0 when $z \mapsto \pm\infty$, so to check the positivity we pick a value between the roots instead. The sum of the roots is $z^+ + z^- = 1$, so $z = 1/2$ always lays between the roots. At $z = 1/2$ sine takes the value

$$\sin^2\tilde{\theta}_{12}|_{z=1/2} = \frac{16y^2(1-4\hat{m}^2-y)}{[(2-y)^2-16\hat{m}^2]^2}. \tag{F.165}$$

The sine segment between the roots is clearly positive for $y < 1-4\hat{m}^2$ and negative for $y > 1-4\hat{m}^2$. Then, accounting for all the results:

- For $y \in (0, 1-4\hat{m}^2)$ the sine is positive between the roots, and $z^\pm$ are also between the roots. Then $z$ must be integrated along $(z^-, z^+)$.

- For $y \in (1-4\hat{m}^2, 1)$ the sine is always negative, so the integration excludes this segment.

- For $y \in (1, 2-4\hat{m})$ the sine is positive outside the roots, but the integration limits are between them. Therefore, this segment is also excluded from the integration.

With this

$$\int_0^{2-4\hat{m}} y\mathrm{d}y \int_{(y-1+2\hat{m})/y}^{(1-2\hat{m})/y} \mathrm{d}z\,\theta(\sin^2\tilde{\theta}_{12}) = \int_0^{1-4\hat{m}^2} y\mathrm{d}y \int_{z^-}^{z^+} \mathrm{d}z, \tag{F.166}$$

which recovers our previous result (F.158).

## F.6 Projection onto thrust

In this section we seek to write the two- and three-particle differential distributions

$$\frac{\mathrm{d}\sigma(a+b\to 1+2)}{\mathrm{d}\cos\theta}, \quad \frac{\mathrm{d}\sigma(a+b\to 1+2+3)}{\mathrm{d}x_i\mathrm{d}x_j\mathrm{d}\cos\theta_i\mathrm{d}\cos\theta_j}, \tag{F.167}$$

where all the angles are measured with respect the $z$-axis in the CM reference frame, in terms of the angle of the thrust axis of the final state with respect to the incoming beam, $\theta_T$.

### F.6.1 Two particles

For two particles and in the CM frame, the thrust axis lays along the direction of the two back-to-back particles. Since the incoming beam lays along the $z$-axis we simply have $\theta=\theta_T$ and from (F.123) we immediately obtain

$$\frac{\mathrm{d}\sigma(a+b\to 1+2)}{\mathrm{d}\cos\theta_T}=\frac{\lambda^{1/2}(s,m_1^2,m_2^2)}{32\pi s\lambda^{1/2}(s,m_a^2,m_b^2)}\frac{(1-\cos^2\theta_T)^{-\epsilon}}{\Gamma(1-\epsilon)}\left[\frac{\lambda(s,m_1^2,m_2^2)}{16\pi s}\right]^{-\epsilon}|\mathcal{M}|^2. \tag{F.168}$$

### F.6.2 Three particles

#### F.6.2.1 Projection

For three particles, result (7.12) establishes the value of thrust is,

$$T=\max\left\{\sqrt{x_1^2-4\hat{m}_1^2},\,\sqrt{x_2^2-4\hat{m}_2^2},\,\sqrt{x_3^2-4\hat{m}_3^2}\right\}, \tag{F.169}$$

and so the thurst axis lies parallel to the tri-momentum of the particle for which the quantity $\sqrt{x_i^2-4\hat{m}_i^2}$ is maximum. To perform the projection one needs to take $\mathrm{d}\Phi_3 \mapsto \mathrm{d}\Phi_3^T \equiv \mathrm{d}\Phi_3\delta_T$ in result (F.134), where

$$\begin{aligned}
&\delta_T \equiv \delta_T^1+\delta_T^2+\delta_T^3, \\
&\delta_T^i \equiv \Theta_i^T\delta(\cos\theta_i-\cos\theta_T)\mathrm{d}\cos\theta_T, \\
&\Theta_i^T \equiv \theta\Big(\sqrt{x_i^2-4\hat{m}_i^2}-\sqrt{x_j^2-4\hat{m}_j^2}\Big)\theta\Big(\sqrt{x_i^2-4\hat{m}_i^2}-\sqrt{x_k^2-4\hat{m}_k^2}\Big),
\end{aligned} \tag{F.170}$$

with $i \neq j \neq k$. This simply changes the notation $\cos\theta_i \to \cos\theta_T$ for the particle defining the thrust axis, and to complete the projection the remaining angle must be integrated over. Note that, since an integration is performed, one must account for the matrix element dependence. Let us perform this procedure for the matrix element in (F.83), which is a polynomial in $\cos\theta_1$ and $\cos\theta_2$.

We start by writing the total cross-section as

$$\begin{aligned}
&\sigma = \frac{s}{512\pi^4\lambda^{1/2}(s,m_a^2,m_b^2)}\frac{1}{\Gamma(1-2\epsilon)}\left(\frac{s}{8\pi}\right)^{-2\epsilon}\bar{\sigma}, \\
&\bar{\sigma} \equiv \int \mathrm{d}x_i\mathrm{d}x_j[(x_i^2-\hat{m}_i^2)(x_j^2-\hat{m}_j^2)]^{-\epsilon}\Theta_{ij}\int\frac{\mathrm{d}\cos\theta_i\mathrm{d}\cos\theta_j\ |M|^2}{(h_{ij})^{\frac{1}{2}+\epsilon}}\theta(h_{ij}).
\end{aligned} \tag{F.171}$$

When (F.170) is inserted into (F.171), the double angular integral over $\cos\theta_i$ and $\cos\theta_j$ is divided into the three regions determined by the three projectors $\Theta_i^T$. In each of these regions, $\cos^m\theta_1\cos^n\theta_2$ must be integrated: the freedom to write the phase-space in terms of any pair of particles allows to choose the outer integral to match the thrust region, and the inner integral to be either over $\cos\theta_1$ or $\cos\theta_2$. Note that this choices affects the integral over $x_i$ and $x_j$, so to easily keep track we define the convenient notation:

$$\int_{ij} \equiv \int \mathrm{d}x_i\mathrm{d}x_j[(x_i^2-\hat{m}_i^2)(x_j^2-\hat{m}_j^2)]^{-\epsilon}\Theta_{ij}f_{i,j}. \tag{F.172}$$

Employing these ideas, the contributions to the first two thrust regions are

$$\begin{aligned}
&\int_{ij}\int\frac{\mathrm{d}\cos\theta_i\mathrm{d}\cos\theta_j\ \cos^m\theta_1\cos^n\theta_2}{(h_{ij})^{\frac{1}{2}+\epsilon}}\theta(h_{ij})\delta_T^{(1)} = \int_{12}\int \mathrm{d}\cos\theta_T\cos^m\theta_T J_n^{\frac{1}{2}+\epsilon}(\cos\theta_T,\cos\tilde{\theta}_{12})\Theta_1^T, \\
&\int_{ij}\int\frac{\mathrm{d}\cos\theta_i\mathrm{d}\cos\theta_j\ \cos^m\theta_1\cos^n\theta_2}{(h_{ij})^{\frac{1}{2}+\epsilon}}\theta(h_{ij})\delta_T^{(2)} = \int_{12}\int \mathrm{d}\cos\theta_T\cos^n\theta_T J_m^{\frac{1}{2}+\epsilon}(\cos\theta_T,\cos\tilde{\theta}_{12})\Theta_2^T.
\end{aligned} \tag{F.173}$$

In the third region we have

$$\begin{aligned}
\int_{ij}\!\!\int \frac{\mathrm{d}\cos\theta_i \mathrm{d}\cos\theta_j \cos^m\theta_1 \cos^n\theta_2}{(h_{ij})^{\frac{1}{2}+\epsilon}}\theta(h_{ij})\delta_T^{(3)} &= \sum_{k=0}^{n}\binom{n}{k}(-1)^n \int_{13} \frac{(x_1^2-4\hat{m}_1^2)^{\frac{k}{2}}(x_3^2-4\hat{m}_3^2)^{\frac{n-k}{2}}}{(x_2^2-4\hat{m}_2^2)^{\frac{n}{2}}} \\
&\quad \times \int \frac{\mathrm{d}\cos\theta_1 \mathrm{d}\cos\theta_3 \cos^{m+k}\theta_1 \cos^{n-k}\theta_3}{(h_{13})^{\frac{1}{2}+\epsilon}}\delta_T^{(3)} \\
&= (-1)^n \sum_{k=0}^{n}\binom{n}{k}\int_{13} \frac{(x_1^2-4\hat{m}_1^2)^{\frac{k}{2}}(x_3^2-4\hat{m}_3^2)^{\frac{n-k}{2}}}{(x_2^2-4\hat{m}_2^2)^{\frac{n}{2}}} \\
&\quad \times \int \mathrm{d}\cos\theta_T \cos^{n-k}\theta_T \\
&\quad \times J_{m+k}^{\frac{1}{2}+\epsilon}(\cos\theta_T, \cos\tilde{\theta}_{13})\Theta_3^T, \qquad \text{(F.174)}
\end{aligned}$$

where in the first step we used

$$\cos\theta_2 = \frac{-\sqrt{x_1^2-4\hat{m}_1^2}\cos\theta_1 - \sqrt{x_3^2-4\hat{m}_3^2}\cos\theta_3}{\sqrt{x_2^2-4\hat{m}_2^2}} \qquad \text{(F.175)}$$

and expanded with the binomial expansion (A.17). The resulting expression is convoluted, but simplifications occur when either $m$ or $n$ are 0:

$$\begin{aligned}
\int_{ij}\!\!\int \frac{\mathrm{d}\cos\theta_i \mathrm{d}\cos\theta_j \cos^m\theta_1}{(h_{ij})^{\frac{1}{2}+\epsilon}}\theta(h_{ij})\delta_T^{(3)} &= \int_{13}\!\!\int \mathrm{d}\cos\theta_T J_m^{\frac{1}{2}+\epsilon}(\cos\theta_T, \cos\tilde{\theta}_{13})\Theta_3^T, \qquad \text{(F.176)} \\
\int_{ij}\!\!\int \frac{\mathrm{d}\cos\theta_i \mathrm{d}\cos\theta_j \cos^n\theta_2}{(h_{ij})^{\frac{1}{2}+\epsilon}}\theta(h_{ij})\delta_T^{(3)} &= \int_{23}\!\!\int \mathrm{d}\cos\theta_T J_n^{\frac{1}{2}+\epsilon}(\cos\theta_T, \cos\tilde{\theta}_{23})\Theta_3^T.
\end{aligned}$$

To combine them all and obtain the differential cross section in terms of $x_1$, $x_2$ and $\cos\theta_T$ we use

$$\int_{13} = \int_{12}\left(\frac{x_3^2-4\hat{m}_3^2}{x_2^2-4\hat{m}_2^2}\right)^{-\epsilon}, \qquad \int_{23} = \int_{12}\left(\frac{x_3^2-4\hat{m}_3^2}{x_1^2-4\hat{m}_1^2}\right)^{-\epsilon}, \qquad \text{(F.177)}$$

where the appropriate changes of variable using $x_1+x_2+x_3=2$ are understood. Thus, defining $\Omega_{m,n}$ as

$$\bar{\sigma} \equiv \int_{12}\!\!\int \mathrm{d}\cos\theta_T \Omega_{m,n}(\Theta_1^T+\Theta_2^T+\Theta_2^T) \qquad \text{(F.178)}$$

the projection

$$\frac{\mathrm{d}\sigma}{\mathrm{d}x_1 \mathrm{d}x_2 \mathrm{d}\cos\theta_1 \mathrm{d}\cos\theta_2} \longrightarrow \frac{\mathrm{d}\sigma}{\mathrm{d}x_1 \mathrm{d}x_2 \mathrm{d}\cos\theta_T} \qquad \text{(F.179)}$$

is simply performed by taking

$$\mathrm{d}x_i\mathrm{d}x_j\mathrm{d}\cos\theta_i\cos\theta_j\cos^m\theta_1\cos^n\theta_2 \longrightarrow \mathrm{d}x_1\mathrm{d}x_2\mathrm{d}\cos\theta_T\Omega_{m,n}. \tag{F.180}$$

As an example, let us consider a matrix element of the form

$$|M|^2 = f_{0,0} + f_{1,1}\cos\theta_1\cos\theta_2 + f_{2,0}\cos^2\theta_1 + f_{0,2}\cos^2\theta_2. \tag{F.181}$$

The relevant values of $\Omega_{m,n}$ are

$$\begin{aligned}
\Omega_{0,0} &= J_0^{1/2+\epsilon}(\cos\theta_T, \cos\tilde{\theta}_{12})(\Theta_1^T + \Theta_2^T) + \left(\frac{x_3^2 - 4\hat{m}_3^2}{x_2^2 - 4\hat{m}_2^2}\right)^{-\epsilon} J_0^{1/2+\epsilon}(\cos\theta_T, \cos\tilde{\theta}_{13})\Theta_3^T,\\
\Omega_{2,0} &= \cos^2\theta_T\Theta_1^T + J_2^{\frac{1}{2}+\epsilon}(\cos\theta_T, \cos\tilde{\theta}_{12})\Theta_2^T + \left(\frac{x_3^2 - 4\hat{m}_3^2}{x_2^2 - 4\hat{m}_2^2}\right)^{-\epsilon} J_2^{\frac{1}{2}+\epsilon}(\cos\theta_T, \cos\tilde{\theta}_{13})\Theta_3^T\\
\Omega_{0,2} &= J_2^{\frac{1}{2}+\epsilon}(\cos\theta_T, \cos\tilde{\theta}_{12})\Theta_1^T + \cos^2\theta_T\Theta_2^T + \left(\frac{x_3^2 - 4\hat{m}_3^2}{x_1^2 - 4\hat{m}_1^2}\right)^{-\epsilon} J_2^{\frac{1}{2}+\epsilon}(\cos\theta_T, \cos\tilde{\theta}_{23})\Theta_3^T\\
\Omega_{1,1} &= \cos\theta_T J_1^{\frac{1}{2}+\epsilon}(\cos\theta_T, \cos\tilde{\theta}_{12})(\Theta_1^T + \Theta_2^T) - \left(\frac{x_3^2 - 4\hat{m}_3^2}{x_2^2 - 4\hat{m}_2^2}\right)^{-\epsilon} \frac{\sqrt{x_3^2 - 4\hat{m}_3^2}}{\sqrt{x_2^2 - 4\hat{m}_2^2}}\\
&\times\left[\cos\theta_T J_1^{\frac{1}{2}+\epsilon}(\cos\theta_T, \cos\tilde{\theta}_{13}) + \frac{\sqrt{x_1^2 - 4\hat{m}_1^2}}{\sqrt{x_3^2 - 4\hat{m}_3^2}} J_2^{\frac{1}{2}+\epsilon}(\cos\theta_T, \cos\tilde{\theta}_{13})\right]\Theta_3^T.
\end{aligned} \tag{F.182}$$

To work with the functions $J_n^{1/2+\epsilon}$ we define

$$\tilde{J}_{ij} \equiv J_0^{1/2+\epsilon}(\cos\theta_T, \cos\tilde{\theta}_{ij}) = \frac{\sqrt{\pi}\Gamma\left(\frac{1}{2} - \epsilon\right)}{\Gamma(1-\epsilon)}\sin^{-2\epsilon}\tilde{\theta}_{ij}(1 - \cos^2\theta_T)^{-\epsilon}, \tag{F.183}$$

which leads to

$$\begin{aligned}
J_1^{1/2+\epsilon}(\cos\theta_T, \cos\tilde{\theta}_{ij}) &= \cos\theta_T\cos\tilde{\theta}_{ij}\,\tilde{J}_{ij}, \qquad \text{(F.184)}\\
J_2^{1/2+\epsilon} &= \frac{1}{4}\left[\left(1 + \frac{\epsilon}{1-\epsilon}\sin^2\tilde{\theta}_{ij}\right)(1 + \cos^2\theta_T) - \left(1 - \frac{2-\epsilon}{1-\epsilon}\sin^2\tilde{\theta}_{ij}\right)(1 - 3\cos^2\theta_T)\right]\tilde{J}_{ij}.
\end{aligned}$$

On top of that for region 3 we also have

$$\left(\frac{x_3^2 - 4\hat{m}_3^2}{x_2^2 - 4\hat{m}_2^2}\right)^{-\epsilon}\tilde{J}_{13} = \tilde{J}_{12}, \quad \left(\frac{x_3^2 - 4\hat{m}_3^2}{x_1^2 - 4\hat{m}_1^2}\right)^{-\epsilon}\tilde{J}_{23} = \tilde{J}_{12}, \tag{F.185}$$

which arise from the identities

$$\frac{x_j^2 - 4\hat{m}_j^2}{x_k^2 - 4\hat{m}_k^2}\sin^2\tilde{\theta}_{ij} = \sin^2\tilde{\theta}_{ik}, \quad \sin^2\tilde{\theta}_{ij} = \sin^2\tilde{\theta}_{ji}, \tag{F.186}$$

and finally

$$\frac{\sqrt{x_3^2-4\hat{m}_3^2}}{\sqrt{x_2^2-4\hat{m}_2^2}}\cos\tilde{\theta}_{13}+\frac{\sqrt{x_1^2-4\hat{m}_1^2}}{\sqrt{x_2^2-4\hat{m}_2^2}}=-\cos\tilde{\theta}_{12}. \tag{F.187}$$

With all these ideas we group the contributions to each thrust region, each solely written in terms of $\cos\tilde{\theta}_{12}$ and $\sin^2\tilde{\theta}_{12}$. To avoid dragging the common factor of the integral over $x_1$ and $x_2$ we define them as

$$\begin{aligned}
\frac{\mathrm{d}\sigma^i}{\mathrm{d}x_1\mathrm{d}x_2\cos\theta_T} &\equiv F(x_1,x_2,\epsilon)\frac{\mathrm{d}\bar{\sigma}^i}{\mathrm{d}x_1\mathrm{d}x_2\cos\theta_T}, &\text{(F.188)}\\
F(x_1,x_2,\epsilon) &\equiv \frac{s\left(\frac{s}{8\pi}\right)^{-2\epsilon}}{512\pi^4\lambda^{1/2}(s,m_a^2,m_b^2)}\frac{[(x_1^2-\hat{m}_1^2)(x_2^2-\hat{m}_2^2)]^{-\epsilon}}{\Gamma(1-2\epsilon)}\Theta_{12}\frac{\tilde{J}_{12}}{4}\\
&= \frac{s\left(\frac{s}{16\pi}\right)^{-2\epsilon}}{2048\pi^3\lambda^{1/2}(s,m_a^2,m_b^2)}\frac{[(x_1^2-4\hat{m}_1^2)(x_2^2-4\hat{m}_2^2)\sin^2\tilde{\theta}_{12}(1-\cos\theta_T)]^{-\epsilon}}{\Gamma^2(1-\epsilon)},
\end{aligned}$$

where we pulled out a factor of $\tilde{J}_{12}$ arising in all the $\Omega_{m,n}$ due to (F.184) and (F.185) and a convenient overall $1/4$. The results are then

$$\begin{aligned}
\frac{\mathrm{d}\bar{\sigma}^1}{\mathrm{d}x_1\mathrm{d}x_2\mathrm{d}\cos\theta_T} &= \left[3f_{0,0}+f_{2,0}+f_{2,0}+\cos\tilde{\theta}_{12}f_{1,1}+\frac{\epsilon}{1-\epsilon}\sin^2\tilde{\theta}_{12}f_{0,2}\right](1+\cos^2\theta_T)\\
&+\left[f_{0,0}-f_{2,0}-f_{0,2}-\cos\tilde{\theta}_{12}f_{1,1}+\frac{2-\epsilon}{1-\epsilon}\sin^2\tilde{\theta}_{12}f_{0,2}\right](1-3\cos^2\theta_T),\\
\frac{\mathrm{d}\bar{\sigma}^2}{\mathrm{d}x_1\mathrm{d}x_2\mathrm{d}\cos\theta_T} &= \left[3f_{0,0}+f_{2,0}+f_{2,0}+\cos\tilde{\theta}_{12}f_{1,1}+\frac{\epsilon}{1-\epsilon}\sin^2\tilde{\theta}_{12}f_{2,0}\right](1+\cos^2\theta_T)\\
&+\left[f_{0,0}-f_{2,0}-f_{0,2}-\cos\tilde{\theta}_{12}f_{1,1}+\frac{2-\epsilon}{1-\epsilon}\sin^2\tilde{\theta}_{12}f_{2,0}\right](1-3\cos^2\theta_T),\\
\frac{\mathrm{d}\bar{\sigma}^3}{\mathrm{d}x_1\mathrm{d}x_2\mathrm{d}\cos\theta_T} &= (1+\cos^2\theta_T)\Bigg\{3f_{0,0}+f_{2,0}+f_{2,0}+\cos\tilde{\theta}_{12}f_{1,1}+\frac{\epsilon}{1-\epsilon}\frac{\sin^2\tilde{\theta}_{12}}{x_3^2-4\hat{m}_3^2}\\
&\times\left[(x_2^2-4\hat{m}_2^2)f_{2,0}+(x_1^2-4\hat{m}_1^2)f_{0,2}+\sqrt{x_1^2-4\hat{m}_1^2}\sqrt{x_2^2-4\hat{m}_2^2}f_{1,1}\right]\Bigg\}\\
&+(1-3\cos^2\theta_T)\Bigg\{f_{0,0}-f_{2,0}-f_{2,0}-\cos\tilde{\theta}_{12}f_{1,1}+\frac{2-\epsilon}{1-\epsilon}\frac{\sin^2\tilde{\theta}_{12}}{x_3^2-4\hat{m}_3^2}\\
&\times\left[(x_2^2-4\hat{m}_2^2)f_{2,0}+(x_1^2-4\hat{m}_1^2)f_{0,2}+\sqrt{x_1^2-4\hat{m}_1^2}\sqrt{x_2^2-4\hat{m}_2^2}f_{1,1}\right]\Bigg\}
\end{aligned} \tag{F.189}$$

In 4 dimensions, we can simply use the values

$$\begin{aligned}
&J_0(\cos\theta_T, \cos\tilde{\theta}_{ij}) = \pi, \\
&J_1(\cos\theta_T, \cos\tilde{\theta}_{ij}) = \pi\cos\theta_T\cos\tilde{\theta}_{ij}, \\
&J_2(\cos\theta_T, \cos\tilde{\theta}_{ij}) = \pi\left[\cos^2\theta_T + \frac{1}{2}\sin^2\tilde{\theta}_{ij}(1 - 3\cos^2\theta_T)\right],
\end{aligned} \tag{F.190}$$

to write

$$\begin{aligned}
\frac{1}{\pi}\Omega_{0,0} &= \Theta_1^T + \Theta_2^T + \Theta_3^T, \\
\frac{1}{\pi}\Omega_{2,0} &= \cos^2\theta_T\Theta_1^T + \left[\cos^2\theta_T + \frac{1}{2}\sin^2\tilde{\theta}_{12}(1 - 3\cos^2\theta_T)\right]\Theta_2^T \\
&\quad + \left[\cos^2\theta_T + \frac{1}{2}\sin^2\tilde{\theta}_{13}(1 - 3\cos^2\theta_T)\right]\Theta_3^T \\
&= \cos^2\theta_T(\Theta_1^T + \Theta_2^T + \Theta_3^T) + \frac{1}{2}(1 - 3\cos^2\theta_T)(\sin^2\tilde{\theta}_{12}\Theta_2^T + \sin^2\tilde{\theta}_{13}\Theta_3^T), \\
\frac{1}{\pi}\Omega_{0,2} &= \left[\cos^2\theta_T + \frac{1}{2}\sin^2\tilde{\theta}_{12}(1 - 3\cos^2\theta_T)\right]\Theta_1^T + \cos^2\theta_T\Theta_2^T \\
&\quad + \left[\cos^2\theta_T + \frac{1}{2}\sin^2\tilde{\theta}_{23}(1 - 3\cos^2\theta_T)\right]\Theta_3^T \\
&= \cos^2\theta_T(\Theta_1^T + \Theta_2^T + \Theta_3^T) + \frac{1}{2}(1 - 3\cos^2\theta_T)(\sin^2\tilde{\theta}_{12}\Theta_1^T + \sin^2\tilde{\theta}_{23}\Theta_3^T), \\
\frac{1}{\pi}\Omega_{1,1} &= \cos^2\theta_T\cos\tilde{\theta}_{12}(\Theta_1^T + \Theta_2^T) - \frac{\sqrt{x_3^2 - 4\hat{m}_3^2}}{\sqrt{x_2^2 - 4\hat{m}_2^2}}\left[\cos^2\theta_T\cos\tilde{\theta}_{13} + \frac{\sqrt{x_1^2 - 4\hat{m}_1^2}}{\sqrt{x_3^2 - 4\hat{m}_3^2}}\cos^2\theta_T\right. \\
&\quad \left. + \frac{1}{2}\frac{\sqrt{x_1^2 - 4\hat{m}_1^2}}{\sqrt{x_3^2 - 4\hat{m}_3^2}}\sin^2\tilde{\theta}_{13}(1 - 3\cos^2\theta_T)\right]\Theta_3^T \\
&= \cos^2\theta_T\cos\tilde{\theta}_{12}(\Theta_1^T + \Theta_2^T) \\
&\quad - \left[-\cos\tilde{\theta}_{12}\cos^2\theta_T + \frac{1}{2}\frac{\sqrt{x_1^2 - 4\hat{m}_1^2}}{\sqrt{x_2^2 - 4\hat{m}_2^2}}\sin^2\tilde{\theta}_{13}(1 - 3\cos^2\theta_T)\right]\Theta_3^T \\
&= \cos^2\theta_T\cos\tilde{\theta}_{12}(\Theta_1^T + \Theta_2^T + \Theta_3^T) - \frac{1}{2}\frac{\sqrt{x_1^2 - 4\hat{m}_1^2}}{\sqrt{x_2^2 - 4\hat{m}_2^2}}\sin^2\tilde{\theta}_{13}(1 - 3\cos^2\theta_T)\Theta_3^T,
\end{aligned} \tag{F.191}$$

thus arriving to

$$\begin{aligned}
\frac{d\bar{\sigma}^1}{dx_1 dx_2 d\cos\theta_T} &= (3f_{0,0} + f_{1,1}\cos\tilde{\theta}_{12} + f_{2,0} + f_{0,2})(1 + \cos^2\theta_T) \\
&\quad + [f_{0,0} - f_{1,1}\cos\tilde{\theta}_{12} - f_{2,0} + (2\sin^2\tilde{\theta}_{12} - 1)f_{0,2}](1 - 3\cos^2\theta_T), \\
\frac{d\bar{\sigma}^2}{dx_1 dx_2 d\cos\theta_T} &= (3f_{0,0} + f_{1,1}\cos\tilde{\theta}_{12} + f_{2,0} + f_{0,2})(1 + \cos^2\theta_T)
\end{aligned} \tag{F.192}$$

$$
\begin{aligned}
&+[f_{0,0}-f_{1,1}\cos\tilde{\theta}_{12}-f_{0,2}+(2\sin^2\tilde{\theta}_{12}-1)f_{2,0}](1-3\cos^2\theta_T),\\
\frac{d\bar{\sigma}^3}{dx_1dx_2d\cos\theta_T} &= (1+\cos^2\theta_T)(3f_{0,0}+f_{1,1}\cos\tilde{\theta}_{12}+f_{2,0}+f_{0,2})+(1-3\cos^2\theta_T)\\
&\times\Bigg[f_{0,0}+(2\sin^2\tilde{\theta}_{13}-1)f_{2,0}-\left(\cos\tilde{\theta}_{12}+2\frac{\sqrt{x_1^2-4\hat{m}_1^2}}{\sqrt{x_2^2-4\hat{m}_2^2}}\sin^2\tilde{\theta}_{13}\right)f_{1,1}\\
&+(2\sin^2\tilde{\theta}_{23}-1)f_{0,2}\Bigg].
\end{aligned}
$$

Results (F.189) reduce to those in (F.192) for $\epsilon=0$. In this case one can sum the contributions and use $\Theta_1^T+\Theta_2^T+\Theta_3^T=1$ to find the more compact expression

$$
\begin{aligned}
\frac{d\bar{\sigma}}{dx_1dx_2d\cos\theta_T} &= (1+\cos^2\theta_T)(3f_{0,0}+f_{1,1}\cos\tilde{\theta}_{12}+f_{2,0}+f_{0,2})\\
&+(1-3\cos^2\theta_T)\Bigg[f_{0,0}-f_{1,1}\cos\tilde{\theta}_{12}-f_{2,0}-f_{0,2}\\
&+2\sin^2\tilde{\theta}_{12}(f_{0,2}\Theta_1^T+f_{2,0}\Theta_2^T)\\
&+2\left(f_{2,0}\sin^2\tilde{\theta}_{13}+f_{0,2}\sin^2\tilde{\theta}_{23}-\frac{\sqrt{x_1^2-4\hat{m}_1^2}}{\sqrt{x_2^2-4\hat{m}_2^2}}f_{1,1}\sin^2\tilde{\theta}_{13}\right)\Theta_3^T\Bigg].
\end{aligned}
\tag{F.193}
$$

#### F.6.2.2 Thrust regions

In the projected result $d\sigma/(dx_1dx_2d\cos\theta_T)$, the Heaviside functions in $\Theta_{12}$ delimit the integration region for $x_1$ and $x_2$ as explained in section F.3.3. In addition, the functions in $\Theta_i^T$ divide the physical region into three sub-regions according to whether the thrust axis lays parallel to $\vec{p}_1$, $\vec{p}_2$ or $\vec{p}_3$. In section F.3.3 we found the physical region for the case $m_1=m_2=m$ and $m_3=0$; we now solve the thrust regions for that case.

To find the curves separating the regions we simply impose the equalities between different thrust values

$$
\begin{aligned}
x_1^2-4\hat{m}^2=x_2^2-4\hat{m}^2 &\implies x_2^{12}(x_1)=x_1,\\
x_1^2-4\hat{m}^2=(2-x_1-x_2)^2 &\implies x_2^{13}(x_1)=2-x_1-\sqrt{x_1^2-4\hat{m}^2},\\
x_2^2-4\hat{m}^2=(2-x_1-x_2)^2 &\implies x_2^{23}(x_1)=\frac{(2-x_1)^2+4\hat{m}^2}{2(2-x_1)}.
\end{aligned}
\tag{F.194}
$$

Also, the intersection points are

$$P_{12}=(1,1),\qquad P_{13}=\left(\frac{1+2\hat{m}(\hat{m}-1)}{1-\hat{m}},2\hat{m}\right),\tag{F.195}$$
$$P_{23}=\left(2\hat{m},\frac{1+2\hat{m}(\hat{m}-1)}{1-\hat{m}}\right),\quad Q=\frac{2}{3}\big(2-\sqrt{1-3\hat{m}^2},2-\sqrt{1-3\hat{m}^2}\big).$$

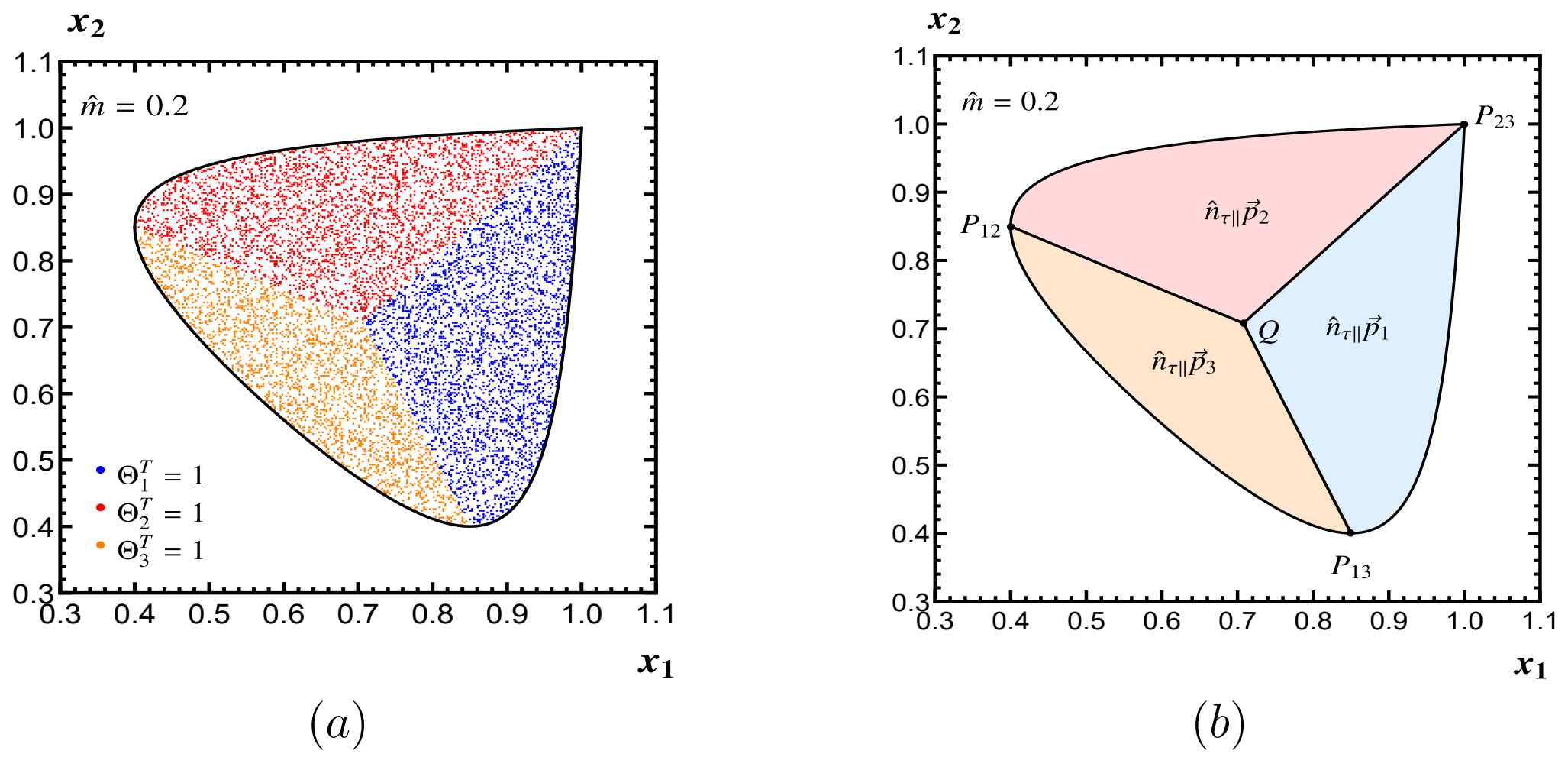

**Figure F.7.** Thrust regions for two particles of equal mass $\hat{m}=0.2$ and one massless particle. Panel $(a)$: $10^5$ random points $(x_1,x_2)$ have been generated in the Dalitz region. Those satisfying $\Theta_1^T=1$ ($\Theta_2^T=1$, $\Theta_3^T=1$) are colored in blue (red, orange). Panel $(b)$: detail of the curves separating the regions and intersection points.

In terms of the variables $y$ and $z$ the results for the curves delimiting the thrust regions are

$$[1+y(z-1)]^2-4\hat{m}^2=x_2^2-4\hat{m}^2 \implies z^{12}=1/2,\tag{F.196}$$
$$[1+y(z-1)]^2-4\hat{m}^2=y^2 \implies y^{13}(z)=\frac{1-z-\sqrt{1-4z(2-z)\hat{m}^2}}{z(2-z)},$$
$$(1-yz)^2-4\hat{m}^2=y^2 \implies y^{23}(z)=\frac{-z+\sqrt{1-4(1-z^2)\hat{m}^2}}{1-z^2},$$

and the coordinates of the points are

$$P_{12}=\left(\frac{1}{2},0\right),\qquad P_{13}=\left(1-\hat{m},\frac{1-2\hat{m}}{1-\hat{m}}\right),\tag{F.197}$$
$$P_{23}=\left(\hat{m},\frac{1-2\hat{m}}{1-\hat{m}}\right),\quad Q=\left(\frac{1}{2},-\frac{2}{3}+\frac{4}{3}\sqrt{1-3\hat{m}^2}\right).$$

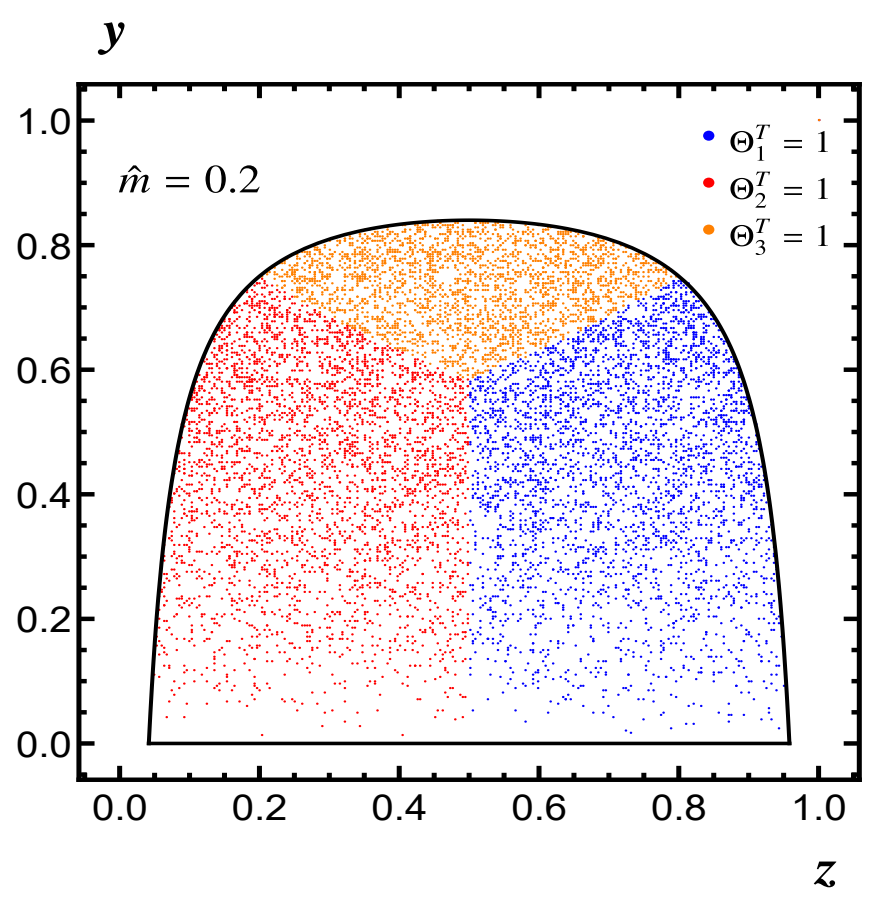

(a)

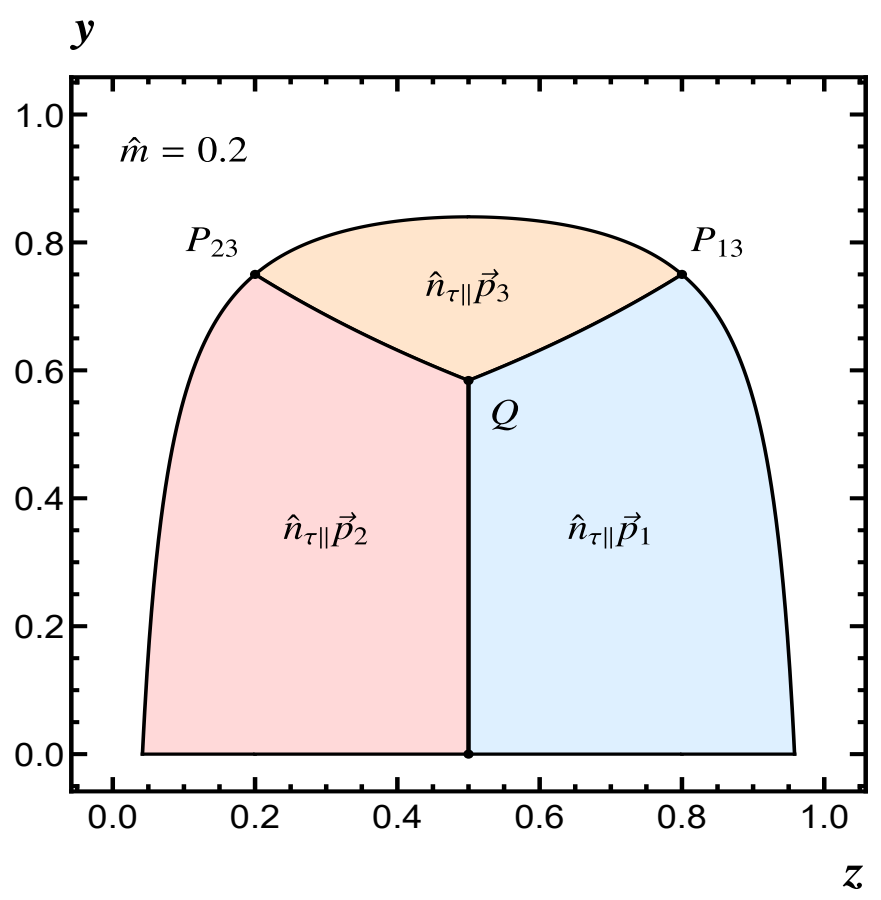

(b)

**Figure F.8.** Thrust regions for two particles of equal mass $\hat{m} = 0.2$ and one massless particle. Panel $(a)$: $10^5$ random points $(z, y)$ have been generated in the Dalitz region. Those satisfying $\Theta_1^T = 1$ ($\Theta_2^T = 1$, $\Theta_3^T = 1$) are colored in blue (red, orange). Panel $(b)$: detail of the curves separating the regions and intersection points.

# Bibliography

[1] N. Bleistein and R. A. Handelsman. *Asymptotic Expansions of Integrals*. Dover Publications, Inc., 1986.

[2] J.P. Boyd. The devil's invention: asymptotic, superasymptotic and hyperasymptotic series. *Acta Applicandae Mathematicae*, 1999.

[3] Jan Fischer. The use of power expansions in quantum field theory. *International Journal of Modern Physics A*, 12(21):3625–3663, aug 1997.

[4] M. Beneke. Renormalons. *Phys. Rept.*, 317:1–142, 1999.

[5] M. Beneke and V.M. Braun. Naive nonabelianization and resummation of fermion bubble chains. *Physics Letters B*, 348(3-4):513–520, apr 1995.

[6] Marcos Mariño. *Instantons and Large N: An Introduction to Non-Perturbative Methods in Quantum Field Theory*. Cambridge University Press, 9 2015.

[7] G. N. Watson. A theory of asymptotic series. *Phil.Trans.Roy.Soc. A*, 211, Jan 1912.

[8] F. Nevanlinna. *Zur Theorie der asymptotichen Potenzreihen*. Alexander University, 1918.

[9] A. D. Sokal. An improvement of Watson's theorem on Borel summabilit. *J. Math. Phys.*, 21:261–263, 1980.

[10] D. Gillam and V. Gurarii. On functions uniquely determined by their asymptotic expansion. *Functional Analysis and its Applications*, 40:273–284, 10 2006.

[11] G. 't Hooft. *Can We Make Sense Out of "Quantum Chromodynamics"?*, pages 943–982. Springer US, Boston, MA, 1979.

[12] N. N. Khuri. Coupling-constant analyticity and the renormalization group. *Phys. Rev. D*, 23:2285–2290, May 1981.

[13] Alexander Moroz. Quantum field theory as a problem of resummation (short guide to using summability methods). 1992.

[14] Andrey Grozin. Lectures on qed and qcd. 2005.

[15] Matthias Jamin. Qcd and renormalisation group methods. *Lecture presented at Herbstschule für Hochenergiephysik*, 2006.

[16] Matthew D. Schwartz. *Quantum Field Theory and the Standard Model*. Cambridge University Press, 3 2014.

[17] N. G. Gracia and V. Mateu. Toward massless and massive event shapes in the large-β0 limit. *Journal of High Energy Physics*, 2021(7), jul 2021.

[18] Andre H. Hoang, Christopher Lepenik, and Vicent Mateu. REvolver: Automated running and matching of couplings and masses in QCD. 2 2021.

[19] Inc Wolfram Research. *Mathematica Edition: Version 10.0*. Wolfram Research, Inc., Champaign, Illinois, 2014.

[20] Guido Rossum. *Python Reference Manual*. CWI, Amsterdam The Netherlands, (1995) [CS-R9525].

[21] Pauli Virtanen et al. SciPy 1.0–Fundamental Algorithms for Scientific Computing in Python. *Nature Meth.*, 17:261, 2020.

[22] Travis E Oliphant. *A guide to NumPy*, volume 1. Trelgol Publishing USA, 2006.

**[23]** A. Palanques-Mestre and P. Pascual. The $1/N_f$ Expansion of the $\gamma$ and $\beta$ functions in QED. *Commun. Math. Phys.*, 95:277, 1984.

**[24]** Andrey G. Grozin. Renormalons: Technical introduction. 11 2003.

**[25]** J. A. M. Vermaseren, S. A. Larin, and T. van Ritbergen. The 4-loop quark mass anomalous dimension and the invariant quark mass. *Phys. Lett.*, B405:327–333, 1997.

**[26]** K. G. Chetyrkin. Quark mass anomalous dimension to $O(\alpha_s^4)$. *Phys. Lett.*, B404:161–165, 1997.

**[27]** P. A. Baikov, K. G. Chetyrkin, and J. H. Kühn. Quark Mass and Field Anomalous Dimensions to $\mathcal{O}(\alpha_s^5)$. *JHEP*, 10:76, 2014.

**[28]** Thomas Luthe, Andreas Maier, Peter Marquard, and York Schröder. Five-loop quark mass and field anomalous dimensions for a general gauge group. *JHEP*, 01:81, 2017.

**[29]** R. L. Workman and Others. Review of Particle Physics. *PTEP*, 2022:83–1, 2022.

**[30]** Andre H. Hoang, Ambar Jain, Ignazio Scimemi, and Iain W. Stewart. Infrared Renormalization Group Flow for Heavy Quark Masses. *Phys. Rev. Lett.*, 101:151602, 2008.

**[31]** Andre H. Hoang, Ambar Jain, Christopher Lepenik, Vicent Mateu, Moritz Preisser, Ignazio Scimemi, and Iain W. Stewart. The MSR mass and the $\mathcal{O}(\Lambda_{\rm QCD})$ renormalon sum rule. *JHEP*, 04:3, 2018.

**[32]** Christian W. Bauer, Sean Fleming, Christopher Lee, and George Sterman. Factorization of $e^+e^-$ event shape distributions with hadronic final states in soft collinear effective theory. *Physical Review D*, 78(3), aug 2008.

**[33]** Vicent Mateu and Germán Rodrigo. "Oriented Event Shapes at N$^3$LL $+O(\alpha_{\rm S}^2)$". *Journal of High Energy Physics*, 2013(11), nov 2013.

**[34]** Christopher Lepenik and Vicent Mateu. NLO massive event-shape differential and cumulative distributions. *Journal of High Energy Physics*, 2020(3), mar 2020.

**[35]** Edward Farhi. A QCD Test for Jets. *Phys. Rev. Lett.*, 39:1587–1588, 1977.

**[36]** Iain W. Stewart, Frank J. Tackmann, and Wouter J. Waalewijn. N-Jettiness: An Inclusive Event Shape to Veto Jets. *Phys. Rev. Lett.*, 105:92002, 2010.

**[37]** Paul E. L. Rakow and B. R. Webber. Transverse Momentum Moments of Hadron Distributions in QCD Jets. *Nucl. Phys. B*, 191:63–74, 1981.

**[38]** L. Clavelli. Jet invariant mass in quantum chromodynamics. *Physics Letters B*, 85(1):111–114, 1979.

**[39]** T. Chandramohan and L. Clavelli. Consequences of Second Order QCD for Jet Structure in $e^+e^-$ Annihilation. *Nucl. Phys. B*, 184:365–380, 1981.

**[40]** L. Clavelli and D. Wyler. Kinematical Bounds on Jet Variables and the Heavy Jet Mass Distribution. *Phys. Lett. B*, 103:383–387, 1981.

**[41]** Christian W. Bauer, Sean Fleming, and Michael Luke. Summing sudakov logarithms in $B\to X_s\gamma$ in effective field theory. *Phys. Rev. D*, 63:14006, Dec 2000.

**[42]** Christian W. Bauer, Sean Fleming, Dan Pirjol, and Iain W. Stewart. An effective field theory for collinear and soft gluons: heavy to light decays. *Physical Review D*, 63(11), may 2001.

**[43]** Christian W. Bauer and Iain W. Stewart. Invariant operators in collinear effective theory. *Physics Letters B*, 516(1-2):134–142, sep 2001.

**[44]** Christian W. Bauer, Dan Pirjol, and Iain W. Stewart. Soft-collinear factorization in effective field theory. *Physical Review D*, 65(5), feb 2002.

**[45]** M. Beneke and Vladimir M. Braun. Heavy quark effective theory beyond perturbation theory: Renormalons, the pole mass and the residual mass term. *Nucl. Phys. B*, 643(1-3):431–476, nov 2002.

**[46]** M. Beneke and Th. Feldmann. Multipole-expanded soft-collinear effective theory with non-abelian gauge symmetry. *Physics Letters B*, 553(3-4):267–276, feb 2003.

**[47]** Richard J. Hill and Matthias Neubert. Spectator interactions in soft-collinear effective theory. *Nuclear Physics B*, 657:229–256, may 2003.

**[48]** Christian W. Bauer and Aneesh V. Manohar. Shape function effects in $B \to X_s \gamma$ and $\to X_u \ell \nu$ decays. *Physical Review D*, 70(3), aug 2004.

**[49]** Christian W. Bauer, Christopher Lee, Aneesh V. Manohar, and Mark B. Wise. Enhanced nonperturbative effects iniz/idecays to hadrons. *Physical Review D*, 70(3), aug 2004.

**[50]** Sean Fleming, Andre H. Hoang, Sonny Mantry, and Iain W. Stewart. Factorization approach for top mass reconstruction at high energies. 2007.

**[51]** Sean Fleming, Andre H. Hoang, Sonny Mantry, and Iain W. Stewart. Jets from massive unstable particles: top-mass determination. *Physical Review D*, 77(7), apr 2008.

**[52]** Sean Fleming, Andre H. Hoang, Sonny Mantry, and Iain W. Stewart. Top jets in the peak region: factorization analysis with next-to-leading-log resummation. *Physical Review D*, 77(11), jun 2008.

**[53]** Aneesh V. Manohar. Deep inelastic scattering as $x \to 1$ using soft collinear effective theory. *Phys. Rev. D*, 68:114019, 2003.

**[54]** Junegone Chay and Chul Kim. Deep inelastic scattering near the endpoint in soft-collinear effective theory. *Physical Review D*, 75(1), jan 2007.

**[55]** Thomas Becher, Alessandro Broggio, and Andrea Ferroglia. *Introduction to Soft-Collinear Effective Theory*. Springer International Publishing, 2015.

**[56]** Iain W. Stewart. *Lectures on the Soft-Collinear Effective Theory*. Massachussets Institue of Technology, 2013.

**[57]** Estia Eichten and Brian Russell Hill. An Effective Field Theory for the Calculation of Matrix Elements Involving Heavy Quarks. *Phys. Lett.*, B234:511–516, 1990.

**[58]** Nathan Isgur and Mark B. Wise. Weak Decays of Heavy Mesons in the Static Quark Approximation. *Phys. Lett. B*, 232:113–117, 1989.

**[59]** Nathan Isgur and Mark B. Wise. Weak transition form factors between heavy mesons. *Physics Letters B*, 237(3):527–530, 1990.

**[60]** Benjamin Grinstein. The Static Quark Effective Theory. *Nucl. Phys.*, B339:253–268, 1990.

**[61]** Howard Georgi. An Effective Field Theory for Heavy Quarks at Low-energies. *Phys. Lett.*, B240:447–450, 1990.

**[62]** Sean Fleming, Andre H. Hoang, Sonny Mantry, and Iain W. Stewart. Jets from massive unstable particles: Top-mass determination. *Phys. Rev.*, D77:74010, 2008.

**[63]** Sean Fleming, Andre H. Hoang, Sonny Mantry, and Iain W. Stewart. Top Jets in the Peak Region: Factorization Analysis with NLL Resummation. *Phys. Rev.*, D77:114003, 2008.

**[64]** Ignazio Scimemi and Alexey Vladimirov. Power corrections and renormalons in Transverse Momentum Distributions. *JHEP*, 03:2, 2017.

**[65]** G. P. Korchemsky and A. V. Radyushkin. Renormalization of the Wilson Loops Beyond the Leading Order. *Nucl. Phys. B*, 283:342–364, 1987.

**[66]** S. Moch, J. A. M. Vermaseren, and A. Vogt. The three-loop splitting functions in QCD:

The non-singlet case. *Nucl. Phys.*, B688:101–134, 2004.

**[67]** Johannes M. Henn, Gregory P. Korchemsky, and Bernhard Mistlberger. The full four-loop cusp anomalous dimension in $\mathcal{N}=4$ super Yang-Mills and QCD. *JHEP*, 04:18, 2020.

**[68]** Tobias Huber, Andreas von Manteuffel, Erik Panzer, Robert M. Schabinger, and Gang Yang. The four-loop cusp anomalous dimension from the $N=4$ Sudakov form factor. *Phys. Lett. B*, 807:135543, 2020.

**[69]** W. L. van Neerven. Dimensional Regularization of Mass and Infrared Singularities in two Loop on-shell Vertex Functions. *Nucl. Phys.*, B268:453, 1986.

**[70]** T. Matsuura, S. C. van der Marck, and W. L. van Neerven. The Calculation of the Second Order Soft and Virtual Contributions to the Drell-Yan Cross-Section. *Nucl. Phys. B*, 319:570–622, 1989.

**[71]** S. Moch, J. A. M. Vermaseren, and A. Vogt. The Quark form-factor at higher orders. *JHEP*, 08:49, 2005.

**[72]** T. Matsuura and W. L. van Neerven. Second Order Logarithmic Corrections to the Drell-Yan Cross-Section. *Z. Phys.*, C38:623, 1988.

**[73]** T. Gehrmann, T. Huber, and D. Maitre. Two-loop quark and gluon form factors in dimensional regularisation. *Phys. Lett.*, B622:295–302, 2005.

**[74]** R. N. Lee, A. V. Smirnov, and V. A. Smirnov. Analytic Results for Massless Three-Loop Form Factors. *JHEP*, 04:20, 2010.

**[75]** P. A. Baikov, K. G. Chetyrkin, A. V. Smirnov, V. A. Smirnov, and M. Steinhauser. Quark and gluon form factors to three loops. *Phys. Rev. Lett.*, 102:212002, 2009.

**[76]** T. Gehrmann, E.W.N. Glover, T. Huber, N. Ikizlerli, and C. Studerus. Calculation of the quark and gluon form factors to three loops in QCD. *JHEP*, 1006:94, 2010.

**[77]** Riccardo Abbate, Michael Fickinger, Andre H. Hoang, Vicent Mateu, and Iain W. Stewart. Precision Thrust Cumulant Moments at $N^3$LL. *Phys. Rev. D*, 86:94002, 2012.

**[78]** André H. Hoang, Daniel W. Kolodrubetz, Vicent Mateu, and Iain W. Stewart. $C$-parameter distribution at N$^3$LL' including power corrections. *Phys. Rev. D*, 91(9):94017, 2015.

**[79]** Brad Bachu, André H. Hoang, Vicent Mateu, Aditya Pathak, and Iain W. Stewart. Boosted Top Quarks in the Peak Region with N$^3$LL Resummation. 12 2020.

**[80]** Enrico Lunghi, Dan Pirjol, and Daniel Wyler. Factorization in leptonic radiative $B\to\gamma e\,\nu$ decays. *Nucl. Phys.*, B649:349–364, 2003.

**[81]** Christian W. Bauer and Aneesh V. Manohar. Shape function effects in $B\to X_s\,\gamma$ and $B\to X_u\,\ell\,\bar\nu$ decays. *Phys. Rev.*, D70:34024, 2004.

**[82]** Thomas Becher and Matthias Neubert. Toward a NNLO calculation of the $\bar B\to X_s\,\gamma$ decay rate with a cut on photon energy. II. Two-loop result for the jet function. *Phys. Lett.*, B637:251–259, 2006.

**[83]** Robin Brüser, Ze Long Liu, and Maximilian Stahlhofen. Three-Loop Quark Jet Function. *Phys. Rev. Lett.*, 121(7):72003, 2018.

**[84]** Pulak Banerjee, Prasanna K. Dhani, and V. Ravindran. Gluon jet function at three loops in QCD. *Phys. Rev. D*, 98(9):94016, 2018.

**[85]** Ambar Jain, Ignazio Scimemi, and Iain W. Stewart. Two-loop Jet-Function and Jet-Mass for Top Quarks. *Phys. Rev.*, D77:94008, 2008.

**[86]** Andre H. Hoang, Aditya Pathak, Piotr Pietrulewicz, and Iain W. Stewart. Hard Matching for Boosted Tops at Two Loops. *JHEP*, 12:59, 2015.

**[87]** E. Braaten, Stephan Narison, and A. Pich. QCD analysis of the tau hadronic width. *Nucl. Phys. B*, 373:581–612, 1992.

**[88]** ALEPH collaboration. S. Sachel et al. Branching ratios and spectral functions of ττ decays: final aleph measurements and physics implications. *Physics Reports*, 421:191–284, 12 2005.

**[89]** ALEPH collaboration. R. Barate et al. Measurement of the spectral functions of axial-vector hadronic $\tau$ decays and determination of $\alpha_s(M_\tau^2)$. *Eur. Phys. J. C*, 4:409–431, 1998.

**[90]** Martin Beneke and Matthias Jamin. $\alpha_s$ and the $\tau$ hadronic width: fixed-order, contour- improved and higher-order perturbation theory. *Journal of High Energy Physics*, 2008(09):44–44, Sep 2008.

**[91]** W. J. Marciano and A. Sirlin. Electroweak Radiative Corrections to tau Decay. *Phys. Rev. Lett.*, 61:1815–1818, 1988.

**[92]** Eric Braaten and Chong-Sheng Li. Electroweak radiative corrections to the semihadronic decay rate of the tau lepton. *Phys. Rev. D*, 42:3888–3891, 1990.

**[93]** M. Davier, A. Höcker, B. Malaescu, C. Z. Yuan, and Z. Zhang. Update of the ALEPH non-strange spectral functions from hadronic $\tau$ decays. *The European Physical Journal C*, 74(3), mar 2014.

**[94]** M. Davier, S. Descotes-Genon, Andreas Hocker, B. Malaescu, and Z. Zhang. The Determination of alpha(s) from Tau Decays Revisited. *Eur. Phys. J. C*, 56:305–322, 2008.

**[95]** Matthias Jamin. Contour-improved versus fixed-order perturbation theory in hadronic tau decays. *JHEP*, 09:58, 2005.

**[96]** André H. Hoang and Christoph Regner. On the difference between FOPT and CIPT for hadronic tau decays. *Eur. Phys. J. ST*, 230(12-13):2625–2639, 2021.

**[97]** André H. Hoang and Christoph Regner. Borel representation of $\tau$ hadronic spectral function moments in contour-improved perturbation theory. *Phys. Rev. D*, 105(9):96023, 2022.

**[98]** Miguel A. Benitez-Rathgeb, Diogo Boito, André H. Hoang, and Matthias Jamin. Reconciling the contour-improved and fixed-order approaches for τ hadronic spectral moments. part i. renormalon-free gluon condensate scheme. *Journal of High Energy Physics*, 2022(7), jul 2022.

**[99]** S. G. Gorishnii, A. L. Kataev, and S. A. Larin. The $O(\alpha_s^3)$-corrections to $\sigma_{tot}\,(e^+\,e^- \to$ hadrons) and $\Gamma\,(\tau^- \to \nu_\tau +$ hadrons) in QCD. *Phys. Lett. B*, 259:144–150, 1991.

**[100]** Levan R. Surguladze and Mark A. Samuel. Total hadronic cross-section in e+ e- annihilation at the four loop level of perturbative QCD. *Phys. Rev. Lett.*, 66:560–563, 1991. [Erratum: Phys.Rev.Lett. 66, 2416 (1991)].

**[101]** P. A. Baikov, K. G. Chetyrkin, and J. H. Kuhn. R(s) and hadronic tau-Decays in Order alpha**4(s): Technical aspects. *Nucl. Phys. B Proc. Suppl.*, 189:49–53, 2009.

**[102]** Matthias Jamin and Markus E. Lautenbacher. TRACER version 1.1 A mathematica package for γ-algebra in arbitrary dimensions. *Computer Physics Communications*, 74:265–288, 1993.

**[103]** *Nist digital library of mathematical functions*. http://dlmf.nist.gov/, Release 1.1.6 of 2022-06-30. F. W. J. Olver, A. B. Olde Daalhuis, D. W. Lozier, B. I. Schneider, R. F. Boisvert, C. W. Clark, B. R. Miller, B. V. Saunders, H. S. Cohl, and M. A. McClain, eds.

**[104]** Wolfram Research. *The mathematical functions site*. https://functions.wolfram.com/.